%% file: 0803.3276.English.tex
\documentclass{amsbook}
\input{0803.3276}

\begin{document}
\title{Lorentz Transformation\texorpdfstring{\\}{ }and General Covariance Principle}


\ShowEq{contents}
\end{document}

%% file: 0803.3276.tex
\def\BookNumber{0803.3276}
\def\PrintBook{}
\def\InputEncoding{off}
\input{Commands}
\renewcommand{\gi}[1]{#1}

\DefEq
{
\Prolog
\RefDefinition{metric-affine manifold}.
\input{\FilePrefix Space.Time.\TheLanguage}
\input{\FilePrefix Representation.\TheLanguage}
\input{\FilePrefix Basis.\TheLanguage}
\input{\FilePrefix Refernce.Frame.\TheLanguage}
\input{\FilePrefix Geom.Object.\TheLanguage}
\input{\FilePrefix Gen.Relativity.\TheLanguage}
\input{\FilePrefix Affine.\TheLanguage}
\input{\FilePrefix Newton.\TheLanguage}

\input{\FilePrefix Tidal.\TheLanguage}
\Epilog
}
{contents}

%% file: Commands.tex
\def\Defined{}
\def\ValueOff{off}%
\def\ValueOn{on}%
\scrollmode
\ifx\FilePrefix\undefined
\newcommand{\FilePrefix}{}
\fi
\ifx\UseRussian\Defined
\usepackage{cmap}
\usepackage[T2A,T2B]{fontenc}
\usepackage[cp1251]{inputenc}
\fi

\usepackage{trace}

\ifx\GJSFRA\Defined
\usepackage[top=1.905cm,bottom=1.905cm,inner=1.65cm,outer=1.65cm]{geometry}
\fi
\ifx\Presentation\Defined
\usepackage[margin=1cm]{geometry}
\fi

\ifx\setCACAA\Defined
\usepackage{times,mathptm}
\usepackage{amsmath,amsfonts,amssymb}
\usepackage{graphicx,graphics,epsfig}
\usepackage{fancyhdr,fancybox}
\usepackage{verbatim}
\usepackage{setspace}
\fi

\ifx\CreateSpace\Defined
\usepackage[top=1.905cm,bottom=1.905cm,inner=1.905cm,outer=1.27cm]{geometry}
\def\Publisher{CreateSpace Independent Publishing Platform}
\fi

\ifx\KDP\Defined
\usepackage[top=1.93cm,bottom=1.93cm,inner=1.93cm,outer=1.52cm]{geometry}
\def\Publisher{Kindle Direct Publishing}
\fi
\ifx\PublishBook\Defined
\def\PrintPaper{}
\usepackage{setspace}
\ifx\UseRussian\undefined
\usepackage{pslatex}
\fi
\usepackage[margin=2cm]{geometry}
\fi
\usepackage{graphicx}
\usepackage{footmisc}
\usepackage[all]{xy}
\usepackage{color}
\ifx\PrintPaper\undefined
\definecolor{CoverColor}{rgb}{.82,.7,.55}
\definecolor{UrlColor}{rgb}{.9,0,.3}
\definecolor{SymbColor}{rgb}{.4,0,.9}
\definecolor{IndexColor}{rgb}{1,.3,.6}
\definecolor{BlueColor}{rgb}{.1,.6,.8}
\newcommand\BlueText[1]{\textcolor{BlueColor}{#1}}
\newcommand\RedText[1]{\textcolor{red}{#1}}
\else
\definecolor{UrlColor}{rgb}{.1,.1,.1}
\definecolor{SymbColor}{rgb}{.1,.1,.1}
\definecolor{IndexColor}{rgb}{.5,.1,.1}
\newcommand\BlueText[1]{#1}
\newcommand\RedText[1]{#1}
\fi

\usepackage{xr-hyper}
\usepackage[unicode]{hyperref}
\hypersetup{pdfdisplaydoctitle=true}
\hypersetup{colorlinks}
\hypersetup{citecolor=UrlColor}
\hypersetup{urlcolor=UrlColor}
\hypersetup{linkcolor=UrlColor}
\hypersetup{pdffitwindow=true}
\hypersetup{pdfnewwindow=true}
\hypersetup{pdfstartview={FitH}}
\ifx\UseRussian\Defined
\usepackage[english,russian]{babel}
\else
\ifx\InputEncoding\ValueOn
\UseRawInputEncoding
\fi
\fi
\newcommand\Prolog
{
\ifx\PrimaryMSC\undefined
\else
\ifx\SecondaryMSC\undefined
\subjclass[2010]{Primary \PrimaryMSC}
\else
\subjclass[2010]{Primary \PrimaryMSC;
Secondary \SecondaryMSC}
\fi
\fi
\input{Preliminary.Eq}
\input{\FilePrefix Abstract.\BookNumber.\TheLanguage}
\maketitle
\tableofcontents
\input{\FilePrefix Preface.\BookNumber.\TheLanguage}
\ePrints{0767-8264}%
\ifx\Semafor\ValueOn%
\input{\FilePrefix Convention.\TheLanguage}
\fi
}
\newcommand\Epilog[1][0]
{
\input{\FilePrefix Biblio.\TheLanguage}
\input{\FilePrefix Index.\TheLanguage}
\input{\FilePrefix Symbol.\TheLanguage}
\def\TempA{PrintCover}%
\def\TempB{#1}%
\ifx\TempA\TempB%
\newpage
\pagestyle{empty}
\includegraphics[scale=.75]{Cover_Front_\BookNumber_\CurrentLanguage}
\newpage
\includegraphics[scale=.75]{Cover_Back_\BookNumber_\CurrentLanguage}
\fi
}
\newcounter{Index}
\newcounter{Symbol}
\newcounter{Symbols}

\def\hyph{\penalty0\hskip0pt\relax-\penalty0\hskip0pt\relax}
\def\Hyph{-\penalty0\hskip0pt\relax}%

\def\Items#1{\ItemList#1,LastItem,}%
\def\LastItem{LastItem}%
\def\ItemList#1,{\def\ViewBook{#1}%
\ifx\ViewBook\LastItem%
\else%
\ifx\ViewBook\BookNumber%
\def\Semafor{on}%
\fi%
\expandafter\ItemList%
\fi%
}%

\newcommand{\ePrints}[1]%
{%
\def\Semafor{off}%
\Items{#1}%
}%

\newcommand{\Basis}[1]{\overline{\overline{#1}}{}}
\newcommand{\Vector}[1]{\overline{#1}{}}
\ifx\PrintPaper\undefined
\newcommand{\gi}[1]{\boldsymbol{\textcolor{IndexColor}{\it #1}}}
\else
\newcommand{\gi}[1]{\boldsymbol{\it #1}}
\fi
\newcommand\gii{\gi i}
\newcommand\giI{\gi I}
\newcommand\gij{\gi j}

\newcommand\gik{\gi k}
\newcommand\gil{\gi l}
\newcommand\gin{\gi n}
\newcommand\gim{\gi m}
\newcommand\giA{\gi 1}

\newcommand{\VX}[1]{\Vector{#1}_{[1]}}
\makeatletter
\newcommand{\NameDef}[1]{%
\expandafter\gdef\csname #1\endcsname%
}%
\newcommand{\xNameDef}[1]{%
\expandafter\xdef\csname #1\endcsname%
}%
\newcommand{\ShowSymbol}[2]{%
\@nameuse{ViewSymbol#1,,,#2}%
}%

\newcommand{\symb}[4][]{%
\def\TempA{}%
\def\TempB{#1}%
\ifx\TempA\TempB%
\def\ThisSymbol{#3}%
\else%
\edef\ThisSymbol{#3(#1)}%
\fi%
\@ifundefined{ViewSymbol\ThisSymbol}{%
\addtocounter{Symbols}{1}%
\edef\SymbolId{\arabic{Symbols}}%
\xNameDef{ViewSymbol\ThisSymbol}{\SymbolId}%
\NameDef{ViewSymbol\ThisSymbol:::\SymbolId}{#2}%
\@namedef{RefSymbol}{:}%
}{%
\edef\Symbols{\@nameuse{ViewSymbol\ThisSymbol}}%
\def\aSymbolId{0}%
\@for\Symbol:=\Symbols\do{%
\protected@edef\TempA{#2}%
\protected@edef\TempB{\@nameuse{ViewSymbol\ThisSymbol:::\Symbol}}%
\ifx\TempA\TempB%
\edef\aSymbolId{\Symbol}%
\fi%
}%
\def\Zero{0}%
\ifx\aSymbolId\Zero%
\addtocounter{Symbols}{1}%
\edef\SymbolIds{\@nameuse{ViewSymbol\ThisSymbol},\arabic{Symbols}}%
\xNameDef{ViewSymbol\ThisSymbol}{\SymbolIds}%
\edef\SymbolId{\arabic{Symbols}}%
\NameDef{ViewSymbol\ThisSymbol:::\SymbolId}{#2}%
\else%
\def\SymbolId{\aSymbolId}%
\fi%
\addtocounter{Symbol}{1}%
\@namedef{RefSymbol}{\arabic{Symbol}}%
}%
\@namedef{LabelSymbol}{\label{symbol: \ThisSymbol:\@nameuse{RefSymbol}}}%
\edef\RefIds{RefSymbol\ThisSymbol===\SymbolId}%
\@ifundefined{\RefIds}{%
\xNameDef{\RefIds}{\@nameuse{RefSymbol}}%
}{%
\xNameDef{\RefIds}{\@nameuse{\RefIds},\@nameuse{RefSymbol}}%
}%
\NameDef{ViewSymbol#3,,,#4}{\textcolor{SymbColor}{#2}}%
\def\Temp{#4}%
\def\One{1}%
\def\Two{2}%
\def\Three{3}%
\ifx\Temp\One%
$\@nameuse{ViewSymbol#3,,,#4}$%
\fi%
\ifx\Temp\Two%
\[\@nameuse{ViewSymbol#3,,,#4}\]%
\fi%
\ifx\Temp\Three%
\@nameuse{ViewSymbol#3,,,#4}%
\fi%
\@nameuse{LabelSymbol}%
}%

\newcommand\AddEq[3][0]%
{%
\@ifundefined{ViewEq #2}{%
\expandafter\newcommand\csname ViewEq #2\endcsname[#1]{#3}%
}{%
\errmessage {second entry of AddEq: #2}%
}%
}%

\newcommand\DefEq[2]{%
\@ifundefined{ViewEq #2}{%
\NameDef{ViewEq #2}{#1}%
}{%
\errmessage {second entry of DefEq: #2}%
}%
}%
\newcommand{\DefEquation}[2]{%
\AddEq{#2}%
{%
\begin{equation}%
#1%
\EqLabel{#2}%
\end{equation}%
}%
}%
\newcommand{\AddEquation}[2]{%
\AddEq{#1}{\begin{equation}#2\EqLabel{#1}\end{equation}}%
}%
\def\ViewParm#1{\protect\getParm#1,endParm,}%
\def\endParm{endParm}%
\def\getParm#1,{\def\temp{#1}%
\ifx\temp\endParm%
\else%
\ShowEq{#1}%
\expandafter\getParm%
\fi%
}%
\newcommand{\DefTheorem}[3][]{%
\AddEq{theorem: #2}
{
\begin{theorem}[#1]
\label{theorem: #2}
#3
\end{theorem}
}
}
\newcommand{\ShowTheorem}[1]{%
\ShowEq{theorem: #1}%
}%
\newcommand{\DefProof}[2]{%
\AddEq{proof: #1}
{%
\begin{proof}
#2
\end{proof}%
}
}
\newcommand{\ShowProof}[1]{\ShowEq{proof: #1}}%
\newcommand{\DefLemma}[2]{%
\AddEq{lemma: #1}
{
\begin{lemma}
{\it
\labelLemma{#1}
#2
}
\end{lemma}
}
}
\newcommand{\ShowLemma}[1]{\ShowEq{lemma: #1}}%
\newcommand{\DefCorollary}[2]{%
\AddEq{corollary: #1}
{
\begin{corollary}
\labelCorollary{#1}
#2
\qed
\end{corollary}
}
}
\newcommand{\DefDefinition}[2]{%
\AddEq{definition: #1}
{
\begin{definition}
\labelDefinition{#1}
{\it
#2
}
\qed
\end{definition}
}
}%
\newcommand\ShowDefinition[1]{\ShowEq{definition: #1}}
\newcommand{\DefExample}[2]{%
\AddEq{example: #1}
{
\begin{example}
\labelExample{#1}
#2
\qed
\end{example}
}
}%
\newcommand{\ShowRemark}[1]{%
\ShowEq{remark: #1}%
}%
\newcommand{\EquationParm}[2]{%
\@ifundefined{ViewEq #1[#2]}%
{%
\ViewParm{#2}%
\DefEquation{\ShowEq{#1}}{#1[#2]}%
}{}%
\ShowEq{#1[#2]}%
}%
\newcommand\DrawEqParm[3]{%
\ViewParm{#2}%
\@ifundefined{ViewEq #1(#2)}{%
\DefEq%
{%
\ShowEq{#1}%
}{#1(#2)}%
}{%
}%
\DrawEq{#1(#2)}{#3}%
}%
\newcommand\EqRef[2][]%
{%
\def\Semafor{on}%
\def\Temp{}%
\edef\Tempa{#1}%
\ifx\Tempa\Temp%
\def\Semafor{off}%
\fi%
\@ifundefined{xRefDef#1}{%
\def\Semafor{off}
}{}%
\ifx\Semafor\ValueOff%
\eqref{eq: #2}%
\else%
\citeBib{#1}-\eqref{#1-\TheLanguage-eq: #2}%
\fi%
}%
\newcommand\eqRef[3][]{\EqRef[#1]{#2(#3)}}%
\newcommand\EqLabel[1]{\label{eq: #1}}%
\newcommand\ShowEq[1]{%
\@ifundefined{ViewEq #1}{%
\message {error: missed ShowEq #1}%
\newline%
\RedText{missed ShowEq #1}%
\newline%
}{%
\csname ViewEq #1\endcsname
}%
}%

\newcommand\DrawEq[3][]{%
\def\Temp{}%
\def\Tempa{#3}%
\ifx\Tempa\Temp%
\[\ShowEq{#2}#1\]%
\else%
\def\Temp{-}%
\ifx\Tempa\Temp%
$\ShowEq{#2}#1$
\else%
\@ifundefined{ViewEq #2(#3)}{%
\AddEquation{#2(#3)}{\ShowEq{#2}#1}%
  }{%
\errmessage {second entry of DrawEq: #2(#3)}%
}%
\ShowEq{#2(#3)}
\fi%
\fi%
}%
\makeatother
\DeclareMathOperator{\Hom}{\mathrm{Hom}} 
\DeclareMathOperator{\End}{\mathrm{End}} 
\DeclareMathOperator{\rank}{\mathrm{rank}}

\newcommand{\subs}{${}_*$\Hyph}
\newcommand{\sups}{${}^*$\Hyph}

\newcommand{\CRstar}{{}^*{}_*}
\newcommand{\RCstar}{{}_*{}^*}
\newcommand{\CRcirc}{{}^{\circ}{}_{\circ}}
\newcommand{\RCcirc}{{}_{\circ}{}^{\circ}}

\newcommand{\RC}{$\RCstar$\Hyph}
\newcommand{\CR}{$\CRstar$\Hyph}
\newcommand{\drc}{$D\RCstar$\Hyph}
\newcommand{\Drc}{$\mathcal D\RCstar$\Hyph}
\newcommand{\dcr}{$D\CRstar$\hyph}
\newcommand{\rcd}{$\RCstar D$\Hyph}
\newcommand{\crd}{$\CRstar D$\Hyph}

\newcommand\sT[1]{$*#1$\Hyph}%
\newcommand\Ts[1]{$#1*$\Hyph}%
\newcommand\sD{$\star D$\Hyph}%
\newcommand\Ds{$D\star$\Hyph}%

\newcommand\VirtVar{\vphantom{\overset{\rightarrow}{1}^1}}
\newcommand\pC[2]{{}_{#1\cdot #2}}%
\newcommand\DcrPartial[1]%
{%
\def\tempa{}%
\def\tempb{#1}%
\ifx\tempa\tempc%
(\partial\CRstar)%
\else%
(\partial_{\gi{#1}}\CRstar)%
\fi%
}%
\newcommand\rcDPartial[1]%
{%
\def\tempa{}%
\def\tempb{#1}%
\ifx\tempa\tempc%
(\RCstar\partial)%
\else%
(\RCstar\partial_{\gi{#1}})%
\fi%
}%
\newcommand\StandPartial[3]%
{%
\frac{d^{\gi{#3}} #1}{d #2}%
}%

\ifx\setCACAA\undefined
\renewcommand{\uppercasenonmath}[1]{}

\makeatletter
\def\l@chapter{\@tocline{0}{8pt plus1pt}{1pc}{}{}}
\newcommand\@dotsep{4.5}
\def\dotfill{%
\hbox{$\m@th\mkern \@dotsep mu\hbox{.}\mkern \@dotsep mu$}\hfill}
\def\@tocline#1#2#3#4#5#6#7{\relax
  \ifnum #1>\c@tocdepth 
  \else
    \def\@toclevel{#1}%
    \par \addpenalty\@secpenalty\addvspace{#2}%
    \begingroup 
        \hyphenpenalty\@M
        \@ifempty{#4}{%
          \@tempdima\csname r@tocindent\number#1\endcsname\relax
        }{%
          \@tempdima#4\relax
        }%
		\parindent .5pc
		\leftskip#3\relax \advance\leftskip\@tempdima\relax
        \rightskip\@pnumwidth plus4em \parfillskip-\@pnumwidth
        #5\leavevmode\hskip-\@tempdima #6\relax
        \leaders\dotfill
		\hbox to\@pnumwidth{\@tocpagenum{#7}}\par
        \nobreak
    \endgroup
  \fi
}
\makeatother 

\fi

\ifx\PrintBook\undefined

\ifx\setCACAA\undefined
\makeatletter
\renewcommand{\@indextitlestyle}{%
\twocolumn[\section*{\indexname}]%
\def\IndexSpace{off}%
}
\makeatother 
\ifx\PrintPaper\undefined
\thanks{\href{mailto:Aleks\_Kleyn@MailAPS.org}{Aleks\_Kleyn@MailAPS.org}}
\ePrints{1102.1776,1201.4158}
\ifx\Semafor\ValueOff
\thanks{\ \ \ \url{http://AleksKleyn.dyndns-home.com:4080/}\ \ \ \ \ \url{http://arxiv.org/a/kleyn\_a\_1}}
\thanks{\ \ \ \url{http://sites.google.com/site/AleksKleyn/}\ \ \ \ \url{http://AleksKleyn.blogspot.com/}}
\fi
\fi
\fi
\else

\makeatletter
\def\@maketitle{%
  \cleardoublepage \thispagestyle{empty}%
  \begingroup \topskip\z@skip
  \null\vfil
  \begingroup
  \LARGE\bfseries \centering
  \openup\medskipamount
  \@title
  \par
  \ifx\subtitle\undefined
  \else
  \centerline{\ }
  \centerline{\emph\subtitle}
  \fi
  \ifx\subtitleA\undefined
  \else
  \centerline{\emph\subtitleA}
  \fi
  \ifx\edition\undefined
  \else
  \centerline{\emph\edition}
  \fi
  \par\vspace{24pt}%
  \def\and{\par\medskip}\centering
  \mdseries\authors\par\bigskip
  \endgroup
  \vfill
\noindent
\href{mailto:Aleks\_Kleyn@MailAPS.org}{Aleks\_Kleyn@MailAPS.org}
\newline
\url{http://AleksKleyn.dyndns-home.com:4080/}
\newline
\url{http://sites.google.com/site/AleksKleyn/}
\newline
\url{http://arxiv.org/a/kleyn\_a\_1}
\newline
\url{http://AleksKleyn.blogspot.com/}
  \newpage\thispagestyle{empty}
  \begin{center}
    \ifx\@empty\@subjclass\else\@setsubjclass\fi
    \ifx\@empty\@keywords\else\@setkeywords\fi
    \ifx\@empty\@translators\else\vfil\@settranslators\fi
    \ifx\@empty\thankses\else\vfil\@setthanks\fi
  \end{center}
  \vfil
  \@setabstract
\vfil
  \def\Temp{0000}
  \ifx\copyrightyear\Temp
  \else
  \begin{center}
\begin{tabular}{@{}c}
Copyright\ \copyright\ \copyrightyear\ \copyrightholder
\\
All rights reserved.
\end{tabular}
  \end{center}
  \fi
  \ifx\Publisher\undefined%
  \else
  \begin{center}
\begin{tabular}{@{}c}
\Publisher
\end{tabular}
  \end{center}
  \fi
  \ifx\ISBN\undefined%
  \else%
 \begin{center}
\begin{tabular}{@{}r@{\ }l}
ISBN:&\ISBN
\\
ISBN-13:&\ISBNa
\end{tabular}
  \end{center}
  \fi%
  \ifx\titleNote\undefined
  \else
  \par\vspace{24pt}%
  \centerline{\mdseries\titleNote}
	  \centerline{\Title}
	  \ifx\Subtitle\undefined
	  \else
	  \centerline{\emph\Subtitle}
	  \fi
	  \ifx\Edition\undefined
	  \else
	  \centerline{\Edition}
	  \fi
	  \centerline{\Authors}
  \fi
  \endgroup}
\def\chapter{%
	\clearpage
  \thispagestyle{plain}\global\@topnum\z@
  \@afterindenttrue \secdef\@chapter\@schapter}
\renewcommand{\@indextitlestyle}{%
\twocolumn[\chapter*{\indexname}]%
\def\IndexSpace{off}%
\let\@secnumber\@empty
\chaptermark{\indexname}%
}
\makeatother 
\email{\href{mailto:Aleks\_Kleyn@MailAPS.org}{Aleks\_Kleyn@MailAPS.org}}
\ePrints{1102.1776,1201.4158}
\ifx\Semafor\ValueOff
\urladdr{\url{http://AleksKleyn.dyndns-home.com:4080/}}
\urladdr{\url{http://sites.google.com/site/alekskleyn/}}
\urladdr{\url{http://arxiv.org/a/kleyn\_a\_1}}
\urladdr{\url{http://AleksKleyn.blogspot.com/}}
\fi
\fi

\ifx\SelectlEnglish\undefined
\ifx\UseRussian\undefined
\def\SelectlEnglish{}
\fi
\fi

\newcommand\arXivOldRef{http://arxiv.org/PS_cache/}
\newcommand\arXivRef{http://arxiv.org/pdf/}
\newcommand\RgRef{https://www.researchgate.net/publication/}
\newcommand\AmazonRef{http://www.amazon.com/s/ref=nb_sb_noss?url=search-alias=aps&field-keywords=aleks+kleyn}
\newcommand\wRefDef[2]
{
\def\Tempa{#1}
\def\Tempb{0405.027}
\ifx\Tempa\Tempb
\def\wRef{\arXivOldRef gr-qc/pdf/0405/0405027v3.pdf}
\fi
\def\Tempb{0405.028}
\ifx\Tempa\Tempb
\def\wRef{\arXivOldRef gr-qc/pdf/0405/0405028v5.pdf}
\fi
\def\Tempb{0412.391}
\ifx\Tempa\Tempb
\def\wRef{\arXivOldRef math/pdf/0412/0412391v4.pdf}
\fi
\def\Tempb{0612.111}
\ifx\Tempa\Tempb
\def\wRef{\arXivOldRef math/pdf/0612/0612111v2.pdf}
\fi
\def\Tempb{0701.238}
\ifx\Tempa\Tempb
\def\wRef{\arXivOldRef math/pdf/0701/0701238v6.pdf}
\fi
\def\Tempb{0702.561}
\ifx\Tempa\Tempb
\def\wRef{\arXivOldRef math/pdf/0702/0702561v3.pdf}
\fi
\def\Tempb{0707.2246}
\ifx\Tempa\Tempb
\def\wRef{\arXivRef 0707.2246v2.pdf}
\fi
\def\Tempb{0803.3276}
\ifx\Tempa\Tempb
\def\wRef{\arXivRef 0803.3276v3.pdf}
\fi
\def\Tempb{0812.4763}
\ifx\Tempa\Tempb
\def\wRef{\arXivRef 0812.4763v7.pdf}
\fi
\def\Tempb{0906.0135}
\ifx\Tempa\Tempb
\def\wRef{\arXivRef 0906.0135v3.pdf}
\fi
 \def\Tempb{0909.0855}
\ifx\Tempa\Tempb
\def\wRef{\arXivRef 0909.0855v5.pdf}
\fi
 \def\Tempb{0912.3315}
\ifx\Tempa\Tempb
\def\wRef{\arXivRef 0912.3315v3.pdf}
\fi
 \def\Tempb{0912.4061}
\ifx\Tempa\Tempb
\def\wRef{\arXivRef 0912.4061v2.pdf}
\fi
 \def\Tempb{1001.4852}
\ifx\Tempa\Tempb
\def\wRef{\arXivRef 1001.4852.pdf}
\fi
 \def\Tempb{1003.3714}
\ifx\Tempa\Tempb
\def\wRef{\arXivRef 1003.3714v2.pdf}
\fi
 \def\Tempb{1003.1544}
\ifx\Tempa\Tempb
\def\wRef{\arXivRef 1003.1544v2.pdf}
\fi
 \def\Tempb{1006.2597}
\ifx\Tempa\Tempb
\def\wRef{\arXivRef 1006.2597v2.pdf}
\fi
 \def\Tempb{1011.3102}
\ifx\Tempa\Tempb
\def\wRef{\arXivRef 1011.3102.pdf}
\fi
 \def\Tempb{1102.1776}
\ifx\Tempa\Tempb
\def\wRef{\arXivRef 1102.1776.pdf}
\fi
 \def\Tempb{1104.5197}
\ifx\Tempa\Tempb
\def\wRef{\arXivRef 1104.5197.pdf}
\fi
 \def\Tempb{1105.4307}
\ifx\Tempa\Tempb
\def\wRef{\arXivRef 1105.4307.pdf}
\fi
 \def\Tempb{1107.1139}
\ifx\Tempa\Tempb
\def\wRef{\arXivRef 1107.1139.pdf}
\fi
 \def\Tempb{1107.5037}
\ifx\Tempa\Tempb
\def\wRef{\arXivRef 1107.5037.pdf}
\fi
 \def\Tempb{1111.6035}
\ifx\Tempa\Tempb
\def\wRef{\arXivRef 1111.6035.pdf}
\fi
 \def\Tempb{1202.6021}
\ifx\Tempa\Tempb
\def\wRef{\arXivRef 1202.6021v2.pdf}
\fi
 \def\Tempb{1211.6965}
\ifx\Tempa\Tempb
\def\wRef{\arXivRef 1211.6965.pdf}
\fi
 \def\Tempb{1302.7204}
\ifx\Tempa\Tempb
\def\wRef{\arXivRef 1302.7204v1.pdf}
\fi
 \def\Tempb{1305.4547}
\ifx\Tempa\Tempb
\def\wRef{\arXivRef 1305.4547.pdf}
\fi
 \def\Tempb{1310.5591}
\ifx\Tempa\Tempb
\def\wRef{\arXivRef 1310.5591.pdf}
\fi
 \def\Tempb{1502.04063}
\ifx\Tempa\Tempb
\def\wRef{\arXivRef 1502.04063v2.pdf}
\fi
 \def\Tempb{1505.03625}
\ifx\Tempa\Tempb
\def\wRef{\arXivRef 1505.03625v1.pdf}
\fi
 \def\Tempb{1506.00061}
\ifx\Tempa\Tempb
\def\wRef{\arXivRef 1506.00061.pdf}
\fi
 \def\Tempb{1601.03259}
\ifx\Tempa\Tempb
\def\wRef{\arXivRef 1601.03259v3.pdf}
\fi
 \def\Tempb{1801.01628}
\ifx\Tempa\Tempb
\def\wRef{\arXivRef 1801.01628.pdf}
\fi
 \def\Tempb{1908.04418}
\ifx\Tempa\Tempb
\def\wRef{\arXivRef 1908.04418.pdf}
\fi
 \def\Tempb{322019412}
\ifx\Tempa\Tempb
\def\wRef{\RgRef 322019412}
\fi
 \def\Tempb{323966352}
\ifx\Tempa\Tempb
\def\wRef{\RgRef 323966352}
\fi
 \def\Tempb{8433-5163}
\ifx\Tempa\Tempb
\def\wRef{\AmazonRef}
\fi
 \def\Tempb{8443-0072}
\ifx\Tempa\Tempb
\def\wRef{\AmazonRef}
\fi
 \def\Tempb{4776-3181}
\ifx\Tempa\Tempb
\def\wRef{\AmazonRef}
\fi
 \def\Tempb{4975-6381}
\ifx\Tempa\Tempb
\def\wRef{\AmazonRef}
\fi
 \def\Tempb{4993-2400}
\ifx\Tempa\Tempb
\def\wRef{\AmazonRef}
\fi
 \def\Tempb{5059-9176}
\ifx\Tempa\Tempb
\def\wRef{\AmazonRef}
\fi
 \def\Tempb{5114-6019}
\ifx\Tempa\Tempb
\def\wRef{\AmazonRef}
\fi
 \def\Tempb{5148-4632}
\ifx\Tempa\Tempb
\def\wRef{\AmazonRef}
\fi
 \def\Tempb{5410-9916}
\ifx\Tempa\Tempb
\def\wRef{\AmazonRef}
\fi
 \def\Tempb{6860-2955}
\ifx\Tempa\Tempb
\def\wRef{\AmazonRef}
\fi
 \def\Tempb{CACAA.01.109}
\ifx\Tempa\Tempb
\def\wRef{http://www.cliffordanalysis.com/}
\fi
 \def\Tempb{CACAA.01.291}
\ifx\Tempa\Tempb
\def\wRef{http://www.cliffordanalysis.com/}
\fi
 \def\Tempb{CACAA.02.097}
\ifx\Tempa\Tempb
\def\wRef{http://www.cliffordanalysis.com/}
\fi
 \def\Tempb{CACAA.04.001}
\ifx\Tempa\Tempb
\def\wRef{http://www.cliffordanalysis.com/}
\fi
 \def\Tempb{CACAA.05.001}
\ifx\Tempa\Tempb
\def\wRef{http://www.cliffordanalysis.com/}
\fi
 \def\Tempb{CACAA.06.121}
\ifx\Tempa\Tempb
\def\wRef{http://www.cliffordanalysis.com/}
\fi
 \def\Tempb{GJSFRA.13.1.39}
\ifx\Tempa\Tempb
\def\wRef{http://www.cliffordanalysis.com/}
\fi
\externaldocument[#1-#2-]{\FilePrefix #1.#2}[\wRef]
}
\newcommand\LanguagePrefix{}%
\makeatletter
\newcommand\StartLabelItem[1][theorem]%
{
\expandafter \def \csname%
 theenumi\expandafter \endcsname \expandafter {\csname the#1\expandafter%
 \endcsname.\expandafter \@arabic \csname c@enumi\endcsname }%
}
\newcommand\StopLabelItem[1][theorem]%
{
\counterwithout{enumi}{#1}%

}
\makeatother
\newcommand\labelConvention[1]{\label{convention: #1}}%
\newcommand\labelTheorem[1]{\label{theorem: #1}}%
\newcommand\labelLemma[1]{\label{lemma: #1}}%
\newcommand\labelCorollary[1]{\label{corollary: #1}}%

\newcommand\labelDefinition[1]{\label{definition: #1}}%
\newcommand\labelExample[1]{\label{example: #1}}%
\newcommand\labelRemark[1]{\label{remark: #1}}%
\newcommand\labelSection[1]{\label{section: #1}}%
\newcommand\labelSubsection[1]{\label{subsection: #1}}%
\newcommand\labelChapter[1]{\label{chapter: #1}}%
\newcommand\labelItem[1]{\label{item: #1}}%

\makeatletter
\newcommand\xRef[2][]%
{%
\def\Semafor{on}%
\def\Temp{}%
\edef\Tempa{#1}%
\ifx\Tempa\Temp%
\def\Semafor{off}%
\fi%
\@ifundefined{xRefDef#1}{%
\def\Semafor{off}%
}{}%
\ifx\Semafor\ValueOff%
\ref{#2}%
\else%
\citeBib{#1}-\ref{#1-\TheLanguage-#2}%
\fi%
}%
\makeatother

\newcommand\RefTheorem[2][]{\xRef[#1]{theorem: #2}}
\newcommand\RefLemma[2][]{\xRef[#1]{lemma: #2}}

\newcommand\RefDefinition[2][]{\xRef[#1]{definition: #2}}

\newcommand\RefSection[2][]{\xRef[#1]{section: #2}}

\newcommand\RefItem[2][]{\xRef[#1]{item: #2}}

\ifx\SelectlEnglish\undefined
\newcommand\TheLanguage{Russian}%
\author{Александр Клейн}
\ifx\setCACAA\undefined
\newtheorem{theorem}{Теорема}[section]
\newtheorem{corollary}[theorem]{Следствие}
\newtheorem{example}[theorem]{Пример}
\newtheorem{definition}[theorem]{Определение}
\newtheorem{remark}[theorem]{Замечание}
\newtheorem{question}[theorem]{Вопрос}
\newtheorem{lemma}[theorem]{Лемма}

\else
\theoremstyle{definition}
\theoremstyle{remark}
\fi
\newtheorem{Statement}[theorem]{Утверждение}
\newtheorem{convention}[theorem]{Соглашение}
\newtheorem{xca}[theorem]{Exercise}
\ifx\PrintBook\undefined
\newcommand{\BibTitle}{%
\section*{Список литературы}%
}
\else
\newcommand{\BibTitle}{%
\chapter*{Список литературы}%
}
\fi
\else
\newcommand\TheLanguage{English}%
\author{Aleks Kleyn}
\ifx\setCACAA\undefined
\newtheorem{theorem}{Theorem}[section]
\newtheorem{corollary}[theorem]{Corollary}
\newtheorem{example}[theorem]{Example}
\newtheorem{definition}[theorem]{Definition}
\newtheorem{remark}[theorem]{Remark}

\newtheorem{lemma}[theorem]{Lemma}
\else
\theoremstyle{definition}
\theoremstyle{remark}
\fi

\newtheorem{convention}[theorem]{Convention}

\ifx\PrintBook\undefined
\newcommand{\BibTitle}{%
\section*{References}%
}
\else
\newcommand{\BibTitle}{%
\chapter*{References}%
}
\fi
\fi
\newcommand\CurrentLanguage{\TheLanguage.}%
\newcommand\xRefDef[1]
{
\wRefDef{#1}{\TheLanguage}
\NameDef{xRefDef#1}{}%
}

\theoremstyle{definition}
\theoremstyle{remark}
\renewenvironment{proof}[1][\proofname]
{\par{\sc #1. }}{\qed}%

\ifx\PrintBook\undefined
\numberwithin{Hfootnote}{section}
\else
\numberwithin{section}{chapter}
\numberwithin{footnote}{chapter}
\numberwithin{Hfootnote}{chapter}
\fi

\ifx\Presentation\undefined
\numberwithin{equation}{section}
\numberwithin{figure}{section}
\numberwithin{table}{section}
\numberwithin{Item}{section}
\fi

\makeatletter
\newcommand\org@maketitle{}
\let\org@maketitle\maketitle
\def\maketitle{%
\hypersetup{pdftitle={\@title}}%
\hypersetup{pdfauthor={\authors}}%
\hypersetup{pdfsubject=\@keywords}%
\ifx\UseRussian\Defined
\pdfbookmark[1]{\@title}{TitleRussian}
\else
\pdfbookmark[1]{\@title}{TitleEnglish}
\fi
\org@maketitle
}
\def\make@stripped@name#1{%
\begingroup
\escapechar\m@ne
\global\let\newname\@empty
\protected@edef\Hy@tempa{\CurrentLanguage #1}%
\edef\@tempb{%
\noexpand\@tfor\noexpand\Hy@tempa:=%
\expandafter\strip@prefix\meaning\Hy@tempa
}%
\@tempb\do{%
\if\Hy@tempa\else
\if\Hy@tempa\else
\xdef\newname{\newname\Hy@tempa}%
\fi
\fi
}%
\endgroup
}%
\newenvironment{enumBib}{%
\BibTitle
\advance\@enumdepth \@ne
\edef\@enumctr{enum\romannumeral\the\@enumdepth}\list
{\csname biblabel\@enumctr\endcsname}{\usecounter
{\@enumctr}\def\makelabel##1{\hss\llap{\upshape##1}}}
}{%
\endlist
}

\makeatletter

\newcommand{\BiblioItem}[2]
{
\def\Semafor{off}
\@ifundefined{\LanguagePrefix ViewCite#1}{}{%
\def\Semafor{on}%
}%
\ifx\Semafor\ValueOff
\@ifundefined{xRefDef#1}{}{%
\def\Semafor{on}%
}%
\fi
\ifx\Semafor\ValueOn
\ifx\IndexState\ValueOff
\begin{enumBib}
\def\IndexState{on}
\fi
\item \label{\LanguagePrefix bibitem: #1}#2%
\fi
}
\makeatother
\newcommand{\OpenBiblio}
{
\def\IndexState{off}
}
\newcommand{\CloseBiblio}
{
\ifx\IndexState\ValueOn
\end{enumBib}
\def\IndexState{off}
\fi
}

\makeatletter
\def\StartCite{[}%
\def\citeBib#1{\protect\showCiteBib#1,endCite,}%
\def\endCite{endCite}%
\def\showCiteBib#1,{\def\temp{#1}%
\ifx\temp\endCite
]%
\def\StartCite{[}%
\else
\StartCite\LanguagePrefix \ref{\LanguagePrefix bibitem: #1}%
\@ifundefined{\LanguagePrefix ViewCite#1}{%
\NameDef{\LanguagePrefix ViewCite#1}{}%
}{%
}%
\def\StartCite{, }%
\expandafter\showCiteBib%
\fi}%
\makeatother

\newcommand{\arp}{\ar @{-->}}

\newcommand\Bundle[1]{{\mathbb #1}}
\newcommand{\bundle}[4]%
{%
\def\tempa{}%
\def\tempb{#3}%
\def\tempc{#1}%
\ifx\tempa\tempb%
\ifx\tempa\tempc%
#2%
\else%
\xymatrix{#2:#1\arp[r]&#4}%
\fi%
\else%
\ifx\tempa\tempc%
#2[#3]%
\else%
\xymatrix{#2[#3]:#1\arp[r]&#4}%
\fi%
\fi%
}%
\makeatletter
\newcommand{\AddIndex}[2]%
{%
\@ifundefined{RefIndex#2}{%
\xNameDef{RefIndex#2}{:}%
\@namedef{LabelIndex}{\label{index: #2::}}%
}{%
\addtocounter{Index}{1}%
\xNameDef{RefIndex#2}{\@nameuse{RefIndex#2},\arabic{Index}}%
\@namedef{LabelIndex}{\label{index: #2:\arabic{Index}}}%
}%
\@nameuse{LabelIndex}%
{\bf #1}%
}%

\newcommand{\Index}[2]%
{%
\@ifundefined{RefIndex#2}{%
\def\Semafor{off}%
}{%
\def\Semafor{on}%
}%
\ifx\Semafor\ValueOn%
\def\tempa{}%
\def\tempb{#2}%
\ifx\IndexState\ValueOff%
\ifx\setCACAA\undefined
\begin{theindex}%
\else
\section*{\indexname}%
\begin{itemize}
\fi
\def\IndexState{on}%
\fi%
\ifx\IndexSpace\ValueOn%
\indexspace%
\def\IndexSpace{off}%
\fi%
\item #1%
\ifx\tempa\tempb%
\else%
\edef\PageRefs{\@nameuse{RefIndex#2}}
\def\Sep{\ }%
\@for\PageRef:=\PageRefs\do{%
\Sep
\pageref{index: #2:\PageRef}%
\def\Sep{,\ }%
}%
\fi%
\fi%
}%

\newcommand{\Symb}[4]%
{%
\def\Semafor{off}%
\@ifundefined{ViewSymbol#2}{%
\@ifundefined{ViewSymbol#2(#3-#4)}{%
}{%
\def\Semafor{on}
\edef\ThisSymbol{#2(#3-#4)}%
}%
}{%
\def\Semafor{on}%
\edef\ThisSymbol{#2}%
}%
\ifx\Semafor\ValueOn%
\ifx\IndexState\ValueOff%
\ifx\setCACAA\undefined
\begin{theindex}%
\else
\section*{\indexname}%
\begin{itemize}
\fi
\def\IndexState{on}%
\fi%
\ifx\IndexSpace\ValueOn%
\indexspace%
\def\IndexSpace{off}%
\fi%
\edef\Symbols{\@nameuse{ViewSymbol\ThisSymbol}}%
\@for\Symbol:=\Symbols\do{%
\edef\Temp{ViewSymbol\ThisSymbol:::\Symbol}%
\item $\displaystyle\textcolor{SymbColor}{\@nameuse{\Temp}}$
\ \ #1
\edef\PageRefs{\@nameuse{RefSymbol\ThisSymbol===\Symbol}}
\def\Sep{}%
\@for\PageRef:=\PageRefs\do{%
\Sep
\pageref{symbol: \ThisSymbol:\PageRef}%
\def\Sep{,\ }%
}%
}%
\fi
}

\newcommand{\Symba}[2]
{
\def\Semafor{off}
\@ifundefined{ViewSymbol#2}{%
}{%
\def\Semafor{on}
}%
\ifx\Semafor\ValueOn
\ifx\IndexState\ValueOff
\begin{theindex}
\def\IndexState{on}
\fi
\ifx\IndexSpace\ValueOn
\indexspace
\def\IndexSpace{off}
\fi
\item $\displaystyle\@nameuse{ViewSymbol#2}$\ \ #1
\edef\PageRefs{\@nameuse{RefSymbol#2}}
\def\Sep{}%
\@for\PageRef:=\PageRefs\do{%
\Sep
\pageref{symbol: #2:\PageRef}%
\def\Sep{,\ }%
}%
\fi
}

\makeatother

\newcommand{\SetIndexSpace}%
{%
\def\IndexSpace{on}%
}%

\newcommand{\OpenIndex}
{
\def\IndexState{off}
}
\newcommand{\CloseIndex}
{
\ifx\IndexState\ValueOn
\ifx\setCACAA\undefined
\end{theindex}
\else
\end{itemize}
\fi
\def\IndexState{off}
\fi
}
\newcommand\CiteQuotation[2]
{
\leftskip=40pt
\Memos{#1}
\begin{flushright}
\begin{tabular}{r}
\hline
\makebox[100pt][r]{#2}\\
\end{tabular}
\end{flushright}
\leftskip=0pt
}
\newcommand\CiteQuote[2]
{
\leftskip=40pt
\BlueText{#1}%
\begin{flushright}
\begin{tabular}{r}
\hline
\makebox[100pt][r]{#2}\\
\end{tabular}
\end{flushright}
\leftskip=0pt
}
\def\epigraphSkip{150pt}
\newcommand\epigraph[2]
{
\leftskip=\epigraphSkip
\Memos{#1}
\begin{flushright}
\begin{tabular}{r}
\hline
\makebox[100pt][r]{#2}\\
\end{tabular}
\end{flushright}
\leftskip=0pt
}
\def\Memos#1{\MemoList#1//LastMemo//}%
\def\LastMemo{LastMemo}%
\def\MemoList#1//{\def\temp{#1}%
\ifx\temp\LastMemo
\else%
\setlength{\parindent}{5mm}
\par
\BlueText{#1}%
\expandafter\MemoList%
\fi%
}     



%% file: Preliminary.Eq.tex
\ifx\Preliminary\undefined
\def\Preliminary{off}
\fi

\input{Preliminary.Ref}
\input{Preliminary.Parm}

\ePrints{MVector,1908.04418,6860-2955,6860-2955,1601.03259,4975-6381,5410-9916,1305.4547,1502.04063,1310.5591}%
\Items{309618526,CACAA.06.121,5059-9176,323236904,322019412,1801.01628,9835-2163,9856-6693,0767-8264}%
\Items{BQuater,MQuater,CACAA.07,323966352,0921-2370,5148-4632,1506.00061,7287-9339}%
\ifx\Semafor\ValueOn
\input{\FilePrefix Stmt.Representation.\TheLanguage}
\fi

\ePrints{MVector,309618526,CACAA.06.121,1601.03259,4975-6381,5410-9916,1502.04063,5114-6019}%
\Items{323236904,1908.04418,6860-2955,322019412,1801.01628,9835-2163,9835-2163,9856-6693,0767-8264,MTensor}%
\Items{323966352,0921-2370,CACAA.07,1506.00061,7287-9339}%
\Items{BQuater,MQuater,0803.3276}
\ifx\Semafor\ValueOn
\input{\FilePrefix Stmt.Module.\TheLanguage}
\fi

\ePrints{1502.04063,4975-6381,5114-6019,1801.01628,9835-2163,0767-8264}
\ifx\Semafor\ValueOn
\input{\FilePrefix Stmt.Module.18.\TheLanguage}
\fi

\ePrints{1908.04418,6860-2955}
\Items{6860-2955,5148-4632,MVector,5410-9916,1601.03259,309618526,CACAA.06.121,4975-6381,322019412}
\ifx\Semafor\ValueOn
\input{\FilePrefix Stmt.Omega.\TheLanguage}
\fi

\ePrints{309618526,CACAA.06.121,4975-6381,1601.03259,323236904,1801.01628,9835-2163,0767-8264,322019412}
\Items{BQuater,MQuater,9856-6693,CACAA.07,330532713}
\ifx\Semafor\ValueOn
\input{\FilePrefix Stmt.Derivative.\TheLanguage}
\fi

\ePrints{323966352,0921-2370,1908.04418,6860-2955,0803.3276,FAlgebra,0405.027}
\ifx\Semafor\ValueOn
\input{\FilePrefix Stmt.Gravity.\TheLanguage}
\fi

\ePrints{324329551}
\ifx\Semafor\ValueOn
\input{Stmt.Gravity.Eq}
\fi

\ePrints{MQuater,BQuater,1601.03259,4975-6381,322019412}
\ifx\Semafor\ValueOn
\input{Algebra.Complex.2016.Eq}
\fi

\ePrints{309618526,323966352,0921-2370,CACAA.06.121,MQuater,BQuater,CACAA.07,322019412}%
\Items{1506.00061,7287-9339,1801.01628,9835-2163,0767-8264,330532713}%
\ifx\Semafor\ValueOn%
\input{Algebra.Quaternion.2016.Eq}
\fi

\ePrints{309618526,323966352,0921-2370,MQuater,BQuater}
\ifx\Semafor\ValueOn
\input{Algebra.Octonion.2016.Eq}
\fi

\ePrints{330532713}
\ifx\Semafor\ValueOn
\input{Stmt.Module.Eq}
\fi

\ePrints{323966352,0921-2370}
\ifx\Semafor\ValueOn
\input{Stmt.Derivative.Eq}
\fi

\ePrints{330532713}
\ifx\Semafor\ValueOn
\input{Stmt.Representation.Eq}
\fi

\def\Ii{i\in I}
\def\iI{$\Ii$}
\def\Times{A_1\times...\times A_n}
\def\LAA{\mathcal L(D;A\rightarrow A)}
\newcommand\Kn[2][k]{$#1=#2$, ..., $n$}
\newcommand\Kb[2][k]{$(#1)=(#2)$, ..., $(n)$}
\def\ATwo{A_2\otimes A_2}
\newcommand\BoxB[1]{$#1\otimes #1$\Hyph}

\newcommand\aUD[3]{#1^{\gi{#2}}_{\gi{#3}}}%
\newcommand\aU[2]{#1^{\gi{#2}}}%
\newcommand\aD[2]{#1_{\gi{#2}}}%
\newcommand\FC[2]{f^{\gi{#1#2}}}%
\newcommand\TensorBasis[1]
{e_{1\cdot\gi{#1_1}}\otimes...\otimes e_{n\cdot\gi{#1_n}}}
\newcommand\re{\mathrm{Re}\,} 
\newcommand\im{\mathrm{Im}\,} 

\AddEq [4]{L(A;B)}
{
$\mathcal L(#1;#2\rightarrow #3)$#4
}

\DefEq
{
\symb{|\sigma|}{parity of permutation}{}
\[
\ShowSymbol{parity of permutation}{}
:\sigma\in S(n)\rightarrow \{-1,1\}
\]
}
{sigma->+-}

\DefEq
{
$\{a_1,...,a_n\}$\Pt
}
{set a1n}

\AddEq[1]{a=b}
{
$a_1=b_1$#1
}

\DefEq
{
\[
\ShowSymbol{parity of permutation}{}=
\left\{
\begin{matrix}
1&\textrm{\permutation}\sigma\textrm{ \even}
\\
-1&\textrm{\permutation}\sigma\textrm{ \odd}
\end{matrix}
\right.
\]
}
{parity of permutation}

\AddEq{quaternion algebra}
{
\symb{H}{quaternion algebra}1
}

\AddEq{quaternion, 3, 1}
{
\[
\ShowEq{quaternion matrix fUD<-fUU}^{-1}
=
\ShowEq{quaternion matrix fUD->fUU}
\]
}

\AddEquation{quaternion fUD<-fUU}
{
\begin{split}
&
\ShowEq{quaternion matrix fUD}
\\=&
\ShowEq{quaternion matrix fUD<-fUU}
\ShowEq{quaternion matrix fUU}f
\end{split}
}

\AddEquation{quaternion fUD->fUU}
{
\begin{split}
&
\ShowEq{quaternion matrix fUU}f
\\=&
\ShowEq{quaternion matrix fUD->fUU}
\ShowEq{quaternion matrix fUD}
\end{split}
}

\DefEq
{
$\mathcal L(C;C\rightarrow C)=CE$.
}
{L(CCC)=CE}

\AddEquation{norm of quaternion}
{
|x|^2=x x^*=(\aU x0)^2+(\aU x1)^2+(\aU x2)^2+(\aU x3)^2
}

\AddEq{inverce quaternion}
{
\[
x^{-1}=|x|^{-2}x^*
\]
}

\DefEq
{
\[
x=x^{\gi 0}+x^{\giA}i+x^{\gi 2}j+x^{\gi 3}k
\]
\[
x^{\gi 0}, x^{\giA}, x^{\gi 2}, x^{\gi 3}\in R
\]
}
{quaternion algebra, element}

\DefEquation
{
\begin{array}{c|ccc}
&i&j&k\\
\hline
i&-1&k&-j\\
j&-k&-1&i\\
k&j&-i&-1\\
\end{array}
}
{product of quaternions}

\DefEq
{
$\Basis e_i$
}
{tensor product of algebras, basis i}

\DefEq
{
\symb{a^{\gi{i_1...i_n}}}{standard component of tensor}{}
}
{standard component of tensor}

\DefEquation
{
\circ:(g,f)\in\mathcal L(D;A\rightarrow A)\times\mathcal L(D;A\rightarrow A)
\rightarrow g\circ f\in\mathcal L(D;A\rightarrow A)
}
{module L(A;A) is algebra}

\DefEq
{
\[
\xymatrix{
&A\ar[rd]^g
\\
A\ar[ru]^f\ar[rr]^{g\circ f}&&A
}
\]
}
{product of linear map, algebra 1}

\DefEquation
{
a=
\ShowSymbol{standard component of tensor}{}
\TensorBasis i
}
{tensor canonical representation, algebra}

\AddEq{k,1n}
{
$k$, $k=1$, ..., $n$,
}

\AddEq{n,1.}
{
$n$, $n=1$, ...,
}

\newcommand{\Tensor}[1]{#1_1\otimes...\otimes #1_n}

\AddEq[3]{set f:A->B}
{
\[
\{#1_i:#2\rightarrow #3,i\in I\}
\]
}

\AddEq[1]{set Bi}
{
$\{#1_i,i\in I\}$
}

\AddEq{product in category diagram}
{
\[
\xymatrix{
P\ar[r]^{f_i}&B_i&f_i\circ h=g_i
\\
R\ar[ur]_{g_i}\ar[u]^h
}
\]
}

\AddEq{coproduct in category diagram}
{
\[
\xymatrix{
P\ar[d]_h&B_i\ar[l]_{f_i}\ar[dl]^{g_i}&h\circ f_i=g_i
\\
R
}
\]
}

\AddEq{product in category, 1 n}
{
\symb[Pi-1]{\prod_{i=1}^nB_i}{product in category}{i 1 n}
\symb[B-0]{B_1\times...\times B_n}{product in category}{1 n}
\[
P=\ShowSymbol{product in category}{i 1 n}
=\ShowSymbol{product in category}{1 n}
\]
}

\AddEq{Cartesian product of sets}
{
\[A=\prod_{\Ii}A_i\]
}

\AddEq{Ai iI}
{
\((A_i,\Ii)\)
}

\AddEq{projection on i factor}
{
\[p_i:A\rightarrow A_i\]
}

\DefEq
{
\[p_i:A\rightarrow A_i\ \ \ \Ii\]
}
{p:A->Ai i in I}

\DefEq
{
\begin{align*}
\re&:A\rightarrow A
\\
\im&:A\rightarrow A
\end{align*}
}
{Re Im A->A}

\DefEq
{
\symb{\re d}{scalar of algebra}{}
\symb{\im d}{vector of algebra}{}
}
{Re Im A->F 0}

\DefEquation
{
\begin{matrix}
\ShowSymbol{scalar of algebra}{}=d^{\gi 0}
&\ShowSymbol{vector of algebra}{}=d-d^{\gi 0}
&d\in D
&d=d^{\gii}e_{\gii}
\end{matrix}
}
{Re Im A->F 1}

\DefEq
{
$\ShowSymbol{scalar of algebra}{}$
}
{Re Im A->F 2}

\DefEq
{
$\ShowSymbol{vector of algebra}{}$
}
{Re Im A->F 3}

\DefEq
{
\[
F=\{d\in A:\re d=d\}
\]
}
{Re Im A->F 4}

\DefEq
{
\symb{\re A}{scalar algebra of algebra}1
}
{scalar algebra of algebra}

\DefEq
{
\symb{\im A}{vector module of algebra}{}
\begin{equation}
\EqLabel{vector module of algebra}
\ShowSymbol{vector module of algebra}{}=\{d\in A:\re d=0\}
\end{equation}
}
{vector module of algebra}

\DefEquation
{
C_{\gi{0k}}^{\gil}=C^{\gil}_{\gi{k0}}=\delta^{\gik}_{\gil}
}
{structural constant re algebra, 1}

\DefEquation
{
d=\re d+\im d
}
{d Re Im}

\DefEq
{
\symb{d^*}{conjugation in algebra}{}
}
{conjugation in algebra}

\DefEquation
{
\ShowSymbol{conjugation in algebra}{}=\re d-\im d
}
{conjugation in algebra, 0}

\DefEquation
{
(cd)^*=d^*\,c^*
}
{conjugation in algebra, 1}

\DefEquation
{
\begin{matrix}
C^{\gi 0}_{\gi{kl}}=C^{\gi 0}_{\gi{lk}}
&
C^{\gi p}_{\gi{kl}}=-C^{\gi p}_{\gi{lk}}
\end{matrix}
}
{conjugation in algebra, 5}

\DefEq
{
\[
\begin{matrix}
\giA\le\gik\le\gin&\giA\le\gil\le\gin&\giA\le\gi p\le\gin
\end{matrix}
\]
}
{conjugation in algebra, 4}

\DefEq
{
$p(x)=p_1\circ x$
}
{p=p circ x}

\AddEq{kernel of map}
{
\symb{\mathrm{ker}\,f}{kernel of map}{}
}

\AddEquation{def kernel of map}
{
\ShowSymbol{kernel of map}{}
=\{
(a,b):a,b\in A,f(a)=f(b)
\}
}

\AddEq{lemmas for ker}
{
\RefLemma{ker - reflexive correspondence},
\RefLemma{ker - symmetric correspondence},
\RefLemma{ker - transitive correspondence}
}

\AddEq [2]{fa=fb}
{
f(#1)=f(#2)
}

\AddEq{x in C=>fx=gx}
{
$x\in C\Rightarrow f(x)=g(x)$.
}

\AddEq[3]{A in B}
{
$#1\subseteq #2$#3
}

\AddEq [2]{ab in ker}
{
(#1,#2)\in\mathrm{ker}\,f
}

\AddEquation{ab in ker -> ba in ker}
{
(a,b)\in\mathrm{ker}\,f\Rightarrow(b,a)\in\mathrm{ker}\,f
}

\AddEquation{ab,ac in ker -> ac in ker}
{
(a,b),(b,c)\in\mathrm{ker}\,f\Rightarrow(a,c)\in\mathrm{ker}\,f
}

\AddEq{ker f}
{
$\mathrm{ker}\,f$
}

\AddEq{fa=fa}
{
\[f(a)=f(a)\]
}

\AddEq{Phi in AxA}
{
$\Phi\in A\times A$
}

\AddEq{transitive correspondence}
{
\labelItem{transitive correspondence}
\[
(a,b),(b,c)\in\Phi\Rightarrow (a,c)\in\Phi
\]
}

\AddEq{symmetric correspondence}
{
\labelItem{symmetric correspondence}
\[
(a,b)\in\Phi\Rightarrow (b,a)\in\Phi
\]
}

\AddEq{reflexive correspondence}
{
\labelItem{reflexive correspondence}
\[
(a,a)\in\Phi
\]
}

\DefEquation
{
\begin{split}
r(x)&=r_0+q_{1\cdot 0} p(x)q_{1\cdot 1}
+q_{2\cdot 0}(x) p(x)q_{2\cdot 1}+...
+q_{k\cdot 0}(x) p(x)q_{k\cdot 1}
\\
&=r_0+(q_{1\cdot 0}\otimes q_{1\cdot 1})\circ p(x)
+(q_{2\cdot 0}(x)\otimes q_{2\cdot 1})\circ p(x)
\\&+...
+(q_{k\cdot 0}(x)\otimes q_{k\cdot 1})\circ p(x)
\end{split}
}
{r=+q circ p}

\DefEquation
{
p(x)=p_0+p_1\circ x
}
{p=p+p circ x}

\DefEq
{
\[
r(x)=r_0+r_1\circ x+...+r_k\circ x^k
\]
}
{r power k}

\DefEquation
{
\begin{split}
r(x)&=r_0
-((r_{1\cdot 0\cdot s}\otimes r_{1\cdot 1\cdot s})\circ
p^{-1}_1)\circ p_0
\\
&-(((r_{2\cdot 0\cdot s}\circ x)\otimes r_{2\cdot 1\cdot s})\circ
p^{-1}_1)\circ p_0
\\
&-...-(((r_{k\cdot 0\cdot s}\circ x^{k-1})
\otimes r_{k\cdot 1\cdot s})\circ
p^{-1}_1)\circ p_0
\\
&+((r_{1\cdot 0\cdot s}\otimes r_{1\cdot 1\cdot s})\circ
p^{-1}_1)\circ p(x)
\\
&+(((r_{2\cdot 0\cdot s}\circ x)\otimes r_{2\cdot 1\cdot s})\circ
p^{-1}_1)\circ p(x)
\\
&+...+(((r_{k\cdot 0\cdot s}\circ x^{k-1})
\otimes r_{k\cdot 1\cdot s})\circ
p^{-1}_1)\circ p(x)
\\
&=r_0
-((r_{1\cdot 0\cdot s}\otimes r_{1\cdot 1\cdot s}
+(r_{2\cdot 0\cdot s}\circ x)\otimes r_{2\cdot 1\cdot s}
\\
&+...+(r_{k\cdot 0\cdot s}\circ x^{k-1})
\otimes r_{k\cdot 1\cdot s})\circ
p^{-1}_1)\circ p_0
\\
&+((r_{1\cdot 0\cdot s}\otimes r_{1\cdot 1\cdot s}
+(r_{2\cdot 0\cdot s}\circ x)\otimes r_{2\cdot 1\cdot s}
\\
&+...+(r_{k\cdot 0\cdot s}\circ x^{k-1})
\otimes r_{k\cdot 1\cdot s})\circ
p^{-1}_1)\circ p(x)
\end{split}
}
{r=+q circ p 1}

\DefEq
{
\[
ax-xa=1
\]
}
{ax-xa=1}

\DefEquation
{
r(x)=q_{i\cdot 0}(x)p(x)q_{i\cdot 1}(x)
=(q_{i\cdot 0}(x)\otimes q_{i\cdot 1}(x))\circ p(x)
}
{r=qpq(x)}

\DefEquation
{
a_1...a_n\omega=(p_i(a_1)...p_i(a_n)\omega,\Ii)
}
{omega(ai)=(omega ai)}

\DefEquation
{
\xymatrix{
A\ar[r]^{p_i}&A_i&p_i\circ \omega=g_i
\\
A^n\ar[ur]_{g_i}\ar[u]^{\omega}
}
}
{operation is defined componentwise, diagram}

\DefEquation
{
a_1...a_n\omega=(a_{1i}...a_{ni}\omega,\Ii)
}
{operation is defined componentwise}

\DefEq
{
\((p_i(a),\Ii)\)
}
{tuple represent A number}

\AddEq[2]{omega in Omega}
{
\(\omega\in\Omega_{#1}\)#2
}

\DefEq
{
h=f_1...f_n\omega
}
{h=f1...fn omega}

\DefEq
{
\(f_X(g_{1\cdot n})(g_{2\cdot n})\)
}
{f(g1n)(g2n)}

\DefEq
{
\(a_1=(a_{1i},\Ii)\), ..., \(a_n=(a_{ni},\Ii)\)
}
{a=ai 1n}

\DefEq
{
\((B_i,\Ii)\)
}
{Bi iI}

\DefEq
{
f(a_i,\Ii)=(f_i(a_i),\Ii)
}
{f:A->B=}

\DefEq
{
\[
g_i(a_1, ..., a_n)=p_i(a_1)...p_i(a_n)\omega
\]
}
{gi()=}

\DefEq
{
\[f_i:A_i\rightarrow B_i\]
}
{f:A->B i}

\DefEquation
{
\xymatrix
{
B\ar[r]^{p'_i}&B_i
\\
A\ar[u]^f\ar[r]_{p_i}&A_i\ar[u]_{f_i}
}
}
{homomorphism of Cartesian product of Omega algebras diagram}

\DefEquation
{
\xymatrix
{
B\ar[rrr]^{p'_i}\ar@{}[dr]^(.6){(1)}&&&B_i
\\
&&
\\
A\ar[uu]^f\ar[uurrr]^{g_i}\ar[rrr]_{p_i}&&&A_i\ar[uu]_{f_i}\ar@{}[ul]^(.8){(2)}
}
}
{homomorphism of Cartesian product of Omega algebras}

\DefEquation
{
b=f(a)\in B
}
{b=f(a)}

\DefEquation
{
\begin{matrix}
b=(b_i,\Ii)&b_i=p'_i(b)\in B_i
\end{matrix}
}
{b=p(b)i}

\DefEquation
{
b_i=g_i(b)
}
{b=g(a)i}

\DefEq
{
\[b_i=f_i(a_i)\]
}
{b=f(a)i}

\DefEq
{
\(b_1=(b_{1i},\Ii)\), ..., \(b_n=(b_{ni},\Ii)\)
}
{b=bi 1n}

\DefEq
{
\begin{align*}
f(a_1...a_n\omega)&=f(a_{1i}...a_{ni}\omega,\Ii)
\\&=(f_i(a_{1i}...a_{ni}\omega),\Ii)
\\&=((f_i(a_{1i}))...(f_i(a_{ni})),\Ii)
\\&=(b_{1i}...b_{ni}\omega,\Ii)
\end{align*}
\[
f(a_1)...f(a_n)\omega=b_1...b_n\omega
=(b_{1i}...b_{ni}\omega,\Ii)
\]
}
{f:A->B omega}

\DefEquation
{
\begin{matrix}
a=(a_i,\Ii)&a_i=p_i(a)\in A_i
\end{matrix}
}
{a=p(a)i}

\DefEq
{
\(p_i\), \(p'_i\)
}
{pi p'i}

\AddEq [2]{f->g}
{
\[#1\rightarrow #2\]
}

\AddEq [3]{A->B}
{
$#1\rightarrow #2$#3
}

\DefEq
{
\[h:S_1\rightarrow S_2\]
}
{f->g 1}

\AddEq{polylinear maps category, diagram}
{
\[
\xymatrix{
&S_1\ar[dd]^h
\\
\Times
\ar[ru]^f\ar[rd]_g
\\
&S_2
}
\]
}

\DefEq
{
$a_1\ne b_1$, $a_1$, $b_1\in A_1$,
}
{a1 ne b1}

\DefEq
{
\[
F:A_1\cup A_2\rightarrow B_1\cup B_2
\]
}
{F:A1+A2->B1+B2}

\DefEq
{
\[
F(A_1)=B_1 \ \ \ \ F(A_2)=B_2
\]
}
{F:A1+A2->B1+B2 1}

\AddEq [3]{map r,R}
{
$(#1,#2)$#3
}

\AddEq [2]{map r12}
{
$#1=(#1_1,#1_2)$#2
}

\DefEq
{
\[
r_i:A_i\rightarrow B_i
\]
}
{ri:A->B}

\AddEq[1]{i=1,2}
{
$i=1$, $2$#1
}

\DefEq
{
\symb{\mathrm{ker}\,f}{kernel of homomorphism}0
$\ShowSymbol{kernel of homomorphism}0
=f\circ f^{-1}$
}
{kernel of homomorphism}

\DefEq
{
$\mathrm{ker}\,t_i
=t_i\circ t_i^{-1}$
}
{kernel of homomorphism i}

\DefEq
{
$A/\mathrm{ker}\,f$
}
{A/ker f}

\DefEq
{
\[
p:a\in A\rightarrow a^{\mathrm{ker}\,f}\in A/\mathrm{ker}\,f
\]
}
{p:A->/ker}

\DefEq
{
$a$, $b\in \pA$\Pt
}
{ab in A}

\DefEq
{
\[f(a)(m)\ne f(b)(m)\]
}
{fam ne fbm}

\DefEq
{
\[f(a_1)(a_2)\ne a_2\]
}
{fam ne m}

\DefEq
{
$B_2\subset A_2$
}
{B2 subset A2}

\DefEq
{
\[f(a_1)(a_2)=f(b_1)(a_2)\]
}
{faa=fba}

\AddEq [2]{a in A}
{
$a_{#1}\in A_{#1}$#2
}

\DefEq
{
\symb{A^{\otimes n}}{tensor power of algebra}{}
\[
\begin{matrix}
\ShowSymbol{tensor power of algebra}{}
=\Tensor A
&
A_1=...=A_n=A
\end{matrix}
\]
}
{tensor power of algebra}

\DefEq
{
\[
g:A_1\times...\times A_n\rightarrow V
\]
}
{map g, algebra, tensor product}

\DefEq
{
\[
h:\Tensor A\rightarrow V
\]
}
{map h, algebra, tensor product}

\DefEquation
{
\xymatrix{
&\Tensor A\ar[dd]^h
\\
\Times
\ar[ru]^f\ar[rd]_g
\\
&V
}
}
{map gh, algebra, tensor product}

\DefEquation
{
h\circ(\Tensor a)=g\circ(a_1,...,a_n)
}
{g=h, algebra, tensor product}

\AddEq{basis of complex field}
{
\[
\begin{array}{rr}
e_{\gi 0}=1&e_{\giA}=i
\end{array}
\]
}

\AddEquation{product of complex field}
{
e_{\giA}^2=-e_{\gi 0}
}

\DefEq
{
\[f:H\rightarrow H\ \ \ y^{\gii}=f^{\gii}_{\gij}x^{\gij}\]
}
{f:H->H ij}

\DefEq
{
\[f:C\rightarrow C\]
}
{f:C->C}

\DefEq
{
\[(aI)\circ (bI)\circ x=(aI)\circ (b\overline x)=a\overline{(b\overline x)}
=(a\overline b)x=((a\overline b)E)\circ x\]
}
{aI o bI=...E}

\DefEq
{
\[CE=\{aE:a\in C\}\]
}
{CE set}

\DefEq
{
\[HE=\{aE:a\in H\}\]
}
{HE set}

\DefEq
{
\[CI=\{aI:a\in C\}\]
}
{CI set}

\AddEquation{structural constants of complex field}
{
\begin{array}{r@{}rr@{}r}
\aUD C0{00}=&1&\aUD C1{01}=&1
\\
\VirtVar
\aUD C1{10}=&1&\aUD C0{11}=&-1
\end{array}
}

\AddEquation{quaternion conjugation}
{
\begin{split}
\overline x&
=-\frac 12(1\otimes 1+i\otimes i+j\otimes j+k\otimes k)\circ x
\\
&=-\frac 12(x+ixi+jxj+kxk)
\end{split}
}

\AddEq [1]{conjugate to the quaternion}
{
\[
#1^*=\aU{#1}0-\aU{#1}1i-\aU{#1}2j-\aU{#1}3k
\]
}

\DefEquation
{
\begin{split}
f_{\gi 0}^{\gi 0}&=\hphantom{-\,}f_{\giA}^{\giA}
\\
f_{\gi 0}^{\giA}&=-f_{\giA}^{\gi 0}
\end{split}
}
{complex field over real field}

\DefEquation
{
\begin{split}
f_{\gi 0}^{\gi 0}&=-f_{\giA}^{\giA}
\\
f_{\gi 0}^{\giA}&=\hphantom{-\,}f_{\giA}^{\gi 0}
\end{split}
}
{matrix of linear map f in CI}

\DefEq
{
\[
(b_{\gi 0}+b_{\giA}i)I(x_{\gi 0}+x_{\giA}i)
=(b_{\gi 0}+b_{\giA}i)(x_{\gi 0}-x_{\giA}i)
=b_{\gi 0}x_{\gi 0}+b_{\giA}x_{\giA}
+(-b_{\gi 0}x_{\giA}+b_{\giA}x_{\gi 0})i
\]
\[
\begin{pmatrix}
b_{\gi 0}&b_{\giA}
\\
b_{\giA}&-b_{\gi 0}
\end{pmatrix}
\begin{pmatrix}
x_{\gi 0}\\x_{\giA}
\end{pmatrix}
=
\begin{pmatrix}
b_{\gi 0}x_{\gi 0}+b_{\giA}x_{\giA}
\\
b_{\giA}x_{\gi 0}-b_{\gi 0}x_{\giA}
\end{pmatrix}
\]
}
{congugate map complex field, 3}

\DefEq
{
\[
x=x^{\gi 0}+x^{\giA}i\rightarrow
f=f^{\gi 0}(x^{\gi 0},x^{\giA})+f^{\giA}(x^{\gi 0},x^{\giA})i
\]
}
{map of complex variable}

\AddEq [1]{Cauchy Riemann equations, complex field, 1}
{
\begin{split}
\frac{\partial #1^{\giA}}{\partial x^{\gi 0}}
&=-\frac{\partial #1^{\gi 0}}{\partial x^{\giA}}
\\[5pt]
\frac{\partial #1^{\gi 0}}{\partial x^{\gi 0}}
&=\hphantom{-\,}\frac{\partial #1^{\giA}}{\partial x^{\giA}}
\end{split}
}

\AddEq [1]{Cauchy Riemann equations, complex field, 2, 1}
{
\frac{\partial #1^{\gi 0}}{\partial x^{\gi 0}}
+i\frac{\partial #1^{\giA}}{\partial x^{\gi 0}}
+i\left(\frac{\partial #1^{\gi 0}}{\partial x^{\giA}}
+i\frac{\partial #1^{\giA}}{\partial x^{\giA}}\right)=
\frac{\partial #1^{\gi 0}}{\partial x^{\gi 0}}
+i\frac{\partial #1^{\giA}}{\partial x^{\gi 0}}
+i\frac{\partial #1^{\gi 0}}{\partial x^{\giA}}
-\frac{\partial #1^{\giA}}{\partial x^{\giA}}=
0
}

\DefEq
{
\[
\begin{matrix}
a\circ E:H\rightarrow H&
a=a^{\gi 0}+a^{\gi 1}i+a^{\gi 2}j+a^{\gi 3}k\in H
\end{matrix}
\]
}
{aE, quaternion}

\AddEquation{df=...dx= CE}
{
\frac{d f}{d x}
=\frac 12\left(\frac{\partial f}{\partial x^{\gi 0}}
-i\frac{\partial f}{\partial x^{\gi 1}}\right)\circ E
}

\AddEq{df=...dx 03}
{
\begin{split}
\frac{d f}{d x}=&-\frac 12
\left(
\frac{\partial f}{\partial x^{\gi 0}}
+\frac{\partial f}{\partial x^{\gi 1}}i
+\frac{\partial f}{\partial x^{\gi 2}}j
+\frac{\partial f}{\partial x^{\gi 3}}k
\right)\circ E
\\&
+\frac 12\left(
\frac{\partial f}{\partial x^{\gi 0}}
+\frac{\partial f}{\partial x^{\gi 1}}i
\right)\circ I
+\frac 12\left(
\frac{\partial f}{\partial x^{\gi 0}}
+\frac{\partial f}{\partial x^{\gi 2}}j
\right)\circ J
\\&
+\frac 12\left(
\frac{\partial f}{\partial x^{\gi 0}}
+\frac{\partial f}{\partial x^{\gi 3}}k
\right)\circ K
\end{split}
}

\DefEquation
{
\begin{split}
dx^{\gi 0}&=\frac 12(E+I)\circ dx
\\
dx^{\gi 1}&=-\frac i2(E-I)\circ dx
\end{split}
}
{dx01=}

\DefEquation
{
\begin{split}
df&=\frac 12\frac{\partial f}{\partial x^{\gi 0}}(E+I)\circ dx
-\frac i2\frac{\partial f}{\partial x^{\gi 1}}(E-I)\circ dx
\\
&=\frac 12\left(\left(\frac{\partial f}{\partial x^{\gi 0}}
-i\frac{\partial f}{\partial x^{\gi 1}}\right)E
+\left(\frac{\partial f}{\partial x^{\gi 0}}
+i\frac{\partial f}{\partial x^{\gi 1}}\right)I\right)\circ dx
\end{split}
}
{df=...dx}

\DefEq
{
\[
df=\frac{\partial f}{\partial x^{\gi 0}}dx^{\gi 0}
+\frac{\partial f}{\partial x^{\gi 1}}dx^{\gi 1}
\]
}
{df=df/dx0+df/dx1}

\DefEq
{
\[dx=dx^{\gi 0}+dx^{\gi 1}i\]
}
{dx=dx0+dx1}

\DefEq
{
\[
y^{\gi i}=x^{\gi j}f_{\gi j}^{\gi i}
\]
}
{yi=xj fji}

\DefEq
{
\symb{B^A}{Cartesian power}1
}
{Cartesian power}

\DefEq
{
$m_1\equiv m_2(\mathrm{mod} S)$
}
{m1 m2 modS}

\DefEq
{
\[f\circ(a_1,...,a_n)=\Tensor a\]
}
{fxa=oxa}

\DefEq
{
\[
\sigma:a\in A\rightarrow \sigma(a)\in A\ \ \ A=\{a_1,...,a_n\}
\]
}
{a->sigma a}

\DefEquation
{
\sigma=
\begin{pmatrix}
a_1&...&a_n
\\
\sigma(a_1)&...&\sigma(a_n)
\end{pmatrix}
}
{permutation as matrix}

\DefEquation
{
\sigma=
\begin{pmatrix}
\sigma(a_1)&...&\sigma(a_n)
\end{pmatrix}
}
{permutation as matrix 2}

\DefEq
{
\[f:A_1\times...\times A_n\rightarrow\Tensor A\]
}
{f:xA->oxA}

\AddEq{tensor product of algebras}
{
\symb{\Tensor A}{tensor product}1
}

\AddEq{set of n-ary operators}
{
\symb{\Omega(n)}{set of n-ary operators}{}
}

\AddEq[4]{omega n ari}
{
$\omega_{#1}\in\Omega_{#2}(#3)$#4
}

\DefEquation
{
ea\omega=a
}
{left neutral element}

\DefEquation
{
ae\omega=a
}
{right neutral element}

\DefEq
{
\[ab\omega=ba\omega\]
}
{commutative operation}

\DefEq
{
\[
a(bc\omega)\omega=(ab\omega)c\omega
\]
}
{associative operation}

\DefEq
{
\symb{A_{\Omega}}{Omega-algebra}1
}
{Omega-algebra}

\DefEq
{
$\omega\in A^{A^n}$.
}
{o in AAn}

\DefEq
{
\[\Omega(n)\rightarrow A^{A^n}\ \ \ n\in N\]
}
{O(n)->AAn}

\AddEq [1]{B subset A}
{
$B\subseteq A$#1
}

\DefEq
{
\[
\ShowSymbol{set of n-ary operators}{}=
\{\omega\in\Omega:a(\omega)=n\}
\]
}
{set of n-ary operators =}

\DefEq
{
$a(\omega)=n$,
}
{a(o)=n}

\DefEq
{
\[\omega_A|B=\omega_B\]
}
{oAB=oB}

\DefEq
{
\[a:\Omega\rightarrow N\]
}
{a:O->N}

\DefEq
{
\symb{\Omega}{operator domain}1
}
{operator domain}

\DefEq
{
$\omega(a_1,...,a_n)$, $a_1...a_n\omega$
}
{a1no=oa1n}

\DefEq
{
$b_1...b_n\omega\in B$,
}
{b1no in B}

\DefEq
{
$\b_1$, ..., $\b_n\in\B$,
}
{b1n in B}

\DefEq
{
\[\omega:A^n\rightarrow A\]
}
{o:An->A}

\DefEq
{
\[
\begin{matrix}
f_1:A\rightarrow S_1&\mathrm{ker}\,f_1\supseteq N
\\
f_2:A\rightarrow S_2&\mathrm{ker}\,f_2\supseteq N
\end{matrix}
\]
}
{maps category}

\DefEq
{
\[
\xymatrix{
&S_1\ar[dd]^h
\\
A
\ar[ru]^{f_1}\ar[rd]_{f_2}
\\
&S_2
}
\]
}
{maps category, diagram}

\DefEq
{
\[
\xymatrix
{
&A/N\ar[dd]^h
\\
A\ar[ur]^{j=\mathrm{nat}\,N}\ar[dr]_f
\\
&S
}
\]
}
{maps category, universal, diagram}

\DefEquation
{
\mathrm{ker}\,f\supseteq N
}
{maps category, universal, ker}

\DefEq
{
\[
\mathrm{nat}\,N:A\rightarrow A/N
\]
}
{maps category, universal}

\DefEq
{
\[j(a_1)=j(a_2)\]
}
{maps category 1}

\DefEq
{
\[f(a_1)=f(a_2)\]
}
{maps category 2}

\DefEq
{
\[h(\BlueText{j(b)})=f(b)\]
}
{maps category, h}

\DefEq
{
\[
(f+g)(a)=f(a)+g(a)
\]
}
{(f+g)(a)=}

\AddEquation{sum of transformations of Abelian group, 1}
{
f(a+b)(x)=f(a)(x)+f(b)(x)
}

\ePrints{1502.04063,5114-6019}
\ifx\Semafor\ValueOff
\DefEquation
{
f(a)(m)=f(b)(m)
}
{representation of ring, 1}

\AddEquation{representation of ring, 2}
{
f(a-b)(m)=0
}
\fi

\AddEq{polynomials p,r}
{
$p\circ F_{[1]}\circ x^n$, $r\circ F_{[2]}\circ x^m$
}

\DefEq
{
\[
(p\circ F_{[1]}\circ x^n)(r\circ F_{[2]}\circ x^m)
=(p\underline{\otimes}r)\circ
(F_{[1](1)},...,F_{[1](n)},F_{[2](1)},...,F_{[2](m)})
\circ x^{n+m}
\]
}
{pr=p circ r}

\DefEquation
{
p(x)=
(a_{k\cdot 0}\circ x^{k-1})
((1\otimes a_{k\cdot 1})\circ x)
}
{p(x)=a k-1 k 1}

\DefEquation
{
p(x)=
((a_{k\cdot 0}\circ x^{k-1})
\otimes a_{k\cdot 1})\circ x
}
{p(x)=a k-1 k 2}

\DefEq
{
$A_{ij}$, $i=1$, ..., $n$, $j=1$, ..., $m$,
}
{Aij}

\DefEq
{
$\Omega_{ij}$\Hyph%
}
{Omegaij}

\DefEq
{
$f_1$, ..., $f_n\in B^A$,
}
{f1n in B**A}

\DefEq
{
$\Hom(\emptyset;A\rightarrow B)=B^A$.
}
{Hom empty A B=B**A}

\DefEq
{
$\End(\emptyset;A)=A^A$.
}
{End empty A=A**A}

\DefEq
{
$\End(\Omega;A)=\Hom(\Omega;A\rightarrow A)$
}
{End A=Hom AA}

\DefEq
{
\symb{\Hom(\Omega;A\rightarrow B)}{set of homomorphisms}1
}
{set of homomorphisms}

\DefEq
{
\symb{\End(\Omega;A)}{set of endomorphisms}1
}
{set of endomorphisms}

\DefEq
{
(f_1...f_n\omega)(x)=f_1(x)...f_n(x)\omega
}
{f1n omega=}

\DefEq
{
\[t:A\rightarrow A\]
}
{t:A->A}

\AddEq{polylinear maps category}
{
\[
\xymatrix@C=15pt{
f:\Times\ar[r]&S_1
&
g:\Times\ar[r]&S_2
}
\]
}

\DefEq
{
\[f:A_1\rightarrow\End(\Omega_2;A_2)\]
}
{representation of algebra}

\DefEq
{
\symb{\mathcal B_f}{lattice of subrepresentations}1
}
{lattice of subrepresentations}

\AddEq[1]{R1 X->}
{
\[R_1:#1\rightarrow #1'\]
}

\DefEq
{
\[
R\circ m=R_1(m)
\]
}
{Rm=R1m}

\DefEq
{
$f(a)(m)\in B_2$
}
{fam in B2}

\ePrints{1502.04063,5114-6019}
\ifx\Semafor\ValueOff
\AddEq[1]{a in A1}
{
$a\in A_1$#1
}
\fi

\DefEq
{
\[
\xymatrix{
&&A_2/s_2\ar[rrrrr]^{q_2}\ar@{}[drrrrr]|{(5)}&&\ar@{}[dddddll]|{(4)}&
\ar@{}[dddddrr]|{(6)}&&t_2A_2\ar[ddddd]^{r_2}\\
&&&&&&&\\
A_1/s_1\ar[r]^{q_1}\ar@/^2pc/@{=>}[urrr]^F&
t_1A_1\ar[d]^{r_1}\ar@{=>}[urrrrr]^(.4)G&&&
A_2/s_2\ar[r]^{q_2}\ar[lluu]_{F(\RedText{p_1(a)})}&
t_2A_2\ar[d]^{r_2}\ar[rruu]_{G(\RedText{t_1(a)})}\\
A_1\ar[r]_{t_1}\ar[u]^{p_1}\ar@{}[ur]|{(1)}\ar@/_2pc/@{=>}[drrr]^f&
B_1\ar@{=>}[drrrrr]^(.4)g&&&
A_2\ar[r]_{t_2}\ar[u]^{p_2}\ar@{}[ur]|{(2)}\ar[ddll]^{f(a)}&
B_2\ar[ddrr]^{g(\RedText{t_1(a)})}\\
&&&&&&&\\
&&A_2\ar[uuuuu]^{p_2}\ar[rrrrr]_{t_2}\ar@{}[urrrrr]|{(3)}&&&&&B_2
}
\]
}
{decompositions of morphism of representations, diagram}

\DefEquation
{
t_i=r_i\circ q_i\circ p_i
}
{morphism of representations of algebra, homomorphism, 1}

\DefEq
{
\[
\RedText{p_1(a)}=a^{\mathrm{ker}\,t_1}
\]
}
{morphism of representations of algebra, p1=}

\DefEq
{
\[
\BlueText{p_2(a)}=a^{\mathrm{ker}\,t_2}
\]
}
{morphism of representations of algebra, p2=}

\DefEquation
{
q_1(\RedText{p_1(a)})=\RedText{t_1(a)}
}
{morphism of representations of algebra, q1=}

\DefEquation
{
q_2(\BlueText{p_2(a)})=\BlueText{t_2(a)}
}
{morphism of representations of algebra, q2=}

\DefEq
{
\[
r_1:\RedText{t_1(a)}\in f(A_1)\rightarrow \RedText{t_1(a)}\in B_1
\]
}
{morphism of representations of algebra, r1=}

\DefEq
{
\[
r_2:\BlueText{t_2(a)}\in f(A_2)\rightarrow \BlueText{t_2(a)}\in B_2
\]
}
{morphism of representations of algebra, r2=}

\DefEquation
{
(t_1,t_2)
=
(r_1,r_2)
\circ
(q_1,q_2)
\circ
(p_1,p_2)
}
{decompositions of morphism of representations}

\DefEq
{
\[
\xymatrix
{
A/\mathrm{ker}\,f\ar[r]^q&f(A)\ar[d]_r
\\
A\ar[u]^p\ar[r]^f&B
}
\ \ \ f=p\circ q\circ r
\]
}
{decomposition of map f}

\DefEq
{
\[
q:p(a)\in A/\mathrm{ker}\,f\rightarrow f(a)\in f(A)
\]
}
{q:A/ker->f(A)}

\DefEq
{
\[
r:f(a)\in f(A)\rightarrow f(a)\in B
\]
}
{r:f(A)->B}

\DefEq
{
$r_1=q_1\circ p_1$, $r_2=q_2\circ p_2$.
}
{r=q*p}

\DefEq
{
\[
r_2:A_2\rightarrow A_2
\]
}
{R:A2->A2}

\AddEq{show closure operator, representation}
{
$\ShowSymbol{subrepresentation generated by set}1$
}

\DefEq
{
\symb{J[f,X]}{subrepresentation generated by set}1
}
{subrepresentation generated by set}

\AddEq{generating set of representation}
{
$J[f,X]=A_2$.
}

\DefEq
{
\[
\xymatrix
{
f_{B_2}:A_1\ar[r]|{*}&B_2
}
\]
}
{representation of algebra A in algebra B}

\DefEq
{
$f_{B_2}(a)=f(a)|_{B_2}$.
}
{fB2(a)=}

\AddEq [3]{r12:A->B}
{
\[
(#1_1:#2_1\rightarrow #3_1,\ #1_2:#2_2\rightarrow #3_2)
\]
}

\DefEq
{
$\BlueText{r_2(f(a)(m))}$.
}
{r2(f(a,m))}

\DefEq
{
$\BlueText{r_2(m)}\in B_2$
}
{r2(m)in B2}

\DefEq
{
$g(\RedText{r_1(a)})$
}
{g(r1(a))}

\DefEq
{
\[
\RedText{r_1(a)}=r_1(a)
\]
}
{r1(a)=r1(a)}

\DefEq
{
\BlueText{r_2(f(a)(m))}=g(\RedText{r_1(a)})(\BlueText{r_2(m)})
}
{morphism of representations of universal algebra, 2m}

\DefEq
{
\xymatrix{
&A_2\ar[dd]_(.3){f(a)}\ar[rr]^{r_2}&&B_2\ar[dd]^(.3){g(\RedText{r_1(a)})}\\
&\ar @{}[rr]|{(1)}&&\\
&A_2\ar[rr]^{r_2}&&B_2\\
A_1\ar[rr]^{r_1}\ar@{=>}[uur]^(.3)f&&B_1\ar@{=>}[uur]^(.3)g
}
}
{morphism of representations of universal algebra, 2m 1}

\DefEquation
{
\xymatrix{
A_2\ar[rr]^{r_2}&&B_2\\
A_1\ar[rr]^{r_1}\ar[u]^f|{*}&&B_1\ar[u]^g|{*}
}
}
{morphism of representations of universal algebra, definition, 2m 2}

\DefEq
{
\[
a\pC i0xa\pC i1
\]
}
{Sum over repeated index}

\DefEq
{
$\Omega=\{\omega\}$.
}
{Omega=omega}

\DefEq
{
\[ab\omega=ab\]
}
{abo=ab}

\DefEq
{
\[ab\omega=a+b\]
}
{abo=a+b}

\DefEq
{
\[f:A_1\times...\times A_n\rightarrow S\]
}
{polylinear map of algebras}

\DefEq
{
\symb{\mathcal L(D;A_1\times...\times A_n\rightarrow S)}{set polylinear maps}1
}
{set polylinear maps}

\DefEquation
{
(A_1\otimes A_2)\otimes A_3=A_1\otimes(A_2\otimes A_3)
=A_1\otimes A_2\otimes A_3
}
{A1xA2xA3}

\DefEq
{
\[
(v_1, ..., v_n)\in V_1\times...\times V_n
\rightarrow v_1\otimes...\otimes v_n\in V_1\otimes...\otimes V_n
\]
}
{V times->V otimes}

\DefEquation
{
\begin{split}
&\,a_1\otimes...\otimes(a_i+b_i)\otimes...\otimes a_n
\\
=&\,a_1\otimes...\otimes a_i\otimes...\otimes a_n+
a_1\otimes...\otimes b_i\otimes...\otimes a_n
\\
&\,a_i, b_i\in A_i
\end{split}
}
{tensors 1, tensor product}

\DefEquation
{
\begin{split}
&a_1\otimes...\otimes(ca_i)\otimes...\otimes a_n=
c(a_1\otimes...\otimes a_i\otimes...\otimes a_n)
\\
&a_i\in A_i\ \ \ c\in D
\end{split}
}
{tensors 2, tensor product}

\DefEq
{
\[f\circ 0=0\]
}
{linear map, 0, D algebra}

\AddEq [3]{f:A->B}
{
\[#1:#2\rightarrow #3\]
}

\DefEquation
{
f:A^n\rightarrow A,
a=f\circ(a_1,...,a_n)
}
{polylinear map, algebra}

\DefEquation
{
a=f\pC{s}{0}^n\ \sigma_s(I_{(s\cdot 1)}\circ a_1)
\ f\pC{s}{1}^n\ ...\ \sigma_s(I_{(s\cdot n)}\circ a_n)\ f\pC{s}{n}^n
}
{polylinear map, algebra, canonical morphism}

\DefEq
{
$\{a_1,...,a_n\}$
\[
\sigma_s=
\begin{pmatrix}
a_1&...&a_n
\\
\sigma_s(a_1)&...&\sigma_s(a_n)
\end{pmatrix}
\]
}
{transposition of set of variables, algebra}

\DefEq
{
$I_{(s\cdot 1)}$, ..., $I_{(s\cdot n)}\in\mathcal L(D;A\rightarrow A)$
}
{I1n in L(A;A)}

\DefEq
{
\labelItem{monomial of power 0}
$p_0(x)=a_0$, $a_0\in A$.
}
{p0(x)=a0}

\AddEq{monomial of power k}
{
\labelItem{monomial of power k}
\[
p_k(x)=p_{k-1}(x)(F\circ x)a_k
\]
}

\AddEq{p1(x)=aFxa}
{
$p_1(x)=a_0(F\circ x)a_1$.
}

\AddEq{Basis F=...}
{
$\Basis F=(F_1,...,F_n)$.
}

\AddEq{Fi=Fj}
{
$F_{(i)}=F_{(j)}$
}

\AddEq{F=...}
{
$F=(F_{(1)},...,F_{(k)})$
}

\AddEq{module of homogeneous polynomials over algebra}
{
\symb{A_k[x]}{module of homogeneous polynomials over algebra}1
}

\DefEq
{
$a\in A^{\otimes (n+1)}$\Pt
}
{a in Aoxn+1}

\DefEq
{
$x_1=...=x_n=x$,
}
{x1=xn=x}

\DefEq
{
\[
a\circ F\circ x^n=a\circ(F_{(1)},...,F_{(n)})\circ(x_1\otimes...\otimes x_n)
\]
}
{a xn=}

\AddEq{F1n in F}
{
$F_{(1)}$, ..., $F_{(n)}\in\Basis F$.
}

\AddEq{F[k]=...}
{
\[
F_{[k]}=(F_{[k](1)},...,F_{[k](n)})
\]
}

\DefEq
{
\[
a=a_{i\cdot 0}\otimes a_{i\cdot 1}\otimes...\otimes a_{i\cdot n}
\ \ \ i\in I
\]
}
{a=oxi}

\DefEq
{
$\sigma=\{\sigma_i\in S(n):i\in I\}$
}
{si in Sn}

\DefEq
{
\[
(a,\sigma):A^{\times n}\rightarrow A
\]
}
{ox:An->A}

\DefEq
{
\begin{align*}
(a,\sigma)\circ (b_1,...,b_n)&=
(a_{i\cdot 0}\otimes a_{i\cdot 1}\otimes...\otimes a_{i\cdot n},\sigma_i)\circ (b_1,...,b_n)
\\&=
a_{i\cdot 0}\sigma_i(b_1)a_{i\cdot 1}...\sigma_i(b_n)a_{i\cdot n}
\end{align*}
}
{ox circ =}

\AddEq{p(x)=a circ xk}
{
\[
\begin{matrix}
p(x)=a_{[s]}\circ F_{[s]}\circ x^k
&a_{[s]}\in A^{\otimes(k+1)}
\end{matrix}
\]
}

\DefEq
{
\symb{A[x]}{algebra of polynomials over algebra}{}
\[
\ShowSymbol{algebra of polynomials over algebra}{}
=\bigoplus_{n=0}^{\infty}A_n[x]
\]
}
{algebra of polynomials over algebra}

\DefEquation
{
}
{derivative in L = +o}

\AddEq[3]{a1n}
{
$#1_1$, ..., $#1_{#2}$#3
}

\AddEq[3]{A1n}
{
$#1^1$, ..., $#1^{#2}$#3
}

\DefEq
{
$\epsilon\in R$, $\epsilon>0$,
}
{epsilon in R}

\DefEq
{
$a_0\in U$.
}
{a0 in U}

\DefEq
{
\[
f:U\rightarrow \mathcal L(D;A^p\rightarrow B)
\]
}
{U->L(Ap,B)}

\DefEq
{
$\partial_x f(x)\in CE$.
}
{dx fx in CE}

\DefEquation
{
\begin{array}{r@{\ }l}
&((a_0,...,a_n,\sigma)\circ(f_1,...,f_n))\circ(x_1,...,x_n)
\\=&
(a_0\sigma(f_1)a_1...a_{n-1}\sigma(f_n)a_n)\circ(x_1,...,x_n)
\\=&
a_0\sigma(f_1\circ x_1)a_1...a_{n-1}\sigma(f_n\circ x_n)a_n
\end{array}
}
{n linear map A LA}

\DefEq
{
\symb{\mathcal L(D;A^n\rightarrow S)}{set polylinear maps An}1
}
{set polylinear maps An}

\DefEq
{
\begin{align*}
f\circ(
a_1, ...,
a_i+ b_i, ...,
a_n)
&=
f\circ(
a_1, ...,
a_i, ...,
a_n)
+
f\circ(
a_1, ...,
b_i, ...,
a_n)
\\
f\circ(
a_1, ...,
pa_i, ...,
a_n)
&=
pf\circ(
a_1, ...,
a_i, ...,
a_n)
\end{align*}
\[
\begin{matrix}
1\le i\le n
&
a_i, b_i \in A_i
&
p\in D
\end{matrix}
\]
}
{polylinear map of algebras, 1}

\DefEquation
{
r_2\circ\BlueText{f(a)}=g(\RedText{r_1(a)})\circ r_2
}
{morphism of representations of universal algebra, definition, 2}

\DefEq
{
\symb{A\cong B}{isomorphic}1.
}
{isomorphic}

\DefEquation
{
f(a_1)...f(a_n)\omega=f(a_1...a_n\omega)
}
{afo=aof}

\DefEquation
{
\begin{array}{r@{\,}l@{\ \ \ }r@{\,}l@{\ \ \ }r@{\,}l}
a_k&=a_{k\cdot 0}\underline{\otimes}(1\otimes a_{k\cdot 1})
&a_{k\cdot 0}&\in A^{\otimes k}&a_{k\cdot 1}&\in A
\end{array}
}
{p(x)=a k-1 k 3}

\AddEq [2]{M M1 Mn}
{
\[#1=\max(#1_1,...,#1_{#2})\]
}

\DefEq
{
\[N=\max(N_1,N_2)\]
}
{N N1 N2}

\DefEq
{
f_i(x)=\lim_{m\rightarrow\infty}f_{i\cdot m}(x)
}
{fi(x)=lim}

\DefEq
{
\delta_1\in R,\ \delta_1>0
}
{delta1 in R}

\DefEq
{
\(m>M_i\)
}
{m>Mi}

\DefEq
{
\[h_m=f_{1\cdot m}...f_{n\cdot m}\omega\]
}
{hm=f1m...fnm omega}

\DefEq
{
\[
\epsilon_1(\delta)<\epsilon
\]
}
{e(d)<e}

\DefEq
{
\[\delta_1=0\Rightarrow\epsilon_1=0\]
}
{d1=0=>e1=0}

\DefEq
{
F_i\ge 0
}
{F>0}

\DefEq
{
F_i=\sup\|f_i(x)\|
}
{F=sup|f|}

\AddEq[3]{a in AA}
{
$#1\in #2\otimes #2$#3
}

%% file: Preliminary.Ref.tex

\ifx\UseRussian\Defined
\def\TheoremFollows{Теорема является следствием теоремы }
\else
\def\TheoremFollows{The theorem follows from the theorem }
\fi

\ePrints{4975-6381,9835-2163,0767-8264,BQuater,7287-9339,9856-6693}%
\ifx\Semafor\ValueOn%
\def\RefLinearMap{5114-6019}%
\xRefDef{5114-6019}
\fi
\ePrints{1601.03259,1506.00061}%
\Items{MQuater,309618526,CACAA.06.121,1801.01628,CACAA.07}%
\ifx\Semafor\ValueOn%
\def\RefLinearMap{1502.04063}%
\xRefDef{1502.04063}
\fi
\ePrints{}%
\ifx\Semafor\ValueOn%
\def\RefLinearMap{MVector}%
\xRefDef{MVector}
\fi

\ePrints{330532713,1801.01628}
\ifx\Semafor\ValueOn
\def\RefQuadratic{1506.00061}%
\xRefDef{1506.00061}
\fi%
\ePrints{0921-2370,0767-8264}
\ifx\Semafor\ValueOn
\def\RefQuadratic{7287-9339}%
\xRefDef{7287-9339}
\fi

\ePrints{330532713}
\ifx\Semafor\ValueOn
\def\RefDiffEq{1801.01628}%
\xRefDef{1801.01628}
\fi

\ePrints{309618526,1801.01628,322019412,323236904,CACAA.07,330532713,323966352,MQuater}
\ifx\Semafor\ValueOn
\def\RefCalculus{1601.03259}%
\xRefDef{1601.03259}
\fi%
\ePrints{9835-2163,0767-8264,9856-6693,7287-9339,0921-2370,BQuater}
\ifx\Semafor\ValueOn
\def\RefCalculus{4975-6381}%
\xRefDef{4975-6381}
\fi

\ePrints{1908.04418,323966352,324329551,0921-2370}
\ifx\Semafor\ValueOn
\def\RefGravity{0803.3276}%
\xRefDef{0803.3276}
\fi%
\ePrints{6860-2955}
\ifx\Semafor\ValueOn
\def\RefGravity{0803.3276}%
\xRefDef{0803.3276}
\fi

\AddEq{SetupRefRepresentation}
{
\ifx\Preliminary\ValueOn%
\ePrints{1506.00061,1601.03259}%
\ifx\Semafor\ValueOn%
\def\RefRepresentation{1502.04063}%
\xRefDef{1502.04063}
\fi
\ePrints{5148-4632,BQuater}
\ifx\Semafor\ValueOn
\def\RefRepresentation{6860-2955}%
\xRefDef{6860-2955}
\fi
\ePrints{4975-6381}
\ifx\Semafor\ValueOn
\def\RefRepresentation{5114-6019}%
\fi
\else
\def\RefRepresentation{}%
\ePrints{5114-6019,1502.04063}
\ifx\Semafor\ValueOn
\def\RefRepresentation{0912.3315}%
\xRefDef{0912.3315}
\fi
\ePrints{MQuater}
\ifx\Semafor\ValueOn
\def\RefRepresentation{1908.04418}%
\xRefDef{1908.04418}
\fi
\ePrints{1801.01628,0767-8264}
\ifx\Semafor\ValueOn
\def\RefRepresentation{1111.6035}%
\xRefDef{1111.6035}
\fi
\ePrints{MVector}
\ifx\Semafor\ValueOn
\def\RefRepresentation{1908.04418}%
\xRefDef{1908.04418}
\fi
\fi
}

\AddEq{SetupRefPolymorphism}
{
\ifx\Preliminary\ValueOn%
\ePrints{9835-2163,9856-6693,0767-8264}%
\ifx\Semafor\ValueOn%
\def\RefPolymorphism{5114-6019}%
\fi
\else
\def\RefPolymorphism{}%
\ePrints{1801.01628,CACAA.07}%
\ifx\Semafor\ValueOn
\def\RefPolymorphism{1502.04063}%
\fi
\ePrints{MVector,323236904}%
\ifx\Semafor\ValueOn
\def\RefPolymorphism{1502.04063}%
\xRefDef{1502.04063}
\fi
\ePrints{9856-6693}%
\ifx\Semafor\ValueOn
\def\RefPolymorphism{5114-6019}%
\fi
\fi
}

\AddEq{SetupRefOmegaNorm}
{
\ifx\Preliminary\ValueOn%
\ePrints{5410-9916,4975-6381}%
\ifx\Semafor\ValueOn%
\def\RefTheoremOmegaNorm{5059-9176}
\xRefDef{5059-9176}
\fi
\else
\ePrints{1102.5168,1502.04063}%
\ifx\Semafor\ValueOn%
\def\RefTheoremOmegaNorm{1305.4547}
\xRefDef{1305.4547}
\fi
\ePrints{CACAA.06.121}%
\ifx\Semafor\ValueOn%
\def\RefTheoremOmegaNorm{CACAA.04.001}
\xRefDef{CACAA.04.001}
\fi
\fi
}

\AddEq{SetupRefMeasure}
{
\ifx\Preliminary\ValueOn%
\ePrints{1601.03259,309618526}
\ifx\Semafor\ValueOn
\def\RefMeasure{1310.5591}
\xRefDef{1310.5591}
\fi
\ePrints{4975-6381}
\ifx\Semafor\ValueOn
\def\RefMeasure{5410-9916}
\xRefDef{5410-9916}
\fi
\ePrints{CACAA.06.121}%
\ifx\Semafor\ValueOn%
\def\RefMeasure{CACAA.04.001}
\fi
\else
\def\RefMeasure{}
\ePrints{9856-6693}
\ifx\Semafor\ValueOn
\def\RefMeasure{5410-9916}
\xRefDef{5410-9916}
\fi
\ePrints{323236904}
\ifx\Semafor\ValueOn
\def\RefMeasure{1310.5591}
\xRefDef{1310.5591}
\fi
\fi
}

\AddEq{SetupRef}
{
\ShowEq{SetupRefRepresentation}
\ShowEq{SetupRefOmegaNorm}
\ShowEq{SetupRefMeasure}
\ShowEq{SetupRefPolymorphism}
}

\ShowEq{SetupRef}

\newcommand\ProofTheorem[2]
{
\begin{proof}
\TheoremFollows
\RefTheorem[#1]{#2}.
\end{proof}
}%


\DefEq
{
\ePrints{1601.03259,4975-6381}%
\ifx\Semafor\ValueOff%
\EqRef[\RefCalculus]{derivative of Order n, algebra},
\else%
\EqRef{derivative of Order n, algebra},
\fi%
}
{ref derivative of Order n, algebra}

\DefEq
{
\ePrints{1310.5591,5059-9176}%
\ifx\Semafor\ValueOff%
\ePrints{5410-9916}%
\ifx\Semafor\ValueOn%
\RefDefinition{representation of algebra}\Pt
\else
\RefDefinition{left-side representation of algebra}\Pt
\fi
\else
\RefDefinition[0912.3315]{left-side representation of algebra}\Pt
\fi
}
{ref definition: left-side representation of algebra}

\DefEq
{
\ePrints{1310.5591}
\ifx\Semafor\ValueOff
\RefDefinition{morphism of representations of universal algebra}.
\else
\RefDefinition[0912.3315]{morphism of representations of universal algebra}.
\fi
}
{ref definition: morphism of representations of universal algebra}

\DefEq
{
\ePrints{1310.5591}%
\ifx\Semafor\ValueOff%
\RefDefinition{closed ball}
\else%
\RefDefinition[1305.4547]{closed ball}
\fi%
}
{ref definition: closed ball}

\DefEq
{
\ePrints{1310.5591}%
\ifx\Semafor\ValueOff%
\RefDefinition{open ball}\Pt
\else%
\RefDefinition[1305.4547]{open ball}\Pt
\fi%
}
{ref definition: open ball}

\DefEq
{
\ePrints{1310.5591}%
\ifx\Semafor\ValueOff%
\RefDefinition{open set},
\else%
\RefDefinition[1305.4547]{open set},
\fi%
}
{ref definition: open set}

\DefEq
{
\ePrints{1310.5591}%
\ifx\Semafor\ValueOff%
\RefDefinition{fundamental sequence}\Pt
\else%
\RefDefinition[1305.4547]{fundamental sequence}\Pt
\fi%
}
{ref definition: fundamental sequence}

\DefEq
{
\ePrints{1310.5591}%
\ifx\Semafor\ValueOff%
\RefDefinition{sequence converges uniformly}\Pt
\else%
\RefDefinition[1305.4547]{sequence converges uniformly}\Pt
\fi%
}
{ref definition: sequence converges uniformly}

\DefEq
{
\ePrints{1310.5591}%
\ifx\Semafor\ValueOff%
\RefDefinition{compact set},
\else%
\RefDefinition[1305.4547]{compact set},
\fi%
}
{ref definition: compact set}

\DefEq
{
\ePrints{1310.5591}%
\ifx\Semafor\ValueOff%
\RefDefinition{Omega group},
\else%
\RefDefinition[1305.4547]{Omega group},
\fi%
}
{ref definition: Omega group}

\DefEq
{
\ePrints{1310.5591}%
\ifx\Semafor\ValueOff%
\RefItem{|a|>=0}
\else%
\RefItem[1305.4547]{|a|>=0}
\fi%
}
{ref item |a|>=0}

\DefEq
{
\ePrints{1310.5591}%
\ifx\Semafor\ValueOff%
\RefItem{|a|=0}.
\else%
\RefItem[1305.4547]{|a|=0}.
\fi%
}
{ref item |a|=0}

\DefEq
{
\ePrints{1310.5591}%
\ifx\Semafor\ValueOff%
\RefItem{|a+b|<=|a|+|b|}\Pt
\else%
\RefItem[1305.4547]{|a+b|<=|a|+|b|}\Pt
\fi%
}
{ref item |a+b|<=|a|+|b|}

\DefEq
{
\ePrints{1310.5591}%
\ifx\Semafor\ValueOff%
\EqRef{|a omega|<|omega||a|1n}.
\else%
\EqRef[1305.4547]{|a omega|<|omega||a|1n}.
\fi%
}
{ref EqRef |a omega|<|omega||a|1n}

\DefEq
{
\ePrints{1310.5591}%
\ifx\Semafor\ValueOff%
\EqRef{|a-b|>|a|-|b|},
\else%
\EqRef[1305.4547]{|a-b|>|a|-|b|},
\fi%
}
{ref EqRef |a-b|>|a|-|b|}

\DefEq
{
\ePrints{1310.5591}%
\ifx\Semafor\ValueOff%
\EqRef{|fab|<|f||a||b|}.
\else%
\EqRef[1305.4547]{|fab|<|f||a||b|}.
\fi%
}
{ref |fab|<|f||a||b|}

\AddEq{ref tensor product and polylinear map}
{
\RefTheorem[\RefLinearMap]{there exists tensor product of modules},
\RefTheorem[\RefLinearMap]{tensor product and polylinear map}.
}

\DefEq
{
\ePrints{1502.04063,5114-6019}
\ifx\Semafor\ValueOff
\RefDefinition{reduced morphism of representations},
\else
\xRef[0912.3315]{remark: reduced morphism of representations},
\fi
}
{ref reduced morphism of representations}

\DefEq
{
\RefTheorem[1202.6021]{expand linear mapping, quaternion}.
}
{ref expand linear mapping, quaternion}

\DefEq
{
\RefTheorem[0912.4061]{ax-xa=1 quaternion algebra}.
}
{ref ax-xa=1 quaternion algebra}

\AddEq{ref theorem derivative, representation in algebra}
{
\ePrints{9835-2163,0767-8264}
\ifx\Semafor\ValueOn
\RefTheorem{derivative, representation in algebra},
\else
\RefTheorem[\RefCalculus]{derivative, representation in algebra},
\fi
}

\AddEq{ref partial derivatives equal}
{
\ePrints{9835-2163,0767-8264}
\ifx\Semafor\ValueOn
\EqRef{partial derivatives equal, Direct Sum of Banach Modules},
\else
\ePrints{CACAA.07}
\ifx\Semafor\ValueOff
\EqRef[0812.4763]{Gateaux partial derivatives equal, D vector space},
\else
\EqRef[CACAA.05.001]{partial derivatives equal, D vector space},
\fi
\fi
\EqRef{dM/dy=dN/dx yx},
\EqRef{d int f=f}.
}

\AddEq{ref aE, quaternion, Jacobian matrix}
{
\ePrints{CACAA.07}
\ifx\Semafor\ValueOn
\RefTheorem[CACAA.01.109]{aE, quaternion, Jacobian matrix},
\else
\RefTheorem[1107.1139]{aE, quaternion, Jacobian matrix},
\fi
}

\AddEq{ref Ea, quaternion, Jacobian matrix}
{
\ePrints{CACAA.07}
\ifx\Semafor\ValueOn
\RefTheorem[CACAA.01.109]{Ea, quaternion, Jacobian matrix},
\else
\RefTheorem[1107.1139]{Ea, quaternion, Jacobian matrix},
\fi
}

\AddEq[1]{ref additive map of complex field}
{
\ePrints{CACAA.07}
\ifx\Semafor\ValueOn
\RefTheorem[CACAA.06.121]{linear map of complex field}#1
\else
\RefTheorem[1003.1544]{additive map of complex field, structure}#1
\fi
}

\DefEq
{
\RefTheorem[1003.1544]{quaternion conjugation}.
}
{ref quaternion conjugation}

\DefEq
{
\RefTheorem[1003.1544]{Quaternion over real field, matrix}.
}
{ref Quaternion over real field, matrix}

\DefEq
{
\RefTheorem[1601.03259]{representation of derivative, algebra A->B}.
}
{ref differentiable map A->B}

\DefEq
{
\RefTheorem[1601.03259]{map is continuous, derivative}.
}
{ref map is continuous, derivative}

\DefEq
{
\RefTheorem[1302.7204]{monomial of power k}.
}
{ref monomial of power k}

\DefEq
{
\RefTheorem[1302.7204]{map A(k+1)->pk otimes is polylinear map}.
}
{ref map A(k+1)->pk otimes is polylinear map}

\DefEq
{
\RefTheorem[1302.7204]{r=+q circ p}.
}
{ref r=+q circ p}

\DefEq
{
\RefTheorem[1302.7204]{r=+q circ p 1}.
}
{ref r=+q circ p 1}

\DefEq
{
\RefTheorem[1105.4307]{Re Im A->F}.
}
{ref Re Im A->F}

\DefEq
{
\RefTheorem[1105.4307]{conjugation in algebra}.
}
{ref conjugation in algebra}

\DefEq
{
\RefTheorem[1105.4307]{structural constant of algebra with unit}.
}
{ref structural constant of algebra with unit}

%% file: Preliminary.Parm.tex

\AddEq{=DA1DA2}
{
\def\DFDT{D1 D2 }%
\def\MF{r1:D1->D2 }%
\def\DF{1}%
\def\DT{2}%
\def\VF{1}%
\def\VT{2}%
\def\MapE{hgf}%
}

\AddEq{=D1D2}
{
\def\DFDT{D1 D2 }%
\def\MF{r1:D1->D2 }%
\def\DF{1}%
\def\DT{2}%
\def\VF{1}%
\def\VT{2}%
\def\MapE{hf}%
}

\AddEq{=DD}
{
\def\DFDT{D }%
\def\MF{}%
\def\DF{}%
\def\DT{}%
\def\VF{1}%
\def\VT{2}%
\def\MapE{f}%
}

\AddEq{=left}
{
\def\SideWS{left }%
\def\SideNS{left}%
\def\HSide{\Hyph side }%
\def\SideA{A*}%
\def\CBase{D}%
\def\Base{A}%
\def\Module{V}%
\def\BaseRings{of commutative ring $D$ and $D$\Hyph algebra $A$ }
\def\Algebra{associative division $D$\Hyph algebra}%
\def\algebra{associative $D$\Hyph algebra}%
\def\algebraa{$D$\Hyph algebra }%
\def\ATransf{g_{3,4}}%
\def\DTransf{g_{1,4}}%
\def\DArg{d}%
\ifx\UseRussian\Defined%
\def\SideRu{ле}%
\def\SideRuC{Ле}%
\def\HSide{востороннее }%
\def\From{вого }%
\def\To{вый }%
\def\ToV{вое }%
\def\What{вым }%
\def\Which{вом }
\def\BaseRings{коммутативного кольца $D$ и $D$\Hyph алгебры $A$ }
\def\Algebra{ассоциативная $D$\Hyph алгебра с делением}%
\def\algebra{ассоциативная $D$\Hyph алгебра}%
\def\algebraa{$D$\Hyph алгебре }%
\def\algebrab{$D$\Hyph алгебры }%
\def\algebrac{$D$\Hyph алгебра }%
\def\algebrad{$D$\Hyph алгебру }%
\def\algebraD{$D$\Hyph алгебра }%
\fi%
}

\AddEq{=right}
{
\def\SideWS{right }%
\def\SideNS{right}%
\def\HSide{\Hyph side }%
\def\SideA{*A}%
\def\CBase{D}%
\def\Base{A}%
\def\Module{V}%
\def\BaseRings{of commutative ring $D$ and $D$\Hyph algebra $A$ }
\def\Algebra{associative division $D$\Hyph algebra}%
\def\algebra{associative $D$\Hyph algebra}%
\def\algebraa{$D$\Hyph algebra }%
\def\ATransf{g_{3,4}}%
\def\DTransf{g_{1,4}}%
\def\DArg{d}%
\ifx\UseRussian\Defined%
\def\SideRu{пра}%
\def\SideRuC{Пра}%
\def\HSide{востороннее }%
\def\From{вого }%
\def\To{вый }%
\def\ToV{вое }%
\def\What{вым }%
\def\Which{вом }
\def\BaseRings{коммутативного кольца $D$ и $D$\Hyph алгебры $A$ }
\def\Algebra{ассоциативная $D$\Hyph алгебра с делением}%
\def\algebra{ассоциативная $D$\Hyph алгебра}%
\def\algebraa{$D$\Hyph алгебре }%
\def\algebrab{$D$\Hyph алгебры }%
\def\algebrac{$D$\Hyph алгебра }%
\def\algebrad{$D$\Hyph алгебру }%
\def\algebraD{$D$\Hyph алгебра }%
\fi%
}

\AddEq{=module}
{
\def\SideWS{}%
\def\SideNS{}%
\def\HSide{}%
\def\SideA{D}%
\def\CBase{Z}%
\def\Base{D}%
\def\Module{A}%
\def\BaseRings{of ring of rational integers $Z$ and commutative ring $D$ }
\def\Algebra{field}%
\def\algebra{ring}%
\def\algebraa{ring }%
\def\Algebrab{module}%
\def\Algebrac{module}%
\def\ATransf{g_2}%
\def\DTransf{g_1}%
\def\DArg{n}%
\ifx\UseRussian\Defined%
\def\SideRu{}%
\def\SideRuC{}%
\def\From{}%
\def\To{}%
\def\ToV{}%
\def\What{}%
\def\Which{}
\def\WhatA{ый }%
\def\BaseRings{кольца целых чисел $Z$ и коммутативного кольца $D$ }
\def\Algebra{поле}%
\def\algebra{кольцо}%
\def\algebraa{кольце }%
\def\algebrab{кольца }%
\def\algebrac{кольцо }%
\def\algebrad{кольцо }%
\def\algebraD{Кольцо }%
\def\Algebrab{модуль}%
\fi%
}

\AddEq{=Omega}
{
\def\SideA{\Omega}%
\def\Algebrab{group}%
\def\Algebrac{Omega group}%
\def\Module{A}%
\ifx\UseRussian\Defined%
\def\WhatA{ая }%
\def\Algebrab{группа}%
\fi%
}

\DefEq%
{%
\def\Pt{.}%
}%
{=.}%

\DefEq%
{%
\def\Pt{;}%
}%
{=.c}%

\DefEq%
{%
\def\Pt{,}%
}%
{=c}%

\DefEq%
{%
\def\Pt{}%
}%
{=z}%

\DefEq
{
\def\PD{}%
\def\PF{}%
\def\PA{}%
}
{D= F= A=}

\DefEq%
{%
\def\b{a}%
\def\B{A}%
}%
{B=A}%

\DefEq%
{%
\def\b{b}%
\def\B{B}%
}%
{B=B}%

\DefEq
{
\def\PX{X}
}
{X}

\AddEq{f=f}
{%
\def\Pf{f}%
}%

\AddEq{f=g}
{%
\def\Pf{g}%
}%

\DefEq%
{%
\def\Pf{m}%
}%
{f=m}%

\DefEq
{
\def\Pf{I}%
}
{f=I}

\DefEq
{
\def\pD{D}%
\def\pA{A}%
\def\pB{A}%
}
{A=A}

\DefEq%
{%
\def\pD{D}%
\def\pA{A}%
\def\pB{B}%
}%
{A=AB}

\DefEq
{
\def\pD{D}%
\def\pA{B}%
\def\pB{C}%
}
{A=BC}

\DefEq%
{%
\def\pD{R}%
\def\pA{C}%
\def\pB{C}%
}%
{A=C}%

\DefEq%
{%
\def\pD{C}%
\def\pA{C}%
\def\pB{C}%
}%
{A=CC}%

\DefEq
{
\def\pD{D}%
\def\pA{A_1}%
\def\pB{A_2}%
}
{A=12}

\DefEq
{
\def\pD{D}%
\def\pA{A_2}%
\def\pB{A_2}%
}
{A=2}

\DefEq
{
\def\pD{D}%
\def\pA{A_1\rightarrow A_2}%
\def\pB{A_3}%
}
{A=123}

\DefEq
{
\def\pD{D}%
\def\pA{A_1}%
\def\pB{A_3}%
}
{A=13}

\DefEq
{
\def\pD{D}%
\def\pA{A}%
\def\pB{C}%
}
{A=AC}

\DefEq
{%
\def\Pn{0}%
}%
{n=0}

\DefEq
{%
\def\Pn{n}%
}%
{n=n}

\DefEq
{%
\def\Pn{k}%
}%
{n=k}

\DefEq
{%
\def\Pn{1}%
\def\Pp{p}%
}%
{n=1}

\DefEq
{%
\def\Pn{2}%
\def\Pp{q}%
}%
{n=2}

\DefEq
{%
\def\Pn{3}%
\def\Pp{r}%
}%
{n=3}

\DefEq
{%
\def\tM{1}%
}%
{m=1}

\DefEq
{%
\def\tM{2}%
}%
{m=2}

%% file: Stmt.Module.English.tex
\input{Stmt.Module.Eq}

\DefExample{Abelian Group is Z module}
{
From the theorem
\RefTheorem{action of ring of rational integers in Abelian group}
and the definition
\RefDefinition{module over commutative ring},
it follows that Abelian group $G$ is module over ring of integers $Z$.
}

\DefEq
{
\begin{definition}
\labelDefinition{module over ring}
Let commutative ring $D$ has unit $1$.
Effective representation
\DrawEq{D->*V}{def}
of ring $D$
in an Abelian group $V$
is called
\AddIndex{module over ring}{module over ring} $D$
or
\AddIndex{$D$\Hyph module}{D module}.
Effective representation
\eqRef{D->*V}{def}
of commutative ring $D$ in an Abelian group $V$
is called
\AddIndex{vector space over field}{vector space over field} $D$
or
\AddIndex{$D$\Hyph vector space}{D vector space}.
\qed
\end{definition}
}
{... definition: module over ring}

\DefTheorem{tensor product of D-algebras is D-algebra}
{
Let
\ShowEq{A1...n}
be $D$\Hyph algebras.
Tensor product
$\Tensor A$
of $D$\Hyph modules
\ShowEq{A1...n}
is $D$\Hyph algebra,
if we define product by the equality
\ShowEq{xA1n*xA1n->oxA1n=}
}

\DefTheorem{representation of algebra A2 in LA}
{
Let $A$ be $D$\Hyph algebra.
Let product in algebra
\ShowEq{AoxA}A{}
be defined according to rule
\DrawEq{product in algebra AA}{}
A representation
\ShowEq{h:AoxA->L(A)}
of $D$\Hyph algebra
\ShowEq{AoxA}A{}
in module
\ShowEq{L(A;B)}DAA{}
defined by the equality
\ShowEq{representation AA in LA}
allows us to identify tensor
\ShowEq{d in AxoA}
and linear map
\ShowEq{product in algebra AA 2}
where
\ShowEq{product in algebra AA 3}
is identity map.
Linear map
\ShowEq{a ox b}
has form
\ShowEq{a ox b c=}
}

\DefDefinition{module over commutative ring}
{
Effective representation of commutative ring $D$
in an Abelian group $V$
\DrawEq{D->*V}{module}
is called
\AddIndex{module over ring}{module over ring} $D$
or
\AddIndex{$D$\Hyph module}{D module}.
$V$\Hyph number is called
\AddIndex{vector}{vector}.
\ePrints{309618526,CACAA.06.121}
\ifx\Semafor\ValueOn
If $D$ is a field, then the Abelian group $V$
is called
\AddIndex{vector space}{vector space} over field $D$
or
\AddIndex{$D$\Hyph vector space}{D vector space}.
\fi
}

\DefDefinition{direct sum of D modules}
{
Coproduct in category of $D$\Hyph modules is called
\AddIndex{direct sum}{direct sum}.\,\footnote{
See also the definition
of direct sum of modules in
\citeBib{Serge Lang},
page 128.
On the same page, Lang proves the existence of direct sum of modules.
}
We will use notation
\ShowEq{direct sum of Abelian groups}
for direct sum of $D$\Hyph modules $A$ and $B$.
}

\AddEq{theorem: definition of module over commutative ring}
{
\begin{theorem}
\labelTheorem{definition of module}
Following conditions hold for $D$\Hyph module:
\StartLabelItem
\begin{enumerate}
\item
\AddIndex{associative law}{associative law}
\DrawEq{associative law, module}2
\item 
\labelItem{distributive law, module}
\AddIndex{distributive law}{distributive law}
\ShowEq{distributive law, module}
\item
\AddIndex{unitarity law}{unitarity law}
\ShowEq{unitarity law, D-module}
\end{enumerate}
for any
\ShowEq{p,q in D, v,w in V}
\end{theorem}
}

\AddEq{definition: module over algebra}
{
\begin{definition}
\labelDefinition{\SideWS module over algebra}
Effective \SideNS\Hyph side representation
\DrawEq{\SideWS A->*V}{def}
of associative $D$\Hyph algebra $A$
in $D$\Hyph module $V$
is called
\AddIndex{\SideWS module}{\SideWS module} over $D$\Hyph algebra $A$.
We will also say that $D$\Hyph module $V$ is
\AddIndex{\SideWS $A$\Hyph module}{\SideWS A module}
or
\AddIndex{$\SideA$\Hyph module}{\SideA-module}.
$V$\Hyph number is called
\AddIndex{vector}{vector}.
\qed
\end{definition}
}

\DefDefinition{left vector space over algebra}
{
Let $A$ be division algebra.
Effective left\Hyph side representation
\DrawEq{left A->*V}{}
of Abelian group $A$
in $D$\Hyph module $V$
is called
\AddIndex{left vector space}{left vector space}
over $D$\Hyph algebra $A$.
We will also say that $D$\Hyph module $V$ is
\AddIndex{left $A$\Hyph vector space}{left A vector space}
or
\AddIndex{$A*$\Hyph vector space}{A*-vector space}.
$V$\Hyph number is called
\AddIndex{vector}{vector}.
}

\DefDefinition{right vector space over algebra}
{
Let $A$ be division algebra.
Effective right\Hyph side representation
\DrawEq{right A->*V}{}
of Abelian group $A$
in $D$\Hyph module $V$
is called
\AddIndex{right vector space}{right vector space}
over $D$\Hyph algebra $A$.
We will also say that $D$\Hyph module $V$ is
\AddIndex{right $A$\Hyph vector space}{right A vector space}
or
\AddIndex{$*A$\Hyph vector space}{*A-vector space}.
$V$\Hyph number is called
\AddIndex{vector}{vector}.
}

\AddEq{theorem: module over algebra}
{
\begin{theorem}
\labelTheorem{\SideWS module over algebra}
The following diagram of representations describes \SideWS $\Base$\Hyph module $V$
\ShowEq{diagram of representations, \SideWS module}
The diagram of representations
\EqRef{diagram of representations, \SideWS module}
holds
\AddIndex{commutativity of representations}{commutativity of representations}
\BaseRings
in Abelian group $V$
\ShowEq{\SideWS module, a d v}
\end{theorem}
}

\DefProof{module over algebra}
{
The diagram of representations
\EqRef{diagram of representations, \SideWS module}
follows from the definition
\def\Temp{}
\ifx\SideNS\Temp
\RefDefinition{module over commutative ring}
and from the theorem
\RefTheorem{action of ring of rational integers in Abelian group}.
\else
\RefDefinition{\SideWS module over algebra}
and the theorem
\RefTheorem{Free Algebra over Ring}.
\fi
Since \SideNS\HSide transformation $\ATransf(a)$
is endomorphism
of $\CBase$\Hyph module $V$,
we obtain the equality
\EqRef{\SideWS module, a d v}.
}

\DefProof{Free Algebra over Ring}
{
The structure of $D$\Hyph module $A$ is generated by effective representation
\ShowEq{f:A->*B}{g_{12}}DA
of ring $D$ in Abelian group $A$.

\begin{lemma}
\labelLemma{structure of D algebra is generated by product}
{\it
Let the structure of $D$\Hyph algebra $A$
defined in $D$\Hyph module $A$,
be generated by product
\DrawEq{product in D algebra}{}
\AddIndex{Left shift of $D$\Hyph module $A$}{left shift of module}
defined by equation
\ShowEq{l(v):w->vw}
generates the representation
\ShowEq{endomorphism of module from product, 1}
of $D$\Hyph module $A$
in $D$\Hyph module $A$
}
\end{lemma}

{\sc Proof.}
According to definitions
\RefDefinition{algebra over ring}
and
\RefDefinition{polylinear map of modules},
left shift of $D$\Hyph module $A$
is linear map.
According to the definition
\RefDefinition{linear map from A1 to A2, commutative module},
the map \(l(v)\)
is endomorphism of $D$\Hyph module $A$.
The equation
\ShowEq{l(v1+v2)w}
follows from the definition
\RefDefinition{polylinear map of modules}
and from the equation
\EqRef{l(v):w->vw}.
\ShowEq{def sum of linear maps}
\ShowEq{ref sum of linear maps}
the equation
\ShowEq{l(v1+v2)}
follows from equation
\EqRef{l(v1+v2)w}.
The equation
\ShowEq{l(dv)w}
follows from the definition
\RefDefinition{polylinear map of modules}
and from the equation
\EqRef{l(v):w->vw}.
\ShowEq{ref sum of linear maps}
the equation
\ShowEq{l(dv)}
follows from equation
\EqRef{l(dv)w}.
The lemma follows from equalities
\EqRef{l(v1+v2)},
\EqRef{l(dv)}.
\hfill\(\odot\)

\begin{lemma}
\labelLemma{representation of D module in D module determines the product}
{\it
The representation
\ShowEq{endomorphism of module from product, 1}
of $D$\Hyph module $A$ in $D$\Hyph module $A$
determines the product in
$D$\Hyph module $A$ according to rule
\ShowEq{endomorphism of module from product, 8}
}
\end{lemma}

{\sc Proof.}
Since map $g_{23}\circ v$ is endomorphism of $D$\Hyph module $A$, then
\ShowEq{endomorphism of module from product, 3}
Since the map $g_{23}$ is
linear map
\ShowEq{g23:A->L}
then,
\ShowEq{ref sum and product over scalar, linear map}
\ShowEq{endomorphism of module from product, 4}
\ShowEq{endomorphism of module from product, 7}
From equations
\EqRef{endomorphism of module from product, 3},
\EqRef{endomorphism of module from product, 4},
\EqRef{endomorphism of module from product, 7}
and the definition
\RefDefinition{polylinear map of modules},
it follows that the map $g_{23}$ is bilinear map.
Therefore, the map $g_{23}$ determines the product in
$D$\Hyph module $A$ according to rule
\ShowEq{endomorphism of module from product, 8}
\hfill\(\odot\)

The theorem follows from lemmas
\RefLemma{structure of D algebra is generated by product},
\RefLemma{representation of D module in D module determines the product}.
}

\AddEq{theorem: A module -> algebra is associative}
{
\begin{theorem}
\labelTheorem{\SideWS A module -> algebra is associative}
Let $g$ be effective left\Hyph side representation of $D$\Hyph algebra $A$
in $D$\Hyph module $V$.
Then $D$\Hyph algebra $A$ is associative.
\end{theorem}
}

\DefProof{A module -> algebra is associative}
{
Let
\ShowEq{abc in A, v in v}
Since \SideNS\Hyph side representation
$g$ is \SideNS\Hyph side representation
of the multiplicative group
of $D$\Hyph algebra $A$,
we obtain the equality
\DrawEq{\SideWS module, associative law}0
The equality
\ShowEq{a(b(cv))=(a(bc))v \SideNS}
follows from the equality
\eqRef{\SideWS module, associative law}0.
Since
\ShowEq{cv \SideNS}
the equality
\ShowEq{a(b(cv))=((ab)c)v \SideNS}
follows from the equality
\eqRef{\SideWS module, associative law}0.
The equality
\ShowEq{(a(bc))v=((ab)c)v \SideNS}
follows from equalities
\EqRef{a(b(cv))=(a(bc))v \SideNS},
\EqRef{(a(bc))v=((ab)c)v \SideNS}.
Since $v$ is any vector of $A$\Hyph module $V$,
the equality
\ShowEq{a(bc)=(ab)c \SideNS}
follows from the equality
\EqRef{(a(bc))v=((ab)c)v \SideNS}.
Therefore, $D$\Hyph algebra $A$ is associative.
}

\AddEq{theorem: definition of A module}
{
\begin{theorem}
\labelTheorem{definition of \SideWS module}
Let $V$ be \SideWS $\Base$\Hyph module.
For any vector $v\in V$,
vector generated by the diagram of representations
\EqRef{diagram of representations, \SideWS module}
has the following form
\ShowEq{\SideWS module, (a+n)v=av+nv}
\StartLabelItem
\begin{enumerate}
\item
The set of maps
\ShowEq{\SideWS module, a+n:V->V}
generates\,\footnote{
See the definition of unital extension also on the pages
\citeBib{McCrimmon: Jordan Algebras}\Hyph 52,
\citeBib{Zharinov: foundation of mathematical physics}\Hyph 64.
}
\algebraa $\Base_{(1)}$
where the sum is defined by the equality
\DrawEq{(a+n)+(b+m)=}{\SideWS module}
and the product is defined by the equality
\DrawEq{(a+n)(b+m)=}{\SideWS module}
The \algebraa $\Base_{(1)}$ is called
\AddIndex{unital extension}{unital extension}
of the \algebraa $\Base$.
\begin{table}[h]
\begin{tabular}{|l|l|l|}
\hline
If \algebraa $\Base$ has unit, then
\ShowEq{A1=A unital extension}
If \algebraa $\Base$ is ideal of $\CBase$, then
\ShowEq{A1=D unital extension}
Otherwise
\ShowEq{A1=A+D unital extension}
\end{tabular}
\end{table}
\item
The \algebraa $\Base$ is \SideWS ideal of \algebraa $\Base_{(1)}$.
\labelItem{Algebra is \SideWS ideal of algebra (1)}
\item
The set of transormations
\EqRef{\SideWS module, (a+n)v=av+nv}
is \SideNS\HSide representation of \algebraa $\Base_{(1)}$ in Abelian group $V$.
\labelItem{\SideWS representation of D1 in Abelian group}
\end{enumerate}
We use the notation
\ShowEq{set of vectors generated by vector \Base}
for the set of vectors generated by vector $v$.
\end{theorem}
}

\AddEq{theorem: definition of A module, property}
{
\begin{theorem}
\labelTheorem{definition of \SideWS module, property}
Following conditions hold for \SideWS $\Base$\Hyph module $V$:
\StartLabelItem
\begin{enumerate}
\item 
\AddIndex{associative law}{associative law}
\labelItem{associative law, \SideWS module}
\DrawEq{associative law, \SideWS module}1
\item 
\AddIndex{distributive law}{distributive law}
\labelItem{distributive law, \SideWS module}
\ShowEq{distributive law, \SideWS module}
\item
\AddIndex{unitarity law}{unitarity law}
\labelItem{unitarity law, \SideWS \Base-module}
\ShowEq{unitarity law, \SideWS \Base-module}
\end{enumerate}
for any
\ShowEq{p,q in \Base 1, v,w in V}
\end{theorem}
}

\AddEq{proof: definition of A module}
{%
{\sc Proof of theorems
\RefTheorem{definition of \SideWS module},
\RefTheorem{definition of \SideWS module, property}.}
Let $v\in V$.

\begin{lemma}
\labelLemma{\SideWS module, map a+n:V->V is endomorphism of Abelian group}
{\it
Let
\ShowEq{module, d in D}
\ShowEq{module, a in A}
The map
\EqRef{\SideWS module, a+n:V->V}
is endomorphism of Abelian group $V$.
}
\end{lemma}

{\sc Proof.}
Statements
\ShowEq{\SideWS module, dv in V}
\ShowEq{\SideWS module, av in V}
follow from the theorems
\RefTheorem[\RefRepresentation]{structure of subrepresentations},
\RefTheorem{\SideWS module over algebra}.
Since $V$ is Abelian group, then
\ShowEq{\SideWS module, dv+av in V}
Therefore,
for any $\CBase$\Hyph number $\DArg$
and for any $\Base$\Hyph number $a$,
we defined the map
\EqRef{\SideWS module, a+n:V->V}.
Since transformation $\DTransf(\DArg)$
and \SideNS\HSide transformation $\ATransf(a)$
are endomorphisms of Abelian group $V$,
then the map
\EqRef{\SideWS module, a+n:V->V}
is endomorphism of Abelian group $V$.
\hphantom{aaaa}\hfill\(\odot\)

Let $\Base_{(1)}$ be the set of maps
\EqRef{\SideWS module, a+n:V->V}.
The equality
\EqRef{distributive law, \SideWS module, 1}
follows from the lemma
\RefLemma{\SideWS module, map a+n:V->V is endomorphism of Abelian group}.

Let
\ShowEq{p,q in D1}
According to the statement
\RefItem{representation of D1 in Abelian group},
we define the sum of $\Base_{(1)}$\Hyph numbers $p$ and $q$ by the equality
\EqRef{distributive law, \SideWS module, 2}.
The equality
\ShowEq{\SideWS module, (a+n)+(b+m)=, 1}
follows from the equality
\EqRef{distributive law, \SideWS module, 2}.
Since representation
$\DTransf$ is homomorphism of the aditive group of ring $\CBase$,
we obtain the equality
\ShowEq{distributive law, \CBase, \SideWS module, 2}
Since \SideNS\HSide representation
$\ATransf$ is homomorphism of the aditive group of \algebraa $\Base$,
we obtain the equality
\ShowEq{distributive law, \Base, \SideWS module, 2}
Since $V$ is Abelian group, then the equality
\ShowEq{\SideWS module, (a+n)+(b+m)=}
follows from equalities
\EqRef{\SideWS module, (a+n)+(b+m)=, 1},
\EqRef{distributive law, \CBase, \SideWS module, 2},
\EqRef{distributive law, \Base, \SideWS module, 2}.
From the equality
\EqRef{\SideWS module, (a+n)+(b+m)=},
it follows that the definition
\eqRef{(a+n)+(b+m)=}{\SideWS module}
of sum on the set $\Base_{(1)}$ does not depend on vector $v$.

Equalities
\eqRef{associative law, \SideWS module}1,
\EqRef{unitarity law, \SideWS \Base-module}
follow from the statement
\RefItem{\SideWS representation of D1 in Abelian group}.
Let
\ShowEq{p,q in D1}
\def\Temp{}
\ifx\SideNS\Temp
Since representation $\DTransf$ is representation
of the multiplicative group of ring $\CBase$,
we obtain the equality
\DrawEq{associative law, \CBase, \SideWS module}1
Since representation $g_2$ is representation
of the multiplicative group of ring $D$,
we obtain the equality
\DrawEq{\SideWS module, associative law}1
Since the ring $D$
is Abelian group,
we obtain the equality
\DrawEq{associative law, \CBase\Base, \SideWS module}1
The equality
\ShowEq{module, (a+n)(b+m)=}
follows from equalities
\ShowEq{ref module, (a+n)(b+m)=}
The equality
\eqRef{(a+n)(b+m)=}{\SideWS module}
follows from the equality
\EqRef{module, (a+n)(b+m)=}.
\else%
Since the product in $D$\Hyph algebra $A$ can be non associative,
then, based on the theorem
\RefTheorem{definition of \SideWS module, property},
we consider product
of $\Base_{(1)}$\Hyph numbers $p$ and $q$
as bilinear map
\ShowEq{f:D1xD1->D1}
such that following equalities are true
\DrawEq{fab=ab}{\SideWS module}
\DrawEq{f1p=p}{\SideWS module}
The equality
\DrawEq{pq=fpq}{\SideWS module}
follows from equalities
\eqRef{fab=ab}{\SideWS module},
\eqRef{f1p=p}{\SideWS module}.
The equality
\eqRef{(a+n)(b+m)=}{\SideWS module}
follows from the equality
\eqRef{pq=fpq}{\SideWS module}.
\fi%

The statement
\RefItem{Algebra is \SideWS ideal of algebra (1)}
follows from the equality
\eqRef{(a+n)(b+m)=}{\SideWS module}.
\qed
}

\AddEq{definition: submodule}
{
\begin{definition}
\labelDefinition{submodule, \SideWS module}
Subrepresentation of \SideWS $D$\Hyph module $V$ is called
\AddIndex{submodule}{submodule}
of $D$\Hyph module $V$.
\qed
\end{definition}
}

\AddEq{theorem: submodule}
{
\begin{theorem}
\labelTheorem{submodule, \SideWS module}
Let
\ShowEq{vi V}{}
be set of vectors of \SideWS $\Base$\Hyph module $V$.
If vectors
\ShowEq{set vi}
belongs submodule $V'$ of \SideWS $\Base$\Hyph module $V$,
then linear combination of vectors
\ShowEq{set vi}
belongs submodule $V'$.
\end{theorem}
}

\DefProof{submodule}
{
The theorem follows from
the theorem
\RefTheorem{set of vectors generated by set of vectors, \SideWS module}
and definitions
\RefDefinition{\SideWS linear combination of vectors},
\RefDefinition{submodule, \SideWS module}.
}

\AddEq{theorem: set of vectors generated by set of vectors}
{
\begin{theorem}
\labelTheorem{set of vectors generated by set of vectors, \SideWS module}
Let $V$ be \SideWS $\Base$\Hyph module.
The set of vectors generated by the set of vectors
\ShowEq{vi V}{}
has form\,\footnote{
For a set $A$,
we denote by $|A|$ the cardinal number of the set $A$.
The notation $|A|<\infty$ means that the set $A$ is finite.
}
\ShowEq{w=sum vi, \SideWS module}
\end{theorem}
}

\DefProof{set of vectors generated by set of vectors}
{
We prove the theorem by induction based on the theorem
\RefTheorem[\RefRepresentation]{structure of subrepresentations},
Acording to the theorem
\RefTheorem[\RefRepresentation]{structure of subrepresentations},
we need to prove following statements:
\ShowEq{vectors generated by set of vectors}

\begin{itemize}
\item
For any
\ShowEq{vk in v}
let
\ShowEq{ci=dik}{\Base_{(1)}}
Then
\ShowEq{vk=sum vi, \SideWS module}
The statement
\RefItem{\SideWS module, vk in Jv}
follows from
\EqRef{w=sum vi, \SideWS module},
\EqRef{vk=sum vi, \SideWS module}.
\item
The statement
\RefItem{\SideWS module, cvk in Jv}
follow from the theorems
\RefTheorem[\RefRepresentation]{structure of subrepresentations},
\RefTheorem{definition of \SideWS module}
and from the statement
\RefItem{\SideWS module, vk in Jv}.
\item
Since $V$ is Abelian group,
then the statement
\RefItem{\SideWS module, sum cvk in Jv}
follows from the statement
\RefItem{\SideWS module, cvk in Jv}
\ePrints{1502.04063}
\ifx\Semafor\ValueOn
and from the theorem
\RefTheorem[\RefRepresentation]{structure of subrepresentations}.
\else
and from theorems
\RefTheorem[\RefRepresentation]{structure of subrepresentations},
\RefTheorem{structure of Abelian group}.
\fi
\item
Let
\ShowEq{w12 in Jv}wv
Since $V$ is Abelian group,
then, according to the statement
\RefItem[\RefRepresentation]{x1n omega in Xk+1},
\DrawEq{w1+w2 in V}{\SideNS}
According to the equality
\EqRef{w=sum vi, \SideWS module},
there exist $\Base_{(1)}$\Hyph numbers
\ShowEq{gi12}w
such that
\DrawEq[w]{w12= \SideNS}{module}
where sets
\DrawEq[w]{|ci12 ne 0|}{\SideWS module}
are finite.
Since $V$ is Abelian group,
then from the equality
\eqRef{w12= \SideNS}{module}
it follows that
\DrawEq[w]{w1+w2= \SideNS}{module}
The equality
\DrawEq[w]{w1+w2= 1 \SideNS}{module}
follows from equalities
\EqRef{distributive law, \SideWS module, 2},
\eqRef{w1+w2= \SideNS}{module}.
From the equality
\eqRef{|ci12 ne 0|}{\SideWS module},
it follows that
the set
\ShowEq{|gi 1+2 ne 0|}w
is finite.
\item
Let
\ShowEq{w in Jv}
According to the statement
\RefItem[\RefRepresentation]{ax in Xk+1},
for any $\Base_{(1)}$\Hyph number $a$,
\ShowEq{aw in Xk \SideWS module}
According to the equality
\EqRef{w=sum vi, \SideWS module},
there exist $\Base_{(1)}$\Hyph numbers
\ShowEq{set au vi}w
such that
\ShowEq{w= \SideWS module}
where
\DrawEq{|wi ne 0|}{\SideWS module}
From the equality
\EqRef{w= \SideWS module}
it follows that
\ShowEq{aw= \SideWS module}
From the statement
\eqRef{|wi ne 0|}{\SideWS module},
it follows that
the set
\ShowEq{\SideWS module, |awi ne 0|}
is finite.
\end{itemize}
From equalities
\eqRef{w1+w2 in V}{\SideNS},
\eqRef{w1+w2= 1 \SideNS}{module},
\EqRef{aw in Xk \SideWS module},
\EqRef{aw= \SideWS module},
it follows that
\ShowEq{Xk+1 in Jv}
}

\AddEq{definition: linear combination of vectors}
{
\begin{definition}
\labelDefinition{\SideWS linear combination of vectors}
Let
\ShowEq{vi V}{}
be set of vectors.
\ShowEq{\SideWS linear combination}%
The expression
\ShowEq{A linear combination =}
is called
\AddIndex{linear combination}{linear combination} of vectors
\ShowEq{vi}
A vector
\ShowEq{w=wi vi\SideNS}
is called
\AddIndex{linearly dependent}{linearly dependent}
on vectors
\ShowEq{vi}
\qed
\end{definition}
}

\AddEq{definition: generating set of module}
{
\begin{definition}
\labelDefinition{generating set of \SideWS module}
$J(v)$
is called
\AddIndex{submodule generated by set}
{submodule generated by set} $v$,
and $v$ is a
\AddIndex{generating set}{generating set}
of submodule $J(v)$.
In particular, a
\AddIndex{generating set}{generating set}
of \SideWS $D$\Hyph module $V$
is a subset $X\subset V$ such that
\ShowEq{generating set of module}
\qed
\end{definition}
}

\AddEq{remark: generating set of module}
{
The following definition follows from the theorems
\RefTheorem{set of vectors generated by set of vectors, \SideWS module},
\RefTheorem[\RefRepresentation]{structure of subrepresentations}
and from the definition
\RefDefinition[\RefRepresentation]{generating set of representation}.
}

\AddEq{remark: basis of module}
{
The following definition follows from the theorems
\RefTheorem{set of vectors generated by set of vectors, \SideWS module},
\RefTheorem[\RefRepresentation]{structure of subrepresentations}
and from the definition
\RefDefinition[\RefRepresentation]{basis of representation}.
}

\AddEq{definition: basis of module}
{
\begin{definition}
\labelDefinition{basis of \SideWS module}
If the set $X\subset V$ is generating set of \SideWS $D$\Hyph module
$V$, then any set $Y$, $X\subset Y\subset V$
also is generating set of \SideWS $D$\Hyph module $V$.
If there exists minimal set $X$ generating
the \SideWS $D$\Hyph module $V$, then the set $X$ is called
\AddIndex{basis}{basis} of \SideWS $D$\Hyph module $V$.
\qed
\end{definition}
}

\AddEq{theorem: linearly depends on rest of vectors}
{
\begin{theorem}
\labelTheorem{linearly depends on rest of vectors, \SideWS module}
Let $\Base$ be \Algebra.
Since the equation
\ShowEq{\SideWS wi vi=0}
implies existence of index
\ShowEq{i=j}
such that
\ShowEq{wj ne 0},
then the vector $v_{\gij}$
linearly depends on rest of vectors $v$.
\end{theorem}
}

\AddEq{proof: linearly depends on rest of vectors}
{
\begin{proof}
The theorem follows from the equality
\ShowEq{\SideWS vj=sum vi}v
and from the definition
\RefDefinition{\SideWS linear combination of vectors}.
\end{proof}
}

\AddEq{remark: 0=0vi}
{
It is evident that for any set of vectors $v_{\gii}$
\ShowEq{\SideWS 0=0vi}
}

\AddEq{remark: scalars and vectors as matrix}
{
We represent the set of $\Base_{(1)}$\Hyph numbers
\ShowEq{set au vi}w
as matrix
\ShowEq{column vector w=wi}w
We represent the set of vectors
\ShowEq{set vi}
as matrix
\ShowEq{row vector w=wi}vn
Then we can represent linear combination of vectors
\ShowEq{w=wi vi\SideNS}
as
\ShowEq{w=w cr v}
}

\AddEq{definition: linearly independent vectors}
{
\begin{definition}
\labelDefinition{\SideWS linearly independent vectors}
The set of vectors\,\footnote{
I follow to the definition in
\citeBib{Serge Lang}, page 130.}
\ShowEq{set vi}
of \SideWS $\Base$\Hyph module $V$ is
\AddIndex{linearly independent}{linearly independent set}
if $w=0$ follows from the equation
\ShowEq{\SideWS wi vi=0}
Otherwise the set of vectors
\ShowEq{set vi}
is \AddIndex{linearly dependent}{linearly dependent set}.
\qed
\end{definition}
}

\AddEq{theorem: basis of module}
{
\begin{theorem}
\labelTheorem{basis of \SideWS module}
The set of vectors
\ShowEq{basis for module over algebra}
is basis of \SideWS $\Base$\Hyph module
$V$, if following statements are true.
\StartLabelItem
\begin{enumerate}
\item
\labelItem{vector is linear combination of set, \SideWS module}
Arbitrary vector $v\in V$
is linear combination of
vectors of the set $\Basis e$.
\item
\labelItem{cannot be represented as a linear combination, \SideWS module}
Vector $e_{\gii}$
cannot be represented as a linear combination
of the remaining vectors of the set $\Basis e$.
\end{enumerate}
\end{theorem}
}

\DefProof{basis of module}
{
According to the statement
\RefItem{vector is linear combination of set, \SideWS module},
the theorem
\RefTheorem{set of vectors generated by set of vectors, \SideWS module}
and the definition
\RefDefinition{\SideWS linear combination of vectors},
the set $\Basis e$ generates \SideWS $\Base$\Hyph module $V$
(the definition
\RefDefinition{generating set of \SideWS module}).
According to the statement
\RefItem{cannot be represented as a linear combination, \SideWS module},
the set $\Basis e$ is minimal set
generating \SideWS $\Base$\Hyph module $V$.
According to the definitions
\RefDefinition{basis of \SideWS module},
the set $\Basis e$ is a basis of \SideWS $\Base$\Hyph module $V$.
}

\AddEq{theorem: division algebra, basis}
{
\begin{theorem}
\labelTheorem{division algebra, \SideWS basis}
Let $\Base$ be \Algebra.
The set of vectors
\ShowEq{basis, module}
is a
\AddIndex{basis of \SideWS $\Base$\Hyph vector space}{basis, vector space} $V$
if vectors $e_{\gii}$ are linearly independent and any vector $v\in V$
linearly depends on vectors $e_{\gii}$.
\end{theorem}
}

\DefProof{division algebra, basis}
{
Let the set of vectors
\ShowEq{ei, i in I}
be linear dependent. Then the equation
\DrawEq[w]{c*e=0, \SideWS module}{}
implies existence of index $\gii=\gij$ such that
\ShowEq{wj ne 0}.
According to the theorem
\RefTheorem{linearly depends on rest of vectors, \SideWS module},
the vector $e_{\gij}$
linearly depends on rest of vectors of the set $\Basis e$.
According to the definition
\RefDefinition{basis of \SideWS module},
the set of vectors
\ShowEq{ei, i in I}
is not a basis for \SideWS $\Base$\Hyph vector space $V$.

Therefore, if the set of vectors
\ShowEq{ei, i in I}
is a basis, then these vectors
are linearly independent.
Since an arbitrary vector $v\in V$
is linear combination of vectors
\ShowEq{ei, i in I},
then the set of vectors $v$,
\ShowEq{ei, i in I}
is not linearly independent.
}

\AddEq{convention: we use separate color for index of element}
{
\begin{convention}
\labelConvention{we use separate color for index of element}
Let $A$ be free algebra
with finite or countable basis.
Considering expansion of element of algebra $A$ relative basis $\Basis e$
we use the same root letter to denote this element and its coordinates.
In expression $a^2$, it is not clear whether this is component
of expansion of element
$a$ relative basis, or this is operation $a^2=aa$.
To make text clearer we use separate color for index of element
of algebra. For instance,
\ShowEq{Expansion relative basis in algebra}
\qed
\end{convention}
}

\AddEq{convention: unit of algebra in basis}
{
\begin{convention}
Let $\Basis e$ be the basis of free algebra $A$ over ring $D$.
If algebra $A$ has unit,
then we assume that $e_{\gi 0}$ is the unit of algebra $A$.
\qed
\end{convention}
}

\DefDefinition{linear map from D1 A1 to D2 A2, module}
{
Morphism of representations
\ShowEq{h:D1->D2 f:A1->A2}h\Base fV
of $D_1$\Hyph module $A_1$
into $D_2$\Hyph module $A_2$
is called
\AddIndex{linear map}{linear map}
of $D_1$\Hyph module $A_1$ into $D_2$\Hyph module $A_2$.
Let us denote
\ShowEq{set linear maps, D12 module}
set of linear maps
of $D_1$\Hyph module $A_1$ into $D_2$\Hyph module $A_2$.
}

\ifx\texFuture\Defined
\AddEq{linear map from D1 A1 to D2 A2, module}
{
\begin{definition}
\labelDefinition{linear map from D1 A1 to D2 A2, \SideNS module}
{\it
Morphism of representations
\ShowEq{h:D1->D2 f:A1->A2}h\Base fV
of \SideWS $D_1$\Hyph module $A_1$
into \SideWS $D_2$\Hyph module $A_2$
is called
\AddIndex{linear map}{linear map}
of \SideWS $D_1$\Hyph module $A_1$ into \SideWS $D_2$\Hyph module $A_2$.

Let us denote
\ShowEq{set linear maps, D12 module}
set of linear maps
of \SideWS $D_1$\Hyph module $A_1$ into \SideWS $D_2$\Hyph module $A_2$.
}
\qed
\end{definition}
}
\fi

\DefDefinition{linear map from A1 to A2, commutative module}
{
Reduced morphism of representations
\ShowEq{f:A->B}f{A_1}{A_2}
of $D$\Hyph module $A_1$
into $D$\Hyph module $A_2$
is called
\AddIndex{linear map}{linear map}
of $D$\Hyph module $A_1$ into $D$\Hyph module $A_2$.
Let us denote
\ShowEq{set linear maps, module}
set of linear maps
of $D$\Hyph module $A_1$ into $D$\Hyph module $A_2$.
}

\DefEq
{
\begin{theorem}
\labelTheorem{set of A->B is D module}
Let $A$ be Banach $D$\Hyph module with norm $|x|_A$.
Let $B$ be Banach $D$\Hyph module with norm $|y|_B$.
\StartLabelItem
\begin{enumerate}
\item
The set
$B^A$
of maps
\ShowEq{f:A->B}fAB
is $D$\Hyph module.
\labelItem{set of A->B is D module}
\item
The map
\ShowEq{f in BA->|f|}
defined by the equality
\ShowEq{norm of map}
\ShowEq{norm of map, algebra}
is the norm in $D$\Hyph module $B^A$
and the value
\ShowEq{show|f|}
is called
\AddIndex{norm of map $f$}{norm of map}.
\labelItem{norm of map}
\end{enumerate}
\end{theorem}
}
{theorem: set of A->B is D module}

\DefDefinition{coordinates of vector 2016}
{
Let $\Basis e$ be the basis of \SideWS $\Base$\Hyph module $V$
and vector
\ShowEq{vv in V}
has expansion
\DrawEq{vv=ve \SideWS module}{}
with respect to the basis $\Basis e$.
$\Base$\Hyph numbers $v^{\gii}$ are called
\AddIndex{coordinates}{coordinates}
of vector $\Vector v$ with respect to the basis $\Basis e$.
\ePrints{309618526,CACAA.06.121}
\ifx\Semafor\ValueOn
Matrix of $\Base$\Hyph numbers
\ShowEq{coordinate matrix of vector}
is called
\AddIndex{coordinate matrix of vector}{coordinate matrix of vector}
$\Vector v$ in basis $\Basis e$.
\fi
}

\AddEq{definition: coordinates of vector}
{
\begin{definition}
\labelDefinition{coordinates of vector, \SideWS module}
Let $\Basis e$ be the basis of \SideWS $\Base$\Hyph module $V$
and vector
\ShowEq{vv in V}
has expansion
\DrawEq{vv=ve \SideWS module}{}
with respect to the basis $\Basis e$.
$\Base_{(1)}$\Hyph numbers $v^{\gii}$ are called
\AddIndex{coordinates}{coordinates}
of vector $\Vector v$ with respect to the basis $\Basis e$.
Matrix of $\Base_{(1)}$\Hyph numbers
\ShowEq{coordinate matrix of vector}
is called
\AddIndex{coordinate matrix of vector}{coordinate matrix of vector}
$\Vector v$ in basis $\Basis e$.
\qed
\end{definition}
}

\AddEq{theorem: linear dependence between vectors of basis}
{
\begin{theorem}
\labelTheorem{linear dependence between vectors of basis, \SideWS module}
Let $\Base$ be \algebra.
Let $\Basis e$ be basis of \SideWS $\Base$\Hyph module $V$.
Let
\DrawEq[w]{c*e=0, \SideWS module}{1}
be linear dependence of vectors of the basis $\Basis e$.
Then
\StartLabelItem
\begin{enumerate}
\item
$\Base_{(1)}$\Hyph number
\ShowEq{wi i in I}
does not have inverse element
in \algebraa $\Base_{(1)}$.
\item
The set $\Base'$ of matrices
\ShowEq{w=wi i in I}
generates \SideWS $\Base$\Hyph module.
\end{enumerate}
\end{theorem}
}

\DefProof{linear dependence between vectors of basis}
{
Let there exist matrix
\ShowEq{w=wi i in I}
such that the equality
\eqRef{c*e=0, \SideWS module}{1}
is true and there exist index
\ShowEq{i=j}
such that
\ShowEq{wj ne 0}.
If we assume that $\Base_{(1)}$\Hyph number $c^{\gij}$
has inverse one, then the equality
\ShowEq{\SideWS vj=sum vi}e
follows from the equality
\eqRef{c*e=0, \SideWS module}{1}.
Therefore, the vector $e_{\gij}$
is linear combination of other vectors of the set $\Basis e$
and the set $\Basis e$ is not basis.
Therefore, our assumption is false,
and $\Base_{(1)}$\Hyph number $c^{\gij}$ does not have inverse.

Let matrices
\ShowEq{b in D'}b,
\ShowEq{b in D'}c.
From equalities
\DrawEq[b]{c*e=0, \SideWS module}{}
\DrawEq[c]{c*e=0, \SideWS module}{}
it follows that
\ShowEq{(b+c)*e, \SideWS module}
Therefore, the set $\Base'$ is Abelian group.

Let matrix
\ShowEq{b in D'}c{}
and $a\in\Base$.
From the equality
\DrawEq[w]{c*e=0, \SideWS module}{}
it follows that
\ShowEq{(ac)*e, \SideWS module}
Therefore, Abelian group $\Base'$ is \SideWS $\Base$\Hyph module.
}

\AddEq{theorem: coordinates of vector with linear dependence}
{
\begin{theorem}
\labelTheorem{coordinates of vector with linear dependence, \SideWS module}
Let \SideWS $\Base$\Hyph module $V$
have the basis $\Basis e$ such that in the equality
\DrawEq[w]{c*e=0, \SideWS module}{2}
there exists index
\ShowEq{i=j}
such that
\ShowEq{wj ne 0}.
Then
\StartLabelItem
\begin{enumerate}
\item
The matrix
\ShowEq{w=wi i in I}
determines coordinates of vector $0\in V$ with respect to basis $\Basis e$.
\labelItem{coordinates of vector 0 with linear dependence, \SideWS module}
\item
Coordinates of vector $\Vector v$ with respect to basis $\Basis e$
are uniquely determined up to a choice of coordinates of vector $0\in V$.
\labelItem{coordinates of vector with linear dependence, \SideWS module}
\end{enumerate}
\end{theorem}
}

\DefProof{coordinates of vector with linear dependence}
{
The statement
\RefItem{coordinates of vector 0 with linear dependence, \SideWS module}
follows from the equality
\eqRef{c*e=0, \SideWS module}{2}
and from the definition
\RefDefinition{coordinates of vector, \SideWS module}.

Let vector $\Vector v$ have expansion
\DrawEq{vv=ve \SideWS module}{2}
with respect to basis $\Basis e$.
The equality
\ShowEq{v=v+0, \SideWS module}
follows from equalities
\eqRef{c*e=0, \SideWS module}{2},
\eqRef{vv=ve \SideWS module}{2}.
The statement
\RefItem{coordinates of vector with linear dependence, \SideWS module}
follows from equalities
\eqRef{vv=ve \SideWS module}{2},
\EqRef{v=v+0, \SideWS module}
and from the definition
\RefDefinition{coordinates of vector, \SideWS module}.
}

\AddEq{definition: free module over ring}
{
\begin{definition}
\labelDefinition{free \SideWS module}
The \SideWS $\Base$\Hyph module $V$ is
\AddIndex{free \SideWS $\Base$\Hyph module}{free module},\,\footnote{
I follow to the
definition in \citeBib{Serge Lang}, page 135.}
if \SideWS $\Base$\Hyph module $V$ has basis
and vectors of the basis are linearly independent.
\qed
\end{definition}
}

\AddEq{theorem: coordinates of vector of free module}
{
\begin{theorem}
\labelTheorem{coordinates of vector of free \SideWS module}
Coordinates of vector $v\in V$ relative to basis $\Basis e$
of free \SideWS $\Base$\Hyph module $V$
are uniquely defined.
\end{theorem}
}

\DefProof{coordinates of vector of free module}
{
The theorem follows from the theorem
\RefTheorem{coordinates of vector with linear dependence, \SideWS module}
and from definitions
\RefDefinition{\SideWS linearly independent vectors},
\RefDefinition{free \SideWS module}.
}

\DefTheorem{standard representation of map A->A, associative algebra}
{
Let $A$ be finite dimensional associative $D$\Hyph algebra.
Let $\Basis e$ be basis of $D$\Hyph module $A$.
Let $\Basis F$
be the basis\,\footnote{
If $D$\Hyph module $A$
is not free $D$\Hyph nodule,
then we may consider the set
\ShowEq{Ik 1n}
of linear independent linear maps. The theorem is true for any linear map
\ShowEq{f:A->B}fAA
generated by the set of linear maps $\Basis F$.
}
of left \BoxB{A}module
\ShowEq{L(A;B)}DAA.
\StartLabelItem
\begin{enumerate}
\item
The linear map
\ShowEq{f:A->B}fAA
has the following expansion
\labelItem{map f generated by basis F}
\DrawEq{map f generated by basis F}{expansion}
where
\ShowEq{fk= in AxA}
\item
The linear map $f$ has the standard representation
\labelItem{standard representation of map A1 A2, associative algebra}
\ShowEq{standard representation of map A->A, associative algebra}
\end{enumerate}
}

\DefTheorem{set FoG generates left module L(A;B), n<m}
{
Let $A$ be $D$\Hyph module,
\ShowEq{n=dim A}nA.
Let $B$ be associative $D$\Hyph algebra,
\ShowEq{n=dim A}mB.
Let $\Basis F$ be basis of left \BoxB{B}module
\ShowEq{L(A;B)}DBB.
Let
\ShowEq{gi n<m}
Let
\ShowEq{f:A->B}GAB
be linear map of maximal rank.
The set
\DrawEq{F o G}1
generates left \BoxB{B}module\,\footnote{
I do not claim that this set is a basis,
because maps
\ShowEq{FijG}
can be linearly dependent.
}
\ShowEq{L(A;B)}DAB.
}

\DefProof{set FoG generates left module L(A;B), n<m}
{
Let
\ShowEq{f:A->B}gAB
be a linear map.
Let $\Basis e_A$ be the basis of $D$\Hyph module $A$.
Let $\Basis e_B$ be the basis of $D$\Hyph module $B$.
According to the theorem
\ShowEq{ref linear map of D1 D2 module, coordinates}
the linear map $G$ has coordinates
\ShowEq{matrix amn}Gmn
with respect to bases $\Basis e_A$, $\Basis e_B$
and the linear map $g$ has coordinates
\ShowEq{matrix amn}gmn
with respect to bases $\Basis e_A$, $\Basis e_B$.
A row $G_{\gii}$ of the matrix $G$,
as well a row $g_{\gii}$ of the matrix $g$,
is coordinates of linear form
\ShowEq{A->B}AD.
Since the matrix $G$ has maximal rank,
then rows of the matrix $G$ generate $D$\Hyph module
\ShowEq{L(A;B)}DAD{}
and rows of the matrix $g$ are linear combination of rows of the matrix $G$
\ShowEq{g=CG}
Since we can consider the matrix $C$
as coordinates of linear map
\ShowEq{f:A->B}CBB
then the equality
\ShowEq{g=(cF)G}
follows from the equality
\EqRef{g=CG}
and from the equality
\ShowEq{C=cF}
Since
\ShowEq{Gkj in D}
then the equality
\ShowEq{g=c(FG)}
follows from the equality
\EqRef{g=(cF)G}.
Therefore, the map $g$ belongs to
linear span of the set of maps
\eqRef{F o G}1.
}

\DefTheorem{set FoG generates left module L(A;B), n>m}
{
Let
\ShowEq{n=dim A}nA,
\ShowEq{n=dim A}mB.
Let $\Basis F$ be basis of left \BoxB{B}module
\ShowEq{L(A;B)}DBB.
Let
\ShowEq{gi n>m}
Let
\ShowEq{f:A->B}GAB
be linear map of maximal rank.
The set
\DrawEq{F o G}{}
generates the set of maps
\ShowEq{set ker G in ker g}
}

\DefProof{set FoG generates left module L(A;B), n>m}
{
The proof of the theorem is similar to the proof of the theorem
\RefTheorem{set FoG generates left module L(A;B), n<m}.
However, since number of rows of the matrix $G$ less then dimention
of $D$\Hyph module $A$, then rows of the matrix $G$ do not generate $D$\Hyph module
\ShowEq{L(A;B)}DAD{}
and the map $G$ has non trivial kernel.
In particular, rows of the matrix $g$ linearly depend on rows of the matrix $G$
iff
\ShowEq{ker G in ker g}g.
}

\DefTheorem{set FoG generates left module L(A;B)}
{
Let $\Basis F$ be basis of left \BoxB{B}module
\ShowEq{L(A;B)}DBB.
Let
\ShowEq{f:A->B}GAB
be linear map of maximal rank.
The set
\DrawEq{F o G}{}
generates the set of maps
\ShowEq{set ker G in ker g}
}

\DefProof{set FoG generates left module L(A;B)}
{
It is easy to see that theorem
\RefTheorem{set FoG generates left module L(A;B), n<m}
is particular case of the theorem
\RefTheorem{set FoG generates left module L(A;B), n>m},
because, in the theorem
\RefTheorem{set FoG generates left module L(A;B), n>m},
$\ker G=\emptyset$.
}

\AddEq{remark: set FoG generates left module L(A;B), n>m}
{
From the theorem
\RefTheorem{set FoG generates left module L(A;B), n>m},
it follows that
choice of the map $G$ depends on the map $g$.
}

\DefTheorem{basis D module L(D->A)}
{
Let $\Basis e$ be the basis of $D$\Hyph module $A$.
Then the set of maps
\ShowEq{d->dei}i
is the basis of $D$\Hyph module
\ShowEq{L(A->B)}DDA.
}

\DefProof{basis D module L(D->A)}
{
The theorem follows from the equality
\ShowEq{f(t)=fit ei}
}

\DefDefinition{component of linear map}
{
Expression
\ShowEq{component of linear map}
in equality
\EqRef{fk= in AxA}
is called
\AddIndex{component of linear map}
{component of linear map} $f$.
Expression
\ShowEq{standard component of linear map}
in the equality
\EqRef{standard representation of map A->A, associative algebra}
is called
\AddIndex{standard component of linear map}
{standard component of linear map} $f$.
}

\DefTheorem{endomorphism of module from product}
{
The representation
\ShowEq{endomorphism of module from product}
of $D$\Hyph module $A$ in $D$\Hyph module $A$
is equivalent to structure of $D$\Hyph algebra $A$.
}

\DefTheorem{direct sum of n Abelian groups}
{
Direct sum of Abelian groups
\ShowEq{a1n}An{}
coincides with their Cartesian product
\ShowEq{A1o+An=A1xAn}
}

\AddEq{remark: direct sum of n Abelian groups}
{
Let
\ShowEq{a1o+.n}An
be direct sum of Abelian groups
\ShowEq{a1n}An.
According to the proof of the theorem
\RefTheorem{direct sum of Abelian groups},
any $A$\Hyph number $a$ has form
$(a_1,...,a_n)$
where $a_i\in A_i$.
We also will use notation
\ShowEq{a=a1 o+ an}
}

\DefTheorem{direct sum of n D modules}
{
Direct sum of $D$\Hyph modules
\ShowEq{a1n}An{}
coincides with their Cartesian product
\ShowEq{A1o+An=A1xAn}
}

\DefTheorem{foa=fi ai}
{
Let
\ShowEq{A1n}An{}
be $D$\Hyph modules and
\ShowEq{A1o+.n}An
Let us represent $A$\Hyph number
\ShowEq{A1o+.n}an
as column vector
\ShowEq{a=(a1.n)}an
Let us represent a linear map
\ShowEq{f:A->B}fAB
as row vector
\ShowEq{f=(f1.n)}
Then we can represent value of the map $f$ in $A$\Hyph number $a$
as product of matrices
\ShowEq{foa=fi ai}
}

\DefTheorem{foa=fi a}
{
Let
\ShowEq{A1n}Bm{}
be $D$\Hyph modules and
\ShowEq{A1o+.n}Bm
Let us represent $B$\Hyph number
\ShowEq{A1o+.n}bm
as column vector
\ShowEq{a=(a1.n)}bm
Then the linear map
\ShowEq{f:A->B}fAB
has representation as column vector of maps
\ShowEq{a=(a1.n)}fm
such way that, if
$b=f\circ a$,
then
\DrawEq{foa=fi a}{}
}

\DefTheorem{map of direct sum of modules}
{
Let
\ShowEq{A1n}An,
\ShowEq{A1n}Bm{}
be $D$\Hyph modules and
\ShowEq{A1o+.n}An
\ShowEq{A1o+.n}Bm
Let us represent $A$\Hyph number
\ShowEq{A1o+.n}an
as column vector
\ShowEq{a=(a1.n)}an
Let us represent $B$\Hyph number
\ShowEq{A1o+.n}bm
as column vector
\ShowEq{a=(a1.n)}bm
Then the linear map $f$
has representation as a matrix of maps
\ShowEq{matrix of maps}
such way that, if
$b=f\circ a$,
then
\ShowEq{b=f rco a}
The map
\ShowEq{fij:->}
is a linear map and is called
\AddIndex{partial linear map}{partial linear map}.
}

\AddEq{remark: map of direct sum of modules}
{
Let
\ShowEq{A1n}Bm{}
be $D$\Hyph algebras.
Then we can represent linear map $f^i_j$
using \BoxB{B^i}number.
}

\DefTheorem{direct sum of D modules}
{
Let
\ShowEq{set Bi}A
be set of $D$\Hyph modules.
Then the representation
\ShowEq{D->*o+A}
of the ring $D$
in direct sum of Abelian groups
\ShowEq{o+Ai}
is direct sum of $D$\Hyph modules
\ShowEq{o+Ai}
}

\AddEq{text: Free Algebra over Ring}
{
Let $D$ be commutative ring and $A$ be Abelian group.
The diagram of representations
\DrawEq[{}{}{}g]{diagram of representations of D algebra}{}
generates the structure of $D$\Hyph algebra $A$.
}

\AddEq{theorem: linear map of A module, coordinates}
{
\begin{theorem}
\labelTheorem{linear map of \SideWS A module, coordinates \Base\DF\Base\DT}
\ShowEq{Let e be basis}{\Base_\DF}iI{}{D_1}{\Base_\DF}
\ShowEq{Let e be basis}{\Base_\DT}jJ{}{D_2}{\Base_\DT}
\ShowEq{Let e be basis}{\Module_1}kK{\SideWS}{\Base_\DF}{\Module_1}
\ShowEq{Let e be basis}{\Module_2}lL{\SideWS}{\Base_\DT}{\Module_2}
Then linear map
\ShowEq{show linear map \MapE}{\Vector g}{\Vector f}
has presentation
\DrawEq[g]{r2:A1->A2, \DFDT module}{1 \SideWS g}
\ShowEq{linear map in \SideWS A module}
\DrawEq{linear map in A module}{\SideNS}
relative to selected bases. Here
\begin{itemize}
\ShowEq{coordinates of the linear map}a{\Base}ghiI{D_2}{}b
\ShowEq{coordinates of the linear map}v{\Module}fgkK{\Base_\DT}{\SideWS}w
\end{itemize}
The map
\ShowEq{fij:-> \SideWS A}
is a linear map and is called
\AddIndex{partial linear map}{partial linear map}.
\end{theorem}
}

\DefProof{linear map of A module, coordinates}
{
The equality
\eqRef{r2:A1->A2, \DFDT module}{1 \SideWS g}
follows from the theorem
\RefTheorem{linear map of D1 D2 module, coordinates}.

\ShowEq{module as direct sum}1kK
\ShowEq{module as direct sum}2lL
The equality
\eqRef{linear map in A module}{\SideNS}
follows from the theorem
\RefTheorem{map of direct sum of modules}.
}

\AddEq[3]{module as direct sum}
{
$A_{#1}$\Hyph module $V_{#1}$ is direct sum
\ShowEq{V=o+Ae}#1#2#3
}

\DefTheorem{representation of algebra An in LAnA}
{
Consider $D$\Hyph algebra $A$.
A representation
\ShowEq{representation An in LAnA}
of algebra $\AOn$
in module $\LAnA$
defined by the equality
\DrawEq{representation An in LAnA, 1}{}
allows us to identify tensor
\ShowEq{product in algebra An 1}
and transposition $\sigma\in S^n$
with map
\ShowEq{product in algebra An 2}
where
\ShowEq{product in algebra AA 3}
is identity map.
}

\DefDefinition{linear map from A1 to A2, algebra}
{
Let $A_1$ and
$A_2$ be algebras over commutative ring $D$.
The linear map
of the $D_{\DF}$\Hyph module $A_\VF$
into the $D_{\DT}$\Hyph module $A_\VT$
is called
\AddIndex{linear map}{linear map}
of $D_{\DF}$\Hyph algebra $A_\VF$ into $D_{\DT}$\Hyph algebra $A_\VT$.

Let us denote
\ShowEq{set linear maps, module}
set of linear maps
of $D$\Hyph algebra
$A_\VF$
into $D$\Hyph algebra
$A_\VT$.
}

\AddEq{remark: notation for linear map}
{
If the map
\ShowEq{f:A->B}f{A_1}{A_2}
is linear map of $D$\Hyph algebra $A_1$ into $D$\Hyph algebra $A_2$,
then I use notation
\ShowEq{f circ a =}
for image of the map $f$.
}

\DefTheorem{linear map times constant, algebra}
{
Let map
\ShowEq{f:A->B}f{A_1}{A_2}
be linear map of $D$\Hyph algebra $A_1$ into $D$\Hyph algebra $A_2$.
Then maps
\ShowEq{linear map times constant, algebra}
defined by equalities
\ShowEq{linear map times constant, 0, algebra}
are linear.
}

\DefTheorem{linear map AA LAA}
{
\ePrints{1502.04063}
\ifx\Semafor\ValueOn
Consider $D$\Hyph algebras $A_1$ and $A_2$.
For given map
\ShowEq{f in L(A->B)}D{A_1}{A_2},
\else
For given map
\ShowEq{f in L(A->B)}D{A_1}{A_2}{}
of $D$\Hyph algebra $A_1$ into $D$\Hyph algebra $A_2$,
\fi
there exists linear map
\ShowEq{linear map AA LAA}
defined by the equality
\ShowEq{linear map AA LAA, 1}
\ShowEq{linear map AA LAA, 2}
}

\DefEq
{
\begin{theorem}
\labelTheorem{conjugation transformation}
Let $A_1$ be free $D$\Hyph module.
Let $A_2$ be free associative $D$\Hyph algebra.
Let $\Basis F$ be the basis of left \BoxB{A_2}module
\ShowEq{L(A;B)}D{A_1}{A_2}.
For any map
\ShowEq{Ik in I}
there exists set of linear maps
\ShowEq{conjugation transformation}
\ShowEq{conjugation transformation:}
of $D$\Hyph module
$A_1\otimes A_1$
into $D$\Hyph module
$A_2\otimes A_2$
such that
\ShowEq{conjugation transformation =}
The map
\ShowEq{show conjugation transformation}
is called
\AddIndex{conjugation transformation}{conjugation transformation}.
\end{theorem}
}
{theorem: conjugation transformation}

\DefEq
{
\begin{proof}
According to the theorem
\RefTheorem{product of linear map, algebra},
for any tensor
$a\in A_1\otimes A_1$,
the map
\ShowEq{x->Ik a x}
is linear.
According to the statement
\RefItem{map f generated by basis F},
there exists expansion
\ShowEq{expansion Ik a x}
Let
\ShowEq{b=Ikl a}
The equality
\EqRef{conjugation transformation =}
follows from equalities
\EqRef{expansion Ik a x},
\EqRef{b=Ikl a}.
From equalities
\ShowEq{Ikl a1+a2}
\ShowEq{Ikl d a}
it follows that the map $I_k^l$ is linear map.
\end{proof}
}
{proof: conjugation transformation}

\DefEq
{
\begin{theorem}
\labelTheorem{representation of composition of linear maps}
Let $A_1$ be free $D$\Hyph module.
Let $A_2$, $A_2$ be free associative $D$\Hyph algebras.
Let $\Basis F$ be the basis of left \BoxB{A_2}module
\ShowEq{L(A;B)}D{A_1}{A_2}.
Let $\Basis G$ be the basis of left \BoxB{A_3}module
\ShowEq{L(A;B)}D{A_2}{A_3}.
\StartLabelItem
\begin{enumerate}
\item
The set of maps
\labelItem{basis of composition of linear maps}
\DrawEq{JlIk}{linear map}
is the basis of left \BoxB{A_3}module
\ShowEq{L(A;B)}D{A_1\rightarrow A_2}{A_3}.
\item
Let
\ShowEq{expansion of f with respect to basis I}
be expansion of linear map
\ShowEq{f:A->B}f{A_1}{A_2}
with respect to the basis $\Basis I$.
Let
\ShowEq{expansion of g with respect to basis J}
be expansion of linear map
\ShowEq{f:A->B}g{A_2}{A_3}
with respect to the basis $\Basis J$.
Then linear map
\DrawEq{h=g o f}{123}
has expansion
\labelItem{hlk=glfk}
\ShowEq{expansion of h with respect to basis K}
with respect to the basis $\Basis K$ where
\ShowEq{hlk=glfk}
\end{enumerate}
\end{theorem}
}
{theorem: representation of composition of linear maps}

\DefEq
{
\begin{proof}
The equality
\ShowEq{h(a)1}
follows from equalities
\EqRef{expansion of f with respect to basis I},
\EqRef{expansion of g with respect to basis J},
\eqRef{h=g o f}{123}.
The equality
\ShowEq{h(a)2}
follows from equalities
\eqRef{JlIk}{linear map},
\EqRef{h(a)1}
and from the theorem
\RefTheorem{conjugation transformation}.
From the equality
\EqRef{h(a)2}
it follows that set of maps $\Basis K$ generates
left \BoxB{A_3}module\newline
\ShowEq{L(A;B)}D{A_1\rightarrow A_2}{A_3}.
From the equality
\ShowEq{aK=aJI}
it follows that
\ShowEq{aJ=0}
and, therefore, $a^{lk}=0$.
Therefore, the set $\Basis K$ is the basis of
left \BoxB{A_3}module
\ShowEq{L(A;B)}D{A_1\rightarrow A_2}{A_3}.
\end{proof}
}
{proof: representation of composition of linear maps}

\DefEq
{
\begin{theorem}
\labelTheorem{representation of composition of linear maps A->A}
Let $A$ be free associative $D$\Hyph algebra.
Let left \BoxB{A}module
\ShowEq{L(A;B)}DAA{}
is generated by the identity map $F_0=\delta$.
Let
\ShowEq{expansion of f A->A}
be expansion of linear map
\ShowEq{f:A->B}fAA
Let
\ShowEq{expansion of g A->A}
be expansion of linear map
\ShowEq{f:A->B}gAA
Then linear map
\DrawEq{h=g o f}{A}
has expansion
\ShowEq{expansion of h A->A}
where
\ShowEq{hlk=glfk 01}
\end{theorem}
}
{theorem: representation of composition of linear maps A->A}

\DefEq
{
\begin{proof}
The equality
\ShowEq{h(a)}
follows from equalities
\EqRef{expansion of f A->A},
\EqRef{expansion of g A->A},
\eqRef{h=g o f}{A}.
The equality
\EqRef{hlk=glfk 01}
follows from the equality
\EqRef{h(a)}.
\end{proof}
}
{proof: representation of composition of linear maps A->A}

\DefEq
{
\begin{theorem}
\labelTheorem{h generated by f, associative algebra}
Consider $D$\Hyph algebra $A_1$
and associative $D$\Hyph algebra $A_2$.
Consider the representation of algebra $\ATwo$
in the module $\mathcal L(D;A_1;A_2)$.
The map
\ShowEq{h:A1->A2}
generated by the map
\ShowEq{f:A->B}f{A_1}{A_2}
has form
\ShowEq{h generated by f, associative algebra}
\end{theorem}
}
{theorem: h generated by f, associative algebra}

\DefTheorem{coordinates of map A->A, algebra}
{
Let $A$ be free finite dimensional associative $D$\Hyph algebra.
Let $\Basis e$ be basis of $D$\Hyph module $A$.
Let
\ShowEq{structural constants, algebra}
be structural constants of algebra $A$.
Let $\Basis F$ be the basis
of left \BoxB{A}module
\ShowEq{L(A;B)}DAA{}
and
\ShowEq{coordinates of map Ik}
be coordinates of map $F_k$ with respect to basis $\Basis e$.
Coordinates
\ShowEq{Coordinates of map f}
of the map
\ShowEq{f in L(A->B)}DAA{}
and its standard components
\ShowEq{standard components of map f}k
are connected by the equation
\DrawEq{coordinates of map A->A, 2}{associative algebra}
}

\DefProof{coordinates of map A->A, algebra}
{
Relative to basis
$\Basis e$, linear maps $f$ and $I_k$ have form
\ShowEq{coordinates of map f, associative algebra}
\ShowEq{coordinates of map Fk, associative algebra}
The equality
\ShowEq{coordinates of map A->A, 3, associative algebra}
follows from equalities
\EqRef{standard representation of map A1 A2, associative algebra},
\EqRef{coordinates of map f, associative algebra},
\EqRef{coordinates of map Fk, associative algebra}.
Since vectors $\aD ek$
are linear independent and $x^{\gi i}$ are arbitrary,
then the equation
\eqRef{coordinates of map A->A, 2}{associative algebra}
follows from the equation
\EqRef{coordinates of map A->A, 3, associative algebra}.
}

\DefTheorem{coordinates of map A1 A2 FoG, algebra}
{
Let $\Basis e_1$ be basis of the finite dimensional
$D$\Hyph module $A_1$.
Let $\Basis e_2$ be basis of the finite dimensional associative
$D$\Hyph algebra $A_2$.
Let
\ShowEq{f in L(A->B)}D{A_1}{A_2}.
Let
\ShowEq{structural constants, algebra}
be structural constants of algebra $A_2$.
Let $\Basis F$ be the basis
of left \BoxB{A_2}module
\ShowEq{L(A;B)}D{A_2}{A_2}{}
and
\ShowEq{coordinates of map Ik}
be coordinates of map $F_k$ with respect to basis $\Basis e_2$.
Let
\ShowEq{f:A->B}GAB
be linear map of maximal rank such that
\ShowEq{ker G in ker g}f{}
and
\ShowEq{coordinates of map G}
be coordinates of map $G$ with respect to bases $\Basis e_1$ and $\Basis e_2$.
Coordinates
\ShowEq{Coordinates of map f}
of the map $f$
and its standard components
\ShowEq{standard components of map f}k
are connected by the equation
\DrawEq[f-]{coordinates of map A1 A2, FoG, associative algebra}f
}

\AddEq[1]{theorem: linear map, maps of conjugation, algebra}
{
\begin{theorem}
\labelTheorem{linear map, maps of conjugation, algebra #1}
A linear map
\ShowEq{f in L(A->B)}R{#1}{#1}{}
have expansion
\ShowEq{linear map of algebra #1, structure, 1}
\ShowEq{linear map of algebra #1, structure, 2}
where $#1$\Hyph numbers
\ShowEq{a... #1}
are defined by the equality
\ShowEq{a...= #1}
\end{theorem}
}

\AddEq [1]{proof: L is left vector space}
{
\begin{proof}
According to the theorem
\RefTheorem{linear map, maps of conjugation, algebra #1},
the expansion
\EqRef{linear map of algebra #1, structure, 1}
of the linear map $f$
exists and is unique.
Therefore, the set
\ShowEq{I=(E,I)}{#1}{}
is basis of left $#1$\Hyph vector space
\ShowEq{L(A;B)}R{#1}{#1}.
\end{proof}
}

\DefTheorem{f=.E+.I+.J+.K}
{
Let
\ShowEq{f:H->H ij}
be linear map of quaternion algebra.
Let
\DrawEq{fi=fij ej}H
Then
\ShowEq{f=.E+.I+.J+.K}
}

\DefProof{f=.E+.I+.J+.K}
{
Equalities
\ShowEq{a0=f03}
\ShowEq{a1=f03}
\ShowEq{a2=f03}
\ShowEq{a3=f03}
follow from the equality
\EqRef{a...= H}.
The equalitiy
\EqRef{f=.E+.I+.J+.K}
follows from equalities
\ShowEq{f=.E+.I+.J+.K ref}
}

\DefTheorem{f=.I0.7}
{
Let
\ShowEq{f:H->H ij}
be linear map of quaternion algebra.
Let
\DrawEq{fi=fij ej}H
Then
\ShowEq{f=.I0.7}
}

\DefProof{f=.I0.7}
{
Equalities
\ShowEq{a0=f0.7}
follow from the equality
\EqRef{a...= H}.
The equalitiy
\EqRef{f=.I0.7}
follows from equalities
\ShowEq{f=.I0.7 ref}
}

\DefDefinition{maps of conjugation, complex field}
{
\ShowEq{C maps of conjugation}
Complex field has following
\AddIndex{maps of conjugation}{map of conjugation}
\ShowEq{C list maps of conjugation}
}

\DefTheorem{HE is algebra isomorphic to quaternion algebra}
{
The set
\ShowEq{HE set}
is $R$\Hyph algebra isomorphic to quaternion algebra.
}

\DefProof{HE is algebra isomorphic to quaternion algebra}
{
The theorem follows from equalities
\ShowEq{aE+bE=(a+b)E}
\ShowEq{aE o bE=(ab)E}
based on the theorem
\RefTheorem{linear map, maps of conjugation, algebra H}.
}

\DefDefinition{algebra, left and right action on L}
{
Let $A$ be $D$\Hyph module
and $B$ be $D$\Hyph algebra.
For any map
\ShowEq{f in L(A->B)}DAB{}
and $b\in B$,
we define left side transformation of the map $f$ using equality
\ShowEq{b o f =}
and right side transformation of the map $f$ using equality
\ShowEq{b * f =}
}

\DefDefinition{projection maps, complex field}
{
Following projection maps are defined in complex field
\ShowEq{projection maps, complex field}
}

\DefTheorem{Expansion of projection maps relative E,I}
{
Expansion of projection maps relative basis
\ShowEq{e=(E,I)}{}
has form
\ShowEq{projection mappings 1, complex field}
}

\DefProof{Expansion of projection maps relative E,I}
{
The theorem follows from the theorem
\RefTheorem{linear map of complex field}
and from the definition
\RefDefinition{projection maps, complex field}.
}

\DefTheorem{f=.E+.I}
{
Let
\ShowEq{f:A->B}fCC
be linear map of complex field.
Let
\DrawEq{fi=fij ej}C
Then
\ShowEq{f=.E+.I}
}

\DefProof{f=.E+.I}
{
Equalities
\ShowEq{a0=f01}
\ShowEq{a1=f01}
follow from the equality
\EqRef{a01=}.
The equalitiy
\EqRef{f=.E+.I}
follows from equalities
\ShowEq{f=.E+.I ref}
}

\DefTheorem{linear map of complex field}
{
A linear map
\ShowEq{f in LRC y=fx}
of complex field has form
\ShowEq{linear map of algebra C, structure, 1}
\ShowEq{linear map of algebra C, structure, 2}
where $C$\Hyph numbers
\ShowEq{a... C}
are defined by the equality
\ShowEq{a01=}
}

\DefDefinition{polylinear map of algebras}
{
Let $A_1$, ..., $A_n$, $S$ be $D$\Hyph algebras.
Polylinear map
\ShowEq{polylinear map of algebras}
of $D$\Hyph modules
$A_1$, ..., $A_n$
into $D$\Hyph module $S$
is called
\AddIndex{polylinear map}{polylinear map} of $D$\Hyph algebras
$A_1$, ..., $A_n$
into $D$\Hyph algebra $S$.
Let us denote
\ShowEq{set polylinear maps}
set of polylinear maps
of $D$\Hyph algebras
$A_1$, ..., $A_n$
into $D$\Hyph algebra
$S$.
Let us denote
\ShowEq{set polylinear maps An}
set of $n$\hyph linear maps
of $D$\Hyph algebra $A$ ($A_1=...=A_n=A$)
into $D$\Hyph algebra
$S$.
}

\DefEq
{
\begin{theorem}
\labelTheorem{sum of linear maps, D module}
Let $A_1$, $A_2$ be $D$\Hyph modules.
The map
\ShowEq{sum of maps,,D module}
\ShowEq{sum of maps, 1, D module}
defined by equation
\ShowEq{sum of maps, D module}
is called
\AddIndex{sum of maps}{sum of maps}
$f$ and $g$
and is linear map.
The set
$\mathcal L(D;A_1;A_2)$
is an Abelian group
relative sum of maps.
\end{theorem}
}
{theorem: sum of linear maps, D module}

\DefEq
{
\begin{theorem}
\labelTheorem{sum of polylinear maps, module}
Let $D$ be the commutative ring.
Let $A_1$, ..., $A_n$, $S$ be $D$\Hyph modules.
The map
\ShowEq{sum of maps,,polylinear}
\ShowEq{sum of maps, 1, polylinear}
defined by the equality
\ShowEq{sum of  maps, polylinear}
is called
\AddIndex{sum of polylinear maps}{sum of maps}
$f$ and $g$
and is polylinear map.
The set
\ShowEq{module of polylinear maps}
is an Abelian group
relative sum of maps.
\end{theorem}
}
{theorem: sum of polylinear maps, module}

\DefCorollary{sum of linear maps, D module}
{
Let $A_1$, $A_2$ be $D$\Hyph modules.
The map
\ShowEq{sum of maps,,D module}
\ShowEq{sum of maps, 1, D module}
defined by equation
\ShowEq{sum of maps, D module}
is called
\AddIndex{sum of maps}{sum of maps}
$f$ and $g$
and is linear map.
The set
$\mathcal L(D;A_1;A_2)$
is an Abelian group
relative sum of maps.
}

\AddEq{definition: linear map A module}
{
\begin{definition}
\labelDefinition{linear map \SideWS A module}
Let
\ShowEq{Ai, i=}A
be algebra over commutative ring $D_i$.
Let
\ShowEq{Ai, i=}V
be
\ShowEq{\SideWS column module}{A_i}
module.
Morphism of diagram of representations
\ShowEq{A module diagram for linear}1
into diagram of representations
\ShowEq{A module diagram for linear}2
is called
\AddIndex{linear map}{linear map}
of
\ShowEq{\SideWS column module}{A_1}
module $V_1$ into
\ShowEq{\SideWS column module}{A_2}
module $V_2$.
Let us denote
\ShowEq{set linear maps, \SideWS A12 module}
set of linear maps
of
\ShowEq{\SideWS column module}{A_1}
module $V_1$ into
\ShowEq{\SideWS column module}{A_2}
module $V_2$.
\qed
\end{definition}
}

\AddEq{definition: homomorphism A module}
{
\begin{definition}
\labelDefinition{homomorphism \SideWS A module}
Let
\ShowEq{Ai, i=}A
be algebra over commutative ring $D_i$.
Let
\ShowEq{Ai, i=}V
be
\ShowEq{\SideWS column module}{A_i}
module.
Morphism of diagram of representations
\ShowEq{A module diagram for homomorphism}1
into diagram of representations
\ShowEq{A module diagram for homomorphism}2
is called
\AddIndex{homomorphism}{homomorphism}
of
\ShowEq{\SideWS column module}{A_1}
module $V_1$ into
\ShowEq{\SideWS column module}{A_2}
module $V_2$.
Let us denote
\ShowEq{set homomorphisms, \SideWS A12 module}
set of homomorphisms
of
\ShowEq{\SideWS column module}{A_1}
module $V_1$ into
\ShowEq{\SideWS column module}{A_2}
module $V_2$.
\qed
\end{definition}
}

\DefTheorem{L(An;B) is free D module}
{
Let $A_1$, ..., $A_n$, $B$ be free modules over commutative ring $D$.
$D$\Hyph module
\ShowEq{L(A;B)}D{A_1\times...\times A_n}B{}
is free $D$\Hyph module.
}

\DefEq
{
\begin{definition}
\labelDefinition{linear map from A1 to A2, module}
Reduced morphism of representations
\ShowEq{f:A->B}f{A_1}{A_2}
of $D$\Hyph module $A_1$
into $D$\Hyph module $A_2$
is called
\AddIndex{linear map}{linear map}
of \SideWS $D$\Hyph module $A_1$ into \SideWS $D$\Hyph module $A_2$.

Let us denote
\ShowEq{set linear maps, module}
set of linear maps
of \SideWS $D$\Hyph module $A_1$ into \SideWS $D$\Hyph module $A_2$.
\qed
\end{definition}
}
{definition: linear map from A1 to A2, module}

\AddEq{theorem: linear map of module}
{
\begin{theorem}
\labelTheorem{linear map, \Base\DF \Module\VF->\Base\DT \Module\VT, \SideWS module}
Linear map
\ShowEq{show linear map \MapE}hf
of \SideWS $\Base_{\DF}$\Hyph module $\Module_\VF$
into \SideWS $\Base_{\DT}$\Hyph module $\Module_\VT$
satisfies to equations\,\footnote{
In some books
(for instance, on page \citeBib{Serge Lang}\Hyph 119) the theorem
\RefTheorem{linear map, \Base\DF \Module\VF->\Base\DT \Module\VT, \SideWS module}
is considered as a definition.}
\ShowEq{property linear map \MapE}
\end{theorem}
}

\DefProof{linear map, DA1->DA2, module}
{
From definitions
\RefDefinition[\RefRepresentation]{reduced morphism of representations},
\RefDefinition{linear map from A1 to A2, commutative module},
it follows that
the map $f$ is a homomorphism of the Abelian group $A_1$
into the Abelian group $A_2$ (the equality
\eqRef{f(a+b)=...}{D }).
The equality
\eqRef{f(da)=h... module}{\DFDT \SideWS module}
follows from the equality
\EqRef[\RefRepresentation]{morphism of representations of universal algebra}.
}

\DefDefinition{polylinear map of modules}
{
Let $D$ be the commutative ring.
Reduced polymorphism of $D$\Hyph modules
$A_1$, ..., $A_n$ into $D$\Hyph module $S$
\ShowEq{polylinear map of algebras}
is called
\AddIndex{polylinear map}{polylinear map} of $D$\Hyph modules
$A_1$, ..., $A_n$
into $D$\Hyph module $S$.
We denote
\ShowEq{set polylinear maps}
the set of polylinear maps
of $D$\Hyph modules
$A_1$, ..., $A_n$
into $D$\Hyph module
$S$.
Let us denote
\ShowEq{set polylinear maps An}
set of $n$\hyph linear maps
of $D$\Hyph module $A$ ($A_1=...=A_n=A$)
into $D$\Hyph module
$S$.
}

\DefTheorem{polylinear map of modules}
{
Let $D$ be the commutative ring.
The polylinear map of $D$\Hyph modules
$A_1$, ..., $A_n$
into $D$\Hyph module $S$
\ShowEq{polylinear map of algebras}
satisfies to equalities
\DrawEq[f]{f(ai+bi)=fai+fbi}{}
\DrawEq[f]{f(pai)=pfai}{}
\ShowEq{polylinear map of algebras, 1}
}

\DefProof{polylinear map of modules}
{
The theorem follows from definitions
\RefDefinition[\RefRepresentation]{reduced polymorphism of representations},
\RefDefinition{linear map from A1 to A2, commutative module},
\RefDefinition{polylinear map of modules}
and from the theorem
\RefTheorem{linear map, DA1->DA2, module}.
}

\DefProof{sum of polylinear maps, module}
{
According to the theorem
\RefTheorem{polylinear map of modules}
\DrawEq[f]{f(ai+bi)=fai+fbi}{sum of maps f}
\DrawEq[f]{f(pai)=pfai}{sum of maps f}
\DrawEq[g]{f(ai+bi)=fai+fbi}{sum of maps g}
\DrawEq[g]{f(pai)=pfai}{sum of maps g}
The equality
\ShowEq{sum of maps, 31, polylinear}
follows from the equalities
\EqRef{sum of maps, polylinear},
\eqRef{f(ai+bi)=fai+fbi}{sum of maps f},
\eqRef{f(ai+bi)=fai+fbi}{sum of maps g}.
The equality
\ShowEq{sum of maps, 32, polylinear}
follows from the equalities
\EqRef{sum of maps, polylinear},
\eqRef{f(pai)=pfai}{sum of maps f},
\eqRef{f(pai)=pfai}{sum of maps g}.
From equalities
\EqRef{sum of maps, 31, polylinear},
\EqRef{sum of maps, 32, polylinear}
and from the theorem
\RefTheorem{polylinear map of modules},
it follows that the map
\EqRef{sum of maps, 1, polylinear}
is linear map of $D$\Hyph modules.

Let
\ShowEq{module of polylinear maps, 1}
For any
\ShowEq{module of polylinear maps, 2}
Therefore, sum of polylinear maps is commutative and associative.

From the equality
\EqRef{sum of maps, polylinear},
it follows that the map
\ShowEq{0:A1n->S}
is zero of addition
\ShowEq{0+f=f 1n}
From the equality
\EqRef{sum of maps, polylinear},
it follows that the map
\ShowEq{-f:A1n->S}
is map inversed to map \(f\)
\ShowEq{f-f=0}
because
\ShowEq{f-f=0 1n}
From the equality
\ShowEq{sum of maps, 4, polylinear}
it follows that sum of maps is commutative.
Therefore, the set
\ShowEq{module of polylinear maps}
is an Abelian group.
}

\DefTheorem{module of polylinear maps}
{
Let $D$ be the commutative ring.
Let $A_1$, ..., $A_n$, $S$ be $D$\Hyph modules.
The map
\ShowEq{product of map over scalar,,polylinear}
\ShowEq{product of map over scalar, 1, polylinear}
defined by equality
\ShowEq{product of map over scalar, polylinear}
is polylinear map
and is called
\AddIndex{product of map $f$ over scalar}
{product of map over scalar} $d$.
The representation
\ShowEq{a:LA1n->LA1n}
of ring $D$ in Abelian group
\ShowEq{module of polylinear maps}
generates structure of $D$\Hyph module.
}

\DefProof{module of polylinear maps}
{
According to the theorem
\RefTheorem{polylinear map of modules}
\DrawEq[f]{f(ai+bi)=fai+fbi}{product of map over scalar}
\DrawEq[f]{f(pai)=pfai}{product of map over scalar}
The equality
\ShowEq{product of map over scalar, 31, polylinear}
follows from equalities
\EqRef{product of map over scalar, polylinear},
\eqRef{f(ai+bi)=fai+fbi}{product of map over scalar}.
The equality
\ShowEq{product of map over scalar, 32, polylinear}
follows from equalities
\EqRef{product of map over scalar, polylinear},
\eqRef{f(pai)=pfai}{product of map over scalar}.
From equalities
\EqRef{product of map over scalar, 31, polylinear},
\EqRef{product of map over scalar, 32, polylinear}
and from the theorem
\RefTheorem{polylinear map of modules},
it follows that the map
\EqRef{product of map over scalar, 1, polylinear}
is polylinear map of $D$\Hyph modules.

The equality
\DrawEq{(p+q)f=pf+qf}{polylinear}
follows from the equality
\ShowEq{(p+q)f=pf+qf 1}
The equality
\DrawEq{p(qf)=(pq)f}{polylinear}
follows from the equality
\ShowEq{p(qf)=(pq)f 1}
From equalities
\eqRef{(p+q)f=pf+qf}{polylinear}
\eqRef{p(qf)=(pq)f}{polylinear}
it follows that the map
\EqRef{a:LA1n->LA1n}
is representation of ring $D$
in Abelian group
\ShowEq{module of polylinear maps}.
Since specified representation is effective,
then, according to the definition
\RefDefinition{module over commutative ring}
and the theorem
\RefTheorem{sum of polylinear maps, module},
Abelian group
\ShowEq{L(A;B)}D{A_1}{A_2}{}
is $D$\Hyph module.
}

\DefCorollary{product of linear map over scalar, D module}
{
Let $A_1$, $A_2$ be $D$\Hyph modules.
The map
\ShowEq{product of map over scalar,,D module}
\ShowEq{product of map over scalar, 1, D module}
defined by equality
\ShowEq{product of map over scalar, D module}
is linear map
and is called
\AddIndex{product of map $f$ over scalar}
{product of map over scalar} $d$.
The representation
\ShowEq{a:LA12->LA12}
of ring $D$ in Abelian group
\ShowEq{L(A;B)}D{A_1}{A_2}{}
generates structure of $D$\Hyph module.
}

\AddEq{theorem: linear map of A module}
{
\begin{theorem}
\labelTheorem{linear map, \Base\DF \Module\VF->\Base\DT \Module\VT, \SideWS module}
Linear map
\ShowEq{show linear map \MapE}gf
of \SideWS $\Base_\DF$\Hyph module $\Module_\VF$
into \SideWS $\Base_\DT$\Hyph module $\Module_\VT$
satisfies to equations\,\footnote{
In some books
(for instance, on page \citeBib{Serge Lang}\Hyph 119) the theorem
\RefTheorem{linear map from \Base\DF \Module\DF to \Base\DT \Module\DT, \SideWS module}
is considered as a definition.}
\ShowEq{property linear map \MapE}
\end{theorem}
}

\DefProof{linear map of module}
{
From definitions
\RefDefinition[\RefRepresentation]{morphism of representations of universal algebra},
\RefDefinition{linear map from D1 A1 to D2 A2, module},
it follows that
the map $h$ is a homomorphism of the ring $D_1$
into the ring $D_2$ (the equalities
\eqRef{h(d1+d2)=...}{\SideWS module},
\eqRef{h(d1d2)=...}{\SideWS module})
and the map $f$ is a homomorphism of the Abelian group $A_1$
into the Abelian group $A_2$ (the equality
\eqRef{f(a+b)=...}{D1 D2 \SideWS module}).
The equality
\eqRef{f(da)=h... \SideWS module}{\DFDT \SideWS module}
follows from the equality
\eqRef[\RefRepresentation]{morphism of representations of universal algebra, 2m}{definition}.
}

\DefProof{linear map of A module}
{
From definitions
\RefDefinition[\RefRepresentation]{morphism of representations of universal algebra},
\RefDefinition{linear map \SideWS A module},
it follows that
\begin{itemize}
\item
the map $h$ is a homomorphism of the ring $D_1$
into the ring $D_2$ (the equalities
\eqRef{h(d1+d2)=...}{\SideWS module},
\eqRef{h(d1d2)=...}{\SideWS module});
\item
the map $g$ is homomorphism of the Abelian group $A_1$
into the Abelian group $A_2$ (the equality
\eqRef{f(a+b)=...}{g \DFDT \SideWS module});
\item
the map $f$ is homomorphism of the Abelian group $V_1$
into the Abelian group $V_2$ (the equality
\eqRef{f(a+b)=...}{f \DFDT \SideWS module}).
\end{itemize}
Equalities
\eqRef{f(da)=h...}{g \DFDT \SideWS module},
\eqRef{f(da)=h...}{f \DFDT \SideWS module}
follow from the equality
\EqRef[\RefRepresentation]{morphism of representations of F algebra, definition, 2m}.
}

\DefTheorem{complex field over real field}
{
Consider complex field $C$ as two-dimensional algebra over real field.
Let
\ShowEq{basis of complex field}
be the basis of algebra $C$.
Then in this basis product has form
\ShowEq{product of complex field}
and structural constants have form
\ShowEq{structural constants of complex field}
}

\DefProof{complex field over real field}
{
Equalities
\EqRef{product of complex field} and
\EqRef{structural constants of complex field}
follow from the equality $i^2=-1$.
}

\AddEq[1]{theorem: maps of conjugation antilinear}
{
\begin{theorem}
\labelTheorem{#1 maps of conjugation antilinear}
Maps of conjugation
\ShowEq{I... #1}
are \DoVerb antilinear homomorphisms.
\end{theorem}
}

\DefProof{H maps of conjugation antilinear}
{
The product of $H$\Hyph numbers
\ShowEq{H number Ea}a
and
\ShowEq{H number Ea}b
has form
\DrawEq[ab]{H product aEx}{ab}
\ShowEq{H items maps of conjugation antilinear}
}

\DefProof{O maps of conjugation antilinear}
{
The product of $O$\Hyph numbers
\ShowEq{O number Ea}a
and
\ShowEq{O number Ea}b
has form
\DrawEq[ab]{O product aEx}{ab}
To prove the theorem, it is enough to consider map $\aU I1$,
because the proof is the same for other maps.
\ShowEq{item maps of conjugation antilinear}O1
}

\AddEq[3]{theorem: ao Jacobian matrix}
{
\begin{theorem}
\labelTheorem{a#1, #2, Jacobian matrix}
The map
\DrawEq[{#1}{#2}]{map ao}{#1#2}
has matrix
\ShowEq{maps of conjugation, Jacobian matrix}#1#2{#3}
\DrawEq[{#1}{#2}]{a o, Jacobian matrix}{#1#2}
\end{theorem}
}

\AddEq[2]{proof: ao Jacobian matrix}
{
\begin{proof}
The product of $#2$\Hyph numbers
\ShowEq{#2 number Ea}a
and
\ShowEq{#2 number #1a}x
has form
\DrawEq[ax]{#2 product a#1x}{}
Therefore, function
\eqRef{map ao}{#1#2}
has Jacobian matrix
\eqRef{a o, Jacobian matrix}{#1#2}.
\end{proof}%
}

\AddEq [2]{item maps of conjugation antilinear}
{

The equality
\ShowEq{#1#2 product ab}
follows from the equalities
\EqRef{#2x= #1},
\eqRef{#1 product aEx}{ab}.
From the equality
\eqRef{#1 product aEx}{ab},
it follows that product of $#1$\Hyph numbers
\ShowEq{#1 number #2a}b
and
\ShowEq{#1 number #2a}a
has form
\ShowEq{#1 #2b*#2a}
From equalities
\EqRef{#1#2 product ab},
\EqRef{#1 #2b*#2a},
it follows that the map $\aU I#2$
is \DoVerb linear antihomomorphism.
}

\AddEq[2]{theorem: ao Jacobian matrix 1}
{
\begin{theorem}
\labelTheorem{a#1, #2, Jacobian matrix 1}
\DrawEq[#1]{ao, #2 matrix 1}{#1}
\end{theorem}
}

\AddEq[2]{proof: ao Jacobian matrix 1}
{
\begin{proof}
The equality
\eqRef{ao, #2 matrix 1}{#1}
follows from the chain of equalities
\ShowEq{a#1, #2 matrix 2}
\end{proof}%
}

\AddEq[2]{theorem: a* Jacobian matrix}
{
\begin{theorem}
\labelTheorem{a*#1, #2, Jacobian matrix}
The map
\DrawEq[{#1}{#2}]{map a*}{#1#2}
has matrix
\ShowEq{maps of conjugation, Jacobian matrix}#1#2r
\DrawEq[{#1}{#2}]{a *, Jacobian matrix}{#1#2}
\end{theorem}
}

\AddEq[2]{proof: a* Jacobian matrix}
{
\begin{proof}
The product of $#2$\Hyph numbers
\ShowEq{#2 number #1a}x
and
\ShowEq{#2 number Ea}a
has form
\ShowEq{#2 product #1xa}
Therefore, function
\eqRef{map a*}{#1#2}
has Jacobian matrix
\eqRef{a *, Jacobian matrix}{#1#2}.
\end{proof}%
}

\AddEq[2]{theorem: a* Jacobian matrix 1}
{
\begin{theorem}
\labelTheorem{a*#1, #2, Jacobian matrix 1}
\DrawEq[#1]{a*, #2 matrix 1}{#1}
\end{theorem}
}

\AddEq[2]{proof: a* Jacobian matrix 1}
{
\begin{proof}
The equality
\eqRef{a*, #2 matrix 1}{#1}
follows from the chain of equalities
\ShowEq{#1a, #2 matrix 2}
\end{proof}%
}

\DefDefinition{antilinear homomorphism}
{
The map
\ShowEq{f in L(A->B)}DAA{}
is called
\AddIndex{antilinear homomorphism}{antilinear homomorphism}
if the map $f$ satisfies the equality
\ShowEq{fab=fbfa}
}

\DefDefinition{quaternion maps of conjugation}
{
\ShowEq{H maps of conjugation}
Quaternion algebra has following
\AddIndex{maps of conjugation}{map of conjugation}
\ShowEq{H list maps of conjugation}
We also use notation
\ShowEq{I0=E}
}

\DefDefinition{octonion maps of conjugation}
{
\ShowEq{O maps of conjugation}
Octonion algebra has following
\AddIndex{maps of conjugation}{map of conjugation}
\ShowEq{O list maps of conjugation}
We also use notation
\ShowEq{I0=E}
}

\AddEq[1]{theorem: L is left vector space}
{
\begin{theorem}
\labelTheorem{L(#1->#1) is left #1-vector space}
$#1\otimes #1$\Hyph module
\ShowEq{L(A;B)}R{#1}{#1}{}
is left $#1$\Hyph vector space
and has the basis
\ShowEq{I=(E,I)}{#1}.
\end{theorem}
}

\AddEq [6]{Let e be basis}
{
Let
\ShowEq{basis e of module}{#1}{#2}{#3}
be a basis of #4 $#5$\Hyph module $#6$.
}

\AddEq [5]{Let e be basis 1n}
{
Let
\ShowEq{basis e of module 1n}{#1}{#2}
be a basis of free #3 $#4$\Hyph module $#5$.
}

\AddEq [9]{coordinates of the linear map}
{
\item $#1$ is coordinate matrix of $#2_1$\Hyph number
$\Vector #1$
relative the basis
\ShowEq{basis e}1{}
\DrawEq [#1{1}]{va=ae1, \SideWS module}{#1 \DFDT\SideWS module}
\def\Temp{D1 D2 }
\ifx\DFDT\Temp
\item
\ShowEq{h(a)=...}{#1}{#4}{#5}{#6}
is a matrix of $#7$\Hyph numbers.
\fi
\item $#9$ is coordinate matrix of vector
\DrawEq[#1#9#3]{vb=f(va)}{#1 \DFDT\SideWS module}
relative the basis
\ShowEq{basis e}2{}
\DrawEq [#9{2}]{va=ae1, #8module}{#9 \DFDT\SideWS module}
\item $#3$ is coordinate matrix of set of vectors
\ShowEq{Vector f(e1) module}#3#2#5#6
relative the basis
\ShowEq{basis e}2.
The matrix $#3$ is
called \AddIndex{matrix of linear map}{matrix of linear map}
$\Vector #3$ relative bases
\ShowEq{basis e}1{}
and
\ShowEq{basis e}2.
}

\AddEq{theorem: linear map of module, coordinates}
{
\begin{theorem}
\labelTheorem{linear map of \SideWS \DFDT module, coordinates}
\ShowEq{Let e be basis}1iI{}{\Base_{\DF}}{\Module_1}
\ShowEq{Let e be basis}2jJ{}{\Base_{\DT}}{\Module_2}
Then linear map
\ShowEq{show linear map \MapE}h{\Vector f}
has presentation
\DrawEq[f]{r2:A1->A2, \DFDT module}{1 \SideWS f}
relative to selected bases. Here
\begin{itemize}
\ShowEq{coordinates of the linear map}a{\Module}fhiI{D_2}{}b
\end{itemize}
\end{theorem}
}

\DefProof{linear map of module, coordinates}
{
Since
\ShowEq{show linear map \MapE}h{\Vector f}
is a linear map, then the equality
\ShowEq{\DFDT vb=h(e1)a \SideNS}
follows from equalities
\eqRef{f(da)=h... \SideWS module}{\DFDT \SideWS module},
\eqRef{va=ae1, \SideWS module}{a \DFDT\SideWS module},
\eqRef{vb=f(va)}{a \DFDT\SideWS module}.
$\Module_2$\Hyph number
\ShowEq{f(e1) \SideNS}
has expansion
\DrawEq{f(e1)= \SideNS}{\DFDT}
relative to basis $\Basis e_2$.
Combining \EqRef{\DFDT vb=h(e1)a \SideNS}
and \eqRef{f(e1)= \SideNS}{\DFDT}, we get
\ShowEq{\DFDT vb=a f e \SideNS}
\eqRef{r2:A1->A2, \DFDT module}{1 \SideWS f}
follows from comparison of
\eqRef{va=ae1, \SideWS module}{b \DFDT\SideWS module}
and \EqRef{\DFDT vb=a f e \SideNS} and
theorem \RefTheorem{coordinates of vector of free \SideWS module}.
}

\DefDefinition{algebra over ring}
{
Let $D$ be commutative ring.
$D$\Hyph module $A$ is called
\AddIndex{algebra over ring}{algebra over ring} $D$
or
\AddIndex{$D$\Hyph algebra}{D algebra},
if we defined product\,\footnote{
I follow the definition
given in
\citeBib{Richard D. Schafer}, page 1,
\citeBib{0105.155}, page 4.  The statement which
is true for any $D$\Hyph module,
is true also for $D$\Hyph algebra.} in $A$
\DrawEq{product in D algebra}{definition}
where $C$ is bilinear map
\ShowEq{product in algebra, definition 1}
If $A$ is free
$D$\Hyph module, then $A$ is called
\AddIndex{free algebra over ring}{free algebra over ring} $D$.
}

\DefDefinition{commutator of algebra}
{
The \AddIndex{commutator}{commutator of algebra}
\ShowEq{commutator of algebra}
measures commutativity in $D$\Hyph algebra $A$.
$D$\Hyph algebra $A$ is called
\AddIndex{commutative}{commutative D algebra},
if
\ShowEq{commutative D algebra}
}

\DefDefinition{nucleus of algebra}
{
The set\,\footnote{
The definition is based on
the similar definition in
\citeBib{Richard D. Schafer}, p. 13}
\ShowEq{nucleus of algebra}
is called the
\AddIndex{nucleus of an $D$\Hyph algebra $A$}{nucleus of algebra}.
}

\DefDefinition{associator of algebra}
{
The \AddIndex{associator}{associator of algebra}
\ShowEq{associator of algebra}
\ShowEq{associator of algebra =}
measures associativity in $D$\Hyph algebra $A$.
$D$\Hyph algebra $A$ is called
\AddIndex{associative}{associative D algebra},
if
\ShowEq{associative D algebra}
}

\DefTheorem{associator of algebra}
{
Let $\Basis e$ be basis of $D$\Hyph algebra $A$. Then
\ShowEq{associator of algebra = A}
where
\AddIndex{coordinates of associator}{coordinates of associator}
are defined by the equality
\ShowEq{coordinates of associator}
\ShowEq{coordinates of associator =}
}

\DefProof{associator of algebra}
{
The equality
\ShowEq{associator of algebra = Ae}
follows from the equality
\EqRef{associator of algebra =}.
The equality
\EqRef{coordinates of associator =}
follows from the equality
\EqRef{associator of algebra = Ae}.
}

\DefEq
{
\begin{definition}
\labelDefinition{center of algebra}
The set\,\footnote{
The definition is based on
the similar definition in
\citeBib{Richard D. Schafer}, page 14}
\ShowEq{center of algebra}
is called the
\AddIndex{center of an $D$\Hyph algebra $A_1$}{center of algebra}.
\qed
\end{definition}
}
{definition: center of algebra}

\DefDefinition{otimes -}
{
Bilinear map
\ShowEq{otimes -}
is defined by the equality
\ShowEq{otimes -, 1}
}

%% file: Stmt.Module.Eq.tex
\ePrints{1506.00061,7287-9339,0803.3276}
\ifx\Semafor\ValueOff
\input{Stmt.Module.Ref}
\fi

\newcommand\eV[1]{$\Basis e_{#1}$}
\newcommand\EV[2]{e_{#1\cdot\gi{#2}}}
\def\AOn{A^{n+1\otimes}}
\def\LAnA{\mathcal L(D;A^n\rightarrow A)}

\newcommand\Ei
{
\begin{pmatrix}
1&0&0&0\\
0&-1&0&0\\
0&0&1&0\\
0&0&0&1
\end{pmatrix}
}

\newcommand\Ej
{
\begin{pmatrix}
1&0&0&0\\
0&1&0&0\\
0&0&-1&0\\
0&0&0&1
\end{pmatrix}
}

\newcommand\Ek
{
\begin{pmatrix}
1&0&0&0\\
0&1&0&0\\
0&0&1&0\\
0&0&0&-1
\end{pmatrix}
}

\newcommand\KR[1]
{
\left(
\begin{array}{rr@{}rr@{}rr@{}r}
\end{array}
\right)
}

\newcommand\PMatrix[3]
{
\begin{pmatrix}
\aUD {#1}11&...&\aUD {#1}1{#2}\\
...&...&...\\
\aUD {#1}{#3}1&...&\aUD {#1}{#3}{#2}
\end{pmatrix}
}

\newcommand\ColMatrix[2]
{
\begin{pmatrix}
\aU {#1}1\\...\\\aU {#1}{#2}
\end{pmatrix}
}

\newcommand\RowMatrix[2]
{
\begin{pmatrix}
\aD {#1}1&...&\aD {#1}{#2}
\end{pmatrix}
}

\AddEq{basis e=1}
{
$\Basis e=\{1\}$.
}

\AddEq[3]{a in Aoxn}
{
$#1\in A^{#2\otimes}$#3
}

\AddEq[2]{Aoxn}
{
$A^{#1\otimes}$#2
}

\AddEq{show definition of A module}
{
\ShowDefinition{module over algebra}

\ShowDefinition{\SideWS vector space over algebra}

\ShowTheorem{module over algebra}
\ShowProof{module over algebra}

\ShowTheorem{A module -> algebra is associative}
\ShowProof{A module -> algebra is associative}

\ShowTheorem{definition of A module}
\ShowTheorem{definition of A module, property}
\ShowProof{definition of A module}
}

\AddEq{show linear combination of vectors}
{
\ShowTheorem{set of vectors generated by set of vectors}
\ShowProof{set of vectors generated by set of vectors}

\ShowDefinition{linear combination of vectors}

\ShowEq{remark: scalars and vectors as matrix}

\ShowTheorem{linearly depends on rest of vectors}
\ShowProof{linearly depends on rest of vectors}

\ShowRemark{0=0vi}

\ShowDefinition{linearly independent vectors}
}

\AddEq{show basis of module}
{
\ShowRemark{generating set of module}

\ShowDefinition{generating set of module}

\ShowRemark{basis of module}

\ShowDefinition{basis of module}

\ShowTheorem{basis of module}
\ShowProof{basis of module}

\ShowTheorem{division algebra, basis}
\ShowProof{division algebra, basis}

\ShowDefinition{coordinates of vector}

\ShowTheorem{linear dependence between vectors of basis}
\ShowProof{linear dependence between vectors of basis}

\ShowTheorem{coordinates of vector with linear dependence}
\ShowProof{coordinates of vector with linear dependence}

\ShowDefinition{free module over ring}

\ShowTheorem{coordinates of vector of free module}
\ShowProof{coordinates of vector of free module}
}

\AddEq{show linear map from D1 A1 to D2 A2, module}
{
\ShowEq{=D1D2}

\ShowDefinition{linear map from D1 A1 to D2 A2, module}

\ShowRemark{notation for linear map}

\ShowTheorem{linear map of module}
\ShowProof{linear map of module}

\ShowTheorem{linear map of module, coordinates}
\ShowProof{linear map of module, coordinates}
}

\AddEq[2]{A1o+.n}
{
\[
#1=#1^1\oplus...\oplus #1^#2
\]
}

\AddEq{D->*o+A}
{
\[
\xymatrix
{
D\ar[r]|(.4){*}&\displaystyle\bigoplus_{i\in I}A_i
}
\ \ \ \ \ \,
d\left(\bigoplus_{i\in I}a_i\right)=\bigoplus_{i\in I}da_i
\]
}

\AddEq[1]{f(pai)=pfai}
{
#1\circ(
a_1, ...,
pa_i, ...,
a_n)
=
p#1\circ(
a_1, ...,
a_i, ...,
a_n)
}

\AddEquation{sum of maps, 31, polylinear}
{
\begin{split}
&(f+g)\circ(x_1,...,x_i+y_i,...,x_n)
\\
=&\,f\circ(x_1,...,x_i+y_i,...,x_n)
+g\circ(x_1,...,x_i+y_i,...,x_n)
\\
=&\,f\circ (x_1,...,x_i,...,x_n)+f\circ (x_1,...,y_i,...,x_n)
\\
+&g\circ (x_1,...,x_i,...,x_n)+g\circ (x_1,...,y_i,...,x_n)
\\
=&(f+g)\circ (x_1,...,x_i,...,x_n)
+(f+g)\circ (x_1,...,y_i,...,x_n)
\end{split}
}

\AddEquation{sum of maps, 32, polylinear}
{
\begin{split}
&(f+g)\circ(x_1,...,px_i,...,x_n)
\\
=&f\circ(x_1,...,px_i,...,x_n)+g\circ(x_1,...,px_i,...,x_n)
\\
=&pf\circ (x_1,...,x_i,...,x_n)+pg\circ (x_1,...,x_i,...,x_n)
\\
=&p(f\circ (x_1,...,x_i,...,x_n)+g\circ (x_1,...,x_i,...,x_n))
\\
=&p(f+g)\circ (x_1,...,x_i,...,x_n)
\end{split}
}

\AddEq{module of polylinear maps, 1}
{
$f$, $g$, $h\in\mathcal L(D;A_1\times...\times A_2\rightarrow S)$.
}

\AddEq{module of polylinear maps, 2}
{
$a=(a_1,...,a_n)$, $a_1\in A_1$, ..., $a_n\in A_n$,
\begin{align*}
(f+g)\circ a=&f\circ a+g\circ a=g\circ a+f\circ a
\\
=&(g+f)\circ a
\\
((f+g)+h)\circ a=&
(f+g)\circ a+h\circ a=(f\circ a+g\circ a)+h\circ a
\\
=&f\circ a+(g\circ a+h\circ a)=f\circ a+(g+h)\circ a
\\
=&(f+(g+h))\circ a
\end{align*}
}

\AddEq{0:A1n->S}
{
\[
0:v\in A_1\times...\times A_n\rightarrow 0\in S
\]
}

\AddEq{0+f=f 1n}
{
\[
(0+f)\circ (a_1,...,a_n)=0\circ (a_1,...,a_n)+f\circ (a_1,...,a_n)=f\circ (a_1,...,a_n)
\]
}

\AddEq{f-f=0 1n}
{
\begin{align*}
(f+(-f))\circ (a_1,...,a_n)&=f\circ (a_1,...,a_n)+(-f)\circ (a_1,...,a_n)
\\&=f\circ (a_1,...,a_n)-f\circ (a_1,...,a_n)
\\&=0
=0\circ (a_1,...,a_n)
\end{align*}
}

\AddEq{sum of maps, 4, polylinear}
{
\begin{align*}
(f+g)\circ (a_1,...,a_n)
&=f\circ (a_1,...,a_n)+g\circ (a_1,...,a_n)\\
&=g\circ (a_1,...,a_n)+f\circ (a_1,...,a_n)\\
&=(g+f)\circ (a_1,...,a_n)
\end{align*}
}

\AddEq{f-f=0}
{
\[
f+(-f)=0
\]
}

\AddEq{-f:A1n->S}
{
\[
-f:(a_1,...,a_n)\in A_1\times...\times A_n\rightarrow -(f\circ(a_1,...,a_n))\in S
\]
}

\AddEq{(p+q)f=pf+qf 1}
{
\begin{align*}
((p+q)f)\circ (x_1,...,x_n)=&(p+q)(f\circ (x_1,...,x_n))
\\
=&p(f\circ (x_1,...,x_n))+q(f\circ (x_1,...,x_n))
\\
=&(pf)\circ (x_1,...,x_n)+(qf)\circ (x_1,...,x_n)
\end{align*}
}

\AddEq{(p+q)f=pf+qf}
{
(p+q)f=pf+qf
}

\AddEq{p(qf)=(pq)f}
{
p(qf)=(pq)f
}

\AddEq{p(qf)=(pq)f 1}
{
\begin{align*}
(p(qf))\circ (x_1,...,x_n)=&p\ (qf)\circ (x_1,...,x_n)
=p\ (q\ f\circ (x_1,...,x_n))
\\
=&(pq)\ f\circ (x_1,...,x_n)=((pq)f)\circ (x_1,...,x_n)
\end{align*}
}

\AddEquation{product of map over scalar, 1, D module}
{
\begin{matrix}
\ShowSymbol{product of map over scalar}{D module}:A_1\rightarrow A_2
&d\in D&f\in\mathcal L(D;A_1\rightarrow A_2)
\end{matrix}
}

\AddEq{endomorphism of module from product, 1}
{
\[
\begin{matrix}
\vcenter
{
\xymatrix
{
A\ar[r]|{*}^{g_{23}}&A
}
}
&
\begin{matrix}
g_{23}:v\rightarrow l\circ v
\\
g_{23}\circ v:w\rightarrow (l\circ v)\circ w
\end{matrix}
\end{matrix}
\]
}

\AddEquation{a:LA12->LA12}
{
a:f\in\mathcal L(D;A_1\rightarrow A_2)\rightarrow af\in\mathcal L(D;A_1\rightarrow A_2)
}

\AddEquation{product of map over scalar, D module}
{
(\ShowSymbol{product of map over scalar}{D module})\circ x=d(f\circ x)
}

\AddEq[1]{f(ai+bi)=fai+fbi}
{
#1\circ(
a_1, ...,
a_i+ b_i, ...,
a_n)
=
#1\circ(
a_1, ...,
a_i, ...,
a_n)
+
#1\circ(
a_1, ...,
b_i, ...,
a_n)
}

\AddEq[1]{w=sum vi}
{
J(v)=\left\{w:w=\sum_{\gii\in\giI}c^{\gii}v_{\gii}, c^{\gii}\in #1,
|\{\gii:c^{\gii}\ne 0\}|<\infty\right\}
}

\AddEq{p,q in D, v,w in V}
{
$p$, $q \in D$, $v$, $w \in V$.
}

\AddEq{o+Ai}
{
\[A=\bigoplus_{i\in I}A_i\]
}

\AddEq{a=a1 o+ an}
{
\[a=a_1\oplus...\oplus a_n\]
}

\AddEq[2]{a1o+.n}
{
\[
#1=#1_1\oplus...\oplus #1_#2
\]
}

\AddEq{A1o+An=A1xAn}
{
\[
A_1\oplus...\oplus A_n=A_1\times...\times A_n
\]
}

\AddEq[2]{a=(a1.n)}
{
\[
#1=
\begin{pmatrix}
#1^1\\...\\#1^#2
\end{pmatrix}
\]
}

\AddEq[2]{vector col}
{
\begin{pmatrix}
\aU {#1}1\\...\\ \aU {#1}{#2}
\end{pmatrix}
}

\AddEq[2]{vector row}
{
\begin{pmatrix}
\aD {#1}1&...&\aD {#1}{#2}
\end{pmatrix}
}

\AddEq[2]{a=(a1.n col)}
{
\[
#1=
\ShowEq{vector col}{#1}{#2}
\]
}

\AddEq[2]{a=(a1.n row)}
{
\[
#1=
\ShowEq{vector row}{#1}{#2}
\]
}

\AddEq{f=(f1.n)}
{
\[
f=
\begin{pmatrix}
f_1&...&f_n
\end{pmatrix}
\]
\ShowEq{f:A->B}{f_i}{A^i}B
}

\AddEquation{foa=fi ai}
{
f\circ a=
\begin{pmatrix}
f_1&...&f_n
\end{pmatrix}
\RCcirc
\begin{pmatrix}
a^1\\...\\a^n
\end{pmatrix}
=f_i\circ a^i
}

\AddEq{foa=fi a}
{
\begin{pmatrix}
b^1\\...\\b^m
\end{pmatrix}
=
\begin{pmatrix}
f^1\\...\\f^m
\end{pmatrix}
\circ a=
\begin{pmatrix}
f^1\circ a\\...\\f^m\circ a
\end{pmatrix}
}

\AddEq{matrix of maps}
{
\[
f=
\begin{pmatrix}
f^1_1&...&f^1_n\\
...&...&...\\
f^m_1&...&f^m_n
\end{pmatrix}
\]
}

\AddEquation{b=f rco a}
{
\begin{pmatrix}
b^1\\...\\b^m
\end{pmatrix}
=
\begin{pmatrix}
f^1_1&...&f^1_n\\
...&...&...\\
f^m_1&...&f^m_n
\end{pmatrix}
\RCcirc
\begin{pmatrix}
a^1\\...\\a^n
\end{pmatrix}
=
\begin{pmatrix}
f^1_i\circ a^i\\
...\\
f^m_i\circ a^i
\end{pmatrix}
}

\AddEq{fij:->}
{
\ShowEq{f:A->B}{f^i_j}{A^j}{B^i}
}

\AddEq{fij:-> left A}
{
\ShowEq{f:A->B}{f^{\gii}_{\gij}}{A_2(g\circ e_{V_1\cdot\gij})}{A_2e_{\gii}}
}

\AddEq{fij:-> right A}
{
\ShowEq{f:A->B}{f^{\gii}_{\gij}}{(g\circ e_{V_1\cdot\gij})A_2}{e_{\gii}A_2}
}

\DefProof{direct sum of n Abelian groups}
{
\TheoremFollows
\RefTheorem{direct sum of Abelian groups}.
}

\DefProof{direct sum of n D modules}
{
\TheoremFollows
\RefTheorem{direct sum of D modules}.
}

\DefEq
{
\begin{matrix}
\xymatrix
{
f:D\ar[r]|{*}&V
}
&
f(d):v\rightarrow d\,v
\end{matrix}
}
{D->*V}

\AddEq{v=veV}
{
$v=v^{\gi i}\EV V i$.
}

\AddEquation{linear map in left A module}
{
f\circ(v\CRstar e_V )=(f_{\gi i}^{\gi j}\circ g\circ v^{\gi i})\ \EV W j
}

\AddEquation{linear map in right A module}
{
f\circ(e_V\RCstar v)=\EV W j(f_{\gi i}^{\gi j}\circ g\circ v^{\gi i})\ 
}

\AddEq{linear map in A module}
{
w=f\RCcirc (g\circ v)
}

\AddEq [1]{Ai, i=}
{
$#1_i$, $i=1$, $2$,
}

\AddEquation{module, (a+n)v=av+nv}
{
(a+n)v=av+nv\ \ \ a\in D\ \ \ n\in Z
}

\AddEquation{left module, (a+n)v=av+nv}
{
(a+n)v=av+nv\ \ \ a\in A\ \ \ n\in D
}

\AddEquation{right module, (a+n)v=av+nv}
{
v(a+n)=va+vn\ \ \ a\in A\ \ \ n\in D
}

\AddEq{(a+n)+(b+m)=}
{
(a+n)+(b+m)=(a+b)+(n+m)
}

\AddEquation{module, (a+n)+(b+m)=}
{
\begin{split}
((a+n)+(b+m))v&=av+nv+bv+mv=av+bv+nv+mv\\
&=(a+b)v+(n+m)v=((a+b)+(n+m))v
\end{split}
}

\AddEquation{g=CG}
{
g^{\gii}_{\gik}=C^{\gii}_{\gij}G^{\gij}_{\gik}
}

\AddEq [3]{matrix amn}
{
\[
#1=
\begin{pmatrix}
#1^{\gi 1}_{\gi 1}&...&#1^{\gi 1}_{\gi #3}\\
...&...&...\\
#1^{\gi #2}_{\gi 1}&...&#1^{\gi #2}_{\gi #3}
\end{pmatrix}
\]
}

\AddEquation{g=(cF)G}
{
g^{\gii}_{\gik}=(c_{1.k.l}F_{k\cdot}^{}{}^{\gii}_{\gij}c_{2.k.l})G^{\gij}_{\gik}
}

\AddEq{C=cF}
{
\[
C=c_{1.k.l}F_kc_{2.k.l}
\]
}

\AddEq{Gkj in D}
{
$G^{\gij}_{\gik}\in D$,
}

\AddEquation{g=c(FG)}
{
g^{\gii}_{\gik}=c_{1.k.l}(F_{k\cdot}^{}{}^{\gii}_{\gij}G^{\gij}_{\gik})c_{2.k.l}
}

\AddEquation{left module, (a+n)+(b+m)=}
{
\begin{split}
((a+n)+(b+m))v&=av+nv+bv+mv=av+bv+nv+mv\\
&=(a+b)v+(n+m)v=((a+b)+(n+m))v
\end{split}
}

\AddEquation{right module, (a+n)+(b+m)=}
{
\begin{split}
v((a+n)+(b+m))&=va+vn+vb+vm=va+vb+vn+vm\\
&=v(a+b)+v(n+m)=v((a+b)+(n+m))
\end{split}
}

\AddEquation{module, (a+n)+(b+m)=, 1}
{
((a+n)+(b+m))v=(a+n)v+(b+m)v
}

\AddEquation{left module, (a+n)+(b+m)=, 1}
{
((a+n)+(b+m))v=(a+n)v+(b+m)v
}

\AddEquation{right module, (a+n)+(b+m)=, 1}
{
v((a+n)+(b+m))=v(a+n)+v(b+m)
}

\AddEq{(a+n)(b+m)=}
{
(a+n)(b+m)=(ab+ma+nb)+(nm)
}

\AddEq [1]{A module diagram for linear}
{
\[
\xymatrix
{
A_{#1}&&V_{#1}\\
&D_{#1}\ar[lu]|{*}\ar[ru]|{*}
}
\]
}

\AddEq [1]{A module diagram for homomorphism}
{
\[
\xymatrix
{
A_{#1}\ar[r]|{*}&A_{#1}\ar[r]|{*}&V_{#1}\\
&D_{#1}\ar[lu]|{*}\ar[u]|{*}\ar[ru]|{*}
}
\]
}

\DefEquation
{
f^k=f^{k\cdot\gi{ij}}e_{\gii}\otimes e_{\gij}
}
{standard representation of map A1 A2, 3, associative algebra}

\AddEquation{A=re+im}
{
A=\re A\oplus\im A
}

\AddEquation{standard representation of map A->A, associative algebra}
{
f=
f^{k\cdot\gi{ij}}(\aU ei\otimes \aU ej)\circ F_k
}

\AddEq{standard representation of map A->A, o, associative algebra}
{
\[
f\circ x=
f^{k\cdot\gi{ij}}\aU ei(F_k\circ x)\aU ej
\]
}

\AddEq{product of map over scalar,,D module}
{
\symb{d\,f}{product of map over scalar}{D module}
}

\AddEq{product of map over scalar,,polylinear}
{
\symb{d\,f}{product of map over scalar}{polylinear}
}

\AddEquation{product of map over scalar, 1, polylinear}
{
\begin{matrix}
\ShowSymbol{product of map over scalar}{polylinear}{}:
A_1\times...\times A_n\rightarrow S
&d\in D&f\in\mathcal L(D;A_1\times...\times A_n\rightarrow S)
\end{matrix}
}

\AddEquation{product of map over scalar, 31, polylinear}
{
\begin{split}
&\,(pf)\circ(x_1,...,x_i+y_i,...,x_n)
\\
=&\,p\ f\circ(x_1,...,x_i+y_i,...,x_n)
\\
=&\,p\ (f\circ (x_1,...,x_i,...,x_n)+f\circ (x_1,...,y_i,...,x_n))
\\
=&\,p(f\circ (x_1,...,x_i,...,x_n))+p(f\circ (x_1,...,y_i,...,x_n))
\\
=&\,(pf)\circ (x_1,...,x_i,...,x_n)+(pf)\circ (x_1,...,y_i,...,x_n)
\end{split}
}

\AddEquation{product of map over scalar, 32, polylinear}
{
\begin{split}
& \,(pf)\circ(x_1,...,qx_i,...,x_n)
\\=& \,p(f\circ(x_1,...,qx_i,...,x_n))
=pq(f\circ (x_1,...,x_i,...,x_n))
\\=& \,qp(f\circ (x_1,...,x_n))
=q(pf)\circ (x_1,...,x_n)
\end{split}
}

\AddEquation{product of map over scalar, polylinear}
{
(\ShowSymbol{product of map over scalar}{polylinear}{})\circ (a_1,...,a_n)
=d(f\circ (a_1,...,a_n))
}

\DefCorollary{Free Algebra over Ring}
{
\ShowEq{text: Free Algebra over Ring}
}

\AddEquation{a:LA1n->LA1n}
{
a:f\in\mathcal L(D;A_1\times...\times A_n\rightarrow S)\rightarrow
af\in\mathcal L(D;A_1\times...\times A_n\rightarrow S)
}

\DefTheorem{Free Algebra over Ring}
{
\ShowEq{text: Free Algebra over Ring}
}

\AddEq{AV->V}
{
\[
(a,v)\in A\times V\rightarrow av\in V
\]
}

\AddEq{VA->V}
{
\[
(v,a)\in V\times A\rightarrow va\in V
\]
}

\AddEquation{l(v1+v2)w}
{
(l\circ (v_1+v_2))\circ w=(v_1+v_2)w=v_1w+v_2w
=(l\circ v_1)\circ w+(l\circ v_2)\circ w
}

\AddEquation{l(v1+v2)}
{
l\circ (v_1+v_2)=l\circ v_1+l\circ v_2
}

\AddEquation{l(dv)w}
{
(l\circ (dv))\circ w=(dv)w=d(vw)
=d((l\circ v)\circ w)
}

\AddEquation{l(dv)}
{
l\circ (dv)=d(l\circ v)
}

\AddEq{g23:A->L}
{
\[g_{23}:A\rightarrow\mathcal L(D;A\rightarrow A)\]
}

\AddEquation{coordinates of map A->A, 3, associative algebra}
{
\begin{split}
\aUD fkl\aU xl\aD ek
&=
f^{k\cdot\gi{ij}} \aD ei
\aUD {F_{k\cdot}^{}{}}ml\aU xl\aD em
\aD ej
\\
&=
f^{k\cdot\gi{ij}}
\aUD {F_{k\cdot}^{}{}}ml\aU xl
\aUD Cp{im}\aUD Ck{pj}
\aD ek
\end{split}
}

\DefEq
{
f=f^k\circ F_k
}
{map f generated by basis F}

\AddEquation{fk= in A2xA2}
{
f^k=f^k_{s_k\cdot 0}\otimes f^k_{s_k\cdot 1}\ \ \ \ f^k\in A_2\otimes A_2
}

\AddEq [3]{n=dim A}
{
$\gi #1=\dim #2$#3
}

\AddEq{gi n<m}
{
$\gi n\le\gi m$.
}

\AddEq{gi n>m}
{
$\gi n>\gi m$.
}

\AddEquation{set ker G in ker g}
{
\{g\in\mathcal L(D;A\rightarrow B):\ker G\subseteq\ker g\}
}

\AddEquation{fk= in AxA}
{
f^k=f^k_{s_k\cdot 0}\otimes f^k_{s_k\cdot 1}\ \ \ \ f^k\in A\otimes A
}

\AddEquation{fk= in (Ax)A}
{
f^k=(f^k_{s_k\cdot 0}\otimes) f^k_{s_k\cdot 1}\ \ \ \ f^k\in A\otimes A
}

\DefEquation
{
f=f_{s\cdot 0}\otimes f_{s\cdot 1}
}
{expansion of f A->A}

\DefEquation
{
f=f^k\circ F_k
}
{expansion of f with respect to basis I}

\DefEquation
{
g=g_{t\cdot 0}\otimes g_{t\cdot 1}
}
{expansion of g A->A}

\DefEquation
{
g=g^l\circ G_l
}
{expansion of g with respect to basis J}

\DefEquation
{
h=h_{ts\cdot 0}\otimes h_{ts\cdot 1}
}
{expansion of h A->A}

\DefEquation
{
h=h^{lk}\circ K_{lk}
}
{expansion of h with respect to basis K}

\DefEquation
{
h\circ a=g\circ f\circ a
=g^l\circ J_l\circ f^k\circ I_k\circ a 
}
{h(a)1}

\DefEquation
{
\begin{split}
h\circ a=g\circ f\circ a
&=g^l\circ(J^m_l\circ f^k)\circ
J_m\circ I_k\circ a
\\&=g^l\circ(J^m_l\circ f^k)\circ K_{mk}\circ a
\end{split}
}
{h(a)2}

\DefEquation
{
\begin{split}
h_{ts\cdot 0}&=g_{t\cdot 0}f_{s\cdot 0}
\\
h_{ts\cdot 1}&=f_{s\cdot 1}g_{t\cdot 1}
\end{split}
}
{hlk=glfk 01}

\DefEq
{
\[a^{lk}K_{lk}=(a^{lk}\circ J_l)\circ I_k=0\]
}
{aK=aJI}

\DefEquation
{
\begin{split}
h\circ a&=g\circ f\circ a
\\&=(g_{t\cdot 0}\otimes g_{t\cdot 1})\circ(f_{s\cdot 0}\otimes f_{s\cdot 1})\circ a
\\&=(g_{t\cdot 0}\otimes g_{t\cdot 1})\circ(f_{s\cdot 0}a f_{s\cdot 1})
\\&=g_{t\cdot 0}f_{s\cdot 0}a f_{s\cdot 1} g_{t\cdot 1}
\end{split}
}
{h(a)}

\DefEq
{
\[a^{lk}\circ J_l=0\]
}
{aJ=0}

\DefEquation
{
h^{lk}=g^l\circ (G^k_m\circ f^m)
}
{hlk=glfk}

\DefEq
{
\Basis H=\{H_{lk}:H_{lk}=G_l\circ F_k, G_l\in\Basis G, F_k\in\Basis F\}
}
{JlIk}

\DefEq
{
h=g\circ f
}
{h=g o f}

\DefEquation
{
\xymatrix
{
h:A\otimes A\ar[r]|{*}&\mathcal L(D;A\rightarrow A)
}
\ \ \ h(p):f\rightarrow p\circ f
}
{h:AoxA->L(A)}

\AddEq [2]{AoxA}
{
$#1\otimes #1$#2
}

\AddEq{representation AA in LA}
{
\[
\begin{matrix}
(a\otimes b)\circ f=afb
&a,b\in A&f\in\mathcal L(D;A\rightarrow A)
\end{matrix}
\]
}

\ePrints{5114-6019}
\ifx\Semafor\ValueOff
\AddEq{component of linear map}
{
\symb{f^k_{s_k\cdot p}}{component of linear map, division ring}1,
$p=0$, $1$,
}

\AddEq{standard component of linear map}
{
\symb{f^{k\cdot\gi{ij}}}{standard component of linear map}1
}
\fi

\AddEq{product in algebra AA}
{
(p_0\otimes p_1)\circ(q_0\otimes q_1)=(p_0q_0)\otimes(q_1p_1)
}

\AddEq{xA1n*xA1n->oxA1n=}
{
\[
(a_1,...,a_n)*(b_1,...,b_n)=(a_1b_1)\otimes...\otimes(a_nb_n)
\]
}

\DefEquation
{
(a\otimes b)\circ c=acb
}
{a ox b c=}

\AddEq{a ox b}
{
$(a\otimes b)\circ\delta$
}

\DefEq
{
$d\in A\otimes A$
}
{d in AxoA}

\AddEq{product in algebra AA 2}
{
$d\circ\delta\in\mathcal L(D;A\rightarrow A)$%
}

\DefEq
{
$A_1$, ..., $A_n$
}
{A1...n}

\AddEq{right A->*V}
{
\begin{matrix}
\xymatrix
{
f:A\ar[r]|{*}&V
}
&
f(a):v\in V\rightarrow va\in V
&a\in A
\end{matrix}
}

\AddEq[2]{diagram of representations, left right module}
{
\begin{matrix}
\vcenter
{
\xymatrix
{
A\ar[r]|{*}^{g_{23}}&
A\ar[r]|{*}^{g_{3,4}}&V
\\
&D\ar[u]|{*}_{g_{12}}\ar@/_1pc/[ur]|{*}_{g_{1,4}}\ar[ul]|(.4){*}^{g_{12}}
}
}
&
\begin{array}{r@{\,}l}
g_{12}(d):a&\rightarrow d\,a
\\
g_{23}(v):w&\rightarrow
C(w, v)
\\
C&\in\mathcal L(A^2\rightarrow A)
\\
g_{3,4}(a):v&\rightarrow v\,a
\\
g_{1,4}(d):v&\rightarrow #1\,#2
\end{array}
\end{matrix}
}

\AddEq{diagram of representations, left module nonassociative}
{
\[
\begin{matrix}
\vcenter
{
\xymatrix
{
A\ar[rd]|{*}^{g_{23}}\ar[rr]|{*}^{g_{34}}&&V
\\
&A
\\
&D\ar[u]|{*}_{g_{12}}\ar@/_1pc/[uur]|{*}_{g_{14}}\ar[uul]|(.4){*}^{g_{12}}
}
}
&
\begin{array}{r@{\,}l}
g_{12}(d):a&\rightarrow d\,a
\\
g_{23}(v):w&\rightarrow
C(w, v)
\\
C&\in\mathcal L(A^2\rightarrow A)
\\
g_{34}(a):v&\rightarrow v\,a
\\
g_{14}(d):v&\rightarrow d\,v
\end{array}
\end{matrix}
\]
}

\AddEquation{diagram of representations, left module}
{
\ShowEq{diagram of representations, left right module}dv
}

\AddEquation{diagram of representations, right module}
{
\ShowEq{diagram of representations, left right module}vd
}

\AddEquation{diagram of representations, module}
{
\xymatrix
{
D\ar[r]|{*}^{g_2}&V\\
&Z\ar[u]|{*}_{g_1}
}
}

\AddEq{A1=A unital extension}
{
&$\CBase\subseteq\Base$&$\Base_{(1)}=\Base$\\
\hline
}

\AddEq{A1=D unital extension}
{
&$\Base\subseteq\CBase$&$\Base_{(1)}=\CBase$\\
\hline
}

\AddEq{A1=A+D unital extension}
{
&\multicolumn{2}{c|}{$\Base_{(1)}=\Base\oplus\CBase$}\\
\hline
}

\AddEq{set of vectors generated by vector A}
{
\symb[A-0]{A_{(1)}v}{set of vectors generated by vector}1
}

\AddEq{set of vectors generated by vector D}
{
\symb[D-0]{D_{(1)}v}{set of vectors generated by vector}1
}

\AddEq{left A->*V}
{
\begin{matrix}
\xymatrix
{
f:A\ar[r]|{*}&V
}
&
f(a):v\in V\rightarrow av\in V
&a\in A
\end{matrix}
}

\AddEq{module, associative law}
{
(ab)v=a(bv)
}

\AddEquation{module, (a+n)(b+m)=}
{
\begin{split}
((a+n)(b+m))v&=(a+n)((b+m)v)=(a+n)(bv+mv)\\
=&a(bv+mv)+n(bv+mv)\\
=&a(bv)+a(mv)+n(bv)+n(mv)\\
=&(ab)v+m(av)++n(bv)+(nm)v\\
=&(ab)v+(ma)v++(nb)v+(nm)v\\
=&((ab+ma+nb)+nm)v
\end{split}
}

\AddEq{fab=ab}
{
f(a,b)=ab\ \ \ a,b\in\Base
}

\AddEq{f1p=p}
{
f(1,p)=f(p,1)=p\ \ \ p\in\Base_{(1)}\ \ \ 1\in\CBase_{(1)}
}

\AddEq{pq=fpq}
{
\begin{split}
(a+n)(b+m)&=f(a+n,b+m)\\
&=f(a,b)+f(a,m)+f(n,b)+f(n,m)\\
&=f(a,b)+mf(a,1)+nf(1,b)+nf(1,m)\\
&=ab+ma+nb+nm
\end{split}
}

\AddEq{f:D1xD1->D1}
{
\[
f:\Base_{(1)}\times\Base_{(1)}\rightarrow\Base_{(1)}
\]
}

\AddEq{distributive law, module}
{
\begin{align}
\EqLabel{distributive law, module, 1}
p(v+w)&=pv+pw\\
\EqLabel{distributive law, module, 2}
(p+q)v&=pv+qv
\end{align}
}

\AddEq{p,q in D1}
{
$p=a+n\in\Base_{(1)}$, $q=b+m\in\Base_{(1)}$.
}

\AddEq [4]{h:D1->D2 f:A1->A2}
{
\[
\begin{pmatrix}
#1:#2_1\rightarrow #2_2&
#3:#4_1\rightarrow #4_2
\end{pmatrix}
\]
}

\AddEq{set linear maps, D12 module}
{
\symb{\mathcal L(D_1\rightarrow D_2;A_1\rightarrow A_2)}{set linear maps}1
}

\AddEq{set linear maps, VW module}
{
\symb{\mathcal L(A;V\rightarrow W)}{set linear maps}1
}

\AddEq{set linear maps, left A12 module}
{
\symb{\mathcal L(D_1\rightarrow D_2;A_1\CRstar\rightarrow A_2\CRstar;V_1\rightarrow V_2)}{set linear maps}1
}

\AddEq{set homomorphisms, left A12 module}
{
\symb{\Hom(D_1\rightarrow D_2;A_1\CRstar\rightarrow A_2\CRstar;V_1\rightarrow V_2)}{set homomorphisms}1
}

\AddEq{set linear maps, right A12 module}
{
\symb{\mathcal L(D_1\rightarrow D_2;\RCstar A_1\rightarrow\RCstar A_2;V_1\rightarrow V_2)}{set linear maps}1
}

\AddEq{set homomorphisms, right A12 module}
{
\symb{\Hom(D_1\rightarrow D_2;\RCstar A_1\rightarrow\RCstar A_2;V_1\rightarrow V_2)}{set homomorphisms}1
}

\AddEq [1]{left column module}
{
$#1\CRstar$\Hyph
}

\AddEq [1]{right column module}
{
$\RCstar #1$\Hyph
}

\AddEquation{unitarity law, D-module}
{
1v=v
}

\DefEq
{
\[
a=a^{\gi i}e_{\gi i}
\]
}
{Expansion relative basis in algebra}

\AddEq{associative law, module}
{
(pq)v=p(qv)
}

\AddEq{associative law, left module}
{
(pq)v=p(qv)
}

\AddEq{associative law, right module}
{
v(pq)=(vp)q
}

\AddEq{left module, associative law}
{
(ab)v=a(bv)
}

\AddEq{associative law, Z, module}
{
(mn)v=m(nv)
}

\AddEq{associative law, ZD, module}
{
(md)v=m(dv)
}

\AddEq{associative law, D, left module}
{
(cd)v=c(dv)
}

\AddEq{associative law, DA, left module}
{
(da)v=d(av)
}

\AddEq{distributive law, left module}
{
\begin{align}
\EqLabel{distributive law, left module, 1}
p(v+w)&=pv+pw\\
\EqLabel{distributive law, left module, 2}
(p+q)v&=pv+qv
\end{align}
}

\AddEquation{distributive law, Z, module, 2}
{
(n+m)v=nv+mv
}

\AddEquation{distributive law, D, module, 2}
{
(a+b)v=av+bv
}

\AddEquation{distributive law, D, left module, 2}
{
(n+m)v=cn+dm
}

\AddEquation{distributive law, A, left module, 2}
{
(a+b)v=av+bv
}

\AddEquation{distributive law, D, right module, 2}
{
v(n+m)=vn+vm
}

\AddEquation{distributive law, A, right module, 2}
{
v(a+b)=va+vb
}

\AddEq{distributive law, right module}
{
\begin{align}
\EqLabel{distributive law, right module, 1}
(v+w)p&=vp+wp\\
\EqLabel{distributive law, right module, 2}
v(p+q)&=vp+vq
\end{align}
}

\AddEquation{aw in Xk module}
{
aw\in X_{k+1}
}

\AddEquation{aw in Xk left module}
{
aw\in X_{k+1}
}

\AddEquation{aw in Xk right module}
{
wa\in X_{k+1}
}

\AddEq[2]{w1+w2 in Jv}
{
$#1_1+#1_2\in J(#2)$.
}

\AddEquation{v=v+0, module}
{
\Vector v=\Vector v+0=v^{\gii}e_{\gii}+c^{\gii}e_{\gii}=(v^{\gii}+c^{\gii})e_{\gii}
}

\AddEquation{v=v+0, left module}
{
\Vector v=\Vector v+0=v^{\gii}e_{\gii}+c^{\gii}e_{\gii}=(v^{\gii}+c^{\gii})e_{\gii}
}

\AddEq{px x-i x-j}
{
\[
p(x)=2(x-i)(x-j)+(x-j)(x-i)
=3x^2-2ix-2xj-jx-xi+k
\]
}

\AddEq{px x-i x-j 1}
{
\[
p(x)=((3-a)(x\otimes 1)+a\otimes x)\circ x-2ix-2xj
-jx-xi+k
\]
}

\AddEq{rx x-i}
{
r(x)=x-i
}

\AddEq{p(x)=sr 2}
{
\begin{align*}
p(x)&=(3-a)(r(x)+i)x+ax(r(x)+i)-2ix-2xj-jx-xi+k
\\
&=(3-a)r(x)x+(3-a)ix+axr(x)+axi-2ix-2xj-jx-xi+k
\\
&=(3-a)r(x)x+axr(x)+(3-a)i(r(x)+i)
\\
&+a(r(x)+i)i-2i(r(x)+i)-2(r(x)+i)j-j(r(x)+i)-(r(x)+i)i+k\\
&=(3-a)r(x)x+axr(x)+3ir(x)-3-air(x)+a
\\
&+ar(x)i-a-2ir(x)+2-2r(x)j-2k-jr(x)+k-r(x)i+1+k
\\
&=(3-a)r(x)x+axr(x)+3ir(x)-air(x)
\\
&+ar(x)i-2ir(x)-2r(x)j-jr(x)-r(x)i
\end{align*}
}

\AddEq{p=s2r}
{
\[p(x)=s_2(a,x)\circ r(x)\]
}

\AddEq{s2(x)=}
{
\begin{align*}
s_2(a,x)&=(3-a)\otimes x+ax\otimes 1+(3-a)i\otimes 1
\\
&+a\otimes i-2i\otimes 1-2\otimes j-j\otimes 1-1\otimes i
\\
&=(3-a)\otimes x+(ax-j-2i+(3-a)i)\otimes 1
+(a-1)\otimes i-2\otimes j
\\
&=(3-a)\otimes x+(ax-j+i-ai)\otimes 1
+(a-1)\otimes i-2\otimes j
\end{align*}
}

\AddEq{s2(1,x)=}
{
\begin{align*}
s_2(1,x)&=2\otimes x+(x-j-2i+2i)\otimes 1
+(1-1)\otimes i-2\otimes j
\\
&=2\otimes (x-j)+(x-j)\otimes 1
\end{align*}
}

\AddEq{s12 in AA}
{
$s_1(x)$, $s_2(x)\in A\otimes A$
}

\AddEq[1]{px=sx o rx}
{
\[
p(x)=s_{#1}(x)\circ r(x)
\]
}

\AddEq{(sx-sx)rx=0}
{
\[
(s_1(x)-s_2(x))\circ r(x)=0
\]
}

\AddEquation{x p0 x*+xQ+Rx*-S=0 3}
{
\begin{split}
p(x)&=(-4\otimes 1\otimes 1-4\otimes i\otimes i-4\otimes j\otimes j
-4\otimes k\otimes k)\circ x^2
\\&+(4\otimes i+2j\otimes 1-2k\otimes i-2\otimes j+2i\otimes k)\circ x+1+2k
\\&=-4x^2-4xixi-4xjxj
-4xkxk 
\\&+4xi+2jx-2kxi-2xj+2ixk+1+2k
\end{split}
}

\AddEquation{q(x)=}
{
q(x)=x-\frac{1}{4}(i+j)
}

\AddEquation{4x=4q+}
{
4x=4q(x)+i+j
}

\AddEquation{rj j-i}
{
r(j)=j-i
}

\AddEq{p(x)=sr}
{
\begin{align*}
p(x)&=3(r(x)+i)x-2ix-2xj-jx-xi+k
\\
&=3r(x)x+ix-2xj-jx-xi+k
\\
&=3r(x)x+i(r(x)+i)-2(r(x)+i)j-j(r(x)+i)-(r(x)+i)i+k
\\
&=3r(x)x+ir(x)-2r(x)j-jr(x)-r(x)i
\\
&=(3\otimes x+i\otimes 1-2\otimes j-j\otimes 1-1\otimes i)\circ r(x)
\end{align*}
}

\AddEq{s(x)=}
{
\[
s(x)=3\otimes x+i\otimes 1-2\otimes j-j\otimes 1-1\otimes i
\]
}

\AddEq{s(j)=}
{
\[
s(j)=1\otimes (j-i)-(j-i)\otimes 1\ne 0\otimes 0
\]
}

\AddEquation{v=v+0, right module}
{
\Vector v=\Vector v+0=e_{\gii}v^{\gii}+e_{\gii}c^{\gii}=e_{\gii}(v^{\gii}+c^{\gii})
}

\AddEq{Xk+1 in Jv}
{
$X_{k+1}\subseteq J(v)$.
}

\AddEq{w1+w2 in V}
{
w_1+w_2\in X_{k+1}
}

\AddEquation{unitarity law, left A-module}
{
1v=v
}

\AddEquation{left module, a d v}
{
a(dv)=d(av)
}

\AddEquation{module, a d v}
{
a(nv)=n(av)
}

\AddEq{abc in A, v in v}
{
$a$, $b$, $c\in A$, $v\in V$.
}

\AddEquation{a(b(cv))=(a(bc))v left}
{
a(b(cv))=a((bc)v)=(a(bc))v
}

\AddEquation{a(b(cv))=(a(bc))v right}
{
((vc)b)a=(v(cb))a=v((cb)a)
}

\AddEquation{a(b(cv))=((ab)c)v left}
{
a(b(cv))=(ab)(cv)=((ab)c)v
}

\AddEquation{a(b(cv))=((ab)c)v right}
{
((vc)b)a=(vc)(ba)=v(c(ba))
}

\AddEquation{(a(bc))v=((ab)c)v left}
{
(a(bc))v=((ab)c)v
}

\AddEquation{(a(bc))v=((ab)c)v right}
{
v((cb)a)=v(c(ba))
}

\AddEquation{a(bc)=(ab)c left}
{
a(bc)=(ab)c
}

\AddEquation{a(bc)=(ab)c right}
{
(cb)a=c(ba)
}

\AddEq{cv left}
{
$cv\in A$,
}

\AddEq{cv right}
{
$vc\in A$,
}

\AddEq{right module, associative law}
{
v(ab)=(va)b
}

\AddEq{associative law, D, right module}
{
v(dc)=(vd)c
}

\AddEq{associative law, DA, right module}
{
v(ad)=(va)d
}

\AddEquation{unitarity law, right A-module}
{
v1=v
}

\AddEquation{right module, a d v}
{
(vd)a=(va)d
}

\AddEq{c,d in Z, a,b in D, v,w in V}
{
$m$, $n\in Z$, $a$, $b \in D$, $v$, $w \in V$.
}

\AddEq{p,q in D1, v,w in V}
{
$p$, $q \in D_{(1)}$, $v$, $w \in V$.
}

\AddEq{p,q in A1, v,w in V}
{
\ePrints{1801.01628,0767-8264}
\ifx\Semafor\ValueOn
$p$, $q \in A$,
\else
$p$, $q \in A_{(1)}$,
\fi
$v$, $w \in V$.
}

\AddEq [1]{vi V}
{
$v=(v_{\gii}\in V,\gii\in\giI)$#1
}

\DefEq
{
$\Vector v\in V$
}
{vv in V}

\AddEq{vv=ve left module}
{
\Vector v=v\CRstar e=v^{\gii}e_{\gii}
}

\AddEq{vv=ve right module}
{
\Vector v=e\RCstar v=e_{\gii}v^{\gii}
}

\AddEq{vv=ve module}
{
\Vector v=v\CRstar e=v^{\gii}e_{\gii}
}

\AddEq{coordinate matrix of vector}
{
$v=(v^{\gii},\Ii)$
}

\AddEquation{w= module}
{
w=\sum_{\gii\in\giI}w^{\gii}v_{\gii}
}

\AddEquation{w= left module}
{
w=\sum_{\gii\in\giI}w^{\gii}v_{\gii}
}

\AddEquation{w= right module}
{
w=\sum_{\gii\in\giI}v_{\gii}w^{\gii}
}

\AddEquation{aw= module}
{
aw=a\sum_{\gii\in\giI}w^{\gii}v_{\gii}
=\sum_{\gii\in\giI}a(w^{\gii}v_{\gii})
=\sum_{\gii\in\giI}(aw^{\gii})v_{\gii}
}

\AddEquation{aw= left module}
{
aw=a\sum_{\gii\in\giI}w^{\gii}v_{\gii}
=\sum_{\gii\in\giI}a(w^{\gii}v_{\gii})
=\sum_{\gii\in\giI}(aw^{\gii})v_{\gii}
}

\AddEquation{aw= right module}
{
wa=\left(\sum_{\gii\in\giI}v_{\gii}w^{\gii}\right)a
=\sum_{\gii\in\giI}(v_{\gii}w^{\gii})a
=\sum_{\gii\in\giI}(v_{\gii}w^{\gii}a)
}

\AddEq[1]{set au vi}
{
$\aU {#1}i$, $\gii\in\giI$,
}

\AddEq[1]{set au v1n}
{
$\aU {#1}1$, ..., $\aU {#1}n$
}

\AddEq{basis for module over algebra}
{
\symb{\Basis{e}=(e_{\gii},\gii\in\giI)}{basis for module}1
}

\AddEq{set vi}
{%
$v_{\gii}$, $\gii\in\giI$,
}

\AddEq[1]{column vector w=wi}
{
\[
#1=
\begin{pmatrix}
\aU {#1}1\\...\\ \aU {#1}n
\end{pmatrix}
\]
}

\AddEq[2]{row vector w=wi}
{
\[
#1=\RowMatrix{#1}{#2}
\]
}

\AddEquation{coordinate transformation, ba}
{
v''^k=b^{-1}{}^k_ia^{-1}{}^i_jv^j=(ab)^{-1}{}^k_jv^j
}

\AddEq[3]{passive coordinate transformation}
{
v#1^i=#3^{-1}{}^i_jv#2^j
}

\AddEq[2]{w cr e}
{
#1\CRstar #2=\aU {#1}i\aD {#2}i=
\begin{pmatrix}
\aU {#1}1\\...\\ \aU {#1}n
\end{pmatrix}
\CRstar
\begin{pmatrix}
\aD {#2}1&...&\aD {#2}n
\end{pmatrix}
}

\AddEq[2]{w rc e}
{
#1\RCstar #2=\aD {#1}i\aU {#2}i=
\begin{pmatrix}
\aD {#1}1&...&\aD {#1}n
\end{pmatrix}
\RCstar
\begin{pmatrix}
\aU {#2}1\\...\\ \aU {#2}n
\end{pmatrix}
}

\AddEq{vector w=w cr e}
{
\[
\Vector w=
\ShowEq{w cr e}we
\]
}

\AddEq{vector w=w rc e}
{
\[
\Vector w=
\ShowEq{w rc e}we
\]
}

\AddEq{vector w=e rc w}
{
\[
\Vector w=
\ShowEq{w rc e}ew
\]
}

\AddEq{vector w=e cr w}
{
\[
\Vector w=
\ShowEq{w cr e}ew
\]
}

\DefEquation
{
\xymatrix
{
g_{23}:A\ar[r]|{*}&A
}
}
{endomorphism of module from product}

\DefEquation
{
l\circ v:w\in A\rightarrow v\,w\in A
}
{l(v):w->vw}

\DefEq
{
\[
\xymatrix
{
h:\AOn\times S_n\ar[r]|{*}&\LAnA
}
\]
}
{representation An in LAnA}

\DefEq
{
\begin{matrix}
(a_0\otimes...\otimes a_n,\sigma)\circ(f_1\otimes...\otimes f_n)
=a_0\sigma(f_1)a_1...a_{n-1}\sigma(f_n)a_n
\\
\begin{matrix}
a_0,...,a_n\in A&\sigma\in S_n&f_1,...,f_n\in\mathcal L(D;A_n\rightarrow A)
\end{matrix}
\end{matrix}
}
{representation An in LAnA, 1}

\DefEq
{
$d\in\AOn$
}
{product in algebra An 1}

\DefEquation
{
\begin{matrix}
(d,\sigma)\circ(f_1,...,f_n)&f_i=\delta\in\mathcal L(D;A\rightarrow A)
\end{matrix}
}
{product in algebra An 2}

\DefEq
{
$\delta\in\mathcal L(D;A\rightarrow A)$
}
{product in algebra AA 3}

\AddEq{wi vi=0}
{
\[w^{\gii}v_{\gii}=0\]
}

\AddEq{left wi vi=0}
{
\[w^{\gii}v_{\gii}=0\]
}

\AddEq{right wi vi=0}
{
\[v_{\gii}w^{\gii}=0\]
}

\AddEq{0=0vi}
{
\[
w^{\gii}=0\Rightarrow w\CRstar v=0
\]
}

\AddEq{left 0=0vi}
{
\[
w^{\gii}=0\Rightarrow w\CRstar v=0
\]
}

\AddEq{right 0=0vi}
{
\[
w^{\gii}=0\Rightarrow v\RCstar w=0
\]
}

\AddEq[1]{vj=sum vi}
{
\[
#1_{\gij}=\sum_{\gii\in \giI\setminus\{\gij\}}
\frac{w^{\gii}}{w^{\gij}}#1_{\gii}
\]
}

\AddEq[1]{left vj=sum vi}
{
\[
#1_{\gij}=\sum_{\gii\in \giI\setminus\{\gij\}}
(w^{\gij})^{-1}w^{\gii}#1_{\gii}
\]
}

\AddEq[2]{b in D'}
{
$#1=(#1^{\gii},\gii\in\giI)\in\Base'$#2
}

\AddEq[1]{right vj=sum vi}
{
\[
#1_{\gij}=\sum_{\gii\in \giI\setminus\{\gij\}}
#1_{\gii}w^{\gii}(w^{\gij})^{-1}
\]
}

\AddEq[1]{wj ne 0}
{
$w^{\gij}\ne 0$#1
}

\AddEq[2]{w12 in Jv}
{
$#1_1$, $#1_2\in X_k\subseteq J(#2)$.
}

\AddEquation{w=sum vi, module}
{
J(v)=\left\{w:w=\sum_{\gii\in \giI}c^{\gii}v_{\gii}, c^{\gii}\in D_{(1)},
|\{\gii:c^{\gii}\ne 0\}|<\infty\right\}
}

\AddEq{generating set of module}
{
$J(X)=V$.
}

\AddEquation{w=sum vi, left module}
{
J(v)=\left\{w:w=\sum_{\gii\in \giI}c^{\gii}v_{\gii}, c^{\gii}\in A_{(1)},
|\{\gii:c^{\gii}\ne 0\}|<\infty\right\}
}

\AddEquation{w=sum vi, right module}
{
J(v)=\left\{w:w=\sum_{\gii\in \giI}v_{\gii}c^{\gii}, c^{\gii}\in A_{(1)},
|\{\gii:c^{\gii}\ne 0\}|<\infty\right\}
}

\AddEq{vectors generated by set of vectors}
{
\StartLabelItem
\begin{enumerate}
\item
\ShowEq{vk in Jv}
\labelItem{\SideWS module, vk in Jv}
\item
\ShowEq{\SideWS module, cvk in Jv}
\item
\ShowEq{\SideWS module, sum cvk in Jv}
\item
\ShowEq{v,w in Jv=>v+w in Jv}
\item
\ShowEq{vector space, v in Jv=>av in Jv}
\end{enumerate}
}

\AddEq{vectors generated by set of vectors 2018}
{
\StartLabelItem
\begin{enumerate}
\item
\ShowEq{vk in Jv}
\labelItem{vector space, vk in Jv}
\item
\ShowEq{vector space, cvk in Jv}
\item
\ShowEq{vector space, sum cvk in Jv}
\item
\ShowEq{v,w in Jv=>v+w in Jv}
\item
\ShowEq{vector space, v in Jv=>av in Jv}
\end{enumerate}
}

\AddEq{vk in Jv}
{%
$v_{\gik}\in X_0\subseteq J(v)$
}

\AddEq{module, d in D}
{
$\DArg\in \CBase$,
}

\AddEq{module, dv in V}
{%
$\DArg v\in V$,
}

\AddEq{left module, dv in V}
{%
$\DArg v\in V$,
}

\AddEq{right module, dv in V}
{%
$v\DArg \in V$,
}

\AddEq{module, a in A}
{
$a\in \Base$.
}

\AddEq{module, av in V}
{%
$av\in V$
}

\AddEq{left module, av in V}
{%
$av\in V$
}

\AddEq{right module, av in V}
{%
$va\in V$
}

\AddEq{module, dv+av in V}
{%
\[\DArg v+av\in V\ \ \ \DArg\in \CBase\ \ \ a\in \Base\]
}

\AddEq{left module, dv+av in V}
{%
\[\DArg v+av\in V\ \ \ \DArg\in \CBase\ \ \ a\in \Base\]
}

\AddEq{right module, dv+av in V}
{%
\[v\DArg +va\in V\ \ \ \DArg\in \CBase\ \ \ a\in \Base\]
}

\AddEq{module, cvk in Jv}
{%
$c^{\gik}v_{\gik}\in J(v)$, $c^{\gik}\in \Base_{(1)}$,
$\gik\in\giI$
\labelItem{module, cvk in Jv}
}

\AddEq{vector space, cvk in Jv}
{%
$c^{\gik}v_{\gik}\in J(v)$, $c^{\gik}\in \Base$,
$\gik\in\giI$
\labelItem{vector space, cvk in Jv}
}

\AddEquation{module, a+n:V->V}
{
a+n:v\in V\rightarrow (a+n)v\in V
}

\AddEquation{left module, a+n:V->V}
{
a+n:v\in V\rightarrow (a+n)v\in V
}

\AddEquation{right module, a+n:V->V}
{
a+n:v\in V\rightarrow v(a+n)\in V
}

\AddEq{left module, cvk in Jv}
{%
$c^{\gik}v_{\gik}\in J(v)$, $c^{\gik}\in \Base_{(1)}$,
$\gik\in\giI$
\labelItem{left module, cvk in Jv}
}

\AddEq{right module, cvk in Jv}
{%
$v_{\gik}c^{\gik}\in J(v)$, $c^{\gik}\in \Base_{(1)}$,
$\gik\in\giI$
\labelItem{right module, cvk in Jv}
}

\AddEq{module, sum cvk in Jv}
{%
$\displaystyle\sum_{\gik\in\giI}c^{\gik}v_{\gik}\in J(v)$,
$c^{\gik}\in\Base_{(1)}$, $|\{\gii:c^{\gii}\ne 0\}|<\infty$
\labelItem{module, sum cvk in Jv}
}

\AddEq{vector space, sum cvk in Jv}
{%
$\displaystyle\sum_{\gik\in\giI}c^{\gik}v_{\gik}\in J(v)$,
$c^{\gik}\in\Base$, $|\{\gii:c^{\gii}\ne 0\}|<\infty$
\labelItem{vector space, sum cvk in Jv}
}

\AddEq{left module, sum cvk in Jv}
{%
$\displaystyle\sum_{\gik\in\giI}c^{\gik}v_{\gik}\in J(v)$,
$c^{\gik}\in\Base_{(1)}$, $|\{\gii:c^{\gii}\ne 0\}|<\infty$
\labelItem{left module, sum cvk in Jv}
}

\AddEq{right module, sum cvk in Jv}
{%
$\displaystyle\sum_{\gik\in\giI}v_{\gik}c^{\gik}\in J(v)$,
$c^{\gik}\in\Base_{(1)}$, $|\{\gii:c^{\gii}\ne 0\}|<\infty$
\labelItem{right module, sum cvk in Jv}
}

\AddEq[1]{|ci12 ne 0|}
{
H_1=\{\gii\in\giI:#1_1^{\gii}\ne 0\}
\ \ \ \,
H_2=\{\gii\in\giI:#1_2^{\gii}\ne 0\}
}

\AddEq{|wi ne 0|}
{
|\{\gii\in\giI:w^{\gii}\ne 0\}|<\infty
}

\AddEq{module, |awi ne 0|}
{
$\{\gii\in\giI:aw^{\gii}\ne 0\}$
}

\AddEq{left module, |awi ne 0|}
{
$\{\gii\in\giI:aw^{\gii}\ne 0\}$
}

\AddEq{right module, |awi ne 0|}
{
$\{\gii\in\giI:w^{\gii}a\ne 0\}$
}

\AddEq{v,w in Jv=>v+w in Jv}
{
$w_1$, $w_2\in J(v)\Rightarrow w_1+w_2\in J(v)$
\labelItem{\SideWS module, v,w in Jv=>v+w in Jv}
}

\AddEq{v in Jv=>av in Jv}
{
$a\in\Base_{(1)}$,
$w\in J(v)\Rightarrow aw\in J(v)$
\labelItem{\SideWS module, v in Jv=>av in Jv}
}

\AddEq{vector space, v in Jv=>av in Jv}
{
$a\in\Base$,
$w\in J(v)\Rightarrow aw\in J(v)$
\labelItem{\SideWS vector space, v in Jv=>av in Jv}
}

\AddEq[1]{gi12}
{
$#1^{\gii}_1$, $#1^{\gii}_2$, $\gii\in\giI$,
}

\AddEq[1]{w1+w2= 1 }
{
#1_1+#1_2=\sum_{\gii\in\giI}(#1^{\gii}_1+#1^{\gii}_2)v_{\gii}
}

\AddEq[1]{w1+w2= 1 left}
{
#1_1+#1_2=\sum_{\gii\in\giI}(#1^{\gii}_1+#1^{\gii}_2)v_{\gii}
}

\AddEq[1]{w1+w2= 1 right}
{
#1_1+#1_2=\sum_{\gii\in\giI}v_{\gii}(#1^{\gii}_1+#1^{\gii}_2)
}

\AddEq[1]{|gi 1+2 ne 0|}
{
\[
\{\gii\in\giI:#1_1^{\gii}+#1_2^{\gii}\ne 0\}\subseteq H_1\cup H_2
\]
}

\AddEq[1]{w1+w2= }
{
#1_1+#1_2=\sum_{\gii\in\giI}#1^{\gii}_1v_{\gii}+\sum_{\gii\in\giI}#1^{\gii}_2v_{\gii}
=\sum_{\gii\in\giI}(#1^{\gii}_1v_{\gii}+#1^{\gii}_2v_{\gii})
}

\AddEq[1]{w1+w2= left}
{
#1_1+#1_2=\sum_{\gii\in\giI}#1^{\gii}_1v_{\gii}+\sum_{\gii\in\giI}#1^{\gii}_2v_{\gii}
=\sum_{\gii\in\giI}(#1^{\gii}_1v_{\gii}+#1^{\gii}_2v_{\gii})
}

\AddEq[1]{w1+w2= right}
{
#1_1+#1_2=\sum_{\gii\in\giI}v_{\gii}#1^{\gii}_1+\sum_{\gii\in\giI}v_{\gii}#1^{\gii}_2
=\sum_{\gii\in\giI}(v_{\gii}#1^{\gii}_1+v_{\gii}#1^{\gii}_2)
}

\AddEq[1]{w12= }
{
#1_1=\sum_{\gii\in\giI}#1^{\gii}_1v_{\gii}
\ \ \ \,
#1_2=\sum_{\gii\in\giI}#1^{\gii}_1v_{\gii}
}

\AddEq[1]{w12= left}
{
#1_1=\sum_{\gii\in\giI}#1^{\gii}_1v_{\gii}
\ \ \ \,
#1_2=\sum_{\gii\in\giI}#1^{\gii}_2v_{\gii}
}

\AddEq[1]{w12= right}
{
#1_1=\sum_{\gii\in\giI}v_{\gii}#1^{\gii}_1
\ \ \ \,
#1_2=\sum_{\gii\in\giI}v_{\gii}#1^{\gii}_2
}

\AddEquation{vk=sum vi, module}
{
v_{\gik}=\sum_{\gii\in\giI}c^{\gii}v_{\gii}
}

\AddEquation{vk=sum vi, left module}
{
v_{\gik}=\sum_{\gii\in\giI}c^{\gii}v_{\gii}
}

\AddEquation{vk=sum vi, right module}
{
v_{\gik}=\sum_{\gii\in\giI}v_{\gii}c^{\gii}
}

\AddEq[1]{ci=dik}
{
$c^{\gii}=\delta^{\gii}_{\gik}\in #1$.
}

\AddEq{vk in v}
{
$v_{\gik}\in v$,
}

\DefEq
{
$v_{\gii}$.
}
{vi}

\AddEq{w=wi vi}
{
$\Vector w=w^{\gii}v_{\gii}$
}

\AddEq{w=wi vileft}
{
$\Vector w=w^{\gii}v_{\gii}$
}

\AddEq{w=wi viright}
{
$\Vector w=v_{\gii}w^{\gii}$
}

\AddEq{w=w cr v}
{
\[\Vector w=\ShowSymbol{linear combination}{\SideWS cr}\]
}

\AddEq{g number linear combination}
{%
\symb{g^is_i}{linear combination}{}%
}

\AddEq{linear combination}
{%
\symb{w^{\gii}v_{\gii}}{linear combination}{i}%
\symb{w\CRstar v}{linear combination}{cr}%
}

\AddEq{left linear combination}
{%
\symb{w^{\gii}v_{\gii}}{linear combination}{left i}%
\symb{w\CRstar v}{linear combination}{left cr}%
}

\AddEq{right linear combination}
{%
\symb{v_{\gii}w^{\gii}}{linear combination}{right i}%
\symb{v\RCstar w}{linear combination}{right cr}%
}

\AddEq{A linear combination =}
{
$\ShowSymbol{linear combination}{\SideWS i}$
}

\DefEq
{
$e_{\gii}$, \iI,
}
{ei, i in I}

\AddEq[1]{c*e=0, module}
{
#1^{\gii}e_{\gii}=0
}

\AddEq[1]{c*e=0, left module}
{
#1^{\gii}e_{\gii}=0
}

\AddEq[1]{c*e=0, right module}
{
e_{\gii}#1^{\gii}=0
}

\AddEq{(b+c)*e, module}
{
\[
(b^{\gii}+c^{\gii})e_{\gii}=0
\]
}

\AddEq{(b+c)*e, left module}
{
\[
(b^{\gii}+c^{\gii})e_{\gii}=0
\]
}

\AddEq{(b+c)*e, right module}
{
\[
e_{\gii}(b^{\gii}+c^{\gii})=0
\]
}

\AddEq{(ac)*e, module}
{
\[
(ac^{\gii})e_{\gii}=0
\]
}

\AddEq{(ac)*e, left module}
{
\[
(ac^{\gii})e_{\gii}=0
\]
}

\AddEq{(ac)*e, right module}
{
\[
e_{\gii}(c^{\gii}a)=0
\]
}

\DefEq
{
\symb{f+g}{sum of maps}{D module}
}
{sum of maps,,D module}

\DefEquation
{
\begin{matrix}
\ShowSymbol{sum of maps}{D module}:A_1\rightarrow A_2
&f,g\in\mathcal L(D;A_1\rightarrow A_2)
\end{matrix}
}
{sum of maps, 1, D module}

\DefEquation
{
(\ShowSymbol{sum of maps}{D module})\circ x=f\circ x+g\circ x
}
{sum of maps, D module}

\AddEq [4]{f in L(A->B)}
{
$f\in\mathcal L(#1;#2\rightarrow #3)$#4
}

\AddEq{fab=fbfa}
{
\[
f(ab)=f(b)f(a)
\]
}

\AddEq{b o f =}
{
\[
(b\circ f)\circ x=b(f\circ x)
\]
}

\AddEq{b * f =}
{
\[
(b*f)\circ x=(f\circ x)b
\]
}

\DefEq
{
$F_k\in\Basis F$,
}
{Ik in I}

\DefEquation
{
x\rightarrow I_k\circ a\circ x
}
{x->Ik a x}

\DefEquation
{
b^l=I^l_k\circ a
}
{b=Ikl a}

\DefEq
{
\begin{align*}
(I_k^l\circ(a_1+\aD a2))\circ I_l\circ x
&=I_k\circ(a_1+\aD a2)\circ x
\\&=I_k\circ a_1\circ x+I_k\circ \aD a2\circ x
\\&=(I_k^l\circ a_1)\circ I_l\circ x+(I_k^l\circ a_1)\circ I_l\circ x
\end{align*}
}
{Ikl a1+a2}

\DefEq
{
\begin{align*}
(I_k^l\circ(da))\circ I_l\circ x
&=I_k\circ(da)\circ x
=I_k\circ(d(a\circ x))
\\&=d(I_k\circ a\circ x)
=d((I_k^l\circ a)\circ I_l\circ x)
\\&=(d(I_k^l\circ a))\circ I_l\circ x
\end{align*}
}
{Ikl d a}

\DefEquation
{
I_k\circ a\circ x=b^l\circ I_l\circ x\ \ \ \ b^l\in A_2\times A_2
}
{expansion Ik a x}

\DefEq
{
\symb{F_k^l}{conjugation transformation}{}
}
{conjugation transformation}

\DefEq
{
\[
\ShowSymbol{conjugation transformation}{}
:A_1\otimes A_1\rightarrow A_2\otimes A_2
\]
}
{conjugation transformation:}

\DefEquation
{
F_k\circ a\circ x=(\ShowSymbol{conjugation transformation}{}\circ a)\circ F_l\circ x
}
{conjugation transformation =}

\DefEq
{
$\ShowSymbol{conjugation transformation}{}$
}
{show conjugation transformation}

\DefEq
{
$\aUD Cp{kl}$
}
{structural constants, algebra}

\AddEq{wi i in I}
{
$w^{\gii}$, $\gii\in\giI$,
}

\AddEq{w=wi i in I}
{
$w=(w^{\gii},\gii\in\giI)$
}

\AddEq{i=j}
{
$\gii=\gij$
}

\DefEq
{
$f^{\gik}_{\gil}$
}
{Coordinates of map f}

\AddEq[1]{standard components of map f}
{
\def\Cdot{}
\def\Temp{-}%
\edef\Tempa{#1}%
\ifx\Tempa\Temp%
\else
\def\Cdot{#1\cdot}
\fi
$\aU {f^{\Cdot}_{}{}}{ij}$
}

\AddEq{linear map over ring, left side}
{
$\aUD flk$
}

\AddEq{left shift algebra, 2}
{
\[
(a,b)_1\circ x=(a,b,x)
\]
}

\AddEquation{left shift algebra}
{
l(a)\circ x=ax
}

\AddEquation{left shift algebra, 3}
{
\begin{split}
(l(a)\circ l(b))\circ x
&=l(a)\circ(l(b)\circ x)
\\
&=a(bx)=(ab)x-(a,b,x)
\\
&=l(ab)\circ x-(a,b)_1\circ x
\end{split}
}

\AddEquation{left shift algebra, 1}
{
l(a)\circ l(b)=l(ab)-(a,b)_1
}

\AddEquation{right shift algebra}
{
r(a)\circ x=xa
}

\AddEquation{right shift algebra, 3}
{
\begin{split}
(r(a)\circ r(b))\circ x
&=r(a)\circ(r(b)\circ x)
\\
&=(xb)a=x(ba)+(x,b,a)
\\
&=r(ba)\circ x+(x,b,a)
\end{split}
}

\AddEquation{right shift algebra, 1}
{
r(a)\circ r(b)=r(ba)+(b,a)_2
}

\AddEq{right shift algebra, 2}
{
\[
(b,a)_2\circ x=(x,b,a)
\]
}

\AddEq{f over A}
{
\[
f:x\in A\rightarrow (ax)b\in A
\]
}

\AddEq{g over A}
{
\[
\begin{matrix}
g:A\rightarrow A
&&
g=(cf)d
\end{matrix}
\]
}

\AddEquation{linear map, standard representation, nonassociative algebra}
{
f\circ x=\aU f{ij}\ (\aD eix)\aD ej
}

\AddEq{linear map, standard representation, nonassociative algebra, 1}
{
\[
f\circ x=\aU f{ij}\aD ei(x\aD ej)
\]
}

\AddEquation{coordinates of map A1 A2, 2, nonassociative algebra}
{
\aUD gkl
=\aUD fml\aU g{ij}
\aUD Cp{im}\aUD Ck{pj}
}

\AddEquation{coordinates of map A1 A2, 21, nonassociative algebra}
{
\aUD gkl
=\aUD fml \aU g{ij}
\aUD Ck{ip}\aUD Cp{mj}
}

\DefEq
{
$F_{k\cdot}^{}{}_{\gii}^{\gij}$
}
{coordinates of map Ik}

\DefEq
{
$G_{\gii}^{\gij}$
}
{coordinates of map G}

\AddEq{f in LRC y=fx}
{
\[
\begin{matrix}
f\in\mathcal L(R;C\rightarrow C)&
y^{\gi i}=x^{\gi j}\aUD fij
\end{matrix}
\]
}

\AddEq{projection maps, complex field}
{
\begin{align*}
P^{\gi 0}\circ a&=\aU a0&P^{\gi 0\cdot}_{}{}^{\gi 0}_{\gi 0}&=1&
R^{\gi 0}\circ a&=\aU a0&R^{\gi 0\cdot}_{}{}^{\gi 0}_{\gi 0}&=1
\\
P^{\giA}\circ a&=\aU a1i&P^{\giA\cdot}_{}{}^{\giA}_{\giA}&=1&
R^{\giA}\circ a&=\aU a1&R^{\giA\cdot}_{}{}^{\gi 0}_{\giA}&=1
\end{align*}
}

\AddEq [1]{e=(E,I)}
{
$\Basis e=(E,I)$#1
}

\AddEq{projection mappings 1, complex field}
{
\begin{align*}
P^{\gi 0}&=\frac 12\circ E+\frac 12\circ I&
R^{\gi 0}&=\frac 12\circ E+\frac 12\circ I
\\
P^{\giA}&=\frac 12\circ E-\frac 12\circ I&
R^{\giA}&=-\frac i2\circ E+\frac i2\circ I
\end{align*}
}

\AddEquation{df/dx= complex field}
{
\frac{df}{dx}=
\frac 12\left(\frac{\partial f}{\partial \aU x0}
-i\frac{\partial f}{\partial x^{\gi 1}}\right)\circ E
+\frac 12\left(\frac{\partial f}{\partial \aU x0}
+i\frac{\partial f}{\partial x^{\gi 1}}\right)\circ I
}

\DefEquation
{
f=\aD a0\circ E+a_1\circ I+\aD a2\circ J+\aD a3\circ K
}
{expand linear mapping, quaternion}

\AddEq{fi=fij ej}
{
f_{\gii}=\aUD fji e_{\gij}
}

\AddEquation{f=.E+.I}
{
f=\frac 12(\aD f0-i\aD f1)\circ E+\frac 12(\aD f0+i\aD f1)\circ I
}

\AddEquation{a0=f01}
{
\begin{split}
\aD a0&=\aUD a00+\aUD a10i
=\frac 12(\aUD f00+\aUD f11+(\aUD f10-\aUD f01)i)
\\&=\frac 12(\aUD f00+\aUD f10i+\aUD f11-\aUD f01i)
\\&=\frac 12((\aUD f00+\aUD f10i)-(\aUD f01+\aUD f11i)i)
\end{split}
}

\AddEquation{a1=f01}
{
\begin{split}
a_1&=\aUD a01+\aUD a11i
=\frac 12(\aUD f00-\aUD f11+(\aUD f10+\aUD f01)i)
\\&=\frac 12(\aUD f00+\aUD f10i-\aUD f11+\aUD f01i)
\\&=\frac 12((\aUD f00+\aUD f10i)+(\aUD f01+\aUD f11i)i)
\end{split}
}

\AddEquation{a01=}
{
\ShowEq{matrix a algebra C}
=
\ShowEq{matrix af algebra C}
\ShowEq{matrix f algebra C}
}

\AddEq [2]{map ao}
{
\def\Map{\aU I#1}
\edef\Tempa{#1}%
\def\Temp{E}
\ifx\Tempa\Temp%
\def\Map{E}
\fi
\def\Temp{I}
\ifx\Tempa\Temp%
\def\Map{I}
\fi
\begin{matrix}
a\circ \Map:#2\rightarrow #2&a\in #2
\end{matrix}
}

\AddEq [2]{map a*}
{
\def\Map{E}
\def\Temp{E}
\edef\Tempa{#1}%
\ifx\Tempa\Temp%
\else
\def\Map{\aU I#1}
\fi
\begin{matrix}
a*\Map:#2\rightarrow #2&a\in #2
\end{matrix}
}

\AddEq [1]{ao, H matrix 1}
{
#1_l(a)=E_l(a)\RCcirc #1
}

\AddEq [1]{a*, H matrix 1}
{
#1_r(a)=E_r(a)\RCcirc #1
}

\AddEq {aI, H matrix 2}
{
\begin{align*}
E_l(a)\RCcirc I
&=
\ShowEq{H aE Jacobian matrix}a
\RCcirc\Ei
\\
&=
\ShowEq{H a1 Jacobian matrix}a
=
I_l(a)
\end{align*}
}

\AddEq {aJ, H matrix 2}
{
\begin{align*}
E_l(a)\RCcirc J
&=
\ShowEq{H aE Jacobian matrix}a
\RCcirc\Ej
\\
&=
\ShowEq{H a2 Jacobian matrix}a
=
J_l(a)
\end{align*}
}

\AddEq{aK, H matrix 2}
{
\begin{align*}
E_l(a)\RCcirc K
&=
\ShowEq{H aE Jacobian matrix}a
\RCcirc\Ek
\\
&=
\ShowEq{H a3 Jacobian matrix}a
=
K_l(a)
\end{align*}
}

\AddEq{Ia, H matrix 2}
{
\begin{align*}
E_r(a)\RCcirc I
&=
\ShowEq{H Ea Jacobian matrix}a
\RCcirc\Ei
\\
&=
\ShowEq{H 1a Jacobian matrix}a
=
I_r(a)
\end{align*}
}

\AddEq{Ja, H matrix 2}
{
\begin{align*}
E_r(a)\RCcirc J
&=
\ShowEq{H Ea Jacobian matrix}a
\RCcirc\Ej
\\
&=
\ShowEq{H 2a Jacobian matrix}a
=
J_r(a)
\end{align*}
}

\AddEq{Ka, H matrix 2}
{
\begin{align*}
E_r(a)\RCcirc K
&=
\ShowEq{H Ea Jacobian matrix}a
\RCcirc\Ek
\\
&=
\ShowEq{H 3a Jacobian matrix}a
=
K_r(a)
\end{align*}
}

\AddEq[3]{maps of conjugation, Jacobian matrix}
{
\def\Map{\aU I#1}
\def\Letter{I}
\def\Temp{E}
\edef\Tempa{#1}%
\ifx\Tempa\Temp%
\def\Map{E}
\def\Letter{E}
\fi
\def\Temp{I}
\ifx\Tempa\Temp%
\def\Map{I}
\def\Letter{I}
\fi
\symb[\Letter-0]{\Map_{#3}(a)}{maps of conjugation, Jacobian matrix}{}
}

\AddEq[2]{a o, Jacobian matrix}
{
\ShowSymbol{maps of conjugation, Jacobian matrix}{}=
\ShowEq{#2 a#1 Jacobian matrix}a
}

\AddEq[2]{a *, Jacobian matrix}
{
\ShowSymbol{maps of conjugation, Jacobian matrix}{}=
\ShowEq{#2 #1a Jacobian matrix}a
}

\AddEq[1]{ao, C, Jacobian matrix}
{
\symb[#1-0]{#1(a)}{maps of conjugation, Jacobian matrix}{}
}

\AddEquation{aE, C, matrix}
{
\ShowSymbol{maps of conjugation, Jacobian matrix}{}=
\begin{pmatrix}
\aU a0&-a^{\gi 1}\\
a^{\gi 1}&\hphantom{-}\aU a0
\end{pmatrix}
}

\AddEquation{aI, C, matrix}
{
\ShowSymbol{maps of conjugation, Jacobian matrix}{}=
\begin{pmatrix}
\aU a0&\hphantom{-}a^{\gi 1}\\
a^{\gi 1}&-\aU a0
\end{pmatrix}
}

\AddEquation{f in L(A,A), 2, nonassociative algebra}
{
f
=a^{k\cdot \gi{ij}}(\aU ei\otimes\aU ej)\circ F_k
=
a^{k\cdot \gi{ij}}(\aU eiI_k)\aU ej
}

\AddEquation{system 1 of linear equations, quaternion}
{
\begin{split}
i\aU x1j+j\aU x2k&=i+j
\\
k\aU x1i+i\aU x2j&=i-j
\end{split}
}

\AddEquation{system 1 of linear equations, quaternion, 1}
{
\begin{split}
(i\otimes j)\circ\aU x1+(j\otimes k)\circ\aU x2&=i+j
\\
(k\otimes i)\circ\aU x1+(i\otimes j)\circ\aU x2&=i-j
\end{split}
}

\AddEquation{system 1 of linear equations, quaternion, 2}
{
\begin{pmatrix}
i\otimes j&j\otimes k\\
k\otimes i&i\otimes j
\end{pmatrix}
\RCcirc
\begin{pmatrix}
c\aU x1\\\aU x2
\end{pmatrix}
=
\begin{pmatrix}
i+j\\i-j
\end{pmatrix}
}

\AddEq{system 1 of linear equations, quaternion, 3}
{
\[
a=
\begin{pmatrix}
i\otimes j&j\otimes k\\
k\otimes i&i\otimes j
\end{pmatrix}
\]
}

\AddEq{system 1 of linear equations, quaternion, 4}
{
\ShowEq{system 1 of linear equations, quaternion, 4 1 1}
\ShowEq{system 1 of linear equations, quaternion, 4 1 2}
\ShowEq{system 1 of linear equations, quaternion, 4 2 1}
\ShowEq{system 1 of linear equations, quaternion, 4 2 2}
}

\AddEquation{system 1 of linear equations, quaternion, 4 1 1}
{
\begin{split}
\aUD{\det\left(a,\RCcirc\right)}11&=\aUD a11
-\aUD a1{[1]}\RCcirc
(\aUD a{[1]}{[1]})^{-1\RCcirc}\RCcirc
\aUD a{[1]}1
=\aUD a11
-\aUD a12\circ
(\aUD a22)^{-1}\circ
\aUD a21
\\
&=i\otimes j
-(j\otimes k)\circ
(i\otimes j)^{-1}\circ
(k\otimes i)
\\
&=i\otimes j
-(j\otimes k)\circ
(i\otimes j)\circ
(k\otimes i)
=i\otimes j
-(-1)\otimes (-1)
\\
&=i\otimes j
-1\otimes 1
\end{split}
}

\AddEquation{system 1 of linear equations, quaternion, 4 1 2}
{
\begin{split}
\aUD{\det\left(a,\RCcirc\right)}12&=\aUD a12
-\aUD a1{[2]}\RCcirc
(\aUD a{[1]}{[2]})^{-1\RCcirc}\RCcirc
\aUD a{[1]}2
=\aUD a12
-\aUD a11\circ
(\aUD a21)^{-1}\circ
\aUD a22
\\
&=j\otimes k
-(i\otimes j)\circ
(k\otimes i)^{-1}\circ
(i\otimes j)
\\
&=j\otimes k
-(i\otimes j)\circ
(k\otimes i)\circ
(i\otimes j)
=j\otimes k
-(-k)\otimes i
\\
&=j\otimes k
+k\otimes i
\end{split}
}

\AddEquation{system 1 of linear equations, quaternion, 4 2 1}
{
\begin{split}
\aUD{\det\left(a,\RCcirc\right)}21&=\aUD a21
-\aUD a2{[1]}\RCcirc
(\aUD a{[2]}{[1]})^{-1\RCcirc}\RCcirc
\aUD a{[2]}1
=\aUD a21
-\aUD a22\circ
(\aUD a12)^{-1}\circ
\aUD a11
\\
&=k\otimes i
-(i\otimes j)\circ
(j\otimes k)^{-1}\circ
(i\otimes j)
\\
&=k\otimes i
-(i\otimes j)\circ
(j\otimes k)\circ
(i\otimes j)
=k\otimes i
-(-j)\otimes k
\\
&=k\otimes i
+j\otimes k
\end{split}
}

\AddEquation{system 1 of linear equations, quaternion, 4 2 2}
{
\begin{split}
\aUD{\det\left(a,\RCcirc\right)}22&=\aUD a22
-\aUD a2{[2]}\RCcirc
(\aUD a{[2]}{[2]})^{-1\RCcirc}\RCcirc
\aUD a{[2]}2
=\aUD a22
-\aUD a21\circ
(\aUD a11)^{-1}\circ
\aUD a12
\\
&=i\otimes j
-(k\otimes i)\circ
(i\otimes j)^{-1}\circ
(j\otimes k)
\\
&=i\otimes j
-(k\otimes i)\circ
(i\otimes j)\circ
(j\otimes k)
=i\otimes j
-(-1)\otimes (-1)
\\
&=i\otimes j
-1\otimes 1
\end{split}
}

\AddEquation{fg=1 e o e}
{
\begin{split}
\aU f{ij}\aUD C0{ik}\aUD C0{lj}\aU g{kl}&=1\\
\aU f{ij}\aUD Cp{ik}\aUD Cq{lj}\aU g{kl}&=0\ \ \ p+q>0
\end{split}
}

\AddEq{g o h=1}
{
\[g\circ h=\aD e0\otimes\aD e0\]
}

\AddEq{f1=...}
{
\[f_1=-1\otimes 1+i\otimes j\]
}

\AddEq{f2=...}
{
\[f_2=k\otimes i+j\otimes k\]
}

\AddEq[2]{g=f-1}
{
$g_{#1}=f_{#1}^{-1}$#2
}

\AddEquation{fg=1 e o e g1}
{
\left\{
\begin{matrix}
-\aUD C0{0k}\aUD C0{l0}\aU g{kl}+
\aUD C0{1k}\aUD C0{l2}\aU g{kl}=1\\
-\aUD C0{0k}\aUD C1{l0}\aU g{kl}+
\aUD C0{1k}\aUD C1{l2}\aU g{kl}=0\\
-\aUD C0{0k}\aUD C2{l0}\aU g{kl}+
\aUD C0{1k}\aUD C2{l2}\aU g{kl}=0\\
-\aUD C0{0k}\aUD C3{l0}\aU g{kl}+
\aUD C0{1k}\aUD C3{l2}\aU g{kl}=0\\
-\aUD C1{0k}\aUD C0{l0}\aU g{kl}+
\aUD C1{1k}\aUD C0{l2}\aU g{kl}=0\\
-\aUD C1{0k}\aUD C1{l0}\aU g{kl}+
\aUD C1{1k}\aUD C1{l2}\aU g{kl}=0\\
-\aUD C1{0k}\aUD C2{l0}\aU g{kl}+
\aUD C1{1k}\aUD C2{l2}\aU g{kl}=0\\
-\aUD C1{0k}\aUD C3{l0}\aU g{kl}+
\aUD C1{1k}\aUD C3{l2}\aU g{kl}=0
\end{matrix}
\ \ \ \ \,
\begin{matrix}
-\aUD C2{0k}\aUD C0{l0}\aU g{kl}+
\aUD C2{1k}\aUD C0{l2}\aU g{kl}=0\\
-\aUD C2{0k}\aUD C1{l0}\aU g{kl}+
\aUD C2{1k}\aUD C1{l2}\aU g{kl}=0\\
-\aUD C2{0k}\aUD C2{l0}\aU g{kl}+
\aUD C2{1k}\aUD C2{l2}\aU g{kl}=0\\
-\aUD C2{0k}\aUD C3{l0}\aU g{kl}+
\aUD C2{1k}\aUD C3{l2}\aU g{kl}=0\\
-\aUD C3{0k}\aUD C0{l0}\aU g{kl}+
\aUD C3{1k}\aUD C0{l2}\aU g{kl}=0\\
-\aUD C3{0k}\aUD C1{l0}\aU g{kl}+
\aUD C3{1k}\aUD C1{l2}\aU g{kl}=0\\
-\aUD C3{0k}\aUD C2{l0}\aU g{kl}+
\aUD C3{1k}\aUD C2{l2}\aU g{kl}=0\\
-\aUD C3{0k}\aUD C3{l0}\aU g{kl}+
\aUD C3{1k}\aUD C3{l2}\aU g{kl}=0
\end{matrix}
\right.
}

\AddEquation{fg=1 e o e g1 1}
{
\left\{
\begin{matrix}
\BlueText{-\aU g{00}+\aU g{12}=1}\\
-\aU g{01}+\aU g{13}=0\\
-\aU g{02}-\aU g{10}=0\\
-\aU g{03}+\aU g{11}=0\\
-\aU g{10}-\aU g{02}=0\\
-\aU g{11}-\aU g{03}=0\\
\BlueText{-\aU g{12}+\aU g{00}=0}\\
-\aU g{13}+\aU g{01}=0
\end{matrix}
\ \ \ \ \,
\begin{matrix}
-\aUD C2{0k}\aUD C0{l0}\aU g{kl}+
\aUD C2{1k}\aUD C0{l2}\aU g{kl}=0\\
-\aUD C2{0k}\aUD C1{l0}\aU g{kl}+
\aUD C2{1k}\aUD C1{l2}\aU g{kl}=0\\
-\aUD C2{0k}\aUD C2{l0}\aU g{kl}+
\aUD C2{1k}\aUD C2{l2}\aU g{kl}=0\\
-\aUD C2{0k}\aUD C3{l0}\aU g{kl}+
\aUD C2{1k}\aUD C3{l2}\aU g{kl}=0\\
-\aUD C3{0k}\aUD C0{l0}\aU g{kl}+
\aUD C3{1k}\aUD C0{l2}\aU g{kl}=0\\
-\aUD C3{0k}\aUD C1{l0}\aU g{kl}+
\aUD C3{1k}\aUD C1{l2}\aU g{kl}=0\\
-\aUD C3{0k}\aUD C2{l0}\aU g{kl}+
\aUD C3{1k}\aUD C2{l2}\aU g{kl}=0\\
-\aUD C3{0k}\aUD C3{l0}\aU g{kl}+
\aUD C3{1k}\aUD C3{l2}\aU g{kl}=0
\end{matrix}
\right.
}

\AddEquation{fg=1 e o e g2}
{
\left\{
\begin{matrix}
\aUD C0{3k}\aUD C0{l1}\aU g{kl}+
\aUD C0{2k}\aUD C0{l3}\aU g{kl}=1\\
\aUD C0{3k}\aUD C1{l1}\aU g{kl}+
\aUD C0{2k}\aUD C1{l3}\aU g{kl}=0\\
\aUD C0{3k}\aUD C2{l1}\aU g{kl}+
\aUD C0{2k}\aUD C2{l3}\aU g{kl}=0\\
\aUD C0{3k}\aUD C3{l1}\aU g{kl}+
\aUD C0{2k}\aUD C3{l3}\aU g{kl}=0\\
\aUD C1{3k}\aUD C0{l1}\aU g{kl}+
\aUD C1{2k}\aUD C0{l3}\aU g{kl}=0\\
\aUD C1{3k}\aUD C1{l1}\aU g{kl}+
\aUD C1{2k}\aUD C1{l3}\aU g{kl}=0\\
\aUD C1{3k}\aUD C2{l1}\aU g{kl}+
\aUD C1{2k}\aUD C2{l3}\aU g{kl}=0\\
\aUD C1{3k}\aUD C3{l1}\aU g{kl}+
\aUD C1{2k}\aUD C3{l3}\aU g{kl}=0
\end{matrix}
\ \ \ \ \,
\begin{matrix}
\aUD C2{3k}\aUD C0{l1}\aU g{kl}+
\aUD C2{2k}\aUD C0{l3}\aU g{kl}=0\\
\aUD C2{3k}\aUD C1{l1}\aU g{kl}+
\aUD C2{2k}\aUD C1{l3}\aU g{kl}=0\\
\aUD C2{3k}\aUD C2{l1}\aU g{kl}+
\aUD C2{2k}\aUD C2{l3}\aU g{kl}=0\\
\aUD C2{3k}\aUD C3{l1}\aU g{kl}+
\aUD C2{2k}\aUD C3{l3}\aU g{kl}=0\\
\aUD C3{3k}\aUD C0{l1}\aU g{kl}+
\aUD C3{2k}\aUD C0{l3}\aU g{kl}=0\\
\aUD C3{3k}\aUD C1{l1}\aU g{kl}+
\aUD C3{2k}\aUD C1{l3}\aU g{kl}=0\\
\aUD C3{3k}\aUD C2{l1}\aU g{kl}+
\aUD C3{2k}\aUD C2{l3}\aU g{kl}=0\\
\aUD C3{3k}\aUD C3{l1}\aU g{kl}+
\aUD C3{2k}\aUD C3{l3}\aU g{kl}=0
\end{matrix}
\right.
}

\AddEquation{fg=1 e o e g2 1}
{
\left\{
\begin{matrix}
\BlueText{\aU g{31}+\aU g{23}=1}\\
-\aU g{30}-\aU g{22}=0\\
-\aU g{33}+\aU g{21}=0\\
\aU g{32}-\aU g{20}=0\\
\aU g{21}-\aU g{33}=0\\
-\aU g{20}+\aU g{32}=0\\
\BlueText{-\aU g{23}-\aU g{31}=0}\\
\aU g{22}+\aU g{30}=0
\end{matrix}
\ \ \ \ \,
\begin{matrix}
\aUD C2{3k}\aUD C0{l1}\aU g{kl}+
\aUD C2{2k}\aUD C0{l3}\aU g{kl}=0\\
\aUD C2{3k}\aUD C1{l1}\aU g{kl}+
\aUD C2{2k}\aUD C1{l3}\aU g{kl}=0\\
\aUD C2{3k}\aUD C2{l1}\aU g{kl}+
\aUD C2{2k}\aUD C2{l3}\aU g{kl}=0\\
\aUD C2{3k}\aUD C3{l1}\aU g{kl}+
\aUD C2{2k}\aUD C3{l3}\aU g{kl}=0\\
\aUD C3{3k}\aUD C0{l1}\aU g{kl}+
\aUD C3{2k}\aUD C0{l3}\aU g{kl}=0\\
\aUD C3{3k}\aUD C1{l1}\aU g{kl}+
\aUD C3{2k}\aUD C1{l3}\aU g{kl}=0\\
\aUD C3{3k}\aUD C2{l1}\aU g{kl}+
\aUD C3{2k}\aUD C2{l3}\aU g{kl}=0\\
\aUD C3{3k}\aUD C3{l1}\aU g{kl}+
\aUD C3{2k}\aUD C3{l3}\aU g{kl}=0
\end{matrix}
\right.
}

\AddEquation{quaternion fUD<-fUU f1 1}
{
\begin{split}
&
\ShowEq{quaternion matrix fUD}
\\=&
\ShowEq{quaternion matrix fUD<-fUU}
\left(
\begin{array}{rrrr}
-1&0&0&0\\
0&0&0&-1\\
0&0&0&0\\
0&0&0&0
\end{array}
\right)
\\=&
\left(
\begin{array}{rrrr}
-1&0&0&1\\
-1&0&0&1\\
-1&0&0&-1\\
-1&0&0&-1
\end{array}
\right)
\end{split}
}

\AddEquation{quaternion fUD<-fUU f2 1}
{
\begin{split}
&
\ShowEq{quaternion matrix fUD}
\\=&
\ShowEq{quaternion matrix fUD<-fUU}
\left(
\begin{array}{rrrr}
0&0&0&0\\
0&0&0&0\\
0&-1&0&0\\
0&0&-1&0
\end{array}
\right)
\\=&
\left(
\begin{array}{rrrr}
0&1&1&0\\
0&-1&-1&0\\
0&1&-1&0\\
0&-1&1&0
\end{array}
\right)
\end{split}
}

\AddEquation{matrix f1}
{
f_1=
\left(
\begin{array}{rrrr}
-1&0&0&1\\
0&-1&-1&0\\
0&-1&-1&0\\
1&0&0&-1
\end{array}
\right)
}

\AddEquation{matrix f2}
{
f_2=
\left(
\begin{array}{rrrr}
0&1&1&0\\
1&0&0&-1\\
1&0&0&-1\\
0&-1&-1&0
\end{array}
\right)
}

\AddEq[1]{f=...ei o ek}
{
#1=\aU {#1}{ij}\aD ei\otimes\aD ej
}

\AddEquation{fg=...ei o ek}
{
f\circ g=\aU {(f\circ g)}{pq}\aD ep\otimes\aD eq
}

\AddEquation{fg=CCfg}
{
\aU {(f\circ g)}{pq}=\aU f{ij}\aU g{kl}\aUD Cp{ik}\aUD Cq{lj}
}

\AddEquation{fg=...e o e}
{
f\circ g=\aU f{ij}\aU g{kl}(\aD ei\aD ek)\otimes(\aD el\aD ej)
}

\AddEquation{fg=...C e o e}
{
f\circ g=\aU f{ij}\aU g{kl}\aUD Cp{ik}\aUD Cq{lj}\aD ep\otimes\aD eq
}

\AddEq[2]{f in AoA}
{
$#1\in A\otimes A$#2
}

\AddEq[2]{texorpdf L(D;A->A)}
{%
\texorpdfstring{%
\ShowEq{L(A;B)}#1#2#2{}%
}{L(#1;#2->#2)}%
}

\AddEq{I0=E}
{
$\aU I0=E$.
}

\AddEq{coordinates of map A->A, 2}
{
\aUD fkl=\aU {f^{k\cdot}_{}{}}{ij}
\aUD {F_{k\cdot}^{}{}}ml
\aUD Cp{im}\aUD Ck{pj}
}

\AddEq [2]{coordinates of map A1 A2, FoG, associative algebra}
{
\def\Cdot{}
\def\Temp{-}%
\edef\Tempa{#2}%
\ifx\Tempa\Temp%
\else
\def\Cdot{#2\cdot}
\fi
#1^{\gik}_{\gil}=#1^{k\cdot\gi{ij}}
F_{k\cdot}^{}{}^{\gim}_{\gi r}G^{\gi r}_{\gil}
C_{\Cdot}{}^{\gi p}_{\gi{im}}C_{\Cdot}{}^{\gik}_{\gi{pj}}
}

\AddEq [2]{coordinates of map A->B}
{
\def\Cdot{}
\def\Temp{-}%
\edef\Tempa{#2}%
\ifx\Tempa\Temp%
\else
\def\Cdot{#2\cdot}
\fi
#1^{\gik}_{\gil}=#1^{k\cdot\gi{ij}}
F_{k\cdot}^{}{}^{\gim}_{\gi r}G^{\gi r}_{\gil}
C_{\Cdot}{}^{\gi p}_{\gi{im}}C_{\Cdot}{}^{\gik}_{\gi{pj}}
}

\AddEquation{a...= H}
{
\begin{split}
&\,
\ShowEq{matrix a algebra H}
\\ = &\,
\ShowEq{matrix af algebra H}
\ShowEq{matrix f algebra H}
\end{split}
}

\AddEquation{a0=f03}
{
\begin{split}
\aD a0&=\aUD a00+\aUD a10i+\aUD a20j+\aUD a30k
\\&=\frac 12
(-\aUD f00+\aUD f11+\aUD f22+\aUD f33
+(-\aUD f10-\aUD f01+\aUD f32-\aUD f23)i
\\&+(-\aUD f20-\aUD f31-\aUD f02+\aUD f13)j
+(-\aUD f30+\aUD f21-\aUD f12-\aUD f03)k)
\\&=\frac 12
((-\aUD f00-\aUD f10i-\aUD f20j-\aUD f30k)
+(-\aUD f01i+\aUD f11+\aUD f21k-\aUD f31j)
\\&+(-\aUD f02j-\aUD f12k+\aUD f22+\aUD f32i)
+(-\aUD f03k+\aUD f13j-\aUD f23i+\aUD f33))
\\&=\frac 12
(-(\aUD f00+\aUD f10i+\aUD f20j+\aUD f30k)
-(\aUD f01+\aUD f11i+\aUD f21j+\aUD f31k)i
\\&-(\aUD f02+\aUD f12i+\aUD f22j+\aUD f32k)j
-(\aUD f03+\aUD f13i+\aUD f23j+\aUD f33k)k)
\end{split}
}

\AddEquation{a1=f03}
{
\begin{split}
\aD a1&=\aUD a01+\aUD a11i+\aUD a21j+\aUD a31k
\\&=\frac 12
(\aUD f00-\aUD f11+(\aUD f10+\aUD f01)i+(\aUD f20+\aUD f31)j+(\aUD f30-\aUD f21)k)
\\&=\frac 12
((\aUD f00+\aUD f10i+\aUD f20j+\aUD f30k)+(\aUD f01i-\aUD f11-\aUD f21k+\aUD f31j))
\\&=\frac 12
((\aUD f00+\aUD f10i+\aUD f20j+\aUD f30k)+(\aUD f01+\aUD f11i+\aUD f21j+\aUD f31k)i)
\end{split}
}

\AddEquation{a2=f03}
{
\begin{split}
\aD a2&=\aUD a02+\aUD a12i+\aUD a22j+\aUD a32k
\\&=\frac 12
(\aUD f00-\aUD f22+(\aUD f10-\aUD f32)i+(\aUD f20+\aUD f02)j+(\aUD f30+\aUD f12)k)
\\&=\frac 12
(\aUD f00-\aUD f22+\aUD f10i-\aUD f32i+\aUD f20j+\aUD f02j+\aUD f30k+\aUD f12k)
\\&=\frac 12
((\aUD f00+\aUD f10i+\aUD f20j+\aUD f30k)+(\aUD f02j+\aUD f12k-\aUD f22-\aUD f32i))
\\&=\frac 12
((\aUD f00+\aUD f10i+\aUD f20j+\aUD f30k)+(\aUD f02+\aUD f12i+\aUD f22j+\aUD f32k)j)
\end{split}
}

\AddEquation{a3=f03}
{
\begin{split}
\aD a3&=\aUD a03+\aUD a13i+\aUD a23j+\aUD a33k
\\&=\frac 12
(\aUD f00-\aUD f33+(\aUD f10+\aUD f23)i+(\aUD f20-\aUD f13)j+(\aUD f30+\aUD f03)k)
\\&=\frac 12
((\aUD f00+\aUD f10i+\aUD f30k+\aUD f20j)+(\aUD f03k-\aUD f13j+\aUD f23i-\aUD f33))
\\&=\frac 12
((\aUD f00+\aUD f10i+\aUD f30k+\aUD f20j)+(\aUD f03+\aUD f13i+\aUD f23j+\aUD f33k)k)
\end{split}
}

\AddEq{a0=f07 1}
{
\begin{align*}
\aD a0&=\aUD a00\aD e0+\aUD a10\aD e1+\aUD a20\aD e2+\aUD a30\aD e3
+\aUD a40\aD e4+\aUD a50\aD e5+\aUD a60\aD e6+\aUD a70\aD e7
\\&=\frac 12
((-5\aUD f00+\aUD f11+\aUD f22+\aUD f33+\aUD f44+\aUD f55+\aUD f66+\aUD f77)\aD e0
\\&+(-5\aUD f10-\aUD f01+\aUD f32-\aUD f23+\aUD f54-\aUD f45-\aUD f76+\aUD f67)\aD e1
\\&+(-5\aUD f20-\aUD f31-\aUD f02+\aUD f13+\aUD f64+\aUD f75-\aUD f46-\aUD f57)\aD e2
\\&+(-5\aUD f30+\aUD f21-\aUD f12-\aUD f03+\aUD f74-\aUD f65+\aUD f56-\aUD f47)\aD e3
\\&+(-5\aUD f40-\aUD f51-\aUD f62-\aUD f73-\aUD f04+\aUD f15+\aUD f26+\aUD f37)\aD e4
\\&+(-5\aUD f50+\aUD f41-\aUD f72+\aUD f63-\aUD f14-\aUD f05-\aUD f36+\aUD f27)\aD e5
\\&+(-5\aUD f60+\aUD f71+\aUD f42-\aUD f53-\aUD f24+\aUD f35-\aUD f06-\aUD f17)\aD e6
\\&+(-5\aUD f70-\aUD f61+\aUD f52+\aUD f43-\aUD f34-\aUD f25+\aUD f16-\aUD f07)\aD e7)
\end{align*}
}

\AddEq{a0=f07 2}
{
\begin{align*}
\aD a0&=\frac 12
(-5\aUD f00\aD e0+\aUD f11\aD e0+\aUD f22\aD e0+\aUD f33\aD e0
+\aUD f44\aD e0+\aUD f55\aD e0+\aUD f66\aD e0+\aUD f77\aD e0
\\&-5\aUD f10\aD e1-\aUD f01\aD e1+\aUD f32\aD e1-\aUD f23\aD e1
+\aUD f54\aD e1-\aUD f45\aD e1-\aUD f76\aD e1+\aUD f67\aD e1
\\&-5\aUD f20\aD e2-\aUD f31\aD e2-\aUD f02\aD e2+\aUD f13\aD e2
+\aUD f64\aD e2+\aUD f75\aD e2-\aUD f46\aD e2-\aUD f57\aD e2
\\&-5\aUD f30\aD e3+\aUD f21\aD e3-\aUD f12\aD e3-\aUD f03\aD e3
+\aUD f74\aD e3-\aUD f65\aD e3+\aUD f56\aD e3-\aUD f47\aD e3
\\&-5\aUD f40\aD e4-\aUD f51\aD e4-\aUD f62\aD e4-\aUD f73\aD e4
-\aUD f04\aD e4+\aUD f15\aD e4+\aUD f26\aD e4+\aUD f37\aD e4
\\&-5\aUD f50\aD e5+\aUD f41\aD e5-\aUD f72\aD e5+\aUD f63\aD e5
-\aUD f14\aD e5-\aUD f05\aD e5-\aUD f36\aD e5+\aUD f27\aD e5
\\&-5\aUD f60\aD e6+\aUD f71\aD e6+\aUD f42\aD e6-\aUD f53\aD e6
-\aUD f24\aD e6+\aUD f35\aD e6-\aUD f06\aD e6-\aUD f17\aD e6
\\&-5\aUD f70\aD e7-\aUD f61\aD e7+\aUD f52\aD e7+\aUD f43\aD e7
-\aUD f34\aD e7-\aUD f25\aD e7+\aUD f16\aD e7-\aUD f07\aD e7)
\end{align*}
}

\AddEq{a0=f07 3}
{
\begin{align*}
\aD a0&=\frac 12
(-5(\aUD f00\aD e0+\aUD f10\aD e1+\aUD f20\aD e2+\aUD f30\aD e3
+\aUD f40\aD e4+\aUD f50\aD e5+\aUD f60\aD e6+\aUD f70\aD e7)
\\&+(-\aUD f01\aD e1+\aUD f11\aD e0+\aUD f21\aD e3-\aUD f31\aD e2
+\aUD f41\aD e5-\aUD f51\aD e4-\aUD f61\aD e7+\aUD f71\aD e6)
\\&+(-\aUD f02\aD e2-\aUD f12\aD e3+\aUD f22\aD e0+\aUD f32\aD e1
+\aUD f42\aD e6+\aUD f52\aD e7-\aUD f62\aD e4-\aUD f72\aD e5)
\\&+(-\aUD f03\aD e3+\aUD f13\aD e2-\aUD f23\aD e1+\aUD f33\aD e0
+\aUD f43\aD e7-\aUD f53\aD e6+\aUD f63\aD e5-\aUD f73\aD e4)
\\&+(-\aUD f04\aD e4-\aUD f14\aD e5-\aUD f24\aD e6-\aUD f34\aD e7
+\aUD f44\aD e0+\aUD f54\aD e1+\aUD f64\aD e2+\aUD f74\aD e3)
\\&+(-\aUD f05\aD e5+\aUD f15\aD e4-\aUD f25\aD e7+\aUD f35\aD e6
-\aUD f45\aD e1+\aUD f55\aD e0-\aUD f65\aD e3+\aUD f75\aD e2)
\\&+(-\aUD f06\aD e6+\aUD f16\aD e7+\aUD f26\aD e4-\aUD f36\aD e5
-\aUD f46\aD e2+\aUD f56\aD e3+\aUD f66\aD e0-\aUD f76\aD e1)
\\&+(-\aUD f07\aD e7-\aUD f17\aD e6+\aUD f27\aD e5+\aUD f37\aD e4
-\aUD f47\aD e3-\aUD f57\aD e2+\aUD f67\aD e1+\aUD f77\aD e0)
\end{align*}
}

\AddEquation{a0=f07}
{
\begin{split}
\aD a0&=-\frac 12
(5(\aUD f00\aD e0+\aUD f10\aD e1+\aUD f20\aD e2+\aUD f30\aD e3
+\aUD f40\aD e4+\aUD f50\aD e5+\aUD f60\aD e6+\aUD f70\aD e7)\aD e0
\\&+(\aUD f01\aD e0+\aUD f11\aD e0+\aUD f21\aD e2+\aUD f31\aD e3
+\aUD f41\aD e4+\aUD f51\aD e5+\aUD f61\aD e6+\aUD f71\aD e7)\aD e1
\\&+(\aUD f02\aD e0+\aUD f12\aD e1+\aUD f22\aD e2+\aUD f32\aD e3
+\aUD f42\aD e4+\aUD f52\aD e5+\aUD f62\aD e6+\aUD f72\aD e7)\aD e2
\\&+(\aUD f03\aD e0+\aUD f13\aD e2+\aUD f23\aD e2+\aUD f33\aD e0
+\aUD f43\aD e4+\aUD f53\aD e5+\aUD f63\aD e6+\aUD f73\aD e7)\aD e3
\\&+(\aUD f04\aD e0+\aUD f14\aD e1+\aUD f24\aD e2+\aUD f34\aD e3
+\aUD f44\aD e4+\aUD f54\aD e5+\aUD f64\aD e6+\aUD f74\aD e7)\aD e4
\\&+(\aUD f05\aD e0+\aUD f15\aD e1+\aUD f25\aD e2+\aUD f35\aD e3
+\aUD f45\aD e4+\aUD f55\aD e5+\aUD f65\aD e3+\aUD f75\aD e7)\aD e5
\\&+(\aUD f06\aD e6+\aUD f16\aD e1+\aUD f26\aD e2+\aUD f36\aD e3
+\aUD f46\aD e4+\aUD f56\aD e5+\aUD f66\aD e6+\aUD f76\aD e7)\aD e6
\\&+(\aUD f07\aD e0+\aUD f17\aD e1+\aUD f27\aD e2+\aUD f37\aD e3
+\aUD f47\aD e4+\aUD f57\aD e5+\aUD f67\aD e6+\aUD f77\aD e7)\aD e7
\end{split}
}

\AddEquation{a1=f07}
{
\begin{split}
\aD a1&=\aUD a01\aD e0+\aUD a11\aD e1+\aUD a21\aD e2+\aUD a31\aD e3
+\aUD a41\aD e4+\aUD a51\aD e5+\aUD a61\aD e6+\aUD a71\aD e7
\\&=\frac 12
((\aUD f00-\aUD f11)\aD e0+(\aUD f10+\aUD f01)\aD e1
+(\aUD f20+\aUD f31)\aD e2+(\aUD f30-\aUD f21)\aD e3
\\ &
+(\aUD f40+\aUD f51)\aD e4+(\aUD f50-\aUD f41)\aD e5
+(\aUD f60-\aUD f71)\aD e6+(\aUD f70+\aUD f61)\aD e7)
\\&=\frac 12
(\aUD f00\aD e0-\aUD f11\aD e0+\aUD f10\aD e1+\aUD f01\aD e1
+\aUD f20\aD e2+\aUD f31\aD e2+\aUD f30\aD e3-\aUD f21\aD e3
\\ &
+\aUD f40\aD e4+\aUD f51\aD e4+\aUD f50\aD e5-\aUD f41\aD e5
+\aUD f60\aD e6-\aUD f71\aD e6+\aUD f70\aD e7+\aUD f61\aD e7)
\\&=\frac 12
((\aUD f00\aD e0+\aUD f10\aD e1+\aUD f20\aD e2+\aUD f30\aD e3+\aUD f40\aD e4
+\aUD f50\aD e5+\aUD f60\aD e6+\aUD f70\aD e7)
\\ &
+(\aUD f01\aD e1-\aUD f11\aD e0-\aUD f21\aD e3+\aUD f31\aD e2
-\aUD f41\aD e5+\aUD f51\aD e4+\aUD f61\aD e7-\aUD f71\aD e6))
\\&=\frac 12
((\aUD f00\aD e0+\aUD f10\aD e1+\aUD f20\aD e2+\aUD f30\aD e3+\aUD f40\aD e4
+\aUD f50\aD e5+\aUD f60\aD e6+\aUD f70\aD e7)
\\ &
+(\aUD f01\aD e0+\aUD f11\aD e1+\aUD f21\aD e2+\aUD f31\aD e3
+\aUD f41\aD e4+\aUD f51\aD e5+\aUD f61\aD e6+\aUD f71\aD e7)\aD e1)
\end{split}
}

\AddEquation{a2=f07}
{
\begin{split}
\aD a2&=\aUD a02\aD e0+\aUD a12\aD e1+\aUD a22\aD e2+\aUD a32\aD e3
+\aUD a42\aD e4+\aUD a52\aD e5+\aUD a62\aD e6+\aUD a72\aD e7
\\&=\frac 12
((\aUD f00-\aUD f22)\aD e0+(\aUD f10-\aUD f32)\aD e1
+(\aUD f20+\aUD f02)\aD e2+(\aUD f30+\aUD f12)\aD e3
\\ &
+(\aUD f40+\aUD f62)\aD e4+(\aUD f50+\aUD f72)\aD e5
+(\aUD f60-\aUD f42)\aD e6+(\aUD f70-\aUD f52)\aD e7)
\\&=\frac 12
(\aUD f00\aD e0-\aUD f22\aD e0+\aUD f10\aD e1-\aUD f32\aD e1
+\aUD f20\aD e2+\aUD f02\aD e2+\aUD f30\aD e3+\aUD f12\aD e3
\\ &
+\aUD f40\aD e4+\aUD f62\aD e4+\aUD f50\aD e5+\aUD f72\aD e5
+\aUD f60\aD e6-\aUD f42\aD e6+\aUD f70\aD e7-\aUD f52\aD e7)
\\&=\frac 12
((\aUD f00\aD e0+\aUD f10\aD e1+\aUD f20\aD e2+\aUD f30\aD e3+\aUD f40\aD e4
+\aUD f50\aD e5+\aUD f60\aD e6+\aUD f70\aD e7)
\\ &
+(\aUD f02\aD e2+\aUD f12\aD e3-\aUD f32\aD e1-\aUD f22\aD e0
-\aUD f42\aD e6-\aUD f52\aD e7+\aUD f62\aD e4+\aUD f72\aD e5))
\\&=\frac 12
((\aUD f00\aD e0+\aUD f10\aD e1+\aUD f20\aD e2+\aUD f30\aD e3+\aUD f40\aD e4
+\aUD f50\aD e5+\aUD f60\aD e6+\aUD f70\aD e7)
\\ &
+(\aUD f02\aD e0+\aUD f12\aD e1+\aUD f22\aD e2+\aUD f32\aD e3
+\aUD f42\aD e4+\aUD f52\aD e5+\aUD f62\aD e6+\aUD f72\aD e7)\aD e2)
\end{split}
}

\AddEquation{a3=f07}
{
\begin{split}
\aD a3&=\aUD a03\aD e0+\aUD a13\aD e1+\aUD a23\aD e2+\aUD a33\aD e3
+\aUD a43\aD e4+\aUD a53\aD e5+\aUD a63\aD e6+\aUD a73\aD e7
\\&=\frac 12
((\aUD f00-\aUD f33)\aD e0+(\aUD f10+\aUD f23)\aD e1
+(\aUD f20-\aUD f13)\aD e2+(\aUD f30+\aUD f03)\aD e3
\\ &
+(\aUD f40+\aUD f73)\aD e4+(\aUD f50-\aUD f63)\aD e5
+(\aUD f60+\aUD f53)\aD e6+(\aUD f70-\aUD f43)\aD e7)
\\&=\frac 12
(\aUD f00\aD e0-\aUD f33\aD e0+\aUD f10\aD e1+\aUD f23\aD e1
+\aUD f20\aD e2-\aUD f13\aD e2+\aUD f30\aD e3+\aUD f03\aD e3
\\ &
+\aUD f40\aD e4+\aUD f73\aD e4+\aUD f50\aD e5-\aUD f63\aD e5
+\aUD f60\aD e6+\aUD f53\aD e6+\aUD f70\aD e7-\aUD f43\aD e7)
\\&=\frac 12
((\aUD f00\aD e0+\aUD f10\aD e1+\aUD f20\aD e2+\aUD f30\aD e3+\aUD f40\aD e4
+\aUD f50\aD e5+\aUD f60\aD e6+\aUD f70\aD e7)
\\ &
+(\aUD f03\aD e3-\aUD f13\aD e2+\aUD f23\aD e1-\aUD f33\aD e0
-\aUD f43\aD e7+\aUD f53\aD e6-\aUD f63\aD e5+\aUD f73\aD e4))
\\&=\frac 12
((\aUD f00\aD e0+\aUD f10\aD e1+\aUD f20\aD e2+\aUD f30\aD e3+\aUD f40\aD e4
+\aUD f50\aD e5+\aUD f60\aD e6+\aUD f70\aD e7)
\\ &
+(\aUD f03\aD e0+\aUD f13\aD e1+\aUD f23\aD e2+\aUD f33\aD e3
+\aUD f43\aD e4+\aUD f53\aD e5+\aUD f63\aD e6+\aUD f73\aD e7)\aD e3)
\end{split}
}

\AddEquation{a4=f07}
{
\begin{split}
\aD a4&=\aUD a04\aD e0+\aUD a14\aD e1+\aUD a24\aD e2+\aUD a34\aD e3
+\aUD a44\aD e4+\aUD a54\aD e5+\aUD a64\aD e6+\aUD a74\aD e7
\\&=\frac 12
((\aUD f00-\aUD f44)\aD e0+(\aUD f10-\aUD f54)\aD e1
+(\aUD f20-\aUD f64)\aD e2+(\aUD f30-\aUD f74\aD e3
\\ &
+(\aUD f40+\aUD f04)\aD e4+(\aUD f50+\aUD f14)\aD e5
+(\aUD f60+\aUD f24)\aD e6+(\aUD f70+\aUD f34)\aD e7)
\\&=\frac 12
(\aUD f00\aD e0-\aUD f44\aD e0+\aUD f10\aD e1-\aUD f54\aD e1
+\aUD f20\aD e2-\aUD f64\aD e2+\aUD f30\aD e3-\aUD f74\aD e3
\\ &
+\aUD f40\aD e4+\aUD f04\aD e4+\aUD f50\aD e5+\aUD f14\aD e5
+\aUD f60\aD e6+\aUD f24\aD e6+\aUD f70\aD e7+\aUD f34\aD e7)
\\&=\frac 12
((\aUD f00\aD e0+\aUD f10\aD e1+\aUD f20\aD e2+\aUD f30\aD e3+\aUD f40\aD e4
+\aUD f50\aD e5+\aUD f60\aD e6+\aUD f70\aD e7)
\\ &
+(\aUD f04\aD e4+\aUD f14\aD e5+\aUD f24\aD e6+\aUD f34\aD e7
-\aUD f44\aD e0-\aUD f54\aD e1-\aUD f64\aD e2-\aUD f74\aD e3))
\\&=\frac 12
((\aUD f00\aD e0+\aUD f10\aD e1+\aUD f20\aD e2+\aUD f30\aD e3+\aUD f40\aD e4
+\aUD f50\aD e5+\aUD f60\aD e6+\aUD f70\aD e7)
\\ &
+(\aUD f04\aD e0+\aUD f14\aD e1+\aUD f24\aD e2+\aUD f34\aD e3
+\aUD f44\aD e4+\aUD f54\aD e5+\aUD f64\aD e6+\aUD f74\aD e7)\aD e4)
\end{split}
}

\AddEquation{a5=f07}
{
\begin{split}
\aD a5&=\aUD a05\aD e0+\aUD a15\aD e1+\aUD a25\aD e2+\aUD a35\aD e3
+\aUD a45\aD e4+\aUD a55\aD e5+\aUD a65\aD e6+\aUD a75\aD e7
\\&=\frac 12
((\aUD f00-\aUD f55)\aD e0+(\aUD f10+\aUD f45)\aD e1
+(\aUD f20-\aUD f75)\aD e2+(\aUD f30+\aUD f65)\aD e3
\\ &
+(\aUD f40-\aUD f15)\aD e4+(\aUD f50+\aUD f05)\aD e5
+(\aUD f60-\aUD f35)\aD e6+(\aUD f70+\aUD f25)\aD e7)
\\&=\frac 12
(\aUD f00\aD e0-\aUD f55\aD e0+\aUD f10\aD e1+\aUD f45\aD e1
+\aUD f20\aD e2-\aUD f75\aD e2+\aUD f30\aD e3+\aUD f65\aD e3
\\ &
+\aUD f40\aD e4-\aUD f15\aD e4+\aUD f50\aD e5+\aUD f05\aD e5
+\aUD f60\aD e6-\aUD f35\aD e6+\aUD f70\aD e7+\aUD f25\aD e7)
\\&=\frac 12
((\aUD f00\aD e0+\aUD f10\aD e1+\aUD f20\aD e2+\aUD f30\aD e3+\aUD f40\aD e4
+\aUD f50\aD e5+\aUD f60\aD e6+\aUD f70\aD e7)
\\ &
+(\aUD f05\aD e5-\aUD f15\aD e4+\aUD f25\aD e7-\aUD f35\aD e6
+\aUD f45\aD e1-\aUD f55\aD e0+\aUD f65\aD e3-\aUD f75\aD e2))
\\&=\frac 12
((\aUD f00\aD e0+\aUD f10\aD e1+\aUD f20\aD e2+\aUD f30\aD e3+\aUD f40\aD e4
+\aUD f50\aD e5+\aUD f60\aD e6+\aUD f70\aD e7)
\\ &
+(\aUD f05\aD e0+\aUD f15\aD e1+\aUD f25\aD e2+\aUD f35\aD e3
+\aUD f45\aD e4+\aUD f55\aD e5+\aUD f65\aD e6+\aUD f75\aD e7)\aD e5)
\end{split}
}

\AddEquation{a6=f07}
{
\begin{split}
\aD a6&=\aUD a06\aD e0+\aUD a16\aD e1+\aUD a26\aD e2+\aUD a36\aD e3
+\aUD a46\aD e4+\aUD a56\aD e5+\aUD a66\aD e6+\aUD a76\aD e7
\\&=\frac 12
((\aUD f00-\aUD f66)\aD e0+(\aUD f10+\aUD f76)\aD e1
+(\aUD f20+\aUD f46)\aD e2+(\aUD f30-\aUD f56\aD e3
\\ &
+(\aUD f40-\aUD f26)\aD e4+(\aUD f50+\aUD f36)\aD e5
+(\aUD f60+\aUD f06)\aD e6+(\aUD f70-\aUD f16)\aD e7)
\\&=\frac 12
(\aUD f00\aD e0-\aUD f66\aD e0+\aUD f10\aD e1+\aUD f76\aD e1
+\aUD f20\aD e2+\aUD f46\aD e2+\aUD f30\aD e3-\aUD f56\aD e3
\\ &
+\aUD f40\aD e4-\aUD f26\aD e4+\aUD f50\aD e5+\aUD f36\aD e5
+\aUD f60\aD e6+\aUD f06\aD e6+\aUD f70\aD e7-\aUD f16\aD e7)
\\&=\frac 12
((\aUD f00\aD e0+\aUD f10\aD e1+\aUD f20\aD e2+\aUD f30\aD e3+\aUD f40\aD e4
+\aUD f50\aD e5+\aUD f60\aD e6+\aUD f70\aD e7)
\\ &
+(\aUD f06\aD e6-\aUD f16\aD e7-\aUD f26\aD e4+\aUD f36\aD e5
+\aUD f46\aD e2-\aUD f56\aD e3-\aUD f66\aD e0+\aUD f76\aD e1))
\\&=\frac 12
((\aUD f00\aD e0+\aUD f10\aD e1+\aUD f20\aD e2+\aUD f30\aD e3+\aUD f40\aD e4
+\aUD f50\aD e5+\aUD f60\aD e6+\aUD f70\aD e7)
\\ &
+(\aUD f06\aD e0+\aUD f16\aD e1+\aUD f26\aD e2+\aUD f36\aD e3
+\aUD f46\aD e4+\aUD f56\aD e5+\aUD f66\aD e6+\aUD f76\aD e7)\aD e6)
\end{split}
}

\AddEquation{a7=f07}
{
\begin{split}
\aD a7&=\aUD a07\aD e0+\aUD a17\aD e1+\aUD a27\aD e2+\aUD a37\aD e3
+\aUD a47\aD e4+\aUD a57\aD e5+\aUD a67\aD e6+\aUD a77\aD e7
\\&=\frac 12
((\aUD f00-\aUD f77)\aD e0+(\aUD f10-\aUD f67)\aD e1
+(\aUD f20+\aUD f57)\aD e2+(\aUD f30+\aUD f47)\aD e3
\\ &
+(\aUD f40-\aUD f37)\aD e4+(\aUD f50-\aUD f27)\aD e5
+(\aUD f60+\aUD f17)\aD e6+(\aUD f70+\aUD f07)\aD e7)
\\&=\frac 12
(\aUD f00\aD e0-\aUD f77\aD e0+\aUD f10\aD e1-\aUD f67\aD e1
+\aUD f20\aD e2+\aUD f57\aD e2+\aUD f30\aD e3+\aUD f47\aD e3
\\ &
+\aUD f40\aD e4-\aUD f37\aD e4+\aUD f50\aD e5-\aUD f27\aD e5
+\aUD f60\aD e6+\aUD f17\aD e6+\aUD f70\aD e7+\aUD f07\aD e7)
\\&=\frac 12
((\aUD f00\aD e0+\aUD f10\aD e1+\aUD f20\aD e2+\aUD f30\aD e3+\aUD f40\aD e4
+\aUD f50\aD e5+\aUD f60\aD e6+\aUD f70\aD e7)
\\ &
+(\aUD f07\aD e7+\aUD f17\aD e6-\aUD f27\aD e5-\aUD f37\aD e4
+\aUD f47\aD e3+\aUD f57\aD e2-\aUD f67\aD e1-\aUD f77\aD e0))
\\&=\frac 12
((\aUD f00\aD e0+\aUD f10\aD e1+\aUD f20\aD e2+\aUD f30\aD e3+\aUD f40\aD e4
+\aUD f50\aD e5+\aUD f60\aD e6+\aUD f70\aD e7)
\\ &
+(\aUD f07\aD e7+\aUD f17\aD e1+\aUD f27\aD e2+\aUD f37\aD e4
+\aUD f61\aD e6+\aUD f47\aD e4+\aUD f57\aD e5+\aUD f77\aD e7)\aD e7)
\end{split}
}

\AddEq{a0=f0.7}
{
\ShowEq{a0=f07 1}
\ShowEq{a0=f07 2}
\ShowEq{a0=f07 3}
\ShowEq{a0=f07}
\ShowEq{a1=f07}
\ShowEq{a2=f07}
\ShowEq{a3=f07}
\ShowEq{a4=f07}
\ShowEq{a5=f07}
\ShowEq{a6=f07}
\ShowEq{a7=f07}
}

\AddEquation{f=.E+.I+.J+.K}
{
\begin{split}
f=&-\frac 12
\left(
\aD f0
+\aD f1i
+\aD f2j
+\aD f3k
\right)\circ E
\\&
+\frac 12\left(
\aD f0
+\aD f1i
\right)\circ I
+\frac 12\left(
\aD f0
+\aD f2j
\right)\circ J
\\&
+\frac 12\left(
\aD f0
+\aD f3k
\right)\circ K
\end{split}
}

\AddEquation{f=.I0.7}
{
\begin{split}
f=&-\frac 12
\left(
5\aD f0+\aD f1i+\aD f2j+\aD f3k
\aD f4l+\aD f5il+\aD f6jl+\aD f7kl
\right)\circ E
\\&
+\frac 12\left(
\aD f0+\aD f1i
\right)\circ\aU I1
+\frac 12\left(
\aD f0+\aD f2j
\right)\circ\aU I2
+\frac 12\left(
\aD f0+\aD f3k
\right)\circ\aU I3
\\&
+\frac 12\left(
\aD f0+\aD f4l
\right)\circ\aU I4
+\frac 12\left(
\aD f0+\aD f5il
\right)\circ\aU I5
+\frac 12\left(
\aD f0+\aD f6jl
\right)\circ\aU I6
\\&
+\frac 12\left(
\aD f0+\aD f7kl
\right)\circ\aU I7
\end{split}
}

\AddEquation{a...= O}
{
\begin{split}
&\,
\ShowEq{matrix a algebra O}
\\ = &\,
\ShowEq{matrix af algebra O}
\\ * &\,
\ShowEq{matrix f algebra O}
\end{split}
}

\AddEq [2]{I=(E,I)}
{
\def\Set{\aD I1,\aD I2,\aD I3}
\def\Temp{O}%
\edef\Tempa{#1}%
\ifx\Tempa\Temp%
\def\Set{\aD Ii,\gi i=\gi 1,...,\gi 7}
\fi
$\Basis I=(E,\Set)$#2
}

\AddEq{F o G}
{
\Basis F\circ G=\{F_k\circ G:F_k\in \Basis F\}
}

\AddEq{FijG}
{
$F_i\circ G$, $i\in I$,
}

\AddEquation{coordinates of map Fk, associative algebra}
{
F_k\circ x=\aUD {F_{k\cdot}^{}{}}ij\aU xj\aD ei
}

\AddEq [1]{linear map over ring, determinant=0, 1}
{
\[
\rank
\begin{pmatrix}
\mathcal C^{\cdot}{}_{\gim}^{\gik}{}_{\cdot\gi{ij}}
&#1_{\gim}^{\gik}
\end{pmatrix}
=\rank\mathcal C
\]
}

\AddEq[1]{linear map over ring, matrix}
{
\def\Cdot{}
\def\Temp{-}%
\edef\Tempa{#1}%
\ifx\Tempa\Temp%
\else
\def\Cdot{#1\cdot}
\fi
\mathcal C=
\left(
\mathcal C^{\cdot}{}_{\gim}^{\gik}{}_{\cdot\gi{ij}}
\right)
=
\left(
C_{\Cdot}{}^{\gi p}_{\gi{im}}C_{\Cdot}{}^{\gik}_{\gi{pj}}
\right)
}

\AddEq{linear map over ring, row of matrix}
{
${}^{\cdot}{}_{\gim}^{\gik}$
}

\AddEq{linear map over ring, column of matrix}
{
${}_{\cdot\gi{ij}}$
}

\AddEquation{coordinates of map f, associative algebra}
{
f\circ x=\aUD fij\aU xj\aD ei
}

\DefEq
{
\[
\Basis F=\{F_k\in\mathcal L(D;A_1\rightarrow A_2): k=1, ..., n\}
\]
}
{Ik 1n}

\DefEquation
{
h=
(
a\pC{s}{0}
\otimes
a\pC{s}{1}
)
\circ f
=a\pC{s}{0}
f
a\pC{s}{1}
}
{h generated by f, associative algebra}

\DefEq
{
\[
h:A_1\rightarrow A_2
\]
}
{h:A1->A2}

\AddEquation{f in L(A,A), 1, associative algebra}
{
f=
(
a_{k\cdot s_k\cdot 0}
\otimes
a_{k\cdot s_k\cdot 1}
)
\circ I_k
=
\sum_{\gik}
a_{k\cdot s_k\cdot 0}
I_k
a_{k\cdot s_k\cdot 1}
}

\DefEq
{
\[
f\in B^A\rightarrow \|f\|\in R
\]
}
{f in BA->|f|}

\DefEq
{
\symb{\|f\|}{norm of map}{}
}
{norm of map}

\DefEquation
{
\ShowSymbol{norm of map}{}=
\text{sup}\frac{\|f(x)\|_B}{\|x\|_A}
}
{norm of map, algebra}

\DefEq
{
$\ShowSymbol{norm of map}{}$
}
{show|f|}

\DefEq
{
\symb{f+g}{sum of maps}{polylinear}
}
{sum of maps,,polylinear}

\DefEquation
{
\begin{matrix}
\ShowSymbol{sum of maps}{polylinear}:A_1\times...\times A_n\rightarrow S
&f,g\in\mathcal L(D;A_1\times...\times A_n\rightarrow S)
\end{matrix}
}
{sum of maps, 1, polylinear}

\DefEquation
{
f^k=f^k_{s_k\cdot 0}\otimes f^k_{s_k\cdot 1}
}
{f=fkxfk}

\AddEq [1]{expansion of linear map with respect to basis}
{
f=f^k\circ I_k\ \ \ f^k\in #1\otimes #1
}

\DefEquation
{
(\ShowSymbol{sum of maps}{polylinear})\circ (a_1,...,a_n)
=f\circ (a_1,...,a_n)+g\circ (a_1,...,a_n)
}
{sum of  maps, polylinear}

\AddEq{module of polylinear maps}
{
$\mathcal L(D;A_1\times...\times A_n\rightarrow S)$
}

\AddEquation{dg h12 xox}
{
dg=(h_2h_1-h_1h_2)x+x(h_1h_2-h_2h_1)
}

\AddEquation{a0=}
{
\aD a0=-\frac 12(\aD f0+\aD f1i+\aD f2j+\aD f3k)
}

\AddEquation{a1=}
{
a_1=\frac 12(\aD f0+\aD f1i)
}

\AddEquation{a2=}
{
\aD a2=\frac 12(\aD f0+\aD f2j)
}

\AddEquation{a3=}
{
\aD a3=\frac 12(\aD f0+\aD f3k)
}

\AddEq{set linear maps, module}
{
\symb{\mathcal L(D;A_1\rightarrow A_2)}{set linear maps}1
}

\AddEq [4]{L(A->B)}
{
$\mathcal L(#1;#2\rightarrow #3)$#4
}

\AddEquation{Am->A(n+m) 1}
{
(p\circ x^n)\circ (q\circ x^m)=(p\underline{\otimes}q)\circ x^{n+m}
}

\AddEq[1]{d->dei}
{
\[
\aD P#1:d\in D\rightarrow d\aD e#1\in A
\]
}

\AddEq{f(t)=fit ei}
{
\[
f\circ t=(\aU fit)\aD ei=\aD Pi\circ(\aU fit)
\]
}

\AddEq{aE+bE=(a+b)E}
{
\[(aE)\circ x+(bE)\circ x=ax+bx=(a+b)x=((a+b)E)\circ x\]
}

\AddEq{aE o bE=(ab)E}
{
\[(aE)\circ (bE)\circ x=(aE)\circ (bx)=a(bx)=(ab)x=((ab)E)\circ x\]
}

\AddEq{g=gE}
{
\[g(x)=g^k(x)\circ E_k\ \ \ \ g^k:A\rightarrow B\]
}

\AddEq{linear map times constant, algebra}
{
$a\circ f$, $b\star f$, $a$, $b\in A_2$,
}

\AddEq{linear map times constant, 0, algebra}
{
\begin{align*}
(a\circ f)\circ x&=a(f\circ x)
\\
(b\star f)\circ x&=(f\circ x)b
\end{align*}
}

\AddEq{linear map AA LAA}
{
\[
h:\ATwo\rightarrow \mathcal L(D;A_1\rightarrow A_2)
\]
}

\AddEquation{linear map AA LAA, 1}
{
(a\otimes b)\circ f=b\star (a\circ f)
}

\AddEq{linear map AA LAA, 2}
{
\[
((a\otimes b)\circ f)\circ x=a(f\circ x)b
\]
}

\AddEq{f circ a =}
{
\[
f\circ a=f(a)
\]
}

\AddEq{norm on D module 3}
{
\item $\|a+b\|\le \|a\|+\|b\|$
\labelItem{norm on D module 3, 1}
\item $\|da\|=|d|\,\|a\|$, $d\in D$, $a\in A$
\labelItem{norm on D module 3, 2}
}

\AddEq [5]{matrix AIJ}
{
\[
#1=(#1^{\gi #2}_{\gi #4},\gi #2\in\gi #3,\gi #4\in\gi #5)
\]
}

\AddEquation{linear map, f a+b=}
{
f\circ(a+b)=f\circ a+f\circ b
}

\AddEquation{linear map, f pa=}
{
f\circ(pa)=p(f\circ a)
}

\AddEq [1]{ab V p D}
{
\[
\begin{matrix}
a,b\in #1
&
p\in D
\end{matrix}
\]
}

\DefEq
{
\symb{[a,b]}{commutator of algebra}{}
\[
\ShowSymbol{commutator of algebra}{}=ab-ba
\]
}
{commutator of algebra}

\DefEq
{
\[
[a,b]=0
\]
}
{commutative D algebra}

\AddEquation{associator of algebra =}
{
\ShowSymbol{associator of algebra}{}=(ab)c-a(bc)
}

\AddEquation{associator of algebra = Ae}
{
\begin{split}
(a,b,c)&=(\aUD Ck{pl}(ab)^{\gi p}c^{\gil}-\aUD Ck{ip}a^{\gii}(bc)^{\gi p})e_{\gik}
\\
&=(\aUD Ck{pl}\aUD Cp{ij}-\aUD Ck{ip}\aUD Cp{jl})a^{\gii}b^{\gij}c^{\gil}e_{\gik}
\end{split}
}

\AddEq{associator of algebra}
{
\symb{(a,b,c)}{associator of algebra}{}
}

\AddEquation{associator of algebra = A}
{
(a,b,c)=A^{\gi k}_{\gi{ijl}}a^{\gii}b^{\gij}c^{\gil}e_{\gik}
}

\AddEq{coordinates of associator}
{
\symb{A^{\gi k}_{\gi{ijl}}}{coordinates of associator}{}
}

\AddEquation{coordinates of associator =}
{
\ShowSymbol{coordinates of associator}{}=
\aUD Ck{pl}\aUD Cp{ij}-\aUD Ck{ip}\aUD Cp{jl}
}

\DefEq
{
\[
(a,b,c)=0
\]
}
{associative D algebra}

\AddEq{nucleus of algebra}
{
\symb{N(A)}{nucleus of algebra}{}
\[
\ShowSymbol{nucleus of algebra}{}=
\{
a\in A:
\forall b, c\in A,
(a,b,c)=(b,a,c)=(b,c,a)=0
\}
\]
}

\AddEq{center of algebra}
{
\symb{Z(A)}{center of algebra}{}
\[
\ShowSymbol{center of algebra}{}=
\{
a\in A:
a\in N(A),
\forall b\in A,
ab=ba
\}
\]
}

\AddEq[3]{f(a+b)=...}
{
#1\circ(#2+#3)=#1\circ #2+#1\circ #3
}

\DefEq
{
\[\Vector r_2:A_1\rightarrow A_2\]
}
{r2:A1->A2}

\AddEq{D ab in A, d in D}
{
\[
\begin{matrix}
a,b\in A_1
&
d\in D
\end{matrix}
\]
}

\AddEq{D1 D2 ab in A, d in D}
{
\[
\begin{matrix}
a,b\in A_1
&d,d_1,d_2\in D_1
\end{matrix}
\]
}

\AddEq{D1 D2 uv in V, ab in A, d in D}
{
\[
\begin{matrix}
u,v\in V_1
&a,b\in A_1
&d,d_1,d_2\in D_1
\end{matrix}
\]
}

\AddEq{property linear map hf}
{
\DrawEq{h(d1+d2)=...}{\SideWS module}
\DrawEq{h(d1d2)=...}{\SideWS module}
\DrawEq[fab]{f(a+b)=...}{\DFDT \SideWS module}
\DrawEq[hfa]{f(da)=h... \SideWS module}{\DFDT \SideWS module}
\ShowEq{D1 D2 ab in A, d in D}
}

\AddEq{property linear map hgf}
{
\DrawEq{h(d1+d2)=...}{\SideWS module}
\DrawEq{h(d1d2)=...}{\SideWS module}
\DrawEq[gab]{f(a+b)=...}{g \DFDT \SideWS module}
\DrawEq[hga]{f(da)=h...}{g \DFDT \SideWS module}
\DrawEq[fuv]{f(a+b)=...}{f \DFDT \SideWS module}
\DrawEq[hfu]{f(da)=h...}{f \DFDT \SideWS module}
\ShowEq{D1 D2 uv in V, ab in A, d in D}
}

\AddEq[2]{show linear map hgf}
{
\[
\begin{pmatrix}
h:D_1\rightarrow D_2&
#1:A_1\rightarrow A_2&
#2:V_1\rightarrow V_2
\end{pmatrix}
\]
}

\AddEq[2]{show linear map hf}
{
\ShowEq{h:D1->D2 f:A1->A2}{#1} \Base {#2} \Module
}

\AddEq[2]{show linear map f}
{
\ShowEq{f:A->B}{#2}{\Module_1}{\Module_2}
}

\AddEq [3]{V=o+Ae}
{
\[
V_{#1}=\bigoplus_{\gi #2\in\gi #3}A_{#1}\EV{V_{#1}}#2
\]
}

\AddEq{w in Jv}
{
$w\in X_k\subseteq J(v)$.
}

\AddEq[3]{f(da)=h... module}
{
\def\Temp{}%
\edef\Tempa{#1}%
\ifx\Tempa\Temp%
\def\Koef{d}%
\else
\def\Koef{#1(d)}
\fi
#2\circ(d#3)=\Koef(#2\circ #3)
}

\AddEq[3]{f(da)=h... left module}
{
#2\circ(d#3)=#1(d)(#2\circ #3)
}

\AddEq[3]{f(da)=h... right module}
{
#2\circ(d#3)=(#2\circ #3)#1(d)
}

\AddEquation{D , f(da)=... module}
{
f\circ(da)=(h\circ d)(f\circ a)
}

\AddEquation{D , f(da)=... left module}
{
f\circ(da)=(h\circ d)(f\circ a)
}

\AddEquation{D , f(da)=... right module}
{
f\circ(ad)=(f\circ a)(h\circ d)
}

\AddEq{h(d1+d2)=...}
{
h(d_1+d_2)=h(d_1)+h(d_2)
}

\AddEq{h(d1d2)=...}
{
h(d_1d_2)=h(d_1)h(d_2)
}

\AddEq{property linear map f}
{
\DrawEq[fab]{f(a+b)=...}{D }
\DrawEq[{}fa]{f(da)=h... module}{\DFDT \SideWS module}
\ShowEq{D ab in A, d in D}
}

\DefEq
{
\symb{\Basis{e}=(e_{\gii},\Ii)}{basis, module}1
}
{basis, module}

\AddEq[2]{basis e}
{
$\Basis e_{#1}$#2
}

\AddEq [3]{basis ei}
{
\[\Basis e_{#1}=(e_{#1\cdot\gi #2},\gi #2\in \gi #3)\]
}

\AddEq [3]{basis e of module}
{
\[\Basis e_{#1}=(\EV{#1}{#2},\gi{#2}\in\gi{#3})\]
}

\AddEq [2]{basis e of module 1n}
{
\[\Basis e_{#1}=(\EV{#1}1,...,\EV{#1}{#2})\]
}

\AddEq[1]{r2:A1->A2, D1 D2 module}
{
b=h(a)\CRstar #1
}

\AddEq[1]{r2:A1->A2, D module}
{
b=a\CRstar #1
}

\AddEq[2]{ker G in ker g}
{
$\ker G\subseteq\ker #1$#2
}

\AddEq [2]{va=ae1}
{
\Vector #1=#1\CRstar e_{#2}
}

\AddEq [2]{va=ae1, left module}
{
\Vector #1=#1\CRstar e_{#2}
}

\AddEq [2]{va=ae1, module}
{
\Vector #1=#1\CRstar e_{#2}
}

\AddEq [2]{va=ae1, right module}
{
\Vector #1=e_{#2}\RCstar #1
}

\AddEq{product in D algebra}
{
v\,w=C\circ(v,w)
}

\AddEq{product in algebra, definition 1}
{
\[
C:A\times A\rightarrow A
\]
}

\DefEq
{
\[
\underline{\otimes}:A^{n\otimes }\times A^{m\otimes }
\rightarrow A^{n+m-1\otimes}
\]
}
{otimes -}

\AddEq{a b in basis of algebra}
{
\[
\begin{matrix}
a=a^{\gii}e_{\gii}&b=b^{\gii}e_{\gii}&a, b\in A
\end{matrix}
\]
}

\AddEquation{product in algebra}
{
(ab)^{\gik}=C^{\gik}_{\gi{ij}}a^{\gii}b^{\gij}
}

\AddEquation{product of basis vectors, algebra}
{
e_{\gii} e_{\gij}=
C^{\gik}_{\gi{ij}}e_{\gik}
}

\AddEq[4]{diagram of representations of D algebra}
{
\begin{matrix}
\vcenter
{
\xymatrix
{
D_{#1}\ar[r]|(.4){*}^{#4_{#2 12}}&
A_{#3}\ar[r]|{*}^{#4_{#2 23}}&
A_{#3}
\\
&&D_{#1}\ar[u]|{*}_{#4_{#2 12}}
}
}
&
\begin{array}{r@{\,}l}
#4_{#2 12}(d):v&\rightarrow d\,v
\\
#4_{#2 23}(v):w&\rightarrow
C_{#3}\circ(v, w)
\\
C_{#3}&\in\mathcal L(D_{#1};A_{#3}^2\rightarrow A_{#3})
\end{array}
\end{matrix}
}

\AddEq{endomorphism of module from product, 8}
{
\[
ab=(g_{23}\circ a)\circ b
\]
}

\AddEquation{endomorphism of module from product, 3}
{
\begin{array}{r@{\,}l}
(g_{23}\circ v)(w_1+w_2)&=(g_{23}\circ v)\circ w_1+(g_{23}\circ v)\circ w_2
\\
(g_{23}\circ v)\circ (dw)&=d((g_{23}\circ v)\circ w)
\end{array}
}

\AddEquation{endomorphism of module from product, 4}
{
(g_{23}\circ (v_1+v_2))\circ w
=(g_{23}\circ v_1+g_{23}\circ v_2)(w)
=(g_{23}\circ v_1)\circ w+(g_{23}\circ v_2)\circ w
}

\AddEquation{endomorphism of module from product, 7}
{
(g_{23}\circ(d\,v))\circ w
=(d\,(g_{23}\circ v))\circ w
=d\,((g_{23}\circ v)\circ w)
}

\AddEq{structural constants of algebra}
{
\symb{C^{\gik}_{\gi{ij}}}{structural constants}1
}

\AddEquation{otimes -, 1}
{
(a_1\otimes...\otimes a_n)\underline{\otimes}(b_1\otimes...\otimes b_n)
=
a_1\otimes...\otimes a_{n-1}\otimes a_nb_1\otimes b_2\otimes...\otimes b_n
}

\AddEq[3]{a=oxai}
{
$#1=#1_0\otimes...\otimes #1_{#2}$#3
}

\AddEq[3]{vb=f(va)}
{
\Vector #2=\Vector #3\circ\Vector #1
}

\AddEquation{D1 D2 vb=h(e1)a }
{
\Vector b=\Vector f\circ\Vector a=\Vector f\circ(a\CRstar e_1)
= h(a)\CRstar(\Vector f\circ e_1)
}

\AddEquation{D1 D2 vb=h(e1)a left}
{
\Vector b=\Vector f\circ\Vector a=\Vector f\circ(a\CRstar e_1)
= h(a)\CRstar(\Vector f\circ e_1)
}

\AddEquation{D vb=h(e1)a }
{
\Vector b=\Vector f\circ\Vector a=\Vector f\circ(a\CRstar e_1)
= a\CRstar(\Vector f\circ e_1)
}

\AddEquation{D1 D2 vb=h(e1)a right}
{
\Vector b=\Vector a\diamond\Vector f=(e_1\RCstar a)\diamond\Vector f
= (e_1\diamond\Vector f)\RCstar h(a)
}

\AddEq{f(e1) }
{
$\Vector f\circ e_{1\cdot\gii}$
}

\AddEq{f(e1) left}
{
$\Vector f\circ e_{1\cdot\gii}$
}

\AddEq{f(e1) right}
{
$e_{1\cdot\gii}\diamond\Vector f$
}

\AddEq{f(e1)= }
{
\Vector f\circ e_{1\cdot\gii}
=f_{\gii}\CRstar e_2
=f_{\gii}^{\gij}e_{2\cdot\gij}
}

\AddEq{f(e1)= left}
{
\Vector f\circ e_{1\cdot\gii}
=f_{\gii}\CRstar e_2
=f_{\gii}^{\gij}e_{2\cdot\gij}
}

\AddEq{f(e1)= right}
{
e_{1\cdot\gii}\diamond\Vector f
=e_2\RCstar f_{\gii}
=e_{2\cdot\gij}f_{\gii}^{\gij}
}

\AddEquation{D1 D2 vb=a f e left}
{
\Vector b=h(a)\CRstar f\CRstar e_2
}

\AddEquation{D vb=a f e }
{
\Vector b=a\CRstar f\CRstar e_2
}

\AddEquation{D1 D2 vb=a f e }
{
\Vector b=h(a)\CRstar f\CRstar e_2
}

\AddEquation{D1 D2 vb=a f e right}
{
\Vector b=e_2\RCstar f\RCstar (h(a))
}

\AddEq [4]{h(a)=...}
{
$#2(#1)=(#2(\aD {#1}{#3}),\gi #3\in\gi #4)$
}

\AddEq[4]{Vector f(e1) module}
{
$(\Vector #1\circ e_{1\cdot\gi #3},\gi #3\in\gi #4)$
}

%% file: Stmt.Gravity.English.tex
\input{Stmt.Gravity.Eq}

\AddEq[2]{note: sum of vectors 1}
{
Sum of vectors
\noindent
\begin{tabular}{@{}lcr}
\ShowEq{Vector v}{#1}
&
and
&
\ShowEq{Vector w}{#2}
\end{tabular}
is defined by the triangle law
\ShowEq{Vector v+w}{#1}{#2}

\noindent
\begin{tabular}{@{}lr}
\begin{minipage}{160pt}
From the parallelogram law
it follows that sum is commutative.
\end{minipage}
&
\begin{minipage}{160pt}
\ShowEq{Vector v+w=w+v}{#1}{#2}
\end{minipage}
\end{tabular}
}

\AddEq[2]{note: sum of vectors 2}
{
Below we consider a model of affine space
in a metric\hyph affine manifold.
\ePrints{0803.3276,0405.027}
\ifx\Semafor\ValueOff
When we consider connection $\Gamma^k_{ij}$ in Riemann space,
we impose a constraint on connection,\,\footnote{
See the definition of affine connection in Riemann space
on the page
\citeBib{Rashevsky}\Hyph 443.
}
that the torsion
\ShowEq{T=G-G}
is $0$ (symmetry of connection)
and parallel transport does not change scalar product of vectors.
If a metric tensor and an arbitrary connection
are defined on a differentiable manifold,
then this manifold is called
\AddIndex{metric\Hyph affine manifold}{metric-affine manifold}.\,\footnote{
See also the definition
\RefDefinition[\RefGravity]{metric-affine manifold}.
}
In particular, connection in metric\Hyph affine manifold has torsion.
\fi

\noindent
\begin{tabular}{@{}lr}
\begin{minipage}{200pt}
\setlength{\parindent}{5mm}
In Riemann space, we use geodesics
instead of straight lines.
So we can represent the vector $#1$
using segment $AB$ of geodesic $L_{#1}$
such that
vector $#1$ is tangent to geodesic $L_{#1}$ at the point $A$
and the length of segment $AB$ equals to the length of the vector $#1$.
\end{minipage}
&
\ShowEq{Vector v Riemann}{#1}
\end{tabular}

\noindent
\begin{tabular}{@{}lr}
\begin{minipage}{200pt}
\setlength{\parindent}{5mm}
This definition allows us to identify the vector $#1$
and the segment $AB$ of geodesic $L_{#1}$.
\end{minipage}
&
\end{tabular}

\noindent
\begin{tabular}{@{}lr}
\begin{minipage}{200pt}
\setlength{\parindent}{5mm}
For given vectors $#1$ and $#2$ in tangent plane at the point $A$,
let $\rho>0$ be the length of the vector $#1$
and $\sigma>0$ be the length of the vector $#2$.
\ePrints{323966352,0921-2370}
\ifx\Semafor\ValueOff
Let $V$ be unit vector collinear to the vector $#1$
\ShowEq{V rho=v}
Let $W$ be unit vector collinear to the vector $#2$
\ShowEq{W sigma=w}
\fi
\end{minipage}
&
\ShowEq{Vector w Riemann}{#2}
\end{tabular}
}

\AddEq{note: sum of vectors 3}
{
\noindent
\begin{tabular}{@{}lr}
\begin{minipage}{240pt}
\setlength{\parindent}{5mm}
We draw geodesic $L_v$ through the point $A$
using the vector $v$ as a tangent vector to $L_v$ in the point $A$.
Let $\tau$ be the canonical parameter on $L_v$ and
\ShowEq{dx/dt=A}{\tau}V
We transfer the vector $w$ along the geodesic $L_v$ from the point $A$
into point $B$ that defined by value of the parameter $\tau=\rho$.
We mark the result as $w'$.
\end{minipage}
&
\ShowEq{Vector v+w= 1 Riemann}vw
\end{tabular}

\noindent
\begin{tabular}{@{}lr}
\begin{minipage}{160pt}
\setlength{\parindent}{5mm}
We draw geodesic $L_{w'}$ through the point $B$
using the vector $W'$ as a tangent vector to $L_{w'}$ in the point $B$.
Let $\varphi'$ be the canonical parameter on $L_{w'}$ and 
\ShowEq{dx/dt=A}{\varphi'}{W'}
We define point $C$ on the geodesic $L_{w'}$
by parameter value $\varphi'=\sigma$
\end{minipage}
&
\ShowEq{Vector v+w= Riemann}vw
\end{tabular}

I assume that length of vectors $v$ and $w$ is small.
Then there exists unique geodesic $L_u$
from point $A$ to point $C$.
I will identify segment $AC$ of geodesic $L_u$
and vector $v+w$.
}

\AddEq{note: sum of vectors 4}
{
The same way, I draw triangle $ADE$
to find vector $w+v$.

\noindent
\begin{tabular}{@{}lr}
\begin{minipage}{160pt}
\setlength{\parindent}{5mm}
We draw geodesic $L_w$ through the point $A$
using the vector $w$ as a tangent vector to $L_w$ in the point $A$.
Let $\varphi$ be the canonical parameter on $L_w$ and
\ShowEq{dx/dt=A}{\varphi}W
We transfer the vector $v$ along the geodesic $L_w$ from the point $A$
into point $D$ that defined by value of the parameter $\varphi=\sigma$.
We mark the result as $v'$.
\end{minipage}
&
\ShowEq{Vector w+v= 1}
\end{tabular}

\noindent
\begin{tabular}{@{}lr}
\begin{minipage}{160pt}
\setlength{\parindent}{5mm}
We draw geodesic $L_{v'}$ through the point $D$
using the vector $v'$ as a tangent vector to $L_{v'}$ in the point $D$.
Let $\tau'$ be the canonical parameter on $L_{v'}$ and 
\ShowEq{dx/dt=A}{\tau'}{V'}
We define point $E$ on the geodesic $L_{v'}$
by parameter value $\tau'=\rho$
\end{minipage}
&
\ShowEq{Vector w+v=}
\end{tabular}

There exists unique geodesic $L_u$
from point $A$ to point $E$.
I will identify segment $AE$ of geodesic $L_u$
and vector $w+v$.
}

\AddEq{note: sum of vectors 5}
{
\noindent
\begin{tabular}{@{}lr}
\begin{minipage}{160pt}
\setlength{\parindent}{5mm}
Formally the lines $AB$ and $DE$ as well as the lines $AD$ and $BC$ are parallel
lines. The lengths of $AB$ and $DE$ are the same, and the lengths
of $AD$ and $BC$ are the same as well. We call this figure
a \AddIndex{parallelogram}{parallelogram} based on vectors
$v$ and $w$ with the origin in the point $A$.
\end{minipage}
&
\ShowEq{Vector v+w=w+v Metric-affine}vw
\end{tabular}
}

\DefLemma{increase of coordinate along geodesic}
{
Let $L_v$ be a geodesic through the point $A$ and the vector $v$
be a tangent vector to $L_v$ in the point $A$.
An increase of coordinate $x^k$ along geodesic $L_v$ is
\ShowEq{increase of coordinate along geodesic}
where $\tau$ is canonical parameter and we take values of derivatives and components
$\Gamma^k_{mn}$ in the initial point. 
}

\DefProof{increase of coordinate along geodesic}
{
The system of differential equations of geodesic $L_v$
has the following form
\ShowEq{differential equation of geodesic}
We write Taylor expansion of solution of
the system of differential equations
\EqRef{differential equation of geodesic}
in the following form
\ShowEq{increase of coordinate along geodesic 1}
The equality
\EqRef{increase of coordinate along geodesic}
follows from the equality
\EqRef{increase of coordinate along geodesic 1}.
}

\DefTheorem{parallelogram is not closed old}
{
Suppose $CBADE$ is a parallelogram with a origin in
the point $A$; then the resulting figure will not be closed \citeBib{torsion}.
The value of the difference of coordinates
of points $C$ and $E$ is equal to surface integral of the torsion
over this parallelogram\,\footnote{Proof of this statement I found in \citeBib{Shilov}}
\ShowEq{CE = surface integral of the torsion}
}

\AddEq{theorem: parallelogram is not closed}
{
\noindent
\begin{tabular}{@{}lr}
\begin{minipage}{160pt}
\setlength{\parindent}{5mm}
\begin{theorem}
\labelTheorem{parallelogram is not closed}
Suppose $CBADE$ is a parallelogram with an origin in
the point $A$; then the resulting figure will not be closed \citeBib{torsion}.
The value of the difference of coordinates
of points $C$ and $E$ is equal to surface integral of the torsion
over this parallelogram
\ShowEq{CE = surface integral of the torsion}
\end{theorem}
\end{minipage}
&
\ShowEq{Vector v+w=w+v Metric-affine}vw
\end{tabular}
}

\DefProof{parallelogram is not closed}
{
According to the lemma
\RefLemma{increase of coordinate along geodesic},
an increase of coordinate $x^k$
along the geodesic $L_v$ has the following form\,\footnote{
Proof of this statement I found in \citeBib{Shilov}}
\ShowEq{CE = surface integral of the torsion}
\ShowEq{Delta AB=}
and an increase of coordinate $x^k$
along the geodesic $L_{b'}$ has the following form
\ShowEq{Delta BC=}
Here
\ShowEq{W'k=Wk-...}
is the result of parallel transport of the vector $w$ from $A$ to $B$ and
\ShowEq{G(B)=G(A)+...}
with precision of small value of first level. Putting
\EqRef{W'k=Wk-...},
\EqRef{G(B)=G(A)+...}
into
\EqRef{Delta BC=}
we will receive
\ShowEq{Delta BC= 1}
Total increase of coordinate $x^K$ along the way $ABC$ has form
\ShowEq{Delta ABC=}
In a similar way, total increase of coordinate $x^K$ along the way $ADE$ has form
\ShowEq{Delta ADE=}
From
\EqRef{Delta ABC=}
and
\EqRef{Delta ADE=},
it follows that
\ShowEq{Delta ADE-ABC=}
and we get integral sum for expression
\[
\Delta_{ADE}x^k-\Delta_{ABC}x^k
=\iint_\Sigma (\Gamma^k_{nm}- \Gamma^k_{mn})dx^m \wedge dx^n
\]

\ePrints{0803.3276,FAlgebra,0405.027}
\ifx\Semafor\ValueOn
However it is not enough to find the difference
\[\Delta_{ADE}x^k-\Delta_{ABC}x^k\]
to find the difference of coordinates of points $C$ and $E$.
Coordinates may be anholonomic and we have to consider that
coordinates along closed loop change \EqRef{change of coordinate along a loop}
\[\Delta x^k=\oint_{ECBADE}dx^k=-\iint_\Sigma c^k_{mn}dx^m \wedge dx^n \]
where $c$ is anholonomity object.

Finally the difference of coordinates of points $C$ and $E$ is
\[\Delta_{CE}x^k=\Delta_{ADE}x^k-\Delta_{ABC}x^k+\Delta x^k=
\iint_\Sigma (\Gamma^k_{nm}- \Gamma^k_{mn}-c^k_{mn})dx^m \wedge dx^n\]
Using \EqRef{Torsion coordinates} we prove the statement.
\fi
}

%% file: Stmt.Gravity.Eq.tex

\AddEq[2]{Vector v+w=w+v Metric-affine}
{
\begin{minipage}{160pt}
\begin{center}
\begin{picture}(160,150)
\put( 20, 25){ $A$ }
\put( 30, 65){ $#1$ }
\put( 65, 50){ $#2$ }
\put( 58, 97){ $#1+#2$ }
\put( 101, 100){ $#2+#1$ }
\put( 55, 130){ $B$ }
\put( 80, 135){ $#2'$ }
\put( 155, 80){ $#1'$ }
\put( 160, 150){ $C$ }
\put( 165, 40){ $D$ }
\put( 180, 125){ $E$ }
\put(-115, -65){\qbezier(150,100)(162,150)(185,190)}
\put(35, 35){\vector(1,4){11}}
\put(-115, -65){\qbezier(185,190)(218,207)(278,210)}
\put(70, 125){\vector(2,1){30}}
\put(-115, -65){\qbezier(150,100)(218,207)(278,210)}
\put(164, 145){\line(3, -2){18}}
\put(-149, -154){\qbezier(315,212)(320,260)(331,287)}
\put(166, 59){\vector(0,1){35}}
\put(-149, -154){\qbezier(185,190)(208,206)(315,212)}
\put(35, 35){\vector(3,2){30}}
\put(-115, -65){\qbezier(150,100)(230,207)(295,197)}
\end{picture}
\end{center}
\end{minipage}
}

\AddEq[2]{Vector v+w= Riemann}
{
\begin{minipage}{160pt}
\begin{center}
\begin{picture}(150,160)
\put( 20, 25){ $A$ }
\put( 30, 65){ $#1$ }
\put( 65, 50){ $#2$ }
\put( 78, 90){ $#1+#2$ }
\put( 55, 130){ $B$ }
\put( 80, 135){ $#2'$ }
\put( 170, 150){ $C$ }
\put(-115, -65){\qbezier(150,100)(162,150)(185,190)}
\put(35, 35){\vector(1,4){11}}
\put(-115, -65){\qbezier(185,190)(218,207)(290,210)}
\put(-115, -65){\qbezier(150,100)(218,207)(290,210)}
\put(70, 125){\vector(2,1){30}}
\put(35, 35){\vector(3,2){30}}
\end{picture}
\end{center}
\end{minipage}
}

\AddEq[2]{Vector v+w= 1 Riemann}
{
\begin{minipage}{80pt}
\begin{center}
\begin{picture}(150,160)
\put( 20, 25){ $A$ }
\put( 30, 65){ $#1$ }
\put( 65, 50){ $#2$ }
\put( 55, 130){ $B$ }
\put( 80, 135){ $#2'$ }
\put(-115, -65){\qbezier(150,100)(162,150)(185,190)}
\put(35, 35){\vector(1,4){11}}
\put(70, 125){\vector(2,1){30}}
\put(35, 35){\vector(3,2){30}}
\end{picture}
\end{center}
\end{minipage}
}

\AddEq[1]{Vector v}
{
\begin{minipage}{30pt}
\[
\xymatrix
{
&\\ \\
\ar[uur]^{#1}
}
\]
\end{minipage}
}

\AddEq[1]{Vector w}
{
\begin{minipage}{80pt}
\[
\xymatrix
{
&&&\\
\ar[urrr]^{#1}
}
\]
\end{minipage}
}

\AddEq[2]{Vector v+w}
{
\[
\xymatrix
{
&&&&\\
&\ar[urrr]^{#2}\\ \\
\ar[uur]^{#1}\ar[uuurrrr]_{#1+#2}
}
\]
}

\AddEq[2]{Vector v+w=w+v}
{
\[
\xymatrix
{
&&&&\\
&\ar[urrr]^{#2}\\
&&&\ar[uur]_{#1}\\
\ar[uur]^{#1}\ar[uuurrrr]_{#1+#2}\ar[urrr]_{#2}
}
\]
}

\AddEquation{Delta ADE=}
{
\begin{split}
\Delta_{ADE}x^k&=\Delta_{AD} x^k+\Delta_{DE} x^k=
\\
&=W^k\sigma+V^k\rho- \Gamma^k_{mn}(A)V^m W^n \rho\sigma-
\\
&-\frac 1 2 \Gamma^k_{mn}(A)V^m V^n\rho^2
-\frac 1 2 \Gamma^k_{mn}(A)W^m W^n\sigma^2+O(dx^2)
\end{split}
}

\AddEq{Delta ADE-ABC=}
{
\begin{align*}
\Delta_{ADE}x^k-\Delta_{ABC}x^k=&
- \Gamma^k_{mn}(A)V^m W^n \rho\sigma
\\&-\frac 1 2 \Gamma^k_{mn}(A)V^m V^n\rho^2
-\frac 1 2 \Gamma^k_{mn}(A)W^m W^n\sigma^2
\\
&+ \Gamma^k_{mn}(A)W^m V^n\sigma\rho
\\&+\frac 1 2 \Gamma^k_{mn}(A)W^m W^n\sigma^2
+\frac 1 2 \Gamma^k_{mn}(A)V^m V^n\rho^2
\end{align*}
}

\AddEquation{Delta ABC=}
{
\begin{split}
\Delta_{ABC}x^k&=\Delta_{AB} x^k+\Delta_{BC} x^k
\\
&=V^k\rho+W^k\sigma- \Gamma^k_{mn}(A)W^m V^n\sigma\rho-
\\
&-\frac 1 2 \Gamma^k_{mn}(A)W^m W^n\sigma^2
-\frac 1 2 \Gamma^k_{mn}(A)V^m V^n\rho^2+O(dx^2)
\end{split}
}

\AddEq{Delta BC= 1}
{
\[
\Delta_{BC} x^k=W^k \sigma- \Gamma^k_{mn}(A)W^m V^n\sigma\rho
-\frac 1 2 \Gamma^k_{mn}(A)W^m W^n\sigma^2+O(\rho^2)
\]
}

\AddEquation{V rho=v}
{
V^k\rho=v^k
}

\AddEquation{W sigma=w}
{
W^k\sigma=w^k
}

\AddEquation{U pi=u}
{
U^k\pi=u^k
}

\AddEq[2]{Vector v+w=w+v Riemann}
{
\begin{center}
\begin{picture}(150,170)
\put( 20, 25){ $A$ }
\put( 30, 65){ $#1$ }
\put( 65, 50){ $#2$ }
\put( 78, 90){ $#1+#2$ }
\put( 55, 130){ $B$ }
\put( 80, 135){ $#2'$ }
\put( 155, 80){ $#1'$ }
\put( 185, 150){ $C$ }
\put( 170, 40){ $D$ }
\put(-115, -65){\qbezier(150,100)(162,150)(185,190)}
\put(35, 35){\vector(1,4){11}}
\put(-115, -65){\qbezier(185,190)(218,207)(305,210)}
\put(-115, -65){\qbezier(150,100)(218,207)(305,210)}
\put(70, 125){\vector(2,1){30}}
\put(-149, -154){\qbezier(320,212)(323,260)(340,300)}
\put(171, 59){\vector(0,1){35}}
\put(-149, -154){\qbezier(185,190)(208,206)(320,212)}
\put(35, 35){\vector(3,2){30}}
\end{picture}
\end{center}
}

\AddEq[1]{Vector v Riemann}
{
\begin{minipage}{120pt}
\begin{center}
\begin{picture}(100,100)
\put( 20, 25){ $A$ }
\put( 30, 65){ $#1$ }
\put( 130, 90){ $B$ }
\put(-115, -65){\qbezier(150,100)(162,150)(250,150)}
\put(35, 35){\vector(1,4){11}}
\end{picture}
\end{center}
\end{minipage}
}

\AddEq[1]{Vector w Riemann}
{
\begin{minipage}{120pt}
\begin{center}
\begin{picture}(180,100)
\put( 20, 25){ $A$ }
\put( 45, 55){ $#1$ }
\put( 135, 45){ $D$ }
\put(-149, -154){\qbezier(185,190)(208,206)(295,212)}
\put(35, 35){\vector(3,2){30}}
\end{picture}
\end{center}
\end{minipage}
}

\AddEq[2]{dx/dt=A}
{
\[\frac {dx^k}{d#1}=#2^k\]
}

\AddEq{Vector w+v=}
{
\begin{minipage}{160pt}
\begin{center}
\begin{picture}(160,150)
\put( 20, 25){ $A$ }
\put( 30, 65){ $v$ }
\put( 60, 60){ $w$ }
\put( 101, 100){ $w+v$ }
\put( 145, 80){ $v'$ }
\put( 155, 45){ $D$ }
\put( 175, 135){ $E$ }
\put(35, 35){\vector(1,4){11}}
\put(-149, -154){\qbezier(315,212)(318,260)(330,287)}
\put(165, 59){\vector(0,1){35}}
\put(-149, -154){\qbezier(185,190)(208,206)(315,212)}
\put(35, 35){\vector(3,2){30}}
\put(-115, -65){\qbezier(150,100)(230,207)(295,197)}
\end{picture}
\end{center}
\end{minipage}
}

\AddEq{Vector w+v= 1}
{
\begin{minipage}{160pt}
\begin{center}
\begin{picture}(160,150)
\put( 20, 25){ $A$ }
\put( 30, 65){ $v$ }
\put( 60, 60){ $w$ }
\put( 145, 80){ $v'$ }
\put( 155, 45){ $D$ }
\put(35, 35){\vector(1,4){11}}
\put(165, 59){\vector(0,1){35}}
\put(-149, -154){\qbezier(185,190)(208,206)(315,212)}
\put(35, 35){\vector(3,2){30}}
\end{picture}
\end{center}
\end{minipage}
}

\AddEq{Schwarzschild metric 1}
{
\begin{align*}
e^{(0)}_0&=c\sqrt{\frac{x^1-r_g} {x^1}}
\\
e^{(1)}_1&=\sqrt{\frac{x^1} {x^1-r_g}}
\\
e^{(2)}_2&=x^1
\\
e^{(3)}_3&=x^1\sin x^2
\end{align*}
}

\AddEquation{Delta BC=}
{
\Delta_{BC} x^k=W'^k\sigma-\frac 1 2 \Gamma^k_{mn}(B)W'^m W'^n\sigma^2+O(\sigma^2)
}

\AddEquation{G(B)=G(A)+...}
{
\begin{split}
\Gamma^k_{mn}(B)
&=\Gamma^k_{mn}(A)
+\partial_p \Gamma^k_{mn}(B)\Delta_{AB} x^p
\\
&=\Gamma^k_{mn}(A)
+\partial_p \Gamma^k_{mn}(B)V^p\rho
\end{split}
}

\AddEquation{W'k=Wk-...}
{
\begin{split}
W'^k&=W^k- \Gamma^k_{mn}(A)W^m \Delta_{AB} x^n+O(dx)
\\
&=W^k- \Gamma^k_{mn}(A)W^m V^n\rho+O(\rho)
\end{split}
}

\AddEquation{Delta AC=}
{
\Delta_{AC} x^k=U^k\pi-\frac 1 2 \Gamma^k_{mn}(A)U^m U^n\pi^2+O(\pi^2)
}

\AddEquation{U pi=V rho +W sigma}
{
U^k\pi=V^k\rho+W^k\sigma
}

\AddEq{Delta AB=}
{
\[
\Delta_{AB} x^k=V^k\rho-\frac 1 2 \Gamma^k_{mn}(A)V^m V^n\rho^2+O(\rho^2)
\]
}

\AddEquation{increase of coordinate along geodesic 1}
{
\begin{split}
\Delta x^k&=\frac{dx^k}{d\tau}\tau+\frac 1 2 \frac{d^2 x^k}{d\tau^2}\tau^2+O(\tau^2)=\\
&=\frac{dx^k}{d\tau}\tau
-\frac 1 2 \Gamma^k_{mn}\frac{dx^m}{d\tau} \frac{dx^n}{d\tau}\tau^2+O(\tau^2)
\end{split}
}

\AddEquation{increase of coordinate along geodesic}
{
\Delta x^k
=\frac{dx^k}{d\tau}\tau
-\frac 1 2 \Gamma^k_{mn}\frac{dx^m}{d\tau} \frac{dx^n}{d\tau}\tau^2+O(\tau^2)
}

\AddEquation{T=G-G}
{
T^i_{kl}=\Gamma^i_{lk}-\Gamma^i_{kl}
}

\AddEquation{differential equation of geodesic}
{
\frac{d^2x^i}{d\tau^2}=-\Gamma^i_{kl}\frac{dx^k}{d\tau}\frac{dx^l}{d\tau}
}

\AddEq{CE = surface integral of the torsion}
{
\[\Delta_{CE}x^k=\iint T^k_{mn}dx^m \wedge dx^n \]
}

\AddEq{e(k)l}
{
$e_{(k)}^l$
}

\AddEq{e(k)lek(k)}
{
\[
e_{(k)}^le^{(k)}_k=\delta^l_k
\]
}

\AddEquation{a(k)=.ak}
{
a^{(k)}=e^{(k)}_ka^k
}

\AddEquation{G(k)=.Gk}
{
\Gamma^{(i)}_{(k)(p)}
=e^i_{(k)}e^p_{(p)}e^{(i)}_j\Gamma^j_{ip}
-e^i_{(k)}e^p_{(p)}\frac{\partial e^{(i)}_i}{\partial x^p}
}

\AddEquation{Connection1}
{
g_{ij,k}+Q_{kij}-S^p_{jk}g_{pi}=\Gamma^p_{ik}g_{pj}+\Gamma^p_{kj}g_{pi}
}

\AddEquation{Connection2}
{
g_{jk,i}+Q_{ijk}-S^p_{ki}g_{pj}=\Gamma^p_{ji}g_{pk}+\Gamma^p_{ik}g_{pj}
}

\AddEquation{Connection3}
{
g_{ki,j}+Q_{jki}-S^p_{ij}g_{pk}=\Gamma^p_{kj}g_{pi}+\Gamma^p_{ji}g_{pk}
}

\AddEq{Connection4}
{
\[
g_{ki,j}+g_{jk,i}-g_{ij,k}+Q_{ijk}+Q_{jki}-Q_{kij}
-S^p_{ij}g_{pk}-S^p_{ki}g_{pj}+S^p_{jk}g_{pi}=2\Gamma^p_{ji}g_{pk}
\]
}

\AddEq{Connection5}
{
\[
\Gamma^p_{ji}=\frac 1 2 g^{pk}
(g_{ki,j}+g_{jk,i}-g_{ij,k}+Q_{ijk}+Q_{jki}-Q_{kij}
-S^r_{ij}g_{rk}-S^r_{ki}g_{rj}+S^r_{jk}g_{ri})
\]
}

\AddEq{anholonomity object 0}
{
\symb{c^{(i)}_{kl}}{anholonomity object}{}
}

\AddEquation{Anholonomity Object}
{
\ShowSymbol{anholonomity object}{} =
\frac{\partial e^{(i)}_k}{\partial x^l} - \frac{\partial e^{(i)}_l}{\partial x^k}
}

\AddEquation{anholonomity object 1}
{
c^{(i)}_{kl} =
\frac{\partial e^{(i)}_k}{\partial x^l} - \frac{\partial e^{(i)}_l}{\partial x^k}
}

\AddEquation{experiment, d1t= indefinite}
{
\int\sqrt{\frac{R_1+vx^0-r_g} {R_1+vx^0}}dx^0
=\frac 1v
\int\sqrt{\frac{r-r_g} r}dr
}

\AddEquation{experiment, d1t= indefinite 1}
{
\int\sqrt{\frac{R_1+vx^0-r_g} {R_1+vx^0}}dx^0
=\frac {2r_g}v
\int\frac{u^2}{(1-u^2)^2}du
}

\AddEquation{experiment, d1t= indefinite 2}
{
\begin{split}
\int\sqrt{\frac{R_1+vx^0-r_g} {R_1+vx^0}}dx^0
&=\frac {r_g}{2v}\int
\frac 1{u-1}du
+\frac {r_g}{2v}\int
\frac 1{(u-1)^2}du\\
&-\frac {r_g}{2v}\int
\frac 1{u+1}du
+\frac {r_g}{2v}\int
\frac 1{(u+1)^2}du\\
&=\frac {r_g}{2v}\left(
\ln\frac{u-1}{u+1}
-\frac 1{u-1}
-\frac 1{u+1}
\right)
\end{split}
}

\AddEquation{experiment, r=..u2}
{
r=\frac{r_g}{1-u^2}
}

\AddEquation{experiment, dr=..du}
{
dr=\frac{2r_gu}{(1-u^2)^2}du
}

\AddEquation{experiment, u2/u2 abcd=}
{
\begin{split}
a+c&=0\\
a-c+b+d&=1\\
b-d&=0\\
b+d-a+c&=0
\end{split}
}

\AddEquation{experiment, u2/u2 abcd= 1}
{
\begin{split}
c&=-a\\
b&=d\\
2a+2b&=1\\
2b-2a&=0
\end{split}
}

\AddEquation{experiment, u2/u2 abcd= 2}
{
a=\frac 14\ \ \ b=\frac 14\ \ \ c=-\frac 14\ \ \ d=\frac 14
}

\AddEquation{experiment, d1t+d2t}
{
\Delta_1x^{(0)}+\Delta_2x^{(0)}=
=\frac {r_g}{v}\left(
\ln\frac{u-1}{u+1}
-\frac 1{u-1}
-\frac 1{u+1}
\right)_{u=u_1}^{u=u_2}
}

\AddEquation{experiment, d3t=}
{
\Delta_3x^{(0)}=\int_0^{2t_1}\sqrt{\frac{R_1-r_g} {R_1}}dx^0
=2t_1\sqrt{\frac{R_1-r_g} {R_1}}
}

\AddEq[1]{experiment, u=u1}
{
\[
u_{#1}=\sqrt{\frac{R_{#1}-r_g}{R_{#1}}}
=\sqrt{1-\frac{r_g}{R_{#1}}}
\]
}

\AddEquation{experiment, u2/u2=abcd}
{
\begin{split}
\frac{u^2}{(1-u^2)^2}
&=
\frac a{u-1}+\frac b{(u-1)^2}+\frac c{u+1}+\frac d{(u+1)^2}
\\&=
\frac {(a+c)u+a-c}{u^2-1}+\frac{(b+d)u^2+2(b-d)u+b+d}{(u^2-1)^2}
\\&=
\frac {(a+c)u(u^2-1)}{(u^2-1)^2}
\\&+
\frac {(a-c+b+d)u^2+2(b-d)u+b+d-a+c}{(u^2-1)^2}
\end{split}
}

\AddEq{experiment, u2=r/r}
{
\[
u^2=\frac{r-r_g} r
\]
}

\AddEq{experiment, r=R1+vt}
{
$r=R_1+vx^0$, $dr=v\,dx^0$
}

\AddEquation{experiment, t1=dR/v}
{
t_1=\frac{R_2-R_1}v
}

\AddEquation{experiment, dt=int}
{
\Delta t_0=\int dx^{(0)}=\int e^{(0)}_0 dx^0
}

\AddEq{experiment, d1t=}
{
\[
\Delta_1x^{(0)}=\int_0^{t_1}\sqrt{\frac{x^1-r_g} {x^1}}dx^0
=\int_0^{t_1}\sqrt{\frac{R_1+vx^0-r_g} {R_1+vx^0}}dx^0
\]
}

\AddEquation{experiment, step 1}
{
x^0=t\ \ \ x^1=R_1+vt\ \ \ 0\le t\le t_1\ \ \ R_1+vt_1=R_2
}

\AddEq{Schwarzschild metric 2}
{
\begin{split}
c^{(0)}_{01}&=c\frac{\sqrt{r_g}}{x^1}
\\
c^{(2)}_{21}&=1
\\
c^{(3)}_{31}&=\sin x^2
\\
c^{(3)}_{32}&=x^1\cos x^2
\end{split}
}

\AddEq{c=0}
{
\[
c^{(i)}_{kl} =0
\]
}

\AddEq{el(k)}
{
$e^{(k)}_l$
}

\AddEquation{dx(k)/dxl}
{
e^{(k)}_l=\frac{\partial x^{(k)}}{\partial x^l}
}

%% file: Biblio.English.tex
\OpenBiblio


\BiblioItem{Doctor Ouch}
{
Kornei Chukovsky. Doctor Ouch.
\\
Translator and illustrator Jan Seabaugh.
\\
Viveca Smith Publishing, 2004, ISBN-10: 0974055107.
}%

\BiblioItem{Einstein: Electrodynamics of Moving Bodies}
{
Albert Einstein,
On the Electrodynamics of Moving Bodies, 1905,
\\
The Principle of Relativity: A Collection of Original
Memoirs on the Special and General Theory of Relativity , 37 - 65,
\\
Courier Dover Publications, 1952; ISBN-13: 978-0486600819
\\
Zur Elektrodynamik der bewegter K\"orper. Ann. Phys., 1905, 17, 891-921. 
}%

\BiblioItem{Einstein: On the Relativity Principle}
{
Albert Einstein,
On the Relativity Principle and the Conclusions Drawn from It, 1907,
\\
The Collected Papers of Albert Einstein, Volume 2:
The Swiss Years: Writings, 1900-1909. English translation. 252 - 311.
\\
Anna Beck, translator; Peter Havas, consultant.
Princeton University Press, 1989; ISBN-13: 9780691085494
\\
\"Uber das Relativit\"atsprinzip und die aus demselben gezogenen Folgerungen. 
Jahrb. d. Radioaktivit\"at u. Elektronik, 1907, 4, 411-462. 
}%

\BiblioItem{Einstein: Foundations of general relativity}
{
Albert Einstein,
Die Grundlage der allgemeinen Relativit\"atstheorie,
Ann. Phys., 1916, {\bf 49}, 769 - 822,\\
Einstein's Annalen Papers: The Complete Collection 1901-1922,
edited by J\"urgen Renn, 517 - 571,\\
Wiley-VCH Verlag GmbH \& Co. KGaA, 2005
}%

\BiblioItem{Einstein: Geometry and Experience}
{
Albert Einstein, Geometry and Experience, (1921)\\
Albert Einstein, Sidelights on Relativity, 25 - 56,\\
Courier Dover Publications, 1983
}%

\BiblioItem{Einstein: Main problems of general relativity}
{
Albert Einstein,
Grundgedanken und Probleme der Relativit\"atstheorie, (1923),\\
Nobelstiftelsen, Les Prix Nobel en 1921 - 1922,
Imprimerie Royale, Stockholm, 1923
}%

\BiblioItem{Einstein: Noneuclidean Geometry and Physics}
{
Albert Einstein,
Nichtenklidische Geometrie in der Physik Neue Rundschan, (1925)
Berlin, S. 16 - 20
}%

\BiblioItem{Einstein: Isaak Newton}
{
Albert Einstein,
Isaak Newton, 1927,
Out of My Later Years, 
Citadel Press, 1995, 219 - 223
}%

\BiblioItem{Einstein: On Science}
{
Albert Einstein,
On Science, 
Cosmic Religion, with Other Opinions and Aphorisms,142 - 146,
New York, 1931, 97 - 103
}%

\BiblioItem{Einstein: Autobiographical Notes}
{
Albert Einstein,
Autobiographical Notes, 1949,\\
Paul A. Schilpp, editor, Albert Einstein: Philosopher-Scientist,
Evanston, 
Illinois, The Library of Living Philosophers, 1949, 1 - 95
}%

\BiblioItem{Feynman 1}
{
Richard Phillips Feynman, Robert B. Leighton, Matthew Linzee Sands.
The Feynman lectures on physics: Volume 1.
Mainly Mechanics, Radiation, and Heat.
Addison\Hyph Wesley, 1965.
}%

\BiblioItem{0538731877}
{
James Shipman, Jerry D. Wilson and Aaron Todd.
Introduction to Physical Science.
Cengage Learning, 2009; ISBN 0538731877.
}%

\BiblioItem{Cite: 104}
{
Cite 104, Source unknown
}%

\BiblioItem{Ghez}
{
Ghez et al.,
The First Measurement of Spectral Lines in a Short-Period Star Bound to the Galaxy's Central Black Hole: A Paradox of Youth,
\href{http://www.journals.uchicago.edu/ApJ/journal/issues/ApJL/v586n2/16990/brief/16990.abstract.html}{ApJL, 586, L127} (2003),
eprint \href{http://arxiv.org/abs/astro-ph/0302299}{arXiv:astro-ph/0302299} (2003)
}%

\BiblioItem{Schodel}
{
R. Sch\"odel et al.,
A star in a 15.2-year orbit around the supermassive black hole at the centre of the Milky Way,
\href{http://www.nature.com/cgi-taf/DynaPage.taf?file=/nature/journal/v419/n6908/abs/nature01121_fs.html}{Nature 419, 694} (2002)
}%

\BiblioItem{Mielke}
{
Eckehard W. Mielke, Affine generalization of the Komar complex of general relativity,
\href{http://prola.aps.org/searchabstract/PRD/v63/i4/e044018}{Phys. Rev. D 63, 044018} (2001)
}%

\BiblioItem{Obukhov}
{
Yu. N. Obukhov and J. G. Pereira, Metric\hyph affine approach to teleparallel gravity,
\href{http://scitation.aip.org/getabs/servlet/GetabsServlet?prog=normal&id=PRVDAQ000067000004044016000001&idtype=cvips&gifs=Yes}
{Phys. Rev. D 67, 044016} (2003),
eprint \href{http://arxiv.org/abs/gr-qc/0212080}{arXiv:gr-qc/0212080} (2002)
}%

\BiblioItem{Sardanashvily}
{
Giovanni Giachetta, Gennadi Sardanashvily, Dirac Equation in Gauge and Affine-Metric Gravitation Theories,
eprint \href{http://arxiv.org/abs/gr-qc/9511035}{arXiv:gr-qc/9511035} (1995)
}%

\BiblioItem{Gauge}
{
Frank Gronwald and Friedrich W. Hehl, On the Gauge Aspects of Gravity, eprint
\href{http://arxiv.org/abs/gr-qc/9602013}{arXiv:gr-qc/9602013} (1996)
}%

\BiblioItem{Neeman}
{
Yuval Neeman, Friedrich W. Hehl, Test Matter in a Spacetime with Nonmetricity, eprint
\href{http://arxiv.org/abs/gr-qc/9604047}{arXiv:gr-qc/9604047} (1996)
}%

\BiblioItem{torsion}
{
F. W. Hehl, P. von der Heyde, G. D. Kerlick, and J. M. Nester,
General relativity with spin and torsion: Foundations and prospects,\\
\href{http://prola.aps.org/abstract/RMP/v48/i3/p393_1}{Rev. Mod. Phys. 48, 393} (1976)
}%

\BiblioItem{Megged}
{
O. Megged, Post-Riemannian Merger of Yang-Mills Interactions with Gravity,
eprint \href{http://arxiv.org/abs/hep-th/0008135}{arXiv:hep-th/0008135} (2001)
}%


\BiblioItem{gr-qc-9604027}
{
Yu.N. Obukhov, E.J. Vlachynsky, W. Esser, R. Tresguerres and F.W. Hehl,
An exact solution of the metric\hyph affine gauge theory with dilation, shear, and spin charges,
eprint \href{http://arxiv.org/abs/gr-qc/9604027}{arXiv:gr-qc/9604027} (1996)
}%

\BiblioItem{4419-7514}
{
Mari\'an Fabian, Petr Habala, Petr H\'ajek, Vicente Montesinos, V\'aclav Zizler.
Banach Space Theory: The Basis for Linear and Nonlinear Analysis.
\\
Springer; New York, 2010; ISBN-13: 978-1441975140
}%

\BiblioItem{Weinberg I}
{
Steven Weinberg.
The Quantum Theory of Fields. Volume I. Foundations.
Cambridge university press, 1995
}%

\BiblioItem{Weinberg II}
{
Steven Weinberg.
The Quantum Theory of Fields. Volume II. Modern applications.
Cambridge university press, 1996
}%

\BiblioItem{Reinhardt}
{
Walter Greiner, Joachim Reinhardt. Field Quantization. Springer.
}%

\BiblioItem{978-3540875604}
{
Walter Greiner, Joachim Reinhardt. Quantum Electrodynamics. Springer, 2009.
}%

\BiblioItem{978-1898563020}
{
H. Robert Mills. Practical Astronomy. Woodhead Publishing, 1994. ISBN-13: 978-1898563020.
}%

\BiblioItem{Landau I}
{
L. D. Landau, E. M. Lifshich.
Course of theoretical physics, volume 1.
Mechanics.
\\
Translated from the Russian by J. B. Sykes and J. S. Bell.
Pergamon Press, 1969
}%

\BiblioItem{Landau}
{
L. D. Landau, E. M. Lifshich, The classical theory of fields.
\\
Translated from the Russian by Morton Hamermesh.
Pergamon Press, 1971
}%

\BiblioItem{Landau III}
{
L. D. Landau, E. M. Lifshich,
Course of Theoretical Physics, Volume 3.
Quantum Mechanics Non-Relativistic Theory, Third Edition.
\\
Translated from the Russian by J. B. Sykes and J. S. Bell.
Butterworth-Heinemann, 1981, ISBN 978-0750635394.
}%

\BiblioItem{Wheeler}
{
Ignazio Ciufolini, John Wheeler. Gravitation and Inertia.
Princeton university press.
}%

\BiblioItem{Gravitation MTW}
{
Charles W. Misner, Kip S. Thorne, John Archibald Wheeler.
Gravitation.
W. H. Freeman and Company, San Francisco, 1973.
}%

\BiblioItem{Anderson98}
{
J. D. Anderson, P. A. Laing, E. L. Lau, A. S. Liu, M. M. Nieto, and S. G. Turyshev,
Indication, from Pioneer 10/11, Galileo, and Ulysses Data, of an Apparent Anomalous, Weak, Long-Range Acceleration,
\href{http://prola.aps.org/abstract/PRL/v81/i14/p2858_1}{Phys. Rev. Lett. 81, 2858}, (1998),
eprint \href{http://arxiv.org/abs/gr-qc/9808081}{arXiv:gr-qc/9808081} (1998)
}%

\BiblioItem{Anderson02}
{
J. D. Anderson, P. A. Laing, E. L. Lau, A. S. Liu, M. M. Nieto, and S. G. Turyshev,
Study of the anomalous acceleration of Pioneer 10 and 11,
\href{http://prola.aps.org/searchabstract/PRD/v65/i8/e082004}{Phys. Rev. D 65, 082004, 50 pp.}, (2002),
eprint \href{http://arxiv.org/abs/gr-qc/0104064}{arXiv:gr-qc/0104064} (2001)
}%


\BiblioItem{H. Aslaksen}
{
H. Aslaksen.  Quaternionic determinants \textit{Math.
Intelligencer} {\bf 18}(3), pp.57-65, (1996).
}%

\BiblioItem{L. Chen: Definition of determinant}
{
L. Chen, Definition of determinant and Cramer solutions over
quaternion field, \textit{Acta Math. Sinica (N.S.)} {\bf 7},
pp.171-180, (1991).
}%

\BiblioItem{L. Chen: Inverse matrix}
{
L. Chen,
Inverse matrix and properties of double determinant over quaternion
field, \textit{Sci. China, Ser. A} {\bf 34}, pp.528-540, (1991).
}%

\BiblioItem{N. Cohen S. De Leo}
{
N. Cohen, S. De Leo, The quaternionic determinant, \textit{The Electronic Journal Linear
Algebra} {\bf 7}, pp.100-111, (2000).
}%

\BiblioItem{Dyson: Quaternion determinants}
{
F. J. Dyson, Quaternion determinants, \textit{Helvetica Phys.
Acta} {\bf 45}, pp. 289-302, (1972).
}%

\BiblioItem{Melvin Hausner}
{
Melvin Hausner,
A Vector Space Approach to Geometry,
Dover Publications, 1998
}%

\BiblioItem{Serge Lang}
{
Serge Lang,
Algebra, Springer, 2002
}%

\BiblioItem{9780534423230}
{
Charles Lanski.
Concepts In Abstract Algebra.
American Mathematical Soc., 2005, ISBN 978-0534423230
}%

\BiblioItem{Burris Sankappanavar}
{
S. Burris, H.P. Sankappanavar,
A Course in Universal Algebra, Springer-Verlag (March, 1982),
\\eprint
\href{http://www.math.uwaterloo.ca/~snburris/htdocs/ualg.html}
{http://www.math.uwaterloo.ca/~snburris/htdocs/ualg.html}
\\(The Millennium Edition)
}%

\BiblioItem{Shilov single 12}
{
G. E. Shilov,
Calculus, Single Variable Functions, Parts 1 - 2,
Moscow, Nauka, 1969
}%

\BiblioItem{Shilov single 3}
{
G. E. Shilov,
Calculus, Single Variable Functions, Part 3,
Moscow, Nauka, 1970
}%

\BiblioItem{Shilov}
{
G. E. Shilov,
Calculus, Multivariable Functions,
Moscow, Nauka, 1972
}%

\BiblioItem{Kolmogorov Fomin}
{
A. N. Kolmogorov and S. V. Fomin.
Introductory Real Analysis.
\\
Translated and edited by Richard A. Silverman.
\\
Dover Publication, 1975, ISBN-13: 978-0486612263
}%

\BiblioItem{Lebedev Vorovich}
{
L. P. Lebedev, I. I. Vorovich,
Functional Analysis in Mechanics,
Springer, 2002
}%

\BiblioItem{8176-4374}
{
Mariano Giaquinta, Giuseppe Modica,
Mathematical Analysis: Linear and Metric Structures and Continuity.
\\
Springer, 2007, ISBN-13: 978-0-8176-4374-4
}%

\BiblioItem
{Rashevsky}
{
P. K. Rashevsky, Riemann Geometry and Tensor Calculus,\\
Moscow, Nauka, 1967
}%

\BiblioItem
{Kurosh: High Algebra}
{
A. G. Kurosh, Higher Algebra,
\\
George Yankovsky translator,
\\
Mir Publishers, 1988, ISBN: 978-5030001319
}%

\BiblioItem
{Kurosh: General Algebra}
{
A. G. Kurosh, Lectures on General Algebra,
Chelsea Pub Co, 1965 
}%

\BiblioItem{Sabinin: Smooth Quasigroups}
{
Lev V. Sabinin, Smooth Quasigroups and Loops,
Kluwer Academic Publisher, 1999 
}%

\BiblioItem{978-0-8176-8384-9}
{
Garret Sobczyk, New Foundations in Mathematics: The Geometric Concept of Number,
\\
Springer, 2013, ISBN: 978-0-8176-8384-9
}%

\BiblioItem{978-0538497817}
{
James Stewart, Calculus,
\\
Cengage Learning, 2012, ISBN: 978-0-538-49781-7
}%

\BiblioItem{470-38334}
{
William E. Boyce, Richard C. DiPrima,
Elementary Differential Equations and Boundary Value Problems,
\\
John Wiley \& Sons, Inc., 2009, ISBN 978-0-470-38334-6
}%

\BiblioItem{Dubrovin Fomenko Novikov part 1}
{
B. A. Dubrovin, A. T. Fomenko, S. P. Novikov,
Modern Geometry - Methods and Applications,\\
Part I, The Geometry of Surfaces, Transformation Groups, and Fields,\\
Translated by Robert G. Burns,\\
Springer - New York, 1992
}%

\BiblioItem{Dubrovin Fomenko Novikov part 2}
{
B. A. Dubrovin, A. T. Fomenko, S. P. Novikov,
Modern Geometry - Methods and Applications,
Part II: The Geometry and Topology of Manifolds,\\
Translated by Robert G. Burns,\\
Springer - New York, 1985
}%

\BiblioItem{Kobayashi Nomizu vol 1}
{
Kobayashi S, Nomizu K,
Foundations of Differential Geometry, volume I,\\
Interscience Publishers, 1963
}%

\BiblioItem{Lichnerowicz}
{
Andre Lichnerowicz,
Global Theory of Connections and Holonomy Groups,\\
Kluwer Academic Publishers, 1976, ISBN-13: 978-9028604964
}%

\BiblioItem{Korn}
{
Granino A. Korn, Theresa M. Korn,
Mathematical Handbook for Scientists and Engineer,
McGraw-Hill Book Company, New York, San Francisco,
Toronto, London, Sydney, 1968
}%

\BiblioItem{Hocking Young Topology}
{
John G. Hocking, Gail S. Young,
Topology,\\
Courier Dover Publications, 1988
}%

\BiblioItem{Olver: Lie groups to differential equations}
{
Peter J. Olver,
Applications of Lie groups to differential equations,\\
Springer, 2000
}%

\BiblioItem{1708.01190}
{
Nathan BeDell,
Doing Algebra over an Associative Algebra,
\\
eprint \href{https://arxiv.org/abs/1708.01190}{arXiv:1708.01190} (2017)
}%

\BiblioItem{Tartaglia}
{
Angelo Tartaglia and Matteo Luca Ruggiero,
Angular Momentum Effects in Michelson\Hyph Morley Type Experiments,
Gen.Rel.Grav. 34, 1371-1382 (2002),\\
eprint \href{http://arxiv.org/abs/gr-qc/0110015}{arXiv:gr-qc/0110015} (2001)
}%

\BiblioItem{Tomozawa}
{
Yukio Tomozawa, Speed of Light in Gravitational Fields, eprint
\href{http://arxiv.org/abs/astro-ph/0303047}{arXiv:astro-ph/0303047} (2004)
}%

\BiblioItem{Magueijo}
{
Joao Magueijo,
Covariant and locally Lorentz-invariant varying speed of light theories,
\href{http://prola.aps.org/abstract/PRD/v62/i10/e103521}{Phys. Rev. D 62, 103521} (2000),
eprint \href{http://arxiv.org/abs/gr-qc/0007036}{arXiv:gr-qc/0007036} (2000)
}%

\BiblioItem{Bassett}
{
Bruce A. Bassett, Stefano Liberati, Carmen Molina-Paris, and Matt Visser,
Geometrodynamics of variable-speed-of-light cosmologies,
\href{http://prola.aps.org/abstract/PRD/v62/i10/e103518}{Phys. Rev. D 62}, 103518 (2000),
eprint \href{http://arxiv.org/abs/astro-ph/0001441}{arXiv:astro-ph/0001441} (2000)
}%

\BiblioItem{C.A. Deavours The Quaternion Calculus}
{
C.A. Deavours, The Quaternion Calculus, 
American Mathematical Monthly, {\bf 80} (1973), pp. 995 - 1008
}%

\BiblioItem{Straumann}
{
Lochlainn O'Raifeartaigh and Norbert Straumann,
Gauge theory: Historical origins and some modern developments,
\href{http://prola.aps.org/abstract/RMP/v72/i1/p1_1}{Rev. Mod. Phys. 72, 1} (2000)
}%

\BiblioItem{Lammerzahl}
{
Claus L\"ammerzahl, Mark P. Haugan,
On the interpretation of Michelson\Hyph Morley experiments,
{Phys. Lett. A282 223-229} (2001),\\
eprint \href{http://arxiv.org/abs/gr-qc/0103052}{arXiv:gr-qc/0103052} (2001)
}%

\BiblioItem{0305117}
{
Holger Mueller, Sven Herrmann, Claus Braxmaier, Stephan Schiller, Achim Peters.
Modern Michelson-Morley Experiment using Cryogenic Optical Resonators.
eprint \href{http://arxiv.org/abs/physics/0305117}{arXiv:physics/0305117} (2003)
\\
Phys. Rev. Lett. 91:020401, 2003
}%

\BiblioItem{0706.2031}
{
Holger Mueller, Paul Louis Stanwix, Michael Edmund Tobar,
Eugene Ivanov, Peter Wolf, Sven Herrmann, Alexander Senger,
Evgeny Kovalchuk, Achim Peters.
Relativity tests by complementary rotating Michelson-Morley experiments.
eprint \href{http://arxiv.org/abs/0706.2031}{arXiv:0706.2031 [physics.class-ph]} (2006)
\\
Phys. Rev. Lett. 99:050401, 2007
}%

\BiblioItem{1008.1205}
{
M. Nagel, K. M\"ohle, K. D\"oringshoff, S. Herrmann, A. Senger, E.V. Kovalchuk, A. Peters.
Testing Lorentz Invariance by Comparing Light Propagation in Vacuum and Matter.
eprint \href{http://arxiv.org/abs/1008.1205}{arXiv:1008.1205 [physics.ins-det]} (2010)
}%

\BiblioItem{1109.4897}
{
The OPERA Collaboration.
Measurement of the neutrino velocity with the OPERA detector in the CNGS beam.
eprint \href{http://arxiv.org/abs/1109.4897}{arXiv:1109.4897 [hep-ex]} (2011)
}%

\BiblioItem{Ranada}
{
Antonio F. Ranada,
Pioneer acceleration and variation of light speed: experimental situation,
eprint \href{http://arxiv.org/abs/gr-qc/0402120}{arXiv:gr-qc/0402120} (2004)
}%

\BiblioItem{Gelfand Minlos: rotation and Lorentz groups}
{
Izrail Moiseevich Gelfand, Robert Adolfovich Minlos,
Representations of the rotation and Lorentz groups and their applications;\\
Engl. transl. ed. H. K. Farahat; Transl. by G. Cummins and T. Boddongton;\\
Pergamon Press, 1963
}%

\BiblioItem{math.QA-0208146}
{
I. Gelfand, S. Gelfand, V. Retakh, R. Wilson,
Quasideterminants,\\
eprint \href{http://arxiv.org/abs/math.QA/0208146}{arXiv:math.QA/0208146} (2002)
}%

\BiblioItem{q-alg-9705026}
{
I. Gelfand, V. Retakh,
Quasideterminants, I,\\
eprint \href{http://arxiv.org/abs/q-alg/9705026}{arXiv:q-alg/9705026} (1997)
}%

\BiblioItem{Gelfand Retakh 1991}
{
I. Gelfand and V. Retakh, Determinants of Matrices over Noncommutative Rings, Funct.
Anal. Appl. 25 (1991), no. 2, 91-102
}%

\BiblioItem{Gelfand Retakh 1992}
{
I. Gelfand and V. Retakh, A Theory of Noncommutative Determinants and Characteristic
Functions of Graphs, Funct. Anal. Appl. 26 (1992), no. 4, 1-20
}%

\BiblioItem{hep-th-9407124}
{
I. M. Gelfand, D. Krob, A. Lascoux, B. Leclerc, V.S. Retakh and J.-Y. Thibon,
Noncommutative symmetric functions,\\
eprint \href{http://arxiv.org/abs/hep-th/9407124}{arXiv:hep-th/9407124} (1994)
}%

\BiblioItem{0911.4454}
{
Vladimir Retakh,
From factorizations of noncommutative polynomials to combinatorial topology,\\
eprint \href{http://arxiv.org/abs/0911.4454}{arXiv:0911.4454} (2009)
}%

\BiblioItem{Naimark Shtern: Theory of group representations}
{
Mark Aronovich Naimark, Aleksandr Isaakovich Shtern,
Theory of group representations;\\
Heidelberg, 1982
}%

\BiblioItem{Barut Raczka: Theory of group representations}
{
Asim Orhan Barut; Ryszard R\c{a}czka;
Theory of group representations and applications;\\
World Scientific Publishing Co. Pre. Ltd., 1986
}%

\BiblioItem{Mihalev Pilz: concise handbook of algebra}
{
Aleksandr Vasilevich Mikhalev; G\"{u}nter Pilz;
The concise handbook of algebra;\\
Kluwer Academic Publishers, 2002
}%

\BiblioItem{McCrimmon: Jordan Algebras}
{
Kevin McCrimmon;
A Taste of Jordan Algebras;\\
Springer, 2004
}%

\BiblioItem{Zharinov: foundation of mathematical physics}
{
V. V. Zharinov,
Algebraic and geometric foundation of mathematical physics,\\
Lecture courses of the scientific and educational center, 9, Steklov Math. Institute of RAS,\\
Moscow, 2008
}%

\BiblioItem{Shafarevich: Basic notions of algebra}
{
I. R. Shafarevich,
Basic notions of algebra,\\
Translated from the Russian by M. Reid,\\
Springer, 2005
}%

\BiblioItem{Coppel: Number Theory}
{
W.A. Coppel,
Number Theory: An Introduction to Mathematics,\\
Springer, 2009
}%

\BiblioItem{978-0486497952}
{
Michael J. Field,
Differential Calculus and Its Applications,\\
Dover Publications, 2012; ISBN-13: 978-0486497952
}%

\BiblioItem{Elsgolts: Differential Equations}
{
Lev Elsgolts,
Differential Equations and the Calculus of Variations,\\
Translated from the Russian by George Yankovsky,\\
MIR Publishers, Moscow, 1977
}%

\BiblioItem{Baez Huerta: algebra of grand unified theories}
{
John Baez; John Huerta;
The algebra of grand unified theories;\\
Bull. Amer. Math. Soc. {\bf 47} (2010), 483-552
}%

\BiblioItem{J. Fan: Determinants}
{
J. Fan, Determinants and multiplicative functionals
on quaternion matrices, \textit{Linear Algebra and Its
Applications} {\bf 369}, pp. 193-201, (2003).
}%

\BiblioItem{Carl Faith 1}
{
Carl Faith, Algebra: Rings, Modules and Categories I,
Springer - Verlag, Berlin - Heidelberg - New York, 1973
}%

\BiblioItem{Gilson Nimmo Ohta}
{
 C.R.Gilson, J.J.C.Nimmo, Y.Ohta, Quasideterminant solutions of a non-Abelian Hirota-Miwa
 equation, \textit{Journal of Physics A: Mathematical and Theoretical} {\bf 40}(42), pp.
 12607-12617,(2007).
}%

\BiblioItem{Haider Hassan}
{
B. Haider, M. Hassan, Quasideterminant solutions of an integrable chiral model in two
 dimensions, \textit{Journal of Physics A: Mathematical and Theoretical} {\bf 42} (35), art. no.
 355211, (2009).
}%



\BiblioItem{0702447}
{
I.I. Kyrchei, Cramer's rule for quaternion systems of linear equations,
\textit{Journal of Mathematical Sciences} {\bf 155}(6), 839-858, (2008).
 Translated from  \textit{Fundamental and Appl. Math.}
 {\bf 13}(4), pp.67-94, (2007). (in Russian)\\
eprint
\href{http://arxiv.org/abs/math/0702447}{arXiv:math.RA/0702447}
(2007)
}%

\BiblioItem{1004.4380}
{
I.I. Kyrchei, Cramer's rule for some quaternion matrix
    equations,  \textit{Applied Mathematics and Computation} {\bf 217}(5), pp.2024-2030, (2010).\\eprint
\href{http://arxiv.org/abs/1004.4380
}{arXiv:math.RA/arXiv:1004.4380 } (2010)
}%

\BiblioItem{1005.0736}
{
I.I. Kyrchei,Determinantal representations of the Moore-Penrose inverse
 over the quaternion skew field and corresponding Cramer's rules,
 \\
eprint
\href{http://arxiv.org/abs/1005.0736}{arXiv:math.RA/1005.0736}
(2010)
}%

\BiblioItem{0412.391}
{
Aleks Kleyn,
Basis Manifold,
eprint \href{http://arxiv.org/abs/math.DG/0412391}{arXiv:math.DG/0412391} (2007)
}%

\BiblioItem{0405.027}
{
Aleks Kleyn,
Reference Frame in General Relativity,\\
eprint \href{http://arxiv.org/abs/gr-qc/0405027}{arXiv:gr-qc/0405027} (2008)
}%

\BiblioItem{0405.028}
{
Aleks Kleyn, Metric\hyph Affine Manifold,\\
eprint \href{http://arxiv.org/abs/gr-qc/0405028}{arXiv:gr-qc/0405028} (2008)
}%

\BiblioItem{0612.111}
{
Aleks Kleyn,
Biring of Matrices,\\
eprint \href{http://arxiv.org/abs/math.OA/0612111}{arXiv:math.OA/0612111} (2007)
}%

\BiblioItem{0701.238}
{
Aleks Kleyn,
Lectures on Linear Algebra over Division Ring,\\
eprint \href{http://arxiv.org/abs/math.GM/0701238}{arXiv:math.GM/0701238} (2010)
}%

\BiblioItem{0702.561}
{
Aleks Kleyn,
Fibered Universal Algebra,\\
eprint \href{http://arxiv.org/abs/math.DG/0702561}{arXiv:math.DG/0702561} (2007)
}%

\BiblioItem{math.RA-0501237}
{
Aleks Kleyn,
Vector Space Over Division Ring,\\
eprint \href{http://arxiv.org/abs/math.RA/0412391}{arXiv:math.RA/0501237} (2007)
}%

\BiblioItem{math.RA-0501237v1}
{
Aleks Kleyn,
Module Over Division Ring, version 1,\\
eprint \href{http://arxiv.org/abs/math/0501237v1}{arXiv:math.RA/0501237v1} (2005)
}%

\BiblioItem{0707.2246}
{
Aleks Kleyn,
Fibered Correspondence,\\
eprint \href{http://arxiv.org/abs/0707.2246}{arXiv:0707.2246} (2007)
}%

\BiblioItem{0803.2620}
{
Aleks Kleyn,
Morphism of \Ts{T}Representations,\\
eprint \href{http://arxiv.org/abs/0803.2620}{arXiv:0803.2620} (2008)
}%

\BiblioItem{0803.3276}
{
Aleks Kleyn,
Lorentz Transformation and General Covariance Principle,\\
eprint \href{http://arxiv.org/abs/0803.3276}{arXiv:0803.3276} (2009)
}%

\BiblioItem{0812.4763}
{
Aleks Kleyn,
Introduction into Calculus over Division Ring,\\
eprint \href{http://arxiv.org/abs/0812.4763}{arXiv:0812.4763} (2010)
}%

\BiblioItem{0906.0135}
{
Aleks Kleyn,
Introduction into Geometry over Division Ring,\\
eprint \href{http://arxiv.org/abs/0906.0135}{arXiv:0906.0135} (2010)
}%

\BiblioItem{0909.0855}
{
Aleks Kleyn,
Quaternion Rhapsody,\\
eprint \href{http://arxiv.org/abs/0909.0855}{arXiv:0909.0855} (2010)
}%

\BiblioItem{0912.3315}
{
Aleks Kleyn,
Representation of Universal Algebra,\\
eprint \href{http://arxiv.org/abs/0912.3315}{arXiv:0912.3315} (2009)
}%

\BiblioItem{0912.4061}
{
Aleks Kleyn,
Linear Equation in Finite Dimensional Algebra,\\
eprint \href{http://arxiv.org/abs/0912.4061}{arXiv:0912.4061} (2010)
}%

\BiblioItem{1001.4852}
{
Aleks Kleyn,
The Matrix of Linear Maps,\\
eprint \href{http://arxiv.org/abs/1001.4852}{arXiv:1001.4852} (2010)
}%

\BiblioItem{1003.1544}
{
Aleks Kleyn,
Linear Maps of Free Algebra,\\
eprint \href{http://arxiv.org/abs/1003.1544}{arXiv:1003.1544} (2010)
}%

\BiblioItem{1006.2597}
{
Aleks Kleyn,
The G\^ateaux Derivative and Integral over Banach Algebra,\\
eprint \href{http://arxiv.org/abs/1006.2597}{arXiv:1006.2597} (2010)
}%

\BiblioItem{1011.3102}
{
Aleks Kleyn,
Polylinear Map of Free Algebra,\\
eprint \href{http://arxiv.org/abs/1011.3102}{arXiv:1011.3102} (2010)
}%

\BiblioItem{1102.1776}
{
Aleks Kleyn, Ivan Kyrchei,
Correspondence between Row\Hyph Column Determinants
and Quasideterminants of Matrices over Quaternion Algebra,\\
eprint \href{http://arxiv.org/abs/1102.1776}{arXiv:1102.1776} (2011)
}%

\BiblioItem{1104.5197}
{
Aleks Kleyn,
$C^*$-Rhapsody,\\
eprint \href{http://arxiv.org/abs/1104.5197}{arXiv:1104.5197} (2011)
}%

\BiblioItem{1105.4307}
{
Aleks Kleyn,
Algebra with Conjugation,\\
eprint \href{http://arxiv.org/abs/1105.4307}{arXiv:1105.4307} (2011)
}%

\BiblioItem{1107.1139}
{
Aleks Kleyn,
Linear Maps of Quaternion Algebra,\\
eprint \href{http://arxiv.org/abs/1107.1139}{arXiv:1107.1139} (2011)
}%

\BiblioItem{1107.5037}
{
Aleks Kleyn,
Orthogonal Basis and Motion in Finsler Geometry,\\
eprint \href{http://arxiv.org/abs/1107.5037}{arXiv:1107.5037} (2011)
}%

\BiblioItem{1111.6035}
{
Aleks Kleyn,
Basis of Representation of Universal Algebra,\\
eprint \href{http://arxiv.org/abs/1111.6035}{arXiv:1111.6035} (2011)
}%

\BiblioItem{1201.4158}
{
Aleks Kleyn, Alexandre Laugier,
Orthonormal Basis in Minkowski Space,\\
eprint \href{http://arxiv.org/abs/1201.4158}{arXiv:1201.4158} (2012)
}%

\BiblioItem{1202.6021}
{
Aleks Kleyn,
Maps of Conjugation of Quaternion Algebra,\\
eprint \href{http://arxiv.org/abs/1202.6021}{arXiv:1202.6021} (2012)
}%

\BiblioItem{1206.0200}
{
Aleks Kleyn,
Algebra of Fractions of Algebra with Conjugation,\\
eprint \href{http://arxiv.org/abs/1206.0200}{arXiv:1206.0200} (2012)
}%

\BiblioItem{1211.6965}
{
Aleks Kleyn,
Free Algebra with Countable Basis,\\
eprint \href{http://arxiv.org/abs/1211.6965}{arXiv:1211.6965} (2012)
}%

\BiblioItem{1302.7204}
{
Aleks Kleyn,
Polynomial over Associative $D$-Algebra,\\
eprint \href{http://arxiv.org/abs/1302.7204}{arXiv:1302.7204} (2013)
}%

\BiblioItem{1305.4547}
{
Aleks Kleyn,
Normed $\Omega$-Group,\\
eprint \href{http://arxiv.org/abs/1305.4547}{arXiv:1305.4547} (2013)
}%

\BiblioItem{1310.5591}
{
Aleks Kleyn,
Integral of Map into Abelian $\Omega$\Hyph group,\\
eprint \href{http://arxiv.org/abs/1310.5591}{arXiv:1310.5591} (2013)
}%

\BiblioItem{1502.04063}
{
Aleks Kleyn,
Linear Map of $D$\Hyph Algebra,\\
eprint \href{http://arxiv.org/abs/1502.04063}{arXiv:1502.04063} (2015)
}%

\BiblioItem{1505.03625}
{
Aleks Kleyn,
Derivative of Map of Banach algebra,\\
eprint \href{http://arxiv.org/abs/1505.03625}{arXiv:1505.03625} (2015)
}%

\BiblioItem{1506.00061}
{
Aleks Kleyn,
Quadratic Equation over Associative $D$\Hyph Algebra,\\
eprint \href{http://arxiv.org/abs/1506.00061}{arXiv:1506.00061} (2015)
}%

\BiblioItem{1601.03259}
{
Aleks Kleyn,
Introduction into Calculus over Banach Algebra,\\
eprint \href{http://arxiv.org/abs/1601.03259}{arXiv:1601.03259} (2016)
}%

\BiblioItem{1801.01628}
{
Aleks Kleyn,
Differential Equation over Banach Algebra,\\
eprint \href{http://arxiv.org/abs/1801.01628}{arXiv:1801.01628} (2018)
}%

\BiblioItem{1908.04418}
{
Aleks Kleyn,
Diagram of Representations of Universal Algebras,\\
eprint \href{http://arxiv.org/abs/1908.04418}{arXiv:1908.04418} (2019)
}%

\BiblioItem{322019412}
{
Aleks Kleyn,
Research Diary, 2017\\
eprint \href{https://www.researchgate.net/publication/322019412}{RG:322019412} (2017)
}%

\BiblioItem{323966352}
{
Aleks Kleyn,
Research Diary, 2018\\
eprint \href{https://www.researchgate.net/publication/323966352}{RG:323966352} (2018)
}%

\BiblioItem{323236904}
{
Aleks Kleyn,
Crash Course in Calculus over  Banach Algebra\\
eprint \href{https://www.researchgate.net/publication/323236904}{RG:323236904} (2018)
}%

\BiblioItem{MVector}
{
Aleks Kleyn,
Linear Map of D-Algebra,\\
eprint \href{http://arxiv.org/abs/MVector}{arXiv:MVector} (2019)
}%

\BiblioItem{8433-5163}
{
Aleks Kleyn,
Linear Maps of Free Algebra: First Steps in Noncommutative Linear Algebra,\\
Lambert Academic Publishing, 2010
}%

\BiblioItem{8443-0072}
{
Aleks Kleyn,
Representation Theory: Representation of Universal Algebra,\\
Lambert Academic Publishing, 2011
}%

\BiblioItem{4776-3181}
{
Aleks Kleyn.\\
Linear Algebra over Division Ring: System of Linear Equations.\\
CreateSpace Independent Publishing Platform, 2012;\\
ISBN-13: 978-1-4776-3181-2
}%

\BiblioItem{4975-6381}
{
Aleks Kleyn.\\
Single Variable Calculus: Noncomutative Banach Algebra.\\
CreateSpace Independent Publishing Platform, 2014;\\
ISBN-13: 978-1-4975-6381-0
}%

\BiblioItem{4993-2400}
{
Aleks Kleyn.\\
Linear Algebra over Division Ring: Vector Space.\\
CreateSpace Independent Publishing Platform, 2014;\\
ISBN-13: 978-1-4993-2400-6
}%

\BiblioItem{5059-9176}
{
Aleks Kleyn.\\
Normed \(\Omega\)-Group.\\
CreateSpace Independent Publishing Platform, 2015;\\
ISBN-13: 978-1-5059-9176-5
}%

\BiblioItem{5114-6019}
{
Aleks Kleyn.\\
Representation of Universal Algebra: Polymorphism.\\
CreateSpace Independent Publishing Platform, 2015;\\
ISBN-13: 978-1-5114-6019-4
}%

\BiblioItem{5148-4632}
{
Aleks Kleyn.\\
Noncommutative Algebra: Introduction.\\
CreateSpace Independent Publishing Platform, 2018;\\
ISBN-13: 978-1-5114-6019-4
}%

\BiblioItem{5410-9916}
{
Aleks Kleyn.\\
Lebesgue Integral in Abelian $\Omega$-Group.\\
CreateSpace Independent Publishing Platform, 2016;\\
ISBN-13: 978-1-5410-9916-6
}%

\BiblioItem{9856-6693}
{
Aleks Kleyn,
Crash Course in Calculus over  Banach Algebra\\
CreateSpace Independent Publishing Platform, 2018;\\
ISBN-13: 978-1-9856-6693-1
}%

\BiblioItem{7287-9339}
{
Aleks Kleyn,
Quadratic Equation over Associative $D$\Hyph Algebra\\
Kindle Direct Publishing, 2018;\\
ISBN-13: 978-1-7287-9339-9
}%

\BiblioItem{9835-2163}
{
Aleks Kleyn,
Differential Equation over Banach Algebra\\
Kindle Direct Publishing, 2018;\\
ISBN-13: 978-1-9835-2163-8
}%

\BiblioItem{6860-2955}
{
Aleks Kleyn,
Diagram of Representations of Universal Algebras,\\
Kindle Direct Publishing, 2019;\\
ISBN-13: 978-1-6860-2955-4
}%

\BiblioItem{CACAA.01.109}
{
Aleks Kleyn,
Mappings of Conjugation of Quaternion Algebra.\\
Clifford Analysis, Clifford Algebras and their applications,
volume 1, Issue 1, pages 109 - 121, 2012
}%

\BiblioItem{CACAA.01.291}
{
Aleks Kleyn,
Introduction into Calculus over Division Ring.\\
Clifford Analysis, Clifford Algebras and their applications,
volume 1, Issue 4, pages 291 - 355, 2012
}%

\BiblioItem{CACAA.02.097}
{
Aleks Kleyn,
Polynomial over Associative $D$-Algebra.\\
Clifford Analysis, Clifford Algebras and their applications,
volume 2, Issue 2, pages 97 - 115, 2013
}%

\BiblioItem{CACAA.04.001}
{
Aleks Kleyn,
Integral of Map into Abelian $\Omega$-group.\\
Clifford Analysis, Clifford Algebras and their applications,
volume 4, Issue 1, pages 1 - 68, 2013
}%

\BiblioItem{CACAA.05.001}
{
Aleks Kleyn,
Introduction into Calculus over Division Ring.\\
Clifford Analysis, Clifford Algebras and their applications,
volume 5, issue 1, pages 1 - 68, 2016 
}%

\BiblioItem{CACAA.06.121}
{
Aleks Kleyn,
Differential Forms in Banach Algebra.\\
Clifford Analysis, Clifford Algebras and their applications,
volume 6, issue 2, pages 121 - 214, 2017 
}%

\BiblioItem{GJSFRA.13.1.39}
{
Aleks Kleyn,
Reference frame and Lorentz transformation,\\
Global Journals of Science Frontier Research A,
volume 13, issue 1, pages 39 - 55, 2013 
}%

\BiblioItem{1506.05848}
{
Rida T. Farouki, Graziano Gentili, Carlotta Giannelli, Alessandra Sestini,
Caterina Stoppato,\\
Solution of a quadratic quaternion equation with mixed coefficients,\\
eprint \href{http://arxiv.org/abs/1506.05848}{arXiv:1506.05848} (2015)
}%

\BiblioItem{1812.03397}
{
Ivan Kyrchei,
Linear differential systems over the quaternion skew field,\\
eprint \href{http://arxiv.org/abs/1812.03397}{arXiv:1812.03397} (2018)
}%

\BiblioItem{Lauve: Quantum coordinates}
{
A. Lauve, Quantum- and quasi-Plucker coordinates,
\textit{Journal of Algebra} {\bf 296}(2), pp.440-461,
(2006).
}%

\BiblioItem{Lewis D. W. Quaternion algebras}
{
Lewis D. W. Quaternion algebras and the algebraic legacy
of Hamilton's quaternions, \textit{Irish Math. Soc. Bulletin} {\bf
57}, pp. 41-64, (2006).
}%

\BiblioItem{0812.2865}
{
Jos\'e Miguel Figueroa-O'Farrill,
Three lectures on 3-algebras,
eprint \href{http://arxiv.org/abs/0812.2865}{arXiv:0812.2865} (2008)
}%

\BiblioItem{1202.0951}
{
Daniel Edward Clark,
Deconvolution of point processes,
eprint \href{http://arxiv.org/abs/1202.0951}{arXiv:1202.0951} (2012)
}%

\BiblioItem{1202.4546}
{
Ming-Liang Hu,
Disentanglement, Bell-nonlocality violation
and teleportation capacity of the decaying tripartite states,
eprint \href{http://arxiv.org/abs/1202.4546}{arXiv:1202.4546} (2012)
}%

\BiblioItem{1203.1629}
{
Borivoje Dakic, Yannick Ole Lipp, Xiaosong Ma, Martin Ringbauer,
Sebastian Kropatschek, Stefanie Barz, Tomasz Paterek, Vlatko Vedral,
Anton Zeilinger, Caslav Brukner, Philip Walther,
Quantum Discord as Optimal Resource for Quantum Communication,
eprint \href{http://arxiv.org/abs/1203.1629}{arXiv:1203.1629} (2012)
}%

\BiblioItem{Li Nimmo: Darboux transformations}
{
C.X.Li, J.J.C. Nimmo, Darboux transformations for a twisted
derivation and quasideterminant solutions to the super KdV
equation, \textit{Proceedings of the Royal Society A:
Mathematical, Physical and Engineering Sciences} {\bf 466} (2120),
pp. 2471-2493, (2010).
}%

\BiblioItem{Schiebold: Cauchy-type determinants}
{
C. Schiebold, Cauchy-type determinants and integrable
systems, \textit{Linear Algebra and Its Applications} {\bf 433}
(2), pp. 447-475, (2010)
}%

\BiblioItem{Suzuki: Noncommutative spectral decomposition}
{
T. Suzuki, Noncommutative
spectral decomposition with qua\-si\-de\-ter\-mi\-nant, \textit{Advances in
Mathematics} {\bf 217}(5), pp. 2141-2158, (2008).
}%

\BiblioItem{1105.3456}
{
C. W. F. Everitt, D. B. DeBra, B. W. Parkinson, J. P. Turneaure, J. W. Conklin,
M. I. Heifetz, G. M. Keiser, A. S. Silbergleit, T. Holmes, J. Kolodziejczak,
M. Al-Meshari, J. C. Mester, B. Muhlfelder, V. Solomonik, K. Stahl, P. Worden,
W. Bencze, S. Buchman, B. Clarke, A. Al-Jadaan, H. Al-Jibreen, J. Li, J. A. Lipa,
J. M. Lockhart, B. Al-Suwaidan, M. Taber, S. Wang,\\
Gravity Probe B: Final Results of a Space Experiment to Test General Relativity,\\
eprint \href{http://arxiv.org/abs/1105.3456}{arXiv:1105.3456[gr-qc]} (2011)
}%

\BiblioItem{0009305}
{
G. S. Asanov.
Can Neutrinos and High-Energy Particles Test Finsler Metric of Space-Time?\\
eprint \href{http://arxiv.org/abs/hep-ph/0009305}{arXiv:hep-ph/0009305} (2000)
}%

\BiblioItem{Asanov 2004}
{
G. S. Asanov.
Finsleroid - space supplemented by angle and scalar product.\\
Hypercomplex Numbers in Geometry and Physics, {\bf 1}, 2004, p. 40 - 62
}%

\BiblioItem{1004.3007}
{
Sergiu I. Vacaru,
Principles of Einstein-Finsler Gravity and Perspectives in Modern Cosmology,\\
eprint \href{http://arxiv.org/abs/1004.3007}{arXiv:1004.3007[math-ph]} (2010)
}%

\BiblioItem{1012.4148}
{
Sergiu I. Vacaru.
Principles of Einstein-Finsler Gravity and Cosmology.\\
eprint \href{http://arxiv.org/abs/1012.4148}{arXiv:1012.4148[physics.gen-ph]} (2010)
}%

\BiblioItem{1112.5641}
{
Christian Pfeifer, Mattias N.R. Wohlfarth.
Finsler geometric extension of Einstein gravity.\\
eprint \href{http://arxiv.org/abs/1112.5641}{arXiv:1112.5641[gr-qc]} (2011)
}%

\BiblioItem{0711.0056}
{
Zhe Chang, Xin Li.
Lorentz Invariance Violation and Symmetry in Randers\Hyph Finsler Spaces.\\
eprint \href{http://arxiv.org/abs/0711.0056}{arXiv:0711.0056[hep-th]} (2011)
}%

\BiblioItem{1510.02224}
{
Kit Ian Kou, Yong-Hui Xia.
Linear Quaternion Differential Equations: Basic Theory and Fundamental Results.\\
eprint \href{http://arxiv.org/abs/1510.02224}{arXiv:1510.02224} (2017)
}%

\BiblioItem{1902.09800}
{
Dong Cheng, Kit Ian Kou, Yong Hui Xia.
Floquet Theory for Quaternion-valued Differential Equations.\\
eprint \href{http://arxiv.org/abs/1902.09800}{arXiv:1902.09800} (2019)
}%

\BiblioItem{Zharinov Kursy NOC, 9}
{
В. В. Жаринов,
Алгебро-геометрические основы математической физики.
\\
Лекц. курсы НОЦ, 9, МИАН, М., 2008, 3–209
}%

\BiblioItem{Rund Finsler geometry}
{
Hanno Rund,
The differential geometry of Finsler spaces.
\\
Springer - Verlag, Berlin - G\"ottingen - Heidelberg, 1959
}%

\BiblioItem{Smirnov vol 1}
{
V. I. Smirnov,
A Course of Higher Mathematics, volume I.
\\
Translated by D. E. Brown.
\\
Translation, edited and additions made by I. N. Sneddon.
\\
Pergamon Press, Addison-Wesley Publishing Company, 1964
}%

\BiblioItem{Beem Dostoglou Ehrlich}
{
John K. Beem, Stamatis A. Dostoglou, Paul E. Ehrlich,
Advances in differential geometry and general relativity.
\\
American Mathematical Society, 2004
}%

\BiblioItem{978-0719033414}
{
Malcolm Pemberton, Nicholas Rau,
Mathematics for economists: an introductory textbook.
\\
Manchester University Press, November 2001; ISBN-13: 978-0719033414
}%

\BiblioItem{0 521 59180 5}
{
Cyrus D. Cantrell,
Modern mathematical methods for physicists and engineers.
\\
Cambridge University Press, 2000
}%

\BiblioItem{Arveson spectral theory}
{
William Arveson,
A short course on spectral theory.
\\
Springer - Verlag, New York, 2002
}%

\BiblioItem{Robert Hermann}
{
Robert Hermann,
Topics in the mathematics of quantum mechanics.
\\
Math Sci Press, 1973
}%

\BiblioItem{9705.009}
{
John C. Baez,
An Introduction to n-Categories,\\
eprint \href{http://arxiv.org/abs/q-alg/9705009}{arXiv:q-alg/9705009} (1997)
}%

\BiblioItem{0105.155}
{
John C. Baez,
The Octonions,\\
eprint \href{http://arxiv.org/abs/math.RA/0105155}{arXiv:math.RA/0105155} (2002)
}%

\BiblioItem{John Baez: Math Blogs}
{
John C. Baez,
What do mathematicians need to know about blogging?,\\
Notices of the American Mathematical Society,
(2010), 3, {\bf 57}, 333,\\
\url{http://www.ams.org/notices/201003/rtx100300333p.pdf}
}%

\BiblioItem{Tolstoi about Anna Karenina}
{
Tolstoi about Anna Karenina,
in book A Karenina Companion, by C. J. G. Turner,
published by Wilfrid Laurier University Press (August 1993)
}%

\BiblioItem
{Cohn: Universal Algebra}
{
Paul M. Cohn,
Universal Algebra,
Springer, 1981
}%

\BiblioItem
{Cohn: Algebra 1}
{
Paul M. Cohn,
Algebra, Volume 1,
John Wiley \& Sons, 1982
}%

\BiblioItem
{Cohn: Algebra 3}
{
Paul M. Cohn,
Algebra, Volume 3,
John Wiley \& Sons, 1991
}%

\BiblioItem
{Cohn: Skew Fields}
{
Paul M. Cohn,
Skew Fields,
Cambridge University Press, 1995
}%

\BiblioItem
{Lam: Noncommutative Rings}
{
T. Y. Lam,
A First Course in
Noncommutative Rings,
Springer-Verlag, 1991
}%

\BiblioItem
{Maunder: Algebraic Topology}
{
C. R. F. Maunder,
Algebraic Topology,
Dover Publications, Inc, Mineola, New York, 1996
}%

\BiblioItem{Pommaret: Partial Differential Equations}
{
J.-F. Pommaret,
Partial Differential Equations and Group Theory,
Springer, 1994
}%

\BiblioItem{Bourbaki: Set Theory}
{
N. Bourbaki,
Theory of sets,
Springer, 2004
}%

\BiblioItem{Bourbaki: Algebra 1}
{
N. Bourbaki,
Algebra 1,
Springer, 2004
}%

\BiblioItem{Bourbaki: Algebra 2}
{
N. Bourbaki,
Algebra II, Chapters 4 - 7,//
Translated by P. M. Cohn & J. Howie,//
Springer, 2004
}%

\BiblioItem
{Bourbaki: General Topology 1}
{
N. Bourbaki,
General Topology, Chapters 1 - 4,
Springer, 1989
}

\BiblioItem{Bourbaki: General Topology: Chapter 5 - 10}
{
N. Bourbaki,
General Topology, Chapters 5 - 10,
Springer, 1989
}

\BiblioItem{Bourbaki: Topological Vector Space}
{
N. Bourbaki,
Topological Vector Spaces, Chapters 1 - 5,
Transl. by H. G. Eggleston $\&$ S. Madan,
Springer, 2003
}

\BiblioItem{Bourbaki: Group Lie}
{
N. Bourbaki,
Lie Groups and Lie Algebras, Chapters 1 - 3,
Springer, 1989
}

\BiblioItem{Bourbaki: Coxeter Group Lie}
{
N. Bourbaki,
Lie Groups and Lie Algebras, Chapters 4 - 6,
Translator Andrew Pressley,
Springer, 2002
}

\BiblioItem{Bourbaki: Real Group Lie}
{
N. Bourbaki,
Lie Groups and Lie Algebras, Chapters 7 - 9,
Translator Andrew Pressley,
Springer, 2005
}

\BiblioItem{Shabat: Complex Analysis}
{
Shabat B. V.,
Introduction to Complex Analysis,
Moscow, Nauka, 1969
}

\BiblioItem{Pontryagin: Topological Group}
{
L. S. Pontryagin,
Selected Works, Volume Two, Topological Groups,
Gordon and Breach Science Publishers, 1986
}

\BiblioItem
{Eisenhart: Riemannian Geometry}
{
Eisenhart,
Riemannian Geometry,
Princeton University Press, Princeton, 1949
}

\BiblioItem
{Eisenhart: Continuous Groups of Transformations}
{
Eisenhart,
Continuous Groups of Transformations,
Dover Publications, New York, 1961
}

\BiblioItem
{Condon Odabasi}
{
Edward Uhler Condon, Halis Odabasi,
Atomic Structure,
CUP Archive, 1980
}

\BiblioItem{Postnikov: Differential Geometry}
{
Postnikov M. M.,
Geometry IV: Differential geometry,
Moscow, Nauka, 1983
}

\BiblioItem{Fikhtengolts: Calculus volume 1}
{
Fikhtengolts G. M.,
Differential and Integral Calculus Course, volume 1,
Moscow, Nauka, 1969
}

\BiblioItem{Fikhtengolts: Calculus volume 2}
{
Fikhtengolts G. M.,
Differential and Integral Calculus Course, volume 2,
Moscow, Nauka, 1969
}

\BiblioItem{Fikhtengolts: Calculus volume 3}
{
Fikhtengolts G. M.,
Differential and Integral Calculus Course, volume 3,
Moscow, Nauka, 1969
}

\BiblioItem{Hatcher: Algebraic Topology}
{
Allen Hatcher,
Algebraic Topology,
Cambridge University Press, 2002
}

\BiblioItem{geometry of differential equations}
{
Krasil'shchik I. S., Lychagin V. V., Vinogradov A. M.,
Geometry of Jet Spaces and Nonlinear Partial Differential Equations,
\\
Translated from the Russian by A. B. Sosinskii,
\\
Gordon and Breach Science Publishers, 1985
}

\BiblioItem{Basic Concepts of Differential Geometry}
{
Alekseyevskii D. V., Vinogradov A. M., Lychagin V. V.,
Basic Concepts of Differential Geometry
\\
VINITI Summary 28
\\
Moscow. VINITI, 1988
}

\BiblioItem{cohomological analysis}
{
A. M. Vinogradov,
Cohomological Analysis of Partial Differential Equations
and Secondary Calculus,
American Mathematical Society, 2001
}

\BiblioItem{0801.1734}
{
Brandon S. DiNunno, Richard A. Matzner,
The Volume Inside a Black Hole,\\
eprint \href{http://arxiv.org/abs/0801.1734v1}{arXiv:0801.1734v1} (2008)
}

\BiblioItem{0702.447}
{
Ivan Kyrchei,
Cramer's rule for some quaternion matrix equations,\\
eprint \href{http://arxiv.org/abs/math/0702447}{arXiv:math.RA/0702447} (2007)
}

\BiblioItem{Izrail M. Gelfand: Quaternion Groups}
{
I. M. Gelfand, M. I. Graev,
Representation of Quaternion Groups over Localy Compact and
Functional Fields,\\
Funct. Anal. Appl. {\bf 2} (1968) 19 - 33;\\
Izrail Moiseevich Gelfand, Semen Grigorevich Gindikin,\\
Izrail M. Gelfand: Collected Papers, volume II, 435 - 449,\\
Springer, 1989
}

\BiblioItem{Richard D. Schafer}
{
Richard D. Schafer,
An Introduction to Nonassociative Algebras,
Dover Publications, Inc., New York, 1995
}

\BiblioItem{Bamberg Sternberg}
{
Paul Bamberg, Shlomo Sternberg,
A course in mathematics for students of physics,
Cambridge University Press, 1991
}

\BiblioItem{Conway Smith}
{
John Horton Conway, Derek Alan Smith,
On quaternions and octonions: their geometry, arithmetic, and symmetry,
A K Peters, Natick, Massachussets, 2003
}

\BiblioItem{Fueter}
{
Fueter, R.
Die Funktionentheorie der Differentialgleichungen $\Delta u = 0$ und
$\Delta \Delta u = 0$ mit vier reellen Variablen.
Comment. Math. Helv. {\bf 7} (1935), 307-330
}

\BiblioItem{Sudbery Quaternionic Analysis}
{
A. Sudbery,
Quaternionic Analysis,\\
Math. Proc. Camb. Phil. Soc. (1979), {\bf 85}, 199 - 225
}

\BiblioItem{Sudbery 2657821}
{
A. Sudbery,
Quaternionic Analysis,\\
eprint \href{https://www.researchgate.net/publication/2657821}{ResearchGate:2657821} (1977)
}

\BiblioItem{0902.4771}
{
Fabrizio Colombo, Graziano Gentili, Irene Sabadini,
A Cauchy kernel for slice regular functions,\\
eprint \href{http://arxiv.org/abs/0902.4771v1}{arXiv:0902.4771v1} (2009)
}

\BiblioItem{Vadim Komkov}
{
Vadim Komkov,
Variational Principles of Continuum Mechanics with Engineering Applications: Critical Points Theory,\\
Springer, 1986
}

\BiblioItem{Alain Connes 1994}
{
Alain Connes,
Noncommutative Geometry,\\
Academic Press, 1994
}

\BiblioItem{Hamilton papers 3}
{
Sir William Rowan Hamilton,
The Mathematical Papers, Vol. III, Algebra,\\
Cambridge at the University Press, 1967
}

\BiblioItem{Hamilton Elements of Quaternions 1}
{
Sir William Rowan Hamilton,
Elements of Quaternions, Volume I,\\
Longmans, Green, and Co., London, New York, and Bombay, 1899
}

\BiblioItem{Cartan geometry in reper}
{
Elie Cartan, Vladislav V. Goldberg, Serge\u{i} Pavlovich Finikov,\\
Riemannian geometry in an orthogonal frame:
from lectures delivered by Elie Cartan at the Sorbonne in 1926-1927,\\
translated by Vladislav V. Goldberg,\\
World Scientific, 2001
}

\BiblioItem{Cartan differential form}
{
Henri Cartan.
Differential forms.\\
Kershaw Publishing Company Limited, London, 1971
}

\BiblioItem{Arnautov Glavatsky Mikhalev}
{
V. I. Arnautov, S. T. Glavatsky, A. V. Mikhalev,\\
Introduction to the theory of topological rings and modules,
Volume 1995,\\
Marcel Dekker, Inc, 1996
}

\BiblioItem{Moore Yaqub}
{
Hal G. Moore, Adil Yaqub,
A first course in linear algebra with applications,
Edition 3, Academic Press, 1998 
}

\BiblioItem{math.CV-0405471}
{
S. V. Ludkovsky,
Differentiable functions of Cayley-Dickson numbers,\\
eprint \href{http://arxiv.org/abs/math.CV/0405471}{arXiv:math.CV/0405471} (2004)
}%

\BiblioItem{W.Bertram H.Glockner K.Neeb}
{
W.Bertram, H.Glockner, K.Neeb,
Differential Calculus over General Base Fields and Rings,
Expositiones Mathematicae (2004), Volume 22, Issue 3, Pages 213-282
}

\CloseBiblio

%% file: Index.English.tex
\OpenIndex
\SetIndexSpace%
\Index
   {$1$\Hyph form}%
   {1-form}%
\SetIndexSpace%
\Index
   {$2$\Hyph ary fibered relation}%
   {2 ary fibered relation}%
\SetIndexSpace%
\Index
   {$A$\Hyph algebra of polynomials over $D$\Hyph algebra $A$}%
   {algebra of polynomials over algebra}%
\Index
   {$A$\Hyph number}%
   {A number}%
\Index
   {$\mathcal A(A)$\Hyph map}%
   {A(A) map}%
\Index
   {$A*$\Hyph module}%
   {A*-module}%
\Index
   {$A*$\Hyph vector space}%
   {A*-vector space}%
\Index
   {$A$\Hyph module}%
   {module over algebra}%
\Index
   {$A$\Hyph valued function}%
   {A valued function}%
\Index
   {$A$\Hyph representation in $\Omega$\Hyph algebra}%
   {A representation of algebra}%
\Index
   {Abelian multiplicative $\Omega$\Hyph group}%
   {Abelian multiplicative Omega group}%
\Index
   {Abelian $\Omega$\Hyph group}%
   {Abelian Omega group}%
\Index
   {Abelian semigroup}%
   {Abelian semigroup}%
\Index
   {absolute value}%
   {absolute value}%
\Index
   {active \sT{G}representation}%
   {active representation, vector space}%
\Index
   {active representation}%
   {active representation}%
\Index
   {active representation in basis manifold}%
   {active representation in basis manifold}%
\Index
   {active representation of group $G(\Vector f)$ in basis manifold of tower of representations}%
   {active representation in basis manifold, tower of representations}%
\Index
   {active transformation of basis manifold}%
   {active transformation of basis}%
\Index
   {active transformation on basis manifold}%
   {active transformation}%
\Index
   {active transformation on the set of \rcd bases}%
   {active transformation, vector space}%
\Index
   {additive map}%
   {additive map}%
\Index
   {affine basis}%
   {Affine Basis}%
\Index
   {affine functional}%
   {affine functional}%
\Index
   {affine representation of Lie group}%
   {affine representation of Lie group}%
\Index
   {affine space}%
   {affine space}%
\Index
   {affine structure on set}%
   {affine structure on set}%
\Index
   {affine transformation}%
   {affine transformation}%
\Index
   {affine transformation group}%
   {affine transformation group}%
\Index
   {affine transformation group}%
   {affine transformation group}%
\Index
   {affine transformation on basis manifold}%
   {affine transformation}%
\Index
   {algebra of fractions of algebra with conjugation}%
   {algebra of fractions of algebra with conjugation}%
\Index
   {algebra of polynomials over $D$\Hyph algebra}%
   {algebra of polynomials over D algebra}%
\Index
   {algebra of rational mappings of algebra}%
   {algebra of rational mappings of algebra}%
\Index
   {algebra of sets}%
   {algebra of sets}%
\Index
   {algebra over ring}%
   {algebra over ring}%
\Index
   {algebra with conjugation}%
   {algebra with conjugation}%
\Index
   {alternation of polylinear map}%
   {alternation of polylinear map}%
\Index
   {alternative representation of matrix}%
   {Alternative representation}%
\Index
   {anholonomic coordinate}%
   {anholonomic coordinate}%
\Index
   {anholonomic coordinates of connection}%
   {anholonomic coordinates of connection}%
\Index
   {anholonomic coordinates of vector}%
   {vector anholonomic coordinates}%
\Index
   {anholonomic coordinates on manifold}%
   {anholonomic coordinates on manifold}%
\Index
   {anholonomity object}%
   {anholonomity object}%
\Index
   {antilinear homomorphism}%
   {antilinear homomorphism}%
\Index
   {antilinear map}%
   {antilinear map}%
\Index
   {antisymmetric $2$\Hyph ary fibered relation}%
   {antisymmetric 2 ary fibered relation}%
\Index
   {$A\RCstar$\Hyph basis for vector space}%
   {Arc basis, vector space}%
\Index
   {arity}%
   {arity}%
\Index
   {arity of operation}%
   {arity of operation}%
\Index
   {associative $D$\Hyph algebra}%
   {associative D algebra}%
\Index
   {associative law}%
   {associative law}%
\Index
   {associative multiplicative $\Omega$\Hyph group}%
   {associative multiplicative Omega group}%
\Index
   {associative $\Omega$\Hyph group}%
   {associative Omega group}%
\Index
   {associative operation}%
   {associative operation}%
\Index
   {associator of $D$\Hyph algebra}%
   {associator of algebra}%
\Index
   {auto parallel line}%
   {auto parallel line}%
\Index
   {automorphism}%
   {automorphism}%
\Index
   {automorphism of diagram of representations}%
   {automorphism of diagram of representations}%
\Index
   {automorphism of representation of $\Omega$\Hyph algebra}%
   {automorphism of representation}%
\Index
   {automorphism of tower of representations}%
   {automorphism of tower of representations}%
\Index
   {automorphism of vector space}%
   {automorphism of vector space}%
\Index
   {$(^j_i)$\hyph \CR quasideterminant}%
   {j i cr-quasideterminant}%
\Index
   {norm of quaternion}%
   {norm of quaternion}%
\SetIndexSpace%
\Index
   {$B$\Hyph set}%
   {B set}%
\Index
   {Banach $D$\Hyph algebra}%
   {Banach algebra}%
\Index
   {Banach $D$\Hyph module}%
   {Banach module}%
\Index
   {base of fibered correspondence}%
   {base of fibered correspondence}%
\Index
   {base of mapping}%
   {base of map}%
\Index
   {basis}%
   {Basis}%
\Index
   {basis dual to basis}%
   {basis dual to basis}%
\Index
   {basis dual to basis}%
   {dual basis}%
\Index
   {basis for \crd vector space}%
   {basis, crd vector space}%
\Index
   {basis for \dcr vector space}%
   {basis, dcr vector space}%
\Index
   {basis for \drc vector space}%
   {basis, drc vector space}%
\Index
   {basis for module}%
   {basis, module}%
\Index
   {basis for \rcd vector space}%
   {basis, rcd vector space}%
\Index
   {basis for vector space}%
   {basis, vector space}%
\Index
   {basis manifold}%
   {basis manifold}%
\Index
   {basis manifold of affine space}%
   {Basis Manifold, Affine Space}%
\Index
   {basis manifold of central affine space}%
   {Basis Manifold, Central Affine Space}%
\Index
   {basis manifold of Euclid space}%
   {Basis Manifold, Euclid Space}%
\Index
   {basis manifold of Euclid space}%
   {Basis Manifold, Euclid Space, division ring}%
\Index
   {basis manifold of \rcd vector space}%
   {basis manifold of rcd vector space}%
\Index
   {basis manifold of tower of representations}%
   {basis manifold tower of representations}%
\Index
   {basis manifold of vector space}%
   {basis manifold of vector space}%
\Index
   {basis of Abelian group}%
   {basis of Abelian group}%
\Index
   {basis of algebra $\mathcal L(A;A)$}%
   {basis of algebra L(A,A)}%
\Index
   {basis of diagram of representations}%
   {basis of diagram of representations}%
\Index
   {basis of representation}%
   {basis of representation}%
\Index
   {basis of tower of representations}%
   {basis of tower of representations}%
\Index
   {basis vector of representation of Lie group over algebra $A$}%
   {basis vector of representation of Lie group over algebra A}%
\Index
   {biring}%
   {biring}%
\Index
   {Borel algebra}%
   {Borel algebra}%
\Index
   {Borel set}%
   {Borel set}%
\Index
   {Borel\Hyph measurable map}%
   {Borel-measurable map}%
\Index
   {bundle of level $2$}%
   {bundle of level 2}%
\Index
   {bundle of level $n$}%
   {bundle of level n}%
\SetIndexSpace%
\Index
   {\subs row of matrix}%
   {c row}%
\Index
   {$c$\hyph row of matrix}%
   {c-row}%
\Index
   {can be embeded}%
   {can be embeded}%
\Index
   {canonical remainder of the division}%
   {canonical remainder of the division}%
\Index
   {canonical representation of division with remainder}%
   {canonical representation of division with remainder}%
\Index
   {carrier of $\Omega$\Hyph algebra}%
   {carrier of Omega-algebra}%
\Index
   {Cartan connection}%
   {Cartan connection}%
\Index
   {Cartan curvature}%
   {Cartan curvature}%
\Index
   {Cartan derivative}%
   {Cartan derivative}%
\Index
   {Cartan equation}%
   {Cartan equation}%
\Index
   {Cartan symbol}%
   {Cartan symbol}%
\Index
   {Cartan transport}%
   {Cartan transport}%
\Index
   {Cartesian power}%
   {Cartesian power}%
\Index
   {Cartesian power $\Bundle A$ of bundle $\Bundle B$}%
   {Cartesian power A of bundle B}%
\Index
   {Cartesian power $A$ of set $B$}%
   {Cartesian power of set}%
\Index
   {Cartesian power $n$ of bundle $\Bundle E$}%
   {Cartesian power n of bundle E}%
\Index
   {Cartesian power $n$ of $\mathfrak{H}$\Hyph algebra}%
   {Cartesian power of algebra}%
\Index
   {Cartesian power of systems of subsets}%
   {Cartesian power of systems of subsets}%
\Index
   {Cartesian product of groups}%
   {Cartesian product of groups}%
\Index
   {Cartesian product of measures}%
   {Cartesian product of measures}%
\Index
   {Cartesian product of \(\Omega\)\Hyph algebras}%
   {Cartesian product of Omega algebras}%
\Index
   {Cartesian product of systems of subsets}%
   {Cartesian product of systems of subsets}%
\Index
   {category of \drc vector spaces}%
   {category of drc vector spaces}%
\Index
   {category of fibered correspondences over diagonal}%
   {category of fibered correspondences over diagonal}%
\Index
   {category of left-side representations}%
   {category of left-side representations}%
\Index
   {category of left-side representations of $\Omega_1$\Hyph algebra $A$}%
   {category of left-side representations of Omega1 algebra}%
\Index
   {category of reduced fibered correspondences}%
   {category of reduced fibered correspondences}%
\Index
   {category of representations}%
   {category of representations}%
\Index
   {Cauchy sequence}%
   {Cauchy sequence}%
\Index
   {center of $D$\Hyph algebra $A$}%
   {center of algebra}%
\Index
   {center of ring $D$}%
   {center of ring}%
\Index
   {central affine basis}%
   {Central Affine Basis}%
\Index
   {closed ball}%
   {closed ball}%
\Index
   {closure of set}%
   {closure of set}%
\Index
   {coefficient of polynomial}%
   {coefficient of polynomial}%
\Index
   {column $D*$\Hyph vector}%
   {column D* vector}%
\Index
   {column determinant}%
   {column determinant}%
\Index
   {column vector}%
   {column vector}%
\Index
   {common factor}%
   {common factor}%
\Index
   {commutative $D$\Hyph algebra}%
   {commutative D algebra}%
\Index
   {commutative diagram of correspondences}%
   {commutative diagram of correspondences}%
\Index
   {commutative diagram of representations of universal algebras}%
   {commutative diagram of representations}%
\Index
   {commutative operation}%
   {commutative operation}%
\Index
   {commutativity of representations}%
   {commutativity of representations}%
\Index
   {commutator of $D$\Hyph algebra}%
   {commutator of algebra}%
\Index
   {compact set}%
   {compact set}%
\Index
   {compact\hyph open topology}%
   {compact open topology}%
\Index
   {complete division ring}%
   {complete division ring}%
\Index
   {complete measure}%
   {complete measure}%
\Index
   {complete normed $\Omega$\Hyph group}%
   {complete Omega group}%
\Index
   {complete ring}%
   {complete ring}%
\Index
   {complete system of linear partial differential equations}%
   {Complete System of Linear Partial Differential Equations}%
\Index
   {completely integrable system}%
   {completely integrable system}%
\Index
   {completion of normed $\Omega$\Hyph group}%
   {completion of normed Omega group}%
\Index
   {completion of representation}%
   {completion of representation}%
\Index
   {component of derivative}%
   {component of derivative}%
\Index
   {component of derivative of Second Order}%
   {component of derivative of Second Order}%
\Index
   {component of linear map}%
   {component of linear map}%
\Index
   {component of polylinear map}%
   {component of polylinear map}%
\Index
   {component of the G\^ateaux derivative}%
   {component of Gateaux derivative}%
\Index
   {component of the G\^ateaux derivative of second order}%
   {component of Gateaux derivative of Second Order}%
\Index
   {composition of fibered correspondences}%
   {composition of fibered correspondences}%
\Index
   {composition of reduced fibered correspondences}%
   {composition of reduced fibered correspondences}%
\Index
   {condition of reducibility of products}%
   {condition of reducibility of products}%
\Index
   {congruence}%
   {congruence}%
\Index
   {conjugate of quaternion $x$}%
   {conjugate of quaternion}%
\Index
   {conjugated affine space}%
   {conjugated affine space}%
\Index
   {conjugated $D$\Hyph  module}%
   {conjugated D module}%
\Index
   {conjugated vector space}%
   {conjugated vector space}%
\Index
   {conjugation in algebra}%
   {conjugation in algebra}%
\Index
   {conjugation in ring}%
   {conjugation in ring}%
\Index
   {conjugation transformation}%
   {conjugation transformation}%
\Index
   {connected set}%
   {connected set}%
\Index
   {connection coefficients in affine space}%
   {connection coefficients, affine space}%
\Index
   {connection in principal fibre bundle}%
   {connection in principal bundle}%
\Index
   {contact point of set}%
   {contact point of set}%
\Index
   {continues basis}%
   {continues basis}%
\Index
   {continuous correspondence}%
   {continuous correspondence}%
\Index
   {continuous map}%
   {continuous map}%
\Index
   {continuous multivariable map}%
   {continuous multivariable map}%
\Index
   {contravariant representation}%
   {contravariant representation}%
\Index
   {convex set}%
   {convex set}%
\Index
   {coordinate isomorphism}%
   {coordinate isomorphism}%
\Index
   {coordinate matrix of set of vectors}%
   {coordinate matrix of set of vectors}%
\Index
   {coordinate matrix of vector}%
   {coordinate matrix of vector}%
\Index
   {coordinate matrix of vector field in \rcD basis}%
   {coordinate matrix of vector field in drc basis}%
\Index
   {coordinate \rcd vector space}%
   {coordinate rcd vector space}%
\Index
   {coordinate reference frame}%
   {coordinate reference frame}%
\Index
   {coordinate representation}%
   {coordinate representation}%
\Index
   {coordinate representation in \rcd vector space}%
   {coordinate representation, rcd vector space}%
\Index
   {coordinate representation in tuple of $\VX\Omega$\Hyph algebras}%
   {coordinate tower of representations, Omega algebra}%
\Index
   {coordinate representation of group in vector space}%
   {coordinate representation, vector space}%
\Index
   {coordinate representation of vector}%
   {coordinate representation of vector}%
\Index
   {coordinate vector bundle}%
   {coordinate vector bundle}%
\Index
   {coordinate vector space}%
   {coordinate vector space}%
\Index
   {coordinates}%
   {coordinates}%
\Index
   {coordinates of $A_2$\Hyph number $m$ relative to set $X$}%
   {coordinates relative to set}%
\Index
   {coordinates of associator}%
   {coordinates of associator}%
\Index
   {coordinates of basis}%
   {coordinates of basis}%
\Index
   {coordinates of basis of representation}%
   {coordinates of basis relative to basis, representation}%
\Index
   {coordinates of endomorphism of representation}%
   {coordinates of endomorphism, representation}%
\Index
   {coordinates of endomorphism of tower of representations}%
   {coordinates of endomorphism, tower of representations}%
\Index
   {coordinates of geometric object}%
   {coordinates of geometric object, vector space}%
\Index
   {coordinates of geometric object}%
   {coordinates of geometric object}%
\Index
   {coordinates of geometric object in coordinate \rcd vector space}%
   {coordinates of geometric object, coordinate rcd vector space}%
\Index
   {coordinates of geometric object in coordinate space of tower of representations}%
   {coordinates of geometric object, coordinate tower of representations}%
\Index
   {coordinates of geometric object in \rcd vector space}%
   {coordinates of geometric object, rcd vector space}%
\Index
   {coordinates of morphism of diagram of representations}%
   {coordinates of morphism, diagram of representations}%
\Index
   {coordinates of point $A$ of affine space $\overset{\circ}{A}$ relative to basis $(O,\Basis e)$}%
   {coordinates in affine space}%
\Index
   {coordinates of reduced morphism of representation}%
   {coordinates of reduced morphism of representation}%
\Index
   {coordinates of representation}%
   {coordinates of representation, drc vector space}%
\Index
   {coordinates of representation}%
   {coordinates of representation}%
\Index
   {coordinates of set of vectors}%
   {coordinates of set of vectors}%
\Index
   {coordinates of vector}%
   {coordinates of vector}%
\Index
   {coordinates of vector field in \Drc basis}%
   {coordinates of vector field in drc basis}%
\Index
   {coordinates of vector relative to Hamel basis}%
   {coordinates of vector, Hamel basis}%
\Index
   {coordinates of vector relative to Schauder basis}%
   {coordinates of vector, Schauder basis}%
\Index
   {coproduct of objects in category}%
   {coproduct in category}%
\Index
   {correspondence continuous on the set}%
   {correspondence continuous on the set}%
\Index
   {correspondence of homomorphism}%
   {correspondence of homomorphism}%
\Index
   {cosine}%
   {cosine}%
\Index
   {covariant representation}%
   {covariant representation}%
\Index
   {\CR eigencolumn}%
   {cr eigencolumn}%
\Index
   {\CR eigenrow}%
   {cr eigenrow}%
\Index
   {\CR eigenvalue}%
   {cr eigenvalue}%
\Index
   {\CR exponent}%
   {CR exponent}%
\Index
   {\CR inverse element of biring}%
   {cr-inverse element}%
\Index
   {\CR matrix group}%
   {cr-matrix group}%
\Index
   {\CR power}%
   {cr power}%
\Index
   {\CR product (product column over row)}%
   {cr-product}%
\Index
   {$\CRcirc$\Hyph product of matrices of maps}%
   {cr product of matrices of maps}%
\Index
   {\crd vector}%
   {crd vector}%
\Index
   {\crd vector space}%
   {crd vector space}%
\Index
   {$C^*$\Hyph algebra}%
   {Cstar-algebra}%
\Index
   {curvilinear coordinates of point in affine space}%
   {curvilinear coordinates of point in affine space}%
\SetIndexSpace%
\Index
   {$D$\Hyph linear functional}%
   {D linear functional}%
\Index
   {$D*$\hyph matrices vector space}%
   {matrices vector space}%
\Index
   {$D*$\hyph  vector space}%
   {D* vector space}%
\Index
   {$D*$\Hyph module}%
   {D*-module}%
\Index
   {$D$\Hyph affine connection on manifold with affine connections}%
   {D affine connection, affine manifold}%
\Index
   {$D$\Hyph algebra}%
   {D algebra}%
\Index
   {$D$\Hyph module}%
   {D-module}%
\Index
   {$D$\Hyph module}%
   {D module}%
\Index
   {$D$\Hyph valued variable}%
   {D valued variable}%
\Index
   {$D$\Hyph vector function}%
   {d vector function}%
\Index
   {$D$\Hyph affine connection coefficients on manifold}%
   {D affine connection coefficients, manifold}%
\Index
   {$D$\hyph vector space}%
   {D vector space}%
\Index
   {\dcr vector}%
   {dcr vector}%
\Index
   {\dcr vector space}%
   {dcr vector space}%
\Index
   {definite integral}%
   {definite integral}%
\Index
   {derivative of map}%
   {derivative of map}%
\Index
   {derivative of order $n$}%
   {derivative of Order n}%
\Index
   {derivative of second order}%
   {derivative of Second Order}%
\Index
   {determinant of matrix}%
   {determinant}%
\Index
   {deviation of trajectories}%
   {deviation of trajectories}%
\Index
   {diagonal in bundle}%
   {diagonal in bundle}%
\Index
   {diagram of correspondences}%
   {diagram of correspondences}%
\Index
   {diagram of representations}%
   {diagram of representations}%
\Index
   {diagram of representations of universal algebras}%
   {diagram of representations of algebras}%
\Index
   {differentiable map}%
   {differentiable map}%
\Index
   {differential equation with separated variables}%
   {differential equation with separated variables}%
\Index
   {differential form of degree $p$}%
   {differential form of degree p}%
\Index
   {differential of independent variable}%
   {differential of independent variable}%
\Index
   {differential of map}%
   {differential of map}%
\Index
   {differential $p$\Hyph form}%
   {differential p form}%
\Index
   {differential separable equation}%
   {differential separable equation}%
\Index
   {dimension of \rcd vector space}%
   {dimension of vector space}%
\Index
   {direct product of bundles}%
   {Cartesian product of bundles}%
\Index
   {direct product of $D$\Hyph vector spaces}%
   {direct product of D vector spaces}%
\Index
   {direct product of division rings}%
   {direct product of division rings}%
\Index
   {direct product of \Ts{G}representations}%
   {direct product of G* representations}%
\Index
   {direct product of \(\Omega\)\Hyph algebras}%
   {direct product of Omega algebras}%
\Index
   {direct product of \rcd vector spaces}%
   {direct product, rcd vector space}%
\Index
   {direct product of representations of fibered group}%
   {direct product of representations of fibered group}%
\Index
   {direct product of representations of group}%
   {direct product of representations of group}%
\Index
   {direct product of total spaces}%
   {Cartesian product of total spaces}%
\Index
   {direct sum}%
   {direct sum}%
\Index
   {direct sum of representations}%
   {direct sum of representations}%
\Index
   {direction over commutative ring}%
   {direction over commutative ring}%
\Index
   {distributive law}%
   {distributive law}%
\Index
   {division algebra}%
   {division algebra}%
\Index
   {division with remainder}%
   {division with remainder}%
\Index
   {division without remainder}%
   {division without remainder}%
\Index
   {divisor of polynomial}%
   {divisor of polynomial}%
\Index
   {double determinant}%
   {double determinant}%
\Index
   {\Drc linear map of vector bundles}%
   {drc linear map of vector bundles}%
\Index
   {\drc vector}%
   {drc vector}%
\Index
   {\drc vector space}%
   {drc vector space}%
\Index
   {$D\star$\Hyph antilinear homomorphism}%
   {Dstar antilinear homomorphism}%
\Index
   {$\mathcal D\star$\Hyph vector bundle}%
   {Dstar vector bundle}%
\Index
   {$\mathcal D\star$\Hyph vector field}%
   {Dstar vector field}%
\Index
   {$\mathcal D\star$\hyph linear composition of vector fields}%
   {linear composition of vector fields}%
\Index
   {$\mathcal D\star$\hyph product of vector field over scalar}%
   {Dstar product of vector field over scalar, vector space}%
\Index
   {dual space of \rcd vector space}%
   {dual space of rcd vector space}%
\Index
   {duality principle for biring}%
   {duality principle for biring}%
\Index
   {duality principle for biring of matrices}%
   {duality principle for biring of matrices}%
\SetIndexSpace%
\Index
   {effective \Ts{G}representation}%
   {effective G* representation}%
\Index
   {effective representation}%
   {effective representation}%
\Index
   {effective representation of division ring}%
   {effective representation of division ring}%
\Index
   {effective representation of fibered $\Omega$\Hyph algebra}%
   {effective representation of fibered Omega-algebra}%
\Index
   {effective representation of group}%
   {effective representation of group}%
\Index
   {effective representation of ring}%
   {effective representation of ring}%
\Index
   {effective \Ts representation of fibered division ring}%
   {effective representation of fibered division ring}%
\Index
   {effective \Ts representation of fibered group}%
   {effective representation of fibered group}%
\Index
   {Einstein equation}%
   {Einstein equation}%
\Index
   {endomorphism}%
   {endomorphism}%
\Index
   {endomorphism of diagram of representations}%
   {endomorphism of diagram of representations}%
\Index
   {endomorphism of representation of $\Omega$\Hyph algebra}%
   {endomorphism of representation}%
\Index
   {endomorphism of representation regular on generating set $X$}%
   {endomorphism of representation, regular on set}%
\Index
   {endomorphism of representation singular on generating set $X$}%
   {endomorphism of representation, singular on set}%
\Index
   {endomorphism of tower of representations}%
   {endomorphism of tower of representations}%
\Index
   {endomorphism of tower of representations regular on tuple of generating sets}%
   {endomorphism of representation, regular on tuple}%
\Index
   {endomorphism of tower of representations singular on tuple of generating sets}%
   {endomorphism of representation, singular on tuple}%
\Index
   {enhanced Lie group}%
   {enhanced Lie group}%
\Index
   {epimorphism}%
   {epimorphism}%
\Index
   {equivalence}%
   {equivalence}%
\Index
   {equivalence generated by representation $f$}%
   {equivalence of representation}%
\Index
   {equivalent norms}%
   {equivalent norms}%
\Index
   {essential parameters in a set of functions}%
   {essential parameters}%
\Index
   {Euclidean metric on division ring}%
   {Euclidean metric on division ring}%
\Index
   {Euclidean scalar product in $D$\Hyph vector space}%
   {Euclidean scalar product, vector space}%
\Index
   {Euclidean scalar product on division ring}%
   {Euclidean scalar product on division ring}%
\Index
   {everywhere dense subset}%
   {everywhere dense subset}%
\Index
   {exact differential equation}%
   {exact differential equation}%
\Index
   {expansion of vector relative to basis converges}%
   {expansion converges}%
\Index
   {expansion of vector relative to basis converges normally}%
   {expansion converges normally}%
\Index
   {exponent}%
   {exponent}%
\Index
   {extended matrix of \drc linear equations}%
   {extended matrix, system of drc linear equations}%
\Index
   {extended matrix of \rcd linear equations}%
   {extended matrix, system of rcd linear equations}%
\Index
   {extension of correspondence}%
   {extension of correspondence}%
\Index
   {extension of measure}%
   {extension of measure}%
\Index
   {exterior differential}%
   {exterior differential}%
\Index
   {exterior product}%
   {exterior product}%
\Index
   {extreme line}%
   {extreme line}%
\SetIndexSpace%
\Index
   {fibered coordinate isomorphism}%
   {fibered coordinate isomorphism}%
\Index
   {fibered correspondence from $\Bundle A$ to $\Bundle B$}%
   {fibered correspondence from A to B}%
\Index
   {fibered correspondence in $\Bundle{A}$}%
   {fibered correspondence in A}%
\Index
   {fibered correspondence of homomorphism}%
   {fibered correspondence of homomorphism}%
\Index
   {fibered equivalence}%
   {fibered equivalence}%
\Index
   {fibered group}%
   {fibered group}%
\Index
   {fibered identification morphism}%
   {fibered identification morphism}%
\Index
   {fibered little group}%
   {fibered little group}%
\Index
   {fibered morphism from bundle $\Bundle A$ into $\Bundle B$}%
   {fibered morphism from A into B}%
\Index
   {fibered natural morphism}%
   {fibered natural morphism}%
\Index
   {fibered $\Omega$\Hyph algebra}%
   {fibered Omega-algebra}%
\Index
   {fibered $\Omega$\Hyph subalgebra}%
   {fibered Omega-subalgebra}%
\Index
   {fibered ordering}%
   {fibered ordering}%
\Index
   {fibered preordering}%
   {fibered preordering}%
\Index
   {fibered ring}%
   {fibered ring}%
\Index
   {fibered stability group}%
   {fibered stability group}%
\Index
   {fibered subset}%
   {fibered subset}%
\Index
   {field equation}%
   {field equation}%
\Index
   {field-strength tensor}%
   {field-strength tensor}%
\Index
   {filter $\mathfrak{F}$ converges to $A$}%
   {filter converges}%
\Index
   {finite expansion of set}%
   {finite expansion of set}%
\Index
   {Finsler metric}%
   {Finsler metric}%
\Index
   {Finsler space}%
   {Finsler space}%
\Index
   {Finsler structure}%
   {Finsler structure}%
\Index
   {first Newton law}%
   {First Newton law}%
\Index
   {frame\Hyph dragging effect}%
   {frame dragging effect}%
\Index
   {free $A$\Hyph module}%
   {free A module}%
\Index
   {free Abelian group}%
   {free Abelian group}%
\Index
   {free algebra over ring}%
   {free algebra over ring}%
\Index
   {free module}%
   {free module}%
\Index
   {free representation}%
   {free representation}%
\Index
   {free representation of group}%
   {free representation of group}%
\Index
   {free \Ts representation of fibered group}%
   {free representation of fibered group}%
\Index
   {Frenet transport}%
   {Frenet transport}%
\Index
   {function homogeneous of degree $k$}%
   {function homogeneous}%
\Index
   {function of division ring \Ds differentiable in the Fr\'echet sense}%
   {function Dstar differentiable in Frechet sense, division ring}%
\Index
   {fundamental sequence}%
   {fundamental sequence}%
\SetIndexSpace%
\Index
   {$G$\Hyph reference frame}%
   {G reference frame}%
\Index
   {$G$\Hyph basis of vector space}%
   {G-basis}%
\Index
   {$G$\Hyph coordinates of basis}%
   {G-coordinates}%
\Index
   {$G$\Hyph space}%
   {GSpace}%
\Index
   {the G\^ateaux \dcr derivative of map $f$ of $D$\Hyph vector space $V$ to $D$\Hyph vector space $W$}%
   {Gateaux dcr derivative of map, D vector space}%
\Index
   {the G\^ateaux derivative of map}%
   {Gateaux derivative of map}%
\Index
   {the G\^ateaux derivative of order $n$}%
   {Gateaux derivative of Order n}%
\Index
   {the G\^ateaux derivative of second order}%
   {Gateaux derivative of Second Order}%
\Index
   {the G\^ateaux \Ds derivative of map $f$ of division ring $D$}%
   {Gateaux Dstar derivative of map, division ring}%
\Index
   {the G\^ateaux mixed partial derivative}%
   {Gateaux partial derivative of Second Order}%
\Index
   {the G\^ateaux partial \dcr derivative of map $f^{\gi b}$ with respect to variable $x^{\gi a}$}%
   {Gateaux partial dcr derivative of map with respect to variable, D vector space}%
\Index
   {the G\^ateaux partial derivative}%
   {Gateaux partial derivative}%
\Index
   {the G\^ateaux partial \rcd derivative of map $f^{\gi b}$ with respect to variable $x^{\gi a}$}%
   {Gateaux partial rcd derivative of map with respect to variable, D vector space}%
\Index
   {the G\^ateaux \rcd derivative of map $f$ of $D$\hyph vector space $V$ to $D$\hyph vector space $W$}%
   {Gateaux rcd derivative of map, D vector space}%
\Index
   {the G\^ateaux \sD derivative of map $f$ of division ring $D$}%
   {Gateaux starD derivative of map, division ring}%
\Index
   {generating set}%
   {generating set}%
\Index
   {generator of linear map}%
   {generator of linear map}%
\Index
   {geodetic effect}%
   {geodetic effect}%
\Index
   {geometric object}%
   {geometric object}%
\Index
   {geometric object defined in \rcd vector space}%
   {geometric object, rcd vector space}%
\Index
   {geometric object defined in tuple of $\VX\Omega$\Hyph algebras $\VX A$}%
   {geometric object, tower of representations g}%
\Index
   {geometric object in coordinate representation}%
   {geometric object, coordinate representation}%
\Index
   {geometric object in coordinate representation defined in \rcd vector space}%
   {geometric object, coordinate rcd vector space}%
\Index
   {geometric object in coordinate representation defined in tuple of $\VX\Omega$\Hyph algebras $\VX A$}%
   {geometric object, coordinate tower of representations g}%
\Index
   {geometric object in vector space}%
   {geometric object, vector space}%
\Index
   {geometric object of type $H$}%
   {geometric object of type H}%
\Index
   {geometric object of type $A$ in vector space}%
   {geometric object of type A, vector space}%
\Index
   {group algebra}%
   {group algebra}%
\Index
   {group of automorphisms of representation}%
   {group of automorphisms of representation}%
\SetIndexSpace%
\Index
   {Hadamard inverse of matrix}%
   {Hadamard inverse of matrix}%
\Index
   {Hamel basis}%
   {Hamel basis}%
\Index
   {hermitian conjugated vector}%
   {hermitian conjugated vector}%
\Index
   {hermitian conjugation in division ring}%
   {hermitian conjugation, division ring}%
\Index
   {hermitian matrix}%
   {hermitian matrix}%
\Index
   {hermitian metric on division ring}%
   {hermitian metric on division ring}%
\Index
   {hermitian scalar product in $D$\Hyph vector space}%
   {hermitian scalar product, vector space}%
\Index
   {hermitian scalar product on division ring}%
   {hermitian scalar product on division ring}%
\Index
   {highest common factor}%
   {highest common factor}%
\Index
   {holomorphic map}%
   {holomorphic map}%
\Index
   {holonomic coordinates of connection}%
   {holonomic coordinates of connection}%
\Index
   {holonomic coordinates of vector}%
   {vector holonomic coordinates}%
\Index
   {homogeneous bundle of fibered group}%
   {homogeneous bundle of fibered group}%
\Index
   {homogeneous linear geometric object}%
   {homogeneous linear geometric object}%
\Index
   {homogeneous map of degree $k$ over field $F$}%
   {homogeneous map of degree over field, D vector space}%
\Index
   {homogeneous polynomial of power $k$}%
   {homogeneous polynomial of power}%
\Index
   {homogeneous space}%
   {homogeneous space}%
\Index
   {homomorphic image}%
   {homomorphic image}%
\Index
   {homomorphism}%
   {homomorphism}%
\Index
   {homomorphism of fibered groups}%
   {homomorphism of fibered groups}%
\Index
   {homomorphism of fibered universal algebras}%
   {homomorphism of fibered universal algebras}%
\Index
   {horizontal component of vector}%
   {horizontal component of vector}%
\Index
   {horizontal subspace}%
   {horizontal subspace}%
\Index
   {horizontal vector}%
   {horizontal vector}%
\Index
   {hyperbolic cosine}%
   {hyperbolic cosine}%
\Index
   {hyperbolic sine}%
   {hyperbolic sine}%
\SetIndexSpace%
\Index
   {ideal of algebra}%
   {ideal of algebra}%
\Index
   {indefinite integral}%
   {indefinite integral}%
\Index
   {independent points}%
   {independent points}%
\Index
   {induction over diagram of representations}%
   {induction over diagram of representations}%
\Index
   {infinitesimal generator of representation}%
   {infinitesimal generator}%
\Index
   {infinitesimal generators of group Lie}%
   {infinitesimal generators of group Lie}%
\Index
   {integrable differential equation}%
   {integrable differential equation}%
\Index
   {integrable form}%
   {integrable form}%
\Index
   {integrable map}%
   {integrable map}%
\Index
   {integral of differential $1$\Hyph form along path}%
   {integral of differential 1 form along path}%
\Index
   {invariance principle}%
   {invariance principle}%
\Index
   {invariance principle in \drc vector space}%
   {invariance principle}%
\Index
   {invariance principle in tower of representations of universal algebras}%
   {invariance principle, tower of representations g}%
\Index
   {inverse fibered correspondence}%
   {inverse fibered correspondence}%
\Index
   {inverse reduced fibered correspondence}%
   {inverse reduced fibered correspondence}%
\Index
   {involution in quaternion algebra}%
   {involution, quaternion algebra}%
\Index
   {isomorphism}%
   {isomorphism}%
\Index
   {isomorphism of fibered $\Omega$\Hyph algebras}%
   {isomorphism of fibered Omega-algebras}%
\Index
   {isomorphism of repesentations of $\Omega$\Hyph algebra}%
   {isomorphism of repesentations of Omega algebra}%
\Index
   {isomorphism of vector spaces}%
   {isomorphism of vector spaces}%
\Index
   {isotropic vector}%
   {isotropic vector}%
\Index
   {Lebesgue integral}%
   {Lebesgue integral}%
\SetIndexSpace%
\Index
   {$(^j_i)$\hyph $\RCcirc$\Hyph quasideterminant}%
   {j i RCcirc-quasideterminant}%
\Index
   {the Jacobi matrix of map}%
   {Jacobi matrix of map}%
\Index
   {Jacobian complete system of differential equations}%
   {Jacobian complete system of differential equations}%
\Index
   {Jacobian complete system of \drv differential equations}%
   {Jacobian complete system of drc differential equations}%
\Index
   {$(ji)$\hyph quasideterminant}%
   {j i quasideterminant}%
\Index
   {the Jacobi\Hyph G\^ateaux matrix of map}%
   {Jacobi Gateaux matrix of map}%
\SetIndexSpace%
\Index
   {kernel of homomorphism}%
   {kernel of homomorphism}%
\Index
   {kernel of inefficiency of \Ts{G}representation}%
   {kernel of inefficiency of G* representation}%
\Index
   {kernel of inefficiency of representation of fibered group}%
   {kernel of inefficiency of representation of fibered group}%
\Index
   {kernel of inefficiency of representation of group}%
   {kernel of inefficiency of representation of group}%
\Index
   {kernel of linear map}%
   {kernel of linear map}%
\Index
   {kernel of map}%
   {kernel of map}%
\Index
   {Killing equation}%
   {Killing equation}%
\Index
   {Killing equation of second type}%
   {Killing equation second type}%
\Index
   {Killing vector of second type}%
   {Killing vector second type}%
\Index
   {Kronecker symbol}%
   {Kronecker symbol}%
\SetIndexSpace%
\Index
   {latitude}%
   {latitude}%
\Index
   {leading coefficient of polynomial}%
   {leading coefficient of polynomial}%
\Index
   {Lebesgue extension of measure}%
   {Lebesgue extension of measure}%
\Index
   {Lebesgue measurable set}%
   {Lebesgue measurable}%
\Index
   {Lebesgue measure}%
   {Lebesgue measure}%
\Index
   {left $A$\Hyph module}%
   {left A module}%
\Index
   {left $A_*$\Hyph vector space}%
   {left A subs vector space}%
\Index
   {left $A^*$\Hyph vector space}%
   {left A sups vector space}%
\Index
   {left $A$\Hyph vector space}%
   {left A vector space}%
\Index
   {left $A$\Hyph column space}%
   {left A-column space}%
\Index
   {left $A$\Hyph row space}%
   {left A-row space}%
\Index
   {left cofactor of entry of matrix}%
   {left cofactor, matrix}%
\Index
   {left $D$\hyph vector space of columns}%
   {left vector space of columns}%
\Index
   {left $D$\hyph vector space of rows}%
   {left vector space of rows}%
\Index
   {left defined Lie algebra of Lie group}%
   {left defined Lie algebra}%
\Index
   {left double cofactor of entry of matrix}%
   {left double cofactor}%
\Index
   {left fraction}%
   {left fraction}%
\Index
   {left ideal of algebra}%
   {left ideal of algebra}%
\Index
   {left invariant vector field}%
   {left invariant vector}%
\Index
   {left linear combination of columns}%
   {left linear combination of columns}%
\Index
   {left linear combination of rows}%
   {left linear combination of rows}%
\Index
   {left module}%
   {left module}%
\Index
   {left principal ideal}%
   {left principal ideal}%
\Index
   {left shift of module}%
   {left shift of module}%
\Index
   {left shift on fibered group}%
   {left shift, fibered group}%
\Index
   {left shift on group}%
   {left shift}%
\Index
   {left shift on group}%
   {left shift, group}%
\Index
   {left structural constant of Lie algebra}%
   {left structural constant of Lie algebra}%
\Index
   {left vector space}%
   {left vector space}%
\Index
   {left zero divisor}%
   {left zero divisor}%
\Index
   {left-ordered cycle notation of permutation}%
   {left-ordered cycle notation of permutation}%
\Index
   {left\Hyph side $A_1$\Hyph representation}%
   {left-side A representation}%
\Index
   {left\Hyph side product}%
   {left-side product}%
\Index
   {left-side product of map over scalar}%
   {left-side product of map over scalar}%
\Index
   {left\Hyph side product of vector over scalar}%
   {left-side product of vector over scalar}%
\Index
   {left-side representation}%
   {left-side representation}%
\Index
   {left-side representation of fibered $\Omega$\Hyph algebra}%
   {left-side representation of fibered Omega-algebra}%
\Index
   {left-side representation of $\Omega_1$\Hyph algebra $A$ in $\Omega_2$\Hyph algebra $M$}%
   {left-side representation of algebra}%
\Index
   {left-side transformation}%
   {left-side transformation}%
\Index
   {left-side transformation on bundle}%
   {left-side transformation of bundle}%
\Index
   {Lie algebra of Lie group}%
   {algebra Lie group Lie}%
\Index
   {Lie derivative}%
   {Lie derivative}%
\Index
   {Lie derivative of connection}%
   {Lie derivative of connection}%
\Index
   {Lie derivative of metric}%
   {Lie derivative of metric}%
\Index
   {Lie group basic operators}%
   {Lie group basic operators}%
\Index
   {lift of correspondence}%
   {lift of correspondence}%
\Index
   {lift of mapping}%
   {lift of map}%
\Index
   {limit of correspondence with respect to the filter}%
   {limit of correspondence with respect to the filter}%
\Index
   {limit of filter}%
   {limit of filter}%
\Index
   {limit of sequence}%
   {limit of sequence}%
\Index
   {limit set of filter}%
   {limit set of filter}%
\Index
   {linear combination}%
   {linear combination}%
\Index
   {linear functional}%
   {linear functional}%
\Index
   {linear \Ts{G}representation}%
   {linear G* representation}%
\Index
   {linear geometric object}%
   {linear geometric object}%
\Index
   {linear homogeneous equation}%
   {linear homogeneous equation}%
\Index
   {linear homomorphism}%
   {linear homomorphism}%
\Index
   {linear map}%
   {linear map}%
\Index
   {linear map generated by map}%
   {linear map generated by map}%
\Index
   {linear map of division ring}%
   {linear map of division ring}%
\Index
   {linear representation of group}%
   {linear representation of group}%
\Index
   {linear representation of Lie group}%
   {linear representation of Lie group}%
\Index
   {linear span}%
   {linear span, vector space}%
\Index
   {linear transformation group}%
   {linear transformation group}%
\Index
   {linear transformation of affine space}%
   {linear transformation, affine space}%
\Index
   {linearly dependent}%
   {linearly dependent}%
\Index
   {linearly dependent set}%
   {linearly dependent set}%
\Index
   {linearly dependent vector fields}%
   {linearly dependent vector fields}%
\Index
   {linearly independent set}%
   {linearly independent set}%
\Index
   {little group}%
   {little group}%
\Index
   {local reference frame}%
   {local reference frame}%
\Index
   {locally compact at point $p$ space}%
   {locally compact at point space}%
\Index
   {locally compact space}%
   {locally compact space}%
\Index
   {longitude}%
   {longitude}%
\Index
   {Lorentz transformation}%
   {Lorentz transformation}%
\SetIndexSpace%
\Index
   {$m$\Hyph dimensional parallelepiped}%
   {m dimensional parallelepiped}%
\Index
   {$m$\Hyph vector}%
   {m-vector}%
\Index
   {manifold with $D$\Hyph affine connections}%
   {manifold with D- affine connections}%
\Index
   {map continuous with respect to set of arguments}%
   {map continuous with respect to set of arguments}%
\Index
   {map differentiable in the G\^ateaux sense}%
   {map differentiable in Gateaux sense}%
\Index
   {map is compatible with operation}%
   {map is compatible with operation}%
\Index
   {map of conjugation}%
   {map of conjugation}%
\Index
   {map of $\gi n$ $D$\Hyph valued variables}%
   {map of n D valued variables}%
\Index
   {map of type $G$ on manifold}%
   {map of type G on manifold}%
\Index
   {map polylinear over finite dimensional algebras}%
   {map polylinear over finite dimensional algebras}%
\Index
   {map projective over commutative ring}%
   {map projective over commutative ring}%
\Index
   {mapping of rings polylinear over commutative ring}%
   {map polylinear over commutative ring, ring}%
\Index
   {mapping space}%
   {mapping space}%
\Index
   {matrix}%
   {matrix}%
\Index
   {matrix of antilinear homomorphism}%
   {matrix of antilinear homomorphism}%
\Index
   {matrix of bilinear function}%
   {matrix of bilinear function}%
\Index
   {matrix of endomorphisms of $\Omega$\Hyph algebra}%
   {matrix of endomorphisms of Omega algebra}%
\Index
   {matrix of fibered \Drc linear map}%
   {matrix of fibered drc linear map}%
\Index
   {matrix of linear homomorphism}%
   {matrix of linear homomorphism}%
\Index
   {matrix of linear map}%
   {matrix of linear map}%
\Index
   {matrix of linear maps}%
   {matrix of linear maps}%
\Index
   {matrix of maps}%
   {matrix of maps}%
\Index
   {matrix of quadratic map}%
   {matrix of quadratic map, division ring}%
\Index
   {Maxwell equation}%
   {Maxwell equation}%
\Index
   {measurable map}%
   {measurable map}%
\Index
   {measure}%
   {measure}%
\Index
   {method of successive differentiation}%
   {method of successive differentiation}%
\Index
   {metric tensor in Minkowski space}%
   {metric tensor, Minkowski space}%
\Index
   {metric-affine manifold}%
   {metric-affine manifold}%
\Index
   {Minkowski space}%
   {Minkowski space, Finsler}%
\Index
   {minor matrix}%
   {minor matrix}%
\Index
   {module over ring}%
   {module over ring}%
\Index
   {monomial of power $k$}%
   {monomial of power}%
\Index
   {monomorphism}%
   {monomorphism}%
\Index
   {morphism from diagram of representations into diagram of representations}%
   {morphism from diagram of representations into diagram of representations}%
\Index
   {morphism from tower of representations into tower of representations}%
   {morphism from tower of representations into tower of representations}%
\Index
   {morphism of fibered \Ts representations from $\Bundle F$ into $\Bundle G$}%
   {morphism of fibered representations from f into g}%
\Index
   {morphism of representation $f$}%
   {morphism of representation f}%
\Index
   {morphism of representations from $f$ into $g$}%
   {morphism of representations from f into g}%
\Index
   {morphism of representations of $\Omega_1$\Hyph algebra in $\Omega_2$\Hyph algebra}%
   {morphism of representations of Omega1 algebra in Omega2 algebra}%
\Index
   {morphism of \Ts representations of fibered $\Omega$\Hyph algebra}%
   {morphism of representations of fibered Omega algebra}%
\Index
   {motion of Minkowski space}%
   {motion, Minkowski space}%
\Index
   {movement on basis manifold}%
   {movement transformation}%
\Index
   {multiplicative map}%
   {multiplicative map}%
\Index
   {multiplicative $\Omega$\Hyph group}%
   {multiplicative Omega group}%
\SetIndexSpace%
\Index
   {$n$\Hyph ary fibered relation}%
   {fibered relation}%
\Index
   {$n$\Hyph ary operation on set}%
   {n-ary operation on set}%
\Index
   {natural homomorphism}%
   {natural homomorphism}%
\Index
   {neutral element of operation}%
   {neutral element of operation}%
\Index
   {nonmetricity}%
   {nonmetricity}%
\Index
   {nonsingular bilinear function}%
   {nonsingular bilinear function}%
\Index
   {nonsingular system of \rcd linear equations}%
   {nonsingular system of linear equations}%
\Index
   {nonsingular tensor}%
   {nonsingular tensor}%
\Index
   {nonsingular transformation}%
   {nonsingular transformation}%
\Index
   {norm in quaternion algebra}%
   {norm, quaternion algebra}%
\Index
   {norm of functional}%
   {norm of functional}%
\Index
   {norm of map}%
   {norm of map}%
\Index
   {norm of octonion}%
   {norm of octonion}%
\Index
   {norm of operation}%
   {norm of operation}%
\Index
   {norm of polylinear map}%
   {norm of polymap}%
\Index
   {norm of representation}%
   {norm of representation}%
\Index
   {norm on $D$\Hyph algebra}%
   {norm on D algebra}%
\Index
   {norm on $D$\Hyph vector space}%
   {norm on D vector space}%
\Index
   {norm on $D$\Hyph module}%
   {norm on D module}%
\Index
   {norm on $\Omega$\Hyph group}%
   {norm on Omega group}%
\Index
   {norm on ring}%
   {norm on ring}%
\Index
   {normal basis}%
   {normal basis}%
\Index
   {normed $D$\Hyph algebra}%
   {normed D algebra}%
\Index
   {normed $D$\Hyph module}%
   {normed D module}%
\Index
   {normed $D$\Hyph vector space}%
   {normed D vector space}%
\Index
   {normed $\Omega$\Hyph group}%
   {normed Omega group}%
\Index
   {normed ring}%
   {normed ring}%
\Index
   {not complete group}%
   {not complete group}%
\Index
   {not complete $\Omega$\Hyph algebra}%
   {not complete Omega algebra}%
\Index
   {nucleus of $D$\Hyph algebra $A$}%
   {nucleus of algebra}%
\SetIndexSpace%
\Index
   {octonion algebra}%
   {octonion algebra}%
\Index
   {open ball}%
   {open ball}%
\Index
   {open set}%
   {open set}%
\Index
   {operation on bundle}%
   {operation on bundle}%
\Index
   {operation on set}%
   {operation on set}%
\Index
   {operator domain}%
   {operator domain}%
\Index
   {opposite algebra to algebra $P$}%
   {opposite algebra}%
\Index
   {opposite fibered preordering}%
   {opposite fibered preordering}%
\Index
   {orbit of linear map}%
   {orbit of linear map}%
\Index
   {orbit of representation}%
   {orbit of representation}%
\Index
   {orbit of representation of fibered group}%
   {orbit of representation of fibered group}%
\Index
   {orbit of representation of group}%
   {orbit of representation of group}%
\Index
   {origin of coordinate system of affine space}%
   {origin of coordinate system of affine space}%
\Index
   {orthogonal basis in Minkowski space}%
   {orthogonal basis, Minkowski space}%
\Index
   {orthogonality in Minkowski space}%
   {Minkowski orthogonality}%
\Index
   {orthonormal basis}%
   {Orthonormal Basis, division ring}%
\Index
   {orthonormal basis in Minkowski space}%
   {orthonormal basis, Minkowski space}%
\Index
   {orthonornal basis}%
   {Orthonornal Basis}%
\Index
   {outer measure}%
   {outer measure}%
\SetIndexSpace%
\Index
   {parallel shift of affine space}%
   {parallel shift, affine space}%
\Index
   {parallelogram}%
   {parallelogram}%
\Index
   {parity of permutation}%
   {parity of permutation}%
\Index
   {partial derivative}%
   {partial derivative}%
\Index
   {partial derivative of second order}%
   {partial derivative of second order}%
\Index
   {partial linear map}%
   {partial linear map}%
\Index
   {passive \sT{G}representation}%
   {passive *G representation}%
\Index
   {passive representation}%
   {passive representation}%
\Index
   {passive representation in basis manifold}%
   {passive representation in basis manifold}%
\Index
   {passive representation of group $G(\Vector f)$ in basis manifold of tower of representations}%
   {passive representation in basis manifold, tower of representations}%
\Index
   {passive transformation of basis manifold}%
   {passive transformation of basis}%
\Index
   {passive transformation of the basis manifold of tower of representations}%
   {passive transformation of basis, tower of representations}%
\Index
   {passive transformation on basis manifold}%
   {passive transformation}%
\Index
   {passive transformation on the set of \rcd bases}%
   {passive transformation, vector space}%
\Index
   {permutability property of trace}%
   {permutability property of trace}%
\Index
   {permutation}%
   {permutation}%
\Index
   {pfaffian derivative}%
   {pfaffian derivative}%
\Index
   {polyadditive map}%
   {polyadditive map}%
\Index
   {polylinear map}%
   {polylinear map}%
\Index
   {polylinear skew symmetric map}%
   {polylinear map skew symmetric}%
\Index
   {polylinear symmetric map}%
   {polylinear map symmetric}%
\Index
   {polymorphism of representations}%
   {polymorphism of representations}%
\Index
   {polynomial}%
   {polynomial}%
\Index
   {polyvector}%
   {polyvector}%
\Index
   {potential energy}%
   {potential energy}%
\Index
   {power of measure}%
   {power of measure}%
\Index
   {prime $A$\Hyph number}%
   {prime A number}%
\Index
   {principal ideal}%
   {principal ideal}%
\Index
   {product in category}%
   {product in category}%
\Index
   {product of geometric object and constant}%
   {product of geometric object and constant}%
\Index
   {product of geometric object and constant in vector space}%
   {product of geometric object and constant, vector space}%
\Index
   {product of measures}%
   {product of measures}%
\Index
   {product of morphisms of diagram of representations}%
   {product of morphisms of diagram of representations}%
\Index
   {product of morphisms of representations of universal algebra}%
   {product of morphisms of representations of universal algebra}%
\Index
   {product of morphisms of tower of representations}%
   {product of morphisms of tower of representations}%
\Index
   {product of morphisms of \Ts representations of fibered $\Omega$\Hyph algebra}%
   {product of morphisms of representations of fibered Omega algebra}%
\Index
   {product of polynomials}%
   {product of polynomials}%
\Index
   {product of rings of sets}%
   {product of rings of sets}%
\Index
   {projection of bundle $\Bundle E$ along fiber $E$}%
   {projection of bundle along fiber}%
\Index
   {projective map is continuous in direction over field}%
   {projective map is continuous in direction over field}%
\Index
   {pseudo\Hyph Euclidean metric on division ring}%
   {pseudo-Euclidean metric on division ring}%
\Index
   {pseudo\Hyph Euclidean scalar product in $D$\Hyph vector space}%
   {pseudo-Euclidean scalar product, vector space}%
\Index
   {pseudo-Euclidean scalar product on division ring}%
   {pseudo-Euclidean scalar product on division ring}%
\SetIndexSpace%
\Index
   {quadratic equation}%
   {quadratic equation}%
\Index
   {quadratic form in division ring}%
   {quadratic form, division ring}%
\Index
   {quadratic map of division ring}%
   {Quadratic Map of Division Ring}%
\Index
   {quasi affine transformation on basis manifold}%
   {quasi affine transformation}%
\Index
   {quasi affine transformation on basis manifold}%
   {quasi affine drc transformation}%
\Index
   {quasi movement on basis manifold}%
   {quasi movement, division ring}%
\Index
   {quasi movement on basis manifold}%
   {quasi movement}%
\Index
   {quasibasis of diagram of representations}%
   {quasibasis of diagram of representations}%
\Index
   {quasibasis of representation}%
   {quasibasis of representation}%
\Index
   {quasiclosed ring of maps}%
   {quasiclosed ring of maps}%
\Index
   {quasideterminant}%
   {quasideterminant definition}%
\Index
   {quasiexponent}%
   {quasiexponent}%
\Index
   {quasimotion of Minkowski space}%
   {Quasimotion, Minkowski space}%
\Index
   {quaternion algebra}%
   {quaternion algebra}%
\Index
   {quaternion algebra $E$ over the field $F$}%
   {quaternion algebra over the field}%
\Index
   {quotient}%
   {quotient divided by}%
\Index
   {quotient bundle}%
   {quotient bundle}%
\SetIndexSpace%
\Index
   {$(\aUD{}ji)$\hyph \RC quasideterminant}%
   {j i RC-quasideterminant}%
\Index
   {\sups row of matrix}%
   {r row}%
\Index
   {$R$\Hyph module}%
   {R- module}%
\Index
   {$r$\hyph row of matrix}%
   {r-row}%
\Index
   {rank of Hermitian matrix by principal minors}%
   {rank of Hermitian matrix by principal minors}%
\Index
   {rank of quadratic map of division ring}%
   {rank of quadratic map, division ring}%
\Index
   {\RC eigencolumn}%
   {rc eigencolumn}%
\Index
   {\RC eigenrow}%
   {rc eigenrow}%
\Index
   {\RC eigenvalue}%
   {rc eigenvalue}%
\Index
   {\RC exponent}%
   {RC exponent}%
\Index
   {\RC inverse element of biring}%
   {rc-inverse element}%
\Index
   {\RC major minor matrix}%
   {RC-major minor}%
\Index
   {\RC matrix group}%
   {rc-matrix group}%
\Index
   {\RC nonsingular matrix}%
   {RC nonsingular matrix}%
\Index
   {\RC power}%
   {rc power}%
\Index
   {\RC product (product of row over column)}%
   {rc-product}%
\Index
   {$\RCcirc$\Hyph product of matrices of maps}%
   {rc product of matrices of maps}%
\Index
   {\RC quasideterminant}%
   {RC-quasideterminant}%
\Index
   {\RC rank of matrix}%
   {rc-rank of matrix}%
\Index
   {\RC singular matrix}%
   {RC singular matrix}%
\Index
   {$\RCcirc$\Hyph nonsingular matrix}%
   {RCcirc nonsingular matrix}%
\Index
   {$\RCcirc$\Hyph nonsingular system of additive equations}%
   {RCcirc nonsingular system of additive equations}%
\Index
   {$\RCcirc$\Hyph quasideterminant}%
   {RCcirc-quasideterminant definition}%
\Index
   {$\RCcirc$\Hyph singular matrix}%
   {RCcirc singular matrix}%
\Index
   {\rcd affine plane}%
   {rcd affine plane}%
\Index
   {\rcd affine space}%
   {rcd affine space}%
\Index
   {\rcd vector}%
   {rcd vector}%
\Index
   {\rcd vector space}%
   {rcd vector space}%
\Index
   {reduced Cartesian product of bundles}%
   {reduced Cartesian product of bundles}%
\Index
   {reduced Cartesian product of total spaces}%
   {reduced Cartesian product of total spaces}%
\Index
   {reduced fibered correspondence from $\Bundle{A}$ to $\Bundle B$}%
   {reduced fibered correspondence from A to B}%
\Index
   {reduced fibered correspondence in $\Bundle{A}$}%
   {reduced fibered correspondence in A}%
\Index
   {reduced morphism of representations}%
   {reduced morphism of representations}%
\Index
   {reduced polymorphism of representations}%
   {reduced polymorphism of representations}%
\Index
   {reduced quadratic equation}%
   {reduced quadratic equation}%
\Index
   {reducible biring}%
   {reducible biring}%
\Index
   {reference frame in event space}%
   {reference frame in event space}%
\Index
   {reference frame manifold}%
   {reference frame manifold}%
\Index
   {reflexive $2$\Hyph ary fibered relation}%
   {reflexive 2 ary fibered relation}%
\Index
   {reflexive correspondence}%
   {reflexive correspondence}%
\Index
   {regular endomorphism}%
   {regular endomorphism}%
\Index
   {regular endomorphism of tower of representations}%
   {regular endomorphism of tower of representations}%
\Index
   {regular quadratic map in division ring}%
   {regular quadratic map, division ring}%
\Index
   {relatively prime $A$\Hyph numbers}%
   {relatively prime A numbers}%
\Index
   {remainder of the division}%
   {remainder of the division}%
\Index
   {representation conjugated to representation}%
   {representation conjugated to representation}%
\Index
   {\Ts{A}representation in $\Omega_2$\Hyph algebra}%
   {A* representation of algebra}%
\Index
   {representation of group}%
   {representation of group}%
\Index
   {representation of $\Omega$\Hyph algebra in representation}%
   {representation of Omega algebra in representation}%
\Index
   {representation of $\Omega$\Hyph algebra in tower of representations}%
   {representation of Omega algebra in tower of representations}%
\Index
   {representation of $\Omega$\Hyph algebra $A$ in category $\mathcal B$}%
   {representation of Omega algebra in category}%
\Index
   {\sT{A}representation of $\Omega_1$\Hyph algebra $A$ in $\Omega_2$\Hyph algebra}%
   {*A representation of algebra}%
\Index
   {representation of $\Omega_1$\Hyph algebra $A$ in $\Omega_2$\Hyph algebra $M$}%
   {representation of algebra}%
\Index
   {representative of geometric object}%
   {representative of geometric object}%
\Index
   {representative of geometric object in \drc vector space}%
   {representative of geometric object, drc vector space}%
\Index
   {representative of geometric object in \rcd vector space}%
   {representative of geometric object, rcd vector space}%
\Index
   {representative of geometric object in tuple of $\VX\Omega$\Hyph algebras}%
   {representative of geometric object, tower of representations g}%
\Index
   {representative of geometric object in vector space}%
   {representative of geometric object, vector space}%
\Index
   {restriction of correspondence $\Phi$ to set $C$}%
   {restriction of correspondence}%
\Index
   {right $A_*$\Hyph vector space}%
   {right A subs vector space}%
\Index
   {right $A^*$\Hyph vector space}%
   {right A sups vector space}%
\Index
   {right $A$\Hyph vector space}%
   {right A vector space}%
\Index
   {right $A$\Hyph column space}%
   {right A-column space}%
\Index
   {right $A$\Hyph row space}%
   {right A-row space}%
\Index
   {right cofactor of entry of matrix}%
   {right cofactor, matrix}%
\Index
   {right $D$\Hyph module}%
   {right D module}%
\Index
   {right $D$\hyph vector space of columns}%
   {right vector space of columns}%
\Index
   {right $D$\hyph vector space of rows}%
   {right vector space of rows}%
\Index
   {right defined Lie algebra of Lie group}%
   {right defined Lie algebra}%
\Index
   {right double cofactor of entry of matrix}%
   {right double cofactor}%
\Index
   {right fraction}%
   {right fraction}%
\Index
   {right ideal of algebra}%
   {right ideal of algebra}%
\Index
   {right invariant vector field}%
   {right invariant vector}%
\Index
   {right linear combination of columns}%
   {right linear combination of columns}%
\Index
   {right linear combination of rows}%
   {right linear combination of rows}%
\Index
   {right module}%
   {right module}%
\Index
   {right module over $D$\Hyph algebra $A$}%
   {right module over algebra}%
\Index
   {right principal ideal}%
   {right principal ideal}%
\Index
   {right shift on group}%
   {right shift}%
\Index
   {right shift on group}%
   {right shift, group}%
\Index
   {right structural constant of Lie algebra}%
   {right structural constant of Lie algebra}%
\Index
   {right vector space}%
   {right vector space}%
\Index
   {right zero divisor}%
   {right zero divisor}%
\Index
   {right-ordered cycle notation of permutation}%
   {right-ordered cycle notation of permutation}%
\Index
   {right\Hyph side $A_1$\Hyph representation}%
   {right-side A representation}%
\Index
   {right\Hyph side product}%
   {right-side product}%
\Index
   {right\Hyph side product of vector over scalar}%
   {right-side product of vector over scalar}%
\Index
   {right-side representation}%
   {right-side representation}%
\Index
   {right-side representation of fibered $\Omega$\Hyph algebra}%
   {right-side representation of fibered Omega-algebra}%
\Index
   {right-side representation of $\Omega_1$\Hyph algebra $A$ in $\Omega_2$\Hyph algebra $M$}%
   {right-side representation of algebra}%
\Index
   {right-side transformation}%
   {right-side transformation}%
\Index
   {ring has characteristic $0$}%
   {ring has characteristic 0}%
\Index
   {ring has characteristic $p$}%
   {ring has characteristic p}%
\Index
   {ring of sets}%
   {ring of sets}%
\Index
   {ring of sets generated by semiring of sets}%
   {ring of sets generated by semiring}%
\Index
   {ring with conjugation}%
   {ring with conjugation}%
\Index
   {root of polynomial}%
   {root of polynomial}%
\Index
   {row $*D$\Hyph vector}%
   {row *D vector}%
\Index
   {row $D*$\Hyph vector}%
   {row D* vector}%
\Index
   {row determinant}%
   {row determinant}%
\Index
   {row vector}%
   {row vector}%
\SetIndexSpace%
\Index
   {$\star A$\Hyph module}%
   {starA-module}%
\Index
   {scalar algebra of algebra}%
   {scalar algebra of algebra}%
\Index
   {scalar algebra of ring}%
   {scalar algebra of ring}%
\Index
   {scalar of element of algebra}%
   {scalar of algebra}%
\Index
   {scalar of element of ring}%
   {scalar of ring}%
\Index
   {scalar potential}%
   {scalar potential}%
\Index
   {Schauder basis}%
   {Schauder basis}%
\Index
   {second axiom of countability}%
   {second axiom of countability}%
\Index
   {second Newton law}%
   {Second Newton law}%
\Index
   {section of bundle}%
   {section of bundle}%
\Index
   {semigroup}%
   {semigroup}%
\Index
   {semiring of sets}%
   {semiring of sets}%
\Index
   {sequence converges}%
   {sequence converges}%
\Index
   {sequence converges almost everywhere}%
   {converges almost everywhere}%
\Index
   {sequence converges uniformly}%
   {sequence converges uniformly}%
\Index
   {series converges normally}%
   {series converges normally}%
\Index
   {set admits operation}%
   {set admits operation}%
\Index
   {set is closed with respect to operation}%
   {set is closed with respect to operation}%
\Index
   {set is dense in set}%
   {dense in set}%
\Index
   {set of coordinates of representation}%
   {coordinate set of representation}%
\Index
   {set of invertible elements of algebra}%
   {set of invertible elements of algebra}%
\Index
   {set of $\Omega_2$\Hyph words of representation}%
   {word set of representation}%
\Index
   {set of tuples of coordinates of diagram of representations}%
   {coordinate set of diagram of representations}%
\Index
   {set of tuples of coordinates of tower of representations}%
   {coordinate set of tower of representations}%
\Index
   {set of tuples of $\Omega$\Hyph words}%
   {set of tuples of Omega words}%
\Index
   {set of tuples of $\Vector\Omega$\Hyph words of tower of representations}%
   {word set of tower of representations}%
\Index
   {set of zeros of algebra}%
   {set of zeros of algebra}%
\Index
   {simple map}%
   {simple map}%
\Index
   {simple polyvector}%
   {simple polyvector}%
\Index
   {simplex}%
   {simplex}%
\Index
   {sine}%
   {sine}%
\Index
   {single transitive representation of fibered $\Omega$\Hyph algebra}%
   {single transitive representation of fibered Omega-algebra}%
\Index
   {single transitive representation of group}%
   {single transitive representation of group}%
\Index
   {single transitive representation of $\Omega$\Hyph algebra $A$}%
   {single transitive representation of algebra}%
\Index
   {singular endomorphism}%
   {singular endomorphism}%
\Index
   {singular linear map}%
   {singular linear map}%
\Index
   {skew product of vectors}%
   {skew product of vectors}%
\Index
   {skew symmetric polylinear map}%
   {skew symmetric polylinear map}%
\Index
   {space of orbits of \Ts{G}representation}%
   {space of orbits of G* representation}%
\Index
   {space of orbits of left\Hyph side representation}%
   {space of orbits of left side representation}%
\Index
   {spacelike vector}%
   {spacelike vector}%
\Index
   {speed of deviation}%
   {speed of deviation}%
\Index
   {spherical coordinates}%
   {spherical coordinates}%
\Index
   {square root}%
   {square root}%
\Index
   {$(\mathcal S\RCstar,\mathcal T\RCstar)$\Hyph linear map of vector bundles}%
   {src trc linear map of vector bundles}%
\Index
   {($S\star$, $\star T$)\hyph bimodule}%
   {(Sstar,starT)-bimodule}%
\Index
   {stability group}%
   {stability group}%
\Index
   {stable set of representation}%
   {stable set of representation}%
\Index
   {standard component of derivative}%
   {standard component of derivative}%
\Index
   {standard component of the G\^ateaux derivative}%
   {standard component of Gateaux derivative}%
\Index
   {standard component of linear map}%
   {standard component of linear map}%
\Index
   {standard component of polylinear map}%
   {standard component of polylinear map}%
\Index
   {standard component of tensor}%
   {standard component of tensor}%
\Index
   {standard component over field $F$ of bilitnear map $f$}%
   {standard component of bilinear map, division ring}%
\Index
   {standard coordinates of basis}%
   {standard coordinates of basis}%
\Index
   {standard coordinates of basis}%
   {standard coordinates of basis}%
\Index
   {standard representation of the derivative}%
   {derivative, standard representation}%
\Index
   {standard representation of the G\^ateaux derivative}%
   {Gateaux derivative, standard representation}%
\Index
   {standard representation of linear map}%
   {linear map, standard representation}%
\Index
   {standard representation of matrix}%
   {Standard representation}%
\Index
   {standard representation of polylinear map}%
   {polylinear map, standard representation}%
\Index
   {standard representation of quadratic map of division ring over field $F$}%
   {quadratic map, standard representation, division ring}%
\Index
   {standard representation over field $F$ of bilinear map of division ring}%
   {bilinear map, standard representation, division ring}%
\Index
   {$\star R$\hyph module}%
   {starR-module}%
\Index
   {$\star D$\hyph product of vector over scalar}%
   {starD product of vector over scalar, vector space}%
\Index
   {starlike set}%
   {starlike set}%
\Index
   {\sT representation of fibered group}%
   {starT representation of fibered group}%
\Index
   {\sT representation of fibered group}%
   {starT representation of fibered group}%
\Index
   {\sT representation of fibered $\Omega$\Hyph algebra}%
   {starT representation of fibered Omega-algebra}%
\Index
   {\sT shift on fibered group}%
   {starT shift, fibered group}%
\Index
   {\sT transformation on bundle}%
   {starT transformation of bundle}%
\Index
   {structural constants}%
   {structural constants}%
\Index
   {subalgebra of $\Omega$\Hyph algebra}%
   {subalgebra of Omega-algebra}%
\Index
   {subbundle}%
   {subbundle}%
\Index
   {subbundle of $\mathcal D\star$\hyph vector space}%
   {subbundle of Dstar vector bundle}%
\Index
   {subgroup of $\Omega$\Hyph group}%
   {subgroup of Omega group}%
\Index
   {submodule}%
   {submodule}%
\Index
   {submodule generated by set}%
   {submodule generated by set}%
\Index
   {subrepresentation}%
   {subrepresentation}%
\Index
   {subrepresentation generated by set $X$}%
   {subrepresentation generated by set}%
\Index
   {subrepresentation of representation}%
   {subrepresentation of representation}%
\Index
   {sum of geometric objects in vector space}%
   {sum of geometric objects, vector space}%
\Index
   {sum of geometric objects}%
   {sum of geometric objects}%
\Index
   {sum of maps}%
   {sum of maps}%
\Index
   {sum of polynomials}%
   {sum of polynomials}%
\Index
   {superposition of coordinates}%
   {superposition of coordinates,}%
\Index
   {superposition of coordinates of the tower of representations $\Vector f$ and the element $\VX a$}%
   {superposition of coordinates, tower of representations}%
\Index
   {symmetric $2$\Hyph ary fibered relation}%
   {symmetric 2 ary fibered relation}%
\Index
   {symmetric bilinear map of $D$\Hyph vector space to division ring}%
   {symmetric bilinear map, vector space to division ring}%
\Index
   {symmetric correspondence}%
   {symmetric correspondence}%
\Index
   {symmetric polylinear map}%
   {symmetric polylinear map}%
\Index
   {symmetric polylinear mapping into associative algebra}%
   {polylinear map symmetric, associative algebra}%
\Index
   {symmetrization of polylinear map}%
   {symmetrization of polylinear map}%
\Index
   {symmetry group}%
   {symmetry group}%
\Index
   {symmetry group}%
   {SymmetryGroup}%
\Index
   {synchronization of reference frame}%
   {synchronization of reference frame}%
\Index
   {system of additive equations}%
   {system of additive equations}%
\Index
   {system of \drc linear equations}%
   {system of drc linear equations}%
\Index
   {system of linear equations}%
   {system of linear equations}%
\Index
   {system of \rcd linear equations}%
   {system of rcd linear equations}%
\SetIndexSpace%
\Index
   {$T_1$\Hyph space}%
   {T1 space}%
\Index
   {Taylor polynomial}%
   {Taylor polynomial, division ring}%
\Index
   {Taylor series}%
   {Taylor series, division ring}%
\Index
   {tensor inverse to tensor}%
   {inverse tensor}%
\Index
   {tensor power}%
   {tensor power}%
\Index
   {tensor product}%
   {tensor product}%
\Index
   {the Fr\'echet \Ds derivative of map $f$ of division ring $D$ at point $x$}%
   {Frechet Dstar derivative of map, division ring}%
\Index
   {timelike vector}%
   {timelike vector}%
\Index
   {topological $D$\Hyph vector space}%
   {topological D vector space}%
\Index
   {topological $D$\Hyph algebra}%
   {topological D algebra}%
\Index
   {topological division ring}%
   {topological division ring}%
\Index
   {topological ring}%
   {topological ring}%
\Index
   {torsion form}%
   {torsion form}%
\Index
   {torsion tensor}%
   {torsion tensor}%
\Index
   {tower of bundles}%
   {tower of bundles}%
\Index
   {tower of effective representations}%
   {tower of effective representations}%
\Index
   {tower of representations of $\Omega$\Hyph algebras}%
   {tower of representations of algebras}%
\Index
   {tower of subrepresentations}%
   {tower of subrepresentations}%
\Index
   {tower of subrepresentations of tower of representations $\Vector f$ generated by tuple of sets $\VX X$}%
   {subrepresentation generated by tuple of sets}%
\Index
   {trace of quaternion}%
   {trace, quaternion algebra}%
\Index
   {transformation coordinated with equivalence}%
   {transformation coordinated with equivalence}%
\Index
   {transformation of universal algebra}%
   {transformation of universal algebra}%
\Index
   {transformation on bundle}%
   {transformation of bundle}%
\Index
   {transitive $2$\Hyph ary fibered relation}%
   {transitive 2 ary fibered relation}%
\Index
   {transitive correspondence}%
   {transitive correspondence}%
\Index
   {transitive representation of fibered $\Omega$\Hyph algebra}%
   {transitive representation of fibered Omega-algebra}%
\Index
   {transitive representation of group}%
   {transitive representation of group}%
\Index
   {transitive representation of $\Omega$\Hyph algebra $A$}%
   {transitive representation of algebra}%
\Index
   {\Ts representation of fibered group}%
   {Tstar representation of fibered group}%
\Index
   {\Ts representation of fibered $\Omega$\Hyph algebra}%
   {Tstar representation of fibered Omega-algebra}%
\Index
   {tuple of equivalence generated by tower of representations $\Vector f$}%
   {tuple of equivalence of tower of representations}%
\Index
   {tuple of generating sets of tower of representations}%
   {tuple of generating sets of tower of representations}%
\Index
   {tuple of $\Omega$\Hyph words}%
   {tuple of Omega words}%
\Index
   {tuple of $\Vector{\Omega}$\Hyph words of element of tower of representations relative to tuple of generating sets}%
   {tuple of words relative to tuple of generating sets, tower of representations}%
\Index
   {tuple of stable sets of diagram of representations}%
   {tuple of stable sets of diagram of representations}%
\Index
   {tuple of stable sets of tower of representation}%
   {tuple of stable sets of tower of representations}%
\Index
   {twin representations}%
   {twin representations}%
\Index
   {twin representations of division ring}%
   {twin representations of division ring}%
\Index
   {twin representations of fibered group}%
   {twin representations of fibered group}%
\Index
   {twin representations of group}%
   {twin representations of group}%
\SetIndexSpace%
\Index
   {unit interval}%
   {unit interval}%
\Index
   {unit of ring of sets}%
   {unit of ring of sets}%
\Index
   {unit sphere in $D$\Hyph algebra}%
   {unit sphere in algebra}%
\Index
   {unit sphere in division ring}%
   {unit sphere in division ring}%
\Index
   {unit vector}%
   {unit vector}%
\Index
   {unital extension}%
   {unital extension}%
\Index
   {unital ring}%
   {unital ring}%
\Index
   {unitarity law}%
   {unitarity law}%
\Index
   {universal algebra}%
   {universal algebra}%
\Index
   {universally attracting object of category}%
   {universally attracting}%
\Index
   {universally repelling  object of category}%
   {universally repelling}%
\SetIndexSpace%
\Index
   {basis for vector  bundle}%
   {basis, vector bundle}%
\Index
   {valued division ring}%
   {valued division ring}%
\Index
   {vector}%
   {vector}%
\Index
   {vector $*A$\Hyph space}%
   {*A-vector space}%
\Index
   {vector bundle}%
   {vector bundle}%
\Index
   {vector module of algebra}%
   {vector module of algebra}%
\Index
   {vector module of ring}%
   {vector module of ring}%
\Index
   {vector of element of algebra}%
   {vector of algebra}%
\Index
   {vector of element of ring}%
   {vector of ring}%
\Index
   {vector potential}%
   {vector potential}%
\Index
   {vector space}%
   {vector space}%
\Index
   {vector space type}%
   {vector space type}%
\Index
   {vertical component of vector}%
   {vertical component of vector}%
\Index
   {vertical subspace}%
   {vertical subspace}%
\Index
   {vertical vector}%
   {vertical vector}%
\SetIndexSpace%
\Index
   {zero divisor}%
   {zero divisor}%
\SetIndexSpace%
\Index
   {$\mu$\Hyph measurable map}%
   {mu measurable map}%
\SetIndexSpace%
\Index
   {$\Omega$\Hyph algebra}%
   {Omega-algebra}%
\Index
   {$\Omega$\Hyph group}%
   {Omega group}%
\Index
   {$\Omega$\Hyph groupoid}%
   {Omega groupoid}%
\Index
   {$\Omega$\Hyph linear mapping}%
   {Omega linear map}%
\Index
   {\(\Omega\)\Hyph ring}%
   {Omega ring}%
\Index
   {$\Omega_2$\Hyph word of element of representation relative to generating set}%
   {word of element relative to generating set, representation}%
\SetIndexSpace%
\Index
   {$\sigma$\Hyph algebra of sets}%
   {sigma algebra of sets}%
\Index
   {$\sigma$\Hyph ring of sets}%
   {sigma ring of sets}%
\Index
   {\(\sigma\)\Hyph additive measure}%
   {sigma-additive measure}%

\CloseIndex

%% file: Symbol.English.tex
\def\indexname{Special Symbols and Notations}
\OpenIndex

\SetIndexSpace
\Symb%
   {direct sum}%
   {direct sum}%
   {0}{0}%
\Symb%
   {unit interval}%
   {unit interval}%
   {0}{0}%

\SetIndexSpace
\Symb%
   {set of vectors whose expansion relative to the basis $\Basis e$ converges normally}%
   {A plus Schauder}%
   {A}{0}%
\Symb%
   {active representation in basis manifold}%
   {active representation in basis manifold}%
   {A}{0}%
\Symb%
   {$A$\Hyph algebra of polynomials over $D$\Hyph algebra $A$}%
   {algebra of polynomials over algebra}%
   {A}{0}%
\Symb%
   {algebra of polynomials over $D$\Hyph algebra $A$}%
   {algebra of polynomials over D algebra}%
   {A}{0}%
\Symb%
   {algebra of rational mappings of algebra $A$}%
   {algebra of rational mappings of algebra}%
   {A}{0}%
\Symb%
   {affine space}%
   {An}%
   {A}{0}%
\Symb%
   {associator of $D$\Hyph algebra}%
   {associator of algebra}%
   {A}{0}%
\Symb%
   {category of left-side representations of $\Omega_1$\Hyph algebra $A$}%
   {category of left-side representations of Omega1 algebra}%
   {A}{0}%
\Symb%
   {category of representations}%
   {category of representations}%
   {A}{0}%
\Symb%
   {commutator of $D$\Hyph algebra}%
   {commutator of algebra}%
   {A}{0}%
\Symb%
   {component of linear map}%
   {component of linear map, vector}%
   {A}{0}%
\Symb%
   {component $p$ of polylinear mapping $\Vector A$}%
   {component of polyadditive map, D vector space}%
   {A}{0}%
\Symb%
   {component of polylinear map}%
   {component of polylinear map, vector}%
   {A}{0}%
\Symb%
   {conjugated $D$\Hyph  module}%
   {conjugated D module}%
   {A}{0}%
\Symb%
   {coordinates of associator}%
   {coordinates of associator}%
   {A}{0}%
\Symb%
   {\CR power of element $A$ of biring}%
   {cr power}%
   {A}{0}%
\Symb%
   {\crd vector}%
   {crd vector}%
   {A}{0}%
\Symb%
   {\CR inverse element of biring}%
   {cr-inverse element}%
   {A}{0}%
\Symb%
   {\CR product}%
   {cr-product}%
   {A}{0}%
\Symb%
   {\dcr vector}%
   {dcr vector}%
   {A}{0}%
\Symb%
   {derivative of left shift}%
   {derivative of left shift}%
   {A}{0}%
\Symb%
   {derivative of left shift in $1$\Hyph parameter Lie group}%
   {derivative of left shift, 1-Parameter Group}%
   {A}{0}%
\Symb%
   {derivative of left shift in $1$\Hyph parameter Lie D group}%
   {derivative of left shift, 1-Parameter Group, algebra}%
   {A}{0}%
\Symb%
   {derivative of right shift}%
   {derivative of right shift}%
   {A}{0}%
\Symb%
   {derivative of right shift in $1$\Hyph parameter Lie group}%
   {derivative of right shift, 1-Parameter Group}%
   {A}{0}%
\Symb%
   {derivative of right shift in $1$\Hyph parameter Lie D group}%
   {derivative of right shift, 1-Parameter Group, algebra}%
   {A}{0}%
\Symb%
   {derivative of left shift}%
   {derivative of Tstar shift}%
   {A}{0}%
\Symb%
   {\drc vector}%
   {drc vector}%
   {A}{0}%
\Symb%
   {coordinates of vector $a$ relative to Hamel basis}%
   {Hamel basis, coordinates}%
   {A}{0}%
\Symb%
   {hermitian conjugation in division ring}%
   {hermitian conjugation, division ring}%
   {A}{0}%
\Symb%
   {tensor inverse to tensor $a$}%
   {inverse tensor}%
   {A}{0}%
\Symb%
   {isomorphic}%
   {isomorphic}%
   {A}{0}%
\Symb%
   {$(^j_i)$\hyph\CR quasideterminant}%
   {j i CR quasideterminant definition}%
   {A}{0}%
\Symb%
   {$(ji)$\hyph quasideterminant of matrix $\bfA$}%
   {j i quasideterminant definition}%
   {A}{0}%
\Symb%
   {$(^j_i)$\hyph $\RCcirc$\Hyph quasideterminant}%
   {j i RCcirc-quasideterminant definition}%
   {A}{0}%
\Symb%
   {$(^j_i)$\hyph \RC quasideterminant}%
   {j i RC-quasideterminant definition}%
   {A}{0}%
\Symb%
   {left fraction}%
   {left fraction}%
   {A}{0}%
\Symb%
   {left principal ideal}%
   {left principal ideal}%
   {A}{0}%
\Symb%
   {left shift in $D$\Hyph algebra}%
   {left shift, D algebra}%
   {A}{0}%
\Symb%
   {linear combination}%
   {linear combination}%
   {A}{0}%
\Symb%
   {little group}%
   {little group}%
   {A}{0}%
\Symb%
   {transformation of matrix}%
   {matrix, replacing its column}%
   {A}{0}%
\Symb%
   {transformation of matrix}%
   {matrix, replacing its row}%
   {A}{0}%
\Symb%
   {minor matrix}%
   {minor matrix}%
   {A}{0}%
\Symb%
   {$A$\Hyph module of homogeneous polynomials over $D$\Hyph algebra $A$}%
   {module of homogeneous polynomials over algebra}%
   {A}{0}%
\Symb%
   {norm on $D$\Hyph module}%
   {norm on D module}%
   {A}{0}%
\Symb%
   {$\Omega$\Hyph algebra}%
   {Omega-algebra}%
   {A}{0}%
\Symb%
   {opposite algebra to algebra $A$}%
   {opposite algebra}%
   {A}{0}%
\Symb%
   {orbit of linear map}%
   {orbit of linear map}%
   {A}{0}%
\Symb%
   {derivative}%
   {overline nabla_l, definition 2}%
   {A}{0}%
\Symb%
   {partial linear map}%
   {partial linear map}%
   {A}{0}%
\Symb%
   {principal ideal}%
   {principal ideal}%
   {A}{0}%
\Symb%
   {quasideterminant of matrix $\bfA$}%
   {quasideterminant definition}%
   {A}{0}%
\Symb%
   {\RC power of element $A$ of biring}%
   {rc power}%
   {A}{0}%
\Symb%
   {$\RCcirc$\Hyph quasideterminant}%
   {RCcirc-quasideterminant definition}%
   {A}{0}%
\Symb%
   {\rcd vector}%
   {rcd vector}%
   {A}{0}%
\Symb%
   {\RC inverse element of biring}%
   {rc-inverse element}%
   {A}{0}%
\Symb%
   {\RC product}%
   {rc-product}%
   {A}{0}%
\Symb%
   {\RC quasideterminant}%
   {RC-quasideterminant definition}%
   {A}{0}%
\Symb%
   {right principal ideal}%
   {right principal ideal}%
   {A}{0}%
\Symb%
   {right shift in $D$\Hyph algebra}%
   {right shift, D algebra}%
   {A}{0}%
\Symb%
   {coordinates of vector $a$ relative to Schauder basis}%
   {Schauder basis, coordinates}%
   {A}{0}%
\Symb%
   {set of additive maps}%
   {set additive maps}%
   {A}{0}%
\Symb%
   {set of invertible elements of algebra $A$}%
   {set of invertible elements of algebra}%
   {A}{0}%
\Symb%
   {set of vectors generated by vector $v$}%
   {set of vectors generated by vector}%
   {A}{0}%
\Symb%
   {set of zeros of algebra $A$}%
   {set of zeros of algebra}%
   {A}{0}%
\Symb%
   {set of polylinear maps of rings $R_1$, ..., $R_n$ into module $S$}%
   {set polylinear maps, ring}%
   {A}{0}%
\Symb%
   {simplex}%
   {simplex}%
   {A}{0}%
\Symb%
   {skew product of vectors $\Vector a_1$, ..., $\Vector a_m$}%
   {skew product of vectors}%
   {A}{0}%
\Symb%
   {space of orbits of left\Hyph side representation}%
   {space of orbits of left side representation}%
   {A}{0}%
\Symb%
   {space of orbits of representation}%
   {space of orbits of representation}%
   {A}{0}%
\Symb%
   {space of orbits of effective right\Hyph side representation}%
   {space of orbits of right-side representation}%
   {A}{0}%
\Symb%
   {square root}%
   {square root}%
   {A}{0}%
\Symb%
   {stability group}%
   {stability group}%
   {A}{0}%
\Symb%
   {\sT shift}%
   {starT shift, fibered group}%
   {A}{0}%
\Symb%
   {tensor power of algebra $A$}%
   {tensor power of algebra}%
   {A}{0}%
\Symb%
   {anholonomic coordinates of vector}%
   {vector anholonomic coordinates}%
   {A}{0}%
\Symb%
   {holonomic coordinates of vector}%
   {vector holonomic coordinates}%
   {A}{0}%

\SetIndexSpace
\Symb%
   {basis manifold}%
   {basis manifold}%
   {B}{0}%
\Symb%
   {basis manifold of \rcd vector space $\Vector V$}%
   {basis manifold of rcd vector space}%
   {B}{0}%
\Symb%
   {basis manifold of vector space}%
   {basis manifold of vector space}%
   {B}{0}%
\Symb%
   {basis manifold of tower of representations $\Vector f$}%
   {basis manifold tower of representations}%
   {B}{0}%
\Symb%
   {basis manifold of affine space}%
   {Basis Manifold, Affine Space}%
   {B}{0}%
\Symb%
   {basis manifold of central affine space}%
   {BCAn}%
   {B}{0}%
\Symb%
   {basis manifold of Euclid space}%
   {BEn}%
   {B}{0}%
\Symb%
   {Borel algebra}%
   {Borel algebra}%
   {B}{0}%
\Symb%
   {canonical remainder of the division}%
   {canonical remainder of the division}%
   {B}{0}%
\Symb%
   {Cartesian power}%
   {Cartesian power}%
   {B}{0}%
\Symb%
   {Cartesian power $\Bundle A$ of bundle $\Bundle B$}%
   {Cartesian power A of bundle B}%
   {B}{0}%
\Symb%
   {Cartesian power $A$ of set $B$}%
   {Cartesian power of set}%
   {B}{0}%
\Symb%
   {closed ball}%
   {closed ball}%
   {B}{0}%
\Symb%
   {closure of set}%
   {closure of set}%
   {B}{0}%
\Symb%
   {coproduct in category}%
   {coproduct in category}%
   {B}{0}%
\Symb%
   {basis manifold of central affine space}%
   {FCAn}%
   {B}{0}%
\Symb%
   {basis manifold of Euclid space}%
   {FEn}%
   {B}{0}%
\Symb%
   {lattice of subrepresentations}%
   {lattice of subrepresentations}%
   {B}{0}%
\Symb%
   {lattice of towers of subrepresentations of tower of representations $\Vector f$}%
   {lattice of subrepresentations, tower of representations}%
   {B}{0}%
\Symb%
   {open ball}%
   {open ball}%
   {B}{0}%
\Symb%
   {product in category}%
   {product in category}%
   {B}{0}%
\Symb%
   {right fraction}%
   {right fraction}%
   {B}{0}%
\Symb%
   {tensor power of representation}%
   {tensor power of representation}%
   {B}{0}%

\SetIndexSpace
\Symb%
   {$\sigma$\Hyph algebra of sets measurable with respect to measure $\mu$}%
   {algebra of sets measurable with respect to measure}%
   {C}{0}%
\Symb%
   {central affine space}%
   {CAn}%
   {C}{0}%
\Symb%
   {central affine space}%
   {central affine space}%
   {C}{0}%
\Symb%
   {continuity class}%
   {class Cn}%
   {C}{0}%
\Symb%
   {$j$th column determinant of matrix $\bfA$}%
   {column determinant}%
   {C}{0}%
\Symb%
   {cosine}%
   {cosine}%
   {C}{0}%
\Symb%
   {$\CRcirc$\Hyph product of matrices of maps}%
   {cr product of matrices of maps}%
   {C}{0}%
\Symb%
   {hyperbolic cosine}%
   {hyperbolic cosine}%
   {C}{0}%
\Symb%
   {left structural constant of Lie algebra}%
   {left structural constant of Lie algebra}%
   {C}{0}%
\Symb%
   {right structural constant of Lie algebra}%
   {right structural constant of Lie algebra}%
   {C}{0}%
\Symb%
   {set of continuous multivariable maps}%
   {set continuous multivariable maps}%
   {C}{0}%
\Symb%
   {structural constants}%
   {structural constants}%
   {C}{0}%

\SetIndexSpace
\Symb%
   {basis vector of representation of Lie group over algebra $A$}%
   {basis vector of representation of Lie group over algebra A}%
   {D}{0}%
\Symb%
   {coordinates of basis vector of representation of Lie group over algebra $A$}%
   {basis vector of representation of Lie group over algebra A, coordinates}%
   {D}{0}%
\Symb%
   {component of derivative of map $f(x)$}%
   {component of derivative}%
   {D}{0}%
\Symb%
   {component of derivative of second order of map $f(x)$}%
   {component of derivative of Second Order}%
   {D}{0}%
\Symb%
   {component of the G\^ateaux derivative of map $f(x)$}%
   {component of Gateaux derivative}%
   {D}{0}%
\Symb%
   {component of the G\^ateaux derivative of map $f(x)$}%
   {component of Gateaux derivative of map, D vector space, short}%
   {D}{0}%
\Symb%
   {component of the G\^ateaux derivative of second order of map $f(x)$}%
   {component of Gateaux derivative of Second Order}%
   {D}{0}%
\Symb%
   {component of the G\^ateaux derivative of second order of map $f(x)$}%
   {component of Gateaux derivative of Second Order, D vector space}%
   {D}{0}%
\Symb%
   {component of the G\^ateaux derivative of map $f(x)$}%
   {component of Gateaux derivative, vector space}%
   {D}{0}%
\Symb%
   {conjugation in algebra}%
   {conjugation in algebra}%
   {D}{0}%
\Symb%
   {conjugation in ring}%
   {conjugation in ring}%
   {D}{0}%
\Symb%
   {coordinate \rcd vector space}%
   {coordinate rcd vector space}%
   {D}{0}%
\Symb%
   {coordinate reference frame}%
   {coordinate reference frame, extensive definition}%
   {D}{0}%
\Symb%
   {coordinate vector bundle}%
   {coordinate vector bundle}%
   {D}{0}%
\Symb%
   {derivative of map $f$}%
   {derivative of map}%
   {D}{0}%
\Symb%
   {derivative of map $f$}%
   {derivative of map inline}%
   {D}{0}%
\Symb%
   {derivative of order $n$}%
   {derivative of Order n}%
   {D}{0}%
\Symb%
   {derivative of order $n$}%
   {derivative of Order n inline}%
   {D}{0}%
\Symb%
   {derivative of second order}%
   {derivative of Second Order}%
   {D}{0}%
\Symb%
   {derivative of second order}%
   {derivative of Second Order inline}%
   {D}{0}%
\Symb%
   {diagonal in bundle $\Bundle A$}%
   {diagonal in bundle, 1}%
   {D}{0}%
\Symb%
   {differential of independent variable}%
   {differential of independent variable}%
   {D}{0}%
\Symb%
   {differential of map $f$}%
   {differential of map}%
   {D}{0}%
\Symb%
   {direct product of division rings $D_1$, ..., $D_n$}%
   {direct product of division rings, 1 n}%
   {D}{0}%
\Symb%
   {double determinant of matrix $\bfA$}%
   {double determinant}%
   {D}{0}%
\Symb%
   {exterior differential}%
   {exterior differential}%
   {D}{0}%
\Symb%
   {the Fr\'echet \Ds derivative of map $f$ of division ring}%
   {Frechet Dstar derivative of map, division ring}%
   {D}{0}%
\Symb%
   {the G\^ateaux \dcr derivative of map $f$ of $D$\Hyph vector space $V$ to $D$\Hyph vector space $W$}%
   {Gateaux dcr derivative of map, D vector space}%
   {D}{0}%
\Symb%
   {the G\^ateaux derivative of map $f$}%
   {Gateaux derivative of map}%
   {D}{0}%
\Symb%
   {the G\^ateaux derivative of map $f$}%
   {Gateaux derivative of map, fraction}%
   {D}{0}%
\Symb%
   {the G\^ateaux derivative of order $n$}%
   {Gateaux derivative of Order n}%
   {D}{0}%
\Symb%
   {the G\^ateaux derivative of order $n$ of map $f$ of division ring}%
   {Gateaux derivative of Order n, division ring}%
   {D}{0}%
\Symb%
   {the G\^ateaux derivative of order $n$ of map $f$ of algebra}%
   {Gateaux derivative of Order n, fraction, algebra}%
   {D}{0}%
\Symb%
   {the G\^ateaux derivative of order $n$ of map $f$ of division ring}%
   {Gateaux derivative of Order n, fraction, division ring}%
   {D}{0}%
\Symb%
   {the G\^ateaux derivative of second order}%
   {Gateaux derivative of Second Order}%
   {D}{0}%
\Symb%
   {the G\^ateaux derivative of second order of mapping $f$ of algebra}%
   {Gateaux derivative of Second Order, fraction, algebra}%
   {D}{0}%
\Symb%
   {the G\^ateaux derivative of second order of map $f$ of division ring}%
   {Gateaux derivative of Second Order, fraction, division ring}%
   {D}{0}%
\Symb%
   {the G\^ateaux differential of map $f$}%
   {Gateaux differential of map, vector}%
   {D}{0}%
\Symb%
   {the G\^ateaux \Ds derivative of map $f$ of division ring $D$}%
   {Gateaux Dstar derivative of map, division ring}%
   {D}{0}%
\Symb%
   {the G\^ateaux Jacobian of map of $D$\Hyph vector space}%
   {Gateaux Jacobian of map, D vector space}%
   {D}{0}%
\Symb%
   {the G\^ateaux partial \dcr derivative of map $f^{\gi b}$ with respect to variable $v^{\gi a}$}%
   {Gateaux partial dcr derivative of map, 1, D vector space}%
   {D}{0}%
\Symb%
   {the G\^ateaux partial \dcr derivative of map $f^{\gi b}$ with respect to variable $v^{\gi a}$}%
   {Gateaux partial dcr derivative of map, 2, D vector space}%
   {D}{0}%
\Symb%
   {the G\^ateaux partial \dcr derivative of map $f^{\gi b}$ with respect to variable $v^{\gi a}$}%
   {Gateaux partial dcr derivative of map, 3, D vector space}%
   {D}{0}%
\Symb%
   {the G\^ateaux partial derivative}%
   {Gateaux partial derivative}%
   {D}{0}%
\Symb%
   {the G\^ateaux mixed partial derivative}%
   {Gateaux partial derivative of Second Order}%
   {D}{0}%
\Symb%
   {the G\^ateaux partial \rcd derivative of map $f^{\gi b}$ with respect to variable $x^{\gi a}$}%
   {Gateaux partial rcd derivative of map, 1, D vector space}%
   {D}{0}%
\Symb%
   {the G\^ateaux partial \rcd derivative of map $f^{\gi b}$ with respect to variable $x^{\gi a}$}%
   {Gateaux partial rcd derivative of map, 2, D vector space}%
   {D}{0}%
\Symb%
   {the G\^ateaux partial \rcd derivative of map $f^{\gi b}$ with respect to variable $x^{\gi a}$}%
   {Gateaux partial rcd derivative of map, 3, D vector space}%
   {D}{0}%
\Symb%
   {the G\^ateaux \rcd derivative of map $f$ of $D$\hyph vector space $V$ to $D$\hyph vector space $W$}%
   {Gateaux rcd derivative of map, D vector space}%
   {D}{0}%
\Symb%
   {the G\^ateaux \sD derivative of map $f$ of division ring $D$}%
   {Gateaux starD derivative of map, division ring}%
   {D}{0}%
\Symb%
   {Jacobi matrix of map}%
   {Jacobi matrix of map}%
   {D}{0}%
\Symb%
   {matrices vector space}%
   {matrices vector space}%
   {D}{0}%
\Symb%
   {Cartan derivative}%
   {overbrace D}%
   {D}{0}%
\Symb%
   {derivative}%
   {overline D}%
   {D}{0}%
\Symb%
   {partial derivative}%
   {partial derivative}%
   {D}{0}%
\Symb%
   {partial derivative of second order}%
   {partial derivative of second order}%
   {D}{0}%
\Symb%
   {derivative $e_{(k)}$}%
   {partial(k)}%
   {D}{0}%
\Symb%
   {product of map over scalar}%
   {product of map over scalar}%
   {D}{0}%
\Symb%
   {set of vectors generated by vector $v$}%
   {set of vectors generated by vector}%
   {D}{0}%
\Symb%
   {speed of deviation}%
   {speed of deviation}%
   {D}{0}%
\Symb%
   {standard component of derivative}%
   {standard component of derivative}%
   {D}{0}%
\Symb%
   {standard component of the G\^ateaux derivative}%
   {standard component of Gateaux derivative}%
   {D}{0}%
\Symb%
   {vector space type}%
   {vector space type}%
   {D}{0}%

\SetIndexSpace
\Symb%
   {affine basis}%
   {Affine Basis}%
   {E}{0}%
\Symb%
   {basis of vector space}%
   {Basis e}%
   {E}{0}%
\Symb%
   {basis for module}%
   {basis for module}%
   {E}{0}%
\Symb%
   {basis in vector space $\Vector V$}%
   {basis in V}%
   {E}{0}%
\Symb%
   {basis of $D$\Hyph module $\mathcal L(D;A_1;A_2)$}%
   {basis L(A1,A2)}%
   {E}{0}%
\Symb%
   {basis for \crd vector space}%
   {basis, crd vector space}%
   {E}{0}%
\Symb%
   {basis for $D$\Hyph vector space}%
   {basis, D vector space}%
   {E}{0}%
\Symb%
   {basis for \dcr vector space}%
   {basis, dcr vector space}%
   {E}{0}%
\Symb%
   {basis for \drc vector space}%
   {basis, drc vector space}%
   {E}{0}%
\Symb%
   {basis for \rcd vector space}%
   {basis, rcd vector space}%
   {E}{0}%
\Symb%
   {basis for vector bundle}%
   {basis, vector bundle}%
   {E}{0}%
\Symb%
   {basis of $(n)$\hyph vector space}%
   {basis,n vector space}%
   {E}{0}%
\Symb%
   {Cartesian power of total spaces}%
   {Cartesian power of total spaces}%
   {E}{0}%
\Symb%
   {Cartesian product of total spaces}%
   {Cartesian product of total spaces, definition 1}%
   {E}{0}%
\Symb%
   {central affine basis}%
   {Central Affine Basis}%
   {E}{0}%
\Symb%
   {\CR exponent}%
   {CR exponent}%
   {E}{0}%
\Symb%
   {form of reference frame}%
   {dual forms, reference frame}%
   {E}{0}%
\Symb%
   {Euclid space}%
   {Euclid space}%
   {E}{0}%
\Symb%
   {Euclid space}%
   {Euclid space, division ring}%
   {E}{0}%
\Symb%
   {exponent}%
   {exponent}%
   {E}{0}%
\Symb%
   {Hamel basis}%
   {Hamel basis}%
   {E}{0}%
\Symb%
   {identical transformation of bundle}%
   {identical transformation of bundle}%
   {E}{0}%
\Symb%
   {map of conjugation}%
   {map of conjugation}%
   {E}{0}%
\Symb%
   {linear automorphism of quaternioin algebra}%
   {mapping E, quaternion}%
   {E}{0}%
\Symb%
   {linear automorphism of quaternioin algebra}%
   {mapping E_1, quaternion}%
   {E}{0}%
\Symb%
   {linear automorphism of quaternioin algebra}%
   {mapping E_2, quaternion}%
   {E}{0}%
\Symb%
   {Jacobian matrix of maps of conjugation}%
   {maps of conjugation, Jacobian matrix}%
   {E}{0}%
\Symb%
   {orthonornal basis}%
   {Orthonornal Basis}%
   {E}{0}%
\Symb%
   {image of basis $\Basis e$ under passive transformation $S$}%
   {passive transformation of basis}%
   {E}{0}%
\Symb%
   {pseudo Euclid space}%
   {pseudo Euclid space}%
   {E}{0}%
\Symb%
   {pseudo Euclid space}%
   {pseudo Euclid space, division ring}%
   {E}{0}%
\Symb%
   {quasiexponent}%
   {quasiexponent}%
   {E}{0}%
\Symb%
   {quaternion algebra over the field $F$}%
   {quaternion algebra over the field}%
   {E}{0}%
\Symb%
   {quaternion division algebra over the field}%
   {quaternion division algebra over the fieldL}%
   {E}{0}%
\Symb%
   {\RC exponent}%
   {RC exponent}%
   {E}{0}%
\Symb%
   {reduced Cartesian product of total spaces}%
   {reduced Cartesian product of total spaces, definition 1}%
   {E}{0}%
\Symb%
   {Schauder basis}%
   {Schauder basis}%
   {E}{0}%
\Symb%
   {set of endomorphisms}%
   {set of endomorphisms}%
   {E}{0}%
\Symb%
   {set of nonsingular \sT transformations of bundle $\Bundle E$}%
   {set of starT nonsingular transformations of bundle}%
   {E}{0}%
\Symb%
   {set of transformations of universal algebra}%
   {set of transformations}%
   {E}{0}%
\Symb%
   {set of nonsingular \Ts transformations of bundle $\Bundle E$}%
   {set of Tstar nonsingular transformations of bundle}%
   {E}{0}%
\Symb%
   {standard coordinates of basis}%
   {standard coordinates of basis}%
   {E}{0}%
\Symb%
   {standard coordinates of reference frame}%
   {standard coordinates of reference frame}%
   {E}{0}%
\Symb%
   {vector field of reference frame}%
   {vector field of reference frame}%
   {E}{0}%
\Symb%
   {vector of basis}%
   {vector of basis}%
   {E}{0}%

\SetIndexSpace
\Symb%
   {alternation of polylinear map}%
   {alternation of polylinear map}%
   {F}{0}%
\Symb%
   {component of linear map $f$ of division ring}%
   {component of linear map, division ring}%
   {F}{0}%
\Symb%
   {component of polylinear map}%
   {component of polylinear map}%
   {F}{0}%
\Symb%
   {conjugation transformation}%
   {conjugation transformation}%
   {F}{0}%
\Symb%
   {exterior product}%
   {exterior product}%
   {F}{0}%
\Symb%
   {fibered morphism from bundle $\Bundle A$ into $\Bundle B$}%
   {fibered morphism from A into B}%
   {F}{0}%
\Symb%
   {filter $\mathfrak{F}$ converges to set $A$}%
   {filter converges}%
   {F}{0}%
\Symb%
   {homomorphism of fibered universal algebras}%
   {homomorphism of fibered universal algebras}%
   {F}{0}%
\Symb%
   {inverse fibered correspondence}%
   {inverse fibered correspondence, 1}%
   {F}{0}%
\Symb%
   {inverse reduced fibered correspondence}%
   {inverse reduced fibered correspondence, 1}%
   {F}{0}%
\Symb%
   {map to Cartesian product}%
   {map to Cartesian product}%
   {F}{0}%
\Symb%
   {norm of functional}%
   {norm of functional}%
   {F}{0}%
\Symb%
   {norm of map}%
   {norm of map}%
   {F}{0}%
\Symb%
   {norm of polylinear map}%
   {norm of polymap}%
   {F}{0}%
\Symb%
   {norm of representation}%
   {norm of representation}%
   {F}{0}%
\Symb%
   {orbit of representation of the group}%
   {orbit of representation}%
   {F}{0}%
\Symb%
   {orthonormal basis}%
   {Orthonormal Basis, division ring}%
   {F}{0}%
\Symb%
   {quaternion algebra  over field ${\rm {\mathbb{F}}}$}%
   {quaternion algebra F a b}%
   {F}{0}%
\Symb%
   {reference frame}%
   {reference frame}%
   {F}{0}%
\Symb%
   {reference frame, extensive definition}%
   {reference frame, extensive definition}%
   {F}{0}%
\Symb%
   {standard component of biadditive map $f$ over field $F$}%
   {standard component of biadditive map, division ring}%
   {F}{0}%
\Symb%
   {standard component of linear map}%
   {standard component of linear map, G}%
   {F}{0}%
\Symb%
   {standard component of polylinear map}%
   {standard component of polylinear map}%
   {F}{0}%
\Symb%
   {standard component of quadratic map $f$ over field $F$}%
   {standard component of quadratic map, division ring}%
   {F}{0}%
\Symb%
   {standard component of tensor}%
   {standard component of tensor}%
   {F}{0}%
\Symb%
   {sum of maps}%
   {sum of maps}%
   {F}{0}%
\Symb%
   {symmetrization of polylinear map}%
   {symmetrization of polylinear map}%
   {F}{0}%

\SetIndexSpace
\Symb%
   {affine transformation group}%
   {affine transformation group}%
   {G}{0}%
\Symb%
   {affine transformation group}%
   {affine transformation group}%
   {G}{0}%
\Symb%
   {Cartesian product of groups $G_1$, ..., $G_n$}%
   {Cartesian product of groups, 1 n}%
   {G}{0}%
\Symb%
   {\CR matrix group}%
   {cr-matrix group}%
   {G}{0}%
\Symb%
   {fibered little group of section $h$}%
   {fibered little group}%
   {G}{0}%
\Symb%
   {fibered stability group of section $h$}%
   {fibered stability group}%
   {G}{0}%
\Symb%
   {group of automorphisms of representation $f$}%
   {group of automorphisms of representation}%
   {G}{0}%
\Symb%
   {group of homomorphisms of vector space $\Vector V$}%
   {GV}%
   {G}{0}%
\Symb%
   {indefinite integral}%
   {indefinite integral}%
   {G}{0}%
\Symb%
   {left defined Lie algebra of Lie group}%
   {left defined Lie algebra of Lie group}%
   {G}{0}%
\Symb%
   {Lie algebra of Lie group}%
   {Lie algebra of Lie group}%
   {G}{0}%
\Symb%
   {linear transformation group}%
   {linear transformation group}%
   {G}{0}%
\Symb%
   {little group}%
   {little group}%
   {G}{0}%
\Symb%
   {orbit of effective Ts representation of group}%
   {orbit of effective starT representation of fibered group}%
   {G}{0}%
\Symb%
   {orbit of effective \Ts representation of fibered group}%
   {orbit of effective Tstar representation of fibered group}%
   {G}{0}%
\Symb%
   {\RC matrix group}%
   {rc-matrix group}%
   {G}{0}%
\Symb%
   {right defined Lie algebra of Lie group}%
   {right defined Lie algebra}%
   {G}{0}%
\Symb%
   {stability group}%
   {stability group}%
   {G}{0}%

\SetIndexSpace
\Symb%
   {Hadamard inverse of matrix}%
   {Hadamard inverse of matrix}%
   {H}{0}%
\Symb%
   {horizontal component of vector}%
   {horizontal component of vector}%
   {H}{0}%
\Symb%
   {horizontal subspace}%
   {horizontal subspace}%
   {H}{0}%
\Symb%
   {quaternion algebra}%
   {quaternion algebra}%
   {H}{0}%
\Symb%
   {quaternion algebra}%
   {quaternion algebra H a b}%
   {H}{0}%
\Symb%
   {set of homomorphisms}%
   {set of homomorphisms}%
   {H}{0}%

\SetIndexSpace
\Symb%
   {infinitesimal generator of representation}%
   {infinitesimal generator i of representation}%
   {I}{0}%
\Symb%
   {infinitesimal generator of representation}%
   {infinitesimal generator of representation}%
   {I}{0}%
\Symb%
   {Lie group infinitesimal generators}%
   {Lie group infinitesimal generators}%
   {I}{0}%
\Symb%
   {map of conjugation}%
   {map of conjugation}%
   {I}{0}%
\Symb%
   {Jacobian matrix of maps of conjugation}%
   {maps of conjugation, Jacobian matrix}%
   {I}{0}%
\Symb%
   {vector module of algebra $A$}%
   {vector module of algebra}%
   {I}{0}%
\Symb%
   {vector module of ring $D$}%
   {vector module of ring}%
   {I}{0}%
\Symb%
   {vector of element $d$ of algebra}%
   {vector of algebra}%
   {I}{0}%
\Symb%
   {vector of element $d$ of ring}%
   {vector of ring}%
   {I}{0}%

\SetIndexSpace
\Symb%
   {closure operator of representation $f$}%
   {closure operator, representation}%
   {J}{0}%
\Symb%
   {closure operator of tower of representations $\Vector f$}%
   {closure operator, tower of representations}%
   {J}{0}%
\Symb%
   {Jacobian matrix of right shift}%
   {Ea, quaternion, Jacobian matrix}%
   {J}{0}%
\Symb%
   {map of conjugation}%
   {map of conjugation}%
   {J}{0}%
\Symb%
   {Jacobian matrix of maps of conjugation}%
   {maps of conjugation, Jacobian matrix}%
   {J}{0}%
\Symb%
   {subrepresentation generated by generating set $X$}%
   {subrepresentation generated by set}%
   {J}{0}%
\Symb%
   {tower of subrepresentations of tower of representations $\Vector f$ generated by tuple of sets $\VX X$}%
   {subrepresentation generated by tuple of sets}%
   {J}{0}%

\SetIndexSpace
\Symb%
   {kernel of homomorphism}%
   {kernel of homomorphism}%
   {K}{0}%
\Symb%
   {kernel of linear map}%
   {kernel of linear map}%
   {K}{0}%
\Symb%
   {kernel of map}%
   {kernel of map}%
   {K}{0}%
\Symb%
   {map of conjugation}%
   {map of conjugation}%
   {K}{0}%
\Symb%
   {Jacobian matrix of maps of conjugation}%
   {maps of conjugation, Jacobian matrix}%
   {K}{0}%

\SetIndexSpace
\Symb%
   {Cartesian power of systems of subsets}%
   {Cartesian power of systems of subsets}%
   {L}{0}%
\Symb%
   {Cartesian product of systems of subsets}%
   {Cartesian product of systems of subsets}%
   {L}{0}%
\Symb%
   {left $ij$th cofactor of entry of matrix}%
   {left cofactor, matrix}%
   {L}{0}%
\Symb%
   {left double $ij$th cofactor of entry of matrix}%
   {left double cofactor}%
   {L}{0}%
\Symb%
   {left shift}%
   {left shift}%
   {L}{0}%
\Symb%
   {Lie derivative of connection}%
   {Lie derivative of connection}%
   {L}{0}%
\Symb%
   {Lie derivative of metric}%
   {Lie derivative of metric}%
   {L}{0}%
\Symb%
   {limit of correspondence $\Phi$ with respect to the filter $\mathfrak{F}$}%
   {limit of correspondence with respect to the filter}%
   {L}{0}%
\Symb%
   {limit of sequence}%
   {limit of sequence}%
   {L}{0}%
\Symb%
   {linear combination}%
   {linear combination}%
   {L}{0}%
\Symb%
   {module of skew symmetric polylinear maps}%
   {module of skew symmetric polylinear maps}%
   {L}{0}%
\Symb%
   {passive transformation}%
   {passive transformation}%
   {L}{0}%
\Symb%
   {$D$\Hyph module of continuous linear mappings of normed $D$\Hyph module $A_1$ into normed $D$\Hyph module $A_2$}%
   {set continuous linear mappings, module}%
   {L}{0}%
\Symb%
   {set of continuous linear maps}%
   {set continuous linear maps, vector}%
   {L}{0}%
\Symb%
   {set of continuous polylinear maps}%
   {set continuous polylinear maps}%
   {L}{0}%
\Symb%
   {set of linear maps}%
   {set linear maps}%
   {L}{0}%
\Symb%
   {set of left-side nonsingular transformations of universal algebra $M$}%
   {set of left-side nonsingular transformations}%
   {L}{0}%
\Symb%
   {set of polylinear maps}%
   {set polylinear maps}%
   {L}{0}%
\Symb%
   {set of $n$\hyph linear maps}%
   {set polylinear maps An}%
   {L}{0}%
\Symb%
   {set of polylinear maps}%
   {set polylinear maps, D vector space}%
   {L}{0}%
\Symb%
   {set of polylinear maps of algebras $A_1$, ..., $A_n$ into algebra $A$}%
   {set polylinear maps, finite dimensional algebra}%
   {L}{0}%

\SetIndexSpace
\Symb%
   {set of left-side transformations of the universal algebra $M$}%
   {set of left-side transformations}%
   {M}{0}%
\Symb%
   {set of maps to $\Omega$\Hyph group $A$}%
   {set of maps to Omega group}%
   {M}{0}%
\Symb%
   {set of right-side transformations of universal algebra $M$}%
   {set of right-side transformations}%
   {M}{0}%
\Symb%
   {space of orbits of \Ts{G}representation}%
   {space of orbits of G* representation}%
   {M}{0}%

\SetIndexSpace
\Symb%
   {norm of quaternion $x$}%
   {norm, quaternion algebra}%
   {N}{0}%
\Symb%
   {nucleus of $D$\Hyph algebra $A$}%
   {nucleus of algebra}%
   {N}{0}%

\SetIndexSpace
\Symb%
   {geometric object}%
   {geometric object}%
   {O}{0}%
\Symb%
   {geometric object in coordinate representation defined in \rcd vector space}%
   {geometric object, coordinate rcd vector space}%
   {O}{0}%
\Symb%
   {geometric object in coordinate representation}%
   {geometric object, coordinate representation}%
   {O}{0}%
\Symb%
   {geometric object in coordinate representation}%
   {geometric object, coordinate vector space}%
   {O}{0}%
\Symb%
   {geometric object defined in \rcd vector space}%
   {geometric object, rcd vector space}%
   {O}{0}%
\Symb%
   {octonion algebra}%
   {octonion algebra}%
   {O}{0}%
\Symb%
   {orbit of representation of fibered group $\Bundle G$}%
   {orbit of representation of fibered group}%
   {O}{0}%
\Symb%
   {tensor product}%
   {tensor product}%
   {O}{0}%

\SetIndexSpace
\Symb%
   {bundle}%
   {bundle}%
   {P}{0}%
\Symb%
   {bundle of level $2$}%
   {bundle of level 2}%
   {P}{0}%
\Symb%
   {bundle of level $n$}%
   {bundle of level n}%
   {P}{0}%
\Symb%
   {Cartesian power $n$ of bundle $\bundle{}{p}{E}{}$}%
   {Cartesian power of bundle}%
   {P}{0}%
\Symb%
   {Cartesian product of bundles}%
   {Cartesian product of bundles, definition 1}%
   {P}{0}%
\Symb%
   {passive representation in basis manifold}%
   {passive representation in basis manifold}%
   {P}{0}%
\Symb%
   {reduced Cartesian product of bundles}%
   {reduced Cartesian product of bundles, definition 1}%
   {P}{0}%
\Symb%
   {set of nonsingular \sT transformations of bundle $\bundle{}pE{}$}%
   {set of starT nonsingular transformations of bundle, projection}%
   {P}{0}%
\Symb%
   {set of nonsingular \Ts transformations of bundle $\bundle{}pE{}$}%
   {set of Tstar nonsingular transformations of bundle, projection}%
   {P}{0}%

\SetIndexSpace
\Symb%
   {active transformation}%
   {active transformation}%
   {R}{0}%
\Symb%
   {Cartan curvature}%
   {Cartan curvature}%
   {R}{0}%
\Symb%
   {\CR rank of matrix}%
   {cr-rank of matrix}%
   {R}{0}%
\Symb%
   {diagonal in bundle  $\bundle{}pA{}$}%
   {diagonal in bundle, 2}%
   {R}{0}%
\Symb%
   {diagonal in bundle $\Bundle A$}%
   {diagonal in reduced bundle, 2}%
   {R}{0}%
\Symb%
   {image of $m$ under endomorphism $R$ of effective representation}%
   {endomorphism image, effective representation}%
   {R}{0}%
\Symb%
   {image of tuple $\VX a$ under endomorphism $\VX r$ of tower of effective representations}%
   {endomorphism image, tower of effective representations}%
   {R}{0}%
\Symb%
   {curvature}%
   {GLn curvature_overline}%
   {R}{0}%
\Symb%
   {product of rings of sets}%
   {product of rings of sets}%
   {R}{0}%
\Symb%
   {$\RCcirc$\Hyph product of matrices of maps}%
   {rc product of matrices of maps}%
   {R}{0}%
\Symb%
   {\RC rank of matrix}%
   {rc-rank of matrix}%
   {R}{0}%
\Symb%
   {right $ij$th cofactor of entry of matrix}%
   {right cofactor, matrix}%
   {R}{0}%
\Symb%
   {right double $ij$th cofactor of entry of matrix}%
   {right double cofactor}%
   {R}{0}%
\Symb%
   {right shift}%
   {right shift}%
   {R}{0}%
\Symb%
   {$i$th row determinant of matrix $\bfA$}%
   {row determinant}%
   {R}{0}%
\Symb%
   {scalar algebra of algebra $A$}%
   {scalar algebra of algebra}%
   {R}{0}%
\Symb%
   {scalar algebra of ring $D$}%
   {scalar algebra of ring}%
   {R}{0}%
\Symb%
   {scalar of element $d$ of algebra}%
   {scalar of algebra}%
   {R}{0}%
\Symb%
   {scalar of element $d$ of ring}%
   {scalar of ring}%
   {R}{0}%
\Symb%
   {set of right-side nonsingular transformations of universal algebra $M$}%
   {set of right-side nonsingular transformations}%
   {R}{0}%
\Symb%
   {spherical coordinates}%
   {spherical coordinates}%
   {R}{0}%

\SetIndexSpace
\Symb%
   {composition of fibered correspondences}%
   {composition of fibered correspondences}%
   {S}{0}%
\Symb%
   {hyperbolic sine}%
   {hyperbolic sine}%
   {S}{0}%
\Symb%
   {inverse fibered correspondence}%
   {inverse fibered correspondence, 2}%
   {S}{0}%
\Symb%
   {inverse reduced fibered correspondence}%
   {inverse reduced fibered correspondence, 2}%
   {S}{0}%
\Symb%
   {Lebesgue integral}%
   {Lebesgue integral}%
   {S}{0}%
\Symb%
   {linear span in vector space}%
   {linear span, vector space}%
   {S}{0}%
\Symb%
   {image of basis $\VX  X$ under passive transformation $\VX s$}%
   {passive transformation of basis, tower of representations}%
   {S}{0}%
\Symb%
   {set of permutations}%
   {set of permutations}%
   {S}{0}%
\Symb%
   {set of transpositions}%
   {set of transpositions}%
   {S}{0}%
\Symb%
   {sine}%
   {sine}%
   {S}{0}%
\Symb%
   {symmetric group}%
   {symmetric group}%
   {S}{0}%

\SetIndexSpace
\Symb%
   {category of left-side representations}%
   {category of left-side representations}%
   {T}{0}%
\Symb%
   {tangent plane to Lie group $G$}%
   {tangent plane to Lie group}%
   {T}{0}%
\Symb%
   {trace of quaternion $x$}%
   {trace, quaternion algebra}%
   {T}{0}%

\SetIndexSpace
\Symb%
   {affine space}%
   {affine space}%
   {V}{0}%
\Symb%
   {conjugated affine space}%
   {conjugated affine space}%
   {V}{0}%
\Symb%
   {conjugated vector space}%
   {conjugated vector space}%
   {V}{0}%
\Symb%
   {coordinate vector space}%
   {coordinate vector space}%
   {V}{0}%
\Symb%
   {coordinates in vector space}%
   {coordinates in vector space}%
   {V}{0}%
\Symb%
   {direct product of $\RCstar D_i$\hyph vector spaces $\Vector V_1$, ..., $\Vector V_n$}%
   {direct product, rcd vector space, 1 n}%
   {V}{0}%
\Symb%
   {dual space of \rcd vector space $\Vector V$}%
   {dual space of rcd vector space}%
   {V}{0}%
\Symb%
   {hermitian conjugated vector}%
   {hermitian conjugated vector}%
   {V}{0}%
\Symb%
   {linear composition of vectors}%
   {linear composition of vectors}%
   {V}{0}%
\Symb%
   {set of vectors generated by vector $v$}%
   {set of vectors generated by vector}%
   {V}{0}%
\Symb%
   {vector space}%
   {V}%
   {V}{0}%
\Symb%
   {vertical component of vector}%
   {vertical component of vector}%
   {V}{0}%
\Symb%
   {vertical subspace}%
   {vertical subspace}%
   {V}{0}%

\SetIndexSpace
\Symb%
   {set of coordinates of representation $J(f,X)$}%
   {coordinate set of representation}%
   {W}{0}%
\Symb%
   {set of tuples of coordinates of tower of representations $\Vector J(\Vector f,\VX X)$}%
   {coordinate set of tower of representations}%
   {W}{0}%
\Symb%
   {coordinates of basis $X'$ relative to basis $X$ of representation}%
   {coordinates of basis relative to basis, representation}%
   {W}{0}%
\Symb%
   {coordinates of element $m$ relative to set $X$}%
   {coordinates of element relative to set, representation}%
   {W}{0}%
\Symb%
   {tuple of coordinates of element $\Vector a*$ relative to tuple of sets $\VX X$}%
   {coordinates of element, tower of representations}%
   {W}{0}%
\Symb%
   {coordinates of element $m$ of representation $f$ relative to set $X$}%
   {coordinates relative to set}%
   {W}{0}%
\Symb%
   {geometric object}%
   {geometric object}%
   {W}{0}%
\Symb%
   {geometric object in coordinate representation}%
   {geometric object, coordinate representation}%
   {W}{0}%
\Symb%
   {geometric object in coordinate representation defined in tuple of $\VX\Omega$\Hyph algebras $\VX A$}%
   {geometric object, coordinate tower of representations g}%
   {W}{0}%
\Symb%
   {geometric object defined in tuple of $\VX\Omega$\Hyph algebras $\VX A$}%
   {geometric object, tower of representations g}%
   {W}{0}%
\Symb%
   {geometric object in vector space}%
   {geometric object, vector space}%
   {W}{0}%
\Symb%
   {set of tuples of $\Omega$\Hyph words}%
   {set of tuples of Omega words}%
   {W}{0}%
\Symb%
   {set of coordinates of set $B\subset J(f,X)$}%
   {subset of coordinates of representation}%
   {W}{0}%
\Symb%
   {coordinates of tuple of sets $\VX B$ relative to tuple of sets $\VX X$}%
   {subset of coordinates of tower of representations}%
   {W}{0}%
\Symb%
   {coordinates of set $B_k$ relative to tuple of sets $\VX X$}%
   {subset of coordinates of tower of representations, k}%
   {W}{0}%
\Symb%
   {set of $\Omega_2$\Hyph words representing set $B\subset J(f,X)$}%
   {subset of words of representation}%
   {W}{0}%
\Symb%
   {superposition of coordinates}%
   {superposition of coordinates}%
   {W}{0}%
\Symb%
   {superposition of coordinates of the tower of representations $\Vector f$ and the element $\VX a$}%
   {superposition of coordinates, tower of representations}%
   {W}{0}%
\Symb%
   {tuple of $\Omega$\Hyph words}%
   {tuple of Omega words}%
   {W}{0}%
\Symb%
   {$\Omega_2$\Hyph word representing element $m\in J(f,X)$}%
   {word of element relative to generating set, representation}%
   {W}{0}%
\Symb%
   {set of $\Omega_2$\Hyph words of representation $J(f,X)$}%
   {word set of representation}%
   {W}{0}%
\Symb%
   {set of tuples of $\VX{\Omega}$\Hyph words of tower of representations $\Vector J(\Vector f,\VX X)$}%
   {word set of tower of representations}%
   {W}{0}%
\Symb%
   {tuple of words of element $\Vector a*$ relative to tuple of sets $\VX X$}%
   {words of element, tower of representations}%
   {W}{0}%

\SetIndexSpace
\Symb%
   {conjugate of quaternion $x$}%
   {conjugate of quaternion}%
   {X}{0}%
\Symb%
   {local basis of affine space}%
   {local basis of affine space}%
   {X}{0}%
\Symb%
   {anholonomic coordinate}%
   {x(k)}%
   {X}{0}%

\SetIndexSpace
\Symb%
   {center of $D$\Hyph algebra $A$}%
   {center of algebra}%
   {Z}{0}%
\Symb%
   {center of ring $D$}%
   {center of ring}%
   {Z}{0}%

\SetIndexSpace
\Symb%
   {deviation of trajectories}%
   {deviation of trajectories}%
   {Delta}{1}%
\Symb%
   {identical transformation}%
   {identical transformation}%
   {Delta}{1}%
\Symb%
   {image of vector $\Vector e_k\in\Basis e$ under isomorphism to coordinate vector space}%
   {image of vector e_k, coordinate vector space}%
   {Delta}{1}%
\Symb%
   {Kronecker symbol}%
   {Kronecker symbol}%
   {Delta}{1}%

\SetIndexSpace
\Symb%
   {anholonomic coordinates of connection}%
   {anholonomic coordinates of connection}%
   {Gamma}{1}%
\Symb%
   {Cartan symbol}%
   {Cartan symbol}%
   {Gamma}{1}%
\Symb%
   {connection}%
   {conection overline}%
   {Gamma}{1}%
\Symb%
   {connection coefficients in $D$\Hyph affine space}%
   {connection coefficients, D affine space}%
   {Gamma}{1}%
\Symb%
   {connection in $D$\Hyph affine manifold}%
   {connection, affine manifold}%
   {Gamma}{1}%
\Symb%
   {$D$\Hyph affine connection coefficients on manifold}%
   {D affine connection coefficients, manifold}%
   {Gamma}{1}%
\Symb%
   {holonomic coordinates of connection}%
   {holonomic coordinates of connection}%
   {Gamma}{1}%
\Symb%
   {Cartan connection}%
   {overbrace Gamma i kl}%
   {Gamma}{1}%
\Symb%
   {set of sections of bundle}%
   {set of sections of bundle}%
   {Gamma}{1}%

\SetIndexSpace
\Symb%
   {inverse operator to operator $\psi_l$}%
   {inverse operator to operator psi l}%
   {Lambda}{1}%
\Symb%
   {inverse operator to operator $\psi_r$}%
   {inverse operator to operator psi r}%
   {Lambda}{1}%

\SetIndexSpace
\Symb%
   {Cartesian product of measures}%
   {Cartesian product of measures}%
   {Mu}{1}%
\Symb%
   {power of measure}%
   {power of measure}%
   {Mu}{1}%
\Symb%
   {product of measures}%
   {product of measures}%
   {Mu}{1}%
\Symb%
   {product of measures}%
   {product of measures, otimes}%
   {Mu}{1}%

\SetIndexSpace
\Symb%
   {anholonomity object}%
   {anholonomity object}%
   {Omega}{1}%
\Symb%
   {definite integral}%
   {definite integral}%
   {Omega}{1}%
\Symb%
   {integral of differential $1$\Hyph form along path}%
   {integral of differential 1 form along path}%
   {Omega}{1}%
\Symb%
   {norm of operation}%
   {norm of operation}%
   {Omega}{1}%
\Symb%
   {operator domain}%
   {operator domain}%
   {Omega}{1}%
\Symb%
   {set of differential $p$\Hyph forms}%
   {set of differential p forms}%
   {Omega}{1}%
\Symb%
   {set of $n$\Hyph ary operations of $\Omega$\Hyph algebra}%
   {set of n-ary operations}%
   {Omega}{1}%
\Symb%
   {set of $n$\Hyph ary operators}%
   {set of n-ary operators}%
   {Omega}{1}%

\SetIndexSpace
\Symb%
   {left basic operator of Lie group over algebra $A$}%
   {L basic operator of Lie group over algebra A}%
   {Psi}{1}%
\Symb%
   {left basic operator of group Lie}%
   {Lie Basic Operator L}%
   {Psi}{1}%
\Symb%
   {left basic operator of Lie 1-parameter group}%
   {Lie Basic Operator L, 1-Parameter Group}%
   {Psi}{1}%
\Symb%
   {left basic operator of Lie 1-parameter group over algebra $A$}%
   {Lie Basic Operator L, 1-Parameter Group, algebra}%
   {Psi}{1}%
\Symb%
   {right basic operator of group Lie}%
   {Lie Basic Operator R}%
   {Psi}{1}%
\Symb%
   {right basic operator of Lie 1-parameter group}%
   {Lie Basic Operator R, 1-Parameter Group}%
   {Psi}{1}%
\Symb%
   {right basic operator of Lie 1-parameter group over algebra $A$}%
   {Lie Basic Operator R, 1-Parameter Group, algebra}%
   {Psi}{1}%
\Symb%
   {right basic operator of Lie group over algebra $A$}%
   {R basic operator of Lie group over algebra A}%
   {Psi}{1}%

\SetIndexSpace
\Symb%
   {fibered subset}%
   {fibered subset}%
   {Sigma}{1}%
\Symb%
   {parity of permutation}%
   {parity of permutation}%
   {Sigma}{1}%
\Symb%
   {subbundle}%
   {subbundle}%
   {Sigma}{1}%

\SetIndexSpace
\Symb%
   {Cartan derivative}%
   {overbrace nabla_l}%
   {Nabla}{2}%
\Symb%
   {derivative}%
   {overline nabla_l, definition 1}%
   {Nabla}{2}%

\SetIndexSpace
\Symb%
   {Lie group composition law}%
   {Lie group composition law}%
   {Phi}{1}%
\Symb%
   {restriction of correspondence $\Phi$ to set $C$}%
   {restriction of correspondence}%
   {Phi}{1}%

\SetIndexSpace
\Symb%
   {Cartesian product of bundles}%
   {Cartesian product of bundles, definition 2}%
   {Pi}{1}%
\Symb%
   {Cartesian product of groups $G_i$, $i\in I$}%
   {Cartesian product of groups}%
   {Pi}{1}%
\Symb%
   {Cartesian product of groups $G_1$, ..., $G_n$}%
   {Cartesian product of groups, i 1 n}%
   {Pi}{1}%
\Symb%
   {Cartesian product of total spaces}%
   {Cartesian product of total spaces, definition 2}%
   {Pi}{1}%
\Symb%
   {coproduct in category}%
   {coproduct in category}%
   {Pi}{1}%
\Symb%
   {direct product of division rings $D_i$, $i\in I$}%
   {direct product of division rings}%
   {Pi}{1}%
\Symb%
   {direct product of division rings $D_1$, ..., $D_n$}%
   {direct product of division rings, i 1 n}%
   {Pi}{1}%
\Symb%
   {direct product of $\RCstar D_i$\hyph vector spaces $\Vector V_i$, $i\in I$}%
   {direct product, rcd vector space}%
   {Pi}{1}%
\Symb%
   {direct product of $\RCstar D_i$\hyph vector spaces}%
   {direct product, rcd vector space, i 1 n}%
   {Pi}{1}%
\Symb%
   {product in category}%
   {product in category}%
   {Pi}{1}%
\Symb%
   {reduced Cartesian product of bundles}%
   {reduced Cartesian product of bundles, definition 2}%
   {Pi}{1}%
\Symb%
   {reduced Cartesian product of total spaces}%
   {reduced Cartesian product of total spaces, definition 2}%
   {Pi}{1}%

\CloseIndex

%% file: Space.Time.English.tex

\chapter{Space and Time in Physics}
\labelChapter{Space and Time in Physics}

\section{Geometry and Physics}
\labelSection{Geometry and Physics}
\epigraph{Reason, of course, is weak,
when measured against its never\Hyph ending task.
Weak, indeed, compared with the follies and passions of mankind,
which, we must admit, almost entirely control our human destinies, in great things and small.
Yet the works of the understanding outlast the noisy bustling generations
and spread light and warmth across the centuries.}
{\citeBib{Einstein: Isaak Newton}, p. 219}
Geometric ideas are not creations of free intellect, but on the contrary, they are product of
human activity. Ideas about space and time changed as knowledge
developed.\footref{footnote: Space and Time in Physics}

\CiteQuotation{Law cannot be exact at least because concepts on the base, which we formulate it,
may develop and may become insufficient. Signs of dogma of impeccability remain on the bottom of
any thesis and any proof.}{\citeBib{Einstein: On Science}}

Mathematics originated from specific needs of human practice and developed along complex and contradictory way of
knowledge. From one hand, this is the most abstract area of science. Mathematician freely operates
with abstract ideas, learns their properties, generalizes, creates new ideas and definitions, and wanders from practice.
This makes more surprising success of applied mathematics. Its success always surprised people, mostly in XX
century when such abstract parts of mathematics like group theory, functional calculus, and differential geometry
became language of modern physics. Contact of mathematics with any other field of science leads to
mutual enrichment of both.

\CiteQuotation{One reason why mathematics enjoys special esteem, above all other sciences, is that its laws
are absolutely certain and indisputable, while those of all other sciences are to some extent
debatable and in constant danger of being overthrown by newly discovered facts. In spite of
this, the investigator in another department of science would not need to envy the mathematician
since the laws of mathematics referred to objects of our mere imagination, and not to objects of reality.
For it cannot occasion surprise that different persons should arrive at the same logical conclusions
when they have already agreed upon the fundamental laws (axioms), as well as the methods by
which other laws are to be deduced therefrom. But there is another reason for the high repute of
mathematics, in that it is mathematics which affords the exact natural sciences a certain measure
of security, to which without mathematics they could not attain.
//
At this point an enigma presents itself which in all ages has agitated inquiring minds.
How can it be that mathematics, being after all a product of human thought which is
independent of experience, is so admirably appropriate to the objects of reality? Is human
reason, then, without experience, merely by taking thought, able to fathom the properties of
real things.}
{\citeBib{Einstein: Geometry and Experience}}

Mathematics has empiric roots. Algebra associated with necessity of calculations. Geometry was created by necessity
of measurement on earth surface. Logic is abstract expression of causal relationship. Calculus studies movement.
This is exactly why on a certain phase of development
it becomes possible to describe real processes
using mathematical models
of different level of complexity.
Gradually statements of theoretical mathematics become instrument of
 applied mathematics; mathematics becomes empiric field of science.

\CiteQuotation{Only main concepts and such called axioms remain in mathematics as evidence of empiric origin of geometry.
People tried to reduce number of this logically irreducible concepts and axioms. Tendency to take out all
geometry from dim area of empiric insensibly leads to erroneous conclusion which we can liken to conversion of heroes of
antiquity into gods. People gradually accustomed to point of view on main concepts as obvious, i.e. objects and
qualities which belong to human mind. According to this point of view, objects of intuition correspond to main concepts
of geometry and negation of any axiom of geometry cannot be done consistently. In this case, possibility of
application of this main concepts and axioms to objects of reality becomes the problem from which Kant's understanding
of space appeared.
//
Physics gave second cause for denial geometry from its empiric foundation. According to more refined point
of view about nature of solid body and light, in nature there are no such objects, which exactly correspond to
main concepts of Euclidean geometry by their property. We cannot assume solid body as steady. Ray of light does not
reproduce nor straight line, nor any one-dimension image. According to modern science view, geometry taken separately
does not correspond ... to any tests. It should be applied to explanation together with mechanics, optics, etc.
Geometry has to appear as science logically preceding any experience and any empirical verification
because geometry has to precede physics as long as laws of physics cannot be expressed without
help of geometry.}{\citeBib{Einstein: Noneuclidean Geometry and Physics}, p 170}

\CiteQuotation{... As far as the laws of mathematics refer to reality, they are not certain; and as far as
they are certain, they do not refer to reality ...}
{\citeBib{Einstein: Geometry and Experience}}

\CiteQuotation{... Question about applicability or no applicability of Euclidean geometry obtains clear meaning
from this point of view. Euclidean geometry as geometry in general keeps nature of mathematical science because
deduction of its theorems from axioms is still logical problem. However, it becomes physical science because its
axioms hold inside statements relative objects of nature and correctness of these statements may be proved
only by experiment.}{\citeBib{Einstein: Noneuclidean Geometry and Physics}, p 181}

\CiteQuotation{Form of object is not something external relative to this object. The form belongs to object
and object defines this form. This is why forms of existence of real world are common structure defined by its
fundamental properties... Efficient theory of spacetime essentially derives properties of spacetime from properties
of matter. This was the source of geometry. First of all it reflected general properties of relations between solid bodies
which were defined in particular by ability to move.}{\citeBib{Cite: 104}, p 117}

\CiteQuotation{... it is certain that mathematics generally, and particularly geometry,
owes its existence to the need which was felt of learning something about the relations
of real things to one another. The very word geometry, which, of course, means
earth measuring, proves this. For earth measuring has to do with the possibilities
of the disposition of certain natural objects with respect to one another,
namely, with parts of the earth, measuring-lines, measuring-wands, etc.
It is clear that the system of concepts of axiomatic geometry alone cannot make any
assertions as to the relations of real objects of this kind, which we will call practically rigid bodies.
To be able to make such assertions, geometry must be stripped of its merely logical-formal
character by the co-ordination of real objects of experience with the empty conceptual
framework of axiomatic geometry. To accomplish this, we need only add the proposition:
//
- Solid bodies are related, with respect to their possible dispositions, as are bodies
in Euclidean geometry of three dimensions. Then the propositions of Euclid
contain affirmations as to the relations of practically rigid bodies.
//
Geometry thus completed is evidently a natural science...
Its affirmations rest essentially on induction from experience, but not on logical inferences only.
We will call this completed geometry practical geometry,
and shall distinguish it in what follows from purely axiomatic geometry.
The question whether the practical geometry of the universe is Euclidean or not has
a clear meaning, and its answer can only be furnished by experience.
All linear measurement in physics is practical geometry in this sense,
so too is geodetic and astronomical linear measurement, if we call to our help
the law of experience that light is propagated in a straight line,
and indeed in a straight line in the sense of
practical geometry.}
{\citeBib{Einstein: Geometry and Experience}}

This discussion shows that mathematics is not separate area of knowledge that develops independently
from cognitive and reformatory  practice of human. In deed, practice is initial source for mathematics and
finally mathematics is indispensable instrument in human practice.

\section{Spacetime}
\labelSection{Spacetime}
\epigraph
{... Space and time are not simple forms of phenomena but objective and real forms of existence.
There is nothing in the world except moving matter and moving matter can move only in space and
time... Changeability of human representations of space and
time deny objective reality of both as little as changeability of scientific
knowledge about structure and forms of movement of matter do not deny of objective reality
of exterior world.}
{Lenin, 110}
\epigraph
{... Not objects assume existence of space and time, but space and time assume existence of objects
because space or extent assumes existence of something that is extensive and time assumes movement.
Time is only idea derived from movement and assumes existence of something that moves.
Everything is spatial and temporal.}
{Fuerbach}

Before study of interaction of geometry and general relativity we have to make clear definition of main concepts.
Main concept of geometry is space. Space filled by physical contents is spacetime.

\CiteQuotation{... Idea "mater object" has to exist before ideas related to space.
This is logically initial idea. We can easily make sure this analyzing such spatial terms as "near", "contact", etc.
and looking for their equivalence in experiment.}{109, p 135}

\CiteQuotation{Physical idea of time responds to idea of intuitive mind. However, such idea traces back
to order in time of sensations of person and we have to accept this order as something initially given.
Somebody feels ... perception at this moment and this perception is connected with remembrance about
(previous) perceptions. This is a reason that perceptions create time series based on estimations
"before" and "after". These series may be repeated and then we can identify them.
In addition, they may be repeated inaccurately,
with replacement of some events by others. Moreover,
we do not lose pattern of repetition. We come this way to introduction
of time like one-dimensional frame that
we can fill different way by sensations. The same series of sensations
respond to the same subjective intervals of time.
//
Transition from this "subjective" time ... to concept of time in prescientific mind
is linked with rise of idea of existence of real world that is independent from subject.
In the sense, they establish correspondence between (objective) event and subjective sensation;
they compare "subjective" time of sensation with "time" of corresponding "objective" event.
External events and their order claim to truth for all subjects contrary to sensations.
//
... At more detailed examination of idea of objective world of external events
it required to establish more complex dependence between events and sensations.
At first, it was done with help
of instinctive rules of mind where  of space plays the most important role.
Process of complication of concepts leads finally to natural fields of science.}{109, p 242, 243}

\CiteQuotation{Form of object ... is set of relations between its parts. Therefore,
we should speak about matter connections
of elements of the world, which in their totality define spacetime.
//
The simplest element of world is something that we call event. It is "point" phenomenon
like flash of bulb. Using visual concepts about spacetime this is phenomenon that we can disregard its
extent in space and time... Event is like point in geometry... Any phenomenon, any process occurs as
connected set of events.
//
Abstracting from all properties of event apart from it exists we represent it as ...
"world point". spacetime is set of all world points.}{\citeBib{Cite: 104}, p 133, 134}

However, such definition is incomplete.
It does not consider that
\CiteQuote{any event ... influences on any other events and it itself is influenced
by other events. In general, an influence is movement that connects one event with other through row
of intermediate events... Using concepts of physics, we can define influence transmission of momentum
and energy.}{\citeBib{Cite: 104}, p 134}

Geometry of spacetime in that way is inseparable from physical processes running in spacetime.

\CiteQuotation{Spacetime is set of all events in the world such that they are abstract from all its properties
except ones that are defined by relation of influence one event on another.
//
Spacetime structure of the world is its causal-investigatory structure acquired in proper abstraction.}
{\citeBib{Cite: 104}, p 135}

Next question that we want to study is question about physical content of concept of coordinates.
To understand all importance of this question let recall how introduced Einstein common time in relativity.
First, each observer has its own clock. To insure that time of each observer is coordinate we
need synchronization of the clocks. However, process of synchronization is based on physical phenomena.

\CiteQuotation{If at the point $A$ of space there is a clock, an observer at $A$ can determine
the time values of events
in the immediate proximity of $A$ by finding the positions of the hands which are simultaneous with these events.
If there is at the point $B$ of space another clock in all respects resembling the one at $A$",
it is possible for an observer at $B$ to determine the time values of events in the immediate
neighbourhood of $B$.
But it is not possible without further assumptions to compare, in respect of time, an event at
$A$ with an events at $B$.
We have not defined a common "time" for $A$ and $B$, for the latter cannot be defined at all
unless we establish by definition that the "time" required by light to travel from $A$ to $B$ equal the "time"
it requires to travel from $B$ to $A$. Let a ray of light start at the "$A$ time" $t_A$ from $A$
toward $B$, let it at the "$B$ time" $t_B$ to be reflected at $B$ in the direction of $A$,
and arrive again at $A$
at "$A$-time" $t'_A$.
//
In accordance with definition the two clocks synchronize if
\[t_B - t_A = t'_A - t_B\]
//
We assume that this definition of synchronism is free from
contradictions, and possible for any number of points; and that the following
relations are universally valid: -
//
1) If the clock at $B$ synchronizes with the clock at $A$, the clock at $A$ synchronizes with the clock at $B$;
//
2) If the clock at $A$ synchronizes with the clock at $B$ and also with the clock at $C$, the clocks at $B$ and $C$
also synchronize with each other.}{\citeBib{Einstein: Electrodynamics of Moving Bodies}, pp. 39, 40}

Similar construction based on measurement of distance between observers (we can do it also with help of light signal)
leads to concept of space coordinates.

Problem of coordinates in general relativity is more complex; however, its main point remains:
coordinates are concentrated expression of interaction of different observers in spacetime.

\section{Covariance Principle}
\labelSection{Covariance Principle}
\epigraph
{Movement is essence of time and space, because it is universal;
to understand it means to tell its essence in form of concept.}
      {Gegel, Lectures about history of philosophy}
\epigraph
{Movement is essence of time and space. Two main concepts express this essence:
(infinite) continuity and "punctuality" (= negation of continuity, discontinuity). Movement is unity
of continuity (of time and space) and discontinuity (of time and space). Movement is contradiction,
is unity of contradictions.}
      {Lenin, Philosophic notebooks}
After definition of space, we can turn to main principles of general relativity.
General covariance principle and equivalence principle are main principles of general relativity.
There is close connection between them. Einstein formulates main principles guided by Mach principle.

\CiteQuotation{Let $K$ is Galiley coordinate system, i.e. certain mass enough remote from others
moves straightforward and uniformly relative this system (at least in considered 4 dimension area).
Let $K'$ is second coordinate system which moves uniformly accelerated relative $K$. Then enough isolated from other mass performs
relative $K'$ accelerated movement; moreover nor acceleration nor direction does not depend from chemical structure
or physical state of this mass.
//
Can observer, rest relative coordinate system $K'$, conclude from this that he is
in "really" accelerated coordinate system? Answer on this question has to be negative because
we can differently; explain behavior of masses free moving relative $K'$. Coordinate system $K'$ does not have
acceleration, however in considered area of spacetime there is gravitational field which causes  accelerated
movement of bodies relative $K'$.}{\citeBib{Einstein: Foundations of general relativity},
pp. 521, 522}

However, Einstein made in this manuscript two mistakes.
Stating foundations of general relativity he writes%

\CiteQuotation{We see from the considerations that creation of general relativity must lead to
theory of gravitation because we can "create" gravitational field simply changing
coordinate system.}{\citeBib{Einstein: Foundations of general relativity}, p. 522}

First, close connection of main principles does not mean their equality. Equivalence principle was
historically first and the simplest principle formulated in general covariant form. Later Einstein realized that
he cannot reduce general relativity to gravitational field. He spent the rest of his life to development of covariance principle.

Second. Einstein by mistake identified Lorentz transformation with general coordinate transformation as result
of identification reference frame and coordinate system. The grossest error was that Einstein supposed that
it is possible to create gravitational field in inertial reference frame using Lorentz transformation even
we did not observe it before.
However this contradicts to covariance principle.
Actually, we have here transfer to noninertial reference frame
expressed in appearance of nonholonomity of coordinates used by observer.

\CiteQuotation{... It is impossible to substitute field of gravity by state of movement of system without
gravitational field as well as it is impossible to transform all points of arbitrarily moving thread to rest
by relativistic transformation.}{107, p 166}

However, isolated observer does not know which coordinate he uses. From other side in inertial reference frame and
in gravitational field free bodies move along trajectory which does not depend on mass.

\CiteQuotation{Equivalence principle is statement about complete equivalence of all physical processes and
phenomena in homogeneous field of gravity and in appropriate uniformly accelerated reference frame.
In general case we speak about enough small spacetime areas. Equality of inertial mass and gravitational mass
follows from equivalence principle because otherwise mechanical movement in accelerated system would be different
from mechanical movement in field of gravity.}{Ginzburg, 105, p 339}

Analysis that I made recently shows that equivalence principle does not relates to equality of two different types of mass.
The main point of equivalence principle is that no one phenomenon
can distinguish one reference frame from another and laws of physics do not depend on which reference frame we use.
Learning of electro magnetic fields shows that mass is less important in general relativity
than we believed before. Not mass but momentum-energy tensor creates gravitational field.
Light does not have mass but it has momentum.
Einstein fixed his errors and formulated main principles next way:
\CiteQuote{Let $K$ is inertial system without field of gravity, $K'$ is coordinate system which moves uniformly accelerated relative $K$.
Then behavior of matter point relative system $K'$ will be the same as when $K'$
is inertial system where homogeneous field of gravity
exists. Thus, definition of inertial system appears unfounded on the base of known from experience properties of gravity.
Arises idea, that from point of view formulating law of nature each any way moving reference frame tantamount any other and
therefore for areas of finite extent there are no physically chosen (privileged) states of movement (general relativity).
Consistent realization of this idea demands more deep then in special relativity
modification of geometric and kinematics fundamentals of theory... Generalizing we come to next result:
field of gravity and metrics represent different forms of manifestation of
the same physical field.}{\citeBib{Einstein: Main problems of general relativity}}

Modern formulating of covariance principle does not depend on equivalence principle. Main point is
that Lorentz transformations create definite group and physical values are invariant structures relative this group.

I want to point out, different erroneous interpretations of statements
of general relativity and Einstein's statements
(in particular, early) deplete content of general relativity.
Thus identifying both principles Fock claimed that general relativity would be
better called gravitation theory. Others authors consider that either
principle like the Mach principle served as midwife and is not
the basic principle of the theory anymore.
Ambiguity of definition of Lorentz transformation
caused wide research. Researchers mistaken simplified problem, choosed privileged reference frame.
However, this contradicts to equivalence principle.

\CiteQuotation{Essential achievement of general relativity consists in escape of physics to introduce "inertial
system". The latter concept is unsatisfactory because it chooses some systems from all conceptually possible coordinate systems
without any proof. Then it assumes that laws of physics apply only for such inertial systems...
On this way, space as such receives
part choosing it from other elements of physical description. It plays certain part in all processes but does not
meet back action. Such theory is possible;
however, it does not appear satisfactory. Newton quite realized this imperfection;
but he as well realized that he does not have another way at his time. Ernst Mach put particular attention on this circumstance
between physicists of posterior time.}{106, p 854, 855}

Ginzburg writes systematizing erroneous conceptions

\CiteQuotation{... Name "general relativity" is completely naturally and we do not have reasons to decline it. Also change
of this name seems impossible because of established custom.
//
Unfortunately, the questions of terminology and word usage interlace such close with existing problems
that they very often prevent discussion of these problems ... and force to argue about words... Question about name
of theory ... such clear terminological and cannot create divergence of the essentials.
However, we cannot tell the same about problem of existing privileged reference frame in general relativity...
Privileged reference frame in general relativity exists for small enough area in spacetime ... i.e.
free falling local inertial reference frame ... where there are no forces of gravity and special relativity is correct.
However, such systems ... do not coincide with inertial systems of classical mechanics...
//
We do not see any analogue of inertial system, any so privileged system for area of finite extend.
Opportunity of terminological disagreement appears here. Research of specific problems is associated with simplification,
approximation, idealization of situation... However quite obviously particular case of such privilege
which is different from privilege of inertial system of classical mechanics.}{105, p 343 - 345}

\section{Spacetime and Quantum}
\labelSection{Spacetime and Quantum}
\epigraph{Every man is prisoner of his own ideas, and everybody must to blow up it in his youth
to try comparing his ideas with reality. However,
it is possible through several centuries that other man will reject his ideas.
It is impossible in case of artist for his peculiarity. It happens only on the way to truth.
And this is not sorrow.}
      {\citeBib{Einstein: On Science}}

Ideas are close associated with physics. In macrophysics, all bodies have accurately definite form
and move along accurately definite trajectories and geometry
 (in the variety of representations and theories)
describes uniquely defined structures.
General relativity is not exception.

In particular simple event is event that we can disregard duration and size.
When we consider quantum processes such definition becomes incorrect.
Point in space, trajectory of movement,
form of body become abstractions, which do not have real analogue. However,
we cannot disregard duration or size in most cases.
Disappearance of border existing in macrophysics between waves and particles leads to washing out accurate geometric forms.
Therefore, geometric forms come to contradiction with physics.

Joint analysis of equations of quantum mechanics and general relativity leads to paradoxical conclusion:
we need to quantize metrics. We still do not have appropriate geometry.
We yet not have corresponding
geometry, However simple constructions allow to say
something definite about geometric structure of space of
general relativistic quantum mechanics.

To understand the kernel of new geometric consepts let us consider visual
geometric example.
Let us consider the manifold of dimention $2$.
Assume that as result of problem solving we have got
two metrics describing spheres of raddii $R$ and $r$. Thus, in any point of
manifold observer will be simultaneously on both spheres.
Assume that we are on north pole. Then we discover that both spheres are tangent 
in north pole. Let us move to south. However the question arises: along which
sphere will we move? Assume that we will move along large sphere.
Then we stop to be tangent to small sphere. However
this contradicts to assumption. Therefore,
during movement along large sphere we carry along small sphere. Similarly, moving
along small sphere, we carry along large sphere.

Undoubtedly, this picture does not depend neither from dimention of manifold nor
from number and
value of metrics. Now we can sum up.

- Quantization of metric tensor leads to foliating of main manifold
(according to value of metric tensor).

- Quantization of metric tensor pushes out moving object from main manifold
into manifold of tangent plains. We saw this process in general relativity: system of
measuring instruments of observer is in tangent plains. During movement of observer the main
manifold moves without sliding along plains and is tangent to them all time.

- Parallel transfer along given line is ambiguous and apparently is irreversible operation.
Geodesic line is no more line as we use it in regular geometry and
disintegrate into set of lines.
Movement along these lines occurs certain probability. This leads to ambiguous
representation of measuring instrument which observer uses.

Terra incognita of mathematical and physical effects follows farther. The time is not far distant
when we are witnesses of new achivements in geometry, physics, phylosophy.

\CiteQuotation{Before I turn to question about completion of general relativity, I must to state
my position relative physical theory that achieved the greatest progress from all physical theories of our time.
I keep in mind statistic quantum mechanics... This is only modern theory giving orderly explanation to our knowledge
of quantum nature of micro-mechanical processes. This theory from one hand and general relativity from other
are considered correct in a sense although merging these theories did not turn out well so far though all efforts.
This may be the reason that between modern physicist-theorist there are absolutely different opinions how
theoretical foundation of future physics will look like. Will it be field
theory?}{\citeBib{Einstein: Autobiographical Notes}, p. 288, 289}

\CiteQuotation{B. Field theory is not yet completely defined by system field equations. Must we acknowledge
existence of singularity? Must we postulate boundary conditions?
//
C. May we think that field theory will allow to understand atomistic and quantum structure?..
I suppose nobody knows know something reliable because we do not know how and to what extent elimination of singularities
reduces set of solutions. We do not have at all any method for methodical getting of solutions that are free from singularities...
Now prevails opinion that before we need transfer field theory to statistic theory of probability using "quantization"...
I see here only attempt to describe relationship of essentially nonlinear nature using linear methods.
//
D. We can conclusively prove that reality cannot be described by continuous field.
It follows from quantum phenomena that finite system with finite energy may be described by finite set of numbers
(quantum numbers). It does not allow combining this with theory of continuum and demand algebraic theory
to describe reality. However now nobody knows how to find basis for such theory.}{106, p 872, 873}

Before we find new geometry we can study what may happens if we join ideas of general relativity and quantum mechanics.

%% file: Representation.English.tex
\input{Representation.Eq}
\ifx\PrintBook\undefined

\section{Introduction}

This paper was written under the great influence of the book \citeBib{Rashevsky}.
The studying of a homogenous space of a group of symmetry of a vector space
leads us to the definition of a basis of this space
and a basis manifold.
We introduce two types of transformation of a basis manifold:
active and passive transformations. The difference between them is
that the active transformation can be expressed as a transformation of
an original space.
As it is shown in \citeBib{Rashevsky} passive transformation
gives ability to define
concepts of invariance and of geometric object.

Two opposite points of view about a geometric object
meet in definition \RefDefinition{geometric object}.
On the one hand we determine coordinates of the geometric object
relative to a given basis and introduce the law of transition of
coordinates during transformation of the basis.
At the same time we study the set of coordinates
of the geometric object relative to different bases
as a single whole. This gives us an opportunity
to study the geometric object without using coordinates.
\else
\chapter{Representation of Group}
\labelChapter{Representation of Group}
\fi
\section{Representation of Group}
\labelSection{Representation of Group}

\begin{definition}
\labelDefinition{nonsingular transformation} 
We call the map
\[
t:M\rightarrow M
\]
\AddIndex{nonsingular transformation}{nonsingular transformation},
if there exists inverse map.
\qed
\end{definition}

\begin{definition}
Let
\ShowEq{product of transformations}
be product of transformations $k$ and $l$.
Transformations is called
\AddIndex{left-side transformation}{left-side transformation}
or it acts from left
\[
u'=t u
\]
if
\ShowEq{left-side transformation}
We denote
\ShowEq{set of left-side nonsingular transformations}
the set of left-side nonsingular transformations of set $M$.
\qed
\end{definition}

\begin{definition}
Let
\ShowEq{product of transformations}
be product of transformations $k$ and $l$.
Transformations is called
\AddIndex{right-side transformation}{right-side transformation}
or it acts from right
\[
u'= ut
\]
if
\ShowEq{right-side transformation}
We denote
\ShowEq{set of right-side nonsingular transformations}
the set of right-side nonsingular transformations of set $M$.
\qed
\end{definition}

We denote
\ShowEq{identical transformation}
identical transformation.

\begin{definition}
\labelDefinition{left-side representation of group}
Let $l(M)$ be a group
and $\delta$ be unit of group $l(M)$.
Let $G$ be group.
We call a homomorphism of group
\[
f:G\rightarrow l(M)
\]
\AddIndex{left-side representation}
{left-side representation} of group\,\footnote{
The theory of group representation is a special case of the theory of
representation of universal algebra
\citeBib{1908.04418}.
}
$G$ in set $M$ if map $f$ holds
\begin{equation}
f(ab)u=f(a)(f(b)u)
\EqLabel{left-side representation of group}
\end{equation}
\qed
\end{definition}

\begin{definition}
\labelDefinition{right-side representation of group} 
Let $r(M)$ be a group
and $\delta$ be unit of group $r(M)$.
Let $G$ be group.
We call a homomorphism of group
\[
f:G\rightarrow r(M)
\]
\AddIndex{right-side representation}
{right-side representation} of group
$G$ in set $M$ if map $f$ holds
\begin{equation}
uf(ab)=(uf(a))f(b)
\EqLabel{right-side representation of group}
\end{equation}
\qed
\end{definition}

Any statement which holds for left-side representation of group
holds also for right-side representation.
For this reason we use the common term
\AddIndex{representation of group}{representation of group}
and use notation for left-side representation
in case when it does not lead to misunderstanding.

\begin{theorem}
\labelTheorem{inverse transformation}
For any $g\in G$
\begin{equation}
\EqLabel{inverse transformation}
f(g^{-1})=f(g)^{-1}
\end{equation}
\end{theorem}
\begin{proof}
Since \EqRef{left-side representation of group} and
\begin{equation}
f(e)=\delta
\EqLabel{right-side identical transformation}
\end{equation}
we have
\[
u=\delta u=f(gg^{-1})u=f(g)(f(g^{-1})u)
\]
This completes the proof.
\end{proof}

\begin{example}
Let $G$ be group.
The group operation determines two different representations on the set $G$:
the \AddIndex{left shift}{left shift, group} which we introduce by the equation
\ShowEq{left shift =}
\ShowEq{left shift}
and the \AddIndex{right shift}{right shift, group} which we introduce by the equation
\ShowEq{right shift =}
\ShowEq{right shift}
\qed
\end{example}

\begin{theorem}
Let representation
\[
u'=f(a)u
\]
be left-side representation. Then representation
\ShowEq{right-side representation h}
is right-side representation.
\end{theorem}
\begin{proof}
Statement follows from chain of equations
\ShowEq{right-side representation h 1}
\end{proof}

\begin{definition}
Let $f$ be left-side representation of the group $G$ in set $M$.
For any $v\in M$ we define
\AddIndex{orbit of representation of the group}
{orbit of representation of group} $G$ as set
\ShowEq{orbit of left representation of group}
\ShowEq{show orbit of left representation of group}
\qed
\end{definition}

\begin{definition}
Let $f$ be right-side representation of the group $G$ in set $M$.
For any $v\in M$ we define
\AddIndex{orbit of representation of the group}
{orbit of representation of group} $G$ as set
\ShowEq{orbit of right representation of group}
\ShowEq{show orbit of right representation of group}
\qed
\end{definition}

Since $f(e)=\delta$ we have
\ShowEq{u in orbit of v}vv.u in orbit of v

\begin{theorem}
\labelTheorem{proper definition of orbit}
Suppose
\ShowEq{orbit, proposition}
Then
\ShowEq{orbit u=orbit v}
\end{theorem}
\begin{proof}
From \EqRef{orbit, proposition} it follows
that there exists $a\in G$ such that
\begin{equation}
v=f(a)u
\EqLabel{orbit, 1}
\end{equation}
Suppose
\ShowEq{u in orbit of v}wv.
Then there exists $b\in G$ such that
\begin{equation}
w=f(b)v
\EqLabel{orbit, 2}
\end{equation}
If we substitute \EqRef{orbit, 1}
into \EqRef{orbit, 2} we get
\begin{equation}
w=f(b)(f(a)u)
\EqLabel{orbit, 3}
\end{equation}
Since \EqRef{left-side representation of group},
we see that from \EqRef{orbit, 3} it follows
that
\ShowEq{u in orbit of v}wu.
Thus
\DrawEq[vu]{orbit v in orbit u}{vu}

Since \EqRef{inverse transformation},
we see that
from \EqRef{orbit, 1} it follows
that
\begin{equation}
u=f(a)^{-1}v=f(a^{-1})v
\EqLabel{orbit, 4}
\end{equation}
From \EqRef{orbit, 4} it follows that
\ShowEq{u in orbit of v}uv{}
and therefore
\DrawEq[uv]{orbit v in orbit u}{uv}
The statement
\EqRef{orbit u=orbit v}
follows from statements
\eqRef{orbit v in orbit u}{vu},
\eqRef{orbit v in orbit u}{uv}.
\end{proof}

\begin{theorem}
\labelTheorem{direct product of representations}
Suppose $f_1$ is representation of group $G$ in set $M_1$ and
$f_2$ is representation of group $G$ in set $M_2$.
Then we introduce
\AddIndex{direct product of representations
$f_1$ and $f_2$ of group}
{direct product of representations of group}
\begin{align*}
f&=f_1\otimes f_2:G\rightarrow M_1\otimes M_2\\
f(g)&=(f_1(g),f_2(g))
\end{align*}
\end{theorem}
\begin{proof}
To show that $f$ is a representation,
it is enough to prove that $f$ satisfies the definition
\RefDefinition{left-side representation of group}.
\[f(e)=(f_1(e),f_2(e))=(\delta_1,\delta_2)=\delta\]
\begin{align*}
f(ab)u&=(f_1(ab)u_1,f_2(ab)u_2)\\
&=(f_1(a)(f_1(b)u_1),f_2(a)(f_2(b)u_2))\\
&=f(a)(f_1(b)u_1,f_2(b)u_2)\\
&=f(a)(f(b)u)
\end{align*}
\end{proof}

\section{Single Transitive Representation of Group}
\labelSection{Single Transitive Representation of Group}

\begin{definition}
We call
\AddIndex{kernel of inefficiency of representation of group}
{kernel of inefficiency of representation of group} $G$
a set \[K_f=\{g\in G:f(g)=\delta\}\]
If $K_f=\{e\}$ we call representation of group $G$
\AddIndex{effective}{effective representation of group}.
\qed
\end{definition}

\begin{theorem}
A kernel of inefficiency is a subgroup of group $G$.
\end{theorem}
\begin{proof}
Assume
$f(a_1)=\delta$ and $f(a_2)=\delta$. Then
\[f(a_1a_2)u=f(a_1)(f(a_2)u)=u\]
\[f(a^{-1})=f^{-1}(a)=\delta\]
\end{proof}

If an action is not effective we can switch to an effective one
by changing group $G_1=G|K_f$
using factorization by the kernel of inefficiency.
This means that we can study only an effective action.

\begin{definition}
We call a representation of group
\AddIndex{transitive}{transitive representation of group}
if for any $a, b \in V$ exists such $g$ that
\[a=f(g)b\]
We call a representation of group
\AddIndex{single transitive}{single transitive representation of group}
if it is transitive and effective.
\qed
\end{definition}

\begin{theorem}
Representation is single transitive if and only if for any $a, b \in V$
exists one and only one $g\in G$ such that $a=f(g)b$
\end{theorem}

\begin{definition}
We call a space $V$
\AddIndex{homogeneous space}{homogeneous space} of group $G$
if we have single transitive representation of group $G$ on $V$.
\qed
\end{definition}

\begin{theorem}%
\labelTheorem{single transitive representation of group}
If we define a single transitive representation $f$ of the group $G$ on the manifold $A$
then we can uniquely define coordinates on $A$ using coordinates on the group $G$.

If $f$ is a covariant representation then $f(a)$ is equivalent to the left shift $L(a)$ on the group $G$.
If $f$ is a contravariant representation then $f(a)$ is equivalent to the right shift $R(a)$ on the group $G$.
\end{theorem}
\begin{proof}
We select a point $v\in A$
and define coordinates of a point $w\in A$
as coordinates of the transformation $a$ such that $w=f(a) v$.
Coordinates defined this way are unique
up to choice of an initial point $v\in A$
because the action is effective.

If $f$ is a covariant representation we will use the notation
\[f(a)v=av\]
Because the notation
\[f(a)(f(b)v)=a(bv)=(ab)v=f(ab)v\]
is compatible with the group structure we see that the covariant representation $f$ is equivalent to the left shift.

If $f$ is a contravariant representation we will use the notation
\[f(a)v=va\]
Because the notation
\[f(a)(f(b)v)=(vb)a=v(ba)=f(ba)v\]
is compatible with the group structure we see that the contravariant representation $f$ is equivalent to the right shift.
\end{proof}

\begin{theorem}%
\labelTheorem{shifts on group commuting}
Left and right shifts on group $G$ are commuting.
\end{theorem}
\begin{proof}
This is the consequence of the associativity on the group $G$
\[(L(a) R(b))c = a(cb)=(ac)b=(R(b) L(a))c\]
\end{proof}

\begin{theorem}%
\labelTheorem{two representations of group}
If we defined a single transitive representation $f$ on the manifold $A$
then we can uniquely define a single transitive representation $h$
such that diagram
\[
\xymatrix{
M\ar[rr]^{h(a)}\ar[d]^{f(b)} & & M\ar[d]^{f(b)}\\
M\ar[rr]_{h(a)}& &M
}
\]
is commutative for any $a$, $b\in G$.\,\footnote{The theorem
\RefTheorem{two representations of group}
is really very interesting. However its
meaning becomes more clear when we apply this theorem to basis manifold,
see section \RefSection{Basis in Vector Space}.}
\end{theorem}
\begin{proof}
We use group coordinates for points $v\in A$.
For the simplicity we assume that $f$ is a covariant representation.
Then according to theorem \RefTheorem{single transitive representation of group}
we can write the left shift $L(a)$ instead of the transformation $f(a)$.

Let points $v_0, v\in A$. Then we can find
one and only one $a\in G$ such that
\[v=v_0 a=R(a) v_0\]
We assume
\[h(a)=R(a)\]
For some $b\in G$ we have
\[w_0=f(b)v_0=L(b) v_0\ \ \ \ w=f(b)v=L(b) v\]
According to theorem \RefTheorem{shifts on group commuting} the diagram
\begin{equation}
\xymatrix{
v_0\ar[rr]^{h(a)=R(a)}\ar[d]^{f(b)=L(b)} & & v\ar[d]^{f(b)=L(b)}\\
w_0\ar[rr]_{h(a)=R(a)}& &w
}
\label{Diagram: two representations of group}
\end{equation}
is commutative.

Changing $b$ we get that $w_0$ is an arbitrary point of $A$.

We see from the diagram that if $v_0=v$ then $w_0=w$ and therefore $h(e)=\delta$.
On other hand if $v_0\neq v$ then $w_0\neq w$ because the representation $f$ is single transitive.
Therefore the representation $h$ is effective.

In the same way we can show that for given $w_0$ we can find $a$
such that $w=h(a)w_0$. Therefore the representation is single transitive.

In general the representation $f$ is not commutative and therefore
the representation $h$ is different from the representation $f$.
In the same way we can create a representation $f$ using the representation $h$.
\end{proof}

\begin{remark}
\labelRemark{one representation of group}
It is clear that transformations $L(a)$ and $R(a)$
are different until the group $G$ is nonabelian.
However they both are maps onto.
Theorem \RefTheorem{two representations of group} states that if both
right and left shift presentations exist on the manifold $A$
we can define two commuting representations on the manifold $A$.
The left shift or the right shift only cannot present both types of representation.
To understand why it is so let us change diagram \eqref{Diagram: two representations of group}
and assume $h(a)v_0=L(a)v_0=v$ instead of $h(a)v_0=R(a)v_0=v$ and let
us see what expression $h(a)$ has at
the point $w_0$. The diagram
\[\xymatrix{
v_0\ar[rr]^{h(a)=L(a)}\ar[d]^{f(b)=L(b)} & & v\ar[d]^{f(b)=L(b)}\\
w_0\ar[rr]_{h(a)}& &w
}\]
is equivalent to the diagram
\[\xymatrix{
v_0\ar[rr]^{h(a)=L(a)} & & v\ar[d]^{f(b)=L(b)}\\
w_0\ar[rr]_{h(a)}\ar[u]_{f^{-1}(b)=L(b^{-1})}& &w
}\]
and we have $w=bv=bav_0=bab^{-1}w_0$. Therefore
\[h(a)w_0=(bab^{-1})w_0\] We see that the representation of $h$
depends on its argument.
\qed  
\end{remark}

\ePrints{0803.3276}
\ifx\Semafor\ValueOff
\section{Linear Representation of Group}
\labelSection{Linear Representation of Group}

Suppose we introduce additional structure on set $M$.
Then we create an additional requirement for the representation of group.

Since we introduce continuity on set $M$,
we suppose that transformation
\[
u'=f(a)u
\]
is continuous in $u$.
Therefore, we get
\[\left|\frac { \partial u'} {\partial u}\right|\neq 0\]

Suppose $M$ is a group. Then representations of
left and right shifts have great importance.

\begin{definition}
Let $M$ be vector space $V$ over field $F$.
We call the representation of group $G$ in vector space $V$
\AddIndex{linear representation}{linear representation of group}
if $f(a)$ is linear map of space $V$ for any $a\in G$.
\qed
\end{definition}

\begin{remark}
\labelRemark{row and column vectors, vector space}

Let transformation $f(a)$ be linear
homogeneous transformation.
$f_\gamma^\beta(a)$ are elements of the matrix of
transformation. We usually assume that the lower index
enumerates rows in the matrix and the upper index enumerates columns.

According to the matrix product rule we can
present coordinates of a vector as a row of a matrix.
We call such vector a \AddIndex{row vector}{row vector}.
We can also study a vector whose coordinates
form a column of a matrix.
We call such vector a \AddIndex{column vector}{column vector}.

Left-side linear representation in column vector space
\[
u'=f(a)u\ \ \ \ u'_\alpha=f_\alpha^\beta(a)u_\beta\ \ \ \ a\in G
\]
is covariant representation
\[
u''_\gamma=f_\gamma^\beta(ba) u_\beta
=f_\gamma^\alpha(b)(f_\alpha^\beta(a) u_\beta)
=(f_\gamma^\alpha(b)f_\alpha^\beta(a)) u_\beta
\]

Left-side linear representation in row vector space
\[
u'=f(a)u\ \ \ \ u'^\alpha=f^\alpha_\beta(a)u^\beta\ \ \ \ a\in G
\]
is contravariant representation
\[
u''^\gamma=f^\gamma_\beta(ba) u^\beta
=f^\gamma_\alpha(b)(f^\alpha_\beta(a) u^\beta)
=(f^\alpha_\beta(a)f^\gamma_\alpha(b)) u^\beta
\]

Right-side representation in column vector space
\[
u'=uf(a)\ \ \ \ u'_\alpha=u_\beta f_\alpha^\beta(a)\ \ \ \ a\in G
\]
is contravariant representation
\[
u''_\gamma=u_\beta f_\gamma^\beta(ab)
=(u_\beta f_\alpha^\beta(a))f_\gamma^\alpha(b)
=u_\beta(f_\gamma^\alpha(b)f_\alpha^\beta(a)) 
\]

Right-side representation in row vector space
\[
u'=uf(a)\ \ \ \ u'^\alpha=u^\beta f^\alpha_\beta(a)\ \ \ \ a\in G
\]
is covariant representation
\[
u''^\gamma=u^\beta f^\gamma_\beta(ab)
=(u^\beta f^\alpha_\beta(a))f^\gamma_\alpha(b)
=u^\beta(f^\alpha_\beta(a)f^\gamma_\alpha(b)) 
\]
\qed
\end{remark}

\begin{remark}
\labelRemark{covariant and contravariant representations}
Studying linear representations
we clearly use tensor notation.
We can use only upper index and notation $u{}^\cdot{}_\alpha^.$ instead of
$u_\alpha$. Then we can write the transformation of this object in the form
\[u'{}^\cdot{}_\alpha^.=f{}^\cdot{}_\alpha^.{}{}_\cdot{}^\beta_. u{}^\cdot{}_\beta^.\]
This way we can hide the difference between covariant and
contravariant representations. This similarity goes
as far as we need.
\qed
\end{remark}
\fi

%% file: Representation.Eq.tex

\DefEq
{
\symb{l(M)}{set of left-side nonsingular transformations}1
}
{set of left-side nonsingular transformations}

\DefEq
{
\symb{r(M)}{set of right-side nonsingular transformations}1
}
{set of right-side nonsingular transformations}

\AddEq{right-side representation h}
{
\[
u'=uh(a)=f(a^{-1})u
\]
}

\AddEq{orbit of left representation of group}
{
\symb{f(G)v}{orbit of representation}l
}

\AddEq{show orbit of left representation of group}
{
\[
\ShowSymbol{orbit of representation}l
=\{w=f(g)v:g\in G\}
\]
}

\AddEquation{orbit, proposition}
{
v\in f(G)u
}

\AddEquation{orbit u=orbit v}
{
f(G)u=f(G)v
}

\AddEq[2]{orbit v in orbit u}
{
f(G)#1\subseteq f(G)#2
}

\AddEq[3]{u in orbit of v}
{
$#1\in f(G)#2$#3
}

\AddEq{orbit of right representation of group}
{
\symb{vf(G)}{orbit of representation}r
}

\AddEq{show orbit of right representation of group}
{
\[
\ShowSymbol{orbit of representation}r
=\{w=vf(g):g\in G\}
\]
}

\AddEq{product of transformations}
{
\[
ts=t\circ s
\]
}

\AddEq{right-side transformation}
{
\[
(vs)t=v(st)
\]
}

\AddEq{left-side transformation}
{
\[
t(sv)=(ts)v
\]
}

\AddEq{right-side representation h 1}
{
\begin{align*}
uh(ab)&=f((ab)^{-1})u=f(b^{-1}a^{-1})u=f(b^{-1})(f(a^{-1})u)
=f(b^{-1})(uh(a))\\&=(uh(a))h(b)
\end{align*}
}

\DefEq
{
\symb{R(a)b}{right shift}{}
}
{right shift =}

\DefEquation
{
b'=\ShowSymbol{right shift}{}=ba
}
{right shift}

\DefEq
{
\symb{L(a)b}{left shift}{}
}
{left shift =}

\DefEquation
{
b'=\ShowSymbol{left shift}{}=ab
}
{left shift}

\DefEq
{
\symb{\delta}{identical transformation}1
}
{identical transformation}

%% file: Basis.English.tex

\input{Basis.Eq}

\ifx\PrintBook\undefined
\else
\chapter{Basis Manifold}
\labelChapter{Basis Manifold}
\fi

\section{Basis in Vector Space}
\labelSection{Basis in Vector Space}

\begin{definition}
\labelDefinition{vector space over field}
Let $D$ be field.
Abelian group \symb{V}{V}1 is called
\AddIndex{vector space}{vector space} over field $D$
if we define the product\,\footnote{
Because product in the field $D$ is commutative,
it doesn't matter to us,
we write the product in the form of $av$ or $va$.
I choose notation $va$ because I want to use
matrix representation of vector.
}
\ShowEq{f:A->B}{*}{V\times D}V
\ShowEq{f:A->B}{*}{D\times V}V
which satisfies to following conditions:
\StartLabelItem
\begin{enumerate}
\item 
\AddIndex{comutative law}{comutative law}
\labelItem{comutative law, vector space}
\ShowEq{comutative law}
\item 
\AddIndex{associative law}{associative law}
\labelItem{associative law, vector space}
\DrawEq{associative law, right module}1
\item 
\AddIndex{distributive law}{distributive law}
\labelItem{distributive law, vector space}
\ShowEq{distributive law, right module}
\item
\AddIndex{unitarity law}{unitarity law}
\labelItem{unitarity law, vector space}
\ShowEq{unitarity law, right A-module}
\end{enumerate}
for any
\ShowEq{p,q in D, v,w in V}
\qed
\end{definition}

\begin{theorem}
\labelTheorem{set of vectors generated by set of vectors, vector space}
Let $V$ be vector space.
The set of vectors generated by the set of vectors
\ShowEq{vi V}{}
has form\,\footnote{
For a set $A$,
we denote by $|A|$ the cardinal number of the set $A$.
The notation $|A|<\infty$ means that the set $A$ is finite.
}
\ShowEq{w=sum vi, vector space}
\end{theorem}

\begin{proof}
We need to prove following statements:
\ShowEq{vector space generated by set of vectors}

\begin{itemize}
\item
For any
\ShowEq{vk in v}
let
\ShowEq{ci=dik}D
Then
\ShowEq{vk=sum vi, right module}
The statement
\RefItem{vector space, vk in Jv}
follows from
\EqRef{w=sum vi, vector space},
\EqRef{vk=sum vi, right module}.
\item
The statement
\RefItem{vector space over D, cvk in Jv}
follows from the definition
\RefDefinition{vector space over field}
and from the statement
\RefItem{vector space, vk in Jv}.
\item
Since $V$ is Abelian group,
then the statement
\RefItem{vector space over D, sum cvk in Jv}
follows from the statement
\RefItem{vector space over D, cvk in Jv}
and from the definition
\RefDefinition{vector space over field}.
\item
Let
\ShowEq{w12 in Jv over D}
Since $V$ is Abelian group,
then the vector $w_1+w_2$ is generated by the set of vectors $v$.
According to the equality
\EqRef{w=sum vi, vector space},
there exist $D$\Hyph numbers
\ShowEq{gi12}w
such that
\DrawEq[w]{w12= right}{module}
where sets
\DrawEq[w]{|ci12 ne 0|}{right module}
are finite.
Since $V$ is Abelian group,
then from the equality
\eqRef{w12= right}{module}
it follows that
\DrawEq[w]{w1+w2= right}{module}
The equality
\DrawEq[w]{w1+w2= 1 right}{module}
follows from equalities
\EqRef{distributive law, right module, 2},
\eqRef{w1+w2= right}{module}.
From the equality
\eqRef{|ci12 ne 0|}{right module},
it follows that
the set
\ShowEq{|gi 1+2 ne 0|}w
is finite.
Therefore, the statement
\RefItem{vector space over D, v,w in Jv=>v+w in Jv}
is true.
\item
Let
\ShowEq{w in Jv over D}
According to the definition
\RefDefinition{vector space over field},
for any $D$\Hyph number $a$,
the vector $wa$ is generated by the set of vectors $v$.
According to the equality
\EqRef{w=sum vi, vector space},
there exist $D$\Hyph numbers
\ShowEq{set au vi}w
such that
\ShowEq{w= right module}
where
\DrawEq{|wi ne 0|}{vector space}
From the equality
\EqRef{w= right module}
it follows that
\ShowEq{aw= right module}
From the statement
\eqRef{|wi ne 0|}{vector space},
it follows that
the set
\ShowEq{right module, |awi ne 0|}
is finite.
Therefore, the statement
\RefItem{vector space over D, v in Jv=>av in Jv}
is true.
\end{itemize}
\end{proof}

\def\SideWS{right }%
\def\SideNS{right}%
\ShowDefinition{linear combination of vectors}

We represent the set of $D$\Hyph numbers
\ShowEq{set au vi}w
as matrix
\ShowEq{column vector w=wi}w
We represent the set of vectors
\ShowEq{set vi}
as matrix
\ShowEq{row vector w=wi}vn
Then we can represent linear combination of vectors
\ShowEq{w=wi vi\SideNS}
as
\ShowEq{w=vw over D}

\begin{theorem}
\labelTheorem{linearly depends on rest of vectors, vector space}
Since the equation
\ShowEq{\SideWS wi vi=0}
implies existence of index
\ShowEq{i=j}
such that
\ShowEq{wj ne 0},
then the vector $v_{\gij}$
linearly depends on rest of vectors $v$.
\end{theorem}

\ShowProof{linearly depends on rest of vectors}

It is evident that for any set of vectors $v_{\gii}$
\ShowEq{right 0=0vi over D}

\begin{definition}
\labelDefinition{linearly independent vectors}
The set of vectors\,\footnote{
I follow to the definition in
\citeBib{Serge Lang}, page 130.}
\ShowEq{set vi}
of vector space $V$ is
\AddIndex{linearly independent}{linearly independent set}
if $w=0$ follows from the equation
\ShowEq{\SideWS wi vi=0}
Otherwise the set of vectors
\ShowEq{set vi}
is \AddIndex{linearly dependent}{linearly dependent set}.
\qed
\end{definition}

Definitions
\RefDefinition{generating set of vector space},
\RefDefinition{basis of vector space}
follow from the theorem
\RefTheorem{set of vectors generated by set of vectors, vector space}.

\begin{definition}
\labelDefinition{generating set of vector space}
$J(v)$
is called
vector subspace generated by set $v$,
and $v$ is a
\AddIndex{generating set}{generating set}
of vector subspace $J(v)$.
In particular, a
\AddIndex{generating set}{generating set}
of vector space $V$
is a subset $X\subset V$ such that
\ShowEq{generating set of module}
\qed
\end{definition}

\begin{definition}
\labelDefinition{basis of vector space}
If the set $X\subset V$ is generating set of vector space
$V$, then any set $Y$, $X\subset Y\subset V$
also is generating set of vector space $V$.
If there exists minimal set $X$ generating
the vector space $V$, then the set $X$ is called
\AddIndex{basis}{basis} of vector space $V$.
\qed
\end{definition}

\begin{theorem}
\labelTheorem{basis of vector space}
The set of vectors
\ShowEq{basis for module over algebra}
is basis of vector space
$V$, if following statements are true.
\StartLabelItem
\begin{enumerate}
\item
\labelItem{vector is linear combination of set, vector space}
Arbitrary vector $v\in V$
is linear combination of
vectors of the set $\Basis e$.
\item
\labelItem{cannot be represented as a linear combination, vector space}
Vector $e_{\gii}$
cannot be represented as a linear combination
of the remaining vectors of the set $\Basis e$.
\end{enumerate}
\end{theorem}

\begin{proof}
According to the statement
\RefItem{vector is linear combination of set, vector space},
the theorem
\RefTheorem{set of vectors generated by set of vectors, vector space}
and the definition
\RefDefinition{\SideWS linear combination of vectors},
the set $\Basis e$ generates vector space $V$
(the definition
\RefDefinition{generating set of vector space}).
According to the statement
\RefItem{cannot be represented as a linear combination, vector space},
the set $\Basis e$ is minimal set
generating vector space $V$.
According to the definitions
\RefDefinition{basis of vector space},
the set $\Basis e$ is a basis of vector space $V$.
\end{proof}

\begin{theorem}
\labelTheorem{field, vector space}
The set of vectors
\ShowEq{basis, module}
is a
\AddIndex{basis of vector space}{basis, vector space} $V$
if vectors $e_{\gii}$ are linearly independent and any vector $v\in V$
linearly depends on vectors $e_{\gii}$.
\end{theorem}

\begin{proof}
Let the set of vectors
\ShowEq{ei, i in I}
be linear dependent. Then the equation
\DrawEq[w]{c*e=0, \SideWS module}{}
implies existence of index $\gii=\gij$ such that
\ShowEq{wj ne 0}.
According to the theorem
\RefTheorem{linearly depends on rest of vectors, vector space},
the vector $e_{\gij}$
linearly depends on rest of vectors of the set $\Basis e$.
According to the definition
\RefDefinition{basis of vector space},
the set of vectors
\ShowEq{ei, i in I}
is not a basis for vector space $V$.

Therefore, if the set of vectors
\ShowEq{ei, i in I}
is a basis, then these vectors
are linearly independent.
Since an arbitrary vector $v\in V$
is linear combination of vectors
\ShowEq{ei, i in I},
then the set of vectors $v$,
\ShowEq{ei, i in I}
is not linearly independent.
\end{proof}

From the theorem
\RefTheorem{field, vector space},
it follows that $\Basis e$ is a maximal set
of linearly independent vectors.
In case when we want to show clearly that this is the basis
in vector space $V$ we use notation
\ShowEq{basis in V}

\begin{definition}
\labelDefinition{coordinates of vector, vector space}
Let $\Basis e$ be the basis of vector space $V$
and vector
\ShowEq{vv in V}
has expansion
\DrawEq{vv=ve vector space}{}
with respect to the basis $\Basis e$.
$D$\Hyph numbers $v^{\gii}$ are called
\AddIndex{coordinates}{coordinates}
of vector $\Vector v$ with respect to the basis $\Basis e$.
\qed
\end{definition}

Let $\Basis e$ be basis of vector space $V$.
We represent the set of vectors
\ShowEq{set ei}
as matrix
\ShowEq{row vector basis e}
We represent the set of coordinates
\ShowEq{set au v1n}v
of vector $\Vector v$ with respect to the basis $\Basis e$
as matrix
\ShowEq{column vector w=wi}v
Matrix of $v$
is called
\AddIndex{coordinate matrix of vector}{coordinate matrix of vector}
$\Vector v$ in basis $\Basis e$.
Then we can represent the vector $\Vector v$
as product of matrices
\DrawEq [v]{vv=vev}{}

\begin{theorem}
\labelTheorem{coordinates of vector of vector space}
Coordinates of vector $v\in V$ relative to basis $\Basis e$
of vector space $V$
are uniquely defined.
\end{theorem}

\begin{proof}
Let vector $\Vector v$ have expansions
\DrawEq[1]{vv=evi}{1}
\DrawEq[2]{vv=evi}{2}
with respect to basis $\Basis e$.
The equality
\ShowEq{ev (1-2)}
follows from equalities
\eqRef{vv=evi}{1},
\eqRef{vv=evi}{2}.
The statement
\ShowEq{vi1=vi2}
follows from the equality
\EqRef{ev (1-2)}
and from the theorem
\RefTheorem{field, vector space}.
\end{proof}

\begin{theorem}
\labelTheorem{coordinates of linear composition}
We represent the set of coordinates of vectors
\ShowEq{row vector w=wi}vm
as matrix
\ShowEq{row vector coordinates}
Then we represent coordinates of linear composition
\ShowEq{vi ci}
of vectors $v$ as product of matrices
\ShowEq{vc=vm cm}
\end{theorem}

\begin{proof}
The theorem follows from the theorem
\RefTheorem{coordinates of vector of vector space}.
\end{proof}

\begin{theorem}
\labelTheorem{coordinate matrix of basis is not singular}
Let $\Basis e$, $\Basis f$ be bases of vector space $V$.
coordinate matrix
\ShowEq{coordinate matrix nn}e
of basis $\Basis e$ with respect to basis $\Basis f$
is not singular matrix.
\end{theorem}

\begin{proof}
According to the definition
\RefDefinition{linearly independent vectors}
and theorems
\RefTheorem{field, vector space},
\RefTheorem{coordinates of linear composition},
the system of linear equations
\ShowEq{matrix of basis *c=0}
has unique solution.
Therefore, the matrix $e$
is not singular matrix.
\end{proof}

\section{Linear Transformation of Vector Space}

\begin{definition}
\labelDefinition{linear transformation of vector space}
Let $V$ be vector space.
The map
\ShowEq{f:A->B}fVV
is called linear transformation of vector space $V$, if
\ShowEq{f linear map}
for any vectors $v$, $w\in V$.
\qed
\end{definition}

\begin{theorem}
\labelTheorem{linear transformation of vector space}
Let
$\Basis e$
be a basis of vector space $V$.
Then linear transformation
\ShowEq{f:A->B}{\Vector f}VV
has presentation
\ShowEq{f:A->A vector space}
\DrawEq[wfv]{w=f v matrix}1
relative to the basis $\Basis e$. Here
\begin{itemize}
\item $v$ is coordinate matrix of vector
$\Vector v$
relative the basis
\ShowEq{basis e}{}{}
\DrawEq [v]{vv=vev}v
\item $w$ is coordinate matrix of vector
\ShowEq{vw=vf(vv)}
relative the basis
\ShowEq{basis e}{}{}
\DrawEq [w]{vv=vev}w
\item $f$ is coordinate matrix of set of vectors
\ShowEq{Vector f(e1) vector space}
relative the basis
\ShowEq{basis e}{}.
The matrix $f$ is
called \AddIndex{matrix of linear map}{matrix of linear map}
$\Vector f$ relative basis
\ShowEq{basis e}{}.
\end{itemize}
\end{theorem}

\begin{proof}
Since the map
$\Vector f$
is a linear map, then the equality
\ShowEq{vw=f(e1)vv}
follows from equalities
\EqRef{f linear map},
\eqRef{vv=vev}v,
\EqRef{vw=vf(vv)}.
The vector
\ShowEq{vf(e1)}
has expansion
\ShowEq{vf(e1)=}
relative to basis $\Basis e$.
Combining \EqRef{vw=f(e1)vv}
and \EqRef{vf(e1)=}, we get
\ShowEq{w=e f v}
\ShowEq{e w=e f v matrix}
The equality
\eqRef{w=f v matrix}1
follows from comparison of
\eqRef{vv=vev}w
and \EqRef{w=e f v}.
\end{proof}

\begin{theorem}
\labelTheorem{non-singular linear transformation}
The matrix $f$ of linear transformation $\Vector f$
is non\Hyph singular
iff
the linear transformation $\Vector f$ maps
a basis of vector space $V$ into basis.
\end{theorem}

\begin{proof}
Let
\ShowEq{Bases egh}
be bases of vector space $V$.
Let linear transformation $\Vector f$
maps the basis $\Basis e$ to the basis $\Basis g$.
Let
\ShowEq{coordinate matrix nn}e
be coordinate matrix of basis $\Basis e$ with respect to basis $\Basis h$.
Let
\ShowEq{coordinate matrix nn}g
be coordinate matrix of basis $\Basis g$ with respect to basis $\Basis h$.
Let
\ShowEq{coordinate matrix nn}f
be coordinate matrix of the map $\Vector f$ with respect to basis $\Basis h$.
Then the equality
\ShowEq{product g=fe}
follows from the theorem
\RefTheorem{linear transformation of vector space}.
According to the theorem
\RefTheorem{coordinate matrix of basis is not singular},
matrices $e$ and $g$ are not singular.
Therefore, the matrix $f$ also is not singular.

On the other hand, let coordinate matrix $f$ of the map $\Vector f$
be non\Hyph singular.
Since coordinate matrix $e$ of basis $\Basis e$ is non\Hyph singular matrix,
then from the equality
\EqRef{product g=fe},
it follows that the matrix $g$ is also non\Hyph singular matrix.
Therefore, the matrix $g$ is
coordinate matrix of basis $\Basis g$
and the map $\Vector f$ maps basis into basis.
\end{proof}

\begin{theorem}
The set of nonsingular matrices
\ShowEq{GV}
forms a group.
We usually call group $G(V)$
\AddIndex{symmetry group}{symmetry group}.
Group $G(V)$
generates left\Hyph side linear representation
in vector space $V$.
\end{theorem}

\begin{proof}
According to the theorem
\RefTheorem{linear transformation of vector space}
linear transformations
\ShowEq{f:A->B}{\Vector f}VV
\ShowEq{f:A->B}{\Vector g}VV
have form
\DrawEq[wfv]{w=f v matrix}{wfv}
\DrawEq[ugw]{w=f v matrix}{ugw}
The equality
\ShowEq{u=gfv}
follows from equalities
\eqRef{w=f v matrix}{wfv},
\eqRef{w=f v matrix}{ugw}.
The theorem follows from the equality
\EqRef{u=gfv}.
The representation is left-side
according to the rule of matrix product.
\end{proof}

According to the theorem
\RefTheorem{linear transformation of vector space},
we identify element $g$ of group $G(V)$
with corresponding transformation of representation
and write its action on vector $v\in V$ as $gv$.
This point of view allows introduction of two types of coordinates
for element $g$ of group $G(V)$. We can either use coordinates
defined on the group, or introduce coordinates as
elements of the matrix of the corresponding transformation.
The former type of coordinates is more effective when we study
properties of group $G(V)$. The latter type of coordinates
contains redundant data;
however, it may be more convenient
when we study representation of group $G(V)$.
The latter type of coordinates is called
\AddIndex{coordinates of representation}{coordinates of representation}.

Depending on considered properties of vector space,
we may consider a group $G\subseteq G(V)$,
whose transformations preserve considered properties.
Considered group $G$ is called symmetry group
of vector space $V$.

From the theorem
\RefTheorem{non-singular linear transformation},
it follows that we can extend linear transformation
of vector space $V$ to the set of bases.
Thus we can extend a left\Hyph side representation of the symmetry group
to the set of bases.
We write the action of element $g$ of group $G$ on basis $\Basis{e}$ as
\ShowEq{active transformation}
However not every two bases can be mapped by a transformation
from the symmetry group
because not every nonsingular linear transformation belongs to
the representation of group $G$. Therefore, we can present
the set of bases
as a union of orbits of group $G$.

Properties of basis depend on the symmetry group.
We can select basis $\Basis{e}$ vectors of which are
in a relationship which is invariant relative to symmetry group.
In this case all bases from orbit
\ShowEq{bases from orbit ge}
have vectors which satisfy the same relationship.
Such a basis we call
\AddIndex{$G$\Hyph basis}{G-basis}.
In each particular case we need to prove the existence of a basis
with certain properties. If such a basis does not exist
we can choose an arbitrary basis.

\begin{definition}
\labelDefinition{basis manifold of vector space}
We call orbit $G\Basis{e}$
of the selected basis $\Basis{e}$
the \AddIndex{basis manifold
\ShowEq{basis manifold of vector space}
of vector space}{basis manifold of vector space}
$V$.
 \qed
 \end{definition}

\begin{theorem}
\labelTheorem{basis manifold of vector space}
Representation of group $G$ on basis manifold
is single transitive representation.
\end{theorem}
\begin{proof}
According to definition \RefDefinition{basis manifold of vector space}
at least one transformation of representation is defined for any two bases.
To prove this theorem it is sufficient to show
that this transformation is unique.

Consider elements $g_1$, $g_2$ of group $G$ and a basis $\Basis{e}$ such that
\ShowEq{two transformations on basis manifold, 1}
From \EqRef{two transformations on basis manifold, 1} it follows that
\ShowEq{two transformations on basis manifold, 2}
Because any vector has a unique expansion
relative to basis $\Basis{e}$ it follows from
\EqRef{two transformations on basis manifold, 2}
that
\ShowEq{two transformations on basis manifold, 3}
is an identical transformation
of vector space $V$. $g_1=g_2$ because representation
of group $G$ is effective on vector space $V$.
Statement of the theorem follows from this.
\end{proof}

Theorem \RefTheorem{basis manifold of vector space}
means that the basis manifold $\mathcal{B}(V)$
is a homogenous space of group $G$.
We constructed left\Hyph side single transitive linear
representation of group $G$ on the basis manifold.
Such representation is called
\AddIndex{active representation}{active representation}.
A corresponding transformation on the basis manifold is called
\AddIndex{active transformation}{active transformation}
(\citeBib{Korn})
because the linear transformation of the vector space
induced the transformation of basis manifold.

According to theorem \RefTheorem{single transitive representation of group}
because basis manifold $\mathcal{B}(V)$
is a homogenous space of group $G$
we can introduce on  $\mathcal{B}(V)$
two types of coordinates defined on group $G$.
In both cases coordinates of basis $\Basis{e}$ are
coordinates of the homomorphism mapping a fixed basis $\Basis{e}_0$
to the basis $\Basis{e}$.
Coordinates of representation are called
\AddIndex{standard coordinates of basis}
{standard coordinates of basis}.
We can show that standard coordinates
\ShowEq{standard coordinates of basis}
of basis $\Basis{e}$ for certain value of $k$
are coordinates of vectors
\ShowEq{vector of basis}
relative to a fixed basis $\Basis{e}_0$.

Basis $\Basis{e}$ creates coordinates in $V$.
In different types of space it may be done in different ways.
In affine space if node of basis is point $A$ then
point $B$ has the same coordinates as vector $\overset{\longrightarrow}{AB}$
relative basis $\Basis{e}$.
In a general case we introduce coordinates of a vector as coordinates
relative to the selected basis.
Using only bases of type $G$ means using of specific coordinates on $\mathcal{A}_n$.
To distinguish them we call this
\AddIndex{$G$\Hyph coordinates}{G-coordinates}.
We also call the space $V$ with such coordinates
\AddIndex{$G$\Hyph space}{GSpace}.

According to theorem \RefTheorem{two representations of group} another
representation, commuting with active, exists on the basis manifold.
As we see from remark \ref{remark: one representation of group}
transformation of this
representation is different from a active transformation and cannot be reduced to
transformation of space $V$.
To emphasize the difference this transformation is called
a \AddIndex{passive transformation}{passive transformation}
of vector space $V$
and the representation is called
\AddIndex{passive representation}{passive representation}.
We write the passive transformation of basis $\Basis{e}$,
defined by element $g\in G$, as
\ShowEq{passive transformation}

\section{Basis in Affine Space}

We identify vectors of the affine space
\ShowEq{An}
with pair of points $\overset{\longrightarrow}{AB}$.
All vectors that have a common beginning $A$ create a vector space
that we call a tangent vector space $T_A\mathcal{A}_n$.

A topology that $\mathcal{A}_n$ inherits from the map $\mathcal{A}_n\rightarrow R^n$
allows us to study smooth transformations of $\mathcal{A}_n$
and their derivatives. More particularly, the derivative
of transformation $f$ maps the vector space $T_A\mathcal{A}_n$ into $T_{f(A)}\mathcal{A}_n$.
If $f$ is linear then its derivative is the same at every point.
Introducing coordinates $A^1,...,A^n$ of a point $A\in\mathcal{A}_n$ we can write
a linear transformation as
\begin{align}
A'^i&=P^i_j A^j + R^i &\det P\ne 0
\EqLabel{AffineTransformation}
\end{align}
Derivative of this transformation is defined by matrix $\|P^i_j\|$.
and does not depend on point $A$. Vector
$(R^1,...,R^n)$ expresses displacement in affine space.
Set of transformations \EqRef{AffineTransformation} is the group Lie
which we denote as
\ShowEq{affine transformation group}
and call
\AddIndex{affine transformation group}{affine transformation group}.

\begin{definition}
\AddIndex{Affine basis}{Affine Basis}
\ShowEq{Affine Basis}
is set of
linear independent vectors $\Vector e_i=\overset{\longrightarrow}{OA_i}=(e^1_i,...,e^n_i)$
with common start point $O=(O^1,...,O^n)$. \qed
\end{definition}

\begin{definition}
\AddIndex{Basis manifold
\ShowEq{Basis Manifold, Affine Space}
of affine space}{Basis Manifold, Affine Space} 
is set of bases of this space.
\qed
\end{definition}

An active transformation is called
\AddIndex{affine transformation}{affine transformation}.
A passive transformation is called
\AddIndex{quasi affine transformation}{quasi affine transformation}.

If we do not concern about starting point of a vector we see little different
type of space which we call central affine space
\symb{\mathcal{CA}_n}{CAn}1.
In the central affine space we can identify all
tangent spaces and denote them $T\mathcal{CA}_n$.
If we assume that the start point
of vector is origin $O$ of coordinate system in space,
then we can identify any
point $A\in\mathcal{CA}_n$ with the vector $a=\overset{\longrightarrow}{OA}$. This leads
to identification of $\mathcal{CA}_n$ and $T\mathcal{CA}_n$.
Now transformation is simply map
\begin{align*}
a'^i&=P^i_j a^j & \det P\ne 0
\end{align*}
and such transformations build up Lie group $GL_n$.

\begin{definition}
\AddIndex{Central affine basis}{Central Affine Basis}
\ShowEq{Central Affine Basis}
is set of
linearly independent vectors $\Vector e_i=(e^1_i,...,e^n_i)$.
\qed
\end{definition}

\begin{definition}
\AddIndex{Basis manifold
\ShowEq{BCAn}
of central affine space}{Basis Manifold, Central Affine Space}
is set of bases of this space. \qed
\end{definition}

\section{Basis in Euclid Space}
\labelSubsection{EuclideSpace}

When we introduce a metric in a central affine space
we get a new geometry because
we can measure a distance and a length of vector. If a metric is
positive defined we call the space Euclid
\ShowEq{Euclid space}
otherwise we call the space pseudo Euclid
\ShowEq{pseudo Euclid space}

Transformations that preserve length
form Lie group $SO(n)$ for Euclid space and Lie group $SO(n,m)$ for pseudo
Euclid space where $n$ and $m$ number of positive and negative terms in
metrics.

\begin{definition}
\AddIndex{Orthonornal basis}{Orthonornal Basis}
\ShowEq{Orthonornal Basis}
is set of
linearly independent vectors $\Vector e_i=(e^1_i,...,e^n_i)$ such that length
of each vector is $1$ and different vectors are orthogonal.
\qed
\end{definition}

We can prove existence of orthonormal basis
using Gram-Schmidt orthogonalization procedure.

\begin{definition}
\AddIndex{Basis manifold
\ShowEq{BEn}
of Euclid space}{Basis Manifold, Euclid Space}
is set of orthonornal bases of this space.
\qed
\end{definition}

An active transformation is called
\AddIndex{movement}{movement transformation}.
A passive transformation is called
\AddIndex{quasi movement}{quasi movement}.

\section{Geometric Object of Vector Space}
\labelSection{Geometric Object of Vector Space}

An active transformation changes bases and vectors uniformly
and coordinates of vector relative basis do not change.
A passive transformation changes only the basis and it leads to change
of coordinates of vector relative to basis.

Let passive transformation $a\in G$ defined by matrix
$(a^i_j)$ maps
basis $\Basis{e}=<e_i>\in\mathcal{B}(V)$
into basis $\Basis{e}'=<e'_i>\in\mathcal{B}(V)$
\ShowEq{passive transformation of vector space}
Let vector $v\in V$ have expansion
\ShowEq{vector expansion in vector space, basis f}
relative to basis $\Basis{e}$ and have expansion
\ShowEq{vector expansion in vector space, basis fprim}
relative to basis $\Basis{e}'$.
From \EqRef{passive transformation of vector space}
and \EqRef{vector expansion in vector space, basis fprim} it follows that
\ShowEq{vector expansion in vector space, basis f, 1}
Comparing \EqRef{vector expansion in vector space, basis f}
and \EqRef{vector expansion in vector space, basis f, 1} we get
\ShowEq{coordinate transformation, 1}
Because $a^i_j$ is nonsingular matrix
we get from \EqRef{coordinate transformation, 1}
\DrawEq['{}a]{passive coordinate transformation}1
Coordinate transformation \eqRef{passive coordinate transformation}1
does not depend on vector $v$ or basis $\Basis{e}$, but is
defined only by coordinates of vector $v$
relative basis $\Basis{e}$.

Suppose we select basis $\Basis{e}$. Then the set of coordinates
\ShowEq{coordinates in vector space}
relative to this basis forms a vector space
\ShowEq{coordinate vector space}
isomorphic to vector space $V$.
This vector space
is called \AddIndex{coordinate vector space}{coordinate vector space}.
This isomorphism
is called \AddIndex{coordinate isomorphism}{coordinate isomorphism}.
Denote by
\ShowEq{image of vector e_k, coordinate vector space}
the image of vector $e_k\in\Basis{e}$ under this .

\begin{theorem}
\labelTheorem{coordinate transformations form representation, vector space}
Coordinate transformations \eqRef{passive coordinate transformation}1
form left\Hyph side effective linear
representation of group $G$ which is called
\AddIndex{contravariant representation}{contravariant representation}.
\end{theorem}
\begin{proof}
Suppose we have two consecutive passive transformations
$a\in G$ and $b\in G$. Coordinate transformation
\DrawEq['{}a]{passive coordinate transformation}a
corresponds to passive transformation $a$.
Coordinate transformation
\DrawEq[{''}'b]{passive coordinate transformation}b
corresponds to passive transformation $b$. Product
of coordinate transformations \eqRef{passive coordinate transformation}a
and \eqRef{passive coordinate transformation}b has form
\ShowEq{coordinate transformation, ba}
and is coordinate transformation
corresponding to passive transformation $ab$.
It proves that coordinate transformations
form contravariant right-side linear
representation of group $G$.

Suppose coordinate transformation does not change vectors $\delta_k$.
Then unit of group $G$ corresponds to it because representation
is single transitive. Therefore,
coordinate representation is effective.
\end{proof}

Let homomorphism of group $G$ to
the group of passive transformations
of vector space $W$ be coordinated with symmetry group
of vector space $V$.
This means that passive transformation $A(a)$ of vector space $W$
corresponds to passive transformation $a\in G$ of vector space $V$
\ShowEq{passive transformation of vector space W}
Then coordinate transformation in $W$ gets form
\ShowEq{coordinate transformation, W}

\begin{definition}
\labelDefinition{geometric object, coordinate representation}
Orbit
\ShowEq{geometric object, coordinate vector space}
\ShowEq{geometric object, coordinate vector space, show}
is called
\AddIndex{geometric object in coordinate representation}
{geometric object, coordinate vector space}
defined in vector space $V$.
For any basis
$\Basis{e}'_V=\Basis{e}_Va$
corresponding point \EqRef{coordinate transformation, W} of orbit defines
\AddIndex{coordinates of geometric object}
{coordinates of geometric object, coordinate vector space}
relative basis $\Basis{e}'_V$.
\qed
\end{definition}

\begin{definition}
\labelDefinition{geometric object}
Orbit
\ShowEq{geometric object, vector space}
\ShowEq{geometric object, vector space, show}
is called
\AddIndex{geometric object}{geometric object, vector space}
defined in vector space $V$.
For any basis
$\Basis{e}'_V=\Basis{e}_Va$
corresponding point \EqRef{coordinate transformation, W} of orbit defines
\AddIndex{coordinates of a geometric object}{coordinates of geometric object, vector space}
relative to basis $\Basis{e}'_V$
and the corresponding vector
\ShowEq{representative of geometric object}{}
is called
\AddIndex{representative of geometric object}
{representative of geometric object, vector space}
in basis $\Basis{e}'_V$.
\qed
\end{definition}

We also say that $w$ is
a \AddIndex{geometric object of type $A$}
{geometric object of type A, vector space}

Since a geometric object is an orbit of representation, we see that
according to theorem \RefTheorem{proper definition of orbit}
the definition of the geometric object is a proper definition.

Definition \RefDefinition{geometric object, coordinate representation}
introduces a geometric object in coordinate space.
We assume in definition \RefDefinition{geometric object}
that we selected a basis in vector space $W$.
This allows using a representative of the geometric object
instead of its coordinates.

\begin{theorem}
(\AddIndex{invariance principle}{invariance principle}).
\labelTheorem{invariance principle}
Representative of geometric object does not depend on selection
of basis $\Basis{e}'_V$.
\end{theorem}
\begin{proof}
To define representative of geometric object,
we need to select basis $\Basis{e}_V$,
basis
$\Basis{e}_W=(e_{W\alpha})$
and coordinates of geometric object
$w^\alpha$. Corresponding representative of geometric object
has form
\ShowEq{representative of geometric object}{}
Suppose we map basis $\Basis{e}_V$
to basis $\Basis{e}'_V$ by passive transformation $a\in G$.
According building this forms passive transformation
\EqRef{passive transformation of vector space W}
and coordinate transformation
\EqRef{coordinate transformation, W}. Corresponding
representative of geometric object has form
\ShowEq{invariance principle}
Therefore representative of geometric object
is invariant relative selection of basis.
\end{proof}

\begin{definition}
\labelDefinition{sum of geometric objects}
Let
\ShowEq{representative of geometric object}1
\ShowEq{representative of geometric object}2
be geometric objects of the same type
defined in vector space $V$.
Geometric object
\ShowEq{sum of geometric objects}
is called \AddIndex{sum
\ShowEq{sum of geometric objects 1}
of geometric objects}{sum of geometric objects, vector space}
$\Vector w_1$ and $\Vector w_2$.
\qed
\end{definition}

\begin{definition}
\labelDefinition{product of geometric object and constant}
Let
\ShowEq{representative of geometric object}1
be geometric object
defined in vector space $V$ over field $F$.
Geometric object
\ShowEq{product of geometric object and constant}
is called \AddIndex{product
\ShowEq{product of geometric object and constant 1}
of geometric object $w_1$ and constant $k\in F$}
{product of geometric object and constant, vector space}.
\qed
\end{definition}

\begin{theorem}
Geometric objects of type $A$
defined in vector space $V$ over field $F$
form vector space over field $F$.
\end{theorem}
\begin{proof}
The statement of the theorem follows from immediate verification
of the properties of vector space.
\end{proof}

%% file: Basis.Eq.tex

\DefEq
{
\symb{\Vector e_k}{vector of basis}{}
$\ShowSymbol{vector of basis}{}\in\Basis{e}$
}
{vector of basis}

\DefEq
{
\symb{\mathcal{A}_n}{An}1
}
{An}

\DefEq
{
\symb{\Basis{e}}{Basis e}{}
$\ShowSymbol{Basis e}{}=<e_{(i)}>$
}
{Basis e}

\DefEq
{
\symb{\tilde{V}}{coordinate vector space}1,
}
{coordinate vector space}

\AddEq[1]{representative of geometric object}
{
\[
\Vector w_{#1}= e'_{W\alpha}w'^\alpha_{#1}
\]
}

\AddEq{product of geometric object and constant}
{
\[\Vector w_2= e_{W\alpha}(kw_1^\alpha)\]
}

\AddEq{product of geometric object and constant 1}
{
\[\Vector w_2=k\Vector w_1\]
}

\AddEq{sum of geometric objects}
{
\[\Vector w=e_{W\alpha}(w_1^\alpha+w_2^\alpha)\]
}

\AddEq{sum of geometric objects 1}
{
\[\Vector w=\Vector w_1+\Vector w_2\]
}

\AddEq{invariance principle}
{
\[
\Vector w'= e'_{W\alpha}w'^\alpha
=  e_{W\gamma}A^\gamma_\alpha(a)A^{-1}_{}{}^\alpha_\beta(a)w^\beta
= e_{W\beta}w^\beta=\Vector w
\]
}

\AddEquation{passive transformation of vector space W}
{
e'_{W\alpha}=e_{W\beta} A^\beta_\alpha(a)
}

\AddEquation{coordinate transformation, W}
{
w'^\alpha=A^\alpha_\beta(a^{-1}) w^\beta=A^{-1}_{}{}^\alpha_\beta(a) w^\beta
}

\AddEq{geometric object, vector space, show}
{
\[
\ShowSymbol{geometric object, vector space}{}
=
(e'_Ww',w'=A(a)^{-1}w,e'_W=e_WA(a),a\in G)
\]
}

\AddEq{geometric object, vector space}
{
\symb{\mathcal{O}(V,W,\Vector w)}{geometric object, vector space}{}
}

\AddEq{geometric object, coordinate vector space}
{
\symb{\mathcal{O}(V,W,\Basis e_V,\Vector w)}{geometric object, coordinate vector space}{}
}

\AddEq{geometric object, coordinate vector space, show}
{
\[
\ShowSymbol{geometric object, coordinate vector space}{}
=A(G)^{-1}w
=(w'=A(a)^{-1}w,e'_V=e_Va,a\in G)
\]
}

\DefEq
{
\symb{\Vector\delta_k=(\delta^i_k)}{image of vector e_k, coordinate vector space}1
}
{image of vector e_k, coordinate vector space}

\DefEq
{
\symb{(v^i)}{coordinates in vector space}1
}
{coordinates in vector space}

\DefEq
{
\symb{\Basis{e}=<\Vector e_i>}{Orthonornal Basis}1
}
{Orthonornal Basis}

\DefEq
{
\symb{\mathcal{B}(\mathcal{E}_n)}{BEn}1
}
{BEn}

\DefEq
{
\symb{\mathcal{E}_{nm}}{pseudo Euclid space}1.
}
{pseudo Euclid space}

\DefEq
{
\symb{\mathcal{E}_n}{Euclid space}1,
}
{Euclid space}

\DefEq
{
\symb{\mathcal{B}(\mathcal{CA}_n)}{BCAn}1
}
{BCAn}

\DefEq
{
\symb{\Basis{e}=<\Vector e_i>}{Central Affine Basis}1
}
{Central Affine Basis}

\DefEq
{
\symb{\mathcal{B}(\mathcal{A}_n)}{Basis Manifold, Affine Space}1
}
{Basis Manifold, Affine Space}

\DefEq
{
\symb{\mathcal{B}(V)}{basis manifold of vector space}1
}
{basis manifold of vector space}

\DefEq
{
\symb{g\Basis{e}}{active transformation}1.
}
{active transformation}

\AddEquation{two transformations on basis manifold, 2}
{
g_2^{-1}(g_1\Basis{e})=(g_2^{-1}g_1)\Basis{e}=\Basis{e}
}

\AddEq{two transformations on basis manifold, 3}
{
$g_2^{-1}g_1$
}

\AddEquation{two transformations on basis manifold, 1}
{
g_1\Basis{e}=g_2\Basis{e}
}

\AddEq{bases from orbit ge}
{
\[
G\Basis{e}=\{g\Basis{e},g\in G\}
\]
}

\AddEq{vv=ve vector space}
{
\Vector v=e_{\gii}v^{\gii}
}

\AddEq{basis in V}
{
\symb{\Basis{e}_V}{basis in V}1.
}

\AddEq{set ei}
{%
$e_i$, $e_i\in\Basis e$,
}

\AddEq{row vector basis e}
{
\[
\Basis e=
\begin{pmatrix}
e_1&...&e_n
\end{pmatrix}
\]
}

\AddEquation{f:A->A vector space}
{
w=fv
}

\AddEq[1]{vv=vev}
{
\Vector{#1}=\Basis e#1=
\begin{pmatrix}
e_1&...&e_n
\end{pmatrix}
\begin{pmatrix}
#1^1\\...\\#1^n
\end{pmatrix}
}

\AddEquation{vw=vf(vv)}
{
\Vector w=\Vector f(\Vector v)
}

\AddEq{Vector f(e1) vector space}
{
$(\Vector f(e_i),e_i\in\Basis e)$
}

\AddEquation{vw=f(e1)vv}
{
\Vector w=\Vector f(\Vector v)=\Vector f(\Basis ev)
= (\Vector f(\Basis e))v
}

\AddEquation{f linear map}
{
f(av+bw)=af(v)+bf(w)
}

\AddEquation{vf(e1)=}
{
\Vector f(e_i)
=\Basis ef_i
=e_jf_i^j=
\begin{pmatrix}
e_1&...&e_n
\end{pmatrix}
\begin{pmatrix}
f_i^1\\...\\f_i^n
\end{pmatrix}
}

\AddEquation{w=e f v}
{
\Vector w=\Basis efv=e_if^i_jv^j
}

\AddEq{e w=e f v matrix}
{
\[
\begin{pmatrix}
e_1&...&e_n
\end{pmatrix}
\begin{pmatrix}
w^1\\...\\w^n
\end{pmatrix}
=
\begin{pmatrix}
e_1&...&e_n
\end{pmatrix}
\begin{pmatrix}
f_1^1&...&f_n^1
\\...&...&...\\
f_1^n&...&f_n^n
\end{pmatrix}
\begin{pmatrix}
v^1\\...\\v^n
\end{pmatrix}
\]
}

\AddEq{vi1=vi2}
{
$v^i_1=v^i_2$, $i=1$, ..., $n$,
}

\AddEq[1]{vv=evi}
{
\Vector v=e_iv^i_{#1}
}

\AddEquation{ev (1-2)}
{
e_iv^i_1=e_iv^i_2\ \ \ \ \ \,
e_i(v^i_1-v^i_2)=0
}

\AddEq[3]{w=f v matrix}
{
\begin{pmatrix}
#1^1\\...\\#1^n
\end{pmatrix}
=
\begin{pmatrix}
#2_1^1&...&#2_n^1
\\...&...&...\\
#2_1^n&...&#2_n^n
\end{pmatrix}
\begin{pmatrix}
#3^1\\...\\#3^n
\end{pmatrix}
}

\AddEquation{u=gfv}
{
\begin{split}
\begin{pmatrix}
u^1\\...\\u^n
\end{pmatrix}
&=
\begin{pmatrix}
g_1^1&...&g_n^1
\\...&...&...\\
g_1^n&...&g_n^n
\end{pmatrix}
\left(
\begin{pmatrix}
f_1^1&...&f_n^1
\\...&...&...\\
f_1^n&...&f_n^n
\end{pmatrix}
\begin{pmatrix}
v^1\\...\\v^n
\end{pmatrix}
\right)
\\
&=
\left(
\begin{pmatrix}
g_1^1&...&g_n^1
\\...&...&...\\
g_1^n&...&g_n^n
\end{pmatrix}
\begin{pmatrix}
f_1^1&...&f_n^1
\\...&...&...\\
f_1^n&...&f_n^n
\end{pmatrix}
\right)
\begin{pmatrix}
v^1\\...\\v^n
\end{pmatrix}
\end{split}
}

\AddEq{w12 in Jv over D}
{
$w_1$, $w_2\in J(v)$.
}

\AddEq{w in Jv over D}
{
$w\in J(v)$.
}

\AddEquation{w=sum vi, vector space}
{
J(v)=\left\{w:w=\sum_{\gii\in \giI}v_{\gii}c^{\gii}, c^{\gii}\in D,
|\{\gii:c^{\gii}\ne 0\}|<\infty\right\}
}

\AddEquation{comutative law}
{
vp=pv
}

\AddEq{vector space generated by set of vectors}
{
\StartLabelItem
\begin{enumerate}
\item
\ShowEq{vk in J(v)}
\labelItem{vector space, vk in Jv}
\item
\ShowEq{vector space over D, cvk in Jv}
\item
\ShowEq{vector space over D, sum cvk in Jv}
\item
\ShowEq{vector space over D, v,w in Jv=>v+w in Jv}
\item
\ShowEq{vector space over D, v in Jv=>av in Jv}
\end{enumerate}
}

\AddEq{w=vw over D}
{
\[\Vector w=vw\]
}

\AddEq{right 0=0vi over D}
{
\[
w^{\gii}=0\Rightarrow vw=0
\]
}

\AddEquation{vc=vm cm}
{
\begin{pmatrix}
v_i^1c^i\\
...\\
v_i^nc^i
\end{pmatrix}
=
\begin{pmatrix}
v^1_1&...&v^1_m\\
...&...&...\\
v^n_1&...&v^n_m
\end{pmatrix}
\begin{pmatrix}
c^1\\...\\c^n
\end{pmatrix}
}

\AddEquation{matrix of basis *c=0}
{
\begin{pmatrix}
e^1_1&...&e^1_n\\
...&...&...\\
e^n_1&...&e^n_n
\end{pmatrix}
\begin{pmatrix}
c^1\\...\\c^n
\end{pmatrix}
=
\begin{pmatrix}
0\\...\\0
\end{pmatrix}
}

\AddEq{Bases egh}
{
$\Basis e$, $\Basis g$, $\Basis h$
}

\AddEq[3]{matrix anm}
{
\begin{pmatrix}
#1_1^1&...&#1_{#2}^1\\
...&...&...\\
#1_1^{#3}&...&#1_{#2}^{#3}
\end{pmatrix}
}

\AddEquation{product g=fe}
{
\ShowEq{matrix anm}gnn
=
\ShowEq{matrix anm}fnn
\ShowEq{matrix anm}enn
}

\AddEq[1]{coordinate matrix nn}
{
\[
#1=
\ShowEq{matrix anm}{#1}nn
\]
}

\AddEq{vi ci}
{
$v_ic^i$
}

\AddEquation{row vector coordinates}
{
\begin{pmatrix}
v^1_1&...&v^1_m\\
...&...&...\\
v^n_1&...&v^n_m
\end{pmatrix}
}

\AddEq{vector space over D, v in Jv=>av in Jv}
{
$a\in D$,
$w\in J(v)\Rightarrow wa\in J(v)$
\labelItem{vector space over D, v in Jv=>av in Jv}
}

\AddEq{vector space over D, v,w in Jv=>v+w in Jv}
{
$w_1$, $w_2\in J(v)\Rightarrow w_1+w_2\in J(v)$
\labelItem{vector space over D, v,w in Jv=>v+w in Jv}
}

\AddEq{vector space over D, sum cvk in Jv}
{%
$\displaystyle\sum_{\gik\in\giI}v_{\gik}c^{\gik}\in J(v)$,
$c^{\gik}\in D$, $|\{\gii:c^{\gii}\ne 0\}|<\infty$
\labelItem{vector space over D, sum cvk in Jv}
}

\AddEq{vector space over D, cvk in Jv}
{%
$v_{\gik}c^{\gik}\in J(v)$, $c^{\gik}\in D$,
$\gik\in\giI$
\labelItem{vector space over D, cvk in Jv}
}

\AddEq{vk in J(v)}
{%
$v_{\gik}\in J(v)$
}

\AddEq{vf(e1)}
{
$\Vector f(e_i)$
}

\DefEq
{
\symb{G(V)}{GV}1
}
{GV}

\DefEq
{
\symb{\Basis{e}=<O,\Vector e_i>}{Affine Basis}1
}
{Affine Basis}

\DefEq
{
\symb{GL(\mathcal{A}_n)}{affine transformation group}1
}
{affine transformation group}

\DefEq
{
\symb{e^i_k}{standard coordinates of basis}1
}
{standard coordinates of basis}

\AddEq{passive transformation}
{
\symb{\Basis{e}g}{passive transformation}1.
}

\AddEquation{vector expansion in vector space, basis f, 1}
{
v=e_ia^i_jv'^j
}

\AddEquation{coordinate transformation, 1}
{
v^i=a^i_jv'^j
}

\AddEquation{vector expansion in vector space, basis f}
{
v=e_iv^i
}

\AddEquation{vector expansion in vector space, basis fprim}
{
v=e'_iv'^i
}

\AddEquation{passive transformation of vector space}
{
e'_j=e_ia^i_j
}

%% file: Refernce.Frame.English.tex
\input{Refernce.Frame.Eq}
\ifx\PrintBook\Defined
\chapter{Reference Frame in Event Space}
\fi

\section{Reference Frame on Manifold}
\labelSection{Reference Frame on Manifold}

\ePrints{GJSFRA.13.1.39}
\ifx\Semafor\ValueOff
As is shown in section
\xRef[0412.391]{section: Basis in Vector Space}
we can identify basis manifold of vector space
and symmetry group of this space.
The details of structure of basis did not interest us 
and this theory can be generalized.
In this section, we generalize the definition of a basis
and introduce a reference
frame on a manifold.
In case of an event space of general relativity it leads us to
\ePrints{4827-2437}
\ifx\Semafor\ValueOn
a natural definition of
$O(3,1)$\Hyph reference frame and the Lorentz transformation.\,\footnote{See
the definition of the Lorentz transformation in the section
\RefSection[GJSFRA.13.1.39]{Reference Frame in Event Space}.
}
\else
a natural definition of a reference frame and the Lorentz transformation.
\fi
We assume that
tangent space to the considered manifold
is vector space $V$ of finite dimension $n$.
\else
When we study manifold $V$ the geometry of tangent space
is one of important factors. 
In this section, we will make the following assumption.
\begin{itemize}
\item All tangent spaces have the same geometry.
\item Tangent space
is vector space $V$
of finite dimension $n$.
\item Symmetry group of tangent space is Lie group $G$.
\end{itemize}

Any homomorphism of the vector space maps one basis into another. 
Thus we can extend a representation of the symmetry group
to the set of bases.
However not every two bases can be mapped by a transformation
from the symmetry group
because not every nonsingular linear transformation belongs to
the representation of group $G$. The basis that belong
to selected orbit of group $G$ is called
\AddIndex{$G$\Hyph basis}{G-basis}.
\fi

\begin{definition}
\labelDefinition{type G reference frame}
\ePrints{GJSFRA.13.1.39}
\ifx\Semafor\ValueOn
Set
\ShowEq{reference frame}
of vector fields
\ShowEq{vector field of reference frame}
is called \AddIndex{$G$\Hyph reference frame}{G reference frame}, if
for any $x\in V$ set
\ShowEq{reference frame at x}
is a $G$\Hyph basis
in tangent space $T_x$.
\else
The set of vector fields
\ShowEq{e(i) 1n}
generates
\AddIndex{$G$\Hyph reference frame}{G reference frame}
\ShowEq{reference frame 1n}
on manifold $V$, if
for any $x\in V$ set
\ShowEq{reference frame at x 1n}
is a $G$\Hyph basis\,\footnote{
According to section
\xRef[0412.391]{section: Basis in Vector Space}
we can identify basis $\Basis e(x)$ with an element of group $G$.}
in tangent space $T_x$.\,\footnote{
In each
particular case we need to prove existence of $G$\Hyph reference frame on manifold.}
We use notation
\ShowEq{show vector field of reference frame}
for vector fields which form
$G$\Hyph reference frame $\Basis e$.
\fi
\qed
\end{definition}

\ePrints{GJSFRA.13.1.39}
\ifx\Semafor\ValueOn
Vector field $a$ has expansion
\ShowEq{Vector field expansion, reference frame}
relative reference frame $\Basis e$.

If we do not limit definition of a reference frame by symmetry group,
then at each point of the manifold
we can select reference frame
\ShowEq{coordinate reference frame}
based on vector fields tangent to lines $x^i=const$.
We call this field of bases
the \AddIndex{coordinate reference frame}{coordinate reference frame}.
Vector field $a$ has expansion
\ShowEq{Vector field expansion, coordinate reference frame}
relative coordinate reference frame.
Then standard coordinates of reference frame $\Basis e$ have form
\ShowEq{standard coordinates of reference frame}
\ShowEq{reference frame expansion relative coordinate reference frame}
Because vectors $e_{(i)}$ are linearly independent at each point matrix
$\|e^k_{(i)}\|$ has inverse matrix $\|e_k^{(i)}\|$
\ShowEq{coordinate reference frame expansion relative reference frame}
\else
We use also a representation for
reference frame on manifold as
\ShowEq{reference frame, extensive definition}
where we use the set of vector
fields $e_{(k)}$ and dual forms
\ShowEq{dual forms, reference frame}
such that at each point
\ShowEq{dual forms and vector fields}
The equality
\EqRef{dual forms and vector fields}
determines forms $e^{(k)}$ uniquely.

At each point of the manifold
we also consider
the \AddIndex{coordinate reference frame}{coordinate reference frame}
\ShowEq{coordinate reference frame}
based on vector fields tangent to lines $x^i=const$.
We use also a representation for
coordinate reference frame on manifold as
\ShowEq{coordinate reference frame, extensive definition}
where we use the set of vector
fields $\partial_i$ and dual forms $dx^i$
such that at each point
\ShowEq{dual forms and coordinate vector fields}
These reference frames are linked by the relationship
\ShowEq{holonomic and anholonomic}
From equalities
\EqRef{holonomic and anholonomic vector field},
\EqRef{holonomic and anholonomic forms},
\EqRef{dual forms and vector fields} it follows
\ShowEq{frame and coframe}

\begin{theorem}
There is the transformation between
reference frames $\Basis e$ and $\Basis{\partial}$
\ShowEq{transformation between reference frames}
where
\ShowEq{transformation between reference frames 4}
\end{theorem}
\begin{proof}
Relative considered reference frames,
vector field $a$ has expansion
\ShowEq{Vector field expansion relative reference frame}
where
\ShowEq{vector holonomic coordinates}
are \AddIndex{holonomic coordinates}{vector holonomic coordinates}
relative coordinate reference frame $\Basis{\partial}$ and
\ShowEq{vector anholonomic coordinates}
are \AddIndex{anholonomic coordinates}{vector anholonomic coordinates}
relative reference frame $\Basis e$.

We get the equality
\EqRef{transformation between reference frames 1},
when we consider coordinates of the vector field $e_{(i)}$
relative to the reference frame $\Basis{\partial}$
and coordinates of the vector field $\partial_k$
relative to the reference frame $\Basis e$.
Because vectors $e_{(i)}$ are linearly independent at each point matrix
$\|e^k_{(i)}\|$ is non singular matrix.
Because vectors $\partial_k$ are linearly independent at each point matrix
$\|e_k^{(i)}\|$ is non singular matrix.
From equalities
\EqRef{transformation between reference frames 1},
it follows that
\ShowEq{transformation between reference frames 3}
The equality
\EqRef{transformation between reference frames 4}
follows from the equality
\EqRef{transformation between reference frames 3}.
From equalities
\EqRef{transformation between reference frames 1},
\EqRef{Vector field expansion relative reference frame},
it follows that
\ShowEq{transformation between reference frames 5}
The equality
\ShowEq{transformation between reference frames 6}
follows from the equality
\EqRef{transformation between reference frames 5}.
The equality
\ShowEq{dual forms and vector fields 1}
follows from equalities
\EqRef{dual forms and vector fields},
\EqRef{Vector field expansion relative reference frame}.
The equality
\ShowEq{dual forms and coordinate vector fields 1}
follows from equalities
\EqRef{dual forms and coordinate vector fields},
\EqRef{Vector field expansion relative reference frame}.
The equality
\EqRef{transformation between reference frames 2}
follows from equalities
\EqRef{transformation between reference frames 6},
\EqRef{dual forms and vector fields 1},
\EqRef{dual forms and coordinate vector fields 1}.
\end{proof}
\fi

\ePrints{GJSFRA.13.1.39}
\ifx\Semafor\ValueOn
We use also a more extensive definition for
reference frame
on manifold, presented in form
\ShowEq{reference frame, extensive definition}
where we use the set of vector
fields $e_{(k)}$ and dual forms
\ShowEq{dual forms, reference frame}
such that
\ShowEq{dual forms and vector fields}
at each point.
Forms $e^{(k)}$ are defined uniquely from
\EqRef{dual forms and vector fields}.

In a similar way, we can introduce a
coordinate reference frame
\ShowEq{coordinate reference frame, extensive definition}.
These reference frames are linked by the relationship
\ShowEq{holonomic and anholonomic}
From equalities
\EqRef{holonomic and anholonomic vector field},
\EqRef{holonomic and anholonomic forms},
\EqRef{dual forms and vector fields} it follows
\ShowEq{frame and coframe}

In particular we assume that we have $GL(n)$-reference frame
$(\partial,dx)$ raised by $n$ differentiable vector fields
$\partial_i$ and 1-forms $dx^i$, that define field of bases $\partial$
and cobases $dx$ dual them.

If we have function $\varphi$ on $V$ then we define \AddIndex{pfaffian derivative}{pfaffian derivative}
\[d\varphi=\partial_i\varphi dx^i\]
\fi

\ePrints{4827-2437}
\ifx\Semafor\ValueOff
\section{Reference Frame in Event Space}
\labelSection{Reference Frame in Event Space}

Starting from this section, we consider orthogonal reference frame
\ShowEq{reference frame in Vn}
in Riemann space with metric $g_{ij}$.
According to definition, at each point of Riemann space
vector fields of orthogonal reference frame satisfy to the equality
\[
g_{ij}e^i_{(k)}e^j_{(l)}=g_{(k)(l)}
\]
where $g_{(k)(l)}=0$, if $(k)\ne(l)$, and $g_{(k)(k)}=1$
or $g_{(k)(k)}=-1$ depending on signature of metric.

We can define the
\AddIndex{reference frame in event space}
{reference frame in event space}
$V$ as $O(3,1)$\Hyph reference frame.
To enumerate vectors, we use index
$k=0, ..., 3$. Index $k=0$ corresponds to time like vector field.

\begin{remark}
\labelRemark{existence of a reference frame}
We can prove the existence of a reference frame
using the orthogonolization procedure at every point of space
time. From the same procedure we get that coordinates of basis
smoothly depend on the point.

A smooth field of time like vectors of each basis defines congruence
of lines that are tangent to this field.
We say that each line is a world line
of an observer or a
\AddIndex{local reference frame}{local reference frame}.
Therefore a reference frame is set of local reference frames.
\qed
\end{remark}

We define the \AddIndex{Lorentz transformation}{Lorentz transformation}
as transformation of a reference frame
$$
{x'}^i = f^i (x^0, x^1, x^2, x^3)
$$
\begin{equation}
{e'}^i_{(k)} = a^i_j b^{(l)}_{(k)} e^j_{(l)}
\label{LorentzTransformation}
\end{equation}
where
$$
a^i_j = \frac {\partial x'^i} {\partial x'^j}
$$
$$
\delta_{(i)(l)} b^{(i)}_{(j)} b^{(l)}_{(k)} = \delta_{(j)(k)}
$$
We call the transformation $a^i_j$ the holonomic part and transformation $b^{(l)}_{(k)}$
the anholonomic part.
\fi

\section{Anholonomic Coordinates}
\labelSection{Anholonomic Coordinates}

\ePrints{4827-2437}
\ifx\Semafor\ValueOff
Let $E(V, G,\pi)$ be the principal bundle,
where $V$ is the differential manifold of dimension $n$ and class
not less than $2$. We also assume that $G$ is symmetry group of tangent plain.

We define connection form on principal bundle 
\begin{equation}
\omega^L=\lambda^L_Nda^N+\Gamma^L_idx^i\ \ \ \omega=\lambda_Nda^N+\Gamma dx
\EqLabel{G connection form}
\end{equation}
We call functions $\Gamma_i$ connection components.

If fiber is group $GL(n)$, then connection has form
\ShowEq{GLn connection form}
\[\Gamma^A_i=\Gamma^a_{bi}\]

A vector field $a$ has two types of coordinates: \AddIndex{holonomic coordinates}{vector holonomic coordinates}
\ShowEq{vector holonomic coordinates}
relative coordinate reference frame
and \AddIndex{anholonomic coordinates}{vector anholonomic coordinates}
\ShowEq{vector anholonomic coordinates}
relative reference frame. These two forms of coordinates also hold the relation
\begin{equation}
a^i(x)=e^i_{(i)}(x)a^{(i)}(x)
\EqLabel{parallel transfer, 3}
\end{equation}
at any point $x$.

We can study parallel transfer of vector fields
using any form of coordinates.
Because \eqref{LorentzTransformation} is a linear transformation
we expect that parallel transfer in
anholonomic coordinates has the same representation
as in holonomic coordinates.
Hence we write
\ShowEq{parallel transfer of vector}
It is required to establish link between
\AddIndex{holonomic coordinate of connection}{holonomic coordinates of connection}
\ShowEq{holonomic coordinates of connection}
and \AddIndex{anholonomic coordinates of connection}{anholonomic coordinates of connection}
\ShowEq{anholonomic coordinates of connection}
\ShowEq{parallel transfer of vector, 1}
Considering \EqRef{parallel transfer, 1},
\EqRef{parallel transfer, 2}, and \EqRef{parallel transfer, 3} we get
\ShowEq{parallel transfer, 4}
It follows from \EqRef{parallel transfer, 4} that 
\begin{align*}
\Gamma^{(i)}_{(k)(p)}e^{(k)}_i(x)e^{(p)}_p(x)a^i(x)dx^p
&=a^{(i)}(x)-e^{(i)}_i(x+dx)\left(a^i(x)-\Gamma^i_{kp}a^k(x)dx^p\right)\\
&=a^i(x)e^{(i)}_i(x)-e^{(i)}_i(x+dx)\left(a^i(x)-\Gamma^i_{kp}a^k(x)dx^p\right)\\
&=a^i(x)\left(e^{(i)}_i(x)-e^{(i)}_i(x+dx)\right)
+e^{(i)}_j(x)\Gamma^j_{ip}a^i(x)dx^p\\
&=e^{(i)}_j(x)\Gamma^j_{ip}a^i(x)dx^p
-a^i(x)\frac{\partial e^{(i)}_i(x)}{\partial x^p}dx^p\\
&=\left(e^{(i)}_j(x)\Gamma^j_{ip}
-\frac{\partial e^{(i)}_i(x)}{\partial x^p}\right)a^i(x)dx^p
\end{align*}
Because $a^i(x)$ and $dx^p$ are arbitrary we get 
\[
\Gamma^{(i)}_{(k)(p)}e^{(k)}_i(x)e^{(p)}_p(x)
=e^{(i)}_j(x)\Gamma^j_{ip}
-\frac{\partial e^{(i)}_i(x)}{\partial x^p}
\]
\begin{equation}
\Gamma^{(i)}_{(k)(p)}
=e^i_{(k)}e^p_{(p)}e^{(i)}_j\Gamma^j_{ip}
-e^i_{(k)}e^p_{(p)}\frac{\partial e^{(i)}_i}{\partial x^p}
\EqLabel{parallel transfer, 5}
\end{equation}
We introduce symbolic operator
\begin{equation}
\frac{\partial}{\partial x^{(p)}}
=e^p_{(p)}\frac{\partial}{\partial x^p}
\EqLabel{anholonomic derivative}
\end{equation}
From \EqRef{frame and coframe} it follows
\ShowEq{parallel transfer, 6}
Equalities
\ShowEq{de(k)i=...dek(i)}
\ShowEq{dek(i)=...de(k)i}
follow from the equality
\EqRef{parallel transfer, 6}.
We substitute \EqRef{anholonomic derivative} and
\EqRef{de(k)i=...dek(i)} into \EqRef{parallel transfer, 5}
\begin{equation}
\Gamma^{(i)}_{(k)(p)}=
e^i_{(k)}e^p_{(p)}e^{(i)}_j\Gamma^j_{ip}
-e^{(i)}_i\frac{\partial e^i_{(k)}}{\partial x^{(p)}}
\EqLabel{Gamma1}
\end{equation}

Equation \EqRef{Gamma1} shows some similarity between holonomic and anholonomic coordinates.
We introduce symbol
\ShowEq{partial(k)}
for the derivative along vector field $e_{(k)}$
$$\partial_{(k)}=e^i_{(k)}\partial_i$$
Then \EqRef{Gamma1} takes the form
$$
\Gamma^{(k)}_{(l)(p)}=
e^i_{(l)}e^r_{(p)}e^{(k)}_j\Gamma^j_{ir}
-e^i_{(l)}\partial_{(p)} e^{(k)}_i
$$

Therefore when we move from holonomic coordinates to anholonomic, the connection transforms
the way similarly to when we move from one coordinate system to another.
This leads us to the model of anholonomic coordinates.

The vector field $e_{(k)}$ generates lines defined by the differential equations
$$
e^j_{(l)} \frac{\partial t} {\partial x^j} = \delta^{(k)}_{(l)}
$$
or the symbolic system 
\begin{equation}
\frac{\partial t} {\partial x^{(l)}} = \delta^{(k)}_{(l)}
\label{DefForX}
\end{equation}
Keeping in mind the symbolic system \eqref{DefForX} we denote
the functional $t$ as
\ShowEq{x(k)}
and call it the
\AddIndex{anholonomic coordinate}{anholonomic coordinate}.
We call the regular coordinate holonomic.

From here we can find derivatives and get
\ShowEq{Derivative For X}
The necessary and sufficient condition of complete integrability
of system \EqRef{Derivative For X} are
\ShowEq{cikl=0}
where we introduced \AddIndex{anholonomity object}{anholonomity object}
\ShowEq{anholonomity object 0}
\ShowEq{Anholonomity Object}
\ShowEq{Anholonomity Object kl=>(k)(l)}

\begin{theorem}
Each reference frame has $n$ vector fields
\DrawEq{partial_(k)}1
which have commutator
\ShowEq{commutator vector fields, reference frame}
\end{theorem}

\begin{proof}
The equality
\ShowEq{commutator vector fields, reference frame, 1}
follows from the equality
\eqRef{partial_(k)}1.
Since
\ShowEq{dij-dji=0}
then the equality
\ShowEq{commutator vector fields, reference frame, 2}
follows from equalities
\EqRef{dek(i)=...de(k)i},
\eqRef{partial_(k)}1,
\EqRef{commutator vector fields, reference frame, 1}.
The equality
\EqRef{commutator vector fields, reference frame}
follows from equalities
\EqRef{Anholonomity Object},
\EqRef{Anholonomity Object kl=>(k)(l)},
\EqRef{commutator vector fields, reference frame, 2}.
\end{proof}

For the same reason we introduce forms
\ShowEq{dx(k)=}
and an exterior differential of this form is
\ShowEq{anholonomity}

Therefore when $c^{(i)}_{(k)(l)} \not =0$, the differential $dx^{(k)}$ is not an exact differential
and the system of differential equations
\EqRef{Derivative For X},
in general, cannot be integrated.
However we can create a meaningful object that
models the solution. We can study how function $x^{(i)}$ changes
along different lines.
We call such coordinates
\AddIndex{anholonomic coordinates on manifold}
{anholonomic coordinates on manifold}.

\begin{remark}
\labelRemark{anholonomic coordinates, model}
Function $x^{(i)}$ is a natural parameter along a flow line
of vector field $\Vector e_{(i)}$.
\ePrints{GJSFRA.13.1.39}
\ifx\Semafor\ValueOn
The time coordinate along a local reference frame
is the observer's proper time.
Because the reference frame consists of local reference frames,
we expect that their proper times
are synchronized.

We introduce the \AddIndex{synchronization of reference frame}{synchronization of reference frame}
as the anholonomic time coordinate.

Because synchronization is the anholonomic coordinate it introduces
new physical phenomena
that we should keep in mind when working
with strong gravitational fields or making
precise measurements. I describe one of these phenomena in sections
\RefSection{Anholonomic Coordinates in Central Body Gravitational Field} -
\RefSection{Doppler Shift in Friedman Space}.
\else
We study an instance of such function in section
\RefSection{Synchronization of Reference Frame}.
The proper time is defined along world line of local
reference frame. As we see in remark
\ref{remark: existence of a reference frame}
world lines of local
reference frames cover spacetime.
To make proper time of local
reference frames as time of reference frame
we expect that proper time smoothly changes
from point to point.
\fi
To synchronize clocks of local
reference frames we use classical procedure of exchange light signals.

Somebody may have impression that we cannot synchronize clock,
however this conflicts with our observation.
We accept that synchronization is possible until
we introduce time along non closed lines.
Synchronization breaks when we try synchronize clocks
along closed line.
\ePrints{GJSFRA.13.1.39}
\ifx\Semafor\ValueOff

From mathematical point of view this is problem
to integrate differential form.
\fi
Namely, a change of function along a loop is
\begin{equation}
\begin{split}
\Delta x^{(i)}
&= \oint dx^{(i)}\\
&=\int\int c^{(i)}_{(k)(l)}dx^{(k)}\wedge dx^{(l)}\\
&=\int\int c^{(i)}_{(k)(l)} e^{(k)}_k e^{(l)}_l dx^k\wedge dx^l
\end{split}
\EqLabel{change of coordinate along a loop}
\end{equation}

This means ambiguity in definition of anholonomic coordinates.
\qed
\end{remark}

From now on we will not make a difference between holonomic and anholonomic coordinates.
Also, we will denote $b^{(l)}_{(k)}$ as ${a^{-1}}^{(l)}_{(k)}$ in the Lorentz transformation
\eqref{LorentzTransformation}.

\ePrints{GJSFRA.13.1.39}
\ifx\Semafor\ValueOff
Even form $dx^{(k)}$ is not exact differential, we can see
that form $d^2x^{(k)}$ is exterior differential of form $dx^{(k)}$.
Therefore
\begin{equation}
\EqLabel{exterior derivative 3}
d^3x^{(k)}=0
\end{equation}

We can represent exterior differential of form, written in anholonomic coordinates, as
\begin{align*}
&d(a_{(i_1)...(i_n)}dx^{(i_1)}\wedge ...\wedge dx^{(i_n)})\\
=&a_{(i_1)...(i_n),p}dx^p\wedge dx^{(i_1)}\wedge ...\wedge dx^{(i_n)}\\
-&a_{(i_1)...(i_n)}ddx^{(i_1)}\wedge ...\wedge dx^{(i_n)}-...
-(-1)^{n-1}a_{(i_1)...(i_n)}dx^{(i_1)}\wedge ...\wedge ddx^{(i_n)}\\
=&a_{(i_1)...(i_n),(p)}e_p^{(p)}e^p_{(r)}dx^{(r)}\wedge dx^{(i_1)}\wedge ...\wedge dx^{(i_n)}\\
-&a_{(i_1)...(i_n)}c^{(i_1)}_{(p)(r)}dx^{(p)}\wedge dx^{(r)}\wedge ...\wedge dx^{(i_n)}-...\\
-&(-1)^{n-1}a_{(i_1)...(i_n)}dx^{(i_1)}\wedge ...\wedge c^{(i_n)}_{(p)(r)}dx^{(p)}\wedge dx^{(r)}\\
=&(a_{(i_1)...(i_n),(p)}
-a_{(r)...(i_n)}c^{(r)}_{(p)(i_1)}-...
-a_{(i_1)...(r)} c^{(r)}_{(p)(i_n)})
dx^{(p)}\wedge dx^{(i_1)}\wedge ...\wedge dx^{(i_n)}
\end{align*}
In case of form $d^3x^{(k)}$ we get equality
\begin{equation}
\EqLabel{exterior derivative, anholonomic form}
\begin{array}{l}
d(c^{(k)}_{(i)(j)}dx^{(i)}\wedge dx^{(j)})\\
=(c^{(k)}_{(i)(j),(p)}
-c^{(k)}_{(r)(j)}c^{(r)}_{(p)(i)}
-c^{(k)}_{(i)(r)} c^{(r)}_{(p)(j)})
dx^{(p)}\wedge dx^{(i)}\wedge dx^{(j)}  
\end{array}
\end{equation}
From equalities \EqRef{exterior derivative 3}
and \EqRef{exterior derivative, anholonomic form}
it follows
\begin{equation}
\EqLabel{exterior derivative, anholonomic form, 1}
(c^{(k)}_{(i)(j),(p)}
-c^{(k)}_{(r)(j)}c^{(r)}_{(p)(i)}
-c^{(k)}_{(i)(r)} c^{(r)}_{(p)(j)})
dx^{(p)}\wedge dx^{(i)}\wedge dx^{(j)}=0  
\end{equation}
It is easy to see that
\begin{equation}
\EqLabel{exterior derivative, anholonomic form, 2}
\begin{array}{l}
(-c^{(k)}_{(r)(j)}c^{(r)}_{(p)(i)}
-c^{(k)}_{(i)(r)} c^{(r)}_{(p)(j)})
dx^{(i)}\wedge dx^{(j)}\\
=(-c^{(k)}_{(r)(j)}c^{(r)}_{(p)(i)}
+c^{(k)}_{(r)(i)} c^{(r)}_{(p)(j)})
dx^{(i)}\wedge dx^{(j)}\\
=-2c^{(k)}_{(r)(j)}c^{(r)}_{(p)(i)}
dx^{(i)}\wedge dx^{(j)}
\end{array}
\end{equation}
Substituting from \EqRef{exterior derivative, anholonomic form, 2}
into \EqRef{exterior derivative, anholonomic form, 1} gives
\begin{equation}
\EqLabel{exterior derivative, anholonomic form, 3}
(c^{(k)}_{(i)(j),(p)}
-2c^{(k)}_{(r)(j)}c^{(r)}_{(p)(i)})
dx^{(p)}\wedge dx^{(i)}\wedge dx^{(j)}=  0
\end{equation}
From \EqRef{exterior derivative, anholonomic form, 3}, it follows
\begin{equation}
\EqLabel{exterior derivative, anholonomic form, 4}
\begin{array}{l}
c^{(k)}_{(i)(j),(p)}+c^{(k)}_{(j)(p),(i)}+c^{(k)}_{(p)(i),(j)}\\
=2c^{(k)}_{(r)(j)}c^{(r)}_{(p)(i)}
+2c^{(k)}_{(r)(p)}c^{(r)}_{(i)(j)}
+2c^{(k)}_{(r)(i)}c^{(r)}_{(j)(p)}
\end{array}
\end{equation}
\fi
\else
We define connection form on event space 
\ShowEq{GLn connection form}
Parallel transfer of vector fields does not depend on choice of coordinates
and the equality
\EqRef{transformation between reference frames 6}
should be invariant relative to parallel transfer.
From this requirement,
according to calculation in section
\RefSection[GJSFRA.13.1.39]{Anholonomic Coordinates},
it follows that we can consider form $e^{(k)}$
as differential of coordinate $x^{(k)}$
\ShowEq{dx(k)=}
and vector field $e_{(k)}$ as differentiation
with respect to coordinate $x^{(k)}$
\DrawEq{partial_(k)}{}
In such case, we consider a matrix of transformation
mapping the reference frame $\Basis{\partial}$
into the reference frame $\Basis e$ as Jacobian matrix
of transformation of coordinates $x^i$ into coordinates $x^{(k)}$
\ShowEq{Derivative For X}
Differential of the form $dx^{(i)}$ is
\ShowEq{d2x(i)}
where
\ShowEq{anholonomity object}
is \AddIndex{anholonomity object}{anholonomity object}
\ShowEq{Anholonomity Object}
Therefore, the necessary and sufficient condition of complete integrability
of system of differential equations
\EqRef{Derivative For X} is
\ShowEq{cikl=0}
In general, the system of differential equations
\EqRef{Derivative For X}
is not complete integrable.
Map
\ShowEq{x(k)}
is called the
\AddIndex{anholonomic coordinate}{anholonomic coordinate}.
Coordinate $x^i$ is called holonomic.

Although we cannot build anholonomic coordinate in a finite domain,
we can define them uniquely along the open curve.\,\footnote{I
described construction of anholonomic coordinates in the section
\RefSection[GJSFRA.13.1.39]{Anholonomic Coordinates}.
}
We also consider anholonomic coordinate $x^{(0)}$
as synchronization of reference frame.
From now on we will not make a difference between holonomic and anholonomic coordinates.
\fi

\ePrints{GJSFRA.13.1.39}
\ifx\Semafor\ValueOn
We define the curvature form for connection \EqRef{G connection form}
\[\Omega=d\omega+[\omega,\omega]\]
\[\Omega^D=d\omega^D+C^D_{AB}\omega^A\wedge \omega^B=R^D_{ij}dx^i\wedge dx^j\]
where we defined a curvature object
\[R^D_{ij}=\partial_i\Gamma^D_j-\partial_j\Gamma^D_i+C^D_{AB}\Gamma^A_i\Gamma^B_j+\Gamma^D_k c^k_{ij}\]

The curvature form for the connection \EqRef{GLn connection form} is
\begin{equation}
\Omega^a_c = d \omega^a_c + \omega^a_b \wedge \omega^b_c
\EqLabel{Curvature}
\end{equation}
where we defined a curvature object
\begin{equation}
R^D_{ij}=R^a_{bij}=\partial_i\Gamma^a_{bj}-\partial_j\Gamma^a_{bi}
+\Gamma^a_{ci}\Gamma^c_{bj}-\Gamma^a_{cj}\Gamma^c_{bi}+\Gamma^a_{bk} c^k_{ij}
\EqLabel{GLn curvature}
\end{equation}
We introduce Ricci tensor
\[R_{bj}=R^a_{baj}=\partial_a\Gamma^a_{bj}-\partial_j\Gamma^a_{ba}
+\Gamma^a_{ca}\Gamma^c_{bj}-\Gamma^a_{cj}\Gamma^c_{ba}+\Gamma^a_{bk} c^k_{aj}\]
\fi

\ePrints{4827-2437,0803.3276}
\ifx\Semafor\ValueOn

\section{Metric-affine Manifold}
\labelSection{Metric-affine Manifold}

In deriving the equations of the gravitational field
(\citeBib{Landau}, \S 93,
\citeBib{Gravitation MTW}, \S 21.2),
we assume that metric tensor
$g^{ij}$ and connection $\Gamma^k_{ij}$ are independent.
However, we assume symmetry
of metric tensor and connection.
In deriving the equations of the gravitational field,
we get relation
\ShowEq{g;k=0}
In quantum mechanics,
measurement of components of
metric tensor and connection
contains an error.
If we assume that connection is not symmetric, then the equality
\EqRef{g;k=0}
may be not true.

Since covariant derivative of the metric tensor
may be different from 0,
we introduce the \AddIndex{nonmetricity}{nonmetricity}
\ShowEq{nonmetricity}
We move derivative of $g$ and torsion to the left-hand side.
\ShowEq{Connection1}
Changing order of indexes we write two more equalities
\ShowEq{Connection2}
\ShowEq{Connection3}
If we subtract equality \EqRef{Connection1} from sum of equalities
\EqRef{Connection2} and \EqRef{Connection1} we get
\ShowEq{Connection4}
Finally we get
\ShowEq{Connection5}

For connection \EqRef{GLn connection form}
we defined the \AddIndex{torsion form}{torsion form}
\begin{equation}
T^a = d^2 x^a + \omega^a_b \wedge dx^b
\EqLabel{Torsion}
\end{equation}
From \EqRef{GLn connection form} it follows
\begin{equation}
\omega^a_b \wedge dx^b=(\Gamma^a_{bc}-\Gamma^a_{cb})dx^c\wedge dx^b
\EqLabel{wedge connection}
\end{equation}
Putting \EqRef{wedge connection} and
\ePrints{4827-2437}
\ifx\Semafor\ValueOn
\EqRef{d2x(i)}
\else
\EqRef{anholonomity}
\fi
into \EqRef{Torsion}
we get
\begin{equation}
T^a=T^a_{cb}dx^c\wedge dx^b =
-c^a_{cb}dx^c\wedge dx^b + (\Gamma^a_{bc}-\Gamma^a_{cb})dx^c\wedge dx^b
\EqLabel{Torsion 1}
\end{equation}
where we defined \AddIndex{torsion tensor}{torsion tensor}
\begin{equation}
T^a_{cb} =
\Gamma^a_{bc}-\Gamma^a_{cb}-c^a_{cb}
\EqLabel{Torsion coordinates}
\end{equation}

The curvature form for the connection \EqRef{GLn connection form} is
\begin{equation}
\Omega^a_c = d \omega^a_c + \omega^a_b \wedge \omega^b_c
\EqLabel{Curvature}
\end{equation}
where we defined a curvature object
\begin{equation}
R^a_{bij}=\partial_i\Gamma^a_{bj}-\partial_j\Gamma^a_{bi}
+\Gamma^a_{ci}\Gamma^c_{bj}-\Gamma^a_{cj}\Gamma^c_{bi}+\Gamma^a_{bk} c^k_{ij}
\EqLabel{GLn curvature}
\end{equation}
We introduce Ricci tensor
\[R_{bj}=R^a_{baj}=\partial_a\Gamma^a_{bj}-\partial_j\Gamma^a_{ba}
+\Gamma^a_{ca}\Gamma^c_{bj}-\Gamma^a_{cj}\Gamma^c_{ba}+\Gamma^a_{bk} c^k_{aj}\]

Commutator of second derivatives has form
\begin{equation}
\EqLabel{commutator second derivative of vector}
\xi^a_{;cb}-\xi^a_{;bc}
=R^a_{d bc}\xi^d
-T^p_{bc}\xi^a_{;p}
\end{equation}

Due to the fact that derivative of the metric tensor is not 0, we cannot raise or lower index
of a tensor under derivative as we do it in regular Riemann space.
Now this operation changes to next
\[a^i_{;k} = g^{ij} a_{j;k} + g^{ij}_{;k} a_j\]
This equality for the metric tensor gets the following form
\[g^{ab}_{;k} = - g^{ai} g^{bj} g_{ij;k}\]

\begin{definition}
\labelDefinition{metric-affine manifold}
Riemannian space with a torsion
and a nonmetricity is called
the \AddIndex{metric\Hyph affine manifold}{metric-affine manifold} \citeBib{Mielke}.
\qed  
\end{definition}

If we study a submanifold $V_n$ of a manifold $V_{n+m}$, we see that the immersion creates
the connection $\Gamma^\alpha_{\beta\gamma}$ that relates to the connection in manifold as
\[
\Gamma^\alpha_{\beta\gamma} e^l_\alpha =
\Gamma^l_{mk} e^m_\beta e^k_\gamma + \frac {\partial e^l_\beta} {\partial u^\gamma}
\]
Therefore there is no smooth immersion
of a space with torsion into the Riemann space.
\fi

\ifx\texFuture\Defined
\section{Covariant exterior Derivative}

exterior derivative of form \EqRef{holonomic and anholonomic forms}
is different from 0. At first sight it looks unusual. However
form \EqRef{holonomic and anholonomic forms} is not
exact differential.

We define covariant exterior derivative of form $t = t_a \theta^a$ as
\[
Dt = t_{a;b} \theta^b \land \theta^a
+ t_a c^a_{bc} \theta^b \land \theta^c
\]

When we look at such equality we ask if $d^2=0$ still is legal.

\section{Moving Basis}

We meet another interesting situation when we study moving of frame along
manifold. If we introduce derivative for each vector of frame as
$$
\frac {De^i_{(k)}} {dx^l}
= e^i_{(k),l} + \Gamma^i_{pl} e^p_{(k)}
$$
then we get transformation for moving frame
$$
\frac {De^i_{(k)}} {dx^l}
= A^{(p)}_{(k)l} e^i_{(p)}
$$
where values $A_{(p)(k)l} = g_{(p)(r)} A^{(r)}_{(k)l}$ satisfy to the equality
\[
A_{(p)(k)l} + A_{(k)(p)l} = - g_{(p)(k);l}
\]
In particular,
\[A_{(k)(k)l}= - \frac 1 2 g_{(k)(k);l}\]
We observed the same type of transformation when we analyzed Frenet transfer.

Therefore, if $g_{pk;l} = 0$, then we get orthogonal transformation of frame.
However this is not so in case of nonzero derivative.
\fi

%% file: Refernce.Frame.Eq.tex

\DefEq
{
\symb{\Basis e=<e_{(i)},i\in I>}{reference frame}1
}
{reference frame}

\DefEq
{
\symb{x^{(k)}}{x(k)}1
}
{x(k)}

\DefEq
{
\symb{\partial_{(k)}}{partial(k)}1
}
{partial(k)}

\DefEq
{
\symb{\Gamma^{(k)}_{(i)(j)}}{anholonomic coordinates of connection}1
}
{anholonomic coordinates of connection}

\DefEq
{
\symb{\Gamma^k_{ij}}{holonomic coordinates of connection}1
}
{holonomic coordinates of connection}

\DefEq
{
\symb{\Basis e=<e_{(i)},i=1, ..., n>}{reference frame}1
}
{reference frame 1n}

\DefEq
{
\symb{e_{(i)}}{vector field of reference frame}1, $i=1$, ..., $n$,
}
{e(i) 1n}

\DefEquation
{
g^{ij}_{,k}+\Gamma^i_{pk}g^{pj}+\Gamma^j_{pk}g^{ip}=0
}
{g;k=0}

\DefEq
{
\symb{\Basis\partial=<\partial_i>}{coordinate reference frame}1
}
{coordinate reference frame}

\DefEq
{
\begin{align}
e_{(i)}&=e^k_{(i)} \partial_k
&\partial_k&=e_k^{(i)} e_{(i)}
\EqLabel{transformation between reference frames 1}
\\
e^{(k)}&=e^{(k)}_idx^i&
dx^i&=e_{(k)}^ie^{(k)}
\EqLabel{transformation between reference frames 2}
\end{align}
}
{transformation between reference frames}

\DefEquation
{
e_{(i)}=e^k_{(i)} \partial_k
=e^k_{(i)} e_k^{(j)} e_{(j)}
}
{transformation between reference frames 3}

\DefEquation
{
e^k_{(i)} e_k^{(j)}
=\delta_{(i)}^{(j)}
}
{transformation between reference frames 4}

\DefEq
{
\symb{e^k_{(i)}}{standard coordinates of reference frame}1
}
{standard coordinates of reference frame}

\DefEquation
{
\partial_k=e_k^{(i)} e_{(i)}
}
{coordinate reference frame expansion relative reference frame}

\AddEq{show vector field of reference frame}
{
$\ShowSymbol{vector field of reference frame}1\in\Basis e$
}

\AddEquation{frame and coframe}
{
e^{(k)}_ie^i_{(l)}=\delta^{(k)}_{(l)}
}

\DefEq
{
$\Basis e=(e_{(k)},e^{(k)})$
}
{reference frame in Vn}

\DefEquation
{
\begin{split}
&a^i(x)-\Gamma^i_{kp}a^k(x)dx^p\\
=&e^i_{(i)}(x+dx)
\left(a^{(i)}(x)-\Gamma^{(i)}_{(k)(p)}e^{(k)}_i(x)a^i(x)e^{(p)}_p(x)dx^p\right)
\end{split}
}
{parallel transfer, 4}

\AddEquation{d2x(i)}
{
d^2x^{(i)}=
-c^{(i)}_{(k)(l)}dx^{(k)}\wedge dx^{(l)}
}

\AddEquation{anholonomity}
{
\begin{split}
d^2x^{(k)}&=d\left( e^{(k)}_i dx^i\right)\\
&=\left(\partial_j e_i^{(k)} -\partial_i e_j^{(k)}\right)dx^i \wedge dx^j\\
&=c^{(k)}_{ij}dx^i \wedge dx^j
=c^{(k)}_{(m)(l)}dx^{(m)}\wedge dx^{(l)}
\end{split}
}

\DefEq
{
\begin{align*}
-Q_{kij}&=g_{ij;k}=g_{ij,k}-\Gamma^p_{ik}g_{pj}-\Gamma^p_{jk}g_{pi}
\\&=g_{ij,k}-\Gamma^p_{ik}g_{pj}-\Gamma^p_{kj}g_{pi}-S^p_{jk}g_{pi}
\end{align*}
}
{nonmetricity}

\DefEq
{
\begin{align}
a^i(x+dx)=a^i(x)+da^i&=a^i(x)-\Gamma^i_{kp}a^k(x)dx^p
\EqLabel{parallel transfer, 1}
\\
a^{(i)}(x+dx)=a^{(i)}(x)+da^{(i)}
&=a^{(i)}(x)-\Gamma^{(i)}_{(k)(p)}a^{(k)}(x)dx^{(p)}
\EqLabel{parallel transfer, 2}
\end{align}
}
{parallel transfer of vector, 1}

\DefEq
{
\begin{align*}
da^k&=-\Gamma^k_{ij}a^idx^j
\\
da^{(k)}&=-\Gamma^{(k)}_{(i)(j)}a^{(i)}dx^{(j)}
\end{align*}
}
{parallel transfer of vector}

\DefEq
{
\begin{align}
\EqLabel{holonomic and anholonomic vector field}
e_{(k)}&=e_{(k)}^i\partial_i\\
\EqLabel{holonomic and anholonomic forms}
e^{(k)}&=e^{(k)}_idx^i
\end{align}
}
{holonomic and anholonomic}

\DefEq
{
\symb{(\partial_i,dx^i)}{coordinate reference frame, extensive definition}1
}
{coordinate reference frame, extensive definition}

\AddEquation{dual forms and vector fields}
{
e^{(k)}(e_{(l)})=\delta^{(k)}_{(l)}
}

\DefEquation
{
e^{(k)}(a)=e^{(k)}(a^{(l)}e_{(l)})
=a^{(l)}e^{(k)}(e_{(l)})=a^{(l)}\delta^{(k)}_{(l)}=a^{(k)}
}
{dual forms and vector fields 1}

\AddEquation{dual forms and coordinate vector fields}
{
dx^k(\partial_l)=\delta^k_l
}

\DefEquation
{
dx^k(a)=dx^k(a^l\partial_l)=a^ldx^k(\partial_l)=a^l\delta^k_l=a^k
}
{dual forms and coordinate vector fields 1}

\DefEq
{
\symb{e^{(k)}}{dual forms, reference frame}1
}
{dual forms, reference frame}

\DefEq
{
\symb{\Basis e=(e_{(k)},e^{(k)})}{reference frame, extensive definition}1
}
{reference frame, extensive definition}

\DefEquation
{
e_{(i)}=\ShowSymbol{standard coordinates of reference frame} \partial_k
}
{reference frame expansion relative coordinate reference frame}

\DefEquation
{
a=a^i \partial_i
}
{Vector field expansion, coordinate reference frame}

\DefEquation
{
a=a^{(i)} e_{(i)}
}
{Vector field expansion, reference frame}

\DefEquation
{
a=a^i \partial_i=a^{(i)} e_{(i)}
}
{Vector field expansion relative reference frame}

\DefEquation
{
a^k \partial_k=a^k e_k^{(i)} e_{(i)}=a^{(i)} e_{(i)}
}
{transformation between reference frames 5}

\AddEq{vector anholonomic coordinates}
{
\symb{a^{(i)}}{vector anholonomic coordinates}1
}

\DefEq
{
\symb{a^i}{vector holonomic coordinates}1
}
{vector holonomic coordinates}

\DefEquation
{
\omega^a_b=\Gamma^a_{bc}dx^c\ \ \ \omega=\Gamma dx
}
{GLn connection form}

\DefEquation
{
a^{(i)}=a^k e_k^{(i)}
}
{transformation between reference frames 6}

\AddEquation{Anholonomity Object kl=>(k)(l)}
{
c^{(i)}_{(k)(l)}=
e_{(k)}^ke_{(l)}^lc^{(i)}_{kl}
}

\AddEquation{commutator vector fields, reference frame}
{
[\partial_{(i)},\partial_{(j)}]
=
-c^{(m)}_{(i)(j)}\partial_{(m)}
}

\AddEq{dij-dji=0}
{
\[
\partial_i\partial_j-\partial_j\partial_i=0
\]
}

\AddEquation{commutator vector fields, reference frame, 1}
{
\begin{split}
[\partial_{(i)},\partial_{(j)}]
&=
e^i_{(i)}\partial_i (e^j_{(j)}\partial_j)-
e^j_{(j)}\partial_j (e^i_{(i)}\partial_i)
\\&=
e^i_{(i)}\partial_ie^j_{(j)}\partial_j-
e^j_{(j)}\partial_je^i_{(i)}\partial_i
+e^i_{(i)}e^j_{(j)}\partial_i\partial_j-
e^j_{(j)}e^i_{(i)}\partial_j\partial_i
\end{split}
}

\AddEquation{parallel transfer, 6}
{
e^i_{(l)}\frac{\partial e^{(k)}_i}{\partial x^p}+
e^{(k)}_i\frac{\partial e^i_{(l)}}{\partial x^p}=0
}

\AddEquation{de(k)i=...dek(i)}
{
\frac{\partial e^{(k)}_i}{\partial x^p}
=-e^{(k)}_je_i^{(l)}\frac{\partial e^j_{(l)}}{\partial x^p}
}

\AddEquation{dek(i)=...de(k)i}
{
\frac{\partial e^j_{(l)}}{\partial x^p}
=-e_{(k)}^je^i_{(l)}\frac{\partial e^{(k)}_i}{\partial x^p}
}

\AddEquation{commutator vector fields, reference frame, 2}
{
\begin{split}
[\partial_{(i)},\partial_{(j)}]
&=
e^i_{(i)}\partial_ie^k_{(j)}\partial_k-
e^j_{(j)}\partial_je^k_{(i)}\partial_k
\\&=
(e^i_{(i)}\partial_ie^k_{(j)}-
e^j_{(j)}\partial_je^k_{(i)})\partial_k
\\&=
(-e^i_{(i)}e_{(k)}^ke^j_{(j)}\partial_ie^{(k)}_j
+e^j_{(j)}e_{(k)}^ke^i_{(i)}\partial_je^{(k)}_i)\partial_k
\\&=e^j_{(j)}e^i_{(i)}
(\partial_je^{(k)}_i-\partial_ie^{(k)}_j)\partial_{(k)}
%
\end{split}
}

\AddEq{anholonomity object}
{
\symb{c^{(i)}_{kl}}{anholonomity object}1
}

\AddEq{cikl=0}
{
\[
c^{(i)}_{kl} =0
\]
}

\DefEquation
{
\frac{\partial x^{(i)}} {\partial x^k} = e^{(i)}_k
}
{Derivative For X}

\AddEq{partial_(k)}
{
\partial_{(k)} = \frac\partial {\partial x^{(k)}} = e^i_{(k)} \partial_i
}

\DefEq
{
$$dx^{(k)}=e^{(k)}=e^{(k)}_l dx^l$$
}
{dx(k)=}

\DefEq
{
$\Basis e(x)=<e_{(i)}(x),i\in I>$
}
{reference frame at x}

\DefEq
{
$\Basis e(x)=<e_{(i)}(x),i=1,...,n>$
}
{reference frame at x 1n}

\DefEq
{
\symb{e_{(i)}}{vector field of reference frame}1
}
{vector field of reference frame}

%% file: Geom.Object.English.tex

\ePrints{0803.3276}
\ifx\Semafor\ValueOff

\ifx\PrintBook\Defined
\chapter{Geometric Object}
\fi

\section{Metric-affine Manifold}
\labelSection{Metric-affine Manifold}

For connection
\EqRef{GLn connection form},
we defined the \AddIndex{torsion form}{torsion form}
\begin{equation}
T^a = d^2 x^a + \omega^a_b \wedge dx^b
\EqLabel{Torsion}
\end{equation}
From \EqRef{GLn connection form} it follows
\begin{equation}
\omega^a_b \wedge dx^b=(\Gamma^a_{bc}-\Gamma^a_{cb})dx^c\wedge dx^b
\EqLabel{wedge connection}
\end{equation}
Putting \EqRef{wedge connection} and \EqRef{anholonomity} into \EqRef{Torsion}
we get
\begin{equation}
T^a=T^a_{cb}dx^c\wedge dx^b =
-c^a_{cb}dx^c\wedge dx^b + (\Gamma^a_{bc}-\Gamma^a_{cb})dx^c\wedge dx^b
\EqLabel{Torsion 1}
\end{equation}
where we defined \AddIndex{torsion tensor}{torsion tensor}
\begin{equation}
T^a_{cb} =
\Gamma^a_{bc}-\Gamma^a_{cb}-c^a_{cb}
\EqLabel{Torsion coordinates}
\end{equation}

Commutator of second derivatives has form
\begin{equation}
u^\alpha_{;kl}-u^\alpha_{;lk}
=R^\alpha_{\beta lk}u^\beta
-T^p_{lk}u^\alpha_{;p}
\EqLabel{commutator second derivative}
\end{equation}
From \EqRef{commutator second derivative} it follows that
\begin{equation}
\EqLabel{commutator second derivative of vector}
\xi^a_{;cb}-\xi^a_{;bc}
=R^a_{d bc}\xi^d
-T^p_{bc}\xi^a_{;p}
\end{equation}

In Rieman space we have metric tensor $g_{ij}$ and connection $\Gamma^k_{ij}$.
One of the features of the Rieman space is symetricity of connection and covariant derivative
of metric is 0. This creates close relation between metric and connection.
However the connection is not necessarily symmetric
and the covariant derivative of the metric tensor may be different from 0.
In latter case we introduce the \AddIndex{nonmetricity}{nonmetricity}
\begin{equation}
Q^{ij}_k=g^{ij}_{;k}=g^{ij}_{,k}+\Gamma^i_{pk}g^{pj}+\Gamma^j_{pk}g^{ip}
\EqLabel{Nonmetricity}
\end{equation}

Due to the fact that derivative of the metric tensor is not 0, we cannot raise or lower index
of a tensor under derivative as we do it in regular Riemann space.
Now this operation changes to next
\[a^i_{;k} = g^{ij} a_{j;k} + g^{ij}_{;k} a_j\]
This equation for the metric tensor gets the following form
\[g^{ab}_{;k} = - g^{ai} g^{bj} g_{ij;k}\]

\begin{definition}
\labelDefinition{metric-affine manifold}
We call a manifold with a torsion
and a nonmetricity
the \AddIndex{metric\Hyph affine manifold}{metric-affine manifold} \citeBib{Mielke}.
\qed  
\end{definition}

\ePrints{GJSFRA.13.1.39}
\ifx\Semafor\ValueOn
Nonmetricity dramatically changes law
how orthogonal basis moves in space time.
However learning of parallel transport in space with nonmetricity
allows us to
introduce the Cartan transport
and the connection compatible with the metric tensor
(section \RefSection[0405.028]{Cartan Transport}). The Cartan transport
holds the basis orthonormal and this makes it valuable tool
because the observer uses an orthonormal basis as his measurement tool.
We will assume that nonmetricity of space
time is equal $0$.
\else

If we study a submanifold $V_n$ of a manifold $V_{n+m}$, we see that the immersion creates
the connection $\Gamma^\alpha_{\beta\gamma}$ that relates to the connection in manifold as
\[
\Gamma^\alpha_{\beta\gamma} e^l_\alpha =
\Gamma^l_{mk} e^m_\beta e^k_\gamma + \frac {\partial e^l_\beta} {\partial u^\gamma}
\]
Therefore there is no smooth immersion
of a space with torsion into the Riemann space.
\fi
\fi

\section{Geometric Meaning of Torsion}

\ShowEq{note: sum of vectors 2}vw

\ShowEq{note: sum of vectors 3}

\ShowEq{note: sum of vectors 4}

\ShowEq{note: sum of vectors 5}

\ShowLemma{increase of coordinate along geodesic}
\ShowProof{increase of coordinate along geodesic}

\ShowTheorem{parallelogram is not closed}
\ShowProof{parallelogram is not closed}

\ePrints{GJSFRA.13.1.39,0803.3276}
\ifx\Semafor\ValueOff
\section{Relation between Connection and Metric}

Now we want to find how we can express connection if we know
metric and torsion. According to definition
\[-Q_{kij}=g_{ij;k}=g_{ij,k}-\Gamma^p_{ik}g_{pj}-\Gamma^p_{jk}g_{pi}\]
\[-Q_{kij}=g_{ij,k}-\Gamma^p_{ik}g_{pj}-\Gamma^p_{kj}g_{pi}-S^p_{jk}g_{pi}\]
We move derivative of $g$ and torsion to the left-hand side.
\ShowEq{Connection1}
Changing order of indexes we write two more equations
\ShowEq{Connection2}
\ShowEq{Connection3}
If we substruct equation \EqRef{Connection1} from sum of equations
\EqRef{Connection2} and \EqRef{Connection1} we get
\ShowEq{Connection4}
Finally we get
\ShowEq{Connection5}
\fi

%% file: Gen.Relativity.English.tex

\input{Gen.Relativity.Eq}

\ifx\PrintBook\Defined
\chapter{Application in General Relativity}
\fi

\ePrints{GJSFRA.13.1.39}
\ifx\Semafor\ValueOff
\section{Synchronization of Reference Frame}
\labelSection{Synchronization of Reference Frame}

Because an observer uses an orthogonal basis for measurement at each point
we can expect that he uses anholonomic coordinates as well.
We also see that the time coordinate along a local reference frame
is the observer's proper time.
Because the reference frame consists of local reference frames, we expect that their proper times
are synchronized.

We introduce the \AddIndex{synchronization of reference frame}{synchronization of reference frame}
as the anholonomic time coordinate.

Because synchronization is the anholonomic coordinate,
it introduces new physical phenomena
that we should keep in mind when working with strong gravitational fields or making
precise measurements. I describe one of these phenomena below.
\fi

\section{Anholonomic Coordinates in Central Body Gravitational Field}
\labelSection{Anholonomic Coordinates in Central Body Gravitational Field}

We will study an observer orbiting around a central body.
The results are only estimation and are good when eccentricity is near $0$
because we study circular orbits.
However, the main goal of this estimation is to show that we have
a measurable effect of anholonomity.

We use the Schwarzschild metric of a central body
\ShowEq{Schwarzschild metric}
$$r_g = \frac {2 G m} {c^2}$$
$G$ is the gravitational constant, $m$ is the mass of the central body,
$c$ is the speed of light.

Connection in this metric is
\ShowEq{Schwarzschild metric, Connection}

The anholonomic basis has form
\ShowEq{Schwarzschild metric 1}
Anholonomity object has form
\DrawEq{Schwarzschild metric 2}{note19}
From the equality
\eqRef{Schwarzschild metric 2}{note19},
it follows that
we should consider radial movement in order
to study the impact of the anholonomity on synchronization of reference frame.

Let John travel in radial direction.
The initial point of his trajectory is at a distance $R_1$ from the center of gravity
(the point $A$).
When John is at a distance $R_2$ from the center of gravity (the point $B$),
he makes a stop and returns into the point $A$.
His friend James remains at the point $A$
and waits when John will return.

We assume that all clocks in reference frame of James synchronized.
John knows that his clock readings is different from clock readings
in reference frame of James.
So John periodically checks clock readings in reference frame of James.
When John meets James again, they verify their time measurement results.

Let $v$ be speed of John's movement in reference frame of James.
Let
\ShowEq{experiment, step 1}
be initial John's movement.
The equality
\ShowEq{experiment, t1=dR/v}
follows from
\EqRef{experiment, step 1}.
The equality
\ShowEq{experiment, d1t=}
follows from the equality
\ShowEq{experiment, dt=int}
and from equalities
\EqRef{experiment, step 1},
\EqRef{experiment, t1=dR/v}.

I need to get indefinite integral
(I changed variable
\ShowEq{experiment, r=R1+vt})
\ShowEq{experiment, d1t= indefinite}
Let
\ShowEq{experiment, u2=r/r}
Then
\ShowEq{experiment, r=..u2}
\ShowEq{experiment, dr=..du}
The equality
\ShowEq{experiment, d1t= indefinite 1}
follows from equalities
\EqRef{experiment, d1t= indefinite},
\EqRef{experiment, r=..u2},
\EqRef{experiment, dr=..du}.
Let
\ShowEq{experiment, u2/u2=abcd}
The system of equations
\ShowEq{experiment, u2/u2 abcd=}
follows from the equality
\EqRef{experiment, u2/u2=abcd}.
The system of equations
\ShowEq{experiment, u2/u2 abcd= 1}
follows from the system of equations
\EqRef{experiment, u2/u2 abcd=}.
The equality
\ShowEq{experiment, u2/u2 abcd= 2}
follows from the system of equations
\EqRef{experiment, u2/u2 abcd= 1}.
The equality
\ShowEq{experiment, d1t= indefinite 2}
follows from equalities
\EqRef{experiment, d1t= indefinite 1},
\EqRef{experiment, u2/u2=abcd},
\EqRef{experiment, u2/u2 abcd= 2}.

If $x^0$ changes from $0$ to $t_1$,
then $r$ changes from $R_1$ to $R_2$
and $u$ changes from
\ShowEq{experiment, u=u1}1
to
\ShowEq{experiment, u=u1}2
Because time of movement in both directions is the same, then we get
\ShowEq{experiment, d1t+d2t}

It remains to calculate the time, how long James waits return of John.
The equality
\ShowEq{experiment, d3t=}
follows from the equality
\EqRef{experiment, dt=int}.

To analyze the results of the experiment,
I assume that, instead of John roundtrip traveling
from the point $A$ into the point $B$,
James will send a beam of light ($v=c-2.998e10$ sm/sec) into the point $B$, where the mirror
will reflect beam of light back into the point $A$.
Let the point $A$ be on the surface of the Earth
($R_1=6.371e8$ santimeters, the distance from the center of the Earth)
and the point $B$ be on the surface of the Moon
($R_2=3.84e10$ santimeters).\,\footnote{
Results of the calculation are an estimate,
because, in the calculation, I do not take into account the gravitational field of the Moon.
}
Time measured by James is $2.5192061357$ seconds.
Time measured by the beam of light is $2.5192002762$ seconds.
Since wave phase of light beam depends on local time
in the reference frame of James, then
James can use interference
of the original beam of light and the returned beam of light
in order to observe interference deviation
from expected picture.

\section{Doppler Shift in Central Body Gravitational Field}

I want to show one more way
to calculate the Doppler shift. The Doppler shift in gravitational field is
well known issue, however the method that I show is useful to
better understand physics of gravitational field.

We can describe the movement of photon in gravitational field
using its wave vector $k^i$. The length of this vector is 0;
$\frac {k^i} {dx^i} = const$; a trajectory is
geodesic and therefore coordinates of this vector satisfy to differential equation
\begin{equation}
dk^i = -\Gamma^i_{kl}k^k dx^l
\EqLabel{Transport}
\end{equation}

We looking for the frequency $\omega$ of light
and $k^0$ is proportional $\omega$.
Let us consider the radial movement of a photon. In this case wave vector has form 
$k=(k^0,k^1,0,0)$.
In the central field with metric
\EqRef{Schwarzschild metric} we can choose
\[k^0=\frac \omega c \sqrt{\frac r {r-r_g}}\]
\[k^1=\omega \sqrt{\frac {r-r_g} r}\]
\[dt=\frac {k^0} {k^1} dr= \frac 1 c \frac r {r-r_g}dr\]
Then the equation \EqRef{Transport} gets form
\[dk^0=-\Gamma^0_{10}(k^1 dt + k^0 dr)\]
\[d\left(\frac\omega c \sqrt{\frac r {r-r_g}}\right)=
-\frac{r_g \omega} {2r(r-r_g)}\left(\sqrt{\frac {r-r_g} r}\frac r {r-r_g}
+\sqrt{\frac r {r-r_g}}\right)\frac {dr} c\]
\[d\omega \sqrt{\frac r {r-r_g}}-\omega\frac 1 2 \sqrt{\frac {r-r_g}r}
\frac {r_gdr} {(r-r_g)^2}
=-\frac {r_g\omega dr} {r(r-r_g)}\sqrt{\frac r {r-r_g}}\]
\[\frac  {d\omega} \omega
=-\frac {r_g} {2r(r-r_g)}dr\]
\[\ln  \omega
=\frac 1 2 \ln \frac r {r-r_g} + \ln C\]
If we define $\omega = \omega_0$ when $r=\infty$, we get finally 
\[\omega
=\omega_0\sqrt{\frac r {r-r_g}}\]

\ePrints{0803.3276}
\ifx\Semafor\ValueOff
\section{Time Delay in Central Body Gravitational Field}
\labelSection{Time Delay in Central Body Gravitational Field}

We will study orbiting around a central body.
The results are only an estimation and are good when eccentricity is near $0$
because we study circular orbits.
However, the main goal of this estimation is to show that we have a measurable effect of anholonomity.

Let us compare the measurements of two observers. The first observer fixed his position in the gravitational field
$$t=\frac s  c \sqrt{\frac r {r-r_g}}$$ $$r=const, \phi=const, \theta=const$$
The second observer orbits the center of the field with constant speed
$$t=s \sqrt{\frac r {(r-r_g) c^2 - \alpha^2 r^3}}$$
$$\phi=\alpha\ s \sqrt{\frac r {(r-r_g) c^2 - \alpha^2 r^3}}$$
$$r=const,\theta=const$$
We choose a natural parameter for both observers.

The second observer starts his travel when $s=0$ and finishes it when returning to the same spatial point.
Because $\phi$ is a cyclic coordinate the second observer finishes his travel when $\phi=2\pi$. We have at this point
$$s_2=\frac{2\pi} \alpha \sqrt{\frac{(r-r_g) c^2 - \alpha^2 r^3} r}$$
$$t=T=\frac{2\pi} \alpha$$
The value of the natural parameter for the first observer at this point is
$$s_1=\frac{2\pi} \alpha c\sqrt{\frac{r-r_g} r}$$
The difference between their proper times is
$$
\Delta s = s_1 - s_2
=\frac{2\pi} \alpha
\left(c\sqrt{\frac{r-r_g} r}-\sqrt{\frac{(r-r_g) c^2 - \alpha^2 r^3} r}\right)
$$
We have a difference in centimeters. To get this difference in seconds we should divide both sides by c.
$$\Delta t = \frac{2\pi} \alpha
\left(\sqrt{\frac{r-r_g} r}-\sqrt{\frac{r-r_g} r - \frac{\alpha^2 r^2} {c^2}}\right)$$

Now we get specific data.

The mass of the Sun is $1.989_{10}33$ g, the Earth orbits the Sun at a
distance of $1.495985_{10}13$ cm from its center during $365.257$ days. In this case we get
$\Delta t = 0.155750625445089$ s. Mercury orbits the Sun at a
distance of $5.791_{10}12$ cm from its center during $58.6462$ days. In this case we get
$\Delta t = 0.145358734930827$ s.

The mass of the Earth is $5.977_{10}27$ g. The spaceship that orbits the Earth at a
distance of $6.916_{10}8$ cm from its center during $95.6$ mins has $\Delta t = 1.8318_{10}-6$ s.
The Moon orbits the Earth at a
distance of $3.84_{10}10$ cm from its center during $27.32$ days. In this case we get
$\Delta t = 1.372_{10}-5$ s.

For better presentation I put these data to
tables \ref{table: Effect of Anholonomic Coordinates, Sun},
\ref{table: Effect of Anholonomic Coordinates, Earth},
and \ref{table: Effect of Anholonomic Coordinates, Sgr A}.

Because clocks of first observer show larger time at meeting,
first observer estimates age of second one
older then real. Hence, if we get parameters of S2 orbit from \citeBib{Ghez},
we get that if first observer estimates age of S2 as 10 Myr then
S2 will be .297 Myr younger.

\begin{table}[ht]
\caption{Sun is central body, mass is $1.989_{10}33$ g}
\label{table: Effect of Anholonomic Coordinates, Sun}
\noindent\[
\begin{tabular}{|c|c|c|}
\hline
Sputnik&Earth&Mercury\\
\hline
distance, cm&$1.495985_{10}13$&$5.791_{10}12$\\
\hline
orbit period, days&$365.257$&$58.6462$\\
\hline
Time delay s&$0.15575$&$0.14536$\\
\hline
\end{tabular}
\]
\end{table}

\begin{table}[ht]
\caption{Earth is central body, mass is $5.977_{10}27$ g}
\label{table: Effect of Anholonomic Coordinates, Earth}
\noindent\[
\begin{tabular}{|c|c|c|}
\hline
Sputnik&spaceship&Moon\\
\hline
distance, cm&$6.916_{10}8$&$3.84_{10}10$\\
\hline
orbit period&$95.6$ mins&$27.32$ days\\
\hline
Time delay, s&$1.8318_{10}-6$&$1.372_{10}-5$\\
\hline
\end{tabular}
\]
\end{table}

\begin{table}[ht]
\caption{Sgr A is central body, S2 is sputnik}
\label{table: Effect of Anholonomic Coordinates, Sgr A}
\noindent\[
\begin{tabular}{|c|c|c|}
\hline
mass, $M_\odot$&$4.1_{10}6$&$3.7_{10}6$\\
\hline
distance sm&$1.4692_{10}16$&$1.1565_{10}16$\\
\hline
orbit period, years&$15.2$&$15.2$\\
\hline
Time delay, min&$164.7295$&$153.8326$\\
\hline
\end{tabular}
\]
\end{table}
\fi

\section{Lorentz Transformation in Orbital Direction}
\labelSection{Lorentz Transformation in Orbital Direction}

The reason for the time delay that we estimated above is in Lorentz transformation between
stationary and orbiting observers.
This means that we have rotation in plain $(e_{(0)}, e_{(2)})$.
The basis vectors for stationary observer are
\[e_{(0)}=(\frac 1 c \sqrt{\frac r {r-a}}, 0, 0, 0)\]
\[e_{(2)}=(0, 0, \frac 1 r, 0)\]
We assume that for orbiting observer changes of $\phi$ and $t$
are proportional and
\[d\phi=\omega dt\]
Unit vector of speed in this case should be proportional to vector
\begin{equation}
(1, 0, \omega, 0)
\label{NewVectorE2}
\end{equation}
The length of this vector is
\begin{equation}
L^2=\frac {r-a} r c^2 - r^2 \omega^2
\label{LengthNewVectorE2}
\end{equation}
We see in this expression very familiar pattern and expect that linear
speed of orbiting observer is $V = \omega r$.

However we have to remember that we make measurement in gravitational
field and coordinates are just tags to label points in spacetime.
This means that we need a legal method to measure speed.

If an object moves from point $(t, \phi)$ to point $(t+dt, \phi+d\phi)$
we need to measure spatial and time intervals between these points.
We assume that in both points there are observers $A$ and $B$.
Observer $A$ sends the same time light signal to $B$ and ball that has
angular speed $\omega$. Whatever observer $B$ receives, he sends light signal
back to $A$.

When $A$ receives first signal he can estimate distance to $B$.
When $A$ receives second signal he can estimate how long ball moved to $B$.

The time of travel of light in both directions is the same.
Trajectory of light is determined by equation $ds^2=0$.
In our case we have
\[\frac {r-r_g} r c^2 dt^2 - r^2 d\phi^2 = 0\]
When light returns back to observer $A$ the change of $t$ is
\[dt=2 \sqrt{\frac r {r-r_g}} c^{-1}r d\phi\]
The proper time of first observer is
\[ds^2= \frac {r-r_g} r c^2 4 \frac r {r-r_g}c^{-2}r^2 d\phi^2\]
Therefore spatial distance is
\[L=r d\phi\]
When object moving with angular speed $\omega$ gets to $B$
change of $t$ is $\frac {d\phi} \omega$.
The proper time at this point is 
\[ds^2 = \frac {r - r_g} r c^2 d\phi^2 \omega^{-2}\]
\[T = \sqrt{\frac {r - r_g} r} \omega^{-1} d\phi\]
Therefore the observer $A$ measures speed
\[V=\frac L T = \sqrt{\frac r {r - r_g}} r\omega\]

We can use speed $V$ as parameter of Lorentz transformation. Then length
\eqref{LengthNewVectorE2} of vector \eqref{NewVectorE2} is
\[L=\sqrt{\frac {r-r_g} r \left(c^2 - \frac r {r-r_g}r^2 \omega^2\right)}=
\sqrt{\frac {r-r_g} r} c\sqrt{\left(1 - \frac {V^2}{c^2}\right)}\]
Therefore time ort of moving observer is
\[e'_{(0)}=\left(\frac 1 L, 0, \frac \omega L, 0\right)\]
\[e'_{(0)}=\left(\sqrt{\frac r {r-r_g}} c^{-1}\frac 1 {\sqrt{\left(1 - \frac {V^2}{c^2}\right)}}, 0,
\omega \sqrt{\frac r {r-r_g}} c^{-1}\frac 1 {\sqrt{\left(1 - \frac {V^2}{c^2}\right)}}, 0\right)\]
Spatial ort $e'_{(2)}=(A,0,B,0)$ is orthogonal $e'_{(0)}$ and has length $-1$. Therefore
\begin{equation}
\frac {r-r_g} r c^2\frac 1 L A - r^2 \frac \omega L B = 0
\label{NewVectorE0_1}
\end{equation}
\begin{equation}
\frac {r-r_g} r c^2 A^2 - r^2 B^2 = -1
\label{NewVectorE0_2}
\end{equation}
We can express $A$ from \eqref{NewVectorE0_1}
\[A = c^{-2}\frac r{r-r_g} r^2 \omega B\]
and substitute into \eqref{NewVectorE0_2}
\[ c^{-2}\frac r{r-r_g} r^4 \omega^2 B^2 - r^2 B^2 = -1\]
\[\frac {V^2}{c^2} r^2 B^2 - r^2 B^2 = -1\]
Finally spatial ort in direction of movement is
\[e'_{(2)}=\left(c^{-2}\frac r{r-r_g} r \omega \frac 1 {\sqrt{1 - \frac {V^2}{c^2}}},
0,\frac 1 r \frac 1 {\sqrt{1 - \frac {V^2}{c^2}}},0\right)\]
\[e'_{(2)}=\left(c^{-2}\sqrt{\frac r{r-r_g}} \frac V {\sqrt{1 - \frac {V^2}{c^2}}},
0,\frac 1 r \frac 1 {\sqrt{1 - \frac {V^2}{c^2}}},0\right)\]

Therefore we get transformation
\begin{equation}
\begin{split}
{e'}_{(0)}=\frac 1 {\sqrt{1 - \frac {V^2}{c^2}}} e_{(0)} +
\frac V  c \frac 1 {\sqrt{1 - \frac {V^2}{c^2}}} e_{(2)}
\\
{e'}_{(2)}=\frac V c \frac 1 {\sqrt{1 - \frac {V^2}{c^2}}} e_{(0)} +
\frac 1 {\sqrt{1 - \frac {V^2}{c^2}}} e_{(2)}
\end{split}
\label{Lorentz}
\end{equation}
If a stationary observer sends light in a radial direction, the orbiting observer observes Doppler shift
$$
\omega'=\frac \omega {\sqrt{1 - \frac {V^2} {c^2}}}
$$
We need to add Doppler shift for gravitational field if the moving observer receives a radial wave that came from
infinity. In this case the Doppler shift will take the form
$$
\omega'=\sqrt{\frac r {r-r_g}}\frac \omega {\sqrt{ 1 - \frac {V^2} {c^2}}}
$$

We see the estimation for dynamics of star S2 that orbits Sgr A
in tables \ref{table:DShift1S2} and \ref{table:DShift2S2}.
The tables are based on two different estimations for mass of Sgr A.

If we get mass Sgr $4.1_{10}6 M_\odot$ \citeBib{Ghez} then
in pericentre (distance $1.868_{10}15$ cm) S2 has speed $738767495.4$ cm/s
and Doppler shift is $\omega'/\omega=1.000628$. In this case we measure length $2.16474 \mu m$
for emitted wave with length $2.1661 \mu m$ (Br $\gamma$).
In apocentre (distance $2.769_{10}16cm$) S2 has speed $49839993.28cm/s$
and Doppler shift is $\omega'/\omega=1.0000232$. We measure length $2.166049 \mu m$
for the same wave.
Difference between two measurements of wave length is $13.098$\AA.

If we get mass Sgr $3.7_{10}6 M_\odot$ \citeBib{Schodel} then
in pericentre (distance $1.805_{10}15cm$) S2 has speed $713915922.3cm/s$
and Doppler shift is $\omega'/\omega=1.000587$. In this case we measure length $2.16483 \mu m$
for emitted wave with length $2.1661 \mu m$ (Br $\gamma$).
In apocentre (distance $2.676_{10}16cm$) S2 has speed $48163414.05cm/s$
and Doppler shift is $\omega'/\omega=1.00002171$. We measure length $2.1666052 \mu m$
for the same wave.
Difference between two measurements of wave length is $12.232$\AA.

\begin{table}[ht]
\caption{Doppler shift on the Earth of a wave emitted from S2;
mass of Sgr A is $4.1_{10}6 M_\odot$ \citeBib{Ghez}}
\label{table:DShift1S2}
\begin{tabular}{|c|c|c|}
\hline
&pericentre&apocentre\\
\hline
distance cm&$1.868_{10}15$&$2.769_{10}16$\\
\hline
speed cm/s&$738767495.4$&$49839993.28$\\
\hline
$\omega'/\omega$&$1.000628$&$1.0000232$\\
\hline
emitted wave $(Br\ \gamma)\ \mu m$& $2.1661$&$2.1661$\\
\hline
observed wave $\mu m$&$2.16474$ &$2.166049$\\
\hline
\end{tabular}

Difference between two measurements of wavelength is $13.098$\AA.
\end{table}

\begin{table}[ht]
\caption{Doppler shift on the Earth of wave emitted from S2;
mass of Sgr A is $3.7_{10}6 M_\odot$ \citeBib{Schodel}}
\label{table:DShift2S2}
\begin{tabular}{|c|c|c|}
\hline
&pericentre&apocentre\\
\hline
distance cm&$1.805_{10}15$&$2.676_{10}16$\\
\hline
speed cm/s&$713915922.3$&$48163414.05$\\
\hline
$\omega'/\omega$&$1.000587$&$1.00002171$\\
\hline
emitted wave $(Br\ \gamma)\ \mu m$& $2.1661$&$2.1661$\\
\hline
observed wave $\mu m$&$2.16483$ &$2.1666052$\\
\hline
\end{tabular}

Difference between two measurements of wavelength is $12.232$\AA.
\end{table}

Difference between two measurements of wavelength in pericentre is $0.9$\AA.
Analyzing this data we can conclude that the use of Doppler shift can help improve estimation
of the mass of Sgr A.

\section{Lorentz Transformation in Radial Direction}

We see that the Lorentz transformation in orbial direction has familiar form.
It is very interesting to see what form this transformation has for radial direction.
We start from procedure of measurement speed and use coordinate speed $v$
\begin{equation}
dr=vdt
\EqLabel{RadialSpeed}
\end{equation}
The time of travel of light in both directions is the same.
Trajectory of light is determined by equation $ds^2=0$.
\[\frac {r-r_g} r c^2 dt^2 - \frac r {r-r_g} dr^2 = 0\]
When light returns back to observer $A$ the change of $t$ is
\[dt=2 \frac r {r-r_g} c^{-1} dr\]
The proper time of observer $A$ is
\ShowEq{proper time, observer A, Radial Direction}
Therefore spatial distance is
\[L=\sqrt{\frac r {r-r_g}} dr\]
When object moving with speed \EqRef{RadialSpeed} gets to $B$
change of $t$ is $\frac {dr} v$.
The proper time of observer $A$ at this point is 
\[ds^2 = \frac {r - r_g} r c^2 dr^2 v^{-2}\]
\[T = \sqrt{\frac {r - r_g} r} v^{-1} dr\]
Therefore the observer $A$ measures speed
\begin{align*}
V&=\frac L T =
=\frac{\sqrt{\frac r {r-r_g}} dr}{\sqrt{\frac {r - r_g} r} v^{-1} dr}=\\
&=\frac r {r-r_g} v
\end{align*}

Now we are ready to find out Lorentz transformation.
The basis vectors for stationary observer are
\[e_{(0)}=\left(\frac 1 c \sqrt{\frac r {r-r_g}}, 0, 0, 0\right)\]
\[e_{(1)}=\left(0,  \sqrt{\frac {r-r_g} r},0, 0\right)\]

Unit vector of speed should be proportional to vector
\begin{equation}
(1, v, 0, 0)
\EqLabel{NewRadialVectorE1}
\end{equation}
The length of this vector is
\begin{equation}
\begin{split}
L^2&=\frac {r-r_g} r c^2 - \frac r {r-r_g} v^2=\\
&=\frac {r-r_g} r c^2\left(1 - \frac {V^2}{c^2}\right)
\EqLabel{LengthNewRadialVectorE1}
\end{split}
\end{equation}
Therefore time ort of moving observer is
\begin{align*}
e'_{(0)}&=\left(\frac 1 L,  \frac v L,0, 0\right)=\\
&=\left(\sqrt{\frac r {r-r_g}} c^{-1}\frac 1 {\sqrt{\left(1 - \frac {V^2}{c^2}\right)}}, 
v \sqrt{\frac r {r-r_g}} c^{-1}\frac 1 {\sqrt{\left(1 - \frac {V^2}{c^2}\right)}}, 0,0\right)\\
&=\left(\sqrt{\frac r {r-r_g}} c^{-1}\frac 1 {\sqrt{\left(1 - \frac {V^2}{c^2}\right)}}, 
\frac V c \sqrt{\frac {r-r_g} r}\frac 1 {\sqrt{\left(1 - \frac {V^2}{c^2}\right)}}, 0,0\right)
\end{align*}
Spatial ort $e'_{(1)}=(A,B,0,0)$ is orthogonal $e'_{(0)}$ and has length $-1$. Therefore
\begin{equation}
\frac {r-r_g} r c^2\frac 1 L A - \frac r {r-r_g} \frac v L B = 0
\label{NewRadialVectorE0_1}
\end{equation}
\begin{equation}
\frac {r-r_g} r c^2 A^2 - \frac r {r-r_g} B^2 = -1
\label{NewRadialVectorE0_2}
\end{equation}
We can express $A$ from \eqref{NewRadialVectorE0_1}
\[A = c^{-2}\frac {r^2}{(r-r_g)^2} v B = c^{-2}\frac r{r-r_g} V B\]
and substitute into \eqref{NewRadialVectorE0_2}
\[\frac {r-r_g} r c^2 c^{-4}\frac {r^2}{(r-r_g)^2} V^2 B^2 - \frac r {r-r_g} B^2 = -1\]
\[\frac r{r-r_g} B^2\left(1- \frac{V^2}{c^2} \right) = 1\]
\[ B^2 = \frac{r-r_g} r\frac 1{1- \frac{V^2}{c^2} }\]
\[ B = \sqrt{\frac{r-r_g} r}\frac 1{\sqrt{1- \frac{V^2}{c^2}} }\]
\begin{align*}
A &= c^{-2}\frac r{r-r_g} V \sqrt{\frac{r-r_g} r}\frac 1{\sqrt{1- \frac{V^2}{c^2}} }\\
&= c^{-2} V \sqrt{\frac r{r-r_g}}\frac 1{\sqrt{1- \frac{V^2}{c^2}} }
\end{align*}
Finally spatial ort in direction of movement is
\[e'_{(1)}=\left(c^{-2} V \sqrt{\frac r{r-r_g}}\frac 1{\sqrt{1- \frac{V^2}{c^2}} },
\sqrt{\frac{r-r_g} r}\frac 1{\sqrt{1- \frac{V^2}{c^2}} },0,0\right)\]

Therefore we get transformationin in familiar form
\begin{equation}
\begin{split}
{e'}_{(0)}=\frac 1 {\sqrt{1 - \frac {V^2}{c^2}}} e_{(0)} +
\frac V  c \frac 1 {\sqrt{1 - \frac {V^2}{c^2}}} e_{(1)}
\\
{e'}_{(1)}=\frac V c \frac 1 {\sqrt{1 - \frac {V^2}{c^2}}} e_{(0)} +
\frac 1 {\sqrt{1 - \frac {V^2}{c^2}}} e_{(1)}
\end{split}
\label{RadialLorentz}
\end{equation}

\section{Doppler Shift in Friedman Space}
\labelSection{Doppler Shift in Friedman Space}

We consider another example in Friedman space. Metric of the space is
$$ds^2 = a^2 (dt^2
- d\chi^2 - \sin^2\chi (d\theta^2 - \sin^2 \theta d\phi^2))
$$
for closed model and
$$ds^2 = a^2 (dt^2
- d\chi^2 - \sinh^2\chi (d\theta^2 - \sin^2 \theta d\phi^2))
$$
for open one. Connection in this space is ($\alpha$, $\beta$ get values 1, 2, 3)
\[\Gamma^0_{00}=\frac {\dot{a}} a\]
\[\Gamma^0_{\alpha\alpha}=-\frac {\dot{a}} {a^2}g_{\alpha\alpha}\]
\[\Gamma^\alpha_{0\beta}=\frac {\dot{a}} a\delta^\alpha_\beta\]

Because space is homogenius we do not care about direction of light.
In this case
\[dk^0=-\Gamma^0_{ij}k^i k^j\]
Because $k$ is isotropic vector tangent to its trajectory we have
\[dx^\alpha=\frac {k^\alpha} {k^0}dt\]
Because $k^0=\frac \omega a$, then
\[d\frac \omega a=-\frac {da} {a^2}\omega+\frac {da} {\omega a}g_{\alpha\alpha}k^\alpha k^\alpha=
-2\frac {da} {a^2}\omega\]
\[ad\omega+\omega da=0\]
\[a\omega=const\]
Therefore when $a$ grows $\omega$ becomes smaller and length of waves grows as well. 

$a$ grows during light travel through spacetime
and this leads to red shift. We observe red shift because geometry changes,
but not because galaxies runs away one from other.

Now we want to see how red shift changes with time
if initial and final points do not move.
For simplicity I will change only $\chi$. Initial value is $\chi_1$
and final value is $\chi_2$. Because $dt=d\chi$ on light trajectory we have
\[\chi = \chi_1 + t - t_1\ \ \ \ t_2 = \chi_2 - \chi_1 + t_1\]
Therefore $a(t_1) \omega_1 = a(t_2) \omega_2$.
Doppler shift is
\[K(t_1) =\frac {\omega_2} {\omega_1} = \frac {a(t_1)} {a(t_2)}\]
If initial time changes
\ShowEq{t'1=t1+dt}
then
\ShowEq{K(t1+dt)=}
Time derivative of $K$ is
\[\dot{K} = \frac {\dot{a}_1 a_2 - a_1 \dot{a}_2} {a_2^2}\]

For closed space $a = \cosh t$. Then $\dot{a} = \sinh t$. 
\[\dot{K} = \frac{\sinh t_1 \cosh t_2 - \sinh t_2 \cosh t_1} {\cosh^2 t_2}
= \frac{\sinh(t_1 - t_2)} {\cosh^2 t_2}\] 
$K$ decreases when $t_1$ increases.

\ePrints{GJSFRA.13.1.39}
\ifx\Semafor\ValueOff
\section{Lorentz Transformation in Friedman Space}
\labelSection{Lorentz Transformation in Friedman Space}

To learn Lorentz transformation in Friedman space I want to use metric in form
\[ds^2=c^2 dt^2 -a^2(d\chi^2 + b^2(d\theta^2 - \sin^2 \theta d\phi^2))\]
Now I have 2 observers. One does not move and has speed $(1,0,0,0)$,
and another moves along $\chi$ and his speed is $C=(1,v,0,0)$ and we assme $V=a v$.

Metric is diagonal and coordinates $\theta, \phi$ do not change.
We have transformation in plane $t, \chi$.

Unit speed of first observer is
\[e_0 = (\frac 1 c , 0, 0, 0)\]
and vector orthonormal this one is
\[e_1 = (0, \frac 1 a , 0, 0)\]

The length of vector $C$ is
\[L=\sqrt{c^2 - a^2 v^2} = c \sqrt{1 - \frac{V^2} {c^2}}\]
Therefore unit vector of speed of second observer is
\[e'_0=(\frac 1 L, \frac v L, 0, 0)\]
We look for vector
\[e'_1=(A, B, 0, 0)\]
that is orthogonal to vector $e'_0$. For this we have
\begin{equation}
c^2 A^2 - a^2 B^2 = -1
\EqLabel{length of vector, e1, 1}
\end{equation}
\begin{equation}
c^2 A \frac 1 L - a^2 B \frac v L = 0
\EqLabel{length of vector, e1, 2}
\end{equation}
We get from the equation \EqRef{length of vector, e1, 2}
\begin{equation}
A = \frac {a^2} {c^2} v B
\EqLabel{length of vector, e1, 3}
\end{equation}
We substitute \EqRef{length of vector, e1, 3}
into \EqRef{length of vector, e1, 1} and get
\[B^2 (\frac {a^4} {c^2} v^2 - a^2) = -1\] 
\[B = \frac 1 {a \sqrt{1 - \frac {V^2} {c^2}}}\]
Therefore basis of second observer is
\[e'_0=(\frac 1 {c \sqrt{1 - \frac{V^2} {c^2}}}, \frac V {c a\sqrt{1 - \frac{V^2} {c^2}}}, 0, 0)\]
\[e'_1=( \frac V {c^2 \sqrt{1 - \frac {V^2} {c^2}}},
\frac 1 {a \sqrt{1 - \frac {V^2} {c^2}}}, 0, 0)\]
Now we can express $e'$ through $e$
\[e'_0= \frac 1 {\sqrt{1 - \frac {V^2} {c^2}}} e_0
+ \frac V c \frac 1 {\sqrt{1 - \frac{V^2} {c^2}}} e_1\]
\[e'_1= \frac V c \frac 1 {\sqrt{1 - \frac{V^2} {c^2}}} e_0
+ \frac 1 {\sqrt{1 - \frac{V^2} {c^2}}} e_1\] 
\fi

%% file: Gen.Relativity.Eq.tex

\AddEquation{Schwarzschild metric}
{
ds^2 = \frac{r-r_g} r c^2 dt^2
- \frac r {r-r_g} dr^2 - r^2 d\phi^2 - r^2 sin^2 \phi d\theta^2
}

\DefEq
{
\begin{align*}
ds^2&= \frac {r-r_g} r c^2 4 \frac {r^2} {(r-r_g)^2}c^{-2}dr^2=\\
&=  4 \frac r {r-r_g}dr^2
\end{align*}
}
{proper time, observer A, Radial Direction}

\DefEq
{
\[K(t_1+dt)=\frac{a(t_1 + dt)}{a(t_2 + dt)}\]
}
{K(t1+dt)=}

\DefEq
{
\[t_1'=t_1+dt\]
}
{t'1=t1+dt}

\DefEquation
{
\left\{
\begin{array}{r@{\,}l@{\ \ \ \ }r@{\,}l}
\Gamma^0_{10}&\displaystyle=\frac {r_g} {2r(r-r_g)}
\\[10pt]
\Gamma^1_{00}&\displaystyle=\frac {r_g(r-r_g)} {2r^3}
&
\Gamma^1_{11}&\displaystyle=-\frac {r_g} {2r(r-r_g)}
\\
\Gamma^1_{22}&=-(r-r_g)
&
\Gamma^1_{33}&=-(r-r_g)\sin^2\phi
\\[10pt]
\Gamma^2_{12}&\displaystyle=-\frac 1 r
&
\Gamma^2_{33}&=-\sin\phi\cos\phi
\\[10pt]
\Gamma^3_{13}&\displaystyle=-\frac 1 r
&
\Gamma^3_{23}&=\cot\phi
\\
\end{array}
\right.
}
{Schwarzschild metric, Connection}

%% file: Affine.English.tex
\input{Affine.Eq}
\ifx\PrintBook\undefined
\else
\chapter{Geometry of Metric-Affine Manifold}
\fi

\section{Line with Extreme Length}
\labelSection{Line with extreme length}

There are two different definitions of a geodesic curve
in the Riemann manifold.
One of them relies on the parallel transport.
We call an appropriate
\AddIndex{line auto parallel}{auto parallel line}.
Another definition depends on the length of trajectory.
We call an appropriate
\AddIndex{line extreme}{extreme line}.
In a metric\hyph affine manifold these lines have different
equations \citeBib{torsion}.
Equation of auto parallel line does not change.
However, the equation of extreme
line changes.\,\footnote{To derive
the equation \EqRef{xtremeLine},
I follow the ideas that Rashevsky \citeBib{Rashevsky}
implemented for the Riemann manifold.}

\begin{theorem}
Let $x^i = x^i(t,\alpha)$ be a line depending on a parameter $\alpha$
with fixed points at $t = t_1$ and
$t = t_2$ and we define its length as
\ShowEq{Line Length}
Then 
\ShowEq{Var Line Length}
where $\delta x^k$ is the change of a line when $\alpha$ changes.
\end{theorem}
\begin{proof}
We have
$$\frac {ds} {dt} = \sqrt{g_{ij} \frac {dx^i} {dt} \frac {dx^j} {dt}}$$
and
\ShowEq{Var Line Length, 1}
We can estimate the numerator of fraction in the equation
\EqRef{Var Line Length, 1}
\ShowEq{Var Line Length, 2}
From equations
\EqRef{Var Line Length, 1},
\EqRef{Var Line Length, 2},
it follows that
\ShowEq{Var Line Length, 3}
First term is $0$ because points, when $t = t_1$ and $t = t_2$, are fixed.
Therefore, we have got the statement of the theorem.
\end{proof}

\begin{theorem}
\labelTheorem{xtremeLine}
An extreme line satisfies equation
\begin{equation}
\EqLabel{xtremeLine}
\frac {D\frac {dx^l} {ds}} {ds} =
\frac 1 2 g^{il} \left(g_{kj;i} - g_{ik;j}
- g_{ij;k}\right) \frac{dx^k}{ds} \frac{dx^j}{ds}
\end{equation}
\end{theorem}
\begin{proof}
To find a line with extreme length,
we use the functional
\EqRef{Line Length}.
Since $\delta s = 0$,
$$
\frac 1 2 \left(g_{kj;i} -
g_{ij;k} -
g_{ik;j}\right) \frac {dx^k} {ds} \frac {dx^j} {ds} ds
- g_{ij} D\frac {dx^j} {ds} = 0$$
follows from \EqRef{Var Line Length}.
\end{proof}

\begin{theorem}
Parallel transport along an extreme line
holds length of tangent vector.
\end{theorem}
\begin{proof}
Let $$v^i=\frac {dx^i}{ds}$$ be the tangent vector to extreme curve.
From theorem \RefTheorem{xtremeLine} it follows that
$$\frac {Dv^l} {ds} =
g^{il} \frac 1 2 \left(g_{kj;i} - g_{ik;j}
- g_{ij;k}\right) v^k v^j
$$
and
\ShowEq{Parallel transport holds length, 1}
Therefore length of the vector $v^i$ does not change along extreme curve.
\end{proof}

\section{Frenet Transport}
\labelSection{Frenet Transport}

\ePrints{4827-2437}
\ifx\Semafor\ValueOff
All equations that we derived before are different,
however they have something common in their structure.
All these equations express movement along a line and in the right side of them
we can see the curvature of this line.

\else
From the equation
\EqRef{xtremeLine},
it follows that an extreme line has curvature equaled $0$.
\fi
By definition curvature of a line is
\[\xi(s) = \left| \frac {D\frac {dx^l} {ds}} {ds} \right|\]
Therefore we can introduce unit vector $e_1$ such that
\[\frac {D\frac {dx^l} {ds}} {ds} = \xi e^l_1\]

Knowledge of the transport of a basis along a line is very important,
because it allows us to study how spacetime changes
when an observer moves through it.
Our task is to discover equations similar to the Frenet transport in
the Riemann space. We design the accompaniment basis $\nu^i_k$ the same
way we do it in the Riemann space.

Vectors
\[\xi^i(t)=\frac{dx^i(t)}{dt}, \ \ \ \frac {D\xi^i} {dt},
\ \ \ ...\ \ \ \frac {D^{n-1}\xi^i} {dt^{n-1}}\]
in general are linearly independent. We call plane that we
create on the base of first $p$ vectors as $p$-th osculating plane $R_p$.
This plane does not depend on choice of parametr $t$.

Our next task is to create orthogonal basis which
shows us how line changes.
We get vector $\nu^i_1\in R_1$ so it is tangent to line.
We get vector $\nu^i_p\in R_p$, $p>1$ such that $\nu^i_p$ is
orthogonal to $R_{p-1}$. If original line is not isotropic
then each $\nu^i_p$ also is not isotropic and we can get unit vector
in the same direction. We call this basis accompaniment.

\begin{theorem}
\labelTheorem{FrenetTransfer}
The \AddIndex{Frenet transport}{Frenet transport} in the metric\hyph affine manifold gets the form
\begin{equation}
\begin{split}
\frac {D\nu^i_p} {dt}
= \frac 1 2 g^{im}(g_{kl;m}-g_{km;l}-g_{ml;k}) \nu^k_1 \nu^l_p - \\
- \epsilon_p \epsilon_{p-1} \xi_{p-1} \nu^i_{p-1} + \xi_p \nu^i_{p+1}
\end{split}
\EqLabel{Frenet}
\end{equation}
\[\epsilon_k = sign(g_{pq} \nu^p_k \nu^q_k)\]
Here $\nu^a_k$ is vector of basis, moving along line,
\[\epsilon_k = sign(g_{pq} \nu^p_k \nu^q_k)\]
\end{theorem}
\begin{proof}
We introduce vectors $\nu^a_k$ in this way that
\begin{equation}
\frac {D\nu^i_p} {dt}
= \frac 1 2 g^{im}(g_{kl;m}-g_{km;l}-g_{ml;k}) \nu^k_1 \nu^l_p
+ a^q_p \nu^i_q
\EqLabel{Frenet_1}
\end{equation}
where $a_p^q = 0$ when $q > p+1$.
Now we can determine coefficients $a^q_p$.
If we get derivative of the equation
\[
g_{ij} \nu^i_p \nu^j_q = const
\]
and substitute \EqRef{Frenet_1} we get the equation
\ShowEq{Frenet 2}

$a_p^q = 0$ when $q > p+1$ by definition.
Therefore $a_p^q = 0$ when $q < p-1$.
Introducing $\xi_p = a^{p+1}_p$ we get
\[
a^p_{p+1} = - \epsilon_p \epsilon_{p+1} \xi_p 
\]
When $q = p$ we get
\[
a^p_p = 0
\]
We get \EqRef{Frenet} when substitute $a^q_p$ in \EqRef{Frenet_1}.
\end{proof}

\section{Cartan Connection}
\labelSection{Cartan Transport}

Theorems \RefTheorem{xtremeLine} and \RefTheorem{FrenetTransfer}
state that
the movement along a line causes a transformation additional
to the parallel transport of a vector.
This transformation is very important
and we call it the \AddIndex{Cartan transport}{Cartan transport}.
We introduce the \AddIndex{Cartan symbol}{Cartan symbol}
\ShowEq{Cartan symbol}
and the \AddIndex{Cartan connection}{Cartan connection}
\ShowEq{overbrace Gamma i kl}
Using the Cartan connection we can write the connection form as
\ShowEq{Cartan connection form}
Respectively we define the
\AddIndex{Cartan derivative}{Cartan derivative}
\ShowEq{overbrace nabla_l}
\ShowEq{overbrace D}

\begin{theorem}
\labelTheorem{LengthTangentVector}
The Cartan transport along an extreme
line holds length of the tangent vector.
\end{theorem}
\begin{proof}
Let $$v^i=\frac {dx^i}{ds}$$ be the tangent vector to an extreme curve.
From theorem \RefTheorem{xtremeLine} it follows that
$$\frac {Dv^l} {ds} =
\frac 1 2 g^{il} \left(g_{kj;i} - g_{ik;j}
- g_{ij;k}\right) v^k v^j
$$
and
\ShowEq{Length of Tangent Vector}
Therefore the length of the vector $v^i$ does not change along the extreme curve.
\end{proof}

We extend the Cartan transport to any geometric object like we do
for the parallel transport.

\begin{theorem}
\labelTheorem{g ij;l=0}
\ShowEq{g ij;l=0}
\end{theorem}
\begin{proof}
\ShowEq{g ij;l=0 proof}
\end{proof}

The Cartan connection $\overbrace{\Gamma^i_{kl}}$
differs from the connection $\Gamma^i_{kl}$
by additional term which is symmetric tensor.
For any connection we introduce standard way derivative and curvature.
Statements of geometry and physics
have the same form independently of whether I use the connection $\Gamma^i_{kl}$ or the Cartan connection.
To show this we can generalize the idea of the Cartan connection
and consider connection defined by equation
\ShowEq{conection overline}
\ShowEq{conection overline =}
where $A$ is $0$, or the Cartan symbol or any other
symmetric tensor.
Respectively we define the derivative
\ShowEq{overline nabla_l, definition 1}
\ShowEq{overline nabla_l, definition 2}
\ShowEq{overline nabla_l}
\ShowEq{overline D}
and curvature
\ShowEq{GLn curvature_overline}
\ShowEq{GLn curvature_overline =}
This connection has the same torsion
\begin{equation}
T^a_{cb} =
\overline{\Gamma^a_{bc}}-\overline{\Gamma^a_{cb}}
\EqLabel{Torsion coordinates_overline}
\end{equation}

In this context theorem \RefTheorem{LengthTangentVector} means
that extreme line is geodesic line
for the Cartan connection.

\begin{theorem}
\labelTheorem{curvature_overline}
Curvature of connection \EqRef{conection overline =} has form
\begin{equation}
\overline{R^a_{bde}}=R^a_{bde}
+A^a_{be;d}-A^a_{bd;e}+A^a_{cd}A^c_{be}-A^a_{ce}A^c_{bd}+S^p_{de}A^a_{bp}
\label{curvature_overline}
\end{equation}
where $R^a_{bde}$ is curvature of connection $\Gamma^i_{kl}$
\end{theorem}
\begin{proof}
\begin{align*}
\overline{R^a_{bde}}&=\overline{\Gamma^a_{be,d}}-\overline{\Gamma^a_{bd,e}}
+\overline{\Gamma^a_{cd}}\ \overline{\Gamma^c_{be}}-\overline{\Gamma^a_{ce}}\ \overline{\Gamma^c_{bd}}\\
&=\Gamma^a_{be,d}+A^a_{be,d}-\Gamma^a_{bd,e}-A^a_{bd,e}\\
&+(\Gamma^a_{cd}+A^a_{cd})(\Gamma^c_{be}+A^c_{be})-(\Gamma^a_{ce}+A^a_{ce})(\Gamma^c_{bd}+A^c_{bd})\\
&=\Gamma^a_{be,d}+A^a_{be,d}-\Gamma^a_{bd,e}-A^a_{bd,e}\\
&+\Gamma^a_{cd}\Gamma^c_{be}+\Gamma^a_{cd}A^c_{be}+A^a_{cd}\Gamma^c_{be}+A^a_{cd}A^c_{be}\\
&-\Gamma^a_{ce}\Gamma^c_{bd}-A^a_{ce}\Gamma^c_{bd}-\Gamma^a_{ce}A^c_{bd}-A^a_{ce}A^c_{bd}\\
&=R^a_{bde}+A^a_{be,d}-A^a_{bd,e}\\
&+\Gamma^a_{cd}A^c_{be}+A^a_{cd}\Gamma^c_{be}+A^a_{cd}A^c_{be}\\
&-A^a_{ce}\Gamma^c_{bd}-\Gamma^a_{ce}A^c_{bd}-A^a_{ce}A^c_{bd}\\
&=R^a_{bde}\\
&+A^a_{be;d}-\underline{\Gamma^a_{pd}A^p_{be}}_2+\underline{\Gamma^p_{bd}A^a_{pe}}_4+\underline{\Gamma^p_{ed}A^a_{bp}}_{1:S}\\
&-A^a_{bd;e}+\underline{\Gamma^a_{pe}A^p_{bd}}_3-\underline{\Gamma^p_{be}A^a_{pd}}_5-\underline{\Gamma^p_{de}A^a_{bp}}_1\\
&+\underline{\Gamma^a_{cd}A^c_{be}}_2+\underline{\Gamma^c_{be}A^a_{cd}}_5+A^a_{cd}A^c_{be}\\
&-\underline{\Gamma^c_{bd}A^a_{ce}}_4-\underline{\Gamma^a_{ce}A^c_{bd}}_3-A^a_{ce}A^c_{bd}
\end{align*}
\end{proof}

\begin{corollary}
\AddIndex{Cartan curvature}{Cartan curvature} has next form
\ShowEq{Cartan curvature}
\ShowEq{Cartan curvature =}
\end{corollary}

\section{Lie Derivative}
\labelSection{Lie Derivative}

Vector field $\xi^k$ on manifold generates infinitesimal transformation
\ShowEq{infinitesimal displacement}
which leads to the \AddIndex{Lie derivative}{Lie derivative}.
Lie derivative tells us how the object changes
when we move along the vector field.

From the equation
\EqRef{infinitesimal displacement},
it follows that
\ShowEq{infinitesimal displacement, derivative}
From the equation
\EqRef{infinitesimal displacement},
it follows as well that
\ShowEq{infinitesimal displacement, 1}
\ShowEq{infinitesimal displacement, 1, derivative}

\begin{theorem}
\AddIndex{Lie derivative of metric}{Lie derivative of metric} has form
\ShowEq{Lie derivative of metric}
\ShowEq{Lie derivative of metric, def}
\end{theorem}
\begin{proof}
We start from transformation \EqRef{infinitesimal displacement}.
Then
\[g_{ab}(x')=g_{ab}(x)+g_{ab,c}\epsilon \xi^c\]
The equation
\begin{align*}
g'_{ab}(x')&=\frac{\partial x^c}{\partial x'^a} \frac{\partial x^d}{\partial x'^b}g_{cd}(x)\\
&=g_{ab}-\epsilon \xi^c_{,a}g_{cb}-\epsilon \xi^c_{,b}g_{ac}
\end{align*}
follows from the equation
\EqRef{infinitesimal displacement, 1, derivative}.
According to definition of Lie derivative we have
\ShowEq{Lie derivative of metric, 1}
\ShowEq{Lie derivative of metric, 2}
\EqRef{Lie derivative of metric, def} follows from
\EqRef{Lie derivative of metric, 2} and \EqRef{Torsion coordinates}.
\end{proof}

\begin{theorem}
\AddIndex{Lie derivative of connection}{Lie derivative of connection} has form
\ShowEq{Lie derivative of connection}
\ShowEq{Lie derivative of connection, def}
\end{theorem}
\begin{proof}
We start from transformation \EqRef{infinitesimal displacement}.
Then
\ShowEq{Lie derivative of connection,1}
The equation
\ShowEq{Lie derivative of connection,2}
follows from equations
\EqRef{infinitesimal displacement, derivative},
\EqRef{infinitesimal displacement, 1, derivative}.
By definition
\ShowEq{Lie derivative of connection 2}
\ShowEq{Lie derivative of connection,3}
Since
\ShowEq{Lie derivative of connection,41}
is tensor, then
\ShowEq{Lie derivative of connection 3}
The equation
\ShowEq{Lie derivative of connection,4}
follows from the equation
\EqRef{Lie derivative of connection 3}.
We substitute \EqRef{Lie derivative of connection,4} and \EqRef{Lie derivative of connection,3}
into \EqRef{Lie derivative of connection,2} and get
\ShowEq{Lie derivative of connection,5}
According definition of Lie derivative we have
using \EqRef{Lie derivative of connection,1} and \EqRef{Lie derivative of connection,5}
\ShowEq{Lie derivative of connection,6}
From \EqRef{Lie derivative of connection,6}
and \EqRef{Torsion coordinates} it follows
\ShowEq{Lie derivative of connection,7}
From \EqRef{Lie derivative of connection,7} and \EqRef{GLn curvature} it follows
\ShowEq{Lie derivative of connection,8}
From \EqRef{Lie derivative of connection,8} and \EqRef{Bianchi 1} it follows
\ShowEq{Lie derivative of connection,9}
We substitute \EqRef{commutator second derivative of vector} into \EqRef{Lie derivative of connection,9}
\ShowEq{Lie derivative of connection,10}
\EqRef{Lie derivative of connection, def}
follows from \EqRef{Lie derivative of connection,10}.
\end{proof}
\begin{corollary}
Lie derivative of connection in Rieman space has form
\begin{equation}
\EqLabel{Lie derivative of connection in Rieman space}
\mathcal{L}_\xi\Gamma^a_{bc}=-R^a_{cbp} \xi^p+\xi^a_{;cb}\\
\end{equation}
\end{corollary}
\begin{proof}
\EqRef{Lie derivative of connection in Rieman space} follows
from \EqRef{Lie derivative of connection, def} when $T^a_{bc}=0$
\end{proof}

\section{Bianchi Identity}
\labelSection{Bianchi Identity}

\begin{theorem}
The first Bianchi identity for the space with torsion has form
\ShowEq{Bianchi 1}
\end{theorem}
\begin{proof}
Differential of equation \EqRef{Torsion 1} has form
\ShowEq{Bianchi 1, 1}
Two forms are equal when their coefficients are equal. Therefore, from
\EqRef{Bianchi 1, 1},
it follows that
\ShowEq{Bianchi 1, 2}
We express derivatives using covariant derivatives
and change order of terms
\ShowEq{Bianchi 1, 3}
The equation
\ShowEq{Bianchi 1, 4}
follows from the equation
\EqRef{Bianchi 1, 3}.
\EqRef{Bianchi 1} follows from \EqRef{Bianchi 1, 4}.
\end{proof}


If we get a derivative of form \EqRef{Curvature}
we will see that the second Bianchi identity does not
depend on the torsion. 

\section{Killing Vector}
\labelSection{Killing Vector}

Invariance of the metric tensor $g$ under the infinitesimal coordinate transformation
\EqRef{infinitesimal displacement}
leads to the \AddIndex{Killing equation}{Killing equation}.

\begin{theorem}
Killing equation in the metric\hyph affine manifold has form
\ShowEq{Killing equation 1}
\end{theorem}
\begin{proof}
Invariance of the metric tensor $g$ means that its Lie derivative equal $0$
\begin{equation}
\mathcal{L}_\xi g_{ab}=0
\EqLabel{Killing equation 1,1}
\end{equation}
\EqRef{Killing equation 1} folows from \EqRef{Killing equation 1,1} and
\EqRef{Lie derivative of metric, def}.
\end{proof}


\begin{theorem}
The condition of invariance of the connection in the metric\hyph affine manifold has form
\ShowEq{Killing equation 2}
\end{theorem}
\begin{proof}
Because connection is invariant under the infinitesimal transformation we have
\ShowEq{Killing equation 2,1}
\EqRef{Killing equation 2} follows from \EqRef{Killing equation 2,1}
and \EqRef{Lie derivative of connection, def}.
\end{proof}

We call equation \EqRef{Killing equation 2}
the \AddIndex{Killing equation of second type}{Killing equation second type}
and vector $\xi^a$
\AddIndex{Killing vector of second type}{Killing vector second type}.

\begin{theorem}
Killing vector of second type satisfies equation
\ShowEq{Killing equation 2, corollary}
\end{theorem}
\begin{proof}
From \EqRef{Killing equation 2}
and \EqRef{commutator second derivative of vector}
it follows that
\ShowEq{Killing equation 2,4}
\EqRef{Killing equation 2, corollary} follows from \EqRef{Killing equation 2,4}.
\end{proof}

\begin{corollary}
The Killing equation of second type in the Riemann space is the identity.
The connection in the Riemann space is invariant under
any infinitesimal transformation \EqRef{infinitesimal displacement}.
\end{corollary}
\begin{proof}
First of all the torsion is 0. The rest is the consequence of the first Bianchi identity.
\end{proof}

%% file: Affine.Eq.tex

\DefEquation
{
s = \int^{t_2}_{t_1} \sqrt{g_{ij} \frac {dx^i} {dt} \frac {dx^j} {dt}} dt
}
{Line Length}

\DefEq
{
\symb{\overline{D} a^i}{overline D}{}
\[\ShowSymbol{overline D}{}=da^i+\overline{\Gamma^i_{kl}}a^k dx^l\]
}
{overline D}

\DefEq
{
\[
\ShowSymbol{GLn curvature_overline}{}
=\partial_i\overline{\Gamma^a_{bj}}-\partial_j\overline{\Gamma^a_{bi}}
+\overline{\Gamma^a_{ci}}\overline{\Gamma^c_{bj}}
-\overline{\Gamma^a_{cj}}\overline{\Gamma^c_{bi}}
\]
}
{GLn curvature_overline =}

\DefEq
{
\symb{\overline{R^a_{bij}}}{GLn curvature_overline}{}
}
{GLn curvature_overline}

\DefEquation
{
\delta s =
\int^{t_2}_{t_1} \left( \frac 1 2 \left(
g_{kj;i}- g_{ik;j}
- g_{ij;k} \right) \frac{dx^k}{ds} \frac{dx^j}{ds} ds -
g_{ij} D\frac {dx^j} {ds} \right) \delta x^i
}
{Var Line Length}

\DefEq
{
\begin{align*}
\frac {Dg_{kl} v^k v^l}{ds}
&= \frac {Dg_{kl}} {ds}v^k v^l
+ g_{kl}\frac {Dv^k} {ds} v^l
+ g_{kl}v^k \frac {Dv^l} {ds}
\\&
\begin{array}{r@{\,}l}
= g_{kl;p}v^p v^k v^l
&\displaystyle
+ g_{kl}g^{ik} \frac 1 2 \left(g_{rj;i} - g_{ir;j}
- g_{ij;r}\right) v^r v^j v^l
\\[8pt]&\displaystyle
+ g_{kl}v^k g^{il} \frac 1 2 \left(g_{rj;i} - g_{ir;j}
- g_{ij;r}\right) v^r v^j
\end{array}
\\&= g_{kl;p}v^p v^k v^l
+ \left(g_{rj;l} - g_{lr;j}
- g_{lj;r}\right) v^r v^j v^l = 0
\end{align*}
}
{Parallel transport holds length, 1}

\DefEq
{
\begin{align*}
\frac {dg_{ij}\nu^i_a \nu^j_b} {ds}&=
\frac {Dg_{ij}} {ds}\nu^i_a \nu^j_b+
g_{ij}\frac {D\nu^i_a} {ds}\nu^j_b+
g_{ij}\nu^i_a\frac {D\nu^j_b} {ds}
\\&
\begin{array}{r@{\,}l}
=g_{ij;k}\nu^k_1 \nu^i_a \nu^j_b&\displaystyle
+g_{ij}(\frac 1 2 g^{im}(g_{kl;m}-g_{km;l}-g_{ml;k}) \nu^k_1 \nu^l_a
+ a^q_a \nu^i_q)\nu^j_b
\\[8pt]&\displaystyle
+g_{ij}\nu^i_a (\frac 1 2 g^{jm}(g_{kl;m}-g_{km;l}-g_{ml;k})
\nu^k_1 \nu^l_b+ a^q_b \nu^i_q)
\end{array}
\\&
\begin{array}{r@{\,}l}
=g_{ij;k}\nu^k_1 \nu^i_a \nu^j_b&\displaystyle
+g_{ij}\frac 1 2 g^{im}(g_{kl;m}-g_{km;l}-g_{ml;k}) \nu^k_1 \nu^l_a\nu^j_b
+ g_{ij}a^q_a \nu^i_q\nu^j_b
\\[8pt]&\displaystyle
+g_{ij}\nu^i_a \frac 1 2 g^{jm}(g_{kl;m}-g_{km;l}-g_{ml;k}) \nu^k_1 \nu^l_b
+ g_{ij}\nu^i_a a^q_b \nu^i_q
\end{array}
\\
&=\frac 1 2 \nu^k_1 \nu^i_a \nu^j_b(2g_{ij;k}+
g_{ki;j}-g_{kj;i}-g_{ji;k}+
g_{kj;i}-g_{ki;j}-g_{ij;k})+
\\&+ \epsilon_b a^b_a+
+ \epsilon_a a^a_b=0
\end{align*}
}
{Frenet 2}

\DefEq
{
\symb{\mathcal{L}_\xi\Gamma^a_{bc}}{Lie derivative of connection}{}
}
{Lie derivative of connection}

\DefEquation
{
\Gamma^a_{bc}(x')
=\Gamma^a_{bc}(x)+\Gamma^a_{bc,p}\epsilon \xi^p
}
{Lie derivative of connection,1}

\DefEquation
{
\xi^a_{,e}
=\xi^a_{;e}
-\Gamma^a_{pe} \xi^p
}
{Lie derivative of connection,3}

\DefEq
{
$\xi^a_{;e}$
}
{Lie derivative of connection 4}

\DefEquation
{
\begin{split}
\mathcal{L}_\xi\Gamma^a_{bc}
&=(\Gamma^a_{bc}(x')-\Gamma'^a_{bc}(x'))\epsilon^{-1}\\
&=(\Gamma^a_{bc}+\Gamma^a_{bc,p}\epsilon \xi^p\\
&-\Gamma^a_{bc}-\epsilon (\xi^a_{;e}T^e_{cb}
-\Gamma^a_{pe} \xi^p\Gamma^e_{bc}
+\xi^e_{;c}T^a_{be}
+\Gamma^e_{pc} \xi^p\Gamma^a_{be}
-\xi^a_{;cb}+\Gamma^a_{pc,b} \xi^p))\epsilon^{-1}
\\
&=\Gamma^a_{bc,p} \xi^p
-\xi^a_{;e}T^e_{cb}
+\Gamma^a_{pe} \xi^p\Gamma^e_{bc}
-\xi^e_{;c}T^a_{be}
-\Gamma^e_{pc} \xi^p\Gamma^a_{be}
+\xi^a_{;cb}-\Gamma^a_{pc,b} \xi^p
\end{split}
}
{Lie derivative of connection,6}

\DefEquation
{
\begin{split}
\mathcal{L}_\xi\Gamma^a_{bc}
&=\Gamma^a_{cb,p} \xi^p-\Gamma^a_{cp,b} \xi^p\\
&+\underline{\Gamma^a_{pe} \Gamma^e_{bc}\xi^p}_{3:T}
-\underline{\Gamma^a_{ep} \Gamma^e_{bc}\xi^p}_3
+\underline{\Gamma^a_{ep} \Gamma^e_{bc}\xi^p}_{4:T}
-\underline{\Gamma^a_{ep} \Gamma^e_{cb}\xi^p}_4
+\Gamma^a_{ep} \Gamma^e_{cb}\xi^p\\
&-\underline{\Gamma^e_{pc} \Gamma^a_{be}\xi^p}_{1:T}
+\underline{\Gamma^e_{pc} \Gamma^a_{eb}\xi^p}_1
- \underline{\Gamma^a_{eb}\Gamma^e_{pc}\xi^p}_{2:T}
+\underline{\Gamma^a_{eb}\Gamma^e_{cp}\xi^p}_2
-\Gamma^a_{eb}\Gamma^e_{cp}\xi^p\\
& -\xi^a_{;e}T^e_{cb}
-\xi^e_{;c}T^a_{be}
+\xi^a_{;cb}-T^a_{cp,b} \xi^p-T^a_{bc,p} \xi^p
\\
&=\Gamma^a_{cb,p} \xi^p-\Gamma^a_{cp,b} \xi^p
+\Gamma^a_{ep} \Gamma^e_{cb}\xi^p
-\Gamma^a_{eb}\Gamma^e_{cp}\xi^p\\
&-\underline{T^a_{pe} \Gamma^e_{bc}\xi^p}_{4:T}
-\underline{\Gamma^a_{ep} T^e_{bc}\xi^p}_1
-\underline{\Gamma^e_{pc} T^a_{eb}\xi^p}_{3:T}
- \underline{\Gamma^a_{eb}T^e_{cp}\xi^p}_2\\
& -\xi^a_{;e}T^e_{cb}
-\xi^e_{;c}T^a_{be}+\xi^a_{;cb}\\
&-T^a_{cp;b} \xi^p
+\underline{\Gamma^a_{eb} T^e_{cp}\xi^p}_2
-\underline{\Gamma^e_{cb} T^a_{ep}\xi^p}_4
-\underline{\Gamma^e_{pb} T^a_{ce}\xi^p}_{5:T}\\
&-T^a_{bc;p} \xi^p
+\underline{\Gamma^a_{ep} T^e_{bc}\xi^p}_1
-\underline{\Gamma^e_{bp} T^a_{ec}\xi^p}_5
-\underline{\Gamma^e_{cp} T^a_{be}\xi^p}_3
\end{split}
}
{Lie derivative of connection,7}

\DefEquation
{
\begin{split}
\mathcal{L}_\xi\Gamma^a_{bc}&=R^a_{cpb} \xi^p\\
&-T^e_{cp} T^a_{eb}\xi^p
-T^a_{pe} T^e_{cb}\xi^p-T^e_{bp} T^a_{ce}\xi^p\\
& -\xi^a_{;e}T^e_{cb}
-\xi^e_{;c}T^a_{be}+\xi^a_{;cb}
-T^a_{cp;b} \xi^p
-T^a_{bc;p} \xi^p
\\
&=-R^a_{cbp} \xi^p\\
&- (T^a_{eb}T^e_{cp}
+T^a_{ep} T^e_{bc}+ T^a_{ec}T^e_{pb})\xi^p\\
& -\xi^a_{;e}T^e_{cb}
-\xi^e_{;c}T^a_{be}+\xi^a_{;cb}
-T^a_{cp;b} \xi^p
-T^a_{bc;p} \xi^p
\end{split}
}
{Lie derivative of connection,8}

\DefEquation
{
\begin{split}
\mathcal{L}_\xi\Gamma^a_{bc}&=\underline{R^a_{cpb} \xi^p}_1\\
&- R^a_{bcp}\xi^p
-R^a_{pbc}\xi^p- \underline{R^a_{cpb}\xi^p}_1\\
&+\underline{T^a_{bc:p}\xi^p}_3
+T^a_{pb;c}\xi^p+ \underline{T^a_{cp;b}\xi^p}_2\\
& -\xi^a_{;e}T^e_{cb}
-\xi^e_{;c}T^a_{be}+\xi^a_{;cb}
-\underline{T^a_{cp;b} \xi^p}_2
-\underline{T^a_{bc;p} \xi^p}_3
\\
&=-R^a_{bcp}\xi^p
-R^a_{pbc}\xi^p
-T^a_{bp;c}\xi^p
-\xi^a_{;e}T^e_{cb}
-\xi^e_{;c}T^a_{be}+\xi^a_{;cb}
\end{split}
}
{Lie derivative of connection,9}

\DefEquation
{
T^k_{ij,m}\theta^m\wedge\theta^i\wedge\theta^j
=(\Gamma^k_{ji,m}-\Gamma^k_{ij,m})\theta^m\wedge\theta^i\wedge\theta^j
}
{Bianchi 1, 1}

\DefEq
{
\[
T^k_{ij,m}+T^k_{mi,j}+T^k_{jm,i}
=\Gamma^k_{ji,m}-\Gamma^k_{ij,m}
+\Gamma^k_{im,j}-\Gamma^k_{mi,j}
+\Gamma^k_{mj,i}-\Gamma^k_{jm,i}
\]
}
{Bianchi 1, 2}

\DefEquation
{
\begin{split}
&T^k_{ij;m}-\underline{\Gamma^k_{pm} T^p_{ij}}_4
+\underline{\Gamma^p_{im} T^k_{pj}}_{2:T}
-\underline{\Gamma^p_{jm} T^k_{pi}}_{3:T}\\
+&T^k_{mi;j}-\underline{\Gamma^k_{pj} T^p_{mi}}_5
+\underline{\Gamma^p_{mj} T^k_{pi}}_3
-\underline{\Gamma^p_{ij} T^k_{pm}}_{1:T}\\
+&T^k_{jm;i}-\underline{\Gamma^k_{pi} T^p_{jm}}_6
+\underline{\Gamma^p_{ji} T^k_{pm}}_1
-\underline{\Gamma^p_{mi} T^k_{pj}}_2\\
=&\Gamma^k_{ji,m}-\Gamma^k_{jm,i}
+\Gamma^k_{pm} \Gamma^p_{ji}
-\Gamma^k_{pi} \Gamma^p_{jm}
-\underline{\Gamma^k_{pm} \Gamma^p_{ji}}_4
+\underline{\Gamma^k_{pi} \Gamma^p_{jm}}_6\\
+&\Gamma^k_{im,j}-\Gamma^k_{ij,m}
+\Gamma^k_{pj} \Gamma^p_{im}
-\Gamma^k_{pm} \Gamma^p_{ij}
-\underline{\Gamma^k_{pj} \Gamma^p_{im}}_5
+\underline{\Gamma^k_{pm} \Gamma^p_{ij}}_4\\
+&\Gamma^k_{mj,i}-\Gamma^k_{mi,j}
+\Gamma^k_{pi} \Gamma^p_{mj}-\Gamma^k_{pj} \Gamma^p_{mi}
-\underline{\Gamma^k_{pi} \Gamma^p_{mj}}_6
+\underline{\Gamma^k_{pj} \Gamma^p_{mi}}_5
\end{split}
}
{Bianchi 1, 3}

\DefEquation
{
\xi^k_{;a}g_{kb}+\xi^k_{;b}g_{ka}
+T^l_{ka}g_{lb}\xi^k+T^l_{kb}g_{la}\xi^k+g_{ab;k}\xi^k=0
}
{Killing equation 1}

\DefEquation
{
\mathcal{L}_\xi\Gamma^a_{bc}=0
}
{Killing equation 2,1}

\DefEquation
{
\begin{split}
R^a_{p bc}\xi^p
-T^p_{bc}\xi^a_{;p}
&=R^a_{cbp}\xi^p
+T^a_{cp;b}\xi^p
+T^a_{cp}\xi^p_{;b}\\
&-R^a_{bcp}\xi^p
-T^a_{bp;c}\xi^p
-T^a_{bp}\xi^p_{;c}
\end{split}
}
{Killing equation 2,4}

\DefEq
{
\[\omega=dx^i-\overbrace{\Gamma^i_{kl}}a^k dx^l\]
}
{Cartan connection form}

\DefEq
{
\symb{\overbrace{D} a^i}{overbrace D}{}
\[\ShowSymbol{overbrace D}{}=da^i+\overbrace{\Gamma^i_{kl}}a^k dx^l\]
}
{overbrace D}

\DefEquation
{
\begin{split}
0&=R^a_{bcp}\xi^p+R^a_{cpb}\xi^p+R^a_{pbc}\xi^p
\\ &+T^a_{bp;c}\xi^p+T^a_{pc;b}\xi^p
+T^p_{cb}\xi^a_{;p}
+T^a_{bp}\xi^p_{;c}
+T^a_{pc}\xi^p_{;b}
\end{split}
}
{Killing equation 2, corollary}

\DefEquation
{
\begin{split}
\xi^a_{;bc}=R^a_{bcp}\xi^p
+T^a_{bp;c}\xi^p
+T^a_{bp}\xi^p_{;c}
\end{split}
}
{Killing equation 2}

\DefEquation
{
\begin{split}
&T^k_{ij;m}
+T^p_{mi} T^k_{pj}+T^p_{jm} T^k_{pi}
+T^k_{mi;j}
+T^p_{ij} T^k_{pm}
+T^k_{jm;i}\\
=&R^k_{jmi}
+R^k_{ijm}
+R^k_{mij}
\end{split}
}
{Bianchi 1, 4}

\DefEquation
{
\frac{\partial x'^l}{\partial x^k}=\delta^l_k+\epsilon\xi^l_{,k}
}
{infinitesimal displacement, derivative}

\DefEquation
{
\frac{\partial x^l}{\partial x'^k}=\delta^l_k-\epsilon\xi^l_{,k}
}
{infinitesimal displacement, 1, derivative}

\DefEquation
{
x'^k=x^k+\epsilon\xi^k
}
{infinitesimal displacement}

\DefEquation
{
x^k=x'^k-\epsilon\xi^k
}
{infinitesimal displacement, 1}

\DefEquation
{
\begin{split}
&T^k_{ij;m}+T^k_{mi;j}+T^k_{jm;i}
+ T^k_{pi}T^p_{jm}+ T^k_{pm}T^p_{ij}+ T^k_{pj}T^p_{mi}\\
=&R^k_{jmi}+R^k_{ijm}+R^k_{mij}
\end{split}
}
{Bianchi 1}

\DefEquation
{
\mathcal{L}_\xi\Gamma^a_{bc}=-R^a_{bcp}\xi^p
-T^a_{bp;c}\xi^p
-\underline{T^e_{cb}\xi^a_{;e}}_1
-T^a_{be}\xi^e_{;c}
-\underline{T^p_{bc}\xi^a_{;p}}_1+\xi^a_{;bc}
}
{Lie derivative of connection,10}

\DefEquation
{
\begin{split}
\Gamma'^a_{bc}(x')
&=\Gamma^a_{bc}+\epsilon (\underline{\xi^a_{;e}}_{4:T}
-\Gamma^a_{pe} \xi^p)\Gamma^e_{bc}
-\epsilon (\underline{\xi^e_{;b}}_2
-\underline{\Gamma^e_{pb} \xi^p}_1)\Gamma^a_{ec}
-\epsilon (\underline{\xi^e_{;c}}_{3:T}
-\Gamma^e_{pc} \xi^p)\Gamma^a_{be}\\
&-\epsilon (\xi^a_{;cb}-\Gamma^a_{pc,b} \xi^p
-\underline{\Gamma^a_{pc} \xi^p_{;b}}_2
+\underline{\Gamma^a_{pc} \Gamma^p_{rb} \xi^r}_1
-\underline{\Gamma^a_{pb} \xi^p_{;c}}_3
+\underline{\Gamma^p_{cb} \xi^a_{;p}}_4)
\\
&=\Gamma^a_{bc}+\epsilon (\xi^a_{;e}T^e_{cb}
-\Gamma^a_{pe} \xi^p\Gamma^e_{bc}
+\xi^e_{;c}T^a_{be}
+\Gamma^e_{pc} \xi^p\Gamma^a_{be}
-\xi^a_{;cb}+\Gamma^a_{pc,b} \xi^p)
\end{split}
}
{Lie derivative of connection,5}

\DefEquation
{
\begin{split}
\xi^a_{;ef}&=\xi^a_{;e,f}
+\Gamma^a_{pf} \xi^p_{;e}
-\Gamma^p_{ef} \xi^a_{;p}\\
&=\xi^a_{,ef}+\Gamma^a_{pe,f} \xi^p+\Gamma^a_{pe} \xi^p_{,f}
+\Gamma^a_{pf} \xi^p_{;e}
-\Gamma^p_{ef} \xi^a_{;p}\\
&=\xi^a_{,ef}+\Gamma^a_{pe,f} \xi^p
+\Gamma^a_{pe} \xi^p_{;f}-\Gamma^a_{pe} \Gamma^p_{rf} \xi^r
+\Gamma^a_{pf} \xi^p_{;e}
-\Gamma^p_{ef} \xi^a_{;p}
\end{split}
}
{Lie derivative of connection 3}

\DefEquation
{
\xi^a_{,ef}
=\xi^a_{;ef}-\Gamma^a_{pe,f} \xi^p
-\Gamma^a_{pe} \xi^p_{;f}+\Gamma^a_{pe} \Gamma^p_{rf} \xi^r
-\Gamma^a_{pf} \xi^p_{;e}
+\Gamma^p_{ef} \xi^a_{;p}
}
{Lie derivative of connection,4}

\DefEq
{
$\xi^a_{;ef}$
}
{Lie derivative of connection,41}

\DefEq
{
\[g_{ij;\{l\}}=0\]
}
{g ij;l=0}

\DefEq
{
\begin{align*}
\overbrace{\nabla_l} g_{ij}&=\partial_l g_{ij}
-\overbrace{\Gamma^k_{il}}g_{kj}
-\overbrace{\Gamma^k_{jl}}g_{ik}=
\\
&=g_{ij;l}
+\frac 1 2 g^{km}(g_{il;m}-g_{im;l}-g_{ml;i})g_{kj}
+\frac 1 2 g^{km}(g_{jl;m}-g_{jm;l}-g_{ml;j})g_{ik}=0
\end{align*}
}
{g ij;l=0 proof}

\DefEq
{
\begin{align*}
\frac {Dg_{kl} v^k v^l}{ds}
&= \frac {Dg_{kl}} {ds}v^k v^l
+ g_{kl}\frac {Dv^k} {ds} v^l
+ g_{kl}v^k \frac {Dv^l} {ds}
\\&
= g_{kl;p}v^p v^k v^l+
\\&
+ g_{kl}g^{ik} \frac 1 2 \left(g_{rj;i} - g_{ir;j}
- g_{ij;r}\right) v^r v^j v^l
\\&
+ g_{kl}v^k g^{il} \frac 1 2 \left(g_{rj;i} - g_{ir;j}
- g_{ij;r}\right) v^r v^j
\\&
= g_{kl;p}v^p v^k v^l
+ \left(g_{rj;l} - g_{lr;j}
- g_{lj;r}\right) v^r v^j v^l = 0
\end{align*}
}
{Length of Tangent Vector}

\DefEq
{
\[
\xi^a_{;e}
=\xi^a_{,e}
+\Gamma^a_{pe} \xi^p
\]
}
{Lie derivative of connection 2}

\DefEquation
{
\begin{split}
\Gamma'^a_{bc}(x')
&=\frac{\partial x'^a}{\partial x^e}\frac{\partial x^f}{\partial x'^b}
\frac{\partial x^g}{\partial x'^c}
\Gamma^e_{fg}(x)
+\frac{\partial x'^a}{\partial x^e}
\frac{\partial^2 x^e}{\partial x'^b\partial x'^c}\\
&=\Gamma^a_{bc}+\epsilon \xi^a_{,e}\Gamma^e_{bc}
-\epsilon \xi^e_{,b}\Gamma^a_{ec}-\epsilon \xi^e_{,c}\Gamma^a_{be}
+(\delta^a_e+\epsilon \xi^a_{,e})(-\epsilon \xi^e_{,cb}))
\\
&=\Gamma^a_{bc}+\epsilon \xi^a_{,e}\Gamma^e_{bc}
-\epsilon \xi^e_{,b}\Gamma^a_{ec}-\epsilon \xi^e_{,c}\Gamma^a_{be}
-\epsilon \xi^a_{,cb}
\end{split}
}
{Lie derivative of connection,2}

\DefEquation
{
\ShowSymbol{Lie derivative of connection}{}=-R^a_{bcp}\xi^p
-T^a_{bp;c}\xi^p
-T^a_{be}\xi^e_{;c}
+\xi^a_{;bc}
}
{Lie derivative of connection, def}

\DefEquation
{
\begin{split}
\mathcal{L}_\xi g_{ab}&=g_{ab;c} \xi^c+\Gamma^d_{ac}g_{db} \xi^c
+\Gamma^d_{bc}g_{ad}\xi^c\\
&+ \xi^c_{;a}g_{cb}- \Gamma^c_{da}\xi^dg_{cb}
+ \xi^c_{;b}g_{ac}- \Gamma^c_{db}\xi^dg_{ac}
\end{split}
}
{Lie derivative of metric, 2}

\DefEq
{
\begin{align*}
\mathcal{L}_\xi g_{ab}&=g_{ab}(x')-g'_{ab}(x')\\
&=g_{ab,c}\epsilon \xi^c+\epsilon \xi^c_{,a}g_{cb}
+\epsilon \xi^c_{,b}g_{ac}\\
&=(g_{ab;c}+\Gamma^d_{ac}g_{db}+\Gamma^d_{bc}g_{ad})\epsilon \xi^c\\
&+\epsilon (\xi^c_{;a}-\Gamma^c_{da}\xi^d)g_{cb}
+\epsilon (\xi^c_{;b}-\Gamma^c_{db}\xi^d)g_{ac}
\end{align*}
}
{Lie derivative of metric, 1}

\DefEquation
{
\ShowSymbol{Lie derivative of metric}{}=\xi^k_{;a}g_{kb}+\xi^k_{;b}g_{ka}
+T^l_{ka}g_{lb}\xi^k+T^l_{kb}g_{la}\xi^k+g_{ab;k}\xi^k
}
{Lie derivative of metric, def}

\DefEq
{
\symb{\overbrace{R^a_{bde}}}{Cartan curvature}{}
}
{Cartan curvature}

\AddEq{overbrace nabla_l}
{
overbrace \symb{\overbrace{\nabla_l} a^i}{overbrace nabla_l}{}
\[\ShowSymbol{overbrace nabla_l}{}= a^i_{;\{l\}}=\partial_l a^i+\overbrace{\Gamma^i_{kl}}a^k\]
}

\AddEquation{conection overline =}
{
\ShowSymbol{conection overline}{}=\Gamma^i_{kl}+A^i_{kl}
}

\AddEq{conection overline}
{
\symb{\overline{\Gamma^i_{kl}}}{conection overline}{}
}

\AddEq{overbrace Gamma i kl}
{
\symb{\overbrace{\Gamma^i_{kl}}}{overbrace Gamma i kl}{}
\[\ShowSymbol{overbrace Gamma i kl}{}
=\Gamma^i_{kl}-\Gamma(C)^i_{kl}
=\Gamma^i_{kl}-\frac 1 2 g^{im}(g_{kl;m}-g_{km;l}-g_{ml;k})\]
}

\DefEq
{
\symb{\Gamma(C)^i_{kl}}{Cartan symbol}{}
\[\ShowSymbol{Cartan symbol}{}=\frac 1 2 g^{im}(g_{kl;m}-g_{km;l}-g_{ml;k})\]
}
{Cartan symbol}

\DefEq
{
\begin{align*}
\ShowSymbol{Cartan curvature}{}&=R^a_{bde}
-\Gamma(C)^a_{be;d}+\Gamma(C)^a_{bd;e}\\
&+\Gamma(C)^a_{cd}\Gamma(C)^c_{be}-\Gamma(C)^a_{ce}\Gamma(C)^c_{bd}-T^p_{de}\Gamma(C)^a_{bp}
\end{align*}
}
{Cartan curvature =}

\DefEq
{
\[\ShowSymbol{overline nabla_l, definition 1}{}=
\ShowSymbol{overline nabla_l, definition 2}{}=
\partial_l a^i+\overline{\Gamma^i_{kl}}a^k\]
}
{overline nabla_l}

\DefEq
{
\symb{\overline{\nabla_l} a^i}{overline nabla_l, definition 1}{}
}
{overline nabla_l, definition 1}

\DefEq
{
\symb{a^i_{;<l>}}{overline nabla_l, definition 2}{}
}
{overline nabla_l, definition 2}

\DefEq
{
\symb{\mathcal{L}_\xi g_{ab}}{Lie derivative of metric}{}
}
{Lie derivative of metric}

\DefEq
{
\begin{align*}
\delta s &= \int^{t_2}_{t_1}
 \frac
 {
  g_{ij;k} \delta x^k dx^i \frac{dx^j}{dt} + 2 g_{ij} D\delta x^i \frac{dx^j}{dt}
 } 
 {2 \frac {ds} {dt}}
\\
&= \int^{t_2}_{t_1}
\left(\frac 1 2 g_{ij;k} \delta x^k dx^i \frac {dx^j} {ds} +
g_{ij} D\delta x^i \frac {dx^j} {ds} \right) 
\\
&= \int^{t_2}_{t_1}
\left(\frac 1 2 g_{kj;i} \delta x^i \frac {dx^k} {ds} {ds} \frac {dx^j} {ds} +
d\left(g_{ij} \delta x^i \frac {dx^j} {ds}\right)
-g_{ij;k} \frac {dx^k} {ds} {ds} \frac {dx^j} {ds} \delta x^i -
g_{ij} D\frac {dx^j} {ds} \delta x^i\right)
\\
&= 
\left.\left(g_{ij} \delta x^i \frac{dx^j}{ds} \right)\right|^{t_2}_{t_1}
+ \int^{t_2}_{t_1} \left(\frac 1 2 \left(
g_{kj;i} -
g_{ij;k} -
g_{ik;j} \right) \frac {dx^k} {ds} \frac{dx^j}{ds} {ds}
- g_{ij} D\frac{dx^j}{ds}\right) \delta x^i
\end{align*}
}
{Var Line Length, 3}

\DefEquation
{
\begin{split}
&g_{ij,k} \delta x^k \frac {dx^i} {dt} \frac {dx^j} {dt} +
2 g_{ij} \delta \frac {dx^i} {dt} \frac {dx^j} {dt}
\\
=& g_{ij;k} \delta x^k {dx^i} {dt} \frac {dx^j} {dt} +
2 g_{ij} \Gamma^i_{lk} \delta x^k \frac {dx^l} {dt} \frac {dx^j} {dt}
+2 g_{ij} d \frac {\delta x^i} {dt} \frac {dx^j} {dt} =
\\
=& g_{ij;k} \delta x^k \frac{dx^i}{dt} \frac{dx^j}{dt} + 2 g_{ij} \frac{D\delta x^i}{dt} \frac{dx^j}{dt}
\end{split}
}
{Var Line Length, 2}

\DefEquation
{
\delta s = \int^{t_2}_{t_1}
\frac
 {\delta \left( g_{ij} \frac{dx^i}{dt} \frac{dx^j}{dt} \right)}
 {2 \frac{ds}{dt}} dt
}
{Var Line Length, 1}

%% file: Newton.English.tex

\ifx\PrintBook\Defined
\chapter{Metric Affine Gravity}
\fi

\section{Newton's Laws: Scalar Potential}
\labelSection{Newton Laws: Scalar Potential}

The knowledge of dynamics of a point particle is important for us
because we can study how the particle interacts with external fields
as well as the properties of the particle itself.

To study the movement of a point particle we can use a potential of a certain field.
The potential may be scalar or vector.

In case of scalar potential we assume that a point particle
has rest mass $m$ and we use lagrangian function in the following form
\[L = - m c ds - U dx^0\]
where $U$ is \AddIndex{scalar potential}{scalar potential}
or \AddIndex{potential energy}{potential energy}.

\begin{theorem}
\labelTheorem{First Newton law}
(\AddIndex{First Newton law}{First Newton law})
If $U = 0$ (therefore we consider free movement)
a body chooses trajectory with extreme length.
\end{theorem}

\begin{theorem}
(\AddIndex{Second Newton law}{Second Newton law})
A trajectory of point particle satisfies the differential equation
\begin{equation}
\frac {\overbrace{D}u^l} {ds} =
\frac {u^0} {m c} F^l
\EqLabel{Newton}
\end{equation}
\[u^j = \frac {dx^l} {ds}\]
where we introduced force
\begin{equation}
F^l = g^{il} \frac{\partial U}{\partial x^i}
\EqLabel{potential force}
\end{equation}
\end{theorem}
\begin{proof}

Using \EqRef{Var Line Length},
we can write variation of the lagrangian as
$$\frac 1 2 m c \left(g_{kl;i} - g_{ik;l}
- g_{il;k}\right) u^k u^j ds
- m c g_{ij} Du^j
+ \frac {\partial U} {\partial x^i} dx^0 = 0$$
The statement of the theorem follows from this.
\end{proof}

\section{Newton's Laws: Vector Potential}
In section \RefSection{Newton Laws: Scalar Potential}
we learned dynamics when potential is scalar.
However in electrodynamics we have \AddIndex{vector potential}
{vector potential} $A^k$.
In this case action is
\[
S = \int^{t_2}_{t_1} \left( - m c ds - \frac e c A_l dx^l \right)
\]
\[
A_c = g_{cd} A^d
\]

\begin{theorem}
The trajectory of a particle moving in the vector field satisfies the differential equation
\[
\frac {\overbrace{D}u^j} {ds} =
\frac e {m c^2} g^{ij} F_{li} u^l
\]
\[u^j = \frac {dx^l} {ds}\]
where we introduce a \AddIndex{field-strength tensor}
{field-strength tensor}
\[F_{dc} = A_{d;c} - A_{c;d} + S^p_{dc} A_p
 = \overbrace{\nabla_c}A_d - \overbrace{\nabla_d}A_c + S^p_{dc} A_p\]
\end{theorem}
\begin{proof}

Using \EqRef{Var Line Length}, we can write the variation of the action as
\[
\delta S =\]
\[=\int^{t_2}_{t_1} \left( - m c \left(
\frac 1 2 \left( g_{kj;i}
- g_{ij;k} - g_{ik;j} \right) u^k u^j ds
- g_{ij} D u^j\right) \delta x^i
- \frac e c \left(\delta A_l dx^l + A_l \delta dx^l \right)
\right)
\]
We can estimate the second term like
\[
- \frac e c \left(A_{l,k} dx^l \delta x^k + A_l d\delta x^l \right) =
\]
\[
=- \frac e c \left(A_{l;k} dx^l \delta x^k
+ \Gamma^p_{lk} A_p dx^l \delta x^k
+ A_l d\delta x^l \right) =
\]
\[
=- \frac e c \left(A_{k;l} dx^l \delta x^k +
\left(A_{l;k} - A_{k;l}\right)dx^l \delta x^k+
S^p_{lk} A_p dx^l \delta x^k + \Gamma^p_{kl} A_p dx^l \delta x^k
+ A_l d\delta x^l \right) =
\]
\[
=- \frac e c \left(DA_k \delta x^k + A_k D\delta x^k +
\left(A_{l;k} - A_{k;l}\right)dx^l \delta x^k
+ S^p_{lk} A_p dx^l \delta x^k \right) =
\]
\[
=- \frac e c \left(\underline{d\left(A_k \delta x^k\right)} +
\left( A_{l;k} - A_{k;l}
+ S^p_{lk} A_p \right) dx^l \delta x^k\right)
\]
The integral of the underlined term is $0$ because points, when $t = t_1$ and $t = t_2$, are fixed.
Therefore
\[
- m c \left(
\frac 1 2 \left( g_{kj;i}
- g_{ij;k} - g_{ik;j} \right) u^k u^j ds
- g_{ij} D u^j\right)-
\frac e c F_{li} dx^l = 0
\]
The statement of the theorem follows from this.
\end{proof}

The dependence of field-strength tensor on derivative of metric follows from
this theorem. It changes form of Einstein equation and momentum of gravitational field
appears in case of vector field.

\begin{theorem}
A field-strength tensor does not change when vector potential changes like
\[A'_j=A_j+\partial_j\Lambda\]
where $\Lambda$ is an arbitrary function of $x$.
\end{theorem}
\begin{proof}
Change in a field-strength tensor is
\[(\partial_d\Lambda){}_{;c} - (\partial_c\Lambda){}_{;d} + S^p_{dc} \partial_p\Lambda =\]
\[\partial_{cd}\Lambda - \Gamma^p_{dc} \partial_p\Lambda
- \partial_{dc}\Lambda+\Gamma^p_{cd} \partial_p\Lambda + S^p_{dc}\partial_p\Lambda =0\]
This proves the theorem.
\end{proof}

%% file: Tidal.English.tex
\input{Tidal.Eq}

\section{Tidal Equation}
\labelSection{Tidal Equation}

Assume  that
considered bodyes perform not geodesic but arbitrary movement.

We assume that both observers start their travel from the same
point\,\footnote{I follow \citeBib{Wheeler}, page 33} and their speed
satisfy to differential equations
\ShowEq{Trajectory}
where $I=1,2$ is the number of the observer and $ds_I$ is infinitesimal arc on geodesic $I$.
Observer $I$ follows the geodesic of connection
\EqRef{conection overline =}
when $a_I=0$.
We assume also that $ds_1=ds_2=ds$.

\AddIndex{Deviation of trajectories}{deviation of trajectories}
\ShowEq{deviation of trajectories}
is vector connecting observers.
The lines are infinitesimally close in the neighborhood of the start point
\[x^i_2(s_2)=x^i_1(s_1)+\delta x^i(s_1)\]
\[v^i_2(s_2)=v^i_1(s_1)+\delta v^i(s_1)\]

Derivative of vector $\delta x^i$ has form
\[\frac{d\delta x^i}{ds}=\frac{d(x^i_2-x^i_1)}{ds}=v_2^i-v_1^i=\delta v^i\]
\AddIndex{Speed of deviation}{speed of deviation}
$\delta x^i$ is covariant derivative
\ShowEq{speed of deviation}
\ShowEq{a speed of deviation}
From \EqRef{a speed of deviation} it follows that
\ShowEq{deviation_extreme_5}
Finally we are ready to estimate second covariant
derivative of vector $\delta x^i$
\ShowEq{deviation_extreme_6 0}
\ShowEq{deviation_extreme_6}
\begin{theorem}
\labelTheorem{deviation_extreme}
Tidal acceleration
\ePrints{GJSFRA.13.1.39}
\ifx\Semafor\ValueOff
of connection \EqRef{conection overline =}
\fi
has form
\ShowEq{deviation_extreme}
\end{theorem}
\begin{proof}
The trajectory of observer $1$ satisfies equation
\ShowEq{deviation_extreme_1}
\ShowEq{deviation_extreme_2}
The same time the trajectory of observer $2$ satisfies equation
\ShowEq{deviation_extreme_3 1}
We can rewrite this equation up to order $1$
\ShowEq{deviation_extreme_3 2}
Using \EqRef{deviation_extreme_1} we get
\ShowEq{deviation_extreme_3 3}
\ShowEq{deviation_extreme_3}
We substitute \EqRef{deviation_extreme_5}, \EqRef{deviation_extreme_2},
and \EqRef{deviation_extreme_3} into \EqRef{deviation_extreme_6}
\ShowEq{deviation_extreme_7}
\ShowEq{deviation_extreme_8}
Terms underscored with symbol $1$ are curvature and terms underscored with symbol $2$ are
covariant derivative of torsion.
\EqRef{deviation_extreme} follows from \EqRef{deviation_extreme_8}.
\end{proof}

\begin{remark}
The body $2$ may be remote from body $1$. In this case we
can use procedure (like in \citeBib{Anderson02}) based on parallel transfer.
For this purpose we transport vector of speed of observer $2$ to the start point of
observer $1$ and then estimate tidal acceleration. This procedure works in case of not
strong gravitational field.
\qed
\end{remark}

\ePrints{GJSFRA.13.1.39}
\ifx\Semafor\ValueOff
\begin{remark}
If in central field observer $1$ has orbital speed $V_\phi$, observer $2$ moves in
radial direction and both observers follow geodesic then tidal acceleration has form
\ShowEq{tidal acceleration in central field}
\qed
\end{remark}

\begin{remark}
If observer $2$ follows geodesic in central field, but observer $1$ fixed his
position at distance $r$ then
\[a^1=\Gamma^1_{kl}v^kv^l
=\frac{r_g}{2r^2c^2}\]
Acceleration follows inverse square law as follows from \EqRef{deviation_extreme}.
\qed
\end{remark}

\begin{remark}
Theorem \RefTheorem{deviation_extreme} has one specific case. If
observer $1$ moves along an extreme line,
we can use Cartan connection. In this case
$a_1^i=0$. If observer $2$ moves along geodesic then
\begin{equation}
\EqLabel{remark deviation_extreme Nongeodesic 1}
a_2^i=-\Gamma(C)^i_{kl}v_2^kv_2^k=-\Gamma(C)^i_{kl}(v_1^kv_1^k+2v_1^l\delta v^k)
\end{equation}
If we substitute \EqRef{deviation_extreme_5}
into \EqRef{remark deviation_extreme Nongeodesic 1} we get
\ShowEq{remark deviation_extreme Nongeodesic 3}
In this case \EqRef{deviation_extreme} gets form
\ShowEq{remark deviation_extreme Nongeodesic 2}
In case of initial conditions
\begin{align*}
\delta x^k&=0\\\frac{D\delta x^k}{ds}&=0
\end{align*}
\EqRef{remark deviation_extreme Nongeodesic 2} is estimation of acceleration \EqRef[0405.028]{Newton}.
\qed
\end{remark}
\fi

\section{Tidal Acceleration and Lie Derivative}

\EqRef{deviation_extreme} reminds expression of Lie derivative
\EqRef[0405.028]{Lie derivative of connection, def}.
To see this similarity we need to write equation \EqRef{deviation_extreme}
different way.

By definition
\ShowEq{Lie 0}
\ShowEq{Lie 1}
Because $\frac{\overline{D}a^k}{ds}$ is vector, we can easy find second derivative
\ShowEq{Lie 2}
On the last step we used \EqRef{Lie 1} when $a^k=v^k$.
When $v^p$ is tangent vector of trajectory of observer $1$ from
\EqRef{Trajectory} it follows that
\ShowEq{Lie 3}
From \EqRef{Lie 2} and \EqRef{Lie 3} it follows that
\ShowEq{Lie 4}
\begin{theorem}
\labelTheorem{Lie deviation_extreme}
Speed of deviation of two trajectories \EqRef{Trajectory} satisfies equation
\ShowEq{Lie and deviation_extreme}
\end{theorem}
\begin{proof}
We substitute \EqRef{Lie 1} and \EqRef{Lie 4} into \EqRef{deviation_extreme}.
\ShowEq{Lie 6}
\ShowEq{Lie 5}
\EqRef{Lie and deviation_extreme} follows from \EqRef{Lie 5}
and \EqRef[0405.028]{Lie derivative of connection, def}.
\end{proof}

At a first glance one can tell that the speed of deviation of
geodesics is the Killing vector of second type. This is an option, however
equation
\ShowEq{Lie deviation geodesic 1}
does not follow from equation
\ShowEq{Lie deviation geodesic}
However equation \EqRef{Lie deviation geodesic} shows a close
relationship between deep symmetry of spacetime and gravitational field.

%% file: Tidal.Eq.tex

\DefEquation
{
\mathcal{L}_{\displaystyle\frac{D\delta x^n}{ds}}\Gamma^i_{kl}v^kv^l=0
}
{Lie deviation geodesic}

\DefEquation
{
\frac{Dv^i_I}{ds_I}=a^i_I
}
{Trajectory}

\DefEquation
{
\delta v^i=\frac{D\delta x^i}{ds}-\Gamma^i_{kl}\delta x^kv_1^l
}
{deviation_extreme_5}

\DefEq
{
\begin{align*}
\frac{D^2\delta x^i}{ds^2}&=\frac{d\frac{D\delta x^i}{ds}}{ds}
+\Gamma^i_{kl}\frac{D\delta x^k}{ds}v_1^l\\
&=\frac{d(\delta v^i+\Gamma^i_{kl}\delta x^kv_1^l)}{ds}
+\Gamma^i_{kl}\frac{D\delta x^k}{ds}v_1^l\\
&=\frac{d\delta v^i}{ds}
+\frac{d\Gamma^i_{kl}}{ds}\delta x^kv_1^l
+\Gamma^i_{kl}\frac{d\delta x^k}{ds}v_1^l
+\Gamma^i_{kl}\delta x^k\frac{dv_1^l}{ds}
+\Gamma^i_{kl}\frac{D\delta x^k}{ds}v_1^l
\end{align*}
}
{deviation_extreme_6 0}

\DefEquation
{
\begin{split}
\frac{D^2\delta x^i}{ds^2}
&=T^i_{ln}\frac{D\delta x^n}{ds}v^l_1\\
&+(\underline{T^i_{km,l}}_2+\underline{\Gamma^i_{km,l}}_1
-\underline{\Gamma^i_{kl,m}}_1\\
&+\underline{T^i_{nl}\Gamma^n_{mk}}_2
+\underline{\Gamma^i_{nl}T^n_{km}}_2
+\underline{\Gamma^i_{nl}\Gamma^n_{km}}_1\\
&-\underline{T^i_{nm}\Gamma^n_{kl}}_2
-\underline{\Gamma^i_{nm}\Gamma^n_{kl}}_1)\delta x^mv_1^kv_1^l\\
&+a^i_2-a^i_1+\Gamma^i_{ml}\delta x^ma^l_1
\end{split}
}
{deviation_extreme_8}

\DefEquation
{
\frac{Da^k}{ds}=a^k_{;p}v^p
}
{Lie 1}

\DefEquation
{
\frac{Dv^i}{ds}=v^i_{;r}v^r=a^i_1
}
{Lie 3}

\DefEq
{
\[
\begin{split}
\delta x^i_{;kl}v^kv^l+\delta x^k_{;p}a^p_1
&=T^i_{ln}\delta x^n_{;k}v^kv^l_1
+(R^i_{klm}+T^i_{km;l})\delta x^mv_1^kv_1^l\\
&+a^i_2-a^i_1+\Gamma^i_{ml}\delta x^ma^l_1
\end{split}
\]
}
{Lie 6}

\DefEq
{
\[
\mathcal{L}_{\displaystyle\frac{D\delta x^n}{ds}}\Gamma^i_{kl}=0
\]
}
{Lie deviation geodesic 1}

\DefEquation
{
\begin{split}
0
&=(T^i_{ln}\delta x^n_{;k}-\delta x^i_{;kl}
+R^i_{klm}\delta x^m+T^i_{km;l}\delta x^m)v_1^kv_1^l\\
&+a^i_2-a^i_1-\delta x^k_{,p}a^p_1
\end{split}
}
{Lie 5}

\DefEquation
{
\mathcal{L}_{\displaystyle\frac{D\delta x^n}{ds}}\Gamma^i_{kl}v^kv^l
=a^i_2-a^i_1+\Gamma^i_{ml}\delta x^ma^l_1
}
{Lie and deviation_extreme}

\DefEquation
{
\frac{D^2a^k}{ds^2}=a^k_{;pr}v^pv^r+a^k_{;p}a^p_1
}
{Lie 4}

\DefEquation
{
\begin{split}
\frac{D^2a^k}{ds^2}&=\frac{D\frac{Da^k}{ds}}{ds}
=\frac{D(a^k_{;p}v^p)}{ds}\\
&=a^k_{;pr}v^pv^r+a^k_{;p}v^p_{;r}v^r
\end{split}
}
{Lie 2}

\DefEq
{
\begin{align*}
\frac{Da^k}{ds}&=\frac{da^k}{ds}+\Gamma^k_{lp}a^l\frac{dx^p}{ds}\\
&=a^k_{,p}v^p+\Gamma^k_{lp}a^lv^p
\end{align*}
}
{Lie 0}

\DefEquation
{
\frac{D^2\delta x^i}{ds^2}
=\frac{d\delta v^i}{ds}
+\Gamma^i_{kl,n}v_1^n\delta x^kv_1^l
+\Gamma^i_{kl}\delta v^k v_1^l
+\Gamma^i_{kl}\delta x^k\frac{dv_1^l}{ds}
+\Gamma^i_{kl}\frac{D\delta x^k}{ds}v_1^l
}
{deviation_extreme_6}

\DefEquation
{
\begin{split}
\frac{D^2\delta x^i}{ds^2}
&=T^i_{ln}\frac{D\delta x^n}{ds}v^l_1
+(R^i_{klm}+T^i_{km;l})\delta x^mv_1^kv_1^l\\
&+a^i_2-a^i_1+\Gamma^i_{ml}\delta x^ma^l_1
\end{split}
}
{deviation_extreme}

\DefEquation
{
\frac{dv^i_1}{ds}=a^i_1-\Gamma^i_{kl}v^k_1v^l_1
}
{deviation_extreme_2}

\DefEq
{
\begin{align*}
\frac{Dv^i_2}{ds}&=\frac{dv^i_2}{ds}+\Gamma^i_{kl}(x_2)v^k_2v^l_2\\
&=\frac{d(v_1^i+\delta v^i)}{ds}
+\Gamma^i_{kl}(x_1+\delta x)(v^k_1+\delta v^k)(v^l_1+\delta v^l)\\
&=\frac{dv_1^i}{ds}+\frac{d\delta v^i}{ds}
+(\Gamma^i_{kl}+\Gamma^i_{kl,m}\delta x^m)
(v^k_1v^l_1+\delta v^kv^l_1
+v^k_1\delta v^l+\delta v^k\delta v^l)\\
&=a^i_2
\end{align*}
}
{deviation_extreme_3 1}

\DefEq
{
\[
\frac{dv_1^i}{ds}+\frac{d\delta v^i}{ds}
+\Gamma^i_{kl}v^k_1v^l_1
+\Gamma^i_{kl}(\delta v^kv^l_1
+v^k_1\delta v^l)
+\Gamma^i_{kl,m}\delta x^m
v^k_1v^l_1=a^i_2
\]
}
{deviation_extreme_3 2}

\DefEq
{
\[
a^i_1+\frac{d\delta v^i}{ds}
+\Gamma^i_{kl}\delta v^kv^l_1
+\Gamma^i_{kl}v^k_1\delta v^l
+\Gamma^i_{kl,m}\delta x^m
v^k_1v^l_1=a^i_2
\]
}
{deviation_extreme_3 3}

\DefEquation
{
\frac{d\delta v^i}{ds}=
-\Gamma^i_{kl}\delta v^kv^l_1
-\Gamma^i_{lk}\delta v^kv^l_1
-\Gamma^i_{kl,m}\delta x^m
v^k_1v^l_1+a^i_2-a^i_1
}
{deviation_extreme_3}

\DefEquation
{
\frac{Dv^i_1}{ds}=\frac{dv^i_1}{ds}+\Gamma^i_{kl}(x_1)v^k_1v^l_1=a^i_1
}
{deviation_extreme_1}

\DefEquation
{
\begin{split}
\ShowSymbol{speed of deviation}{}
&=\frac{d\delta x^i}{ds}+\Gamma^i_{kl}\delta x^kv_1^l\\
&=\delta v^i+\Gamma^i_{kl}\delta x^kv_1^l
\end{split}
}
{a speed of deviation}

\DefEq
{
\EqRef{Trajectory}
\symb{\delta x^k}{deviation of trajectories}1
}
{deviation of trajectories}

\DefEq
{
\symb{\frac{D\delta x^i}{ds}}{speed of deviation}{}
}
{speed of deviation}

\DefEq
{
\begin{align*}
\frac{D^2\delta x^i}{ds^2}
&=-\underline{\Gamma^i_{kl}\delta v^kv^l_1}_1
-\Gamma^i_{ln}
\left(
\frac{D\delta x^n}{ds}-\Gamma^n_{mk}\delta x^mv_1^k
\right)v^l_1
-\Gamma^i_{kl,m}\delta x^mv^k_1v^l_1\\
&+a^i_2-a^i_1\\
&+\Gamma^i_{mk,l}v_1^k\delta x^mv_1^l
+\underline{\Gamma^i_{kl}\delta v^k v_1^l}_1\\
&+\Gamma^i_{mn}\delta x^m(a^n_1-\Gamma^n_{kl}v^k_1v^l_1)
+\Gamma^i_{nl}\frac{D\delta x^n}{ds}v_1^l
\\
\frac{D^2\delta x^i}{ds^2}
&=(\Gamma^i_{mk,l}-\Gamma^i_{kl,m}
+\Gamma^i_{ln}\Gamma^n_{mk}
-\Gamma^i_{mn}\Gamma^n_{kl})\delta x^mv^k_1v^l_1\\
&+\Gamma^i_{nl}\frac{D\delta x^n}{ds}v_1^l
-\Gamma^i_{ln}\frac{D\delta x^n}{ds}v^l_1\\
&+a^i_2-a^i_1+\Gamma^i_{mn}\delta x^ma^n_1
\\
\frac{D^2\delta x^i}{ds^2}
&=(\Gamma^i_{mk,l}-\Gamma^i_{km,l}+\Gamma^i_{km,l}
-\Gamma^i_{kl,m}\\
&+\Gamma^i_{ln}\Gamma^n_{mk}
-\Gamma^i_{nl}\Gamma^n_{mk}
+\Gamma^i_{nl}\Gamma^n_{mk}
-\Gamma^i_{nl}\Gamma^n_{km}
+\Gamma^i_{nl}\Gamma^n_{km}\\
&-\Gamma^i_{mn}\Gamma^n_{kl}
+\Gamma^i_{nm}\Gamma^n_{kl}
-\Gamma^i_{nm}\Gamma^n_{kl})\delta x^mv_1^kv_1^l\\
&+T^i_{ln}\frac{D\delta x^n}{ds}v_1^l
+a^i_2-a^i_1+\Gamma^i_{mn}\delta x^ma^n_1
\end{align*}
}
{deviation_extreme_7}

\DefEq
{
\[
a_2^i=-\Gamma(C)^i_{kl}v_1^kv_1^k
-2\Gamma(C)^i_{kl}v_1^l\frac{D\delta x^k}{ds}
+2\Gamma(C)^i_{ml}\Gamma(C)^m_{kn}v_1^nv_1^l\delta x^k
\]
}
{remark deviation_extreme Nongeodesic 3}

\DefEquation
{
\begin{split}
\frac{\overbrace{D^2}\delta x^i}{ds^2}
&=(\overbrace{R^i_{lnk}}
+\overbrace{\nabla_n}T^i_{lk})v_1^lv_1^n\delta x^k
+T^i_{lk}\frac{\overbrace{D}\delta x^k}{ds} v_1^l\\
&-\Gamma(C)^i_{kl}v_1^kv_1^k
-2\Gamma(C)^i_{kl}v_1^l\frac{\overbrace{D}\delta x^k}{ds}
+2\Gamma(C)^i_{ml}\Gamma(C)^m_{kn}v_1^nv_1^l\delta x^k
\end{split}
}
{remark deviation_extreme Nongeodesic 2}

\DefEq
{
\begin{align*}
\frac{D^2\delta x^1}{ds^2}
&=R^1_{lnk}\delta x^kv^nv^l\\
&=(R^1_{001}v^0v^0+R^1_{221}v^2v^2)\delta x^1\\
&=\left(\frac{r_g } {r^3c^2}\frac 1 {1 - \frac {V_\phi^2}{c^2}}-\left(-1
+\frac{r_g} {2 r}
- \frac {r - r_g} r\right)V_\phi^2\right)\delta x^1
\end{align*}
}
{tidal acceleration in central field}